\documentclass[10pt,phd,a4paper,oneside]{ucl_thesis}

\usepackage{emptypage}
\usepackage{url} 
\usepackage[hidelinks]{hyperref}
\usepackage{graphicx}
\usepackage{float}
\usepackage{amsmath}

\usepackage{gensymb}
\usepackage{textcomp}

\usepackage{multirow}
\usepackage{color}
\usepackage{mathptmx}
\usepackage{booktabs}
\usepackage{subfigure}
\usepackage{amsfonts}
\usepackage{stackengine}

\usepackage[OT1]{fontenc} 
\usepackage{xspace}
\newcommand{\descr}[1]{\smallskip\noindent\textbf{#1}}
\newcommand{\descrit}[1]{\vspace{0.05cm}\noindent{#1}}
\usepackage[font=small,labelfont=bf]{caption}
\usepackage{paralist}
\newcommand{\reduce}{\vspace{-0.5cm}}

\newcommand{\dspol}{{\sf{\fontsize{9}{10}\selectfont /pol/}}\xspace}
\newcommand{\dssp}{{\sf{\fontsize{9}{10}\selectfont /sp/}}\xspace}
\newcommand{\dsint}{{\sf{\fontsize{9}{10}\selectfont /int/}}\xspace}
\newcommand{\dssci}{{\sf{\fontsize{9}{10}\selectfont /sci/}}\xspace}
\newcommand{\tdshort}{T\textunderscore D\xspace}
\newcommand{\td}{The\textunderscore Donald\xspace}

\usepackage[OT2,OT1]{fontenc}
\newcommand\cyr
{
\renewcommand\rmdefault{wncyr}
\renewcommand\sfdefault{wncyss}
\renewcommand\encodingdefault{OT2}
\normalfont
\selectfont
}
\DeclareTextFontCommand{\textcyr}{\cyr}

\usepackage{titlesec}
\usepackage{hyperref}

\titleclass{\subsubsubsection}{straight}[\subsection]

\newcounter{subsubsubsection}[subsubsection]
\renewcommand\thesubsubsubsection{\thesubsubsection.\arabic{subsubsubsection}}

\titleformat{\subsubsubsection}
  {\normalfont\normalsize\bfseries}{\thesubsubsubsection}{1em}{}
\titlespacing*{\subsubsubsection}
{0pt}{3.25ex plus 1ex minus .2ex}{1.5ex plus .2ex}

\makeatletter
\renewcommand\paragraph{\@startsection{paragraph}{5}{\z@}%
  {3.25ex \@plus1ex \@minus.2ex}%
  {-1em}%
  {\normalfont\normalsize\bfseries}}
\renewcommand\subparagraph{\@startsection{subparagraph}{6}{\parindent}%
  {3.25ex \@plus1ex \@minus .2ex}%
  {-1em}%
  {\normalfont\normalsize\bfseries}}
\def\toclevel@subsubsubsection{4}
\def\toclevel@paragraph{5}
\def\toclevel@paragraph{6}
\def\l@subsubsubsection{\@dottedtocline{4}{7em}{4em}}
\def\l@paragraph{\@dottedtocline{5}{10em}{5em}}
\def\l@subparagraph{\@dottedtocline{6}{14em}{6em}}
\makeatother

\title{Towards Understanding the Information Ecosystem Through the Lens of Multiple Web Communities}
\author{Savvas Zannettou}
\department{Department of Electrical Engineering, Computer Engineering and Informatics}

\begin{document}

\maketitle

\begin{abstract} 

The Web consists of numerous Web communities, news sources, and services, which are often exploited by various entities for the dissemination of false or otherwise malevolent information.
 Yet, we lack tools and techniques to effectively track the propagation of information across the multiple diverse communities, and to capture and model the interplay and influence between them. 
 Furthermore, we lack a basic understanding of what the role and impact of some emerging communities and services on the Web information ecosystem are, and how such communities are exploited by bad actors (e.g., state-sponsored trolls) that spread false and weaponized information.

In this thesis, we shed some light on the complexity and diversity of the information ecosystem on the Web by presenting a typology  that includes the various types of false information, the involved actors as well as their possible motives. 
Then, we follow a data-driven cross-platform quantitative approach to analyze billions of posts from Twitter, Reddit, 4chan's Politically Incorrect board (/pol/), and Gab, to shed light on: 
1) how news and image-based memes travel from one Web community to another and how we can model and quantify the influence between the various Web communities; 
2) characterizing the role of emerging Web communities and services on the information ecosystem, by studying Gab and two popular Web archiving services, namely the Wayback Machine and archive.is; and 
3) how popular Web communities are exploited by state-sponsored actors for the purpose of spreading disinformation and sowing public discord.

In a nutshell, our analysis reveal that small fringe Web communities like 4chan's /pol/ and The\_Donald subreddit have a disproportionate influence on mainstream communities such as Twitter with regard to the dissemination of news and image-based memes. 
We find that Gab acts as the new hub for the alt-right community, while for Web archiving services we find that they are popular on fringe Web communities and that they can be misused by Reddit moderators in order to penalize ad revenue from news sources with conflicting ideology. 
Finally, when studying state-sponsored actors, we find that they exhibit substantial differences compared to random users, that their tactics change and evolve over time, and that they were particularly influential in spreading news on popular mainstream communities like Twitter and Reddit.
 
\end{abstract}

\begin{acknowledgements}
First, I would like to acknowledge my advisor, Michael Sirivianos, and my de facto co-advisor, Jeremy Blackburn, for their continuous support and feedback throughout my PhD journey.
Their support and guidance was instrumental in turning me into an independent and competent researcher.
Second, I would like to thank Emiliano De Cristofaro and \emph{The Legendary} Gianluca Stringhini, who I consider my ``second advisors.''
Their expertise and feedback complemented the one received by my two advisors, hence helping me in further expanding my research and writing skills.
More importantly, I am grateful to these four individuals mainly because they ensured that this journey was fun while undertaking important and cool research.

Also, I want to thank various colleagues from Cyprus University of Technology, Telefonica Research, University College London, Princeton University, and University of Illinois at Urbana-Champaign: their help and feedback was pivotal for undertaking the studies presented in this thesis.

Furthermore, I would like to thank my family for their support, patience, and encouragement, which ensured that I was mentally strong to overcome all the obstacles faced during my PhD journey.

Finally, I would like to thank anonymous users on 4chan for their comments on our papers, as well as for creating some high quality memes about our research: their comments and ``meme magic'' made this journey much more enjoyable. 
\end{acknowledgements}

\tableofcontents
\listoffigures
\listoftables

\chapter{Introduction}

Over the past decades, the Web became the predominant medium for the rapid acquisition of information.
Unfortunately, the Web has also become a medium where false information, hateful content, and weaponized information is disseminated.
Recently, we have seen extensive anecdotal evidence suggesting that users on the Web are exposed to such information.
Some examples include the spread of false and weaponized information by state-sponsored actors, namely Russian trolls were disseminating weaponized information related to the vaccine debate~\cite{broniatowski2018weaponized} and the 2016 US elections~\cite{facebook_russians_elections} to sow public discord and likely change voting preferences of people.
Another example is the one of the now infamous Cambridge Analytica: during the 2016 US elections the company targetted millions of people in the US on Facebook by exposing them to political weaponized information with the goal to shift their voting decision~\cite{cambridge_analytica}.
On top of this, regular users on the Web are also involved in the dissemination of false, hateful, or weaponized information, which further compounds these emerging problems on the Web.
Therefore, there is a pressing need to understand how information is shared on the Web, who the main entities involved are, and how information shared on the Web can alter real-world behavior.

At the same time, the information ecosystem on the Web has become an enormous and complex medium.
It involves various entities that contribute to the dissemination of information: ranging from Web communities like Twitter, where users can share information, to news sources that disseminate articles to users.
On top of this, the barrier of entry of new communities and news sources is minimal, hence the Web is becoming more complex as more communities and news sources are added.
Due to the increased complexity of the ecosystem a lot of its aspects are relatively unstudied by the research community.

In this thesis, we focus on the following aspects of the information ecosystem: 1)~how we can track the propagation of information across multiple Web communities and how to study the interplay and influence between these communities; 2)~characterizing the role of emerging Web communities and services; and 3) understanding the exploitation of Web communities from bad actors for the purpose of advancing an agenda or sowing public discord.
We focus on these mainly because we argue that providing knowledge and tools to analyze these aspects is a significant step towards understanding and mitigating emerging socio-technological phenomena on the Web.
Such phenomena include studying the spread of hateful, fake, and weaponized information that can have great impact on the world (e.g., excessive spread of weaponized and targeted information can lead to shift in voting results of major elections).
We elaborate on each of these aspects below.

\descr{Spread of information across multiple Web communities.} As the number and diversity of communities and news sources grow, so does the opportunity for the production and dissemination of hateful or fake content.
Nevertheless, previous work (see Chapter~\ref{chapter:background}) only examined the propagation of information on the Web, to the best of our knowledge, by looking at specific communities in isolation.
In reality, however, the various communities on the Web do not exist in vacuum. 
Users are members of multiple communities and they can share information seen on one community to another, possibly mutating it along the way.
Such interactions, indicate that information travels from one Web community to another, hence denoting influence from the \emph{source} to the \emph{destination} Web community.
Furthermore, anecdotal evidence emerged suggesting that fake news dissemination might start on fringe Web communities, eventually reaching mainstream communities and likely affecting the opinion of a vast amount of people~\cite{bbc_macron,bbc_pizzagate,bbc_wwe}.
Nevertheless, as a research community, we lack tools to effectively track the propagation of information across multiple Web communities, and more importantly, we lack knowledge on understanding the interplay and influence between multiple Web communities.
Gaining this understanding will be extremely important for the research community and the public, allowing us to understand and mitigate emerging pressing issues of our era like the spread of hateful, weaponized, and fake information across the Web.

\descr{Characterizing the Role of Emerging Web Communities and Services on the Information Ecosystem.} 
As new communities are added on the Web, we have limited knowledge of their role on the ecosystem.
One example is the Gab social network~\cite{gab_landing}, which was introduced back in 2016 as an alternative to Twitter.
This specific Web community claims to be all about free speech and welcomes users banned from other Web communities.
However, anecdotal evidence suggests that this community has become the new hub for the alt-right community and likely it is used for the dissemination of false or hateful content~\cite{gab_racism}.
Therefore, it is important to gain knowledge on what content is disseminated in these emerging Web communities, what users are attracted, and how such emerging communities affect the Web.

Information can be extremely diverse: the same piece of information can be disseminated via text, images, and URLs, hence constituting the tracking of information on the Web a non-straightforward task.
In particular, URLs have several aspects that need to be considered.
First, the information provided by the URL can change when the source updates the page. 
Second, URLs can get inaccessible after some time, a problem known as ~\emph{link rot}~\cite{koehler2004longitudinal}, which can affect references across the Web.
Third, there are several services that work with URLs that add complexity to studies that use URLs. 
An example is URL shorteners that generate a shortened URL, which redirects the user to the source page (i.e., different URLs point to the same information).
Another, more interesting example, is the one of archiving services. 
These services, archive the content of the URL at a specific point in time and provide a new URL. 
However, in contrast with URL shorteners, the page is served by the archiving service itself without redirection to the source.
This aspect can have important implications to the Web, as the content will not be able to be changed by the original author and because traffic is taken away from the original source (i.e., content is served by the archiving service).
Despite these interesting aspects, Web archiving services are relatively unstudied:
we lack a general understanding of how these services are used on the Web and what is their role and impact on the information ecosystem.

\descr{Exploitation of Web communities from bad actors.} The Web provides an ideal environment for the diffusion of information to a vast amount of people in a short period of time. 
Clearly, this aspect of the Web can become an extremely powerful and dangerous tool when exploited by bad actors.
Recently, anecdotal evidence emerged that highlights how popular mainstream Web communities like Twitter were exploited by state-sponsored actors that disseminated disinformation on a wide-variety of subjects ranging from health issues~\cite{broniatowski2018weaponized} to politics~\cite{facebook_russians_elections}.
These actors are employed by governments and they possess several online personas that disseminate specific information that helps in pushing the agenda of their government.
Motivated by the real-world impact that these actors can have, popular mainstream communities like Twitter and Reddit, started working on identifying and removing such actors from their platforms.
However, as a research community, we lack an understanding on the behavior of these actors on the Web and how they impact and disrupt the Web's information ecosystem.

Motivated by the above aspects of the information ecosystem on the Web, we focus on providing answers to the following research questions (RQs):
\begin{itemize}
    \item \textbf{RQ1: } What are the various types of false or otherwise malevolent information (e.g., propaganda) that exist on the Web, what are the main actors that contribute to the dissemination of false information, and what are their possible underlying motives? 
    \item \textbf{RQ2: } How information propagates across multiple Web communities and how can we quantify the influence between Web communities? 
    \item \textbf{RQ3: } What type of content is disseminated in small fringe Web communities like Gab, what user base they attract, and what is the influence and impact of these communities to the rest of the Web?
    \item \textbf{RQ4: } What is the role of Web archiving services and how are these services exploited by users on various Web communities. Also, how do such services impact the information ecosystem on the Web? 
    \item \textbf{RQ5: } How are state-sponsored actors exploiting mainstream Web communities in order to disseminate weaponized and possibly fake content? Do these actors have substantial differences when compared to random users? More importantly, how do these actors evolve over time, and what is their influence on the Web.
\end{itemize}

To provide answers to these research questions, we follow a large-scale cross-platform data-driven quantitative approach.
To do so, we first implement a data collection infrastructure that consists of various crawlers, which allow us to collect large-scale datasets from the Web.
Then, we apply various statistical analysis and machine learning techniques to extract meaningful insights from the large-scale datasets.
Specifically, we use the main following techniques:
\begin{itemize}
\item \textbf{Hawkes Processes~\cite{hawkes1971spectra}}: A statistical analysis framework that enable us to investigate possible causalities between events. We use this technique to assess the influence that various Web communities have to each other by modeling and fitting our datasets with Hawkes Processes. More details regarding this technique can be found in Section~\ref{sec:hawkes_background}.
\item \textbf{Changepoint Analysis~\cite{changepoint}}: A statistical analysis technique that allows us to extract points in a time series where statistically significant changes occur. This is particularly useful as it allows us to isolate significant days in a time series and investigate why these changes occur on the various Web communities we study, and possibly link them to real-world events. More details on the methodology and application of this technique can be found in Section~\ref{sec:gab}.
\item \textbf{Neural Networks:} We apply neural networks for various purposes. 
For instance, we use word2vec~\cite{mikolov2013distributed}, which are shallow neural networks, to understand the use of language in the various communities we study.
Also, we use neural networks to build custom classifiers: e.g., we use Convolutional Neural Networks to build a custom screenshot classifier (see Section~\ref{subsec:pipeline}).
\item \textbf{Graph Analysis \& Visualization:} We leverage several graph analysis and visualization techniques to analyze data that can be modeled with graphs. Among other things, we use community detection techniques (e.g., the Louvain method~\cite{blondel2008fast}) to detect meaningful communities from the underlying graph structure, and graph layout techniques (e.g., OpenOrd~\cite{martin2011openord} and ForceAtlas2~\cite{jacomy2014forceatlas2}), which allow us to lay out graphs in the space where the distance between nodes represents something useful (e.g., nodes that are layed out closer means they are more similar). 
\item \textbf{Clustering Algorithms:} We use traditional clustering algorithms for the purpose of creating groups of similar information. For instance, we use the DBSCAN algorithm~\cite{ester1996density} to cluster images that are visually similar with the ultimate goal to track the propagation of memes across the Web (more details can be found in Section~\ref{sec:methodology}).
\end{itemize}

\section{Contributions}
This thesis makes several contributions towards understanding the information ecosystem on the Web.
We make contributions in three main lines of work: 1)~understanding the spread of information across multiple Web communities; 
2)~characterizing emerging Web communities like Gab and assessing the role of Web archiving services on the information ecosystem; 
and 3)~understanding the behavior and influence of state-sponsored actors on the Web's information ecosystem.
In more detail, we make the following contributions:
\begin{itemize}
    \item We provide a comprehensive overview of the information ecosystem on Web.
    To do this, we present a typology that sheds light into the types of false information on the Web, the actors that are involved as well as their possible underlying motives (\textbf{RQ1}).
    \item We introduce a novel methodology, based on Hawkes Processes, to quantify the influence between multiple Web communities. We applied this methodology to several datasets with the goal to quantify the influence that each community has on other communities with respect to the dissemination of news and image-based memes (\textbf{RQ2}).
    \item We present the first study of mainstream and alternative news shared on Twitter, Reddit, and 4chan, measuring how mainstream and alternative news flow between these platforms, and demonstrating how alt-right communities have surprisingly high influence on Twitter (\textbf{RQ2}).
    \item We design and implement a highly scalable processing pipeline that is able to track the propagation of image-based memes across multiple Web communities.\footnote{We make the memes processing pipeline publicly available so it can be used by other researchers~\cite{memes_repo}} By applying the proposed pipeline to 160M images posted on Twitter, Reddit, 4chan, and Gab, we study the memes ecosystem and characterize each community with respect to the memes their users share (\textbf{RQ2}). 
    \item We provide some exploratory analyses on some relatively unknown, by the time we started looking at them, communities and services. Specifically, we provide the first study on Gab, finding that it is becoming the new alt-right's hub despite the fact that it started as a social network promoting free speech (\textbf{RQ3}). Furthermore, we study two Web archiving services, the Wayback Machine and \url{archive.is}, and their use on Twitter, Reddit, 4chan, and Gab, finding several ``nuggets'': 1) these services are used to archive news content and are extensively used by fringe Web communities like 4chan; 2) these services are exploited to a large extent by Reddit bots; 3) these services can be used to deprive ad revenue from the original source and we find evidence that Reddit moderators actually ``force'' users to share archived content from sources with opposing ideology in order to deprive them of ad revenue (\textbf{RQ4}).
    \item We study the behavior of state-sponsored actors on the Web, finding that they exhibit substantial differences when compared to a set of random users.
We find that they change their behavior and that they target different populations over time. 
Also, we quantified the influence that these actors had to Twitter, Reddit, 4chan, and Gab, finding that these actors had a disproportionate influence to the rest of the platforms, with respect to the dissemination of news URLs (\textbf{RQ5}).
\end{itemize}

\section{Peer-Reviewed Papers}

A large body of work presented in this thesis is already published in peer-reviewed journal, conference, and workshop papers. Specifically, some aspects of our work (in collaboration with other researchers and academics) appear in the following papers:
\begin{itemize}
    \item Zannettou, S., Sirivianos, M., Blackburn, J. and Kourtellis, N., 2019. The Web of False Information: Rumors, Fake News, Hoaxes, Clickbait, and Various Other Shenanigans. Journal of Data and Information Quality (JDIQ), 11(3), p.10.
    \item Zannettou, S., Caulfield, T., De Cristofaro, E., Kourtellis, N., Leontiadis, I., Sirivianos, M., Stringhini, G. and Blackburn, J., 2017, November. The Web Centipede: Understanding How Web Communities Influence Each Other Through the Lens Of Mainstream and Alternative News Sources. In Proceedings of the 2017 Internet Measurement Conference (pp. 405-417). ACM.
    \item Zannettou, S., Caulfield, T., Blackburn, J., De Cristofaro, E., Sirivianos, M., Stringhini, G. and Suarez-Tangil, G., 2018, October. On the Origins of Memes By Means of Fringe Web Communities. In Proceedings of the Internet Measurement Conference 2018 (pp. 188-202). ACM  \textbf{(Distinguished Paper Award)}.
    \item Zannettou, S., Bradlyn, B., De Cristofaro, E., Kwak, H., Sirivianos, M., Stringini, G. and Blackburn, J., 2018, April. What is Gab: A Bastion of Free Speech or An Alt-Right Echo Chamber. In Companion of the The Web Conference 2018 on The Web Conference 2018 (pp. 1007-1014). 
    \item Zannettou, S., Blackburn, J., De Cristofaro, E., Sirivianos, M. and Stringhini, G., 2018, June. Understanding Web Archiving Services and Their (Mis) Use on Social Media. In Twelfth International AAAI Conference on Web and Social Media.
    \item Zannettou, S., Caulfield, T., De Cristofaro, E., Sirivianos, M., Stringhini, G. and Blackburn, J., 2019, May. Disinformation Warfare: Understanding State-Sponsored Trolls on Twitter and Their Influence on the Web. In Companion Proceedings of The 2019 World Wide Web Conference (pp. 218-226). ACM \textbf{(Best Paper Award)}.
    \item Zannettou, S., Caulfield, T., Setzer, W., Sirivianos, M., Stringhini, G. and Blackburn, J., 2018. Who Let The Trolls Out? Towards Understanding State-Sponsored Trolls. In Proceedings of the 2019 ACM on Web Science Conference. ACM.

\end{itemize}

\section{Thesis Organization} 

The remainder of this thesis is structured as follows:  
Chapter~\ref{chapter:background} provides the required  background.
Chapter~\ref{chapter:related} describes previous work on: 1) user perception and interaction with false information on the Web; 2) propagation of information on the Web; 3) detection and containment of false information on the Web;  and 4) false information on the political stage.
In Chapter~\ref{chapter:spread_information} we present our work that focuses on understanding how information spreads from one Web community to another and how to measure the influence between multiple communities.
In Chapter~\ref{chapter:various_communities} we present our work related to characterizing various communities and services on the information ecosystem; namely, we study Gab and Web archiving services.
Chapter~\ref{chapter:trolls} describes our work on understanding the role and impact of state-sponsored actors on the Web.
Finally, we conclude in Chapter~\ref{chapter:conclusions}.

\chapter{Background} \label{chapter:background}
In this chapter, we present useful background information regarding the Web communities we study, as well as some statistical techniques that we use to analyze data from various Web communities.
Specifically, for the former, we briefly describe Twitter, Reddit, 4chan, and Gab, while for the latter, we overview Hawkes Processes for estimating influence between Web communities.

\section{Web Communities}
In this section, we briefly describe the Web communities that we use in our work, as well as the methodology for collecting data from each community.

\subsection{Twitter}

\descr{General.} Twitter is a popular microblogging social network.
Users can broadcast 280-character messages, called ``tweets'', to their followers. 
By default, tweets are publicly available, however, users are able to restrict tweets to be available only to their followers.
Twitter includes several traditional social networking features like sharing other tweets (i.e., retweet), liking tweets, as well as posting tweets in reply to other tweets.
On top of this, users can use hashtags (\#) in their tweets, which can help other users to find and weight in on tweets with specific content.
Also, users can refer to other users by mentioning them in tweets (i.e., by using the @ character).

\descr{Moderation.} Twitter moderates content on their site with the goal to remove hateful content and ban users that incite violence or share hateful content. 
Some examples include the permanent ban of Milo Yiannopoulos after continuing hateful abuse against Leslie Jones\footnote{\url{http://fxn.ws/2zshTl8}} and the permanent ban of several accounts linked to the alt-right because of targeted abuse and harassment of others.\footnote{\url{https://wapo.st/2fYdQRG}}
Also, Twitter employs a demoting system which automatically hides content that is likely to be abusive and the user can only see the possibly hateful tweets by pressing a button.\footnote{\url{https://cnnmon.ie/2smPCaF}}

\subsection{Reddit}

\descr{General.} Reddit is called the ``front page of the Internet'' and it is a popular news aggregator.
Users can create threads (called ``submissions'') by posting a URL along with a title.
Other users can reply below in a structured manner (e.g., reply to submission or reply to specific reply).
Popularity of content within the platform is determined via a voting system: each comment or submission can be up-voted or down-voted, hence a score can be calculated.
Submissions and comments with higher score appears on top of submissions and comments with lower score.
Also, there is a user-based ``score'' called \emph{karma} that is basically the sum of scores for all of the user's comments and submissions.
Note that on Reddit the community structure is not defined by the friendship/follower relation like Twitter (a user can list another user as a friend but it does not change anything in the structure or use of the platform).

\descr{Subreddits.}
Reddit is divided into millions of communities called ``subreddits''.\footnote{According to statista as of 2017 there are nearly 1.2M subreddits (\url{https://www.statista.com/chart/11882/number-of-subreddits-on-reddit/}).}
Subreddits are created from users of the platform and this has lead to a plethora of communities discussing a wide variety of topics ranging from video games, to politics, pornography, and even meta-communities that summarize interactions of users on other subreddits/social networks.
Subreddits are monitored by Reddit's administrators and they are removed when they share ``extremely inappropriate'' content.
For instance, in the past, Reddit removed subreddits related to the promotion of conspiracy theories (e.g.,  /r/greatawakening, which promoted the Qanon conspiracy theory~\cite{wapo_qanon}), subreddits that shared suggestive photos of underage girls (e.g., /r/jailbait), as well as subreddits sharing ``deepfakes'' (e.g., /r/deepfakes).\footnote{For a list of controversial subreddits that were removed see \url{https://en.wikipedia.org/wiki/Controversial_Reddit_communities}.}

\subsection{4chan}

\descr{General.} 4chan is a discussion forum known as an imageboard.\footnote{\url{http://4chan.org/}} 
Users can create a new thread by creating a post that must include an image. 
Other users can reply to the thread (images are optional in replies) and possibly add references or quotes to previous posts within the thread.
4chan is an anonymous community: users are not required to have an account in order to create a post.
At the same time, users can add a pseudonym when posting, however, their pseudonym is bounded to the specific post and they can use a different one for other posts.
On top of this, each post is associated with a \emph{flag}. 
Usually, the flag is determined based on the location of the user, however, it can be tricked by the use of Virtual Private Networks (VPN).
Furthermore, there are communities within 4chan that introduce custom flags.
For instance, 4chan's Politically Incorrect board (\dspol), allow users to either add the flag based on their location or from a set of 23 pre-defined flags. 
Examples of such flags include the flag of Kekistan, the Nazi flag, confederate flag, etc.

\descr{Boards.} 4chan is divided into multiple communities called \emph{boards}, which are defined by 4chan.
Each board has its own general theme and topic, ranging from politics, to sports and pornography, and likely attracts different user bases.
For instance, 4chan's Politically Incorrect board (\dspol) focuses on discussions of politics and world news, while the Video games board (/v/) focuses on discussions around video games.
As of April 2019, 4chan has 70 different boards.
In our work, we mainly focus on 4chan \dspol board, which is mainly used for the discussion of news and real-world events that are happening.
Also, we are particularly interested in this community since previous work showed that it exhibits a high degree of hate speech and racism~\cite{hine2016longitudinal}, and because of anecdotal evidence that suggests the community's influence and impact both on the online and offline world (e.g., spread of Pizzagate conspiracy theory~\cite{bbc_4chan_pizzagate}).

\descr{Moderation.} 4chan has an extremely lax moderation: each board has a handful of volunteers called \emph{janitors} that are moderating each board.
Janitors can remove posts and threads and recommend user bans to 4chan employees.
Generally, Janitors pretty much allow everything to be posted as long as it is relevant to the general topic and theme of the board. 
Due to this lax moderation and anonymous nature of the community, 4chan users can use whatever tone and language to express themselves, hence 4chan's high degree of hateful content.

\descr{Ephemerality.} 4chan is an ephemeral community. Each board has a finite amount of active threads. 
Threads are removed after a relatively short period based on a ``bumping system'' that considers the posting activity within the thread.
That is, creating a new thread, results in the \emph{archival} of the thread with the least recent post.
A new post within a thread can help ``bump'' the thread as it keeps the thread alive and makes the thread appear at the top of the board. 
To avoid having a thread alive forever, 4chan has \emph{bump} and \emph{image} limits, which determine the maximum number of images and bumps that a thread can receive.
Once a thread is archived, it remains to the community for 7 days before getting deleted forever.

\subsection{Gab}

\descr{General.} Gab is a new social network, launched in August 2016, that ``champions free speech, individual liberty, and the free flow of information online.\footnote{\url{http://gab.ai}}''
It combines social networking features that exist in popular social platforms like Reddit and Twitter.
A user can broadcast 300-character messages, called ``gabs,'' to their followers (akin to Twitter).
From Reddit, Gab takes a modified voting system (which we discuss later).
Gab allows the posting of pornographic and obscene content, as long as users label it as Not-Safe-For-Work (NSFW).\footnote{What constitutes NSFW material is not well defined.}
Posts can be reposted, quoted, and used as replies to other gabs. 
Similar to Twitter, Gab supports hashtags, which allow indexing and querying for gabs, as well as mentions, which allow users to refer to other users in their gabs.

\descr{Topics and Categories.} Gab posts can be assigned to a specific \emph{topic} or \emph{category}.
\emph{Topics} focus on a particular event or timely topic of discussion and can be created by Gab users themselves; all topics are publicly available and other users can post gabs related to topics.
\emph{Categories} on the other hand, are defined by Gab itself, with 15 categories defined at the time of this writing.
Note that assigning a gab to a category and/or topic is \emph{optional}, and Gab moderates topics, removing any that do not comply with the platform's guidelines.

\descr{Voting system.} Gab posts can get up- and down-voted; a feature that determines the popularity of the content in the platform (akin to Reddit).
Additionally, each user has its own score, which is the sum of up-votes minus the sum of down-votes that it received to all his posts (similar to Reddit's user karma score~\cite{reddit_karma}).
This user-level score determines the popularity of the user and is used in a way unique to Gab: a user must have a score of at least 250 points to be able to down-vote other users' content, and every time a user down-votes a post a point from his user-level score is deducted.
In other words, a user's score is used as a form of currency expended to down-vote content.

\descr{Moderation.} Gab has a lax moderation policy that allows most things to be posted, with a few exceptions.
Specifically, it only forbids posts that contain ``illegal pornography'' (legal pornography is permitted), posts that promote terrorist acts, threats to other users, and doxing other users' personal information~\cite{snyder2017fifteen}.\footnote{For more information on Gab's guidelines, see \url{https://gab.ai/about/guidelines}.}

\descr{Monetization.} Gab is ad-free and relies on direct user support.
On October 4, 2016 Gab's CEO Andrew Torba announced that users were able to donate to Gab~\cite{gab_donate}.
Later, Gab added ``pro'' accounts as well.
``Pro'' users pay a per-month fee granting additional features like live-stream broadcasts, account verification, extended character count (up to 3K characters per gab), special formatting in posts (e.g., italics, bold, etc.), as well as premium content creation.
The latter allows users to create ``premium'' content that can only be seen by subscribers of the user, which are users that pay a monthly fee to the content creator to be able to view his posts.
The premium content model allows for crowdfunding particular Gab users, similar to the way that Twitch and Patreon work.
Finally, Gab is in the process of raising money through an Initial Coin Offering (ICO) with the goal to offer a ``censorship-proof'' peer-to-peer social network that developers can build application on top~\cite{gab_ico}.

\subsection{Remarks}
In this section, we presented the data sources that we use in this thesis.
We select these specific data sources for various reasons.
First, our data sources comprise of an interesting mix of both mainstream Web communities (i.e., Twitter and Reddit), as well as fringe Web communities (i.e., Gab and 4chan). 
This enables us to understand how small fringe Web communities influence large mainstream Web communities.
Second, we select these specific fringe Web communities mainly because of extensive anecdotal evidence that suggest that 4chan and Gab are involved in the dissemination of false information~\cite{bbc_4chan_pizzagate,4chan_disruptions_politics} and hateful content~\cite{gab_racism,gab_hate_speech}.
Third, we avoid using other popular social networks like Facebook mainly due to limits imposed on their APIs by the company itself, hence constituting the task of obtaining data non-straightforward.

Obviously, it is likely that on other Web communities we can find important differences that might affect the results presented in this thesis, however, this thesis sheds light into the information ecosystem through the lens of multiple Web communities highlighting the need to shift focus into understanding the various Web communities on the Web and study the interplay between them.
Despite this fact, as we mention in the next section, our influence estimation experiments via Hawkes Processes allow us to also capture the creation of events from external sources, hence we argue that the Influence Estimation results presented throughout this thesis can be treated as general, as they shed light into the influence that each Web community have to the others by also considering external sources (i.e., communities that we do not study like Facebook).

\section{Hawkes Processes}\label{sec:hawkes_background}
In this section, we provide necessary background for Hawkes Processes and how we use them in order to assess the interplay of multiple Web communities and, more importantly, quantify the influence that specific small fringe Web communities (e.g., 4chan) have to mainstream ones like Twitter.
In a nutshell, Hawkes Processes is a statistical framework that allow us to assess the causality of events that occur on the Web, and find the possible root causes (i.e., Web community that is responsible for the creation of the events) along with their respective probabilities.

\descr{General.} Hawkes Processes are self-exciting temporal point processes~\cite{hawkes1971spectra} that describe how events (e.g., posting of a URL or an image) occur on a set of processes (i.e., Web communities).
Generally, a Hawkes model consists of a number, $K$, of point processes, each with a ``background rate'' of events $\lambda_{0,k}$.
The background rate is the expected rate at which events will occur on a process {\em without} influence from the processes modeled or previous events; this captures events created for the first time, or those seen on a process we do not model and then created on a process we do.

An event on one process can cause an \emph{impulse response} on other processes, which increases the probability of an event occurring above the processes' background rates.
The shape of the impulse determines how the probability of these events occurring is distributed over time; typically the probability of another event occurring is highest soon after the original event and decreases over time.

\begin{figure}[]
\centering
\includegraphics[width=.67\columnwidth]{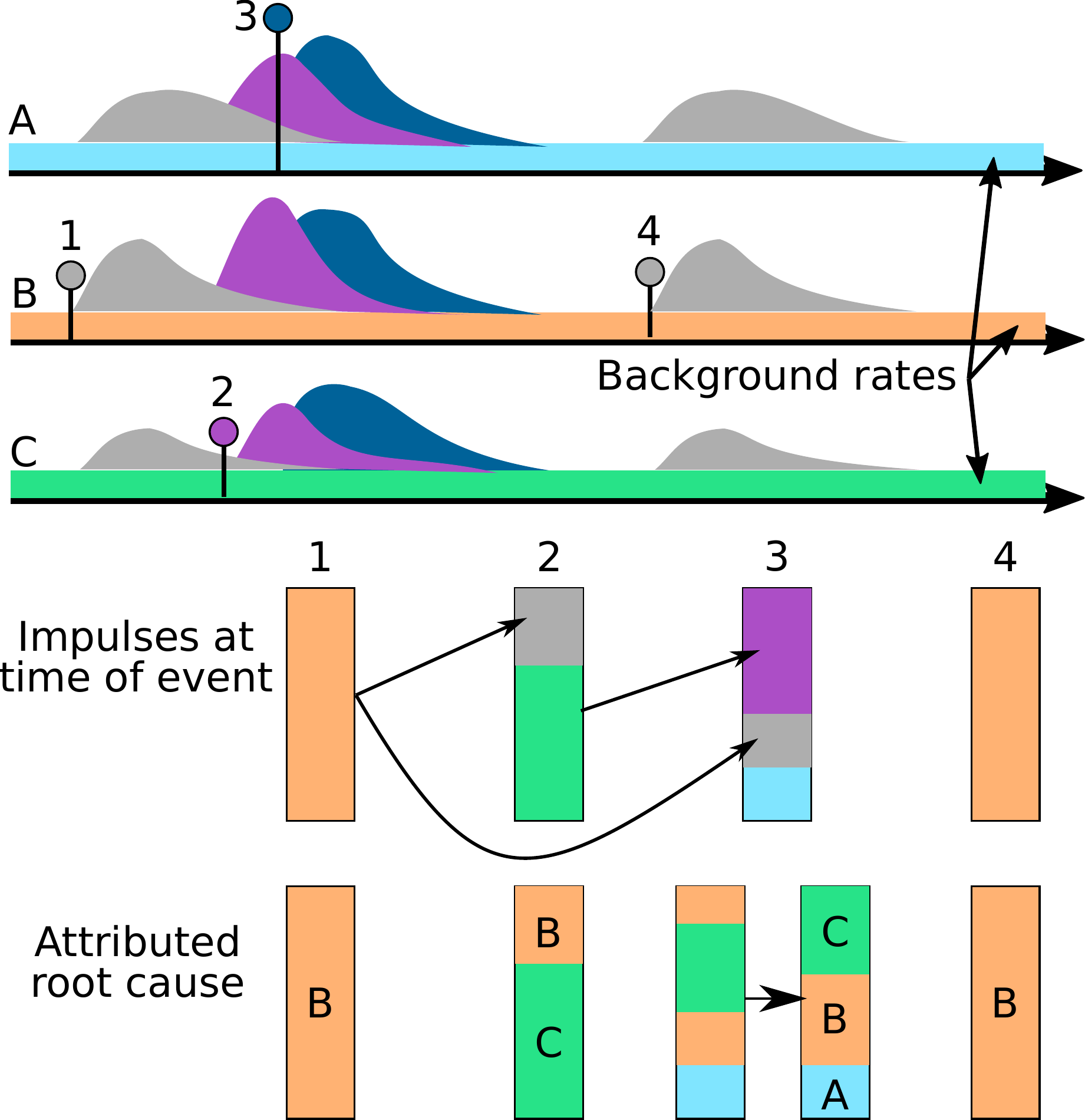}
\caption{Representation of a Hawkes model with three processes.  Events cause impulses that increase the rate of subsequent events in the same or other processes.  By looking at the impulses present when events occur, the probability of a process being the root cause of an event can be determined. Note that on the second part of the Figure, colors represent events while arrows represent impulses between the events. }
\label{fig:hawkes_explanation}
\end{figure}

Fig.~\ref{fig:hawkes_explanation} illustrates a Hawkes model with three processes.
The first event occurs on process B, which causes an increase in the rate of events on all three processes.
The second event then occurs on process C, again increasing the rate of events on the processes.
The third event occurs soon after, on process A.
The fourth event occurs later, again caused by the background arrival rate on process B, after the increases in arrival rate from the other events have disappeared.

To understand the influence different processes have on the creation of specific events, we want to be able to attribute the cause of an event being posted back to a specific process.
For example, if an image is posted on \dspol and then someone sees it there and posts it on Twitter where it is shared several times, we would like to be able to say that \dspol was the \emph{root cause} of those events.
Obviously, we do not actually know where someone saw something and decided to share it, but we can, using the Hawkes models, determine the \emph{probability} of each community being the root cause of an event.

Looking again at Fig.~\ref{fig:hawkes_explanation}, we see that events 1 and 4 are caused directly by the background rate of process B.
This is because, in the case of event 1, there are no previous events on other processes, and in the case of event 4, the impulses from previous events have already stopped.
Events 2 and 3, however, occur when there are multiple possible causes: the background rate for the process and the impulses from previous events.
In these cases, we assign the probability of being the root cause in proportion to the magnitudes of the impulses (including the background rate) present at the time of the event.
For event 2, the impulse from event 1 is smaller than the background rate of process C, so the background rate has a higher probability of being the cause of event 2 than event 1.
Thus, most of the cause for event 2 is attributed to process C, with a lesser amount to B (through event 1).
Event 3 is more complicated: impulses from both previous events are present, thus the probability of being the cause is split three ways, between the background rate and the two previous events.
The impulse from event 2 is the largest, with the background rate and event 1 impulse smaller.
Because event 2 is attributed both to processes B and C, event 3 is partly attributed to process B through both event 1 and event 2.

For our purposes, fitting a Hawkes model to a series of events on the different processes gives us values for the background rates for each process along with the probability of an event on one process causing events on other processes.
We emphasize that the background rates of the Hawkes processes allows us to also account for the probability of an event caused by external sources of information. 
Thus, while we are only able to model the specific influences for a limited number of Web communities we study, the resulting probabilities are affirmatively attributable to each of them; the influence of the greater Web is captured by the background rates.

For a discrete-time Hawkes model, time is divided into a series of bins of duration $ \Delta t$, and events occurring
within the same time bin do not interact with each other.
The rate of each $k$-th process,
$\lambda_{t,k}$ is given by:
\[
  \lambda_{t,k} = \lambda_{0,k} + \sum_{k'=1}^K \sum_{t'=1}^{t-1} s_{t',k'} \cdot h_{k' \to k}[t - t']
\]
where $ s \in \mathbb{N}^{T \times K}$ is the matrix of event counts (how many events occur for process $k$ at time $t$) and $h_{k' \to k}[ t-t']$ is an impulse response function that describes the amplitude of influence that events on process $k'$ have on the rate of process $k$.

Following~\cite{linderman2014}, the impulse response function $h_{k \to k'}[t-t']$ can be decomposed into a scalar weight $W_{k \to k'}$ and a probability mass function $G_{k \to k'}[d]$.
The weight specifies the strength of the interaction from process $k$ to process $k'$ and the probability mass function specifies how the interaction changes over time:
\[
  h_{k \to k'}[d] = W_{k \to k'} G_{k \to k'}[d]
\]
The weight value $W_{k \to k'}$ can be interpreted as the expected number of child events that will be caused on process $k'$ after an event on process $k$.
The probability mass function $G_{k \to k'}$ specifies the probability that a child event will occur at each specific time lag $d \Delta t $, up to a maximum lag $\Delta t_{max}$.
This interpretation of $W_{k \to k'}$ is useful because it allows us to compare how much influence processes have on each other.
For instance, we can examine whether an image posted on Twitter or on Reddit is more likely to cause the same image to be posted on 4chan, or if there is a difference in influence from one platform to another between various images.

\descr{Experimental Setup.} We assume a Hawkes model that is is fully connected, i.e., it is possible for each process to influence all the others, as well as itself, which describes behavior where participants on a platform see an image and re-post it on the same platform.
For example, with Twitter, this value ($W_{\mathrm{Twitter} \to \mathrm{Twitter}}$) would likely be quite high, given that tweets are commonly re-tweeted a number of times: the initial tweet containing an image is likely to cause a number of re-tweets, also containing the image, on the same platform.

For each type of event, we create a matrix $s \in \mathbb{N}^{T \times K}$ containing the number of events per minute for each of the processes (i.e., Web communities).
Here, $T$ is the number of minutes from the first recorded post of the event on any process, to the last recorded post of the event on any process (\textbf{NB:} this value can be different for each type of event).
We select $ \Delta t = 1~\mathrm{minute}$ as a reasonable compromise between accuracy and computational cost.

Next, we fit a Hawkes model for each type of event using the approach described in~\cite{linderman2014,lindermanArxiv}, which uses Gibbs sampling to infer the parameters of the model from the data, including the weights, background rates, and shape of the impulse response functions between the different processes.
By setting $\Delta t_{max} = 60 \cdot 12 = 720$ minutes, we say that a given event can cause other events within a 12-hour time window.
Experiments with other values (6, 12, 24, and 48 hours) gave similar results.
After fitting the models, we have the values for the $W$ matrix -- i.e., the weights of the interactions between events on different processes for each type of event.
These weights can then be interpreted as the expected number of events.
For example, $W_{Twitter \to /pol/} = 0.1$ would mean that an event on Twitter will cause $n$ events on \dspol, where $n$ is drawn from a Poisson distribution with rate parameter $0.1$.
Finally, for each type of event, we also get the $\lambda_{0,k}$ values for each process, which are the background rates for event arrivals that are \emph{not} caused by other events in the system we model.
Again, these background rates capture events due to some \emph{other} process, e.g., someone posting an image after seeing it on the original site or seeing the image on another site not included in the model, like Facebook.

\descr{Metrics.}
Having obtained the weight matrix $W$, which specifies the strength of connections between processes for each type of event, we report our influence estimation results using two metrics.
First, we measure the absolute influence, which can be interpreted as the expected number of events that are created on a destination process because of events previously seen on a source process. 
Since the weight values can be interpreted as the expected number of additional events that will be caused a consequence of an event, we can estimate the percentage of events on each process that were caused by each of the other processes by multiplying the weight by the actual number of events that occurred on the source process (e.g., Process A) and dividing by the number of events that occurred on the destination process (e.g., Process B):

\[
  \mathrm{Influence}_{A \to B} = \frac{ \sum_{e \in \mathrm{events}} \left( W_{A \to B} \cdot \sum_{t = 1}^T s_{t,A} \right) }{  \sum_{e \in \mathrm{events}} \sum_{t = 1}^T s_{t,B} } 
\]

Second, we measure the \emph{efficiency} of each process in pushing events to the rest of the processes.
To do this, we normalize the influence values by the total number of events in the source process (e.g., Process A).
This metric allow us to see how much influence each process has, relative to the number of events that are created in the process and is given by:

\[
  \mathrm{Efficiency}_{A \to B} = \frac{ \sum_{e \in \mathrm{events}} \left( W_{A \to B} \cdot \sum_{t = 1}^T s_{t,A} \right) }{  \sum_{e \in \mathrm{events}} \left(\sum_{t = 1}^T s_{t,A} \right)} 
\]

\descr{Remarks.} In this section, we described how we can use Hawkes Processes for modeling the interplay between multiple Web communities and how to quantify the influence that each Web community have to the others.
To do this, we leverage Bayesian Inference techniques and data that describes the appearances of events in a set of processes. 
This allow us to assess the causality of events that happen on multiple Web communities and assess the possible root causes (i.e., the responsible Web community for the creation of the event) for each event.
Note that by tweaking what an event is, the proposed framework can be applied to a wide variety of use cases (e.g., an event can be text referring to a specific news story, a specific video, etc.).

\chapter{Literature Review}
\label{chapter:related}

In this chapter, we provide an extensive literature review of work that focus on the false information ecosystem on the Web. 
First, we present a typology of the various types and actors that are involved in the spread of information on the Web. 
Then we review the following lines of work: 1) user perception of false information; 2) propagation of false information; 3) detection and containment of false information;  4) false information on the political stage and 5) various other studies that are relevant.

\section{Typology of the False Information Ecosystem}
\label{chapter:taxonomy}

In this section we present our typology, which we believe it will provide a succinct roadmap for future work.
The typology is based on~\cite{taxonomy_firstdraftnews} and extended to build upon the existing literature. 
Specifically, we describe the various types of false information that can be found in OSNs (Section~\ref{subsec:types}), the various types of actors that contribute in the distribution of false information (Section~\ref{subsec:actors}), as well as their motives (Section~\ref{subsec:motives}).
Note that our typology is different from concurrent work by Kumar and Shah~\cite{kumar2018false} as we provide a fine-grained distinction for the types of false information, the actors, and their motives.
Also, note that we make a best effort to cover as many aspects of the false information as per our knowledge; however, the typology should not be treated as an exhaustive representation of the false information ecosystem.

\subsection{Types of False Information} \label{subsec:types}
False information on the Web can be found in various forms, hence we propose the categorization of false information into eight types as listed below:
    \begin{itemize}
         \item \textbf{Fabricated (F)~\cite{rubin2015towards}.} Completely fictional stories disconnected entirely from real facts. 
   This type is not new and it exists since the birth of journalism. 
    Some popular examples include fabricated stories about politicians and aliens~\cite{hillary_fake_news} (e.g., the story that Hillary Clinton adopted an alien baby).

    \item \textbf{Propaganda (P)~\cite{jowett2014propaganda}.} This is a special instance of the fabricated stories that aim to harm the interests of a particular party and usually has a political context.
This kind of false news is not new, as it was widely used during World War II and the Cold War.
Propaganda stories are profoundly utilized in political contexts to mislead people with the overarching goal of inflicting damage to a particular political party or nation-state.
Due to this, propaganda is a consequential type of false information as it can change the course of human history (e.g., by changing the outcome of an election).
Some recent examples of propaganda include stories about the Syria air strikes in 2018 or about specific movements like the BlackLivesMatter (see~\cite{medium_propaganda} for more examples).
            \item \textbf{Conspiracy Theories (CT)~\cite{fenster1999conspiracy}.} Refer to stories that try to explain a situation or an event by invoking a conspiracy without proof.
    Usually, such stories are about illegal acts that are carried out by governments or powerful individuals.
    They also typically present unsourced information as fact or dispense entirely with an ``evidence'' based approach, relying on leaps of faith instead.
    Popular recent examples of conspiracy theories include the Pizzagate theory (i.e., Clinton's campaign running a pedophile ring)~\cite{pizzagate} and conspiracies around the murder of Seth Rich~\cite{seth_rich} (e.g., Seth Rich was involved in the DNC email leaks).
        \item \textbf{Hoaxes (H)~\cite{kumar2016disinformation}.} News stories that contain facts that are either false or inaccurate and are presented as legitimate facts. This category is also known in the research community either as half-truth~\cite{half-truth} or factoid~\cite{factoid} stories.
        Popular examples of hoaxes are stories that report the false death of celebrities (e.g., the Adam Sadler death hoax~\cite{snopes_hoax_sadler}).
        \item \textbf{Biased or one-sided (B).} Refers to stories that are extremely one-sided or biased.
        In the political context, this type is known as Hyperpartisan news~\cite{potthast2017stylometric} and are stories that are extremely biased towards a person/party/situation/event.
    Some examples include the wide spread diffusion of false information to the alt-right community from small fringe Web communities like 4chan's /pol/ board~\cite{hine2016longitudinal} and Gab, an alt-right echo chamber~\cite{zannettou2018gab}.
    \item \textbf{Rumors (R)~\cite{peterson1951rumor}.} Refers to stories whose truthfulness is ambiguous or never confirmed. 
    This kind of false information is widely propagated on OSNs, hence several studies have analyzed this type of false information.
    Some examples of rumors include stories around the 2013 Boston Marathon Bombings like the story that the suspects became citizens on 9/11 or that a Sandy Hook child was killed during the incident~\cite{boston_bombing_rumors}.
    \item \textbf{Clickbait (CL)~\cite{chen2015misleading}.} Refers to the deliberate use of misleading headlines and thumbnails of content on the Web. 
    This type is not new as it appeared years before, during the ``newspaper era,'' a phenomenon known as yellow journalism~\cite{campbell2001yellow}.
    However, with the proliferation of OSNs, this problem is rapidly growing, as many users add misleading descriptors to their content with the goal of increasing their traffic for profit or popularity~\cite{politifact_clickbait_profit}. 
    This is one of the least severe types of false information because if a user reads/views the whole content then he can distinguish if the headline and/or the thumbnail was misleading.
    \item \textbf{Satire News (S)~\cite{burfoot2009automatic}.} Stories that contain a lot of irony and humor. 
    This kind of news is getting considerable attention on the Web in the past few years. 
    Some popular examples of sites that post satire news are TheOnion~\cite{theonion} and SatireWire~\cite{satirewire}.
Usually, these sites disclose their satyric nature in one of their pages (i.e., About page). 
However, as their articles are usually disseminated via social networks, this fact is obfuscated, overlooked, or ignored by users who often take them at face value with no additional verification.
\end{itemize}

It is extremely important to highlight that there is an overlap in the aforementioned types of false information, thus it is possible to observe false information that may fall within multiple categories.
Here, we list two indicative examples to better understand possible overlaps: 1) a rumor may also use clickbait techniques to increase the audience that will read the story; and 2) propaganda stories, which are a special instance of a fabricated story, may also be biased towards a particular party.
These examples highlight that the false information ecosystem is extremely complex and the various types of false information need to be considered to mitigate the problem.

\subsection{False Information Actors} \label{subsec:actors}

In this section, we describe the different types of actors that constitute the false information propagation ecosystem.
We identified a handful of different actors that we describe below. %
\begin{itemize}
    \item \textbf{Bots \cite{boshmaf2011socialbot}.} In the context of false information, bots are programs that are part of a bot network (Botnet) and are responsible for controlling the online activity of several fake accounts with the aim of disseminating false information. Botnets are usually tied to a large number of fake accounts that are used to propagate false information in the wild. A Botnet is usually employed for profit by 3rd party organizations to diffuse false information for various motives (see Section \ref{motives_subsection} for more information on their possible motives).
    Note that various types of bots exist, which have varying capabilities; for instance, some bots only repost content, promote content (e.g., via vote manipulation on Reddit or similar platforms), and others post ``original'' content. However, this distinction is outside of the scope of this work, which provides a general overview of the information ecosystem on the Web.
    \item \textbf{Criminal/Terrorist Organizations \cite{al2015examining}.} Criminal gangs and terrorist organizations are exploiting OSNs as the means to diffuse false information to achieve their goals.
     A recent example is the ISIS terrorist organization that diffuses false information in OSNs for propaganda purposes~\cite{al2015examining}. 
    Specifically, they widely diffuse ideologically passionate messages for recruitment purposes.
    This creates an extremely dangerous situation for the community as there are several examples of individuals from European countries recruited by ISIS that ended-up perpetrating terrorist acts.
    \item \textbf{Activist or Political Organizations.} Various organizations share false information in order to either promote their organization, demote other rival organizations, or for pushing a specific narrative to the public. 
    A recent example include the National Rifle Association, a non-profit organization that advocates gun rights, which disseminated false information to manipulate people about guns~\cite{nra_fake_news}.
    Other examples include political parties that share false information, especially near major elections~\cite{allcott2017social}.
    \item \textbf{Governments~\cite{goverments_fake_news}.} Historically, governments were involved in the dissemination of false information for various reasons. More recently, with the proliferation of the Internet, governments utilize the social media to manipulate public opinion on specific topics. 
    Furthermore, there are reports that foreign goverments share false information on other countries in order to manipulate public opinion on specific topics that regard the particular country. Some examples, include the alleged involvement of the Russian government in the 2016 US elections~\cite{us_elections_russians} and Brexit referendum~\cite{brexit_russians}.
    \item \textbf{Hidden Paid Posters \cite{chen2013battling} and State-sponsored Trolls~\cite{zannettou2018disinformation,zannettou2018let}.} They are a special group of users that are paid in order to disseminate false information on a particular content or targeting a specific demographic.
    Usually, they are employed for pushing an agenda; e.g., to influence people to adopt certain social or business trends. 
    Similar to bots, these actors disseminate false information for profit.
    However, this type is substantially harder to distinguish than bots because they exhibit characteristics similar to regular users.
    \item \textbf{Journalists \cite{lee2004lying}.} Individuals that are the primary entities responsible for disseminating information both to the online and to the offline world.
    However, in many cases, journalists are found in the center of controversy as they post false information for various reasons.
    For example, they might change some stories so that they are more appealing, in order to increase the popularity of their platform, site, or newspaper.
    \item \textbf{Useful Idiots~\cite{useful_idiot}.} The term originates from the early 1950s in the USA as a reference to a particular political party's members that were manipulated by Russia in order to weaken the USA. 
    Useful idiots are users that share false information mainly because they are manipulated by the leaders of some organization or because they are naive.
    Usually, useful idiots are normal users that are not fully aware of the goals of the organization, hence it is extremely difficult to identify them.
    Like hidden paid posters, useful idiots are hard to distinguish and there is no study that focuses on this task. 
    
    \item \textbf{``True Believers'' and Conspiracy Theorists.} Refer to individuals that share false information because they actually believe that they are sharing the truth and that other people need to know about it. 
    For instance, a popular example is Alex Jones, which is a popular conspiracy theorist that shared false information about the Sandy Hook shooting~\cite{sandy_hook_wiki}.
    
    \item \textbf{Individuals that benefit from false information.} Refer to various individuals that will have a personal gain by disseminating false information. This is a very broad category ranging from common persons like an owner of a cafeteria to popular individuals like political persons.
    \item \textbf{Trolls \cite{mihaylov2015finding}.} The term troll is used in great extend by the Web community and refers to users that aim to do things to annoy or disrupt other users, usually for their own personal amusement. 
    An example of their arsenal is posting provocative or off-topic messages in order to disrupt the normal operation or flow of discussion of a website and its users.
    In the context of false information propagation, we define trolls as users that post controversial information in order to provoke other users or inflict emotional pressure.
    Traditionally, these actors use fringe communities like Reddit and 4chan to orcherstrate organized operations for disseminating false information to mainstream communities like Twitter, Facebook, and YouTube~\cite{zannettou2017web, hine2016longitudinal}.
    \end{itemize}
    
Similarly to the types of false information, overlap may exist in actors too. 
Some examples include: 
1) Bots can be exploited by criminal organizations or political persons to disseminate false information~\cite{isis_bots}; and
2) Hidden paid posters and state-sponsored trolls can be exploited by political persons or organizations to push false information for a particular agenda~\cite{facebook_russians_elections}.

\subsection{Motives behind false information propagation} \label{subsec:motives}
\label{motives_subsection}

False information actors and types have different motives behind them. 
Below we describe the categorization of motives that we distinguish:
\begin{itemize}
    \item \textbf{Malicious Intent.} 
    Refers to a wide spectrum of intents that drive actors that want to hurt others in various ways.
    Some examples include inflicting damage to the public image of a specific person, organization, or entity. 
    \item \textbf{Influence.} 
    This motive refers to the intent of misleading other people in order to influence their  decisions, or manipulate public opinion with respect to specific topics. 
    This motive can be distinguished into two general categories; 1)~aiming to get leverage or followers (\emph{power}) and 2)~changing the norms of the public by disseminating false information.
    This is particularly worrisome on political matters~\cite{fake_news_politics}, where individuals share false information to enhance an individuals' public image or to hurt the public image of opposing politicians, especially during election periods.
    \item \textbf{Sow Discord.} In specific time periods individuals or organizations share false information to sow confusion or discord to the public. 
    Such practices can assist in pushing a particular entity's agenda; we have seen some examples on the political stage where foreign governments try to seed confusion in another country's public for their own agenda~\cite{russia_discord}.
    \item \textbf{Profit.}
    Many actors in the false information ecosystem seek popularity and monetary profit for their organization or website. 
    To achieve this, they usually disseminate false information that increases the traffic on their website. 
    This leads to increased ad revenue that results in monetary profit for the organization or website, at the expense of manipulated users.
    Some examples include the use of clickbait techniques, as well as fabricated news to increase views of articles from fake news sites that are disseminated via OSNs~\cite{politifact_clickbait_profit, fake_news_profit}
    \item \textbf{Passion.} A considerable amount of users are passionate about a specific idea, organization, or entity.
    This affects their judgment and can contribute to the dissemination of false information.
    Specifically, passionate users are blinded by their ideology and perceive the false information as correct, and contribute in its overall propagation~\cite{fake_news_passion}.
    \item \textbf{Fun.} As discussed in the previous section, online trolls are usually diffusing false information for their amusement.
  Their actions can sometimes inflict considerable damage to other individuals (e.g., see Doxing~\cite{snyder2017fifteen}), and thus should not be taken lightly.
\end{itemize}

Again, similarly to Sections~\ref{subsec:types} and \ref{subsec:actors}, we have overlap among the presented motives. 
For instance, a political person may disseminate false information for political influence and because he is passionate about a specific idea.

\section{User Perception of False Information}
\label{sec:user_perception}

In this section, we describe work that study how users perceive and interact with false information on OSNs.
Existing work use the following methodologies in understanding how false information is perceived by users:
(i) by analyzing large-scale datasets obtained from OSNs; and
(ii) by receiving input from users either from questionnaires, interviews, or through crowdsourcing marketplaces (e.g., Amazon Mechanical Turk, AMT~\cite{mechanical_turk}).
Table~\ref{tbl:user_perception_papers_summary} summarizes the studies on user perception, as well as their methodology and the considered OSN.
Furthermore, we annotate each entry in Table~\ref{tbl:user_perception_papers_summary} with the type of false information that each work considers.
The remainder of this section provides an overview of the studies on understanding users' perceptions on false information.

\begin{table}[]
\centering
\resizebox{.8\columnwidth}{!}{
\begin{tabular}{@{}cccc@{}}
\toprule
\textbf{Platform} & \textbf{OSN data analysis} & \textbf{Questionnaires/Interviews}& \textbf{Crowdsourcing platforms}\\ \midrule
Twitter & \begin{tabular}[c]{@{}c@{}}Kwon et al.~\cite{kwon2013aspects} \textbf{(R)},\\ Zubiaga et al.~\cite{zubiaga2016analysing} \textbf{(R)},\\
Thomson et al.~\cite{thomson2012trusting} \textbf{(R)}\end{tabular} & Morris et al.~\cite{morris2012tweeting} \textbf{(CA)}
& \begin{tabular}[c]{@{}c@{}}Ozturk et al.~\cite{ozturk2015combating}~\textbf{(R)},\\McCreadie et al.~\cite{mccreadie2015crowdsourced}~\textbf{(R)}\end{tabular}\\ \midrule 

Facebook & \begin{tabular}[c]{@{}c@{}}Zollo et al.~\cite{zollo2015emotional}~\textbf{(CT)},\\ Zollo et al.~\cite{zollo2015debunking}~\textbf{(CT)},\\ Bessi et al.~\cite{bessi2015science}~\textbf{(CT)}\end{tabular} & Marchi~\cite{marchi2012facebook} \textbf{(B)}& X \\ \midrule
Other & Dang et al.~\cite{dang2016toward} \textbf{(R)} & \begin{tabular}[c]{@{}c@{}}Chen et al.~\cite{chen2015why}~\textbf{(F)},\\ Kim and Bock~\cite{kim2011study}~\textbf{(R)},\\ Feldman~\cite{feldman2011partisan}~\textbf{(B)},\\
Brewer et al.~\cite{brewer2013impact}~\textbf{(S)}\\
Winerburg and McGrew~\cite{wineburg2017lateral} \textbf{(CA)}\end{tabular} & X\\ \bottomrule
\end{tabular}
}
\caption{Studies of user perception and interaction with false information on OSNs. The table depicts the main methodology of each paper as well as the considered OSN (if any). Also, where applicable, we report the type of false information that is considered (see bold markers and cf. with Section~\ref{subsec:types}).
}
\label{tbl:user_perception_papers_summary}
\end{table}

\subsection{OSN data analysis}
Previous work focuses on extracting meaningful insights by analyzing data obtained from OSNs.
From Table~\ref{tbl:user_perception_papers_summary} we observe that previous work, leverages data analysis techniques to mainly study how users perceive and interact with rumors and conspiracy theories.

\descr{Rumors.} Kwon et al.~\cite{kwon2013aspects} study the propagation of rumors in Twitter, while considering findings from social and psychological studies. 
By analyzing 1.7B tweets, obtained from~\cite{cha2010measuring}, they find that: 
1) users that spread rumors and non-rumors have similar registration age and number of followers; 
2) rumors have a clearly different writing style; 
3) sentiment in news depends on the topic and not on the credibility of the post; and 
4) words related to social relationships are more frequently used in rumors.
Zubiaga et al.~\cite{zubiaga2016analysing} analyze 4k tweets related to rumors by using journalists to annotate rumors in real time.
Their findings indicate that true rumors resolved faster than false rumors and that the general tendency for users is to support every unverified rumor. 
However, the latter is less prevalent to reputable user accounts (e.g., reputable news outlets) that usually share information with evidence.
Thomson et al.~\cite{thomson2012trusting} study Twitter's activity regarding the Fukushima Daiichi nuclear power plant disaster in Japan.
The authors undertake a categorization of the messages according to their user, location, language, type, and credibility of the source.
They observe that anonymous users, as well as users that live far away from the disaster share more information from less credible sources.
Finally, Dang et al.~\cite{dang2016toward} focus on the users that interact with rumors on Reddit by studying a popular false rumor (i.e., Obama is a Muslim). 
Specifically, they distinguish users into three main categories: the ones that support false rumors, the ones that refute false rumors and the ones that joke about a rumor.
To identify these users they built a Naive Bayes classifier that achieves an accuracy of 80\% and find that more than half of the users joked about this rumor, 25\% refuted the joke and only 5\% supported this rumor.

\descr{Conspiracy Theories.} Zollo et al.~\cite{zollo2015emotional} study the emotional dynamics around conversations regarding science and conspiracy theories. 
They do so by collecting posts from 280k users on Facebook pages that post either science or conspiracy theories posts. 
Subsequently, they use Support Vector Machines (SVMs) to identify the sentiment values of the posts, finding that sentiment is more negative on pages with conspiracy theories.
Furthermore, they report that as conversations grow larger, the overall negative sentiment in the comments increases.
In another work, Zollo et al.~\cite{zollo2015debunking} perform a quantitative analysis of 54M Facebook users. finding the existence of well-formed communities for the users that interact with science and conspiracy news. 
They note that users of each community interact within the community and rarely outside of it. 
Also, debunking posts are rather inefficient and user exposure to such content increases the overall interest in conspiracy theory posts.
Similarly, Bessi et al.~\cite{bessi2015science} study how conspiracy theories and news articles are consumed on Facebook, finding that polarized users contribute more in the diffusion of conspiracy theories, whereas this does not apply for news and their respective polarized users.

\subsection{Questionnaires/Interviews}
To get insights on how users perceive the various types of false information, some of the previous work conducted questionnaires or interviews. 
The majority of the work aims to understand how younger users (students or teenagers) interact and perceive false information.

\descr{Credibility Assessment.} Morris et al.~\cite{morris2012tweeting} highlight that users are influenced by several features related to the author of a tweet like their Twitter username when assessing the credibility of information. 
Winerburg and McGrew~\cite{wineburg2017lateral} study whether users with different backgrounds have differences in their credibility assessments. 
To achieve this they conducted experiments with historians, fact-checkers, and undergraduate students, finding that historians and students can easily get manipulated by official-looking logos and domain names.

\descr{Biased.} Marchi~\cite{marchi2012facebook} focus on how teenagers interact with news on Facebook by conducting interviews with 61 racially diverse teenagers.
The main findings of this study is that teenagers are not very interested in consuming news (despite the fact that their parents do) and that they demonstrate a preference to news that are opinionated when compared to objective news.
Similarly, Feldman~\cite{feldman2011partisan} focus on biased news and conduct 3 different studies with the participants randomly exposed to 2 biased and 1 non-biased news.
The participants were asked to provide information about the news that allowed the authors to understand the perceived bias.
They find that participants are capable of distinguishing bias in news articles; however, participants perceived lower bias in news that agree with their ideology/viewpoints.

\descr{Fabricated.} Chen et al.~\cite{chen2015why} use questionnaires on students from Singapore with the goal to unveil the reasons that users with no malicious intent share false information on OSNs. 
They highlight that female students are more prone in sharing false information, and that students are willing to share information of any credibility just to initiate conversations or because the content seems interesting.

\descr{Rumors.} Kim and Bock~\cite{kim2011study} study the rumor spreading behavior in OSNs from a psychological point of view by undertaking questionnaires on Korean students. 
They find that users' beliefs results in either positive or negative emotion for the rumor, which affects the attitude and behavior of the users towards the rumor spreading.

\descr{Satire.} Brewer et al.~\cite{brewer2013impact} indicate that satirical news programs can affect users' opinion and political trust, while at the same time users tend to have stronger opinion on matters that they have previously seen in satirical programs.

\subsection{Crowdsourcing platforms}
Other related work leverages crowdsourcing platform to get feedback from users about false information. We note that, to the best of our knowledge, previous work that used crowdsourcing platforms focused on rumors.
\descr{Rumors.} Ozturk et al.~\cite{ozturk2015combating} study how users perceive health-related rumors and if their are willing to share them on Twitter.
For acquiring the rumors, they crawl known health-related websites such as Discovery, Food Networks and National Institute of Health websites. 
To study the user perceptions regarding these rumors, they use AMT where they query 259 participants about ten handpicked health-related rumors.
The participants were asked whether they will share a specific rumor or a message that refutes a rumor or a rumor that had a warning on it (i.e., ``this message appeared in a rumor website''). 
Their results indicate that users are less likely to share a rumor that is accompanied with a warning or a message that refutes a rumor. 
Through simulations, they demonstrate that this approach can help in mitigating the spread of rumors on Twitter.
Finally, McCreadie et al.~\cite{mccreadie2015crowdsourced} use crowdsourcing on three Twitter datasets related to emergency situations during 2014, in order to record users' identification of rumors. 
Their results note that users were able to label most of the tweets correctly, while they note that tweets that contain controversial information are harder to distinguish.

\subsection{User Perception - Remarks}
The studies discussed in this section aim to shed light on how users \emph{perceive} false information on the Web.
Overall the main take-away points from the reviewed related work are:
1) teenagers are not interested in consuming news;
2) students share information of any credibility just to initate conversations;
3) in most cases, adults can identify bias in news and this task is harder when the news are biased towards the reader's ideology;
and 4) users can mostly identify rumors except the ones that contain controversial information.

\section{Propagation of False Information}
\label{sec:propagation}

Understanding the dynamics of false information is of paramount importance as it gives useful insights regarding the problem. 
Table~\ref{tbl:propagation_summary} summarizes the studies of false information propagation at OSNs, their methodology, as well as the corresponding type of false information according to the typology in Section~\ref{subsec:types}. 
The research community focuses on studying the propagation by either employing data analysis techniques or mathematical and statistical approaches. 
Furthermore, we note the efforts done on providing systems that visualize the propagation dynamics of false information.  
Below, we describe the studies that are mentioned in Table~\ref{tbl:propagation_summary} by dedicating a subsection for each type of methodology.

\begin{table}[]
\centering
\resizebox{\columnwidth}{!}{
\begin{tabular}{cccc}
\hline
\textbf{Platform} & \textbf{OSN data analysis}                                                                              & \textbf{Epidemic \& Statistical Modeling}                                              & \textbf{Systems} \\ \hline
Twitter           & \begin{tabular}[c]{@{}c@{}}Mendoza et al.~\cite{mendoza2010twitter}~\textbf{(R)},\\Oh et al.~\cite{oh2010exploration}~\textbf{(R)},\\
Andrews et al.~\cite{andrews2016keeping}~\textbf{(R)},\\
Gupta et al.~\cite{gupta20131}~\textbf{(F)},\\
Starbird et al.~\cite{starbird2014rumors}~\textbf{(R)},\\ 
Arif et al.~\cite{arif2016information}~\textbf{(R)},\\
Situngkir~\cite{situngkir2011spread}~\textbf{(H)},\\ 
Nadamoto et al.~\cite{nadamoto2013analysis}~\textbf{(R)},\\
Vosoughi et al.~\cite{vosoughi2018spread}~\textbf{(F)}\end{tabular}                                         & \begin{tabular}[c]{@{}c@{}}Jin et al.~\cite{jin2013epidemiological}~\textbf{(R)}, \\ Doerr et al.~\cite{doerr2012why}~\textbf{(R)},\\
Jin et al.~\cite{jin2014misinformation}~\textbf{(R)}\\\end{tabular}   
 & \begin{tabular}[c]{@{}c@{}}Finn et al.~\cite{finn2014investigating}~\textbf{(R)},\\ Shao el at.~\cite{shao2016hoaxy}~\textbf{(F)}\end{tabular}\\ \midrule
Facebook          & \begin{tabular}[c]{@{}c@{}}Friggeri et al.~\cite{friggeri2014rumor}~\textbf{(R)},\\ Del Vicario et al.~\cite{del2016spreading}~\textbf{(CT)},\\ Anagnostopoulos et al.~\cite{anagnostopoulos2014viral}~\textbf{(CT)}\end{tabular} & Bessi~\cite{bessi2017statistical}~\textbf{(CT)}                                                                                    & X                \\ \midrule
Other            & \begin{tabular}[c]{@{}c@{}}Ma and Li~\cite{ma2016rumor}~\textbf{(R)},\\
Zannettou et al.~\cite{zannettou2017web}~\textbf{(B)}                                                                                             \end{tabular}
 & \begin{tabular}[c]{@{}c@{}}Shah et al.~\cite{shah2011rumors}~\textbf{(R)}, \\ Seo et al.~\cite{seo2012identifying}~\textbf{(R)} ,\\ Wang et al.~\cite{wang2014rumor}~\textbf{(R)}\end{tabular} &    Dang et al.~\cite{dang2016what}~\textbf{(R)}             \\ \midrule
Sina Weibo        & X & Nguyen et al.~\cite{nguyen2012sources}~\textbf{(R)}                                                                                    & X                \\ \hline
\end{tabular}
}
\caption{Studies that focus on the propagation of false information on OSNs. The table summarizes the main methodology of each paper as well as the considered OSNs. Also, we report the type of false information that is considered (see bold markers and cf. with Section~\ref{subsec:types}
}
\label{tbl:propagation_summary}
\end{table}

\subsection{OSN Data Analysis}

\descr{Rumors.} Mendoza et al.~\cite{mendoza2010twitter} study the dissemination of false rumors and confirmed news on Twitter the days following the 2010 earthquake in Chile. 
They analyze the propagation of tweets for confirmed news and for rumors finding that the propagation of rumors differs from the confirmed news and that an aggregate analysis on the tweets can distinguish the rumors from the confirmed news.
Similarly, Starbird et al.~\cite{starbird2014rumors} study rumors regarding the 2013 Boston Bombings on Twitter and confirm both findings from Mendoza et al.~\cite{mendoza2010twitter}. %
In a similar notion, Nadamoto et al.~\cite{nadamoto2013analysis} analyze the behavior of the Twitter community during disasters (Great East Japan Earthquake in 2011) when compared to a normal time period; finding that the spread of rumors during a disaster situation is different from a normal situation. That is in disaster situations, the hierarchy of tweets is shallow whereas in normal situations the tweets follow a deep hierarchy. 

Others focused on understanding how rumors can be controlled and shed light on which types of accounts can help stop the rumor spread.
Oh et al.~\cite{oh2010exploration} study Twitter data about the 2010 Haiti Earthquake and find that credible sources contribute in rumor controlling, while Andrews et al.~\cite{andrews2016keeping} find that official accounts can contribute in stopping the rumor propagation by actively engaging in conversations related to the rumors.

Arif et al.~\cite{arif2016information} focus on the 2014 hostage crisis in Sydney. 
Their analysis include three main perspectives; (i) volume (i.e., number of rumor-related messages per time interval); (ii) exposure (i.e., number of individuals that were exposed to the rumor) and (iii) content production (i.e., if the content is written by the particular user or if it is a share).
Their results highlight all three  perspectives are important in understanding the dynamics of rumor propagation.
Friggeri et al.~\cite{friggeri2014rumor} use known rumors that are obtained through Snopes~\cite{snopes}, a popular site that covers rumors, to study the propagation of rumors on Facebook.
Their analysis indicates that rumors' popularity is bursty and that a lot of rumors change over time, thus creating rumor variants.
These variants aim to reach a higher popularity burst.
Also, they note that rumors re-shares which had a comment containing a link to Snopes had a higher probability to be deleted by their users. 

Finally, Ma and Li~\cite{ma2016rumor} study the rumor propagation process when considering a two-layer network; one layer is online (e.g., Twitter) and one layer is offline (e.g., face-to-face). 
Their simulations indicate that rumor spread is more prevalent in a two-layer network when compared with a single-layer offline network. 
The intuition is that in an offline network the spread is limited by the distance, whereas this constraint is eliminated in a two-layer network that has an online social network. 
Their evaluation indicates that in a two-layer network the spreading process on one layer does not affect the spreading process of the other layer; mainly because the interlayer transfer rate is less effective from an offline to an online network when compared with that from an OSN.

\descr{Fabricated.} Gupta et al.~\cite{gupta20131} study the propagation of false information on Twitter regarding the 2013 Boston Marathon Bombings. 
To do so, they collect 7.9M unique tweets by using keywords about the event.
Using real annotators, they annotate 6\% of the whole corpus that represents the 20 most popular tweets during this crisis situation (i.e., the 20 tweets that got retweeted most times).
Their analysis indicate that 29\% of the tweets were false and a large number of those tweets were disseminated by reputable accounts.
This finding contradicts with the findings of Oh et al.~\cite{oh2010exploration}, which showed that credible accounts help stop the spread of false information, hence highlighting that reputable accounts can share bad information too.
Furthermore, they note that out of the 32K accounts that were created during the crisis period, 19\% of them were deleted or suspended by Twitter, indicating that accounts were created for the whole purpose of disseminating false information.

Vosoughi et al.~\cite{vosoughi2018spread} study the diffusion of false and true stories in Twitter over the course of 11 years. 
They find that false stories propagate faster, farther, and more broadly when compared to true stories.
By comparing the types of false stories, they find that these effects were more intensive for political false stories when compared to other false stories (e.g., related to terrorism, science, urban legends, etc.).

\descr{Hoaxes.} Situngkir~\cite{situngkir2011spread} observe an empirical case in Indonesia to understand the spread of hoaxes on Twitter. 
Specifically, they focus on a case where a Twitter user with around 100 followers posted a question of whether a well-known individual is dead.
Interestingly, the hoax had a large population spread within 2 hours of the initial post and it could be much larger if a popular mainstream medium did not publicly deny the hoax.
Their findings indicate that a hoax can easily spread to the OSN if there is collaboration between the recipients of the hoax.
Again, this work highlights, similarly to Oh et al.~\cite{oh2010exploration} that reputable accounts can help in mitigating the spread of false information.

\descr{Conspiracy Theories.} Del Vicario et al.~\cite{del2016spreading} analyze the cascade dynamics of users on Facebook when they are exposed to conspiracy theories and scientific articles.
They analyze the content of 67 public pages on Facebook that disseminate conspiracy theories and science news.
Their analysis indicates the formulation of two polarized and homogeneous communities for each type of information.
Also, they note that despite the fact that both communities have similar content consumption patterns, they have different cascade dynamics.
Anagnostopoulos et al.~\cite{anagnostopoulos2014viral} study the role of homophily and polarization on the spread of false information by analyzing 1.2M Facebook users that interacted with science and conspiracy theories. 
Their findings indicate that user's interactions with the articles correlate with the interactions of their friends (homophily) and that frequent exposure to conspiracy theories (polarization) determines how viral the false information is in the OSN.

\descr{Biased.} Zannettou et al.~\cite{zannettou2017web}, motivated by the fact that the information ecosystem consists of multiple Web communities, study the propagation of news across multiple Web communities.
To achieve this, they study URLs from 99 mainstream and alternative news sources on three popular Web communities: Reddit, Twitter, and 4chan. 
Furthermore, they set out to measure the influence that each Web community has to each other, using a statistical model called Hawkes Processes.
Their findings indicate that small fringe communities within Reddit and 4chan have a substantial influence to mainstream OSNs like Twitter. 

\subsection{Epidemic and Statistical Modeling}
\descr{Rumors.} Jin et al.~\cite{jin2013epidemiological} use epidemiological models to characterize cascades of news and rumors in Twitter. Specifically, they use the SEIZ model~\cite{bettencourt2006power} which divides the user population in four different classes based on their status; 
(i) Susceptible; 
(ii) Exposed; 
(iii) Infected 
and (iv) Skeptic.
Their evaluation indicates that the SEIZ model is better than other models and it can be used to distinguish rumors from news in Twitter.
In their subsequent work, Jin et al.~\cite{jin2014misinformation} perform a quantitative analysis on Twitter during the Ebola crisis in 2014. By leveraging the SEIZ model, they show that rumors spread in Twitter the same way as legitimate news.

Doerr et al.~\cite{doerr2012why} use a mathematical approach to prove that rumors spread fast in OSNs (similar finding with Vosoughi et al.~\cite{vosoughi2018spread}).
For their simulations they used real networks that represent the Twitter and Orkut Social Networks topologies obtained from \cite{cha2010measuring} and SNAP~\cite{snap_datasets}, respectively. 
Intuitively, rumors spread fast because of the combinations of few large-degree nodes and a large number of small-degree nodes. 
That is, small-degree nodes learn a rumor once one of their adjacent nodes knows it, and then quickly forward the rumor to all adjacent nodes. 
Also, the propagation allows the diffusion of rumors between 2 large-degree nodes, thus the rapid spread of the rumor in the network.

Several related work focus on finding the source of the rumor.
Specifically, Shah et al.~\cite{shah2011rumors} focus on detecting the source of the rumor in a network by defining a new rumor spreading model and by forming the problem as a maximum likelihood estimation problem.
Furthermore, they introduce a new metric, called \textit{rumor centrality}, which essentially specifies the likelihood that a particular node is the source of the rumor. 
This metric is evaluated for all nodes in the network by using a simple linear time message-passing algorithm, hence the source of the rumor can be found by selecting the node with the highest rumor centrality.
In their evaluation, they used synthetic small-world and scale-free real networks to apply their rumor spreading model and they show that they can distinguish the source of a rumor with a maximum error of 7-hops for general networks, and with a maximum error of 4-hops for tree networks. 
Seo et al.~\cite{seo2012identifying} aim to tackle the same problem by injecting monitoring nodes on the social graph.
They propose an algorithm that considers the information received by the monitoring nodes to identify the source.
They indicate that with sufficient number of monitoring nodes they can recognize the source with high accuracy.
Wang et al.~\cite{wang2014rumor} aim to tackle the problem from a statistical point of view. 
They propose a new detection framework based on rumor centrality, which is able to support multiple snapshots of the network during the rumor spreading.
Their evaluation based on small-world and scale-free real networks note that by using two snapshots of the network, instead of one, can improve the source detection. 
Finally, Nguyen et al.~\cite{nguyen2012sources} aim to find the $k$ most suspected users where a rumor originates by proposing the use of a reverse diffusion process in conjunction with a ranking process.

\descr{Conspiracy Theories.} Bessi~\cite{bessi2017statistical} perform a statistical analysis of a large corpus (354k posts) of conspiracy theories obtained from Facebook pages. 
Their analysis is based on the Extreme Value Theory branch of statistics~\cite{extreme_value_theory} and they find that extremely viral posts (greater than 250k shares) follow a Poisson distribution. 

\subsection{Systems}

\descr{Rumors.} Finn et al.~\cite{finn2014investigating} propose a web-based tool, called TwitterTrails, which enables users to study the propagation of rumors in Twitter. 
TwitterTrails demonstrates indications for bursty activity, temporal characteristics of propagation, and visualizations of the re-tweet networks. 
Furthermore, it offers advanced metrics for rumors such as level of visibility and community's skepticism towards the rumor (based on the theory of h-index~\cite{h_index}).
Similarly, Dang et al.~\cite{dang2016what} propose RumourFlow, which visualizes rumors propagation by adopting modeling and visualization tools.
It encompasses various analytical tools like semantic analysis and similarity to assist the user in getting a holistic view of the rumor spreading and its various aspects. 
To demonstrate their system, they collect rumors from Snopes and conversations from Reddit.

\descr{Fabricated.} Shao et al.~\cite{shao2016hoaxy} propose Hoaxy, a platform that provides information about the dynamics of false information propagation on Twitter as well as the respective fact checking efforts.

\subsection{Propagation of False Information - Remarks}
In this section, we provided an overview of the existing work that focuses on the propagation of false information on the Web. 
Some of the main take-aways from the literature review on the propagation of false information are:
1) Accounts on social networks are created with the sole purpose of disseminating false information;
2) False information is more persistent than corrections;
3) The popularity of false information follow a bursty activity;
4) Users on Web communities create polarized communities that disseminate false information;
5) Reputable or credible accounts are usually useful in stopping the spread of false information; however we need to pay particular attention as previous work (see Gupta et al.~\cite{gupta20131}) has showed that they also share false information;
6) Being able to detect the source of false information is a first step towards stopping the spread of false information on Web communities and several approaches exist that offer acceptable performance.

\section{Detection and Containment of False Information}
\label{sec:detection_containment}

\subsection{Detection of false information}
Detecting false information is not a straightforward task, as it appears in various forms, as discussed in Section~\ref{chapter:taxonomy}.
Table~\ref{tbl:detection_summary} summarizes the studies that aim to solve the false information detection problem, as well as their considered OSNs and their methodology.
Most studies try to solve the problem using handcrafted features and conventional machine learning techniques.
Recently, to avoid using handcrafted features, the research community used neural networks to solve the problem (i.e., Deep Learning techniques).
Furthermore, we report some systems that aim to inform users about detected false information. 
Finally, we also note a variety of techniques that are proposed for the detection and containment of false information, such as epidemiological models, multivariate Hawkes processes, and clustering.
Below, we provide more details about existing work grouped by methodology and the type of information, according to Table~\ref{tbl:detection_summary}.

\begin{table}[t]
\centering
\resizebox{\textwidth}{!}{
\begin{tabular}{@{}cccc@{}}
\toprule
\textbf{Platform}                                                          & \textbf{Machine Learning}                                                                                                                                                                                                                                              & \textbf{Systems}                                                                          & \textbf{Other models/algorithms}                                                                                                                       \\ \midrule
Twitter                                                                    & \begin{tabular}[c]{@{}c@{}}Castillo et al.~\cite{castillo2011information}~\textbf{(CA)},\\ Gupta and Kumaraguru~\cite{gupta2012credibility}~\textbf{(CA)},\\ Kwon et al.~\cite{kwon2013prominent}~\textbf{(R)},\\ Yang et al.~\cite{yang2015emerging}~\textbf{(R)},\\ Liu et al.~\cite{liu2015real}~\textbf{(R)},\\ Wu et al.~\cite{wu2017gleaning}~\textbf{(R)},\\ Gupta et al.~\cite{gupta2014tweetcred}~\textbf{(CA)}, \\ AlRubaian et al.~\cite{alrubaian2015multistage}~\textbf{(CA)},\\ Hamidian and Diab~\cite{hamidian2016rumor}~\textbf{(R)},\\ Giasemidis et al.~\cite{giasemidis2016determining}~\textbf{(R)} ,\\ Kwon et al.~\cite{kwon2017rumor}~\textbf{(R)},\\
Volkova et al.~\cite{volkova2017separating}~\textbf{(CA)}  \end{tabular} 
                    & \begin{tabular}[c]{@{}c@{}}Resnick et al.~\cite{resnick2014rumorlens}~\textbf{(R)},\\ Vosoughi et al.~\cite{vosoughi2015human}~\textbf{(R)},\\ Jaho et al.~\cite{jaho2014alethiometer}~\textbf{(CA)}\end{tabular} & \begin{tabular}[c]{@{}c@{}}Qazvinian et al.~\cite{qazvinian2011rumor}~\textbf{(R)} \\(rumor retrieval model),\\ Zhao el al.~\cite{zhao2015enquiring}~\textbf{(R)} \\(clustering),\\ Farajtabar et al.~\cite{farajtabar2017fake}~\textbf{(F)}\\(hawkes process),\\
Kumar and Geethakumari~\cite{kumar2014detecting}~\textbf{(F)}\\
(algorithm with psychological cues)\end{tabular}       \\ \midrule
Sina Weibo                                                                 & \begin{tabular}[c]{@{}c@{}}Yang et al.~\cite{yang2012automatic}~\textbf{(R)},\\ Wu et al.~\cite{wu2015false}~\textbf{(R)},\\ Liang et al.~\cite{liang2015rumor}~\textbf{(R)},\\ Zhang et al.~\cite{zhang2015automatic}~\textbf{(R)},\end{tabular}                                                                                                                                                                      & Zhou et al.~\cite{zhou2015real}~\textbf{(CA)}                                                                          & X                                                                                                                                                      \\ \midrule
\begin{tabular}[c]{@{}c@{}}Twitter and \\ Sina Weibo\end{tabular}          & \begin{tabular}[c]{@{}c@{}}Ma et al.~\cite{ma2015detect}~\textbf{(CA)}\\Ma et al.~\cite{ma2016detecting}~\textbf{(R)} \end{tabular}                                                                                                                                                                                                                                                                    & X                                                                                         & \begin{tabular}[c]{@{}c@{}}Jin et al.~\cite{jin2016news}~\textbf{(CA)}\\(graph optimization)\end{tabular}                                                                                                                          \\ \midrule
Facebook                                                                   & \begin{tabular}[c]{@{}c@{}}Tacchini et al.~\cite{tacchini2017some}~\textbf{(H)},\\ Conti et al.~\cite{conti2017s}~\textbf{(CT)}\end{tabular}                                                                                                                                                                                                                    & X                                                                                         & X                                                                                                                                                      \\ \midrule
\begin{tabular}[c]{@{}c@{}}Wikipedia and/or \\ other articles\end{tabular} & \begin{tabular}[c]{@{}c@{}}Qin et al.~\cite{qin2016spotting}~\textbf{(R)},\\ Rubin et al.~\cite{rubin2016fake}~\textbf{(S)},\\ Kumar et al.~\cite{kumar2016disinformation}~\textbf{(H)},\\ Chen et al.~\cite{chen2015misleading}~\textbf{(CL)},\\ Chakraborty et al.~\cite{chakraborty2016stop}~\textbf{(CL)},\\ Potthast et al.~\cite{potthast2016clickbait}~\textbf{(CL)},\\ Biyani et al.~\cite{Biyani2016}~\textbf{(CL)},\\Wang~\cite{wang2017liar}~\textbf{(F)},\\ Anand et al. ~\cite{anand2016we}~\textbf{(CL)}\end{tabular}                                                                 & X                                                                                         & \begin{tabular}[c]{@{}c@{}}Potthast et al.~\cite{potthast2017stylometric}~\textbf{(B)}\\(unmasking)\end{tabular}                                                                                                                            \\ \midrule
Other                                                                      & \begin{tabular}[c]{@{}c@{}}
Afroz et al.~\cite{afroz2012detecting}~\textbf{(H)},\\
Maigrot et al.~\cite{maigrot2016mediaeval}~\textbf{(H)},\\
Zannettou et al.~\cite{zannettou2018good}~\textbf{(CL)}\end{tabular}                                                                                                                                                                                                                                                         & Vukovic et al.~\cite{vukovic2009intelligent}~\textbf{(H)}                                                                                        & \begin{tabular}[c]{@{}c@{}}Jin et al.~\cite{jin2014news}~\textbf{(CA)}\\(hierarchical propagation model),\\ Chen et al.~\cite{chen2014email}~\textbf{(H)}\\(Levenshtein Distance)\end{tabular} \\ \bottomrule
\end{tabular}
}
\caption{Studies that focus on the detection of false information on OSNs. The table demonstrates the main methodology of each study, as well as the considered OSNs.  Also, we report the type of false information that is considered (see bold markers and cf. with Section~\ref{subsec:types}, \textbf{CA} corresponds to Credibility Assessment and refers to work that aim to assess the credibility of information).}
\label{tbl:detection_summary}
\end{table}

\subsubsection{Machine Learning}

\descr{Credibility Assessment.} 
Previous work leverage machine learning techniques to assess the credibility of information. 
Specifically, Castillo et al.~\cite{castillo2011information} analyze 2.5k trending topics from Twitter during 2010 to determine the credibility of information.
For labeling their data they utilize crowdsourcing tools, namely AMT, and propose the use of conventional machine learning techniques (SVM, Decision Trees, Decision Rules, and Bayes Networks) that take into account message-based, user-based, topic-based and propagation-based features. 
Gupta and Kumaraguru~\cite{gupta2012credibility} analyze tweets about fourteen high impact news events during 2011.
They propose the use of supervised machine learning techniques with a relevance feedback approach that aims to rank the tweets according to their credibility score.
AlRubaian et al.~\cite{alrubaian2015multistage} propose the use of a multi-stage credibility assessment platform that consists of a relative importance component, a classification component, and an opinion mining component.
The relative importance component requires human experts and its main objective is to rank the features according to their importance. 
The classification component is based on a Naive Bayes classifier, which is responsible for classifying tweets by taking the output of the relative importance component (ranked features), while the opinion mining component captures the sentiment of the users that interact with the tweets. 
The output of the three components is then combined to calculate an overall assessment.
Ma et al.~\cite{ma2015detect} observe that typically the features of messages in microblogs vary over time and propose the use of an SVM classifier that is able to consider the messages features in conjunction with how they vary over time. 
Their experimental evaluation, based on Twitter data provided by~\cite{castillo2011information} and on a  Sina Weibo dataset, indicate that the inclusion of the time-varying features increase the performance between 3\% and 10\%.

All of the aforementioned work propose the use of supervised machine learning techniques. 
In contrast, Gupta et al.~\cite{gupta2014tweetcred} propose a semi-supervised model that ranks tweets according to their credibility in real-time.
For training their model, they collect 10M tweets from six incidents during 2013, while they leverage CrowdFlower~\cite{crowdflower} to obtain groundtruth.
Their system also includes a browser extension that was used by approx. 1.1k users in a 3-month timespan, hence computing the credibility score of 5.4M tweets.
Their evaluation indicates that 99\% of the users were able to receive credibility scores under 6 seconds.
However, feedback from users for approx. 1.2k tweets indicate that 60\% of the users disagreed with the predicted score.

Volkova et al.~\cite{volkova2017separating} motivated by the performance gains of deep learning techniques, propose the use of neural networks to distinguish news into satire, hoaxes, clickbait, and propaganda news.
They collect 130k news posts from Twitter and propose the use of neural networks that use linguistic and network features.
Their findings indicate that Recurrent and Convolutional neural networks exhibit strong performance in distinguishing news in the aforementioned categories.

\descr{Rumors.} 
Kwon et al.~\cite{kwon2013prominent} propose the use of Decision Trees, Random Forest, and SVM for detecting rumors on Twitter.
Their models leverage temporal, linguistics, and structural features from tweets and can achieve precision and recall scores between 87\% and 92\%.
Yang et al.~\cite{yang2015emerging} propose the use of a hot topic detection mechanism that work in synergy with conventional machine learning techniques (Naive Bayes, Logistic Regression and Random Forest).
Liu et al.~\cite{liu2015real} demonstrate the feasibility of a real-time rumoring detection system on Twitter. 
To achieve real-time debunking of rumors, they propose the use of an SVM classifier that uses beliefs from the users in conjunction with traditional rumor features from~\cite{castillo2011information, yang2012automatic}.
Their evaluation demonstrates that for new rumors (5-400 tweets), the proposed classifier can outperform the models from~\cite{castillo2011information, yang2012automatic}. 
Furthermore, they compare their approach with human-based rumor debunking services (Snopes and Emergent), showing that they can debunk 75\% of the rumors earlier than the corresponding services.
Similarly, Kwon et al.~\cite{kwon2017rumor} study the rumor classification task with a particular focus on the temporal aspect of the problem, by studying the task over varying time windows on Twitter. 
By considering user, structural, linguistic, and temporal features, they highlight that depending on the time window, different characteristics are more important than others. 
For example, at early stages of the rumor propagation, temporal and structural are not available. 
To this end, they propose a rumor classification algorithm that achieves satisfactory accuracy both on short and long time windows.

Hamidian and Diab~\cite{hamidian2016rumor} propose a supevised model that is based on the Tweet Latent Vector (TLV), which is an 100-dimensional vector, proposed by the authors, that encapsulates the semantics behind a particular tweet. 
For the classification task, they use an SVM Tree Kernel model that achieves 97\% on two Twitter datasets. 
Giasemidis et al.~\cite{giasemidis2016determining} study 72 rumors in Twitter by identifying 80 features for classifying false and true rumors. 
These features include diffusion and temporal dynamics, linguistics, as well as user-related features. 
For classifying tweets, they use several machine learning techniques and conclude that Decision Trees achieve the best performance with an accuracy of 96\%.
Yang et al.~\cite{yang2012automatic} study the rumor detection problem in the Sina Weibo OSN. 
For the automatic classification task of the posts they use SVMs that take as input various features ranging from content-based to 
user- and location-based features.
Their evaluation shows that the classifier achieves an accuracy of approximately 78\%. 
Similarly to the aforementioned work, Wu et al.~\cite{wu2015false} try to tackle the rumor detection problem in the Sina Weibo OSN by leveraging SVMs.
Specifically, they propose an SVM classifier which is able to combine a
normal radial basis function, which captures high level semantic features, and a random walk graph kernel, which captures the similarities between propagation trees. 
These trees encompass various details such as temporal behavior, sentiment of re-posts, and user details.
Liang et al.~\cite{liang2015rumor} study the problem of rumor detection using machine learning solutions that take into account users' behavior in the Sina Weibo OSN. 
Specifically, they introduce 3 new features that are shown to provide up to 20\% improvement when compared with baselines. 
These features are: 
1) average number of followees per day;
2) average number of posts per day; 
and 3) number of possible microblog sources.
Zhang et al.~\cite{zhang2015automatic} propose various implicit features that can assist in the detection of rumors. 
Specifically, they evaluate an SVM classifier against the Sina Weibo dataset proposed in \cite{yang2012automatic} with the following features: 
1) content-based implicit features (sentiment polarity, opinion on comments and content popularity);
2) user-based implicit features (influence of user to network, opinion re-tweet influence, and match degree of messages) 
and 3) shallow message features that are proposed by the literature.
Their evaluation shows that the proposed sets of features can improve the precision and recall of the system by 7.1\% and 6.3\%, respectively.
Qin et al.~\cite{qin2016spotting} propose the use of a new set of features for detecting rumors that aim to increase the detection accuracy; namely novelty-based and pseudo-feedback features. 
The novelty-based features consider reliable news to find how similar is a particular rumor with reliable stories.
The pseudo-feedback features take into account information from historical confirmed rumors to find similarities. 
To evaluate their approach, they obtain messages from the Sina Weibo OSN and news articles from Xinhua News Agency~\cite{xinhuanet}. They compare an SVM classifier, which encompasses the aforementioned set of features and a set of other features (proposed by the literature), with the approaches proposed by~\cite{yang2012automatic, liu2015real}. 
Their findings indicate that their approach provides an improvement between 17\% and 20\% in terms of accuracy.
Similarly to~\cite{qin2016spotting}, Wu et al.~\cite{wu2017gleaning} propose a system that uses historical data about rumors for the detection task. 
Their system consists of a feature selection module, which categorizes and selects features, and a classifier.
For constructing their dataset they use Snopes and the Twitter API to retrieve relevant tweets, acquiring in total 10k tweets, which are manually verified by annotators.
In their evaluation, they compare their system with various baselines
finding that the proposed system offers enhanced performance in rumor detection with an increase of 12\%-24\% for precision, recall, and F1-score metrics.
Ma et al.~\cite{ma2016detecting} leverage Recurrent neural networks  to solve the problem of rumor detection in OSNs. 
Such techniques are able to learn hidden representations of the input without the need for hand-crafted features.
For evaluating their model, they construct two datasets; one from Twitter and one from Sina Weibo. 
For the labeling of their messages they use Snopes for Twitter and the official rumor-busting service of Sina Weibo's OSN.
Their evaluation shows an accuracy of 91\% on the Sina Weibo dataset and 88\% on the Twitter dataset.

\descr{Hoaxes.} Tacchini et al.~\cite{tacchini2017some} study hoaxes in Facebook and argue that they can accurately discern hoax from non-hoax posts by simply looking at the users that liked the posts. 
Specifically, they propose the use of Logistic Regression that classifies posts with features based on users' interactions. 
Their evaluation demonstrate that they can identify hoaxes with an accuracy of 99\%.
Kumar et al.~\cite{kumar2016disinformation} study the presence of hoaxes in Wikipedia articles by considering 20k hoax articles that are explicitly flagged by Wikipedia editors. 
They find that most hoaxes are detected quickly and have little impact, however, a small portion of these hoaxes have a significant life-span and are referenced a lot across the Web. 
By comparing the "successful" hoaxes with failed hoaxes and legitimate articles, the authors highlight that the successful hoaxes have notable differences in terms of structure and content.
To this end, they propose the use of a Random Forest classifier to distinguish if articles are hoaxes. 
Their evaluation reports that their approach achieves an accuracy of 92\% and that is able to outperform human judgments by a significant margin (20\%). 
Maigrot et al.~\cite{maigrot2016mediaeval} propose the use of a multi-modal hoax detection system that fuses the diverse modalities pertaining to a hoax.
Specifically, they take into consideration the text, the source, and the image of tweets.
They observe higher performance when using only the source or text modality instead of the combination of all modalities.

\descr{Conspiracy Theories.} Conti et al.~\cite{conti2017s} focus on identifying conspiracy theories in OSNs by considering only the structural features of the information cascade.
The rationale is that such features are difficult to be tampered by malicious users, which aim to avoid detection from classification systems.   
For their dataset they use data from~\cite{bessi2015science}, which consist of scientific articles and conspiracy theories. 
For classifying their Facebook data they propose conventional machine learning techniques and they find that it is hard to distinguish a conspiracy theory from a scientific article by only looking at their structural dynamics (F1 -score not exceeding 65\%). 

\descr{Satire.} Rubin et al.~\cite{rubin2016fake} propose the use of satirical cues for the detection of false information on news articles.
Specifically, they propose the use of five new set of features, namely absurdity, humor, grammar, negative affect, and punctuation.
Their evaluation shows that by using an SVM algorithm with the aforementioned set of features and others proposed by the literature, they can detect satirical news with 90\% precision and 84\% recall. 

\descr{Clickbait.} Several studies focus on the detection of clickbait on the Web using machine learning techniques. Specifically, Chen et al.~\cite{chen2015misleading} propose tackling the problem using SVMs and Naive Bayes. 
Also, Chakraborty et al.~\cite{chakraborty2016stop} propose the 
use of SVM and a browser add-on to offer a system to users for news articles. 
Potthast et al.~\cite{potthast2016clickbait} proposes the use of Random Forest for detecting clickbait tweets.
Moreover, Biyani et al.~\cite{Biyani2016} propose the use of Gradient Boosted Decision Trees for clickbait detection in news articles and show that the degree of informality in the content of the landing page can help in finding clickbait articles. 
Anand et al.~\cite{anand2016we} is the first work that suggests the use of deep learning techniques for mitigating the clickbait problem.
Specifically, they propose the use of Recurrent Neural Networks in conjunction with word2vec embeddings~\cite{mikolov2013distributed} for identifying clickbait news articles.
 Similarly, Zannettou et al.~\cite{zannettou2018good} use deep learning techniques to detect clickbaits on YouTube. Specifically, they  propose a semi-supervised model based on variational autoencoders (deep learning). Their evaluation indicates that they can detect clickbaits with satisfactory performance and that YouTube's recommendation engine does not consider clickbait videos in its recommendations. 

\descr{Fabricated.} Wang~\cite{wang2017liar} presents a dataset that consists of 12.8k manually annotated short statements obtained from PolitiFact.
They propose the use of Convolutional neural networks for fusing linguistic features with metadata (e.g., who is the author of the statement). 
Their evaluation demonstrates that the proposed model outperforms SVM and Logistic Regression algorithms.

\subsubsection{Systems}
\descr{Rumors.} Resnick et al.~\cite{resnick2014rumorlens} propose a system called RumorLens, which aims to discover rumors in a timely manner, provide insights regarding the rumor's validity, and visualize a rumor's propagation. 
To achieve the aforementioned, RumorLens leverages data mining techniques alongside with a visual analysis tool.
However, their system raises scalability issues as it highly depends on users' labor, which provide labeling of tweets that are subsequently used for classifying tweets related to a particular rumor.
Vosoughi et al.~\cite{vosoughi2015human} propose a human-machine collaborative system that aims to identify rumors by disposing irrelevant data and ranking the relevant data.
Their system consists of two components; the assertion detector and the hierarchical clustering module. 
The assertion detector is a classifier that uses semantic and syntactic features to find tweets that contain assertions. 
These tweets are then presented to the clustering module, which clusters the tweets according to the similarity of the assertions.
During their evaluation, the authors state that for a particular incident (Boston Marathon Bombings) from a dataset of 20M tweets, their system managed to discard 50\% of them using the assertion detector. 
Furthermore, the 10M relevant tweets are clustered somewhere between 100 and 1000 clusters, something that enables users to quickly search and find useful information easier.

\descr{Credibility Assessment.} Jaho et al.~\cite{jaho2014alethiometer} undertake a statistical analysis by crawling Twitter for 3 months and retrieve a dataset that includes 10M users.
They propose a system that is based on contributor-related features (e.g., reputation, influence of source, etc.), content features (e.g., popularity, authenticity, etc.) and context features (e.g., coherence, cross-checking, etc.). 
Their system combines all the features and outputs a single metric that corresponds to the truthfulness of the message.
Zhou et al.~\cite{zhou2015real} note that calculating credibility in the granularity of message is not scalable, therefore they propose the calculation of credibility score per event. 
To this end, they propose a system that is able to collect related data from Sina Weibo using keywords and detect the credibility of a particular event.
The credibility score is calculated by the combination of 3 sub-models; the user model, the propagation model, and the content model. 
Each one of the sub-models considers one aspect of the news and the overall score is calculated using weighted combination.
The system is trained on a dataset that contains 73 real news and 73 fake news from approximately 50k posts.
Their evaluation shows that the proposed system provides an accuracy close to 80\% and that credibility scores are calculated within 35 seconds.

\descr{Hoaxes.} Vukovic et al.~\cite{vukovic2009intelligent} focus on hoaxes and  propose the use of a detection system for email.
The proposed system consists of a feed-forward neural network and a self-organizing map (SOM) and it is trained on a corpus of 298 hoax and 1370 regular emails.
The system achieves an accuracy of 73\% with a ratio of false positives equal to 4.9\%.
Afroz et al.~\cite{afroz2012detecting} focus on detecting hoaxes by observing changes in writing style.
The intuition is that people use different linguistic features when they try to obfuscate or change information from users.
To detect hoaxes they propose the use of an SVM classifier that takes into account the following set of features: 
1) lexical features;
2) syntactic features;
3) content features 
and 4) lying detection features obtained from~\cite{burgoon2003detecting, hancock2007lying}.
Their evaluation on various datasets indicates that the proposed system can detect hoaxes with an accuracy of 96\%.

\subsubsection{Other models/algorithms}

\descr{Rumors.} Qazvinian et al.~\cite{qazvinian2011rumor} study the rumor detection problem on Twitter by retrieving tweets regarding rumors and leveraging manual inspectors to annotate it. 
Specifically, the annotators were asked whether tweets contained rumors or not and whether a user endorsed, debunked or was neutral about the rumors. 
The resulted dataset consists of approximately 10k annotated tweets and was analyzed to demonstrate the effectiveness of the following feature sets in identifying rumors: 1) content-based features; 2) network-based features and 3) Twitter-specific memes (hashtags and URLs). 
Furthermore, the paper proposes a rumor retrieval model that achieves 95\% precision. %
Zhao et al.~\cite{zhao2015enquiring} are motivated by the fact that identifying false factual claims in each individual message is intractable. 
To overcome this, they adapt the problem in finding whole clusters of messages that their topic is a disputed factual claim. 
To do so, they search within posts to find specific phrases that are used from users who want to seek more information or to express their skepticism. For example, some enquiry phrases are "Is this true?", "Really?" and "What?". 
Their approach uses statistical features of the clusters in order to rank them according to the likelihood of including a disputed claim.
Their evaluations on real Twitter data indicate that among the top 50 ranked clusters, 30\% of them are confirmed rumors.

\descr{Fabricated.} Farajtabar et al.~\cite{farajtabar2017fake} propose a framework for tackling false information that combines a multivariate Hawkes process and reinforcement learning. 
Their evaluation highlights that their model shows promising performance in identifying false information in real-time on Twitter.
Kumar and Geethakumari~\cite{kumar2014detecting} measure the diffusion of false information by exploiting cues obtained from cognitive psychology. 
Specifically, they consider the consistency of the message, the coherency of the message, the credibility of the source, and the general acceptability of the content of the message. 
These cues are fused to an algorithm that aims to detect the spread of false information as soon as possible.
Their analysis on Twitter reports that the proposed algorithm has a 90\% True positive rate and a False positive rate less than 10\%.

\descr{Credibility Assessment.} Jin et al.~\cite{jin2016news} aim to provide verification of news by considering conflicting viewpoints on Twitter and Sina Weibo. 
To achieve this, they propose the use of a topic model method that identifies conflicting viewpoints.
Subsequently they construct a credibility network with all the viewpoints and they formulate the problem as a graph optimization problem, which can be solved with an iterative approach.
They compare their approach with baselines proposed in~\cite{castillo2011information, kwon2013prominent}, showing that their solution performs better.
Jin et al.~\cite{jin2014news} propose a hierarchical propagation model to evaluate information credibility in microblogs by detecting events, sub-events, and messages.
This three-layer network assists in revealing vital information regarding information credibility. 
By forming the problem as a graph optimization problem, they propose an iterative algorithm, that boosts the accuracy by 6\% when compared to an SVM classifier that takes into account only features obtained from the event-level network only.

\descr{Biased.} Potthast et al.~\cite{potthast2017stylometric} study the writing style of hyperpartisan news (left-wing and right-wing) and mainstream news and how this style can be applied in hyperpartisan news detection.
Their dataset consists of 1.6k news articles from three right-wing, three left-wings, and three mainstream news sites. 
For annotating the dataset they used journalists from Buzzfeed, who rated each article according to its truthfulness.
By leveraging the Unmasking approach \cite{koppel2007measuring}, the paper demonstrates that right-wing and left-wing hyperpartisan news exhibit similar writing style that differentiates from the mainstream news.
To this end, they propose the use of Random Forest classifier that aims to distinguish hyperpartisanship.
Their evaluation indicates that their style-based classifier can distinguish hyperpartisan news with an accuracy of 75\%. 
However, when the same classifier is used to discern fake or real news, then the accuracy is 55\%. 

\descr{Hoaxes.} Chen et al.~\cite{chen2014email} propose an email hoax detection system by incorporating a text matching method using the Levenshtein distance measure. 
Specifically, their system maintains a database of hoaxes that is used to calculate the distance between a potential hoax email and the stored hoaxes.

\subsection{Containment of false information}

Several studies focus on containing the diffusion of false information. 
Our literature review reveals that the majority of previous work on containment of rumors, while we also find one that focus on Hoaxes (see Tambuscio et al.~\cite{tambuscio2015fact}). 
Below we provide a brief overview of the studies that try to contain the spread of false information, while ensuring that the solutions are scalable.

\descr{Rumors.}
Tripathy et al. ~\cite{tripathy2010study} propose a process, called "anti-rumor", which aims to mitigate the spreading of a rumor in a network. 
This process involves the dissemination of messages, which contradict with a rumor, from agents. 
The authors make the assumption that once a user receives an anti-rumor message, then he will never believe again the rumor, thus the spreading of a rumor is mitigated. 
Their evaluation, based on simulations, indicates the efficacy of the proposed approach.
Budak et al.~\cite{budak2011limiting} formulate the problem of false information spreading as an optimization problem.
Their aim is to identify a subset of the users that need to be convinced to spread legitimate messages in contrast with the bad ones that spread rumors.
The paper shows that this problem is NP-hard and they propose a greedy solution as well as some heuristics to cope with scalability issues. 
Fan et al.~\cite{fan2013least} try to tackle the problem of false information propagation under the assumption that rumors originate from a particular community in the network. 
Similarly to other work, the paper tries to find a minimum set of individuals, which are neighbors with the rumor community to stop the rumor diffusion in the rest of the network. 
To achieve this, they propose the use of two greedy-based algorithms, which are evaluated in two real-world networks (Arxiv Hep and Enron). 
Their experimental results show that the proposed algorithms outperform simple heuristics in terms of the number of infected nodes in the network. 
However, as noted, the greedy algorithms are time consuming and are not applicable in large-scale networks.
Kotnis et al.~\cite{kotnis2014cost} propose a solution for stopping the spread of false information by training a set of individuals in a network that aim to distinguish and stop the propagation of rumors. 
This set of individuals is selected based on their degree in the network with the goal to minimize the overarching training costs. 
For evaluating their solution they create a synthetic network, which takes into account a calculated network degree distribution, based on ~\cite{molloy1995critical}.
Ping et al.~\cite{ping2014sybil} leverage Twitter data to demonstrate that sybils presence in OSNs can decrease the effectiveness of community-based rumor blocking approaches by 30\%. 
To this end, they propose a Sybil-aware rumor blocking approach, which finds a subset of nodes to block by considering the network structure in conjunction with the probabilities of nodes being sybils. 
Their evaluation, via simulations on Twitter data, show that the proposed approach significantly decreases the number of affected nodes, when compared to existing approaches.
He et al.~\cite{he2015modeling} argue that existing false information containment approaches have different costs and efficiencies in different OSNs.
To this end, they propose an optimization method that combines the spreading of anti-rumors and the block of rumors from influential users.
The goal of their approach is to minimize the overarching cost of the method while containing the rumor within an expected deadline.
To achieve this, they use the Pontryagin's maximum principle~\cite{kopp1962pontryagin} on the Digg2009 dataset \cite{hogg2012social}. 
They find that spreading the truth plays a significant role at the start of the rumor propagation, whereas close to the deadline of containment the blocking of rumors approach should be used extensively.
Huang et al.~\cite{huang2015connected} aim to contain the false information spread by finding and decontaminating with good information,
the smallest set of influential users in a network. 
To do so, they propose a greedy algorithm and a community-based heuristic, which takes into consideration the community structure of the underlying network. 
For evaluating their approach, they used traces from three networks; NetHEPT, NetHEPT\_WC and Facebook.
Previous studies on false information containment~\cite{budak2011limiting, nguyen2012containment} assumed that when true and false information arrive the same time at a particular node, then the true information dominates.
Wang et al.~\cite{wang2014containment} state that the dominance of the information should be based on the influence of the neighbors in the network. 
With this problem formulation in mind, the paper proposes two approaches to find the smallest number of nodes that are required to stop the false information spread. 
Their evaluation is based on three networks obtained from Twitter, Friendster, and a random synthetic network. 
Evaluation comparisons with simple heuristics (random and high degree) demonstrate the performance benefits of the proposed approaches.
In a similar notion, Tong et al.~\cite{tong2017efficient} aim to increase performance motivated by the fact that greedy solutions, which include Monte Carlo simulations, are inefficient as they are computationally intensive. 
To overcome this, the paper proposes a random-based approach, which utilizes sampling with the aim to be both effective and efficient. 
The performance evaluations on real-world (obtained from Wikipedia and Epinions~\cite{epinions}) and synthetic networks demonstrate that the proposed solution can provide a 10x speed-up without compromising performance when compared to state-of-the-art approaches.  
Wang et al.~\cite{wang2016drimux} propose a model, called DRIMUX, which aims to minimize the influence of rumors by blocking a subset of nodes while considering users' experience.
User experience is defined as a time threshold that a particular node is willing to wait while being blocked.
Their model utilizes survival theory and takes into account global rumor popularity features, individual tendencies (how likely is a rumor to propagate between a pair of nodes) as well as the users' experience.
Their evaluations on a Sina Weibo network, which consists of 23k nodes and 183k edges, indicate that the proposed model can reduce the overarching influence of false information.

\descr{Hoaxes.} Tambuscio et al.~\cite{tambuscio2015fact} simulate the spread and debunking of hoaxes on networks. 
Specifically, they model the problem as a competition between believers (acknowledge the hoax) and fact checkers which reveal the hoax with a specific probability. 
To study their model they performed simulations on scale-free and random networks finding that a specific threshold for the probability of fact checkers exists and this indicates that the spread can be stopped with a specific number of fact checkers. 
However, the paper oversimplifies the problem by assuming all the nodes to have the same probability.

\subsection{Detection and Containment of False Information - Remarks}

The main findings from the literature review of the detection and containment of false information are:
1) Machine learning techniques can assist in identifying false information. However, they heavily rely on handcrafted set of features and it is unclear if they generalize well on other datasets;
2) Containment of false information can be achieved by adding a set of good nodes that disseminate good information or information that refute false; and
3) The problem of detection of false information requires human-machine collaboration for effectively mitigating it.

\section{False Information in the political stage}
\label{sec:political}

Recently, after the 2016 US elections, the problem of false information dissemination got extensive interest from the community. Specifically, Facebook got openly accused for disseminating false information and that affected the outcome of the elections~\cite{facebook_fake_news_us_elections}. 
It is evident that dissemination of false information on the Web is used a lot for political influence.
Therefore in this section we review the most relevant studies on the political stage. Table~\ref{tbl:political_overview} reports the reviewed work as well as the main methodology and considered OSN.

\subsection{Machine Learning}

\descr{Propaganda.} Ratkiewicz et al.~\cite{ratkiewicz2011detecting} study political campaigns on Twitter that use multiple controlled accounts to disseminate support for an individual or opinion. 
They propose the use of a machine learning-based framework in order to detect the early stages of the spreading of political false information on Twitter.
Specifically, they propose a framework that takes into consideration topological, content-based and crowdsourced features of the information diffusion in Twitter.
Their experimental evaluation demonstrates that the proposed framework achieves more than 96\% accuracy in the detection of political campaigns for data pertaining to the 2010 US midterm elections.
Conover et al.~\cite{conover2011political} study Twitter on a six-week period leading to the 2010 US midterm elections and the interactions between right and left leaning communities. 
They leverage clustering algorithms and manually annotated data to create the re-tweets and mentions networks.
Their findings indicate that the re-tweet network has limited connectivity between the right and left leaning communities, whereas this is not the case in the mentions networks.
This is because, users try to inject different opinions on users with different ideologies, by using mentions on tweets, so that they change their stance towards a political individual or situation.
Ferrara et al.~\cite{ferrara2016detection} propose the use of a k-nearest neighbor algorithm with a dynamic warping classifier in order to capture promoted campaigns in Twitter. 
By extracting a variety of features (user-related, timing-related, content-related and sentiment-related features) from a large corpus of tweets they demonstrate that they can distinguish promoted campaigns with an AUC score close to 95\% in a timely manner.
 
 \begin{table}[]
\centering
\resizebox{\columnwidth}{!}{
\begin{tabular}{@{}cccc@{}}
\toprule
\textbf{Platform} & \textbf{Machine Learning}                                                                                                                                                               & \textbf{OSN Data Analysis}                                                                                                                                                                                                                                                                                                                                                                   & \textbf{Other models/algorithms}                                                                                                                                                                                                                                                                                                                                                                                        \\ \midrule
Twitter           & \begin{tabular}[c]{@{}c@{}} Ratkiewicz et al.~\cite{ratkiewicz2011detecting}~\textbf{(P)},\\ Conover et al.\cite{conover2011political}~\textbf{(P)},\\ Ferrara et al.\cite{ferrara2016detection}~\textbf{(P)}\end{tabular} & \begin{tabular}[c]{@{}c@{}} Wong et al.~\cite{wong2013quantifying}~\textbf{(B)},\\ Golbeck and Hansen~\cite{golbeck2014twitterpolpref}~\textbf{(B)},\\ Jackson and Welles~\cite{jackson2015hijacking}~\textbf{(P)},\\ Hegelich and Janetzko\cite{hegelich2016social}~\textbf{(P)},\\Zannettou et al.~\cite{zannettou2018disinformation}~\textbf{(P)}\\ Howard and Kollanyi\cite{howard2016botsa}~\textbf{(P)},\\ Shin et al.\cite{shin2016political}~\textbf{(R)}\end{tabular} & \begin{tabular}[c]{@{}c@{}}An et al.~\cite{an2012visualizing}~\textbf{(B)} \\(distance model),\\ Al-khateeb and Agarwal~\cite{al2015examining}~\textbf{(P)} \\(social studies)\\ Ranganath et al.\cite{ranganath2016understanding}~\textbf{(P)} \\(exhaustive search),\\ Jin et al.~\cite{jin2017rumor}~\textbf{(R)} \\(text similarity),\\ Yang et al.~\cite{yang2016social}~\textbf{(B)} \\(agenda-setting tool)\end{tabular} \\ \midrule
Digg              & Zhou et al.\cite{zhou2011classifying}~\textbf{(B)}                                                                                                                                                 & X                                                                                                                                                                                                                                                                                                                                                                                            & X                                                                                                                                                                                                                                                                                                                                                                                                                       \\ \midrule
Sina Weibo        & X                                                                                                                                                                                       & \begin{tabular}[c]{@{}c@{}}King et al.~\cite{king2016chinese}~\textbf{(P)},\\ Yang et al.~\cite{yang2015penny}~\textbf{(P)}\end{tabular}                                                                                                                                                                                                                                                                             & X                                                                                                                                                                                                                                                                                                                                                                                                                       \\ \midrule
News articles     & Budak et al.~\cite{budak2016fair}~\textbf{(B)}                                                                                                                                                      & Woolley\cite{woolley2016automating}~\textbf{(P)}                                                                                                                                                                                                                                                                                                                                                        & X                                                                                                                                                                                                                                                                                                                                                                                                                       \\ \midrule
Facebook          & X                                                                                                                                                                                       & Allcot and Gentzkow\cite{allcott2017social}~\textbf{(P)}                                                                                                                                                                                                                                                                                                                                                & X                                                                                                                                                                                                                                                                                                                                                                                                                       \\ \bottomrule
\end{tabular}
}
\caption{Studies on the false information ecosystem on the political stage. The table demonstrates the main methodology of each study as well as the considered OSNs.}
\label{tbl:political_overview}
\end{table}

\descr{Biased.}
Zhou et al.~\cite{zhou2011classifying} study Digg, a news aggregator site, and aim to classify users and articles to either liberal or conservative.
To achieve this, they propose three semi-supervised propagation algorithms that classify users and articles based on users' votes.
The algorithms make use of a few labeled users and articles to predict a large corpus of unlabeled users and articles.
The algorithms are based on the assumption that a liberal user is more likely to vote for a liberal article rather than a conservative article. 
Their evaluations demonstrate that the best algorithm achieves 99\% and 96\% accuracy on the dataset of users and articles, respectively.
Budak et al.~\cite{budak2016fair} use Logistic Regression to identify articles regarding politics from a large corpus of 803K articles obtained from 15 major US news outlets.
Their algorithm filtered out 86\% of the articles as non-political related, while a small subset of the remainder (approx. 11\%) were presented to workers on AMT. 
The workers were asked to answer questions regarding the topic of the article, whether the article was descriptive or opinionated, the level of partisanship, and the level of bias towards democrats or republicans.
Their empirical findings are that on these articles there are no clear indications of partisanship, some articles within the same outlet are left-leaning and some have right-leaning, hence reducing the overall outlet bias. 
Also, they note that usually bias in news articles is expressed by criticizing the opposed party rather than promoting the supporting party.

\subsection{OSN Data Analysis}
\descr{Biased.}
Wong et al.~\cite{wong2013quantifying} collect and analyze 119M tweets pertaining to the 2012 US presidential election to quantify political leaning of users and news outlets. 
By formulating the problem as an ill-posed linear inverse problem, they propose an inference engine that considers tweeting behavior of articles.
Having demonstrated their inference engine, the authors report results for the political leaning scores of news sources and users on Twitter.
Golbeck and Hansen~\cite{golbeck2014twitterpolpref} provide a technique to estimate audience preferences in a given domain on Twitter, with a particular focus on political preferences.
Different from methods that assess audience preference based on citation networks of news sources as a proxy, they directly measure the audience itself via their social network.
Their technique is composed of three steps: 1)~apply ground truth scores (they used Americans for Democratic Action reports as well as DW-Nominate scores) to a set of seed nodes in the network, 2)~map these scores to the seed group's followers to create ``P-scores'', and 3)~map the P-scores to the target of interest (e.g., government agencies or think tanks).
One important take away from this work is that \emph{Republicans are over-represented on Twitter with respect to their representation in Congress}, at least during the 2012 election cycle.
To deal with this, they built a balanced dataset by randomly sampling from bins formed by the number of followers a seed group account had.

\descr{Propaganda.} Jackson and Welles~\cite{jackson2015hijacking} demonstrate how Twitter can be exploited to organize and promote counter narratives.
To do so, they investigate the misuse of a Twitter hashtag (\#myNYPD) during the 2014 New York City Police Department public relations campaign.
In this campaign, this hashtag was greatly disseminated to promote counter narratives about racism and police misconduct.
The authors leverage network and qualitative discourse analysis to study the structure and strategies used for promoting counterpublic narratives.

Hegelich and Janetzko~\cite{hegelich2016social} investigate whether bots on Twitter are used as political actors.
By exposing and analyzing 1.7K bots on Twitter, during the Russian/Ukrainian conflict, they find that the botnet has a political agenda and that bots exhibit various behaviors.
Specifically, they find that bots try to hide their identity, to be interesting by promoting topics through the use of hashtags and retweets.
Howard and Kollanyi~\cite{howard2016botsa} focus on the 2016 UK referendum and the role of bots in the conversations on Twitter.
They analyze 1.5M tweets from 313K Twitter accounts collected by searching specific hashtags related to the referendum.
Their analysis indicates that most of the tweets are in favor of exiting the EU, there are bots with different levels of automation and that 1\% of the accounts generate 33\% of the overall messages.
They also note that among the top sharers, there are a lot of bot accounts that are mostly retweeting and not generating new content. 
In a similar work, Howard et al.~\cite{howard2016botsb} study Twitter behavior during the second 2016 US Presidential Debate. They find that Twitter activity is more pro-Trump and that a lot of activity is driven by bots.
However, they note that a substantial amount of tweets is original content posted from regular Twitter users.
Woolley~\cite{woolley2016automating} analyzes several articles regarding the use of bots in OSNs for political purposes.
Specifically, he undertakes a qualitative content analysis on 41 articles regarding political bots from various countries obtained from the Web. %
One of his main findings is that the use of bots varies from country to country and that some countries (e.g., Argentina, China, Russia, USA, etc.) use political bots on more than one type of event.
For example, they report the use of Chinese political bots for elections, for protests and for security reasons.

In the Chinese political stage, during December 2014, an anonymous blogger released an archive of emails pertaining to the employment of Wumao, a group of people that gets paid to disseminate propaganda on social media, from the Chinese government.
King et al.~\cite{king2016chinese} analyzed these leaks and found out 43K posts that were posted by Wumao. 
Their main findings are:
1) by analyzing the time-series of these posts, they find bursty activity, hence signs of coordination of the posters;
2) most of the posters are individuals working for the government; and 
3) by analyzing the content of the message, they note that posters usually post messages for distraction rather than discussions of controversial matters (i.e., supporting China's regime instead of discussing an event).
Similarly to the previous work, Yang et al.~\cite{yang2015penny} study the Wumao by analyzing 26M posts from 2.7M users on the Sina Weibo OSN, aiming to provide insights regarding the behavior and the size of Wumao.
Due to the lack of ground truth data, they use clustering and topic modeling techniques, in order to cluster users that post politics-related messages with similar topics.
By manually checking the users on the produced clusters, they conclude that users that post pro-government messages are distributed across multiple clusters, hence there is no signs of coordination of the Wumao on Sina Weibo for the period of their dataset (August 2012 and August 2013).

Zannettou et al.~\cite{zannettou2018disinformation} study Russian state-sponsored troll accounts and measure the influence they had on Twitter and other Web communities. They find that Russian trolls were involved in the discussion of political events, and that they exhibit different behavior when compared to random users.
Finally, they show that their influence was not substantial, with the exception of the dissemination of articles from state-sponsored Russian news outlets like Russia Today (RT).
Allcot and Gentzkow~\cite{allcott2017social} make a large scale analysis on Facebook during the period of the 2016 US election.
Their results provide the following interesting statistics about the US election:
1) 115 pro-Trump fake stories are shared 30M times, whereas 41 pro-Clinton fake stories are shared 7.6M times. This indicates that fake news stories that favor Trump are more profound in Facebook.
2) The aforementioned 37.6M shares translates to 760M instances of a user clicking to the news articles. This indicates the high reachability of the fake news stories to end-users. 
3) By undertaking a 1200-person survey, they highlight that a user's education, age and overall media consumption are the most important factors that determine whether a user can distinguish false headlines.

\descr{Rumors.} Shin et al.~\cite{shin2016political} undertake a content-based analysis on 330K tweets pertaining to the 2012 US election.
Their findings agree with existing literature, noting that users that spread rumors are mostly sharing messages against a political person.
Furthermore, they highlight the resilience of rumors despite the fact that rumor debunking evidence was disseminated in Twitter;  however, this does not apply for rumors that originate from satire websites.

\subsection{Other models/algorithms}

\descr{Biased.}
An et al.~\cite{an2012visualizing} study the interactions of 7M followers of 24 US news outlets on Twitter, in order to identify political leaning.
To achieve this, they create a distance model, based on co-subscription relationships, that maps news sources to a dimensional dichotomous political spectrum.
Also, they propose a real-time application, which utilizes the underlying model, and visualizes the ideology of the various news sources.
Yang et al.~\cite{yang2016social} investigate the topics of discussions on Twitter for 51 US political persons, including President Obama.
The main finding of this work is that Republicans and Democrats are similarly active on Twitter with the difference that Democrats tend to use hashtags more frequently.
Furthermore, by utilizing a graph that demonstrates the similarity of the agenda of each political person, they highlight that Republicans are more clustered.
This indicates that Republicans tend to share more tweets regarding their party's issues and agenda.

\descr{Propaganda.}
Al-khateeb and Agarwal~\cite{al2015examining} study the dissemination of propaganda on Twitter from terrorist organizations ( namely ISIS).
They propose a framework based on social studies that aim to identify social and behavioral patterns of propaganda messages disseminated by a botnet. 
Their main findings are that bots exhibit similar behaviors (i.e., similar sharing patterns, similar usernames, lot of tweets in a short period of time) and that they share information that contains URLs to other sites and blogs.
Ranganath et al.~\cite{ranganath2016understanding} focus on the detection of political advocates (individuals that use social media to strategically push a political agenda) on Twitter. 
The authors note that identifying advocates is not a straightforward task due to the nuanced and diverse message construction and propagation strategies.
To overcome this, they propose a framework that aims to model all the different propagation and message construction strategies of advocates.
Their evaluation on two datasets on Twitter regarding gun rights and elections demonstrate that the proposed framework achieves good performance with a 93\% AUC score.

\descr{Rumors.} Jin et al.~\cite{jin2017rumor} study the 2016 US Election through the Twitter activity of the followers of the two presidential candidates.
For identifying rumors, they collect rumor articles from Snopes and then they use text similarity algorithms based on:
1) Term frequency-inverse document frequency (TF-IDF);
2) BM25 proposed in~\cite{robertson2009probabilistic}
3) Word2Vec embeddings~\cite{mikolov2013distributed};
4) Doc2Vec embeddings~\cite{le2014distributed};
5) Lexicon used in ~\cite{zhao2015enquiring}.
Their evaluation indicates that the best performance is achieved using the BM25-based approach.
This algorithm is subsequently used to classify the tweets of the candidates' followers.
Based on the predictions of the algorithm, their main findings are:
1) rumors are more prevalent during election period;
2) most of the rumors are posted by a small group of users;
3) rumors are mainly posted to debunk rumors that are against their presidential candidate, or to inflict damage on the other candidate; and
4) rumor sharing behavior increases in key points of the presidential campaign and in emergency events.

\subsection{False information in political stage - Remarks}
The main insights from the review of work that focus on the political stage are:
1) Temporal analysis can by leveraged to assess coordination of bots, state-sponsored actors, and orchestrated efforts on disseminating political false information;
2) Bots are extensively used for the dissemination of political false information;
3) Machine learning techniques can assist in detecting political false information and political leaning of users. However, there are concerns about the generalization of such solutions on other datasets/domains; and
4) Political campaigns are responsible for the substantial dissemination of political false information in mainstream Web communities.

\section{Other related work}
\label{sec:other}

In this section we shall present work that is relevant to the false information ecosystem that does not fit in the aforementioned lines of work.
Specifically, we group these studies in the following categories:
1) General Studies;
2) Systems; and
3) Use of images on the false information ecosystem.

\subsection{General Studies}

\descr{Credibility Assessment.} Buntain and Golbeck~\cite{buntain2017want} compare the accuracy of models that use features based on journalists assessments and crowdsourced assessments.
They indicate that there is small overlap between the two features sets despite the fact that they provide statistically correlated results. 
This indicates that crowdsourcing workers discern different aspects of the stories when compared to journalists.
Finally, they demonstrate that models that utilize features from crowdsourcing outperform the models that utilize features from journalists.
Zhang et al.~\cite{zhang2018structured} present a set of indicators that can used to assess the credibility of articles. 
To find these indicators they use a diverse set of experts (coming from multiple disciplines), which analyzed and annotated 40 news articles.
Despite the low number of annotated articles, this inter-disciplinary study is important as it can help in defining standards for assessing the credibility of content on the Web.
Mangolin et al.~\cite{margolin2018political} study the interplay between fact-checkers and rumor spreaders on social networks finding that users are more likely to correct themselves if the correction comes from a user they follow when compared to a stranger.

\descr{Conspiracy Theories.} Starbird~\cite{starbird2017examining} performs a qualitative analysis on Twitter regarding shooting events and conspiracy theories.
Using graph analysis on the domains linked from the tweets, she provides insight on how various websites work to promote conspiracy theories and push political agendas.

\descr{Fabricated.} Horne and Adah~\cite{horne2017just} focus on the headline of fake and real news. 
Their analysis on three datasets of news articles highlight that fake news have substantial differences in their structure when compared with real news. 
Specifically, they report that generally the structure of the content and the headline is different.
That is, fake news are smaller in size, use simple words, and use longer and ``clickbaity'' headlines. 
Potts et al.~\cite{potts2013interfaces} study Reddit and 4chan and how their interface is a part of their culture that affects their information sharing behavior.
They analyze the information shared on these two platforms during the 2013 Boston Marathon bombings. 
Their findings highlight that users on both sites tried to find the perpetrator of the attack by creating conversations for the attack, usually containing false information.
Bode and Vraga~\cite{bode2015related} propose a new function on Facebook, which allow users to observe related stories that either confirm or correct false information; they highlight that using this function users acquire a better understanding of the information and its credibility.
Finally, Pennycook and Rand~\cite{pennycook2017implied} highlight that by attaching warnings to news articles can help users to better assess the credibility of articles, however news articles that are not attached with warnings are considered as validated, which is not always true, hence users are tricked.

\descr{Propaganda.} Chen et al.~\cite{chen2013battling} study the behavior of hidden paid posters on OSNs. 
To better understand how these actors work, an author of this work posed as a hidden paid poster for a site\cite{shuijunwang} that gives users the option to be hidden paid posters.
This task revealed valuable information regarding the organization of such sites and the behavior of the hidden paid posters, who are assigned with missions that need to be accomplished within a deadline. 
For example, a mission can be about posting articles of a particular content on different sites.
A manager of the site can verify the completion of the task and then the hidden paid poster gets paid.
To further study the problem, they collect data ,pertaining to a dispute between two big Chinese IT companies, from users of 2 popular Chinese news sites (namely Sohu \cite{sohu} and Sina \cite{sina}). 
During this conflict there were strong suspicions that both companies employed hidden paid posters to disseminate false information that aimed to inflict damage to the other company.  
By undertaking statistical and semantic analysis on the hidden paid posters' content they uncover a lot of useful features that can be used in identifying hidden paid posters.
To this end, they propose the use of SVMs in order to detect such users by taking into consideration statistical and semantic features; their evaluation show that they can detect users with 88\% accuracy.

\descr{Rumors.} Starbird et al.~\cite{starbird2016could} study and identify various types of expressed uncertainty within posts in OSN during a rumor's lifetime.
To analyze the uncertainty degree in messages, the paper acquires 15M tweets related to two crisis incidents (Boston Bombings and Sydney Siege). 
They find that specific linguistic patterns are used in rumor-related tweets.
Their findings can be used in future detection systems in order to detect rumors effectively in a timely manner.
Zubiaga et al.~\cite{zubiaga2015towards} propose a different approach in collecting and preparing datasets for false information detection. 
Instead of finding rumors from busting websites and then retrieving data from OSNs, they propose the retrieval of OSN data that will subsequently annotated by humans.
In their evaluation, they retrieve tweets pertaining to the Ferguson unrest incident during 2014.
They utilize journalists that act as annotators with the aim to label the tweets and their conversations. Specifically, the journalists annotated 1.1k tweets, which can be categorized into 42 different stories. Their findings show that 24.6\% of the tweets are rumorous.
FInally, Spiro et al.~\cite{spiro2012rumoring} undertake a quantitative analysis on tweets pertaining to the 2010 Deepwater Horizon oil spill. 
They note that media coverage increased the number of tweets related to the disaster. 
Furthermore, they observe that retweets are more commonly transmitted serially when they have event-related keywords.

\subsection{Systems}
\descr{Biased.} Park et al.~\cite{park2009newscube} note that biased information is profoundly disseminated in OSNs.
To alleviate this problem, they propose NewsCube: a service that aims to provide end-users with all the different aspects of a particular story. 
In this way, end-users can read and understand the stories from multiple perspectives hence assisting in the formulation of their own unbiased view for the story.
To achieve this, they perform structure-based extraction of the different aspects that exist in news stories. 
These aspects are then clustered in order to be presented to the end-users.
To evaluate the effectiveness of their system, they undertake several user studies that aim to demonstrate the effectiveness in terms of the ability of the users to construct balanced views when using the platform. 
Their results indicate that 16 out of 33 participants stated that the platform helped them formulate a balanced view of the story, 2 out of 33 were negative, whereas the rest were neutral.

\descr{Credibility Assessment.} Hassan et al.~\cite{hassan2014data} propose FactWatcher, a system that reports facts that can be used as leads in stories. 
Their system is heavily based on a database and offers useful features to it's users such as ranking of the facts, keyword-based search and fact-to-statement translation.
Ennals et al.~\cite{ennals2010highlighting} describe the design and implementation of Dispute Finder, which is a browser extension that allows users to be warned about claims that are disputed by sources that they might trust.
Dispute Finder maintains a database with well-known disputed claims which are used to inform end-users in real-time while they are reading stories. 
Users are also able to contribute to the whole process by explicitly flagging content as disputed, or as evidence to dispute other claims.
In the case of providing evidence, the system requires a reference to a trusted source that supports the user's actions, thus ensuring the quality of user's manual annotations.
Mitra and Gilbert~\cite{mitra2015credbank} propose CREDBANK that aims to process large datasets by combining machine and human computations. 
The former is used to summarize tweets in events, while the latter is responsible for assessing the credibility of the content.
Pirolli et al.~\cite{pirolli2009so} focus on Wikipedia and develop and system that presents users an interactive dashboard, which includes the history of article content and edits.
The main finding is that users can better judge the credibility of an article, given that they are presented with the history of the article and edits through an interactive dashboard.

\subsection{Use of images on the false information ecosystem}

Information can be disseminated via images on the Web. The use of images increases the credibility of the included information, as users tend to believe more information that is substantiated with an image.
However, nowadays, images can be easily manipulated, hence used for the dissemination of false information. 
In this section, we provide an overview of the papers that studied the problem of false information on the Web, while considering images.

\descr{Fabricated.} Boididou et al.~\cite{boididou2014challenges, boididou2015verifying} focus on the use of multimedia in false information spread in OSNs. 
In~\cite{boididou2015verifying} they prepare and propose a dataset of 12K tweets, which are manually labeled as fake, true, or unknown.
A tweet is regarded as true if the image is referring to a particular event and fake if the image is not referring to a particular event. 
The authors argue that this dataset can help researchers in the task of automated identification of fake multimedia within tweets. 
In~\cite{boididou2014challenges} they study the challenges that exist in providing an automated verification system for news that contain multimedia.
To this end, they propose the use of conventional classifiers with the aim to discern fake multimedia pertaining to real events. 
Their findings demonstrate that generalizing is extremely hard as their classifiers perform poorly (58\% accuracy) when they are trained with a particular event and they are tested with another. 
Diego Saez-Trumper~\cite{saez2014fake} proposes a Web application, called Fake Tweet Buster, that aims to warn users about tweets that contain false information through images or users that habitually diffuse false information. 
The proposed approach is based on the reverse image search technique (using Google Images) in order to determine the origin of the image, its age and its context.
Furthermore, the application considers user attributes and crowdsourcing data in order to find users that consistently share tweets that contain false information on images.
Pasquini et al.~\cite{pasquini2015towards} aim to provide image verification by proposing an empirical system that seeks visually and semantically related images on the web.
Specifically, their system utilizes news articles metadata in order to search, using Google's search engine, for relevant news articles. 
These images are then compared with the original's article images in order to identify whether the images were tampered.
To evaluate their approach, they created dummy articles with tampered images in order to simulate the whole procedure.

Jin et al.~\cite{jin2016novel} emphasize the importance of images in news articles for distinguishing its truthfulness. 
They propose the use of two sets of features extracted from images in conjunction with features that are proposed by~\cite{castillo2011information,kwon2013prominent}. 
For the image features, they define a set of visual characteristics as well as overall image statistics.
Their data is based on a corpus obtained from the Sina Weibo that comprises 50K posts and 26K images. 
For evaluating the image feature set, they use conventional machine learning techniques: namely SVM, Logistic Regression, KStar, and Random Forest.
They find that the proposed image features increase the accuracy by 7\% with an overall accuracy of 83\%. 
In a follow-up work, Jin et al.~\cite{jin2016image} leverage deep neural networks with the goal of distinguishing the credibility of images. 
They note that this task is extremely difficult as images can be misleading in many ways.
Specifically, images might be outdated (i.e., old images that are falsely used to describe a new event), inaccurate, or even manipulated.
To assess the image credibility, they train a Convolutional Neural Network (CNN) using a large-scale auxiliary dataset that comprises 600K labeled fake and real images.
Their intuition is that the CNN can extract useful hyperparameters that can be used to detect eye-catching and visually striking images, which are usually used to describe false information. 
Their evaluation indicates that the proposed model can outperform several baselines in terms of the precision, recall, F1, and accuracy scores.
Gupta et al.~\cite{gupta2013faking} focus on the diffusion of fake images in Twitter during Hurricane Sandy in 2012. 
They demonstrate that the use of automated techniques (i.e., Decision Trees) can assist in distinguishing fake images from real ones. 
Interestingly, they note that the 90\% of the fake images came from the top 0.3\% of the users.

\chapter{Understanding the Spread Of Information Through The Lens Of Multiple Web Communities}\label{chapter:spread_information}
In this chapter, we present our work that helps in better understanding the spread of information across the Web and how web communities influence each other. 
We focus on understanding the spread of news and image-based memes across multiple Web communities, namely, Twitter, Reddit, 4chan, and Gab.

\section[How Web Communities Influence Each Other Through the Lens of News Sources]{Understanding How Web Communities Influence Each Other Through the Lens of News Sources}

\subsection{Motivation}
Over the past few years, several conspiracy theories and false stories have spread on the Web. 
Some examples include the Boston Marathon bombings in 2013, where a large number of tweets started to claim that the bombings were a ``false  flag'' perpetrated by the goverment of the United States. More recently, the Pizzagate conspiracy~\cite{pizzagate} – a debunked theory connecting a restaurant and members of the US Democratic Party to a child sex ring – led to a shooting in a Washinghton DC restaurant~\cite{nytimes_pizzagate_shooter}. These stories were all propagated, in no small part, via the use of ``alternative'' news sites like Infowars and ``fringe'' Web communities like 4chan. 
This is mainly because the barrier of entry for such alternative news sources has been greatly reduced by the Web and large social networks. Due to the negligible cost of distributing information over social media, fringe sites can quickly gain traction with large audiences. 

Although previous works have studied the dissemination of false information on the Web, as discussed in Chapter~\ref{chapter:related}, very little work provides a holistic view of the modern information ecosystem.
This knowledge, however, is crucial for understanding the alternative news world and for designing appropriate detection and mitigation strategies. Anecdotal evidence and press coverage suggest that alternative news dissemination might start on fringe sites, eventually reaching mainstream online social networks and news outlets~\cite{bbc_4chan_pizzagate, bbc_macron}. Nevertheless, this phenomenon has not been measured and no thorough analysis has focused on how news moves from one online service to another, sort of forming an interconnected centipede of Web Communities.

In this work, we address this gap by performing the first thorough large-scale measurement on how mainstream and alternative news flows through three Web Communities; namely Twitter, Reddit, and 4chan. We focus on these three platforms because of: 1) they are fundamentally different and they drive substantial portions of the online world; 2) there is anecdotal evidence that suggests that specific communities within Reddit and 4chan act as generators~\cite{bbc_4chan_pizzagate} and incubators~\cite{guardian_reddit} of false information; and 3) they are able to have a substantial impact in forming and manipulating peoples' opinions by constantly circulating false information~\cite{nytimes_pizzagate_shooter}.

\descr{Contributions. }
First, we undertake a large-scale measurement and comparison of the occurrence of mainstream and alternative news sources across three social media platforms (4chan, Reddit, and Twitter). Then, we provide an understanding of the temporal dynamics of how URLs from news sites are posted on the different social networks. Finally, we present a measurement of the influence between the platforms that provides insight into how information spreads throughout the greater Web.
Overall, our  findings indicate that Twitter, Reddit, and 4chan are used quite extensively for the dissemination of both alternative and mainstream news. Using a statistical model for influence – namely, Hawkes processes – we show that each of the platforms have varying degrees of in influence on each other, and this influence differs with respect to mainstream and alternative news sources. 

\subsection{Datasets} \label{sec:data_collection}
Our analysis uses a set of news websites that can confidently be labeled as either ``mainstream'' or ``alternative'' news.
More specifically, we create a list of 99 news sites including 45 mainstream and 54 alternative ones.\footnote{The complete list of the 99 sites is available at \url{https://drive.google.com/open?id=0ByP5a__khV0dM1ZSY3YxQWF2N2c}}
For the former, we select 45 from the Alexa top 100 news sites, leaving out those based on user-generated content, those serving specialized content
(e.g., finance news), as well as non-English sites.
For the latter, we use Wikipedia~\cite{fake_news_list} and
FakeNewsWatch~\cite{fakenewswatch}.
We also add two state-sponsored alternative news domains: \url{sputniknews.com} and \url{rt.com}, as they have recently attracted public attention due to their posting of controversial, and seemingly agenda-pushing stories~\cite{sputnik_spotlight2}.

We gather information from posts, threads, and comments on Twitter, Reddit, and 4chan that contain URLs from the 99 news sites. %
With a few gaps (see below), our datasets cover activity on the three platforms between June 30, 2016 and February 28, 2017.
Table~\ref{tbl:post_percentage} shows the total number of posts/comments crawled and the percentage of posts that contains links to URLs from the aforementioned news domains.
We observe that mainstream news URLs are present in a greater percentage of posts on 4chan and Reddit than on Twitter,
while alternative ones are about twice as likely to appear in posts on 4chan than on Twitter or Reddit.
Table~\ref{tbl:datasets} provides a summary of our datasets, which we present in more detail below.
Note that we break Reddit and 4chan datasets into two different instances, as further discussed.

\begin{table}[h!]
\centering
\resizebox{0.67\columnwidth}{!}{
\begin{tabular}{lrrr}
\hline
\textbf{Platform}         & \textbf{Total Posts} & \textbf{\% Alt.} & \textbf{\% Main.} \\ \hline
Twitter                   & 587M   & 0.022\%     & 0.070\%   \\
Reddit (posts + comments) & 332M   & 0.023\%     & 0.181\%   \\
4chan                     & 42M   & 0.050\%     & 0.197\%   \\ \hline
\end{tabular}
}
\caption{Total number of posts crawled and percentage of posts that contain URLs to our list of alternative and mainstream news sites.}
\label{tbl:post_percentage}
\end{table}

\begin{table}[h!]
\centering
\resizebox{0.7\columnwidth}{!}{
\begin{tabular}{lrrr}
\hline
\textbf{Platform}    & \textbf{Posts/Comments} & \textbf{Alt. URLs} & \textbf{Main. URLs} \\ \hline
Twitter              & 486,700                       & 42,550                    & 236,480                  \\
Reddit (six selected subreddits)\hspace{-1cm}      & 620,530                       & 40,046                    & 301,840                  \\
Reddit (all other subreddits)\hspace{-1cm}  & 1,228,105                     & 24,027                    & 726,948                  \\
4chan (\dspol)        & 90,537                        & 8,963                     & 40,164                   \\
4chan (\dsint, \dssci, \dssp) & 7,131                         & 615                       & 5,513                    \\ \hline
\end{tabular}
}
\caption{Overview of our datasets with the number of posts/comments that contain a URL to one of our information sources, as well as the number of unique URLs linking to alternative and mainstream news sites in our list.}
\label{tbl:datasets}
\end{table}

\descr{Twitter.} We collect the 1\% of all publicly available tweets with URLs from the aforementioned news domains between June 30, 2016 and February 28, 2017 using the Twitter Streaming
API~\cite{twitter_streaming_api}.
In total, we gather 487K tweets containing 279K unique URLs pointing to mainstream or alternative news sites. 
Since tweets are retrieved at the time they are posted, we do not get information such as the number of times they are re-tweeted or liked.
Therefore, between March and May 2017, we re-crawled each tweet to retrieve this data.
Basic statistics are summarized in Table~\ref{table-tweets-stats}.
Due to a failure in our collection infrastructure, we have some gaps in the Twitter dataset, specifically between Oct 28--Nov 2 and Nov 5--16, 2016, as well as Nov 22, 2016 -- Jan 13, 2017, and Feb 24--28, 2017.

\begin{table}[h!]
\centering
\resizebox{0.7\columnwidth}{!}{%
\begin{tabular}{@{}lrrrrrr@{}}
\toprule
   & Tweets & Retrieved (\%)  &  Avg. Retweets & Avg. Likes   \\ \midrule
\textbf{Alternative} & 110,629  & 92,104 (83.2\%)  & 341 $\pm$ 1,228           & 0.82 $\pm$ 15.6                           \\
\textbf{Mainstream}  & 376,071  & 329,950 (87.7\%) & 404 $\pm$ 2,146          & 0.96 $\pm$ 55.6                         \\ \bottomrule
\end{tabular}
}
\caption{Basic statistics of the occurrence of alternative and mainstream news URLs in the tweets in our dataset.}
\label{table-tweets-stats}
\end{table}

\descr{Reddit.} We obtain all posts and comments on Reddit between June 30, 2016 and February 28, 2017, using data made available on Pushshift~\cite{pushshift}.
We collect approximately 42M posts, 390M comments, and 300K subreddits.
Once again, we filter posts and comments that contain URLs from one of the 99 news sites, which yields a dataset of 1.8M posts/comments and approximately 1.1M URLs.

\descr{4chan.} For 4chan, we use all threads and posts made on the Politically Incorrect (\dspol)  board, as well as \dssp (Sports), \dsint (International), and \dssci (Science) boards for comparison,
using the same methodology as~\cite{hine2016longitudinal}.
We opt to select both not safe for work boards (i.e., \dspol) and safe for work boards (i.e., \dssp, \dsint, and \dssci) to observe how these compare to each other with respect to the dissemination of news.
The resulting dataset includes 97K posts and replies, including 56K alternative and mainstream news URLs, between June 30, 2016 and February 28, 2017. We have some small gaps due to our crawler failing, specifically, Oct 15--16 and Dec 16--25, 2016 as well as Jan 10--13, 2017.

\subsection{General Characterization}
In this section, we present a general characterization of the mainstream and alternative news URLs found on the three platforms.

\begin{table}[]
\centering
\resizebox{\columnwidth}{!}{%
\begin{tabular}{@{}rlrlrlrl@{}}
\toprule
\multicolumn{1}{l}{\textbf{Subreddit (Alt.)}} & \textbf{(\%)} & \textbf{Subreddit (Alt.)} & \multicolumn{1}{c}{\textbf{(\%)}} & \textbf{Subreddit (Main.)} & \multicolumn{1}{r}{\textbf{(\%)}} & \multicolumn{1}{l}{\textbf{Subreddit (Main.)}} & \textbf{(\%)} \\ \midrule
The\_Donald                                   & 35.37 \%      & KotakuInAction            & \multicolumn{1}{r|}{1.04 \%}      & politics                   & 12.9 \%                           & EnoughTrumpSpam                                & 1.20 \%       \\
politics                                      & 8.21 \%       & HillaryForPrison          & \multicolumn{1}{r|}{0.94 \%}      & worldnews                  & 6.24 \%                           & NoFilterNews                                   & 1.16 \%       \\
news                                          & 3.85 \%       & TheOnion                  & \multicolumn{1}{r|}{0.94 \%}      & The\_Donald                & 4.53 \%                           & BreakingNews24hr                               & 1.07 \%       \\
conspiracy                                    & 3.84 \%       & AskTrumpSupporters        & \multicolumn{1}{r|}{0.84 \%}      & news                       & 4.23 \%                           & conspiracy                                     & 0.89 \%       \\
Uncensored                                    & 2.66 \%       & POLITIC                   & \multicolumn{1}{r|}{0.81 \%}      & TheColorIsBlue             & 3.06 \%                           & todayilearned                                  & 0.83 \%       \\
Health                                        & 2.10 \%       & rss\_theonion             & \multicolumn{1}{r|}{0.67 \%}      & TheColorIsRed              & 2.48 \%                           & thenewsrightnow                                & 0.78 \%       \\
PoliticsAll                                   & 1.54 \%       & the\_Europe               & \multicolumn{1}{r|}{0.67 \%}      & willis7737\_news           & 2.27 \%                           & europe                                         & 0.77 \%       \\
Conservative                                  & 1.45 \%       & new\_right                & \multicolumn{1}{r|}{0.6 \%}       & news\_etc                  & 1.94 \%                           & ReddLineNews                                   & 0.75 \%       \\
worldnews                                     & 1.41 \%       & AskReddit                 & \multicolumn{1}{r|}{0.59 \%}      & AskReddit                  & 1.37 \%                           & hillaryclinton                                 & 0.73 \%       \\
WhiteRights                                   & 1.21 \%       & AnythingGoesNews          & \multicolumn{1}{r|}{0.51 \%}      & canada                     & 1.31 \%                           & nottheonion                                    & 0.73 \%       \\ \bottomrule
\end{tabular}
}
\caption{Top 20 subreddits w.r.t. mainstream and alternative news URLs occurrence and their percentage in Reddit (all subreddits).}
\label{top_subreddits}
\end{table}

\descr{Reddit.} We start by identifying news and politics communities.
In Table~\ref{top_subreddits}, we report the top 20 subreddits with the most URLs, along with their percentage.
Note that we omit automated ones (e.g., /r/AutoNewspaper/) where news articles are posted without user intervention.
Many of the subreddits are indeed related to news and politics -- e.g.,
`The\_Donald' is mostly a community of Donald Trump supporters, while `worldnews' is focused around globally relevant events.
We also find the presence of the `conspiracy' subreddit, which has been involved in disinformation campaigns including Pizzagate, %
as well as  `AskReddit,' where both mainstream and alternative news sources are used
to answer questions submitted by users. Although the latter is intended for open-ended questions that spark discussion, it is evident that commenters often try to push their agenda even in non-political threads.
In the end, based on their propensity to include news URLs of both types, we single out the follow top six subreddits for further exploration:
The\_Donald, politics, conspiracy, news, worldnews, and AskReddit.

\begin{table}[h!]
\centering
\resizebox{\columnwidth}{!}{%
\begin{tabular}{@{}rlrlrlrl@{}}
\toprule
\multicolumn{1}{l}{\textbf{Domain (Alt.)}} & \textbf{(\%)} & \multicolumn{1}{l}{\textbf{Domain (Alt.)}} & \multicolumn{1}{r}{\textbf{(\%)}} & \multicolumn{1}{c}{\textbf{Domain (Main.)}} & \multicolumn{1}{r}{\textbf{(\%)}} & \multicolumn{1}{l}{\textbf{Domain (Main.)}} & \textbf{(\%)} \\ \midrule
breitbart.com                              & 55.58 \%      & prntly.com                                 & \multicolumn{1}{l|}{0.49 \%}      & nytimes.com                                 & 14.07 \%                          & nbcnews.com                                 & 2.86 \%       \\
rt.com                                     & 19.18 \%      & dccclothesline.com                         & \multicolumn{1}{l|}{0.4 \%}       & cnn.com                                     & 11.23 \%                          & time.com                                    & 2.57 \%       \\
infowars.com                               & 8.99 \%       & worldnewsdailyreport.com                   & \multicolumn{1}{l|}{0.36 \%}      & theguardian.com                             & 8.86 \%                           & washinghtontimes.com                        & 2.52 \%       \\
sputniknews.com                            & 3.95 \%       & therealstrategy.com                        & \multicolumn{1}{l|}{0.3 \%}       & reuters.com                                 & 6.67 \%                           & bloomberg.com                               & 2.5 \%        \\
beforeitsnews.com                          & 2.34 \%       & disclose.tv                                & \multicolumn{1}{l|}{0.23 \%}      & huffingtonpost.com                          & 5.67 \%                           & wsj.com                                     & 2.31 \%       \\
lifezette.com                              & 2.28 \%       & clickhole.com                              & \multicolumn{1}{l|}{0.2 \%}       & thehill.com                                 & 5.15 \%                           & cbsnews.com                                 & 2.26 \%       \\
naturalnews.com                            & 1.54 \%       & libertywritersnews.com                     & \multicolumn{1}{l|}{0.2 \%}       & foxnews.com                                 & 4.89 \%                           & thedailybeast.com                           & 2.05 \%       \\
activistpost.com                           & 1.45 \%       & worldtruth.tv                              & \multicolumn{1}{l|}{0.14 \%}      & bbc.com                                     & 4.76 \%                           & forbes.com                                  & 1.87 \%       \\
veteranstoday.com                          & 1.11 \%       & thelastlineofdefence.org                   & \multicolumn{1}{l|}{0.07 \%}      & abcnews.go.com                              & 2.94 \%                           & nypost.com                                  & 1.85 \%       \\
redflagnews.com                            & 0.63 \%       & nodisinfo.com                              & \multicolumn{1}{l|}{0.05 \%}      & usatoday.com                                & 2.87 \%                           & cncb.com                                    & 1.54 \%       \\ \bottomrule
\end{tabular}
}
\caption{Top 20 mainstream and alternative domains and their percentage in the six selected subreddits.}
\label{tbl:reddit_top_domains}
\end{table}

In order to get a better view of the popularity of news sites on the six subreddits, we study the occurrence of each news outlet.
Specifically, we find 76K URLs (40K unique) from alternative news
and 600K (301K unique) from mainstream news domains.
Table~\ref{tbl:reddit_top_domains} reports the top 20 mainstream/alternative news sites and their percentage in the six subreddits.
The top 20 domains for mainstream news account for 89\% of all mainstream news URLs in our data, while for alternative domains the percentage is 99\%.
Known alt-right news outlets, such as \url{breitbart.com} and \url{infowars.com}, are predominantly present, as well as state-sponsored
alternative domains like \url{sputniknews.com} and \url{rt.com}, which have recently been in the spotlight for disseminating false information
and propaganda~\cite{sputnik_spotlight2}. 
The fact that many such URLs appear in our dataset may indeed be an indication that the six subreddits significantly contribute to the dissemination of controversial stories.

\begin{table}[t]
\centering
\resizebox{\columnwidth}{!}{
\begin{tabular}{@{}llrlrlrl@{}}
\toprule
\textbf{Domain (Alt.)} & \textbf{(\%)} & \textbf{Domain (Alt.)} & \multicolumn{1}{r}{\textbf{(\%)}} & \textbf{Domain (Main.)} & \multicolumn{1}{c}{\textbf{(\%)}} & \textbf{Domain (Main.)} & \textbf{(\%)} \\ \midrule
breitbart.com                     & 46.04 \%                 & activistpost.com                  & \multicolumn{1}{l|}{0.41 \%}                 & theguardian.com                    & 19.04 \%                                     & usatoday.com                       & 2.02 \%                  \\
rt.com                            & 17.56 \%                 & disclose.tv                       & \multicolumn{1}{l|}{0.39 \%}                 & nytimes.com                        & 10.07 \%                                     & thedailybeast.com                  & 2.02 \%                  \\
infowars.com                      & 17.25 \%                 & prntly.com                        & \multicolumn{1}{l|}{0.26 \%}                 & bbc.com                            & 8.99 \%                                      & nbcnews.com                        & 1.96 \%                  \\
therealstrategy.com               & 5.63 \%                  & worldtruth.tv                     & \multicolumn{1}{l|}{0.25 \%}                 & forbes.com                         & 6.24 \%                                      & nypost.com                         & 1.95 \%                  \\
sputniknews.com                   & 4.11 \%                  & libertywriternews.com             & \multicolumn{1}{l|}{0.15 \%}                 & thehill.com                        & 4.95 \%                                      & cbsnews.com                        & 1.89 \%                  \\
beforeitsnews.com                 & 2.26 \%                  & worldnewsdailyreport.com          & \multicolumn{1}{l|}{0.06 \%}                 & cbc.ca                             & 4.82 \%                                      & abcnews.go.com                     & 1.78 \%                  \\
redflagnews                       & 2.04 \%                  & mediamass.net                     & \multicolumn{1}{l|}{0.04 \%}                 & foxnews.com                        & 4.79 \%                                      & time.com                           & 1.71 \%                  \\
dccclothesline.com                & 1.37 \%                  & newsbiscuit.com                   & \multicolumn{1}{l|}{0.03 \%}                 & wsj.com                            & 4.04 \%                                      & cnbc.com                           & 1.40 \%                  \\
naturalnews.com                   & 1.29 \%                  & react365.com                      & \multicolumn{1}{l|}{0.02 \%}                 & bloomberg.com                      & 3.48 \%                                      & washingtontimes.com                & 1.34 \%                  \\
clickhole.com                     & 0.53 \%                  & the-daily.buzz                    & \multicolumn{1}{l|}{0.02 \%}                 & reuters.com                        & 2.85 \%                                      & washingtonexaminer.com             & 1.33 \%                  \\ \bottomrule
\end{tabular}
}
\caption{Top 20 mainstream and alternative news sites in the Twitter dataset and their percentage.}
\label{twitter_top_domains}
\end{table}

\descr{Twitter.} In our Twitter dataset, we find 129K (42K unique) URLs of alternative news domains and 413K (236K unique) URLs of
mainstream ones. Recall that we re-crawl
tweets to get the number of retweets and likes, and a small percentage of them are no longer available as they were either deleted
or the associated account was suspended.
This percentage is slightly higher for tweets with URLs from alternative news, possibly due to the fact that some users tend to remove
controversial content when a particular false story is debunked~\cite{friggeri2014rumor}.
Also, alternative and mainstream news tend to get a significant number of retweets, at about the same rate (on average, 404 and 341 retweets
per tweet, respectively). A similar pattern is observed for likes (see Table~\ref{table-tweets-stats}).

In Table~\ref{twitter_top_domains}, we report the top 20 mainstream and alternative news domains, and their percentage, in our Twitter dataset.
These cover, respectively, 86\% and 99\% of all URLs.
Similar to Reddit, there are many popular alt-right and state-sponsored news outlets.

\begin{table}[t]
\centering
\resizebox{\columnwidth}{!}{
\begin{tabular}{@{}rlrlrlrl@{}}
\toprule
\textbf{Domain (Alt.)}   & \textbf{(\%)} & \textbf{Domain (Alt.)} & \textbf{(\%)}                & \textbf{Domain (Main.)} & \textbf{(\%)} & \textbf{Domain (Main.)} & \textbf{(\%)} \\ \midrule
breitbart.com            & 53.00 \%      & activistpost.com       & \multicolumn{1}{l|}{0.38 \%} & theguardian.com         & 14.10 \%      & wsj.com                 & 2.82 \%       \\
rt.com                   & 28.22 \%      & dccclothesline.com     & \multicolumn{1}{l|}{0.29 \%} & nytimes.com             & 10.07 \%      & washinghtontimes.com    & 2.77 \%       \\
infowars.com             & 9.12 \%       & redflagnews.com        & \multicolumn{1}{l|}{0.20 \%} & cnn.com                 & 9.90 \%       & bloomberg.com           & 2.75 \%       \\
sputniknews.com          & 3.36 \%       & libertywritersnews.com & \multicolumn{1}{l|}{0.16 \%} & bbc.com                 & 5.45 \%       & cbc.ca                  & 2.66 \%       \\
veteranstoday.com        & 1.07 \%       & therealstrategy.com    & \multicolumn{1}{l|}{0.16 \%} & foxnews.com             & 5.35 \%       & nypost.com              & 2.65 \%       \\
beforeitsnews.com        & 0.91 \%       & clickhole.com          & \multicolumn{1}{l|}{0.11 \%} & reuters.com             & 5.10 \%       & cbsnews.com             & 2.44 \%       \\
lifezette.com            & 0.86 \%       & disclose.tv            & \multicolumn{1}{l|}{0.10 \%} & time.com                & 3.42 \%       & nbcnews.com             & 2.32 \%       \\
naturalnews.com          & 0.61 \%       & now8news.com           & \multicolumn{1}{l|}{0.06 \%} & abcnews.go.com          & 3.40 \%       & usatoday.com            & 2.25 \%       \\
worldnewsdailyreport.com & 0.46 \%       & firebrandleft.com      & \multicolumn{1}{l|}{0.05 \%} & huffingtonpost.com      & 3.29 \%       & cnbc.com                & 2.13 \%       \\
prntly.com               & 0.41 \%       & nodisinfo.com          & \multicolumn{1}{l|}{0.05 \%} & thehill.com             & 3.04 \%       & forbes.com              & 1.68 \%       \\ \bottomrule
\end{tabular}
}
\caption{Top 20 mainstream and alternative news sites in the /pol/ dataset and their percentage.}
\label{tbl:4chan_top_domains}
\end{table}

\descr{4chan.}
In our \dspol dataset, we find 21K (9K unique) URLs to alternative news outlets and 82K (40K unique) to mainstream news.
Table~\ref{tbl:4chan_top_domains} reports the percentage of URLs of the top 20 domains for each type of news.
These cover 87\% and 99\% of mainstream and alternative news URLs, respectively.
Again, we observe that, by far, the most popular alternative news domains are \url{breitbart.com}, \url{rt.com},
\url{infowars.com}, and \url{sputniknews.com}.
For the mainstream news, we observe that \url{theguardian.com} is the most frequently posted, followed by \url{nytimes.com}, \url{cnn.com}, and \url{bbc.com}.
We also obtained similar statistics for domain popularity in the other boards of 4chan, but we omit them for brevity.

\begin{figure*}[h]
\center
\subfigure[]{\includegraphics[width=0.49\textwidth]{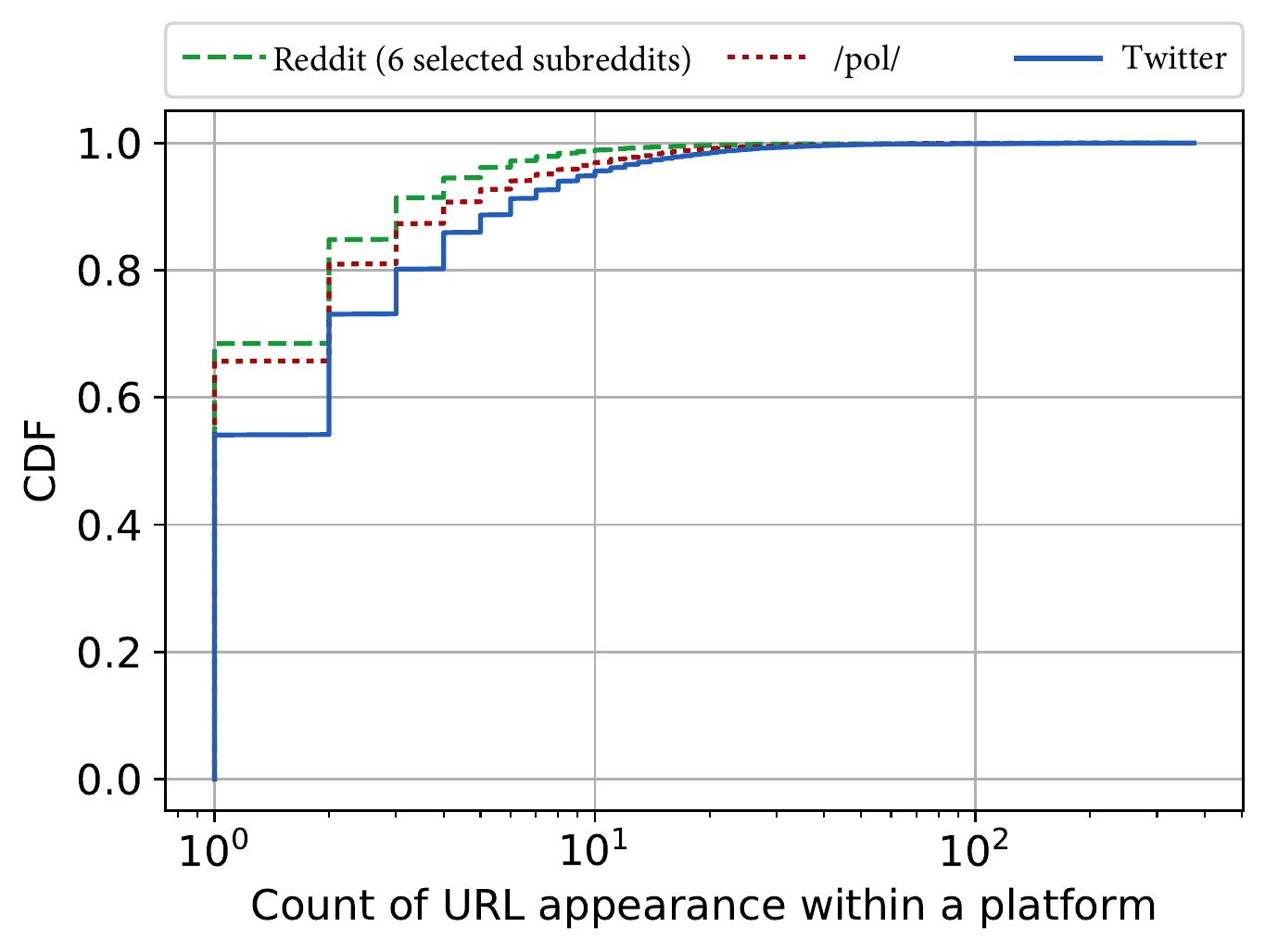}\label{url-occurence-fake}}
\subfigure[]{\includegraphics[width=0.49\textwidth]{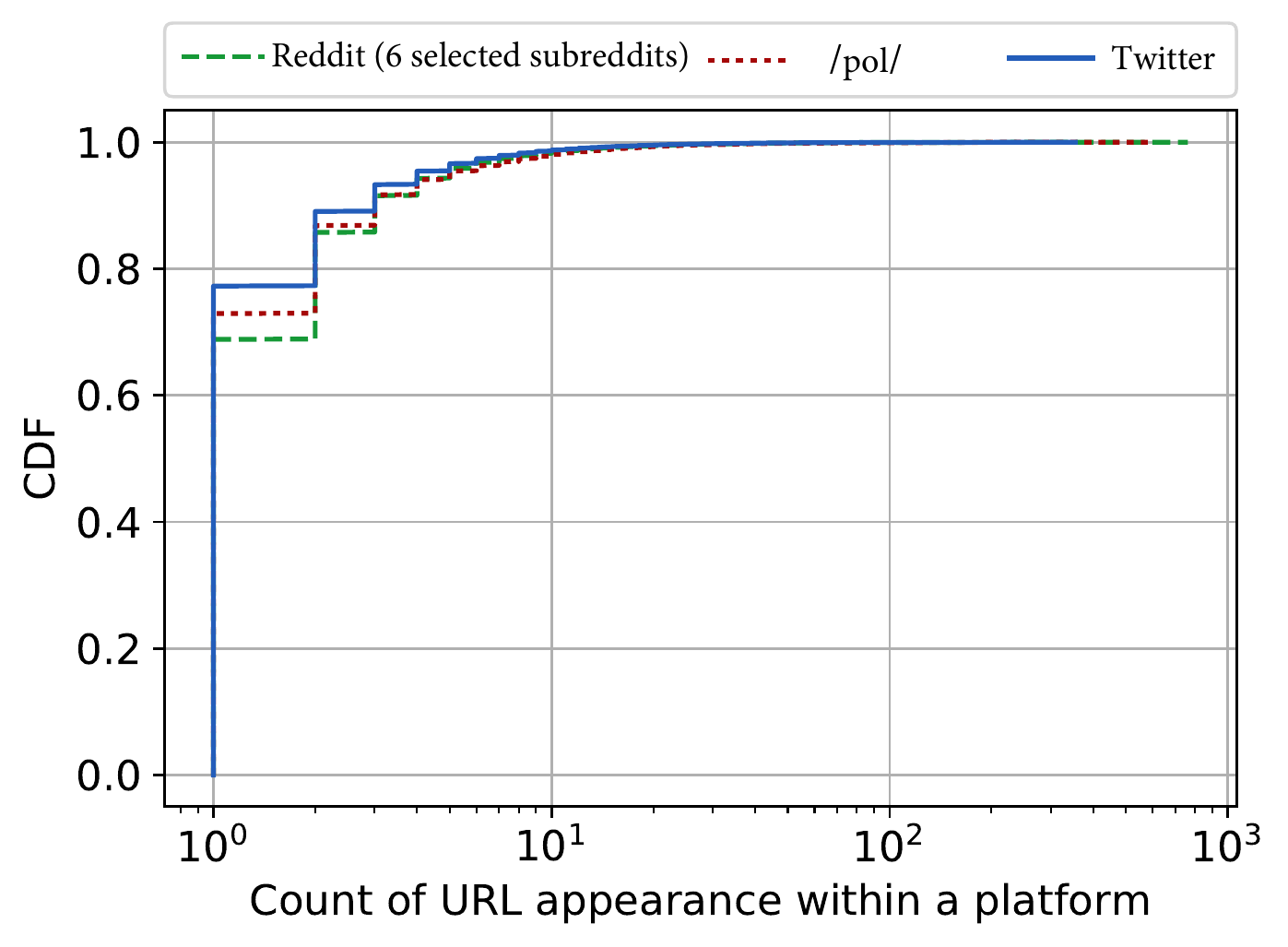}\label{url-occurence-real}}
\reduce
\caption{CDF of the counts of URL appearance within a particular platform: (a) alternative news and (b) mainstream news. }
\label{cdf_url_occurence}
\end{figure*}

To get a better view of the platforms' URL posting behavior, Fig.~\ref{cdf_url_occurence} plots the CDF of URL appearances (i.e., how many times a specific URL appears) within a particular platform.
We observe that a substantial portion of the URLs appear only once for both alternative and mainstream news, and that, on Twitter, alternative news tends to appear more times than  mainstream news. For \dspol and the six subreddits, we observe a similar behavior for both mainstream and alternative news.

\begin{figure*}[h]
\center
\subfigure[]{\includegraphics[width=0.47\textwidth]{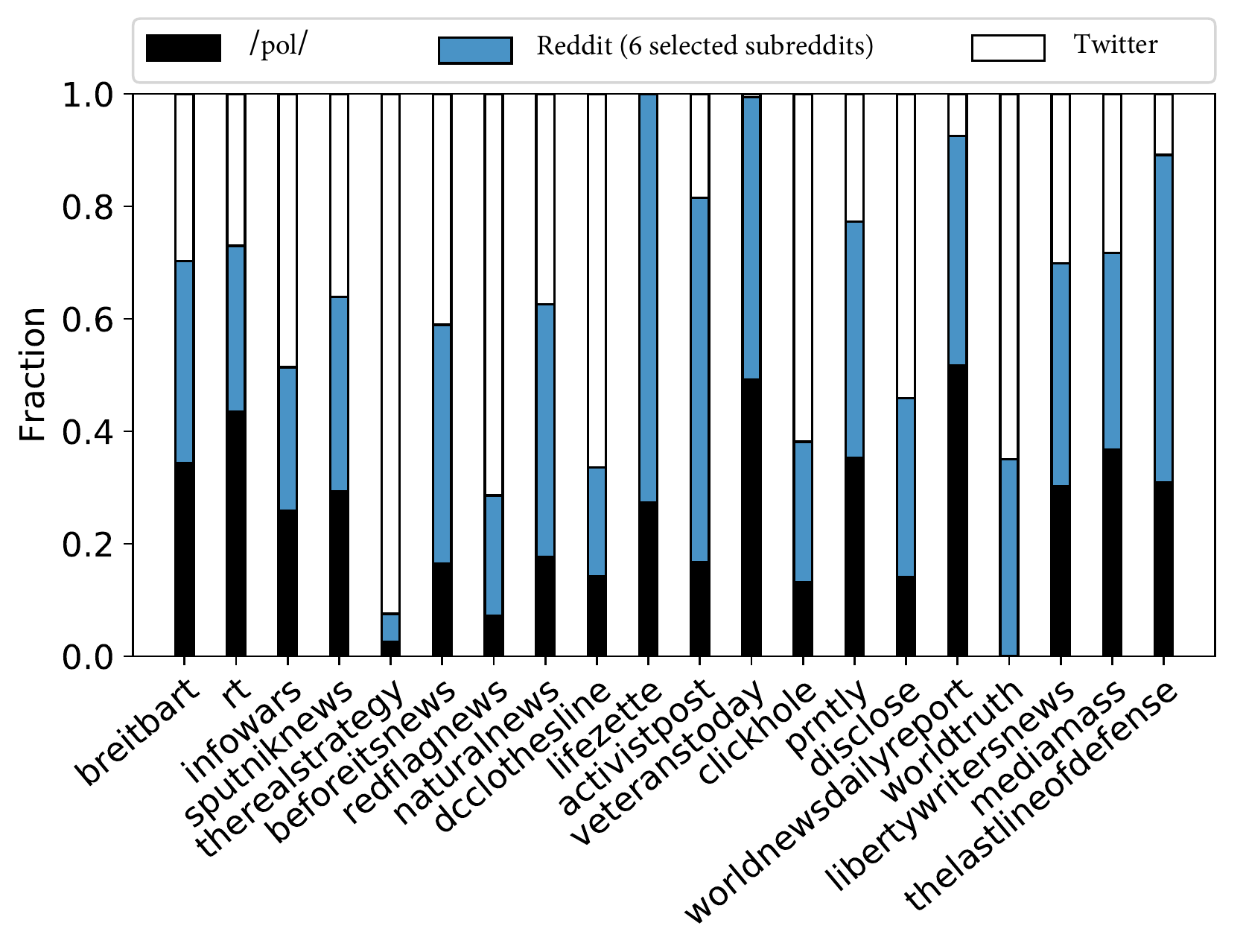}\label{domain_ratio_platforms-fake}}
\subfigure[]{\includegraphics[width=0.495\textwidth]{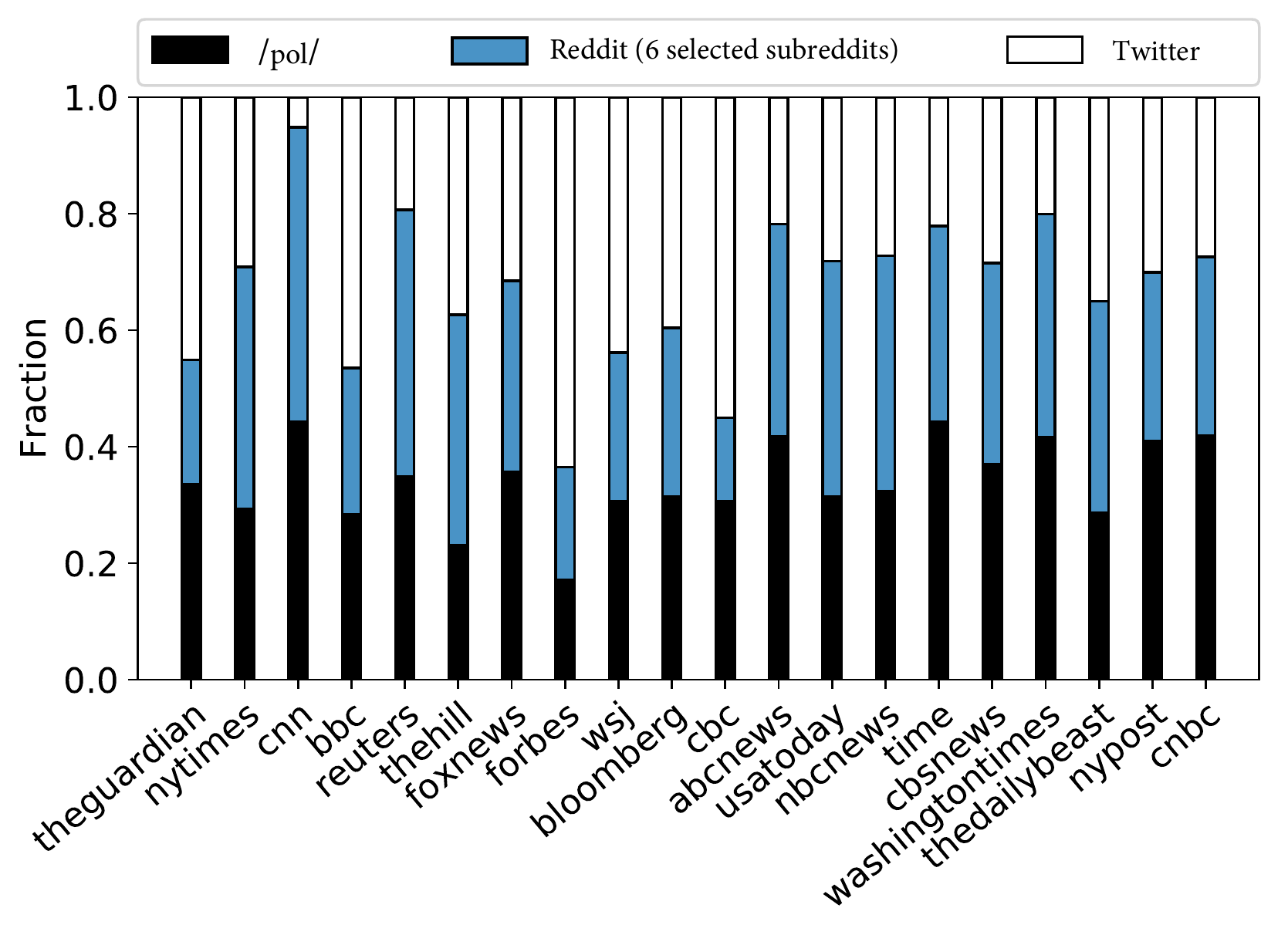}\label{domain_ratio_platforms-real}}
\reduce
\caption{Top 20 domains and each platform's fraction for (a) alternative and (b) mainstream news.}
\label{domain_ratio_platforms}
\end{figure*}

\begin{figure*}[t]
\center
\subfigure[ ]{\includegraphics[width=0.49\textwidth]{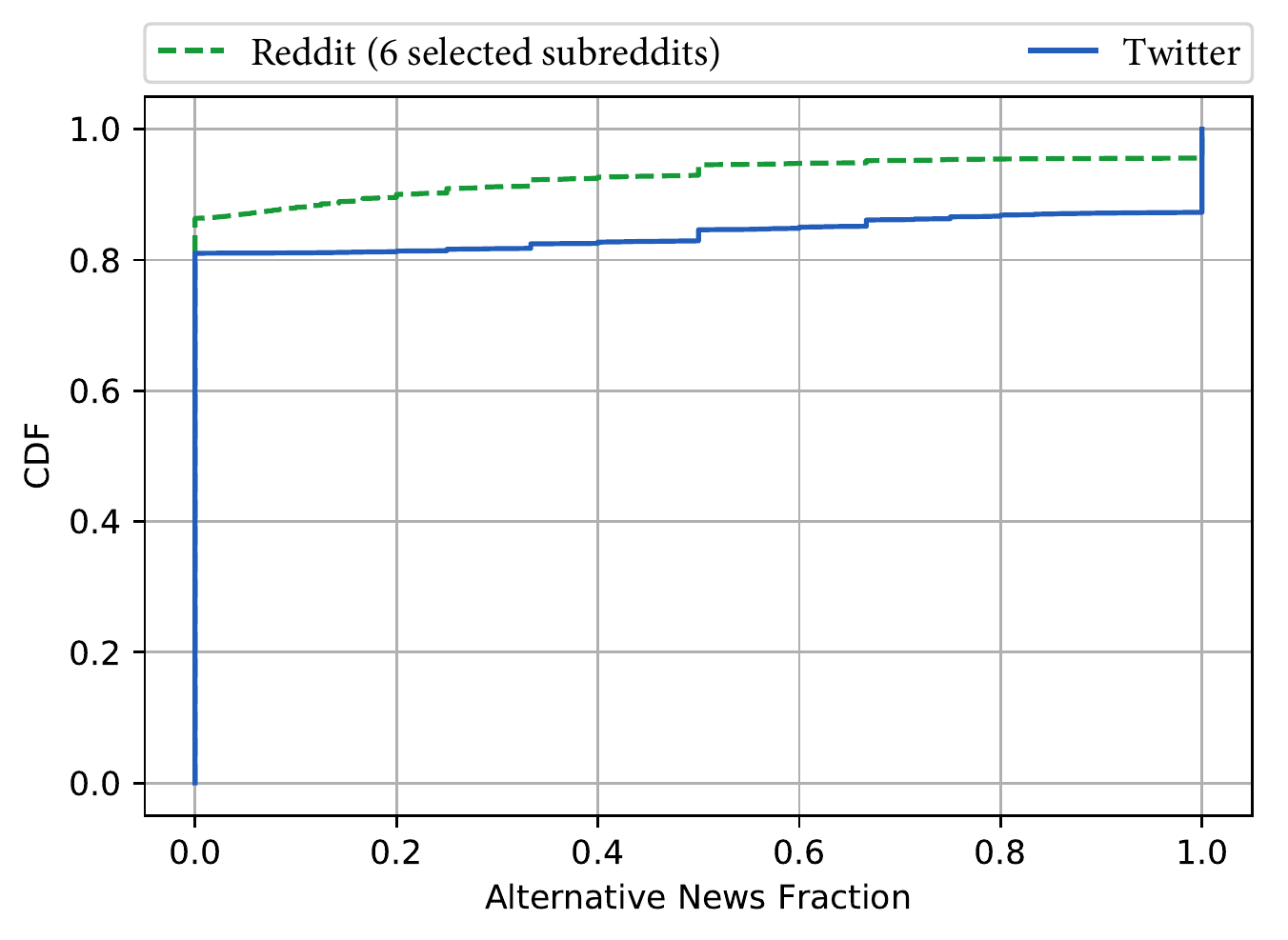}\label{fake_real_ratio_cdf_all}}
\subfigure[]{\includegraphics[width=0.49\textwidth]{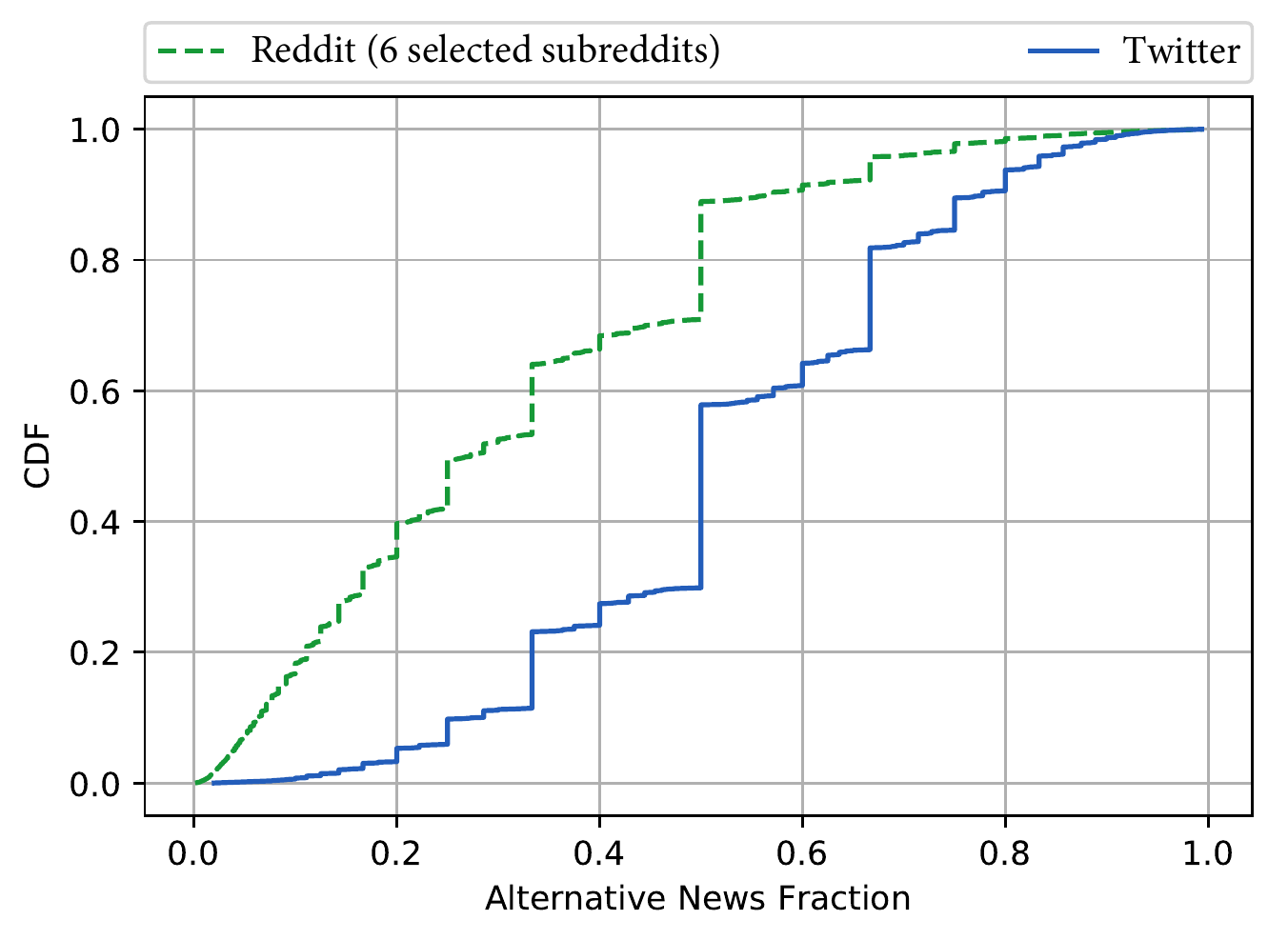}\label{fake_real_ratio_cdf_common}}
\reduce
\caption{CDF of the fraction of URLs from alternative news and overall news URLs for (a) all users in our Twitter and Reddit datasets, and (b) users that shared URLs from both mainstream and alternative news.}
\label{fake_real_ratio_cdf}
\end{figure*}

\begin{figure*}[t]
\center
\subfigure[]{\includegraphics[width=0.49\textwidth]{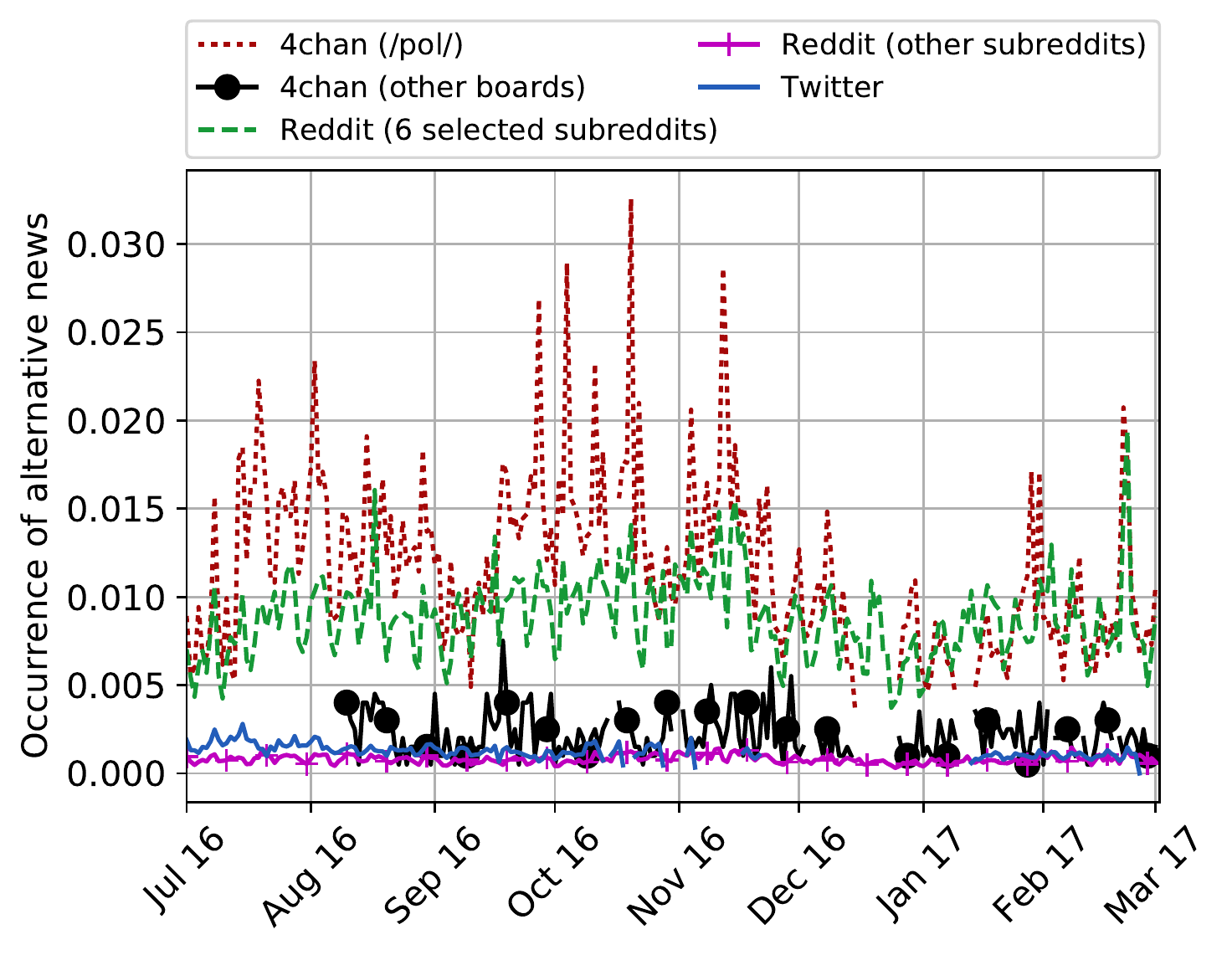}\label{aggregate_month_counts-fake}}
\subfigure[]{\includegraphics[width=0.49\textwidth]{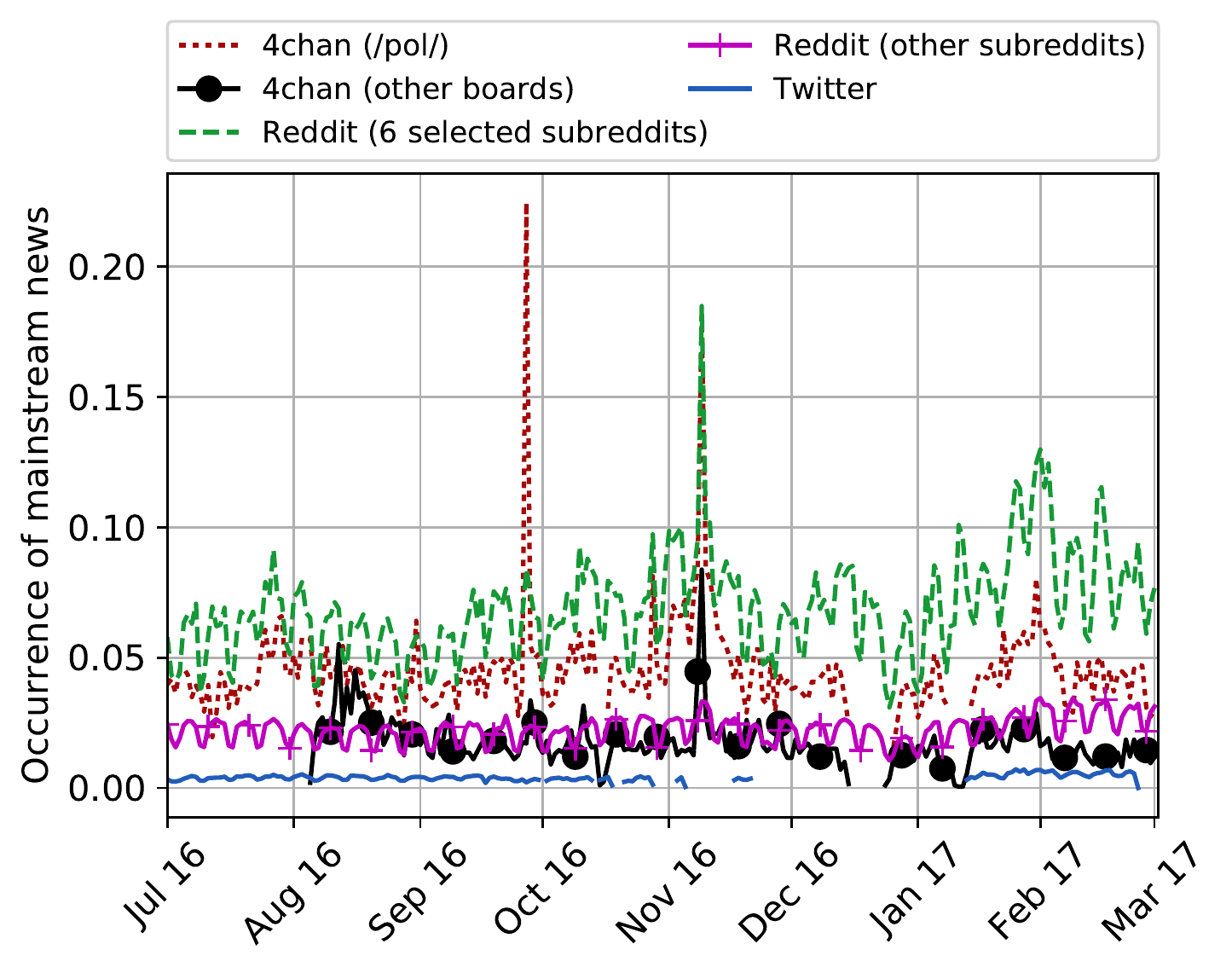}\label{aggregate_month_counts-real}}
\subfigure[]{\includegraphics[width=0.49\textwidth]{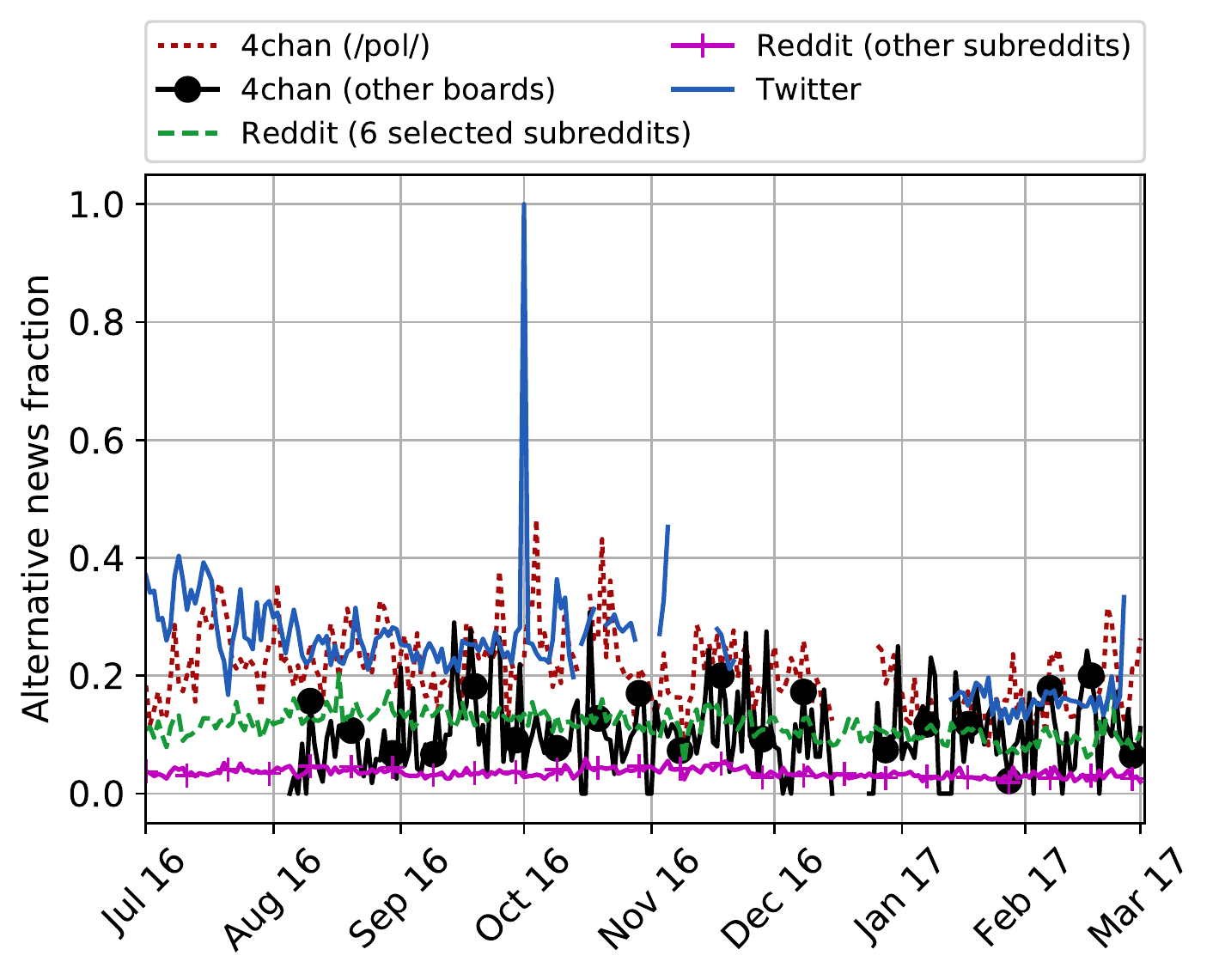}\label{aggregate_month_counts-ratio}}
\reduce
\caption{Normalized daily occurrence of URLs for (a) alternative news, (b) mainstream news, and (c) fraction of alternative news over all news.}
\label{aggregate_month_counts}
\end{figure*}
Next, in Fig.~\ref{domain_ratio_platforms}, we compare how popular domains, in both categories, appear on the three platforms
(i.e., Twitter, the six subreddits, and \dspol).
We find that the top 4 alternative domains -- \url{breitbart.com}, \url{rt.com}, \url{infowars.com}, \url{sputniknews.com} -- influence the three platforms more or less in the same way.
However, some outlets appear predominantly in some platforms but not in others; e.g., \url{therealstrategy.com} is popular only on Twitter,
while \url{lifezette.com} and \url{veteranstoday.com} are popular on the 6 subreddits and \dspol, but not on Twitter.
We believe the primary reason for this has to do with %
Twitter bots.
We cannot exclude with certainty that bots do not exist on 4chan, while bots are actually acceptable on Reddit (as long as they follow the rules of Reddit's API~\cite{reddit_api_rules}), however, they are certainly more prevalent on Twitter.
Thus, if a particular domain is popular on Twitter because of the influence of bots, then it might not be popular on Reddit and 4chan.
We have also considered ways to factor out posting behavior from bots, especially for Twitter, such as the one proposed in~\cite{davis2016botornot}. However, we have not removed this activity due to: 
1) posting behavior from bots can affect real users' posting behavior, hence this activity is part of the overall news dissemination ecosystem and needs to be accounted for; 
and 2) the satisfactory performance of such approaches is yet to be proven.

We also measure the fraction of news URLs that are alternative, {\em per user}, in Fig.~\ref{fake_real_ratio_cdf}.
We report this fraction only for Reddit and Twitter users, since on 4chan posts are anonymous.
We find that 80\% of the users of both platforms share only URLs from mainstream news,
while, 13\% of Twitter users -- which are likely bots~\cite{varol2017online} -- exclusively post URLs to alternative news.
We observe from Fig.~\ref{fake_real_ratio_cdf_common}, which shows the fraction for users sharing URLs from both categories, that there is a wide distribution, especially on the six selected subreddits, between people that rarely share alternative news (fraction close to 0) and those who share them almost all the time (fraction close to 1).
Moreover, we find that Twitter users share more alternative news: just 5\% of these users have a fraction below 0.2, 
which might be also attributed to the presence of bots.

\subsection{Temporal Analysis}
In this section, we present the results of a cross-platform temporal analysis of the way news are posted on Twitter, Reddit, and 4chan.

\subsubsection{URL Occurrence}
In Fig.~\ref{aggregate_month_counts}, we measure the daily occurrence of news URLs over the three platforms normalized by the average daily number of URLs shared in each community.\footnote{Gaps in the plot correspond to gaps in our dataset due to crawler failure.}
We find that \dspol and the six selected subreddits exhibit a much higher percentage of occurrences of alternative news compared to the other communities (Fig.~\ref{aggregate_month_counts-fake}), whereas, for mainstream news, the sharing behavior is more similar across platforms (Fig.~\ref{aggregate_month_counts-real}).
There are also some interesting spikes, likely related to the 2016 US elections, on the date of the first presidential debate and election day itself.
These findings indicate that the selected sub-communities are heavily utilized for the dissemination of alternative news.
We also study the fraction of alternative news URLs with respect to overall news URLs (Fig.~\ref{aggregate_month_counts-ratio}), highlighting that mainstream news URLs are overall more ``popular'' than the alternative news URLs.
Note that the Twitter spike in Fig.~\ref{aggregate_month_counts-ratio} appears to be an artifact of a failure in our collection infrastructure.

\begin{figure*}[t]
\centering
\subfigure[]{\includegraphics[width=0.49\textwidth]{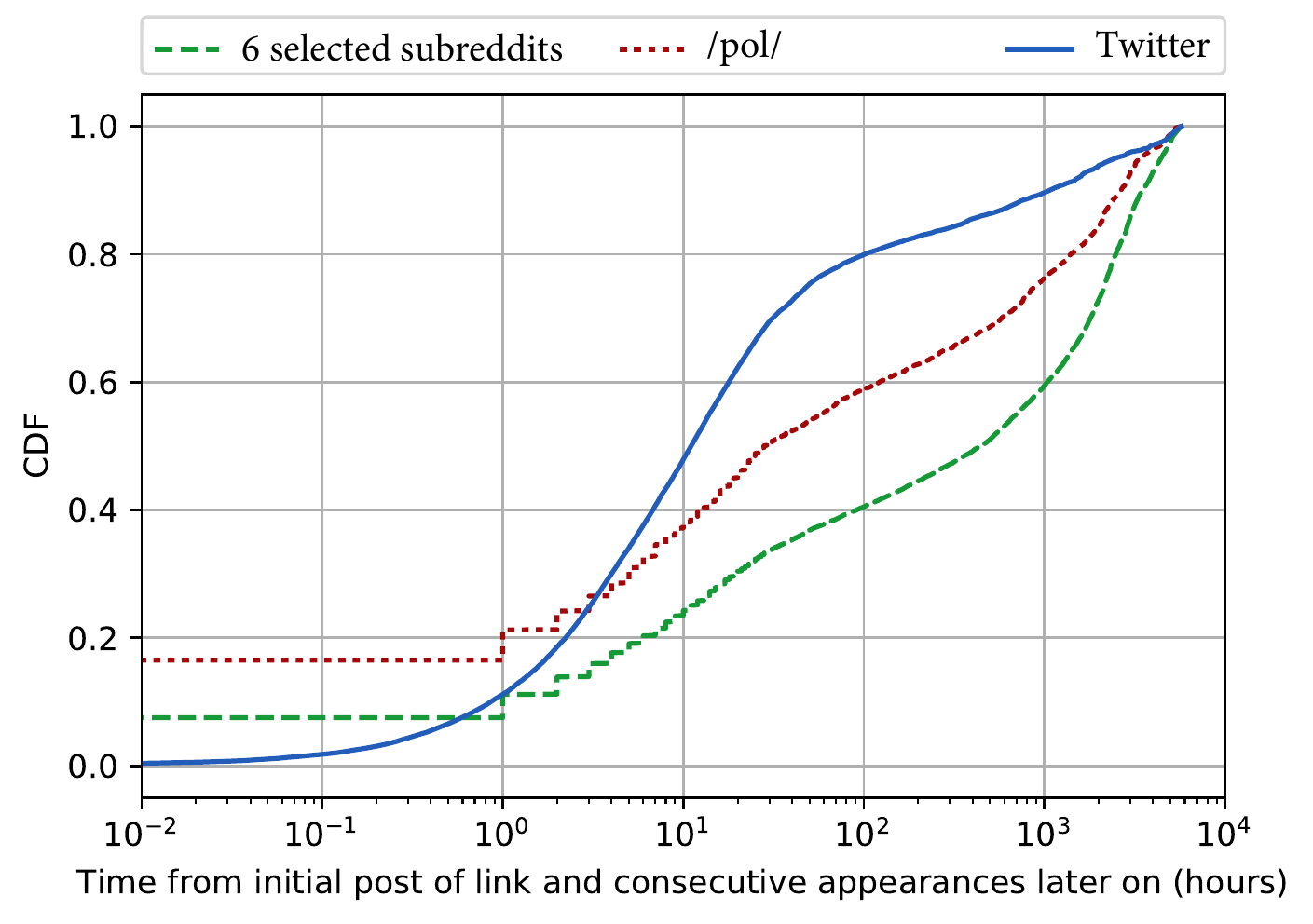}}
\subfigure[]{\includegraphics[width=0.49\textwidth]{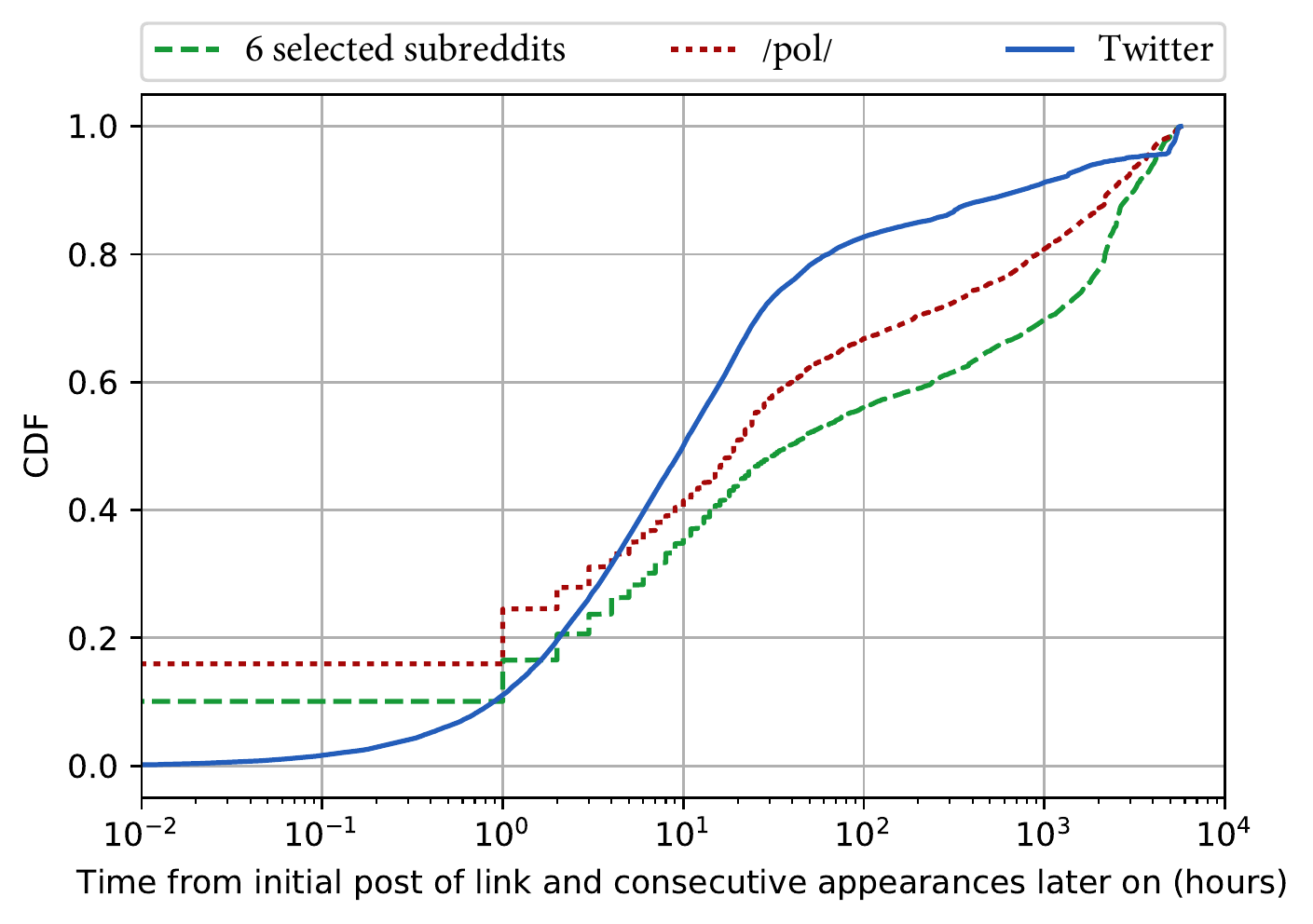}}
\reduce
\caption{CDF of time difference (in hours) between the first occurrence of a URL and its next occurrences on each platform for (a) alternative and (b) mainstream news.}
\label{timeline_for_each_platform}
\end{figure*}

As some users repost the same URL many times within the same platform,
we next study such reposting behavior and extract insights while comparing platforms.
In Fig.~\ref{timeline_for_each_platform}, we plot the CDF of the time difference between the first occurrence of a URL and its next occurrences on the same platform.
Both alternative and mainstream news URLs are recycled over time within the platform (even after several months), but Twitter exhibits a smaller lag between the first occurrence and later ones compared to the other two platforms.
In all three platforms, there is an inflection point at the 24h period, which probably signifies the day-to-day behavior of news propagation within a platform, and this is true for both alternative and mainstream news.
Finally, mainstream news seem to propagate faster in these platforms than alternative news, especially on the six subreddits; for Twitter and \dspol the difference is not evident.

\begin{figure*}[t]
\subfigure[ ]{\includegraphics[width=0.49\textwidth]{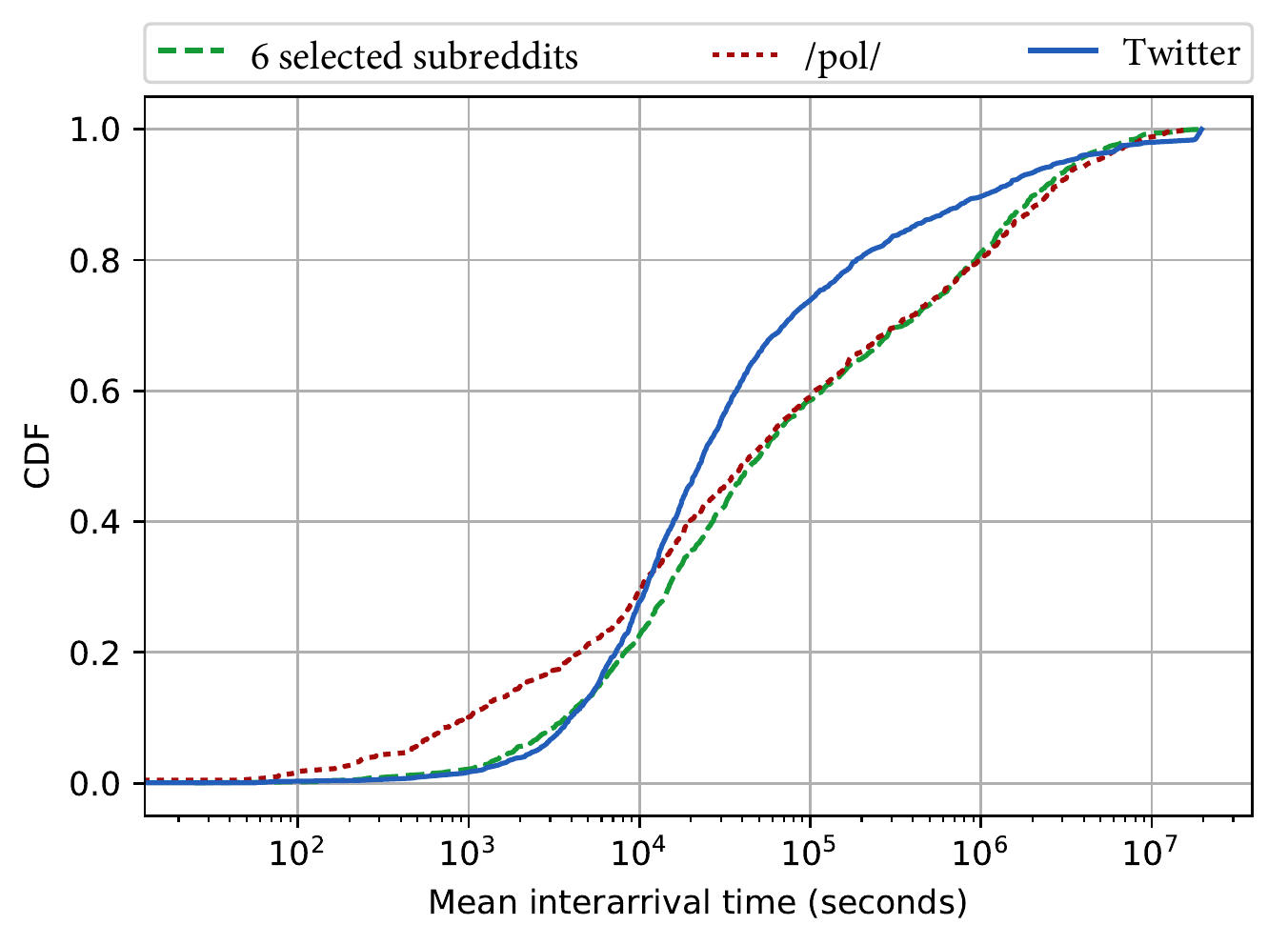}}
\subfigure[]{\includegraphics[width=0.49\textwidth]{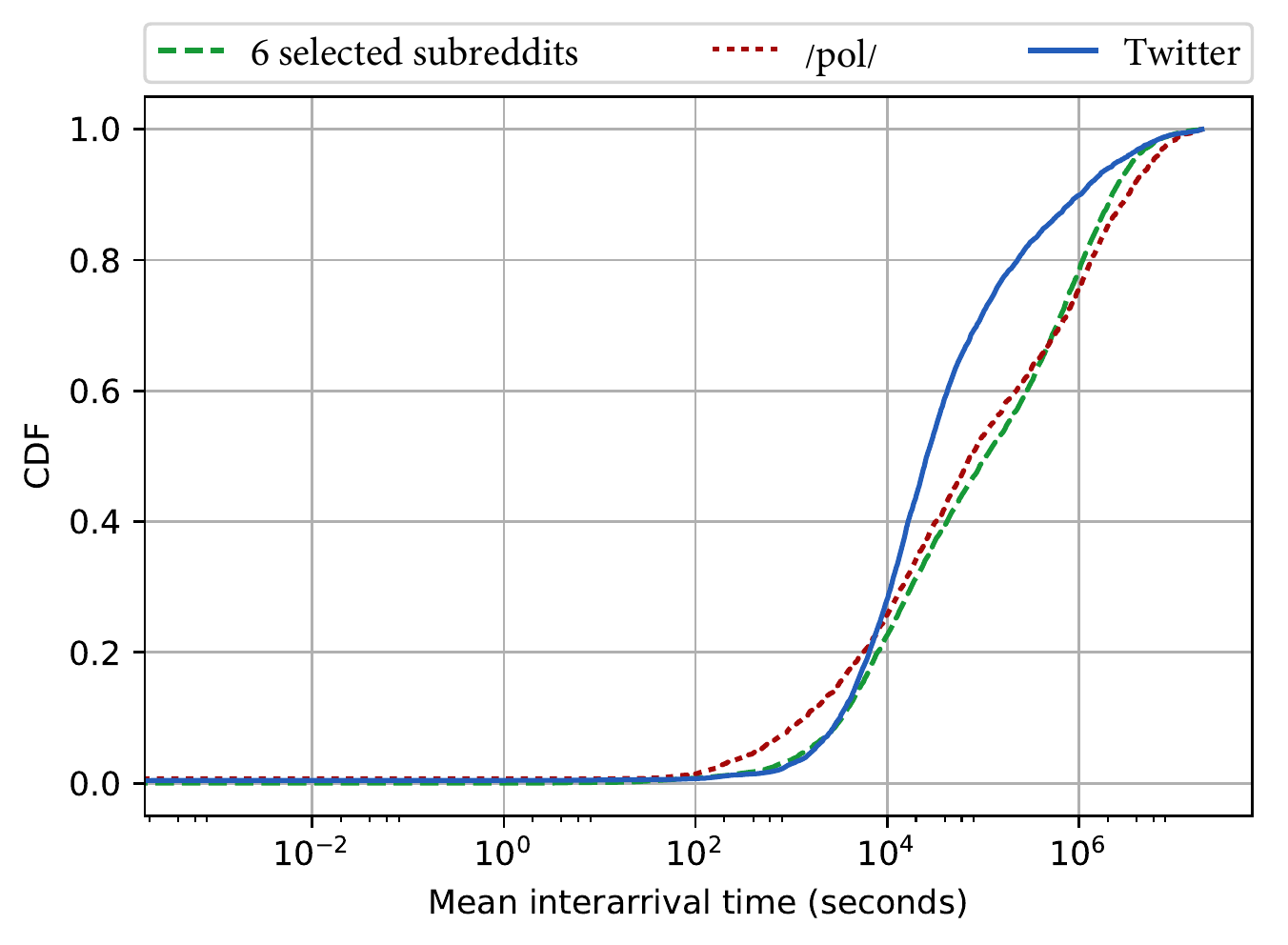}}
\subfigure[ ]{\includegraphics[width=0.49\textwidth]{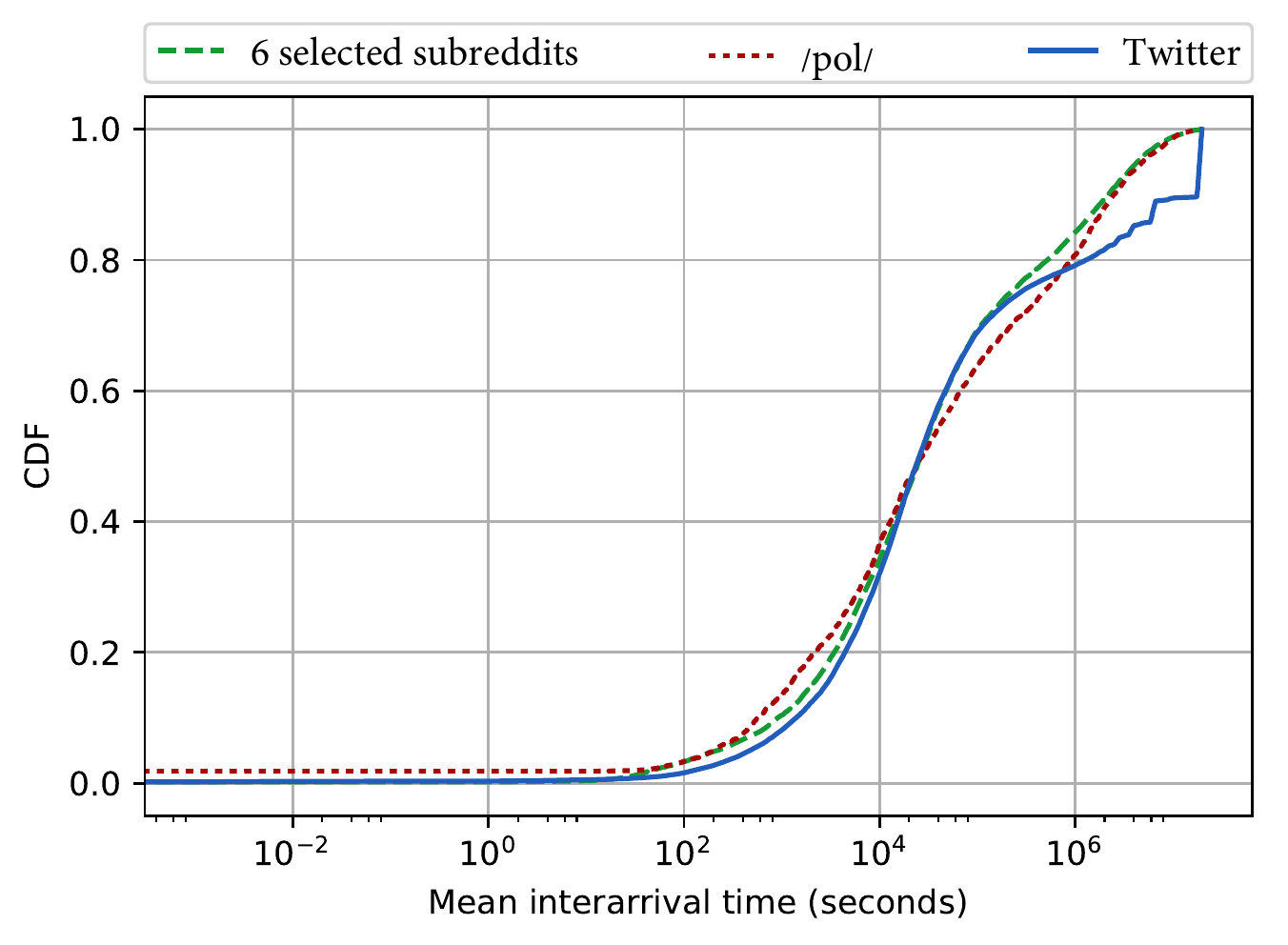}}
\subfigure[]{\includegraphics[width=0.49\textwidth]{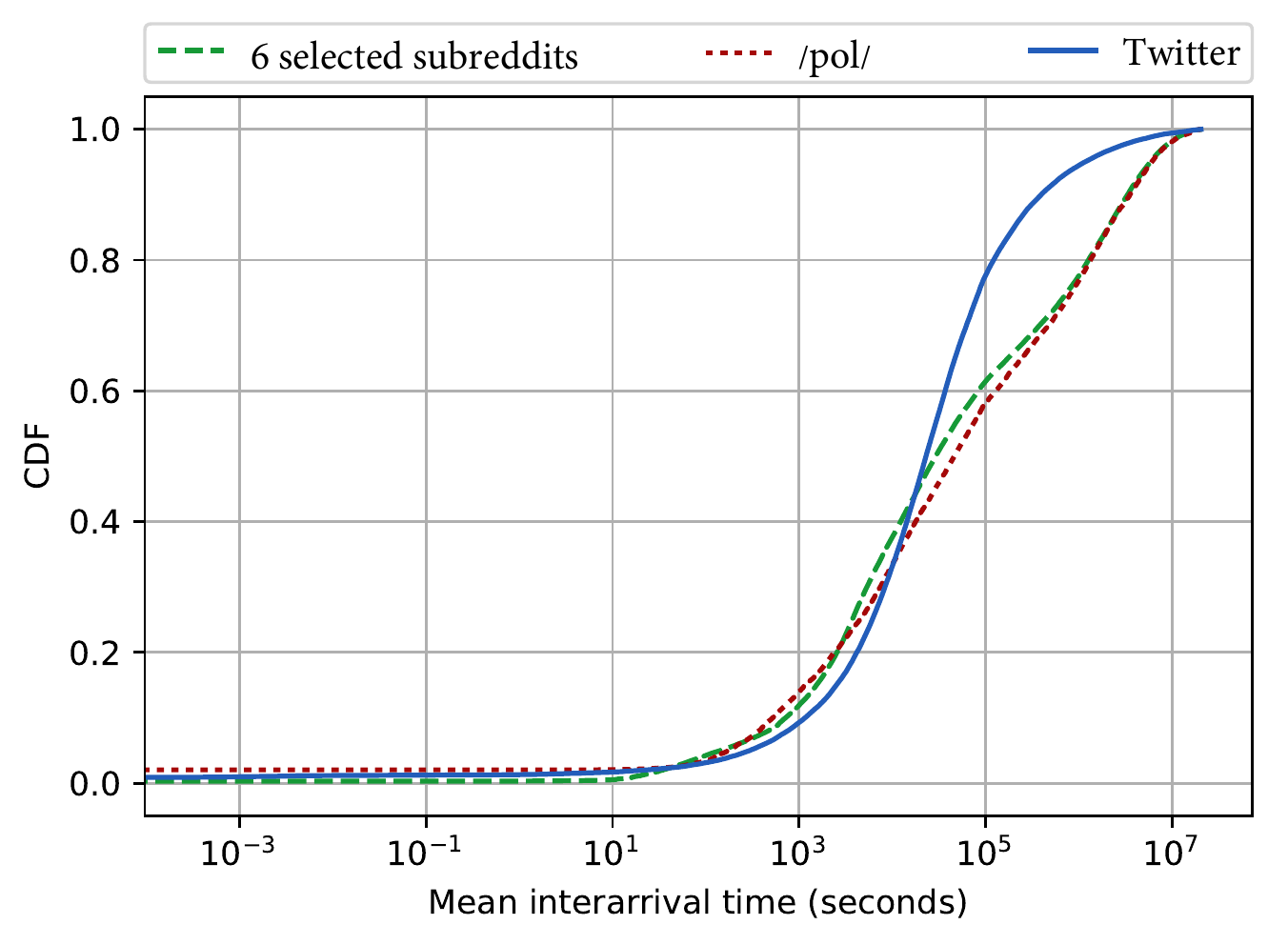}}
\caption{CDF for mean inter-arrival time for the URLs that occur more than once for (a) common alternative news URLs; (b) common mainstream news URLs; (c) all alternative news URLs, and (d) all mainstream news URLs.}
\label{cdf_interarrival_time}
\end{figure*}

\begin{figure*}[t]
\centering
\subfigure[]{\includegraphics[width=0.49\textwidth]{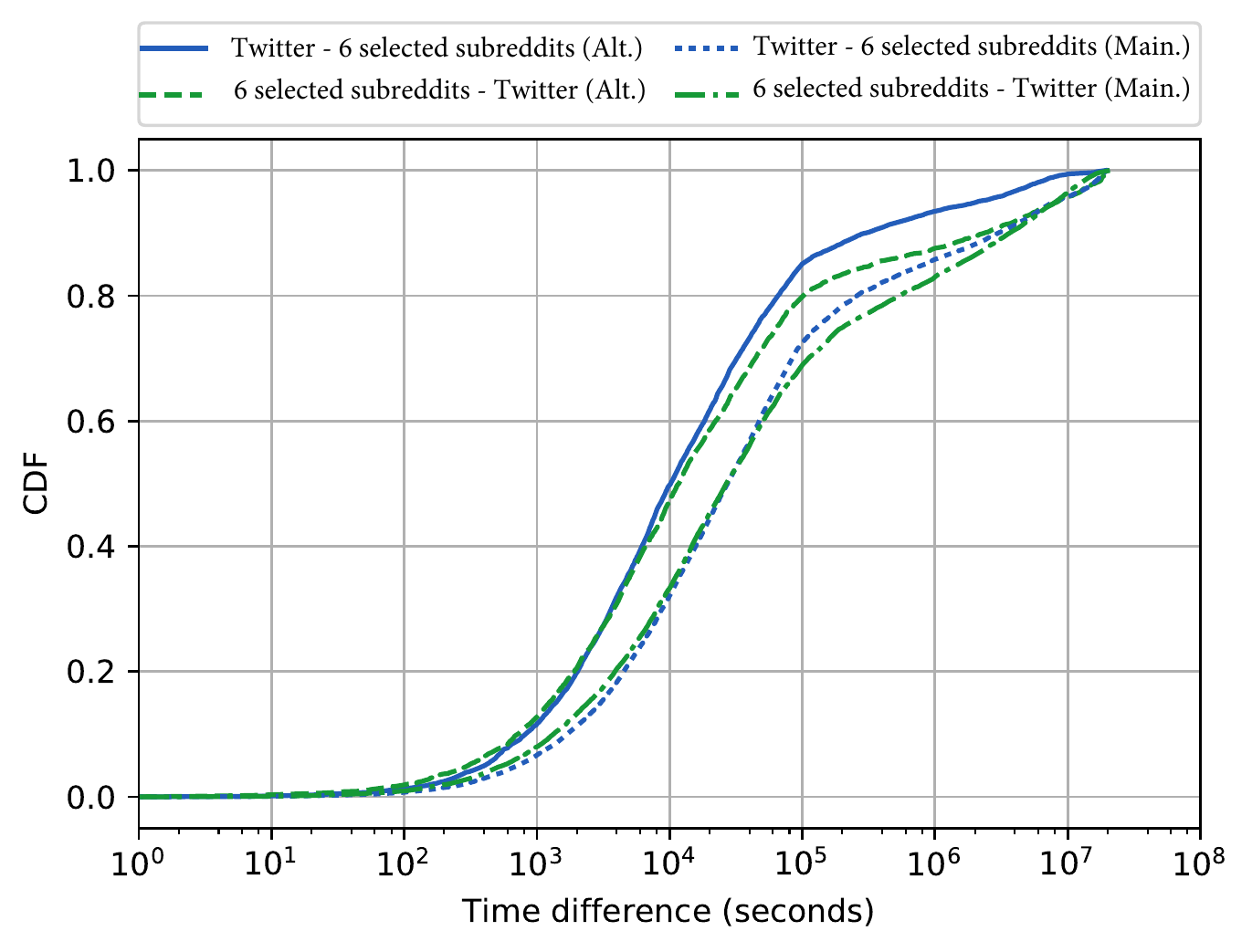}\label{cdf_time_difference-rt}}
\subfigure[]{\includegraphics[width=0.49\textwidth]{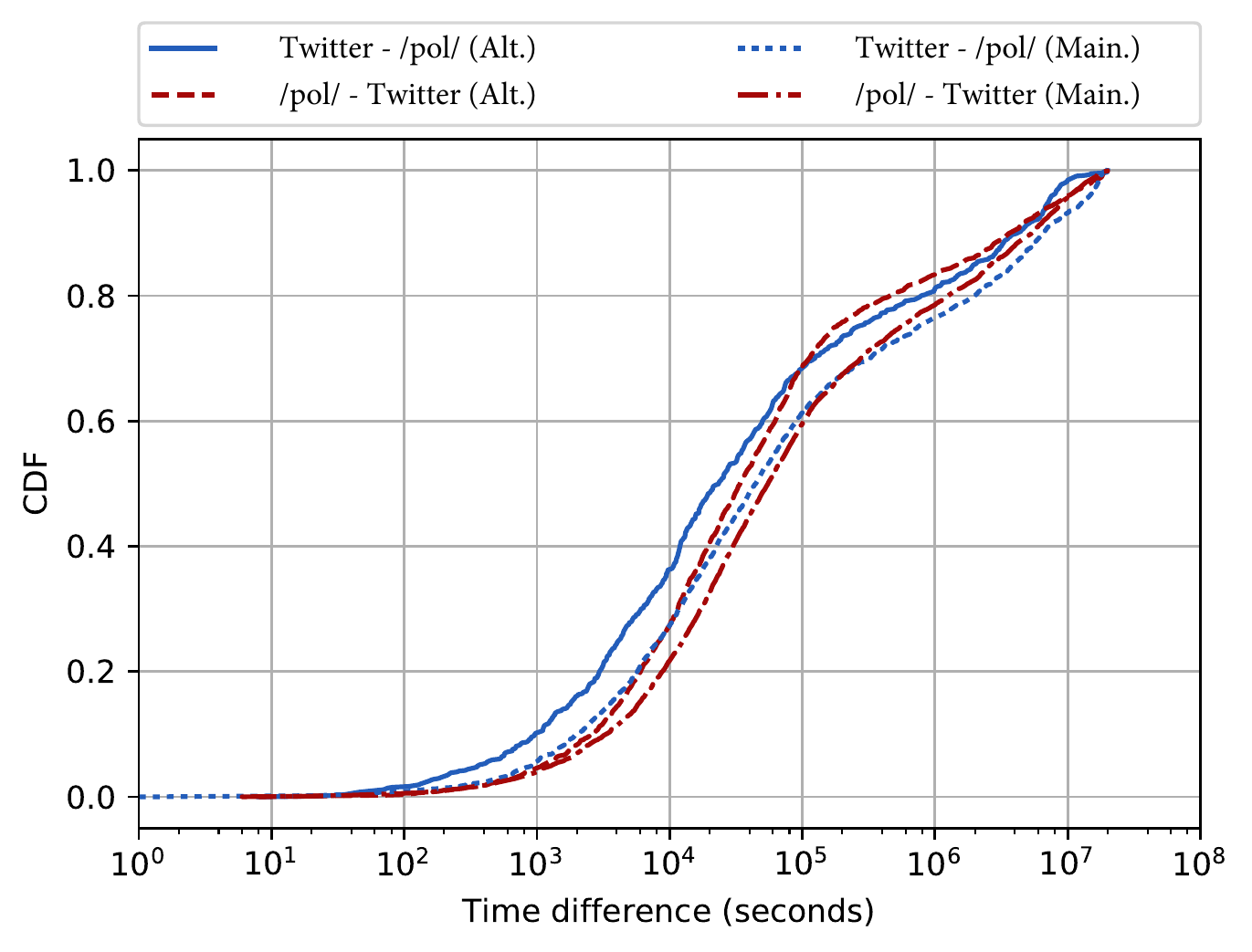}\label{cdf_time_difference-ct}}
\subfigure[]{\includegraphics[width=0.49\textwidth]{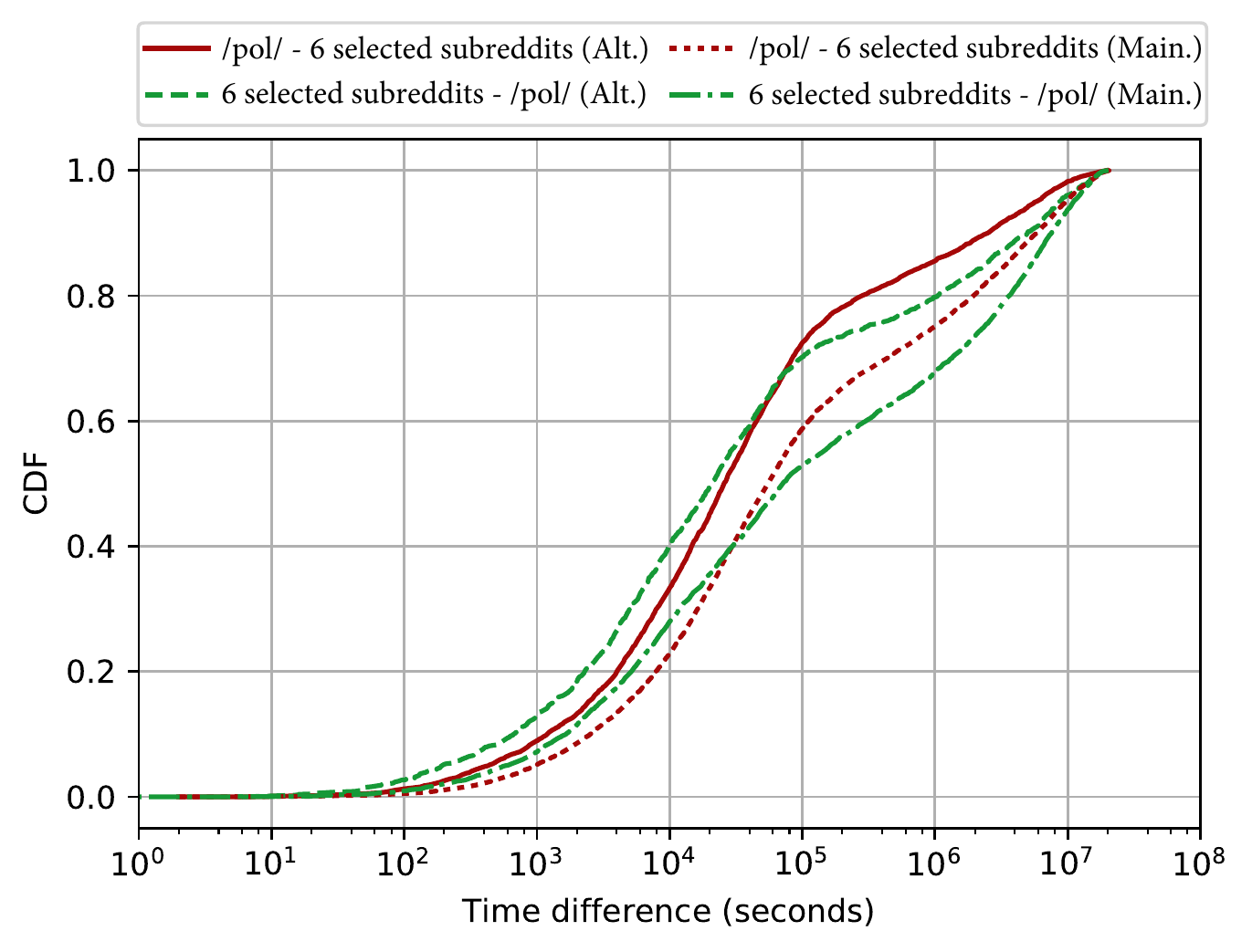}\label{cdf_time_difference-cr}}
\reduce
\caption{CDF of the difference between the first occurrence of a URL between (a) six selected subreddits and Twitter, (b) \dspol and Twitter, and (c) \dspol and six selected subreddits.}
\label{cdf_time_difference}
\end{figure*}

\begin{table}[t]
\centering
\resizebox{0.7\columnwidth}{!}{
\begin{tabular}{@{}llrr@{}}
\toprule
\textbf{Comparison} & \textbf{Type of  News} & \textbf{\begin{tabular}[c]{@{}c@{}}\#URLs where \\ platform 1 is faster\end{tabular}} & \textbf{\begin{tabular}[c]{@{}c@{}}\#URLs where \\ platform 2 is faster\end{tabular}} \\ \midrule
Reddit vs Twitter   & Mainstream         & 18,762                                                                                 & 11,416                                                                                 \\
                    & Alternative        & 5,232                                                                                  & 4,301                                                                                  \\ \midrule
\dspol vs Twitter    & Mainstream         & 2,938                                                                                  & 4,700                                                                                  \\
                    & Alternative        & 778                                                                                    & 2,099                                                                                  \\ \midrule
\dspol vs Reddit     & Mainstream         & 5,382                                                                                  & 14,662                                                                                 \\
                    & Alternative        & 1,455                                                                                  & 3,695                                                                                   \\ \bottomrule
\end{tabular}
}
\caption{Statistics of URLs for the comparisons of time difference between platforms. Reddit refers to the six selected subreddits.}
\label{tbl:time_difference_statistics}
\end{table}

We also study the inter-arrival time of reposted URLs.
Fig.~\ref{cdf_interarrival_time} shows the CDF of the mean inter-arrival time of URLs that appear more than one time in each platform.
Each platform exhibits unique behavior, confirmed by a two sample Kolmogorov-Smirnov test showing significant differences between the distributions ($p < 0.01$ for each pairwise comparison).
However, \dspol and the six subreddits exhibit similar time-related sharing behavior for both mainstream and alternative news URLs, and Twitter has smaller mean inter-arrival time overall.
Interestingly, the six subreddits appear to have a duality in reposting behavior: for URLs with small inter-arrival time, it follows the faster pace of Twitter, whereas, for URLs with longer inter-arrival times, it follows \dspol.

\subsubsection{Cross Platform Analysis}

We now look at URLs that appear on more than one platform and study the time at which they are shared.
Fig.~\ref{cdf_time_difference} plots the CDF of the time difference (in seconds) between the first occurrence of a
URL on pairs of platforms, while Table~\ref{tbl:time_difference_statistics} reports the numbers of URLs involved in each comparison.

We make the following observations: first, when comparing pairs of distributions for a given category of URLs, they are
statistically different (a two sample Kolmogorov-Smirnov test rejects the null hypothesis with $p<10^{-4}$).
Second, alternative news appear on multiple platforms faster than mainstream news.
This is consistent regardless of the pair of platforms we consider, and the sequence of appearances
(i.e., first in platform A and then B, vs. first in B and then in A).
Third, we notice the presence of a ``turning point'' with respect to the delay between URL appearance on
each platform, which seems to be consistent across all pairs of platforms and types of news, and matches the 24h period observed earlier.
Finally, there is a cross point when comparing URLs first posted on platform A and then on B, and URLs which were posted first in B and
then on A (i.e., when the lines for the same type of URLs cross).
Such a point represents which portion of URLs appear faster in one platform than the other.
For the Twitter-six selected subreddits comparison, alternative (mainstream) news appear faster on Twitter than
the six subreddits $80\%$ of the time ($50\%$), with these URLs exhibiting slower propagation, since the turning point is at $\sim$1 hour (5 hours).
Similarly, for the Twitter-\dspol comparison, alternative (mainstream) news appear faster on Twitter than \dspol $70\%$ ($5\%$) of the time,
with the turning point at 1 day (2 days).
Finally, for the six selected subreddits-\dspol comparison, alternative (mainstream) news appear faster on the six subreddits than \dspol for $65\%$ ($40\%$) of the time,
with the turning point around 18 hours (12 hours).

Next, given the set of unique URLs across all platforms and the time they appear for the first time, we analyze their appearance in one,
two, or three platforms, and the order in which this happens.
For each URL, we find the first occurrence on each platform and build corresponding ``sequences,'' e.g., if a URL first appears on the six  subreddits (Reddit) and subsequently on \dspol (4chan), the sequence is Reddit$\rightarrow$ 4chan (R$\rightarrow$4).
Table~\ref{links_sequences_all} reports the distribution of the sequences of appearances considering only the first hop, i.e.,
up to the first two platforms in the sequence.
The majority of URLs only appear on one platform: 82\% of alternative URLs and 89\% of mainstream news URLs.
Also, both alternative and mainstream news URLs tend to appear on the six subreddits first and later appear on either Twitter or \dspol, %
and
on Twitter before \dspol.

\begin{table}[t]
\centering
\small
\begin{tabular}{@{}lrr@{}}
\toprule
Sequence & Alternative (\%) & Mainstream (\%) \\ \midrule
4 only       & 3,236\hfill (4.4\%)     & 18,654\hfill (3.7\%)   \\
4$\rightarrow$R      & 1,118\hfill (1.5\%)      & 4,606\hfill (0.9\%)    \\
4$\rightarrow$T      & 315\hfill (0.5\%)      & 861\hfill (0.17\%)     \\ \midrule
R only       & 24,292\hfill (33.3\%)    & 230,602\hfill (46.1\%)  \\
R$\rightarrow$4      & 2,181\hfill (3.0\%)     & 11,307\hfill (2.3\%)    \\
R$\rightarrow$T      & 4,769\hfill (6.5\%)     & 16,685\hfill (3.35\%)   \\ \midrule
T only       & 32,443\hfill (44.5\%)    & 204,836\hfill (41\%)   \\
T$\rightarrow$4      & 585\hfill (0.8\%)      & 1,345\hfill (0.26\%)     \\
T$\rightarrow$R      & 3,964\hfill (5.5\%)     & 10,640\hfill (2.12\%)    \\ \bottomrule
\end{tabular}
\caption{Distribution of URLs according to the sequence of first appearance within platforms for all URLs, considering only the first hop. ``4'' stands for \dspol (4chan), ``R'' for the six selected subreddits (Reddit), and ``T'' for Twitter.}
\label{links_sequences_all}
\end{table}

We also study the temporal dynamics of URLs that appear on all three platforms, with triplets of sequences.
Table~\ref{links_sequences_common} reports the distribution of these sequences.
The most common sequences are similar for both alternative and mainstream news URLs:
R$\rightarrow$T$\rightarrow$4, R$\rightarrow$4$\rightarrow$T, and T$\rightarrow$R$\rightarrow$4 are the top three sequences.
As already mentioned, the six selected subreddits ``outperform'' both other platforms in terms of the speed of sharing mainstream and alternative news URLs, as evidenced by the fact that it is at the head of the sequence for 51\% and 59\% of alternative and mainstream news URLs, respectively.

\begin{table}[t]
\centering
\small
\begin{tabular}{@{}lrr@{}}
\toprule
Sequence & Alternative (\%) & Mainstream (\%) \\ \midrule
4$\rightarrow$R$\rightarrow$T    & 128\hfill (5.5\%)      & 552\hfill (8.9\%)     \\
4$\rightarrow$T$\rightarrow$R    & 145\hfill (6.2\%)      & 290\hfill (4.7\%)     \\
R$\rightarrow$4$\rightarrow$T    & 335\hfill (14.4\%)      & 1,525\hfill (24.5\%)    \\
R$\rightarrow$T$\rightarrow$4    & 841\hfill (36.3\%)      & 2,189\hfill (35.3\%)    \\
T$\rightarrow$4$\rightarrow$R    & 192\hfill (8.2\%)      & 486\hfill (7.8\%)     \\
T$\rightarrow$R$\rightarrow$4    & 673\hfill (29\%)       & 1,166\hfill (18.8\%)    \\ \bottomrule
\end{tabular}
\caption{Distribution of URLs according to the sequence of first appearance within a platform for URLs common
to all platforms. ``4'' stands for \dspol (4chan), ``R'' for the six selected subreddits (Reddit), and ``T'' for Twitter.}
\label{links_sequences_common}
\end{table}

Finally, we analyze the source of the URLs for each of the three platforms, as follows.
We create two directed graphs, one for each type of news, $\boldsymbol{G}=(\boldsymbol{V},\boldsymbol{E})$, where
$\boldsymbol{V}$ represents alternative or mainstream domains, as well as the three platforms, and $\boldsymbol{E}$ the set
of sequences that consider only the first-hop of the platforms.
For example, if a \url{breitbart.com} URL appears first on Twitter and later on the six selected subreddits, we add an edge
from \url{breitbart.com} to Twitter, and from Twitter to the six selected subreddits.
We also add weights on these edges based on the number of such unique URLs.
By examining the paths %
we can discern which domains' URLs tend to appear first on each of the platforms.

\begin{figure}[t]
\centering
\subfigure[]{\includegraphics[width=0.49\columnwidth]{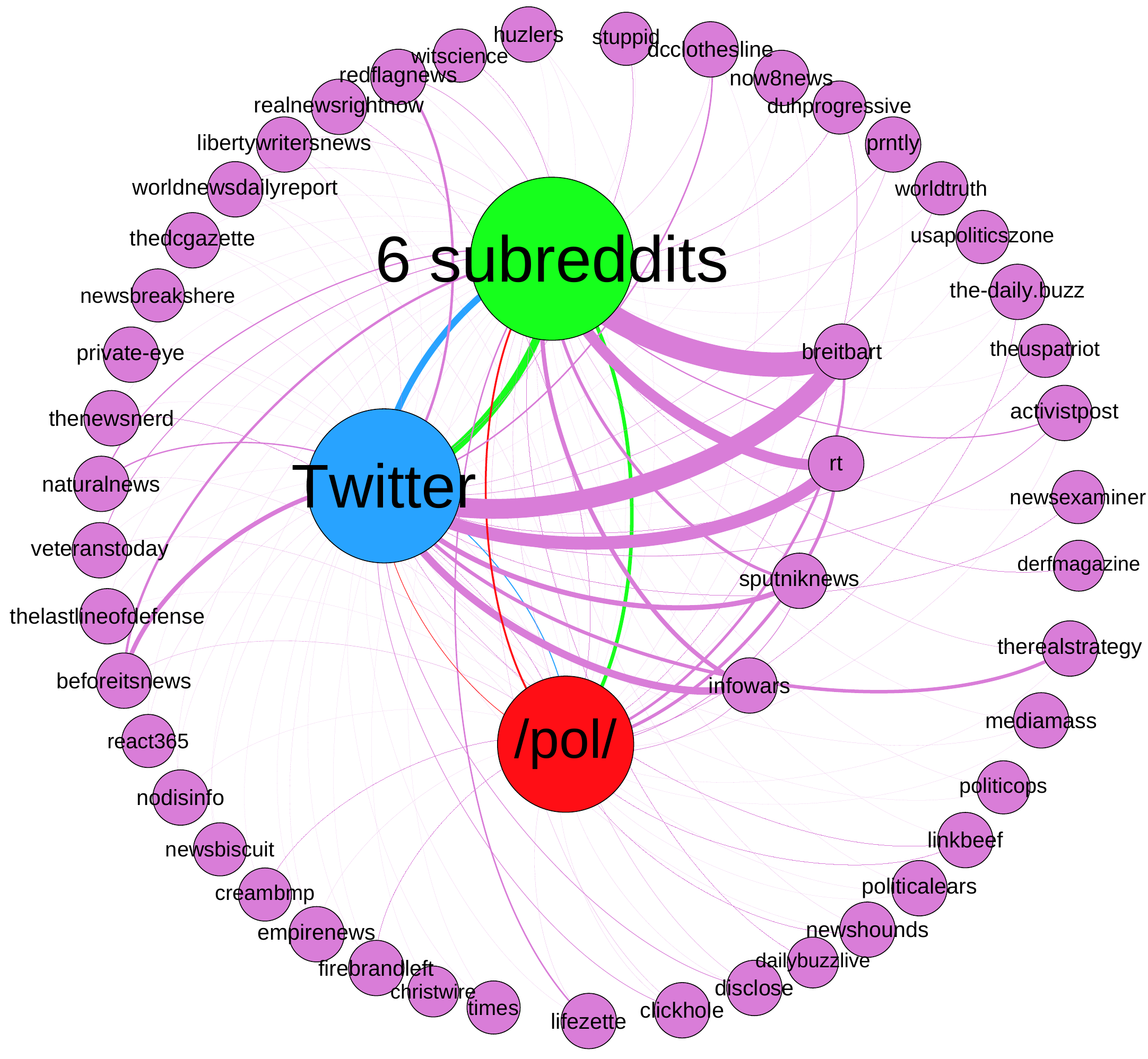}\label{timeline_all_fake-a}}
\subfigure[]{\includegraphics[width=0.49\columnwidth]{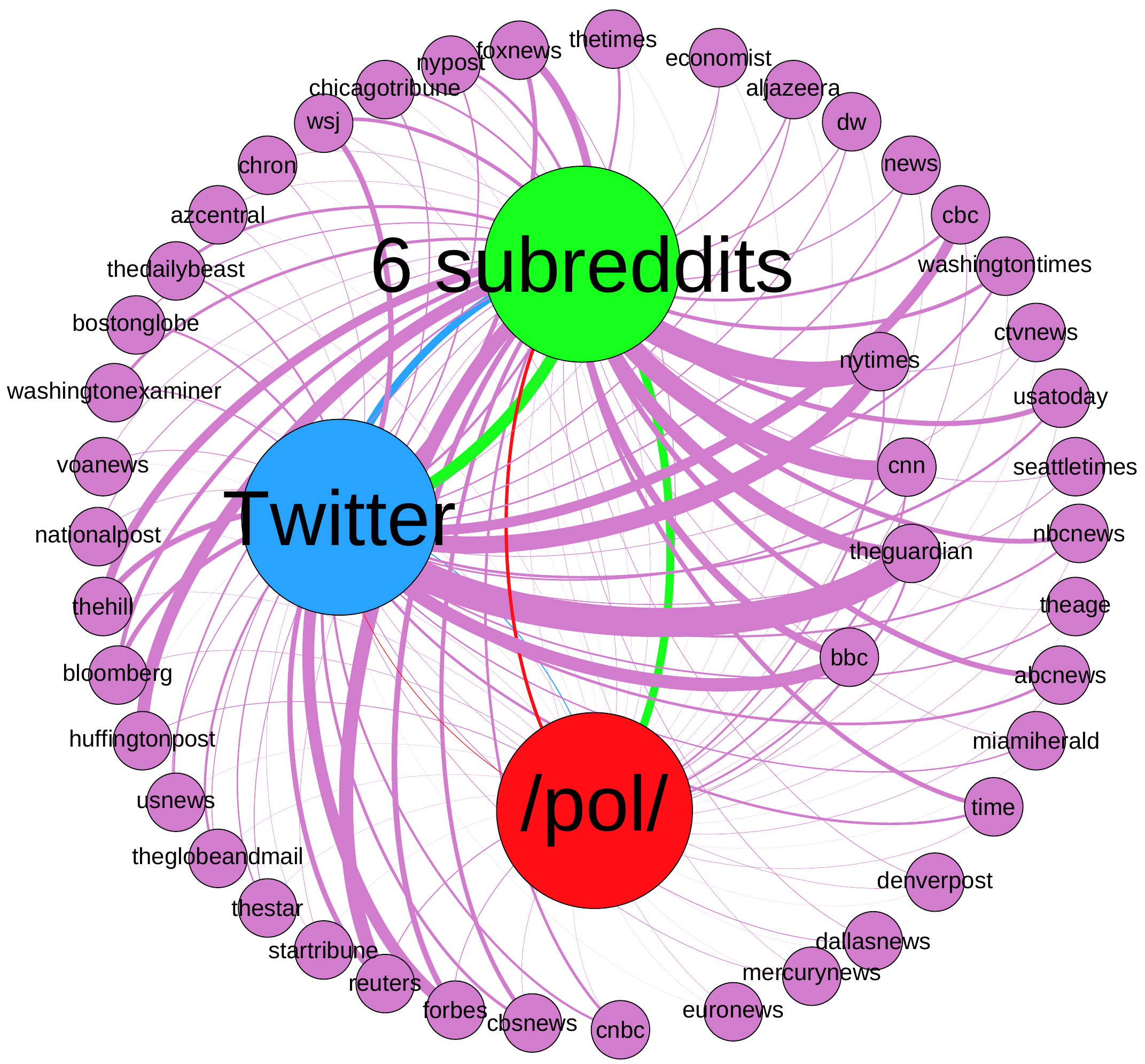}\label{timeline_all_fake-b}}
\reduce
\caption{Graph representation of news ecosystem (a) alternative news domains and (b) mainstream news domains. Edges are colored the same as their source node.}
\label{timeline_all_fake}
\end{figure}

Fig.~\ref{timeline_all_fake} shows the graphs built for alternative and mainstream domains.
Comparing the thickness of the outgoing edges, one can see that \url{breitbart.com} URLs appear first in
the six selected subreddits more often than on Twitter and more frequently than on \dspol.
However, for other popular alternative domains, such as \url{infowars.com}, \url{rt.com}, and \url{sputniknews.com}, URLs
appear first on Twitter more often than the six selected subreddits and \dspol. Also, \dspol is rarely the platform where a URL first shows up.
For the mainstream news domains, we note that URLs from \url{nytimes.com} and \url{cnn.com} tend to appear first more often on the selected subreddits
than Twitter and \dspol,
however, URLs from other domains like \url{bbc.com} and \url{theguardian.com} tend to appear first more often on Twitter than the selected subreddits.
Similar to the alternative domains graph, there is no domain where \dspol dominates in terms of first URL appearance.

\subsection{Influence Estimation}
Thus far, our measurements have shown relative differences in how news media is shared on Reddit, Twitter, and 4chan.
In this section, we provide meaningful evidence of how the individual platforms influence the media shared on other platforms.
We do so by using a mathematical technique known as Hawkes processes.
These statistical models can be used for modeling the dissemination of information in Web communities~\cite{farajtabar2017fake} as well as measuring social influence~\cite{guo2015bayesian}.
For more details regarding the Hawkes Processes and the general methodology used we refer the interested reader to Section~\ref{sec:hawkes_background}.

\subsubsection{Methodology}
We now provide more details about our experiments, once again, considering 4chan (\dspol), Twitter, and the six subreddits.
We study Hawkes processes at the subreddit granularity to get a better understanding of the various platforms and particular subreddits.

We aim to examine how these platforms and subreddits influence each other, so we model the arrival of URLs, in posts or tweets, with a Hawkes model with $K=8$ point processes---one for Twitter, one for \dspol, and one for each of the subreddits.
The model is fully connected, i.e., it is possible for each process to influence all the others, as well as itself, which describes behavior where participants on a platform see a URL and re-post it on the same platform.

We select URLs that have at least one event in Twitter, \dspol, and at least one of the subreddits, and we model each URL individually.
The missing Twitter data affects 3,1K (37\%)  URLs.
One way to mitigate the impact of this missing data is to remove events for which it has a larger impact.
E.g., if an event spans 100 days, the missing Twitter data has less of an effect than if the event only spanned two days.
Thus, we examine URLs from  other platforms that overlap with any of the missing days and remove the 10\% of URLs (895) with the shortest total duration from the first event recorded until the last event recorded.
This results in the missing data making up a smaller portion of the overall duration of the events.

\begin{table*}[h]
\centering

\resizebox{\textwidth}{!}{%
  \begin{tabular}{llrrrrrrrr}
  \toprule
      &        &     \textbf{The\_Donald} & \textbf{worldnews} &   \textbf{politics} &  \textbf{news} &   \textbf{conspiracy} &   \textbf{AskReddit} &  \textbf{/pol/} &  \textbf{Twitter} \\
  \midrule
\textbf{URLs}  & Mainstream &  3,097 &   2,523 &  3,578 &  2,584 &   907 &  841 &   5,589 &     5,589 \\
      & Alternative &  2,008 &    252 &   813 &   362 &   321 &  100 &   2,136 &     2,136 \\
      &   Total &  5,105 &   2,775 &  4,391 &  2,946 &  1,228 &  941 &   7,725 &     7,725 \\

  \midrule

\textbf{Events} & Mainstream &  12,312 &   7,517 &  26,160 &  5,794 &  1,995 &  2,302 &  19,746 &    36,250 \\
       & Alternative &   7,797 &    458 &   2,484 &   586 &   497 &   176 &   7,322 &    23,172 \\
       &Total &  20,109 &   7,975 &  28,644 &  6,380 &  2,492 &  2,478 &  27,068 &    59,422 \\

\midrule
\textbf{Mean $\lambda_0$ } & Mainstream &  0.001502 &  0.001382 &  0.001265 &  0.001392 &  0.000501 &  0.000107 &  0.001564 &  0.002330 \\
                  & Alternative &  0.001627 &  0.000619 &  0.000696 &  0.000553 &  0.000423 &  0.000034 &  0.001525 &  0.002803 \\
\bottomrule
\end{tabular}
}
\caption{Total URLs with at least one event in Twitter, \dspol, and at least one of the subreddits; total events for mainstream and alternative URLs, and the mean background rate ($\lambda_0$) for each platform/subreddit. %
}
\label{tbl:hawkes_centipede}
\end{table*}

The number of remaining URLs and events included for each platform are shown in Table~\ref{tbl:hawkes}.
Next, we fit a Hawkes model for each URL and calculate the influence results using the approach described in Section~\ref{sec:hawkes_background}.

\subsubsection{Results}
\begin{figure}[t]
\centering
\includegraphics[width=0.8\columnwidth]{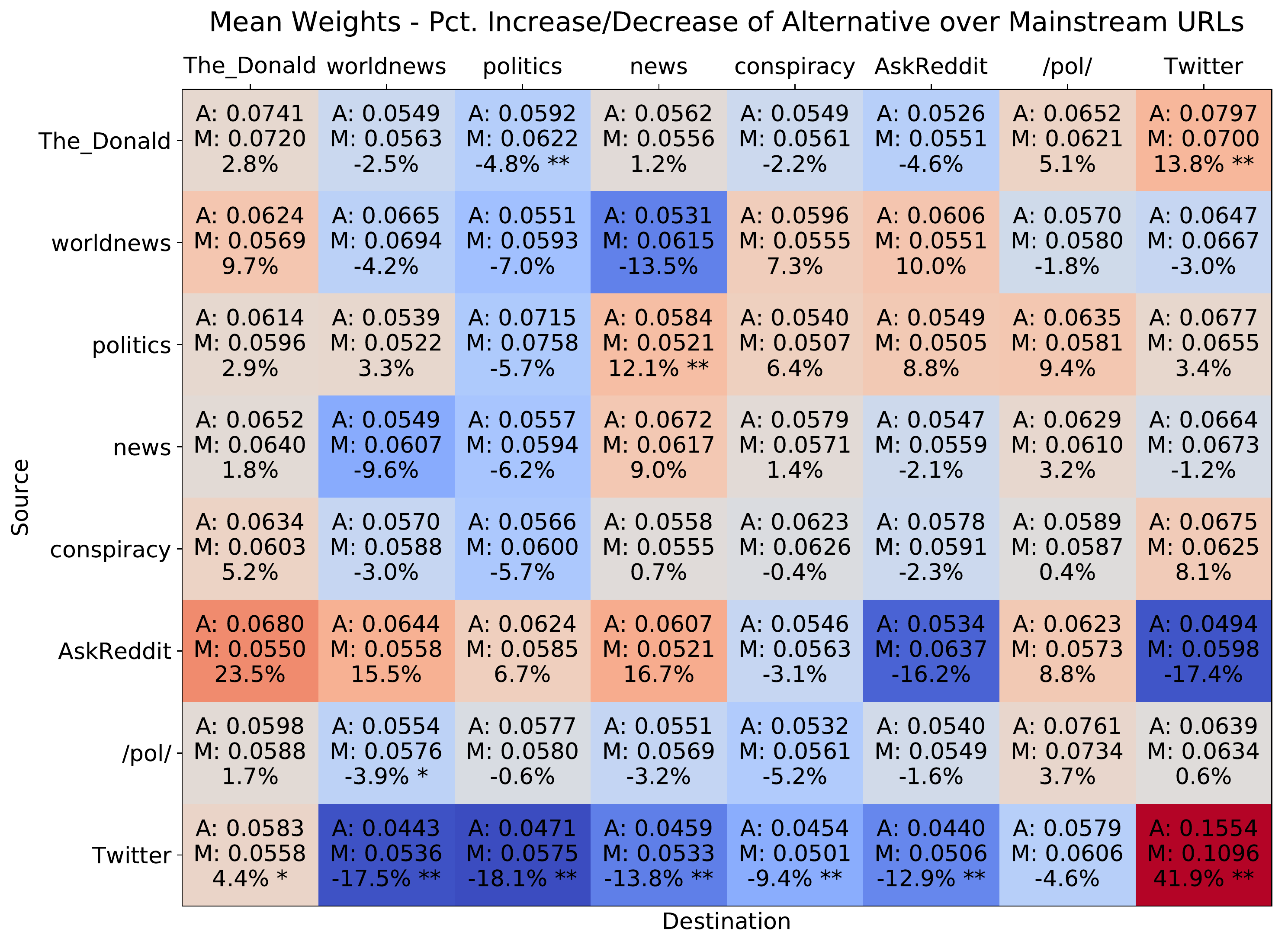}
\caption{The mean weights for alternative URLs (A), the mean weights for mainstream URLs (M), and the percent increase/decrease between mainstream and alternative (also indicated by the coloration). 
The stars on the cells indicate the level of statistical significance (p-value) between the weight distributions: no stars indicate no statistical significance, whereas * and ** indicate statistical significance with $p < 0.05$ and $p < 0.01$ respectively. }
\reduce
\label{fig:mean-weights-difference}
\end{figure}

Looking at the number of URLs in Table~\ref{tbl:hawkes_centipede}, we note that there are substantially more events for mainstream than alternative news URLs.
However, for Twitter, \dspol, and The\_Donald, the ratios of events to URLs for alternative news URLs are similar to or greater than the ratios for mainstream ones.
These high ratios explain the high background rates (see~Table~\ref{tbl:hawkes_centipede}) for alternative news URLs for these platforms despite the lower number of events.

From the Hawkes models for each URL, we obtain the weight matrix $W$ which specifies the strength of the connections between the different platforms and subreddits.
The mean weight values over all URLs for alternative and mainstream news URLs, as well as the percentage difference between them are presented in Fig.~\ref{fig:mean-weights-difference}.
First, we look at Twitter.
Background rates are high for both mainstream and alternative news URLs, which is not surprising given the large number of users on the platform.
The values for $W_{\mathrm{Twitter} \to \mathrm{Twitter}}$ are also substantially higher than all other weights: $0.1096$ for mainstream news URLs and $0.1554$ for alternative news URLs.
This reflects the ease and common practice of re-tweeting: a URL in a tweet is likely to generate other events as users re-tweet it.
There are different possible explanations for why the Twitter to Twitter rate for alternative news URLs is much greater than the rate for mainstream news URLs.
The first is bot activity---if automated Twitter bots are used to spread alternative news URLs, it could result in a much higher rate of tweeting and re-tweeting.
Another possible explanation is the behavior of users who read news stories from alternative sources; they might be more inclined to re-tweet the URL~\cite{gupta2013faking}.

Looking at the weights for Twitter to the other platforms, except The\_Donald, they are all greater for mainstream news URLs, meaning that the average tweet containing a mainstream URL is more likely to cause a subsequent post on the other platforms than the average tweet containing an alternative URL.
The next communities most likely to cause events on others are The\_Donald and \dspol.
It is worth noting that The\_Donald is the only platform/subreddit that has greater alternative URL weights for all of its inputs.
Assuming that the population of The\_Donald users that also read, say, worldnews is the same for both alternative and mainstream news URLs---which is reasonable---then the difference in weights implies that the users have a stronger preference for re-posting alternative news URLs back to The\_Donald than for mainstream news URLs.
The opposite can be seen for worldnews and politics, where most of the input weights are stronger for mainstream news.
However, despite the higher weights for alternative news URLs, The\_Donald is also, interestingly, influenced more strongly by mainstream news URLs than alternative news URLs on all platforms, with the exception of Twitter.  This is in part because of the greater number of mainstream URL events, but The\_Donald also has a higher background rate for alternative news URLs than mainstream news URLs, which implies that a lot of the alternative news URLs on the platform are coming from other sources.

\begin{figure}[t]
\centering
\includegraphics[width=0.8\columnwidth]{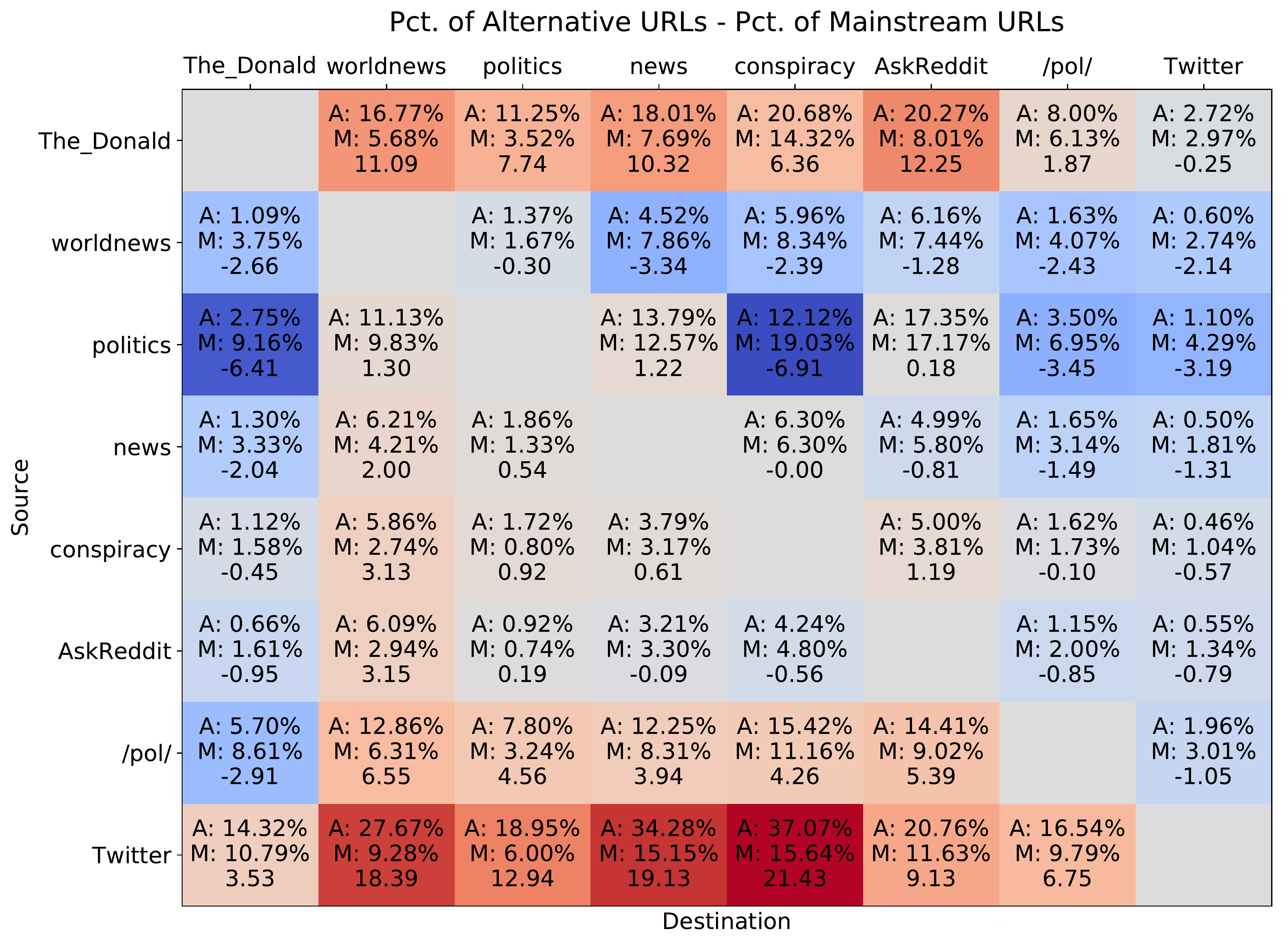}
\reduce
\caption{The estimated mean percentage of alternative URL events caused by alternative news URL events (A), the estimated mean percentage of mainstream news URL events caused by mainstream news URL events (M), and the difference between alternative and mainstream news (also indicated by the coloration).}
\label{fig:percent-events-difference}
\end{figure}

To assess the statistical significance of the results, we perform two sample Kolmogorov-Smirnov tests on the weight distributions of mainstream and alternative news URLs for each source-destination pair (depicted as stars in Fig.~\ref{fig:mean-weights-difference}).
This allow us to assess whether the distributions of the weights for mainstream and alternative news URLs have statistically significant differences, hence indicating whether mainstream news URLs spread differently compared to alternative news URLs across the Web communities we study. 
Unsurprisingly, many of the source-destination pairs have no significant difference. 
However, in most cases where Twitter is the source community there \emph{is} a significant statistical difference with $p < 0.01$.
I.e., for some communities, Twitter is used not just to disseminate news, but to disseminate news from a specific \emph{type} of source.

Fig.~\ref{fig:percent-events-difference} illustrates the estimated total influence of the different platforms on each other, for both mainstream and alternative news URLs.
Twitter contributes heavily to both types of events on the other platforms---and is in fact the most influential single source for most of the other platforms.
Despite Twitter's lower weights for alternative news URLs, it actually has a greater influence on alternative than mainstream news URLs, in terms of percentage of events caused, on all the other platforms/subreddits.
This is due to the fact that, even though it has lower weights, the largest proportion of alternative URL events are on Twitter.
After Twitter, The\_Donald and \dspol also have a strong influence on the alternative news URLs that get posted on other platforms.  The\_Donald has a stronger effect for alternative news URLs on all platforms except Twitter---although it still
has the largest alternative influence on Twitter, causing an estimated 2.72\% of alternative news URLs tweeted.
Interestingly, The\_Donald causes 8\% of \dspol's alternative news URLs, while \dspol's influence on The\_Donald is less, at 5.7\%.
For the mainstream news URLs the strength of influence is reversed.
Specifically, \dspol's influence on The\_Donald is 8.61\% whereas The\_Donald's influence on \dspol is 6.13\%.

In descending order, the influences on Twitter for mainstream news URLs are politics (4.29\%), \dspol (3.01\%), The\_Donald (2.97\%), worldnews (2.74\%), news (1.81\%), AskReddit (1.34\%), and conspiracy (1.04\%).
The strongest influences for alternative news URLs are, unsurprisingly, The\_Donald (2.72\%) and \dspol (1.96\%), followed by politics (1.10\%), worldnews (0.60\%), AskReddit (0.55\%), news (0.50\%), and conspiracy (0.46\%).  Twitter influences the alternative news URLs on other platforms to a large degree---but the largest alternative URL inputs to Twitter are The\_Donald and \dspol.
While we are only looking at a closed system of 8 different platforms and subreddits, we note that Twitter is undoubtedly effective at propagating information.
Thus the influence these two communities have on Twitter is likely to have a disproportional impact on the greater Web compared to their relatively minuscule userbase.

\subsection{Remarks}
In this work, we explored how mainstream and fringe Web communities share mainstream and alternative news sources with a particular focus on how communities influence each other.
We collected millions of posts from Twitter, Reddit, and 4chan, and analyzed the occurrence and temporal dynamics of news shared from 45 mainstream and 54 alternative news sites.
We found that users on various platforms prefer distinct news sources, especially when it comes to alternative ones.
We also explored complex temporal dynamics and we discovered, for example, that Twitter and Reddit users tend to post the same stories within a relatively short period of time, with 4chan posts lagging behind both of them.
However, when a story becomes popular after a day or two, it is usually the case it was posted on 4chan first, lending some credence to 4chan's supposed influence on the Web.

Using Hawkes processes, we also modeled the influence the individual platforms have on each other, while also taking into account influence that comes from external sources of information.
We found that the interplay between platforms manifests in subtle, yet meaningful ways.
For example, of all the platforms and subreddits, Twitter by far has the most influence in terms of the number of URLs it causes to be posted to other platforms, and contributes to the share of alternative news URLs on the other platforms to a much greater degree than to the share of mainstream news URLs.
After Twitter, The\_Donald subreddit and \dspol are the next most influential when it comes to alternative news URLs.
For such URLs, The\_Donald is less influenced by the other platforms than \dspol, and has a higher background rate, i.e., more of the URLs posted there come from other sources.

To the best of our knowledge, our analysis constitutes the first attempt to characterize the dissemination of mainstream and alternative news across multiple social media platforms, and to estimate a quantifiable influence between them.
Overall, our findings shed light on how Web communities influence each other and can be extremely useful to better understand and detect false information as well as informing the design of systems that aim to trace the origins of fake stories and mitigate their dissemination.

\section{Detecting and Understanding the Spread of Memes Across Multiple Web Communities}

\subsection{Motivation}

The Web has become one of the most impactful vehicles for the propagation of ideas and culture.
Images, videos, and slogans are created and shared online at an unprecedented pace.
Some of these, commonly referred to as \emph{memes}, become viral, evolve, and eventually enter popular culture.
The term ``meme'' was first coined by Richard Dawkins \cite{dawkins1976selfish}, who framed them as cultural analogues to genes,
as they too self-replicate, mutate, and respond to selective pressures~\cite{graham2005genes}.
Numerous memes have become integral part of Internet culture, with well-known examples
including the Trollface~\cite{trollface_meme}, Bad Luck Brian~\cite{bad_luck_brian_meme}, and Rickroll~\cite{rickroll_meme}.

While most memes are generally ironic in nature, used with no bad intentions, others have assumed negative and/or hateful connotations, including outright racist and aggressive undertones~\cite{yoon2016not}.
These memes, often generated by fringe communities, are being ``weaponized'' and even becoming part of political and ideological propaganda~\cite{salon_weaponization}. 
For example, memes were adopted by candidates during the 2016 US Presidential Elections as part of their iconography~\cite{guardian_memes_election}; in October 2015, then-candidate Donald Trump retweeted an image depicting him as Pepe The Frog, 
a controversial character considered a hate symbol~\cite{adl_pepe_frog}.
In this context, polarized communities within 4chan and Reddit have been working hard to create new memes and make them go viral, aiming to increase the visibility of their ideas---a phenomenon known as ``attention hacking''~\cite{marwick2017media}.

Despite their increasingly relevant role, we have very little measurements and computational tools to understand the origins and the influence of memes.
The online information ecosystem is very complex; social networks do not operate in a vacuum but rather influence each other as to how information spreads~\cite{zannettou2017web}.
However, previous work 
has mostly focused on social networks in an isolated manner.

In this work, we aim to bridge these gaps by identifying and addressing a few research questions, which are oriented towards fringe Web communities:
1)~How can we characterize memes, and how do they evolve and propagate?
2)~Can we track meme propagation across multiple communities and measure their influence?
3)~How can we study variants of the same meme?
4)~Can we characterize Web communities through the lens of memes?

Our work focuses on four Web communities: Twitter, Reddit, Gab, and 4chan's Politically Incorrect board (\dspol), because of their impact on the information ecosystem~\cite{zannettou2017web} and anecdotal evidence of them disseminating weaponized memes~\cite{scott2018information}. %
We design a processing pipeline and use it over 160M images posted between July 2016 and July 2017. %
Our pipeline relies on perceptual hashing (pHash) and clustering techniques; the former extracts representative feature vectors from the images encapsulating their visual peculiarities, while the latter allow us to detect groups of images that are part of the same meme.
We design and implement a custom distance metric, based on both pHash and meme metadata, obtained from Know Your Meme (KYM), and use it to understand the interplay between the different memes.
Finally, using Hawkes processes, we quantify the reciprocal influence of each Web community with respect to the dissemination of image-based memes.

\descr{Findings.}
Some of our findings (among others) include: \smallskip
\begin{compactenum}
\item Our influence estimation analysis reveals that \dspol and \td are influential actors in the meme ecosystem, despite their modest size.
We find that \dspol substantially influences the meme ecosystem by posting a large number of memes, while \td is the most \emph{efficient} community in pushing memes to both fringe and mainstream Web communities.

\item Communities within 4chan, Reddit, and Gab use memes to share hateful and racist content. For instance, among the most popular cluster of memes, we find variants of the anti-semitic ``Happy Merchant'' meme~\cite{happy_merchant_meme} and the controversial Pepe the Frog~\cite{pepe_frog_meme}.

\item Our custom distance metric effectively reveals the phylogenetic relationships of clusters of images.
This is evident from the graph that shows the clusters obtained from \dspol, Reddit's \td subreddit, and Gab available for
exploration at~\cite{memes_graph_site}.

\end{compactenum}

\descr{Contributions.} 
First, we develop a robust processing pipeline for detecting and tracking memes across multiple Web communities.
Based on pHash and clustering algorithms, it supports large-scale measurements of meme ecosystems, while minimizing processing power and storage requirements.
Second, we introduce a custom distance metric,
geared to highlight hidden correlations between memes and better understand the interplay and overlap between them.
Third, we provide a characterization of multiple Web communities (Twitter, Reddit, Gab, and \dspol) with respect to the memes they share, and an analysis of their reciprocal influence using the Hawkes Processes statistical model.
Finally, we release our processing pipeline and datasets\footnote{\url{https://github.com/memespaper/memes_pipeline}}, in the hope to support further measurements in this space.

\subsection{Methodology}
\label{sec:methodology}
In this section, we present our methodology for measuring the propagation of memes across Web communities.

\subsubsection{Overview}
\label{sec:methodology:overview}

Memes are high-level concepts or ideas that spread within a culture~\cite{dawkins1976selfish}.
In Internet vernacular, a {\em meme} usually refers to variants of a particular image, video, clich\'e, etc.~that share a
common theme and are disseminated by a large number of users.
In this thesis, we focus on their most common incarnation: \emph{static images}.

\begin{figure}[t]
\centering
\includegraphics[width=0.99\columnwidth]{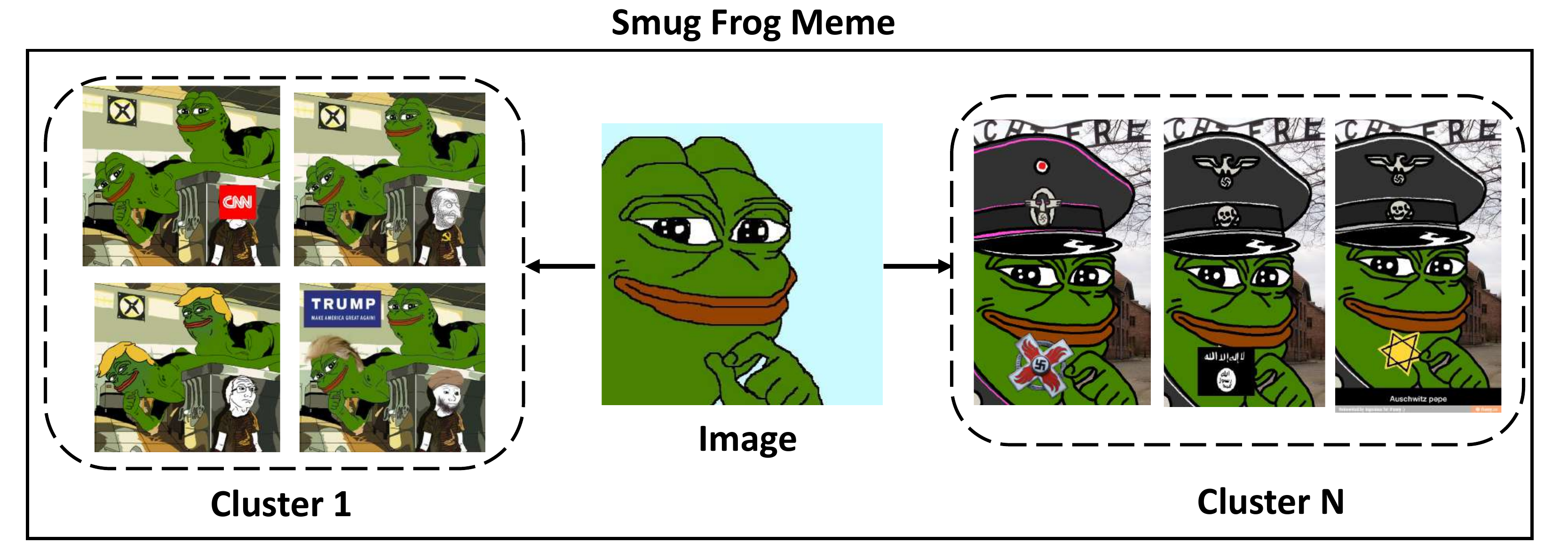}
\caption{An example of a meme (Smug Frog) that provides an intuition of what an image, a cluster, and a meme is.}
\label{fig:smug_frog_example}
\end{figure}

To gain an understanding of how memes propagate across the Web, with a particular focus on discovering the communities that are most
influential in spreading them, our intuition is to build \emph{clusters} of visually similar images, allowing us to track variants of a meme.
We then group clusters that belong to the same meme to study and track the meme itself.
In Figure~\ref{fig:smug_frog_example}, we provide a visual representation of the Smug Frog meme~\cite{smug_frog_meme}, which includes many variants of the same image (a ``smug'' Pepe the Frog) and several clusters of variants.
Cluster 1 has variants from a Jurassic Park scene, where one of the characters is hiding from two velociraptors behind a kitchen counter: the frogs are stylized to look similar to velociraptors, and the character hiding varies to express a particular message.
For example, in the image in the top right corner, the two frogs are searching for an anti-semitic caricature of a Jew (itself a meme known as the Happy Merchant~\cite{happy_merchant_meme}).
Cluster N shows variants of the smug frog wearing a Nazi officer military cap with a photograph of the infamous ``Arbeit macht frei'' slogan from the distinctive curved gates of Auschwitz in the background.
In particular, the two variants on the right display the death's head logo of the SS-Totenkopfverb{\"a}nde organization responsible for
running the concentration camps during World War II.
Overall, these clusters represent the branching nature of memes: as a new variant of a meme becomes prevalent, it often branches into its own sub-meme, potentially incorporating imagery from other memes.

\subsubsection{Processing Pipeline}\label{subsec:pipeline}

\begin{figure}[t]
\hspace*{-0.3cm}
\includegraphics[width=\columnwidth]{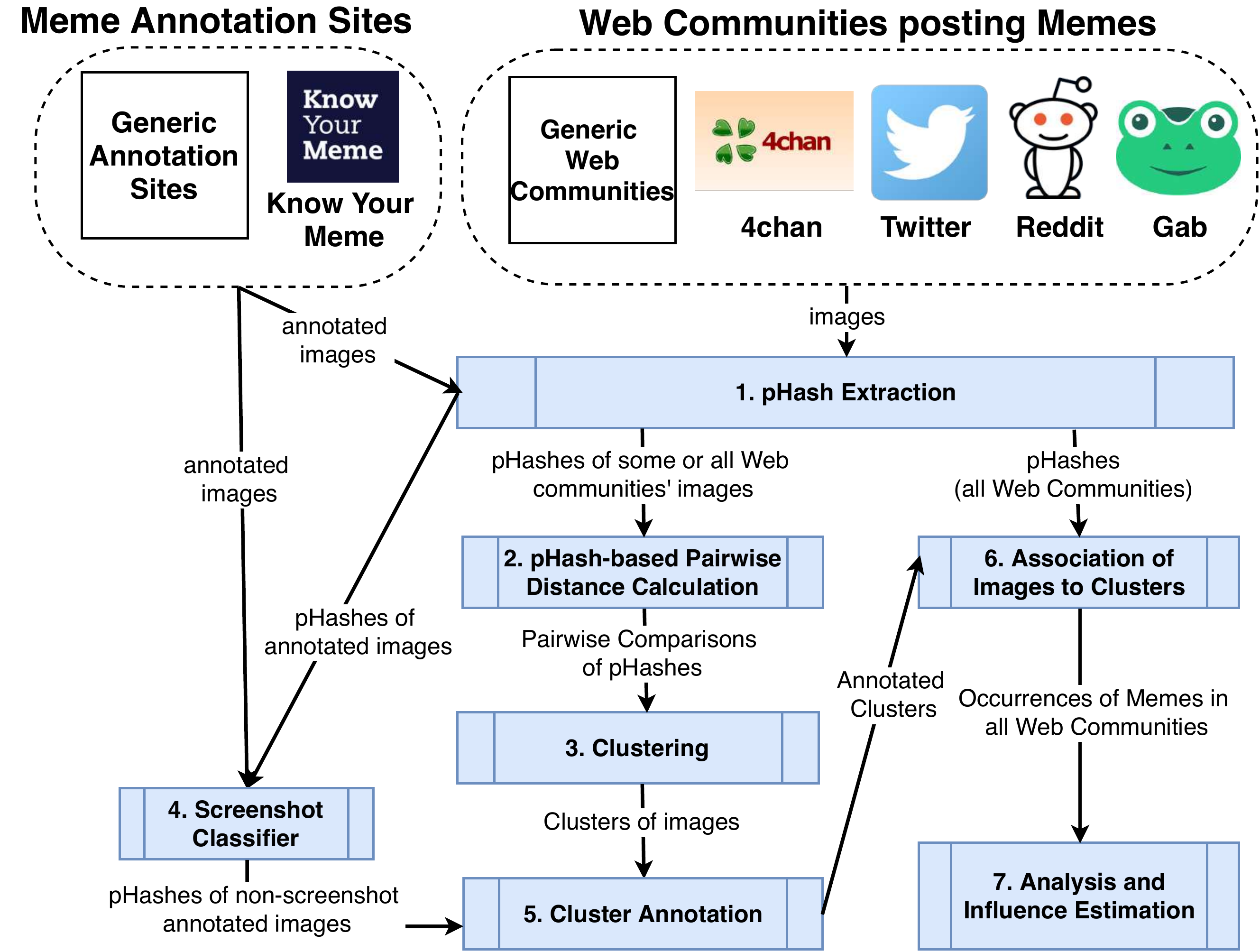}
\caption{High-level overview of our processing pipeline.}
\label{fig:pipeline}
\end{figure}

Our processing pipeline is depicted in Figure~\ref{fig:pipeline}.
As discussed above, our methodology aims at identifying clusters of similar images and assign them to higher level groups, which are the actual memes.
Note that the proposed pipeline is not limited to image macros and can be used to identify any image. 
We first discuss the types of data sources needed for our approach, i.e., meme annotation sites and Web communities that post memes
(dotted rounded rectangles in the figure).
Then, we describe each of the operations performed by our pipeline (Steps 1-7, see regular rectangles).

\descr{Data Sources.} Our pipeline uses two types of data sources: 1)~sites providing meme annotation and 2)~Web communities that disseminate memes.
In this thesis, we use Know  Your Meme for the former, and Twitter, Reddit, \dspol, and Gab for the latter.
We provide more details about our datasets in Section~\ref{sec:memes_dataset}.
Note that our methodology supports any annotation site and any Web community, and this is why we add the ``{\em Generic}'' sites/communities notation in Figure~\ref{fig:pipeline}.

\descr{pHash Extraction (Step 1).} We use the Perceptual Hashing (pHash) algorithm~\cite{monga2006perceptual} to calculate a fingerprint of each image in such a way that any two images that look similar to the human eye map to a ``similar'' hash value. pHash generates a feature vector of 64 elements that describe an image, computed from the Discrete Cosine Transform among the different frequency domains of the image.
Thus, visually similar images have minor differences in their vectors, hence allowing to search for and detect visually similar images.
For example, the string representation of the pHashes obtained from the images in cluster N (see Figure~\ref{fig:smug_frog_example}) are 55352b0b8d8b5b53, 55952b0bb58b5353, and 55952b2b9da58a53, respectively.
The algorithm is also robust against changes in the images, e.g., signal processing operations and direct
manipulation~\cite{zauner2011rihamark}, and effectively reduces the dimensionality of the raw images.

\descr{Clustering via pairwise distance calculation (Steps 2-3).}
Next, we cluster images from one or more Web Communities using the pHash values.
We perform a pairwise comparison of all the pHashes using Hamming distance (Step 2).
To support large numbers of images, we implement a highly parallelizable system on top of TensorFlow~\cite{abadi2016tensorflow}, which uses multiple GPUs to enhance performance. 
Images are clustered using a density-based algorithm (Step 3).
Our current implementation uses DBSCAN~\cite{ester1996density}, mainly because it can discover clusters of arbitrary shape
and performs well over large, noisy datasets.
Nonetheless, our architecture can be easily tweaked to support any clustering algorithm and distance metric.

We also perform an analysis of the clustering performance and the rationale for selecting the clustering threshold. Our implementation uses the DBSCAN algorithm with a clustering threshold equal to 8.
To select this threshold, we perform the clustering step while varying the distances.
Table~\ref{tbl:clustering_statistics_evaluation} shows the number of clusters and the percentage of images that are regarded as noise by the clustering algorithm for varying distances.
We observe that, for distances 2-4, we have a substantially larger percentage of noise, while with distance 10 we have the least percentage of noise.
With distances between 6 and 8 we observe that we get a larger number of clusters than the other distances, while the noise percentages are 73\% and 63\%, respectively.

\begin{figure}[t]
\centering
\includegraphics[width=0.57\columnwidth]{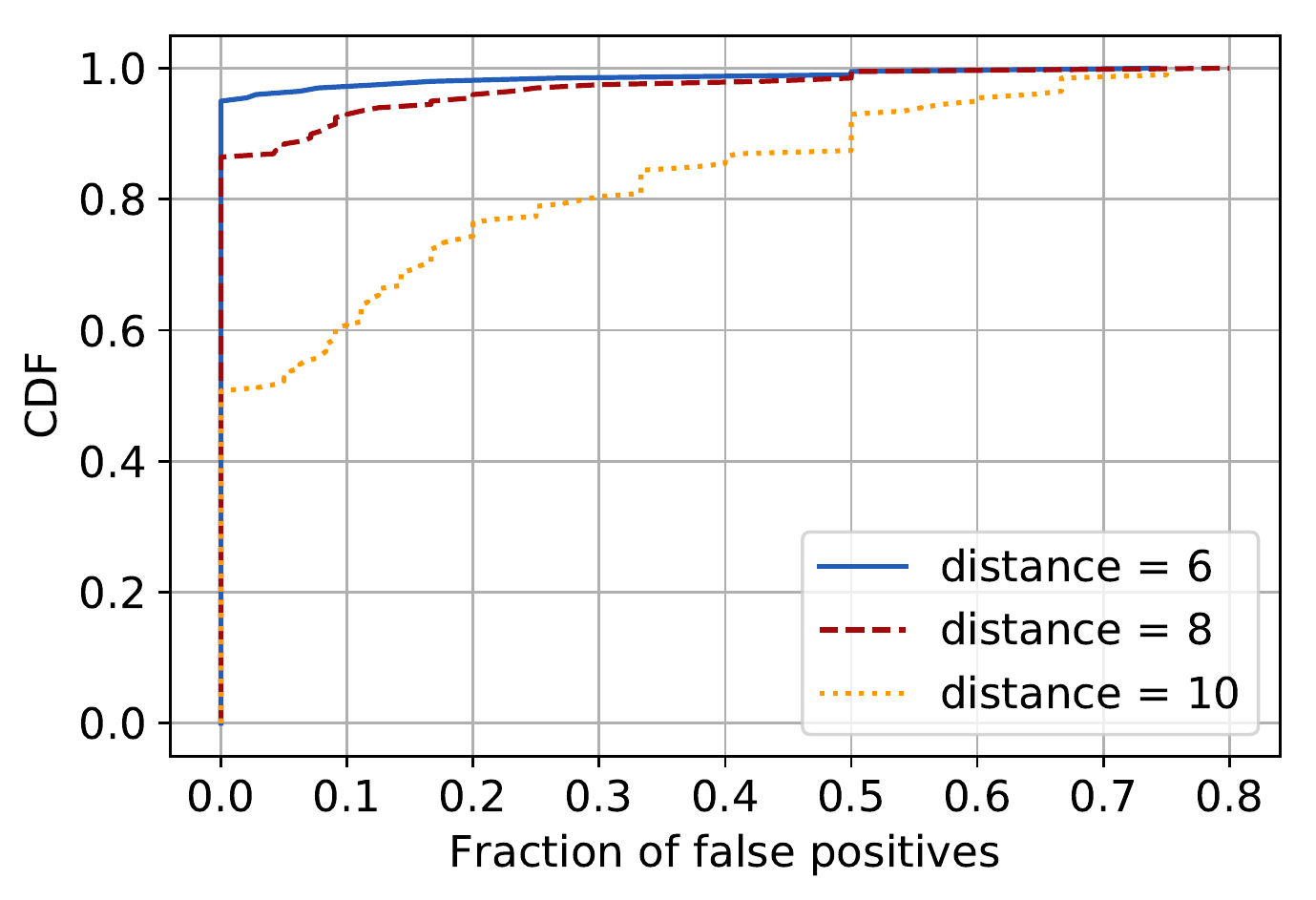}
\caption{Fraction of false positives in clusters with varying clustering distance. }
\label{fig:clustering_evaluation}
\end{figure}

To further evaluate the clustering performance for varying distances, we randomly select 200 clusters and manually calculate the number of images that are false positives within each cluster. Figure~\ref{fig:clustering_evaluation} shows the CDF of the false positive fraction in the random sample of clusters for distances 6, 8, and 10 (we disregard distances 2-4 due to the high percentage of noise).
Distance 10 yields a high number of false positives, while distances 6-8 the overall false positives are below 3\%.
Therefore, we investigate the impact of these false positives in the overall dataset, looking at all posts that contain false and true positives in the random sample of 200 clusters, using distance 8.
We find that the false positives have little impact as they occur substantially fewer times than true positives: the percentage of true positives over the set of false positives and true positives is 99.4\%.
Thus, due to the larger number of clusters, the acceptable false positive performance, and the smaller percentage of noise (when compared to distances 2-6), we elect to use as a threshold the perceptual distance that is equal to 8.

\begin{table}[t]
\centering
\small
\begin{tabular}{@{}rrr@{}}
\toprule
\multicolumn{1}{l}{\textbf{Distance}} & \multicolumn{1}{c}{\textbf{\#Clusters}} & \multicolumn{1}{c}{\textbf{\%Noise}} \\ \midrule
2 & 30,327 & 82.9\% \\
4 & 34,146 & 78.5\% \\
6 & 37,292 & 73.0\% \\
8 & 38,851 & 62.8\% \\
10 & 30,737 & 27.8\% \\ \bottomrule
\end{tabular}%
\caption{Number of clusters and percentage of noise for varying clustering distances.}
\label{tbl:clustering_statistics_evaluation}
\end{table}

\begin{figure*}[t]
\includegraphics[width=\textwidth]{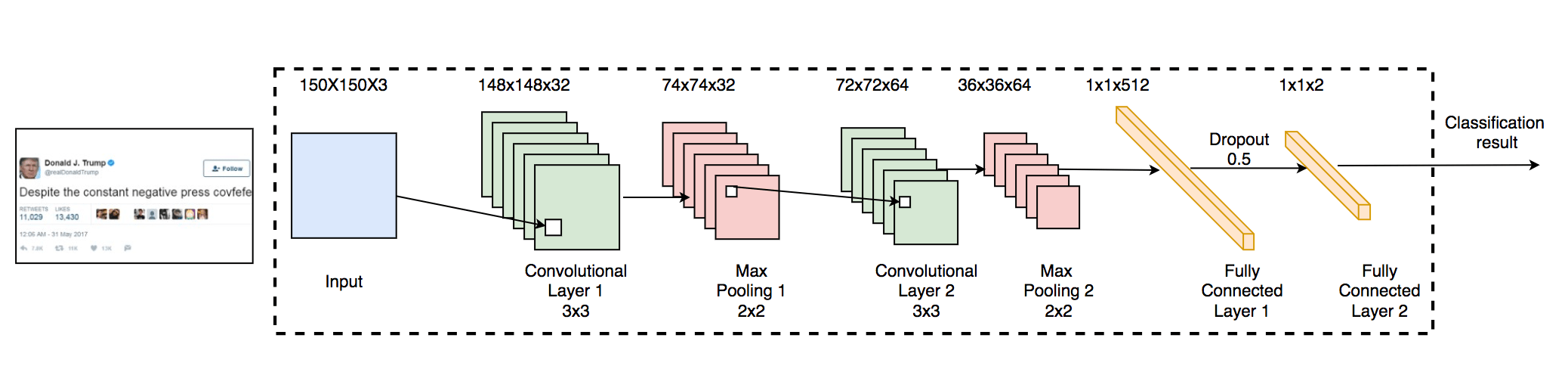}
\caption{Architecture of the deep learning model for detecting screenshots from Twitter, /pol/, Reddit, Instagram, and Facebook.}
\label{fig:ml_architecture}
\end{figure*}

\descr{Screenshots Removal (Step 4).} Meme annotation sites like KYM often include, in their image galleries, screenshots of social network posts that are not variants of a meme but just comments about it.
Hence, we discard social-network screenshots from the annotation sites data sources using a deep learning classifier. 
Below, we provide more details about our screenshot removal classifier.

\descr{Dataset}. Table~\ref{tbl:curated_dataset} summarizes the dataset used for training the classifier.
It includes 28.8K images that depict posts from Twitter, 4chan, Reddit, Facebook, and Instagram, which we collect from public sources.
First, we download images from specific subreddits that only allow screenshots from a particular community.
For example, the 4chan subreddit require all submissions to be of a screenshot of a 4chan thread.
Next, we use the Pinterest platform to download specific boards that contain mostly screenshots from the communities we study.
Also, we search and obtain image datasets that are publicly available on Web archiving services like the Wayback Machine.
We then manually filter out images that were misplaced.
Finally, we include 10K random images posted on \dspol (i.e., a subset of the 4.3M images collected for our measurements).

\begin{table}[t]
\centering
\setlength{\tabcolsep}{5pt}
\resizebox{0.6\columnwidth}{!}{%
\begin{tabular}{@{}lllllll@{}}
\toprule
\textbf{Platform}  & Twitter                    & 4chan                      & Reddit                    & Facebook               & Instagram                  & Other                      \\ \midrule
\textbf{\# images} & \multicolumn{1}{r}{14,602} & \multicolumn{1}{r}{10,127} & \multicolumn{1}{r}{2,181} & \multicolumn{1}{r}{1,414} & \multicolumn{1}{r}{497} &\multicolumn{1}{r}{10,630} \\ \bottomrule
\end{tabular}%
}
\caption{Curated dataset used to train the screenshot classifier.}
\label{tbl:curated_dataset}
\end{table}

\descr{Classifier.} To detect screenshots that contain images from one of the social networks included in our dataset, we use Convolutional Neural Networks.
Figure~\ref{fig:ml_architecture} provides an overview of our classifier's architecture.
It includes two Convolutional Neural Networks, each followed by a max-pooling layer.
The output of these layers is fed to a fully-connected dense layer comprising 512 units.
Finally, we have another fully-connected layer with two units, which outputs the probability that a particular image is a screenshot from one of the five social networks and the probability that an image is a random one.
To avoid overfitting on the two last fully-connected layers, we apply Dropout with $d=0.5$~\cite{srivastava2014dropout}.
This means that, while training, 50\% of the units are randomly omitted from updating their parameters. %

\begin{figure}[t]
\centering
\includegraphics[width=0.6\columnwidth]{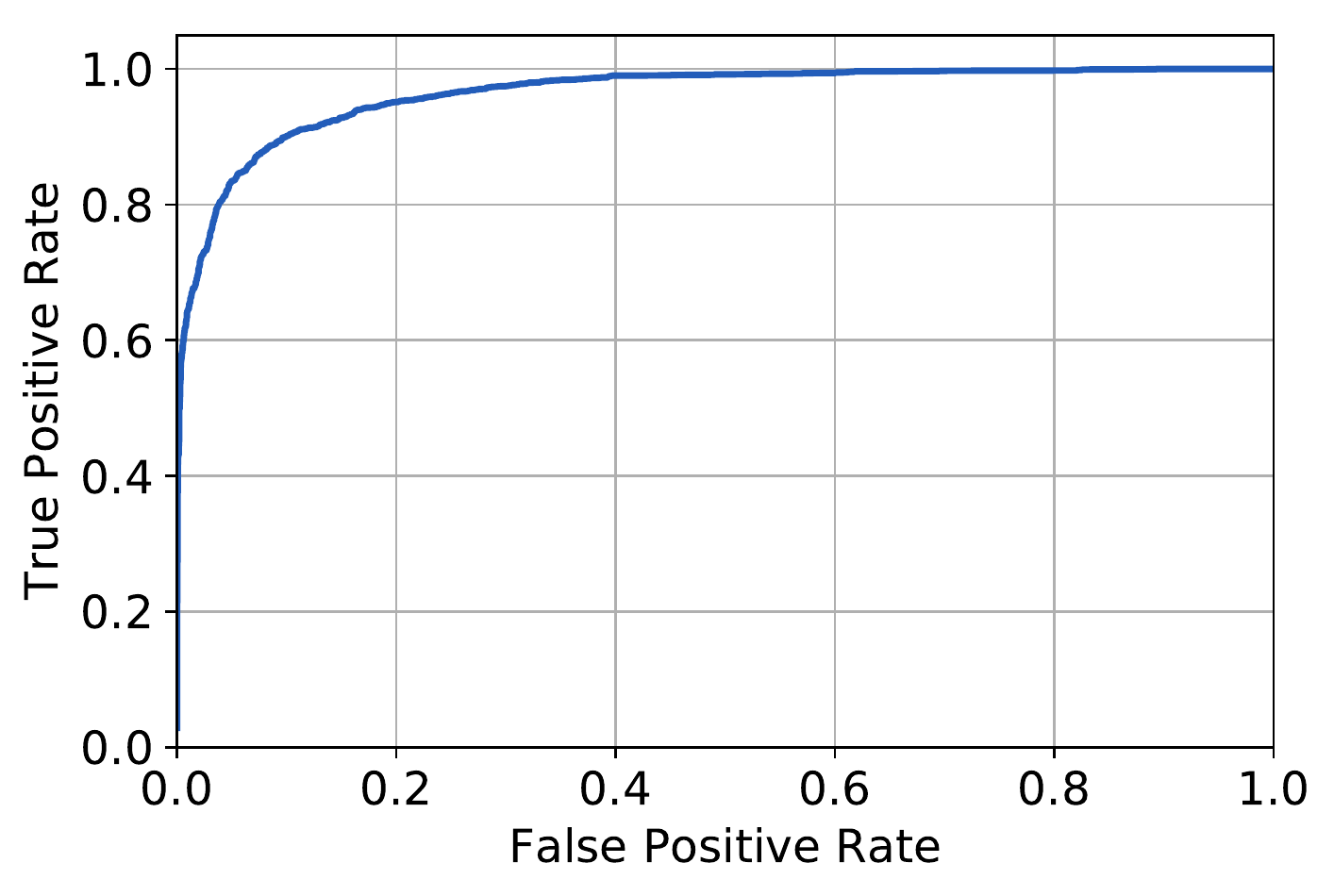}
\caption{ROC curve of the screenshot classifier.}
\label{fig:model_performance}
\end{figure}

\descr{Experimental Evaluation.} Our implementation uses Keras~\cite{chollet2015keras} with TensorFlow as the backend~\cite{abadi2016tensorflow}.
To train our model, we randomly select 80\% of the images and evaluate based on the rest 20\% out-of-sample dataset.
Figure~\ref{fig:model_performance} shows the ROC curve of the model.
We observe that the devised classifier exhibits acceptable performance with an Area Under the Curve (AUC) of 0.96.
We also evaluate our model in terms of accuracy, precision, recall, and F1-score, which amount to 91.3\%, 94.3\%, 93.5\%, and 93.9\%, respectively.

\descr{Cluster Annotation (Steps 5).} Clustering annotation uses the \emph{medoid} of each cluster, i.e., the element with the minimum square average distance from all images in the cluster. 
In other words, the medoid is the image that best represents the cluster.
The clusters' medoids are compared with all images from meme annotation sites, by calculating the Hamming distance between each pair of pHash vectors. We consider that an image matches a cluster if the distance is less than or equal to a threshold $\theta$, which we set to $8$, as it allows us to capture the diversity of images that are part of the same meme while maintaining a low number of false positives.

As the annotation process considers all the images of a KYM entry's image gallery, it is likely we will get multiple annotations for a single cluster.
To find the representative KYM entry for each cluster, we select the one with the largest proportion of matches of KYM images with
the cluster medoid. In case of ties, we select the one with the minimum average Hamming distance.

While KYM might not be a household name, the site is seemingly the largest curated collection of memes on the Web, i.e., KYM is as close to an ``authority'' on memes as there is.
That said, crowdsourcing \emph{is} an aspect of how KYM works, and thus there might be questions as to how ``legitimate'' some of the content is.
To this end, we set out to measure the quality of KYM by sampling a number of pages and manually examining them.
This is clearly a subjective task, and a fully specified definition of what makes a valid meme is approximately as difficult as defining ``art.''
Nevertheless, the authors of this work have, for better or worse, collectively spent thousands of hours immersed in the communities we explore; thus, while we are not confident in providing a strict definition of a meme, we are in claiming that we know a meme when we see it.

Using the same randomly selected 200 clusters as mentioned in Steps2-3 above, we visited each KYM page the cluster was tagged with and noted whether or not it properly documented what we consider an ``actual'' meme.
The 200 clusters were mapped to 162 unique KYM pages, and of these 162 pages, 3 (1.85\%) we decided were ``bad.''
This is mainly due to the lack of completeness and relatively high number of random images in the gallery (see ~\cite{maxvidya_meme,xy_meme} for some examples of ``bad'' KYM entries).

Next, we set out to determine whether the label (i.e., KYM page) assigned to each of our randomly sampled clusters was appropriate.
Using three annotators, for each cluster we examined the KYM page, the medoid of the cluster, and the images in the cluster itself and noted whether the label does in fact apply to the cluster.
Here, again, there is a great degree of subjectivity.
To reign some of the subjectivity in, we used the following guidelines:
\begin{compactenum}
  \item If the exact image(s) in the cluster appear in the KYM gallery, then the label is correct.
  \item For images that do not appear in the KYM gallery, if the label is \emph{appropriate}, then it is a correct labeling.
\end{compactenum}

There are some important caveats with these guidelines.
First, KYM galleries are crowdsourced, and while curated to some extent, the possibility for what amounts to random images in a gallery \emph{does} exist; however, based on our assessment of KYM page validity, this occurs with low probability.
Second, we considered a label correct if it was \emph{appropriate}, even if it was not necessarily the \emph{best} possible label.
For example, as our results show, many memes are related, and many images mix and match pieces of various memes.
While it is definitely true that there might be better labels that exist for a given cluster, this straightforward and comprehensible labeling process is sufficient for our purposes.
We leave a more in-depth study of the subjective nature of memes for future work.
Finally, it is important to note that memes are a \emph{cultural} phenomenon, and thus the potential for cultural bias in our annotation is possible.
Note that our annotators were born in three different countries (USA, Italy, and Cyprus), only one is a native English speaker, and two have spent substantial time in the US.

After annotating clusters, we compute the Fleis agreement score ($\kappa$).
With our cluster samples, we achieve $\kappa {=} 0.67$, which is considered ``substantial'' agreement.
Finally, for each cluster we obtain the majority agreement of all annotators to assess the accuracy of our annotation process; we find that 89\% of the clusters had a legitimate annotation to a specific KYM entry.

\descr{Association of images to memes (Step 6).}
To associate images posted on Web communities (e.g., Twitter, Reddit, etc.) to memes, we compare them with the clusters' medoids, using the
same threshold $\theta$.
This is conceptually similar to Step 5, but uses images from Web communities instead of images from annotation sites.
This lets us identify memes posted in generic Web communities and collect relevant metadata from the posts (e.g., the timestamp of a tweet). 
Note that we track the propagation of memes in generic Web communities (e.g., Twitter) using a {\em seed} of memes obtained by clustering images from other (fringe) Web communities.
More specifically, our seeds will be memes generated on three fringe Web communities (\dspol, \td subreddit, Gab); nonetheless, our methodology can be applied to any community.

\descr{Analysis and Influence Estimation (Step 7).} We analyze all relevant clusters and the occurrences of memes, 
aiming to assess: 1)~their popularity and diversity in each community; 2)~their temporal evolution; and 3)~how communities influence each other with respect to meme dissemination. %

\subsubsection{Distance Metric}
\label{sec:methodology:distance}

To better understand the interplay and connections between the clusters, we introduce a custom distance metric, which relies on both the visual peculiarities of the images (via pHash) and  data available from annotation sites.
The distance metric supports one of two modes: 1)~one for when both clusters are annotated ({\it full-mode}), and 2)~another for when one or none of the clusters is annotated ({\it partial-mode}). 

\descr{Definition.}
Let $c$ be a cluster of images and $\mathsf{F}$ a set of features extracted from the clusters.
The custom distance metric between cluster $c_i$ and $c_j$ is defined as:

\begin{equation}\label{eq:distance}
\mathsf{distance}(c_i, c_j) = 1 - \sum_\mathsf{f \in \mathsf{F}} w_\mathsf{f} \times \texttt{r}_\mathsf{f}(c_i, c_j) %
\end{equation}

\noindent where $\texttt{r}_\mathsf{f}(c_i, c_j)$ denotes the similarity between the features of type $\mathsf{f} \in \mathsf{F}$ of cluster $c_i$ and $c_j$, and $w_f$ is a weight that represents the relevance of each feature.
Note that $\sum_\mathsf{f} w_\mathsf{f} = 1$ and $\mathsf{r}_f(c_i, c_j) = \{x \in \Bbb{R} \mid 0 \leq x \leq 1\}$.
Thus, $\mathsf{distance}(c_i, c_j)$ is a number between 0 and 1. %

\descr{Features.} We consider four different features for $\texttt{r}_\mathsf{f \in \mathsf{F}}$, specifically, $\mathsf{F} = \{perceptual, meme, people, culture\}$; see below.

\descrit{${r_{perceptual}}$:} this feature is the similarity between two clusters from a perceptual viewpoint.
Let $h$ be a pHash vector for an image $m$ in cluster $c$, where $m$ is the medoid of the cluster, and $d_{ij}$ the Hamming distance between vectors $h_i$ and $h_j$ (see Step 5 in Section~\ref{subsec:pipeline}).
We compute $d_{ij}$ from $c_i$ and $c_j$ as follows.
First, we obtain  obtain the medoid $m_i$ from cluster $c_i$.
Subsequently, we obtain $h_i{=}$\mbox{pHash}$(m_i)$.
Finally, we compute $d_{ij} {=} \mbox{Hamming}(h_i, h_j)$.
We simplify notation and use $d$ instead of $d_{ij}$ to denote the distance between two medoid images and refer to this distance as the Hamming {\it score}.

\begin{figure}[t]
\centering
\includegraphics[width=.65\columnwidth]{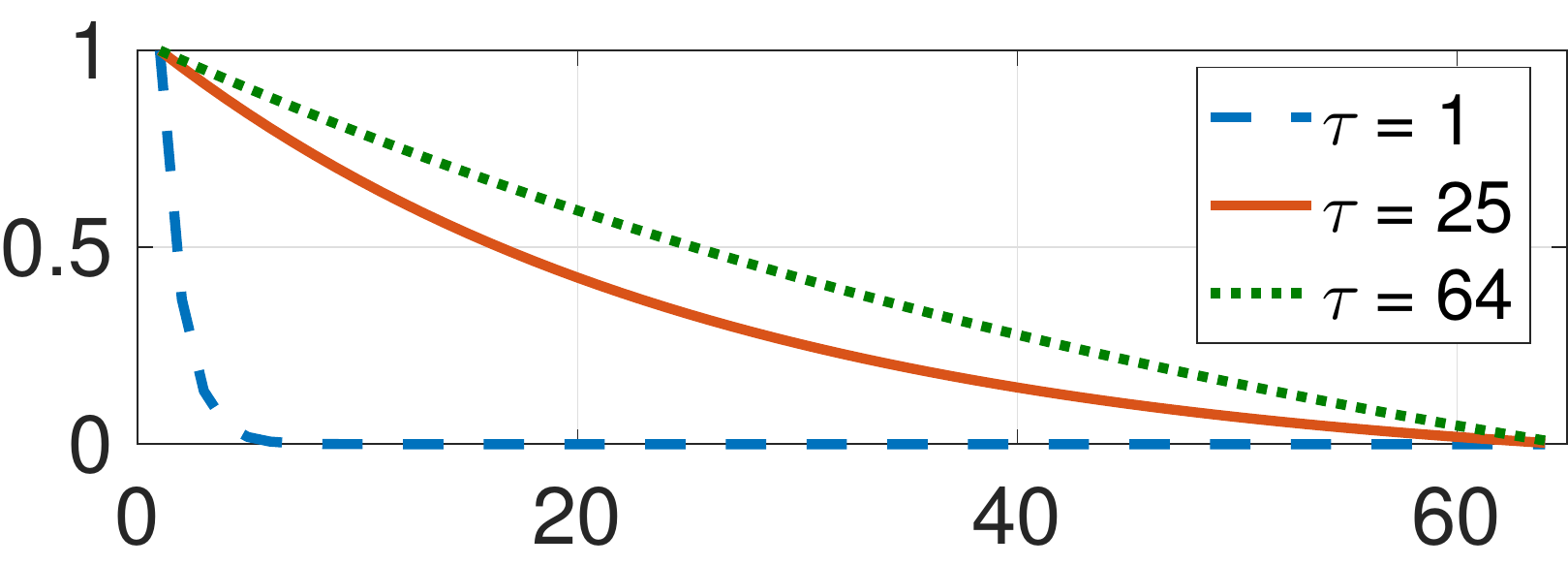}
\caption{Different values of $r_{perceptual}$ (y-axis) for all possible inputs of $d$ (x-axis) with respect to the smoother $\tau$.}
\label{fig:perceptual-relevance}
\end{figure}

We define the perceptual similarity between two clusters as an exponential decay function over the Hamming score $d$:
\begin{equation}\label{eq:score_perceptual}
r_{perceptual}(d) = 1 - \frac{d}{\tau \times e^{\texttt{max} / \tau}} 
\end{equation}

\noindent where $\texttt{max}$ represents the maximum pHash distance between two images and $\tau$ is a constant parameter, or {\it smoother}, that controls how fast the exponential function decays for all values of $d$ (recall that $\{d \in \Bbb{R} \mid 0 \leq d \leq \texttt{max}\}$).
Note that $\texttt{max}$ is bound to the precision given by the pHash algorithm.
Recall that each pHash has a size of $|d| {=} 64$, hence $\texttt{max} {=} 64$.
Intuitively, when $\tau << 64$, $r_{perceptual}$ is a high value only with perceptually indistinguishable images, e.g., for $\tau {=} 1$, two images with $d {=} 0$ have a similarity $r_{perceptual} {=} 1.0$.
With the same $\tau$, the similarity drops to $0.4$ when $d {=} 1$.
By contrast, when $\tau$ is close to $64$, $r_{perceptual}$ decays almost linearly.
For example, for $\tau {=} 64$, $r_{perceptual}(d{=}0) {=} 1.0$ and $r_{perceptual}(d{=}1) {=} 0.98$.
Figure~\ref{fig:perceptual-relevance} shows how $r_{perceptual}$ performs for different values of $\tau$.
As mentioned above, we observe that pairs of images with scores between $d{=}0$ and $d{=}8$ are usually part of the same variant (see Step 5 in Section~\ref{subsec:pipeline}).
In our implementation, we set $\tau {=} 25$ as $r_{perceptual}$ returns high values up to $d{=}8$, and rapidly decays thereafter.

\descrit{$r_{meme}$, $r_{culture}$, and $r_{people}$:} the annotation process (Step 5) provides  contextualized information about the cluster medoid, including the name (i.e., the main identifier) given to a meme, the associated culture (i.e., high-level group of meme), and people that are included in a meme. Note that we use all the annotations for each category and not only the representative one (see Step 5).
Therefore, we model a different similarity for each of the these categories, by looking at the overlap of all the annotations among the medoids of both clusters ($m_i$, $m_j$, for $c_i$ and $c_j$, respectively).
Specifically, for each category, we calculate the Jaccard index between the annotations of both medoids, for memes, cultures, and people, thus acquiring $r_{meme}$, $r_{culture}$, $r_{people}$, respectively.

\descr{Modes.}
Our distance metric measures how similar two clusters are.
If both clusters are annotated, we operate in ``full-mode,'' and in ``partial-mode'' otherwise.
For each mode, we use different weights for the features in Eq.~\ref{eq:distance}, which we set empirically as we lack the ground-truth data needed to automate the computation of the optimal set of thresholds.

\descr{Full-mode.}  In full-mode, we set weights as follows.
1)~The features from the perceptual and meme categories should have higher relevance than people and culture, as they are intrinsically related to the definition of meme (see Section~\ref{sec:methodology:overview}).
The last two are non-discriminant features, yet are informative and should contribute to the metric.
Also, 2)~$r_{meme}$ should not outweigh $r_{perceptual}$ because of the relevance that visual similarities have on the different variants of a meme.
Likewise, $r_{perceptual}$ should not dominate over $r_{meme}$ because of the branching nature of the memes.
Thus, we want these two categories to play an equally important weight.
Therefore, we choose
$w_{perceptual} {=} 0.4$,
$w_{meme} {=} 0.4$,
$w_{people} {=} 0.1$,
$w_{culture} {=} 0.1.$

This means that when two clusters belong to the same meme and their medoids are perceptually similar, the distance between the clusters will be small.
In fact, it will be at most $0.2 = 1 - (0.4 + 0.4)$ if people and culture do not match, and $0.0$ if they also match.
Note that our metric also assigns small distance values for the following two cases:
1)~when two clusters are part of the same meme variant,
and 2)~when two clusters use the same image for different memes.

\descr{Partial-mode.} In this mode, we associate unannotated images with any of the known clusters.
This is a critical component of our analysis (Step 6), allowing us to study images from generic Web communities where annotations are unavailable.
In this case, we rely entirely on the perceptual features.
We once again use Eq.~\ref{eq:distance}, but simply set all weights to 0, except for $w_{perceptual}$ (which is set to 1).
That is, we compare the image we want to test with the medoid of the cluster and we apply Eq.~\ref{eq:score_perceptual} as described above.

\subsection{Datasets}
\label{sec:memes_dataset}

We now present the datasets used in our measurements.

\subsubsection{Web Communities}\label{sec:communities}
As mentioned earlier, our data sources are Web communities that post memes and meme annotation sites.
For the former, we focus on four communities: Twitter, Reddit, Gab, and 4chan (more precisely, 4chan's Politically Incorrect board, \dspol).
This provides a mix of mainstream social networks (Twitter and Reddit) as well as fringe communities that are often associated with the alt-right and have an impact on the information ecosystem (Gab and \dspol)~\cite{zannettou2017web}.

There are several other platforms playing important roles in spreading memes, however, many are ``closed'' (e.g., Facebook) or do not involve memes based on static images (e.g., YouTube, Giphy).
In future work, we plan to extend our measurements to communities like Instagram and Tumblr, as well as to GIF and video memes.
Nonetheless, we believe our data sources already allow us to elicit comprehensive insights into the meme ecosystem.

\begin{table}[t]
\centering
\setlength{\tabcolsep}{3pt}
\small
\begin{tabular}{lrrrr}
\toprule
{\bf Platform} & {\bf \#Posts}    & {\bf \#Posts with} & {\bf \#Images} & {\bf \#Unique} \\
& & {\bf Images} &  & {\bf pHashes}\\
\midrule
\textbf{Twitter} &     1,469,582,378               & 242,723,732  & 114,459,736                               & 74,234,065                                                                                   \\
\textbf{Reddit}  & 1,081,701,536               & 62,321,628 & 40,523,275                                & 30,441,325                                                                                   \\
\textbf{/pol/}    & 48,725,043    & 13,190,390               & 4,325,648                                 & 3,626,184                                                                                    \\
\textbf{Gab}      &   12,395,575       &  955,440         & 235,222                                   & 193,783                                                                                      \\
\textbf{KYM}      & 15,584             & 15,584      & 706,940                                   & 597,060                                                                                      \\
 \bottomrule
\end{tabular}
\caption{Overview of our datasets.}
\label{tbl:datasets_summary}

\end{table}

Table~\ref{tbl:datasets_summary} reports the number of posts and images processed for each community.
Note that the number of images is lower than the number of posts with images because of duplicate image URLs and because some images get deleted.
Next, we discuss each dataset.

\descr{Twitter.} Our Twitter dataset is based on tweets made available via the 1\% Streaming API,  between July 1, 2016 and July 31, 2017.
In total, we parse 1.4B tweets: 242M of them have at least one image.
We extract all the images, ultimately collecting 114M images yielding 74M unique pHashes.

\descr{Reddit.} We gather images from Reddit using publicly available data from Pushshift~\cite{reddit_data_pushshift}.
We parse all submissions and comments between July 1, 2016 and July, 31 2017, and extract 62M posts that contain at least one image.
We then download 40M images producing 30M unique pHashes.

\descr{4chan.} We obtain all threads posted on \dspol, between July 1, 2016 and July 31, 2017, using the same methodology of \cite{hine2016longitudinal}.
Since all threads (and images) are removed after a week, we use a public archive service called 4plebs~\cite{4plebs_site} to collect 4.3M images, thus yielding 3.6M unique pHashes.

\descr{Gab.} We collect 12M posts, posted on Gab between August 10, 2016 and July 31, 2017, and 955K posts have at least one image, using the same methodology as in~\cite{zannettou2018gab}.
Out of these, 235K images are unique, producing 193K unique pHashes.
Note that our Gab dataset starts one month later than the other ones, since Gab was launched in August 2016.

\subsubsection{Meme Annotation Site}
\label{sec:analysis:kym}

\begin{figure*}[t]
\centering
\subfigure[Categories]{\includegraphics[width=0.49\textwidth]{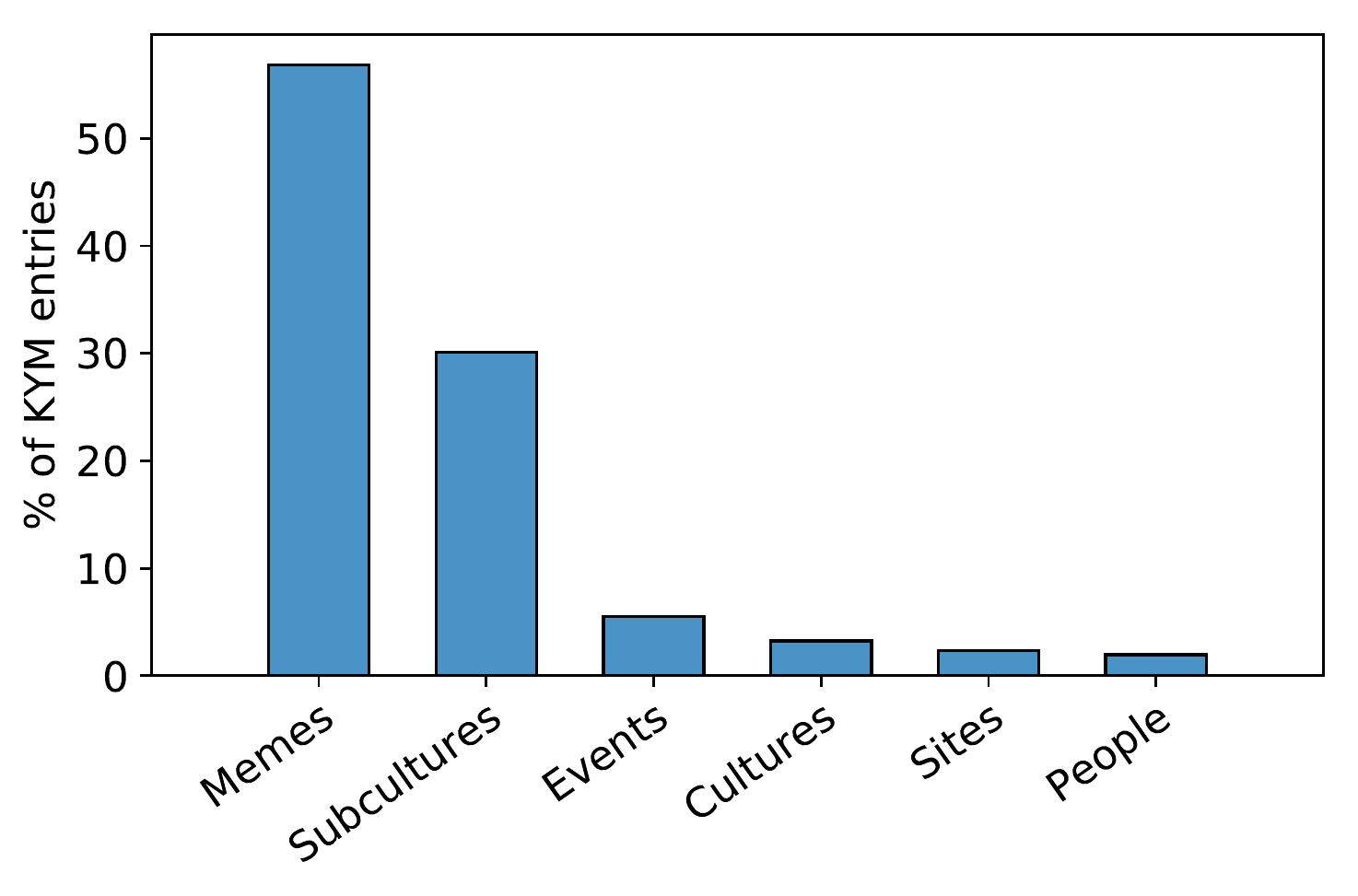}\label{fig:bc_kym_entry_categories}}
\subfigure[Images]{\includegraphics[width=0.48\textwidth]{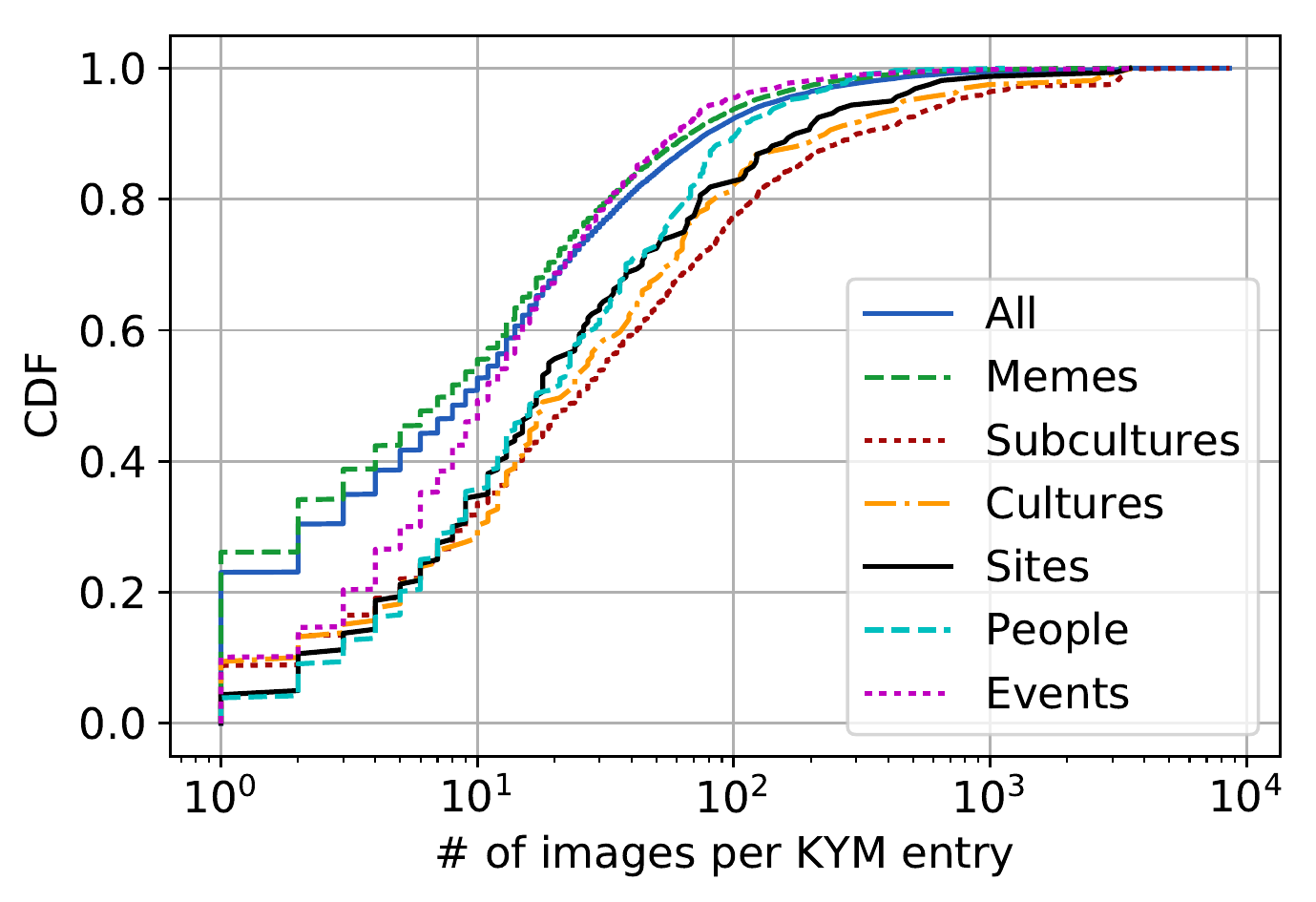}\label{fig:stats_kym_images}}
\subfigure[Origins]{\includegraphics[width=0.49\textwidth]{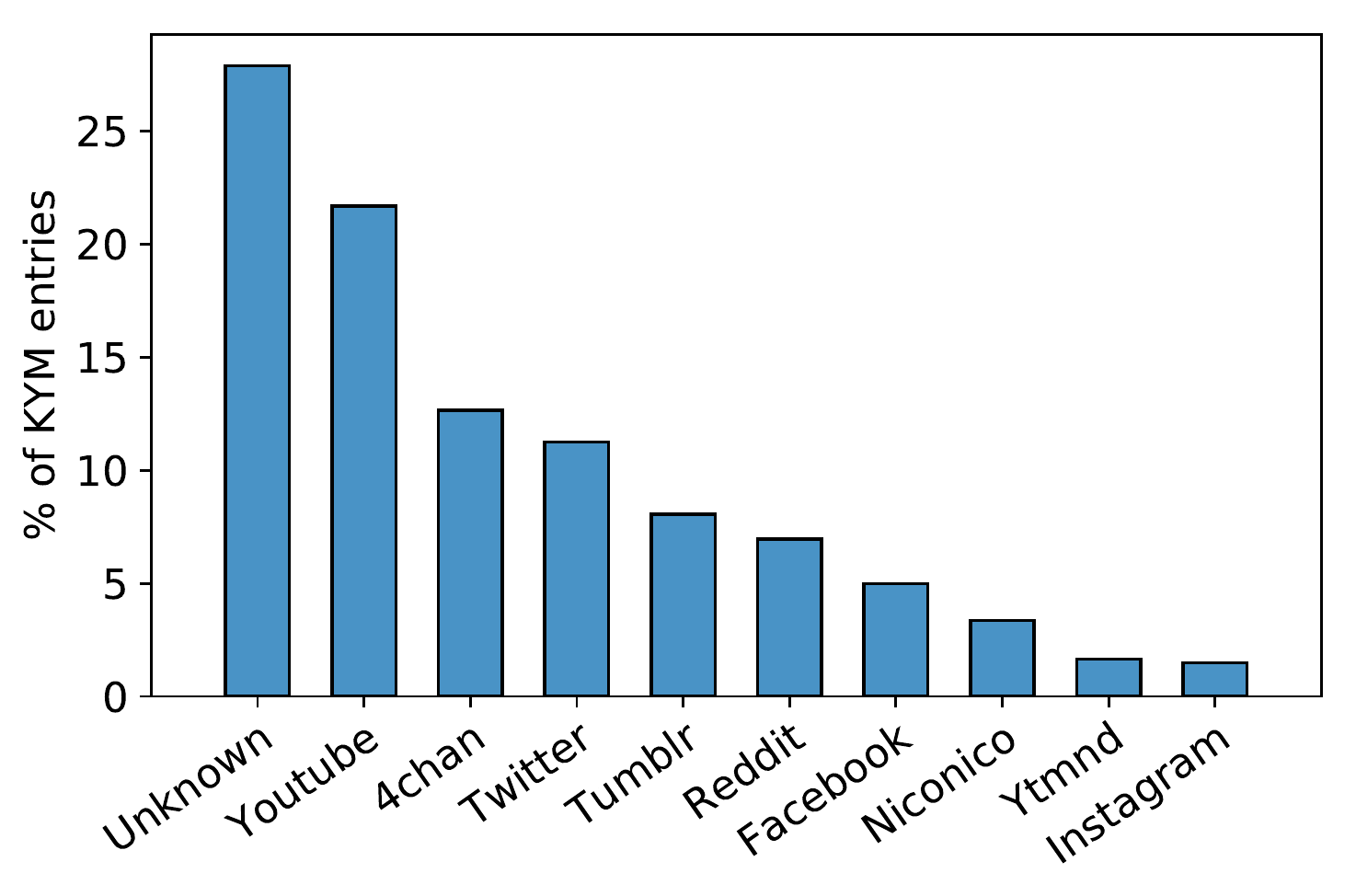}\label{fig:bc_origins_memes}}
\caption{Basic statistics of the KYM dataset.}
\label{fig:kym_overview}
\end{figure*}

\descr{Know Your Meme (KYM).} We choose KYM as the source for meme annotation as it offers a comprehensive database of memes.
KYM is a sort of encyclopedia of Internet memes: for each meme, it provides information such as its origin (i.e., the platform on which it was first observed), the year it started, as well as descriptions and examples.
In addition, for each entry, KYM provides a set of keywords, called {\em tags}, that describe the entry.
Also, KYM provides a variety of higher-level categories that group meme entries; namely, cultures, subcultures, people, events, and sites.
``Cultures'' and ``subcultures'' entries refer to a wide variety of topics ranging from video games to various general categories.
For example, the Rage Comics {\em subculture}~\cite{kym_rage_comics_subculture} is a higher level category associated with memes related to comics like Rage Guy~\cite{rage_guy} or LOL Guy~\cite{lol_guy}, while the Alt-right {\em culture}~\cite{alt_right_culture} gathers entries from a loosely defined segment of the right-wing community.
The rest of the categories refer to specific individuals (e.g., Donald Trump~\cite{donald_trump_meme}), specific \em{events} (e.g.,\#CNNBlackmail~\cite{cnnblackmail_meme}), and \em{sites}  (e.g., \dspol~\cite{pol_kym}), respectively. It is also worth noting that KYM moderates all entries, hence entries that are wrong or incomplete are marked as so by the site.

As of May 2018, the site has 18.3K entries, specifically, 14K memes, 1.3K subcultures,  1.2K people, 1.3K events, and 427 websites~\cite{kym_memes_summary}.
We crawl KYM between October and December 2017, acquiring data for 15.6K entries; %
for each entry, we also download all the images related to it by crawling all the pages of the image gallery.
In total, we collect 707K images corresponding to 597K unique pHashes.
Note that we obtain 15.6K out of 18.3K entries, as we crawled the site several months before May 2018.

\descr{Getting to know KYM.} We also perform a general characterization of KYM. %
First, we look at the distribution of entries across categories: as shown in Figure~\ref{fig:bc_kym_entry_categories}, %
as expected, the majority (57\%) are memes, followed by subcultures (30\%), cultures (3\%), websites (2\%), and people (2\%).

Next, we measure the number of images per entry: as shown in Figure~\ref{fig:stats_kym_images}, this varies considerably (note log-scale on x-axis).
KYM entries have as few as 1 and as many as 8K images, with an average of 45 and a median of 9 images.
Larger values may be related to the meme's popularity, but also to the ``diversity'' of image variants it generates.
Upon manual inspection, we find that the presence of a large number of images for the same meme happens either when images are visually very similar to each other (e.g., Smug Frog images {\em within} the two clusters in Figure~\ref{fig:smug_frog_example}), or if there are actually remarkably different variants of the same meme (e.g., images in `cluster 1' {\em vs.} images in `cluster N' in the same figure).
We also note that the distribution varies according to the category: e.g., higher-level concepts like cultures include more images than more specific entries like memes.

We then analyze the origin of each entry: see Figure~\ref{fig:bc_origins_memes}.
Note that a large portion of the memes (28\%) have an unknown origin, while YouTube, 4chan, and Twitter are the most popular platforms with, respectively, 21\%, 12\%, and 11\%, followed by Tumblr and Reddit with 8\% and 7\%.
This confirms our intuition that 4chan, Twitter, and Reddit, which are among our data sources, play an important role in the generation and dissemination of memes.
As mentioned, we do not currently study video memes originating from YouTube, due to the inherent complexity of video-processing tasks as well as scalability issues. However, a large portion of YouTube memes actually end up being morphed into image-based memes (see, e.g., the Overly Attached Girlfriend meme~\cite{overly_attached_girlfriend}).

\subsubsection{Running the pipeline on our datasets}
For all four Web communities (Twitter, Reddit, \dspol, and Gab), we perform Step 1 of the pipeline (Figure~\ref{fig:pipeline}), using the ImageHash library.\footnote{\url{https://github.com/JohannesBuchner/imagehash}}%
After computing the pHashes, we delete the images (i.e., we only keep the associated URL and pHash) due to space limitations of our infrastructure. 
We then perform Steps 2-3 (i.e., pairwise comparisons between all images and clustering), for all the images from \dspol, \td subreddit, and Gab, as we treat them as fringe Web communities.
Note that, we exclude mainstream communities like the rest of Reddit and Twitter as our main goal is to obtain clusters of memes from fringe Web communities and later characterize all communities by means of the clusters.
Next, we go through Steps 4-5 using all the images obtained from meme annotation websites (specifically, Know Your Meme, see Section~\ref{sec:analysis:kym}) and the medoid of each cluster from \dspol, \td, and Gab.
Finally, Steps 6-7 use all the pHashes obtained from Twitter, Reddit (all subreddits), \dspol, and Gab to find posts with images matching the annotated clusters.
This is an integral part of our process as it allows to characterize and study mainstream Web communities not used for clustering (i.e., Twitter and Reddit).

\subsection{Analysis} \label{sec:analysis}
In this section, we present a cluster-based measurement of memes and an analysis of a few Web communities from the ``perspective'' of memes.
We measure the prevalence of memes across the clusters obtained from fringe communities: \dspol, \td subreddit (\tdshort), and Gab.
We also use the distance metric introduced in Eq.~\ref{eq:distance} to perform a \emph{cross-community} analysis,
then, we group clusters into broad, but related, categories to gain a macro-perspective understanding of larger communities, including  Reddit and Twitter.

\subsubsection{Cluster-based Analysis}
\label{sec:analysis:clusters}
We start by analyzing the 12.6K annotated clusters consisting of 268K images from \dspol, \td, and Gab (Step 5 in Figure~\ref{fig:pipeline}).
We do so to understand the \emph{diversity} of memes in each Web community, as well as the interplay between \emph{variants} of memes. 
We then evaluate how clusters can be grouped into higher structures using hierarchical clustering and graph visualization techniques.

\begin{table}[t]
\centering
\small
\begin{tabular}{@{}lrrrr@{}}
\toprule
\textbf{Platform} & \textbf{\#Images} & \textbf{Noise} & \textbf{\#Clusters} & \hspace*{-0.2cm}\textbf{\#Clusters with}\\
& & & & \hspace*{-0.1cm}\textbf{KYM tags (\%)}\\
\midrule
\textbf{/pol/}    & 4,325,648                                                                            & 63\%                                                                               & 38,851                                                                                 & 9,265 (24\%)                                                                                              \\
\textbf{\tdshort}     & 1,234,940                                                                            & 64\%                                                                               & 21,917                                                                                 & 2,902 (13\%)                                                                                                \\
\textbf{Gab}      & 235,222                                                                              & 69\%                                                                               & 3,083                                                                                  & 447 (15\%)                                                                                                  \\ \bottomrule
\end{tabular}
\caption{Statistics obtained from clustering images from \dspol, \td, and Gab.}
\label{tbl:clustering-statistics}
\end{table}

\begin{figure*}[t]
\center
\subfigure[]{\includegraphics[width=0.49\textwidth]{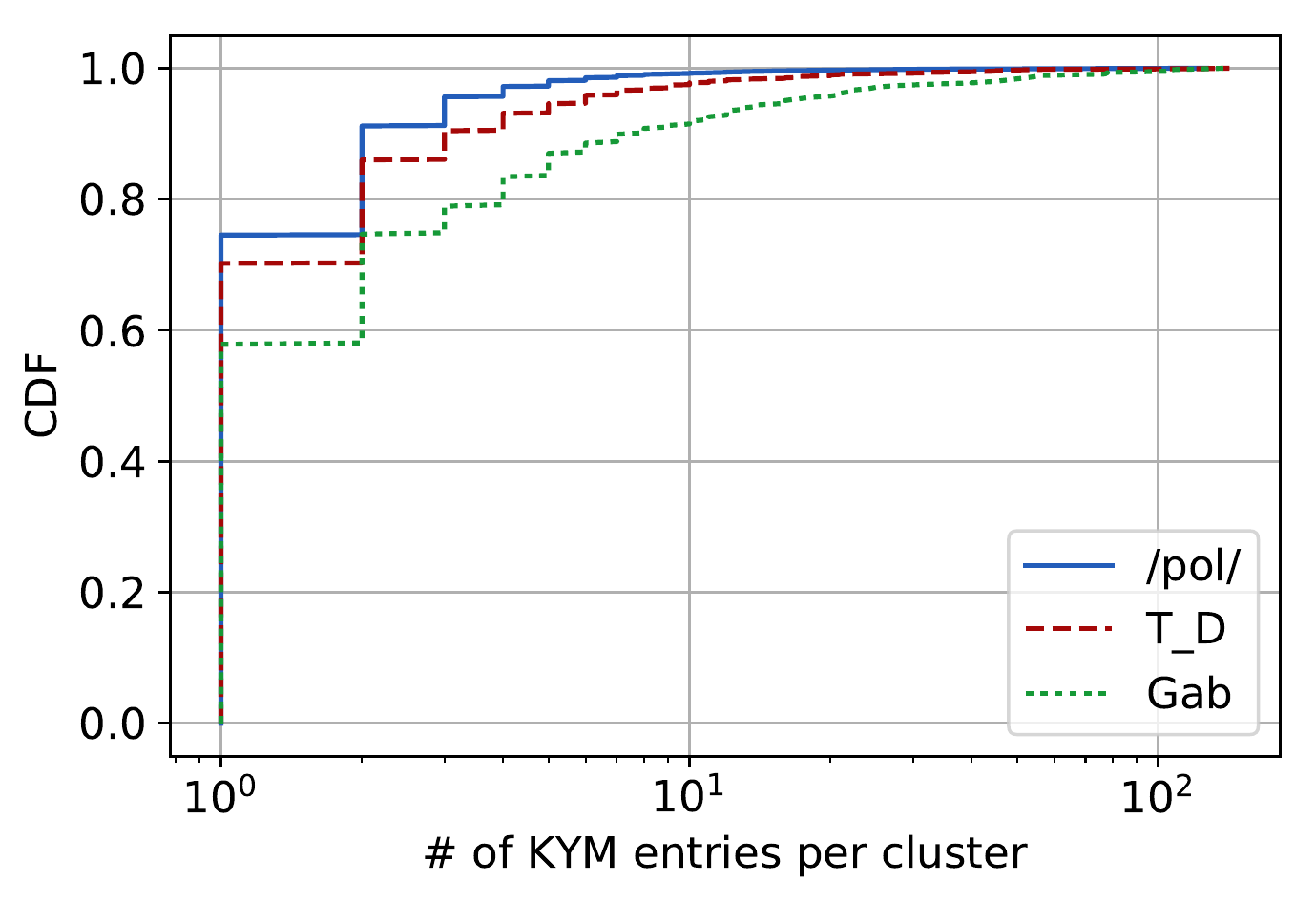}\label{subfig:cdf_kym_entries_per_cluster}}
\subfigure[]{\includegraphics[width=0.49\textwidth]{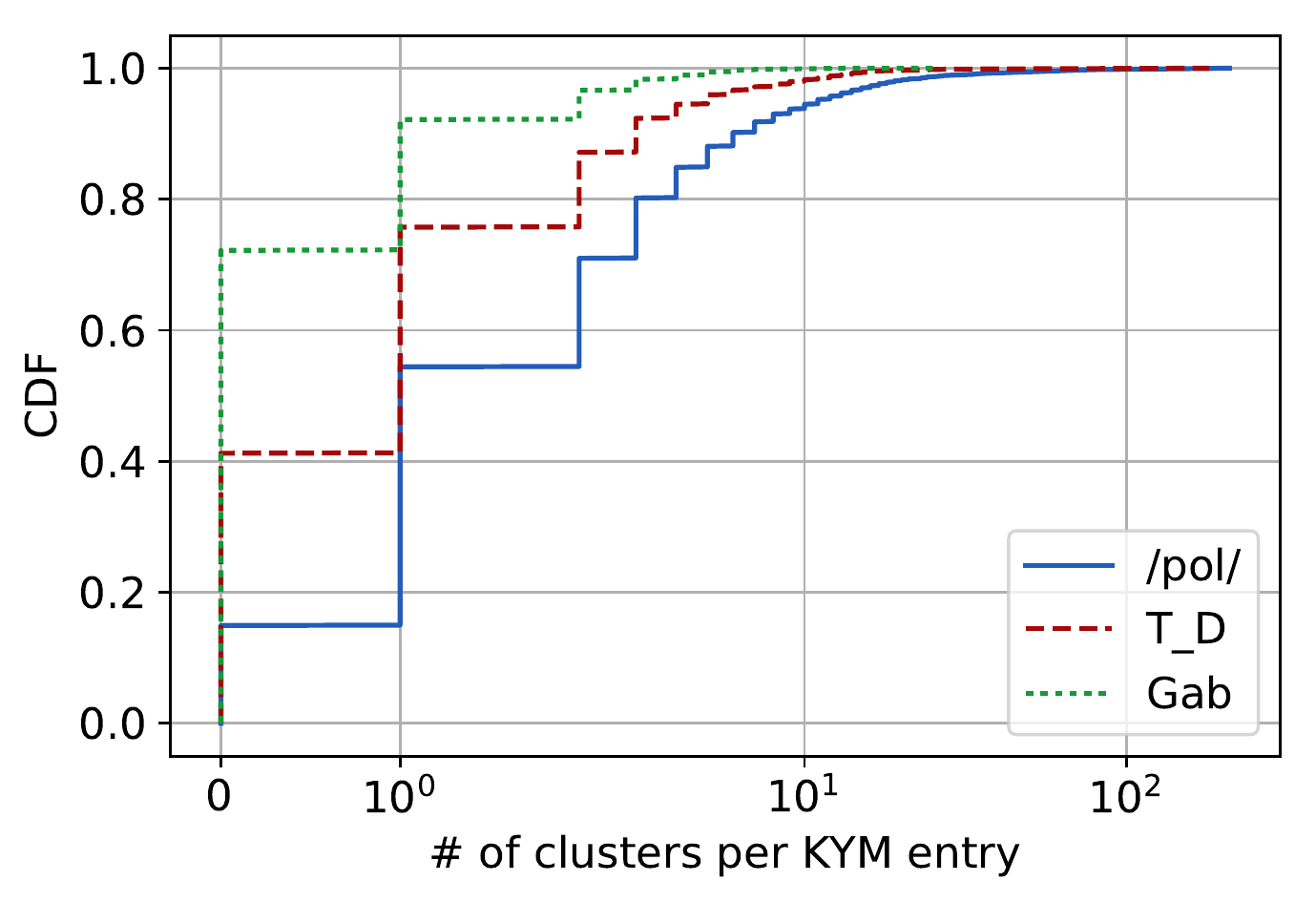}\label{subfig:cdf_clusters_per_kym_entry}}
\caption{CDF of (a) KYM entries per cluster and (b)clusters per KYM entry. }
\label{fig:cdf_clusters_kym_entries}
\end{figure*}

\begin{figure}[t]
\centering
\includegraphics[width=0.5\columnwidth]{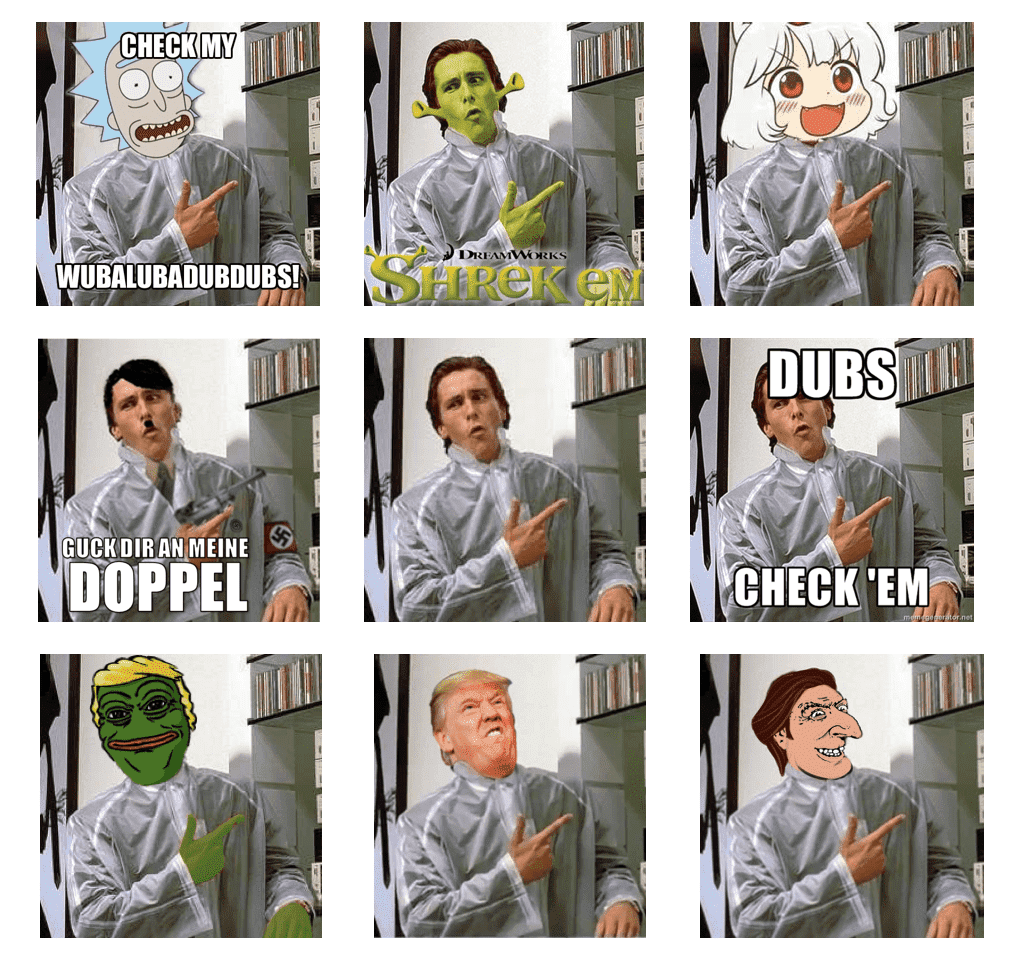}
\caption{Images that are part of the Dubs Guy/Check Em Meme.}
\label{fig:check_em_example}
\end{figure}
\begin{figure}[t]
\centering
\includegraphics[width=0.5\columnwidth]{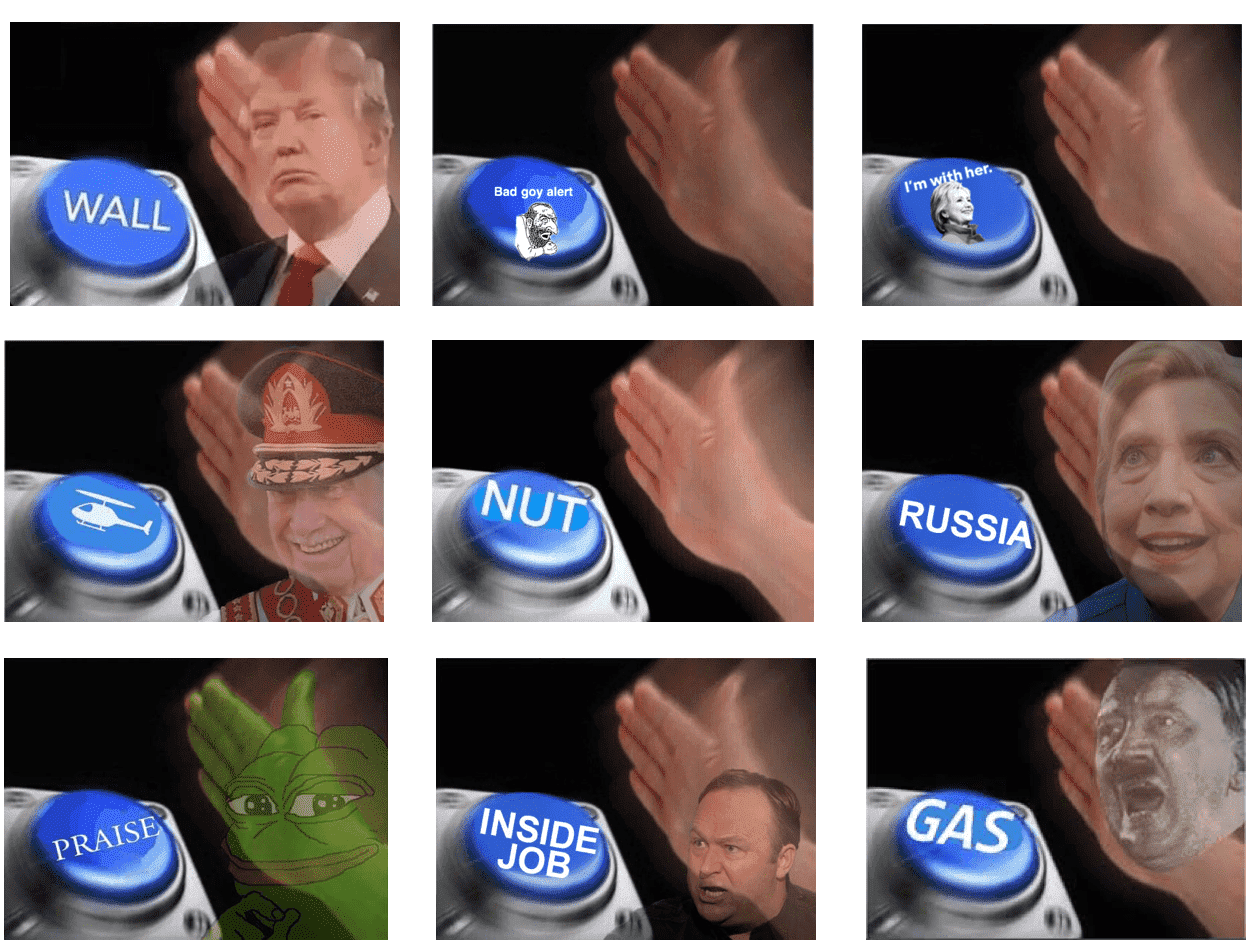}
\caption{Images that are part of the Nut Button Meme.}
\label{fig:nut_button}
\end{figure}
\begin{figure}[t]
\centering
\includegraphics[width=0.5\columnwidth]{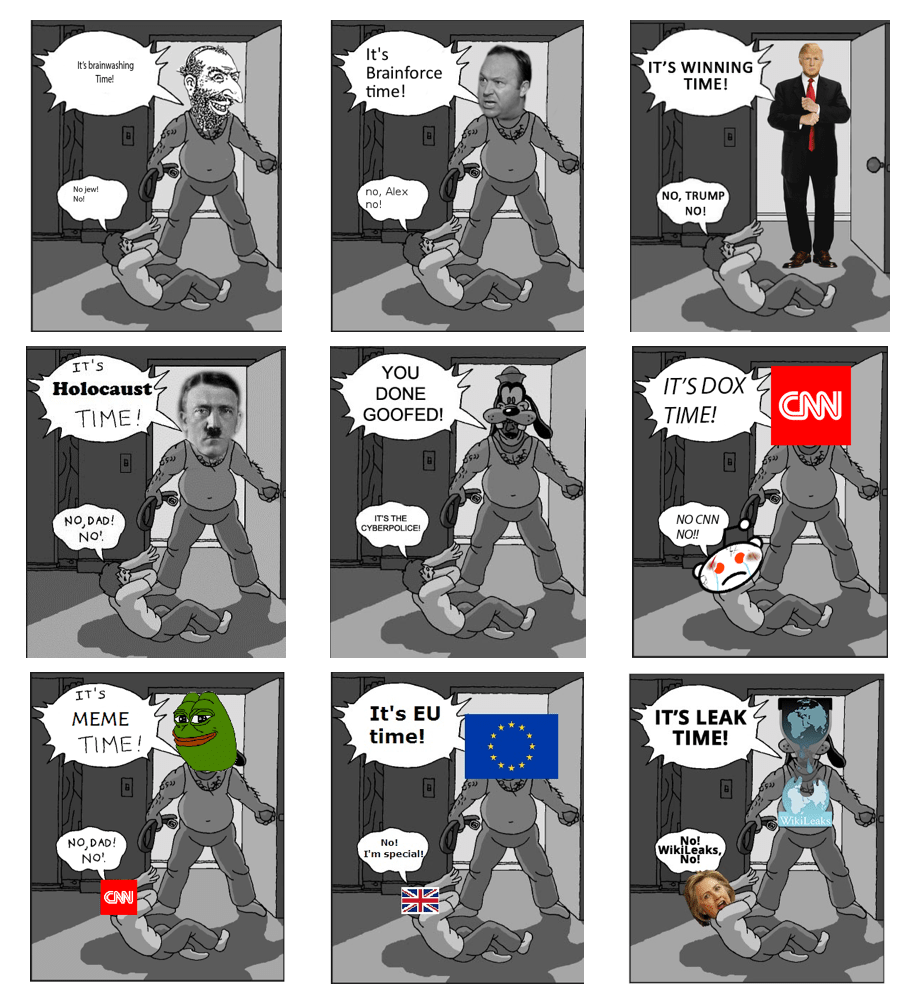}
\caption{Images that are part of the Goofy's Time Meme.}
\label{fig:goofys}
\end{figure}

\subsubsubsection{Clusters}

\noindent\textbf{Statistics.} In Table~\ref{tbl:clustering-statistics}, we report some basic statistics of the clusters obtained for each Web community.
A relatively high percentage of images (63\%--69\%) are not clustered, i.e., are labeled as noise.
While in DBSCAN ``noise'' is just an instance that does not fit in any cluster (more specifically, there are less than 5 images with perceptual distance $\leq 8$ from that particular instance), we note that this likely happens as these images are not memes, but rather ``one-off images.''
For example, on \dspol there is a large number of pictures of random people taken from various social media platforms.

Overall, we have 2.1M images in 63.9K clusters: 38K clusters for \dspol, 21K for \td, and 3K for Gab.
12.6K of these clusters are successfully annotated using the KYM data: 
9.2K from \dspol (142K images), 2.9K from \td (121K images), and 447 from Gab (4.5K images).

We also present some examples of clusters showcasing how the proposed pipeline can effectively detect and group images that belong to the same meme.

Specifically, Figure~\ref{fig:check_em_example} shows a subset of the images from the Dubs Guy/Check Em meme~\cite{dubs_guy_meme}, Figure~\ref{fig:nut_button} a subset of images that belong to the Nut Button meme~\cite{nut_button_meme}, while Figure~\ref{fig:goofys} -- to the Goofy's Time meme~\cite{goofy_meme}.
Note that all these images are obtained from /pol/ clusters.

In all clusters, we observe similar variations, i.e., variations of Donald Trump, Adolf Hitler, The Happy Merchant, and Pepe the Frog appear in all examples.
Once again, this emphasizes the overlap that exists among memes.

As for the un-annotated clusters, manual inspection confirms that many include miscellaneous images unrelated to memes, e.g., similar screenshots of social networks posts (recall that we only filter out screenshots from the KYM image galleries), images captured from video games, etc.

\descr{KYM entries per cluster.} Each cluster may receive multiple annotations, depending on the KYM entries that have at least one image matching that cluster's medoid.
As shown in Figure~\ref{subfig:cdf_kym_entries_per_cluster}, the majority of the annotated clusters (74\% for \dspol, 70\% for \td, and 58\% for Gab) only have a single matching KYM entry.
However, a few clusters have a large number of matching entries, e.g., the one matching the Conspiracy Keanu meme~\cite{conspiracy_keanu} is annotated by 126 KYM entries (primarily, other memes that add text in an image associated with that meme).
This highlights that memes do overlap and that some are highly influenced by other ones.

\descr{Clusters per KYM entry.} We also look at the number of clusters annotated by the \emph{same} KYM entry.
Figure~\ref{subfig:cdf_clusters_per_kym_entry} plots the CDF of the number of clusters per entry.
About 40\% only annotate a single \dspol cluster, while 34\% and 20\% of the entries annotate a single \td and a single Gab cluster, respectively.
We also find that a small number of entries are associated to a large number of clusters: for example, the Happy Merchant meme~\cite{happy_merchant_meme} annotates 124 different clusters on \dspol.
This highlights the \emph{diverse} nature of memes, i.e., memes are mixed and matched, not unlike the way that genetic traits are combined in biological reproduction.

\begin{table*}[t!]
\centering
\setlength{\tabcolsep}{0.18em} %
\resizebox{0.999\textwidth}{!}{%
\begin{tabular}{@{}llrllrllr@{}}
\toprule
\multicolumn{3}{c|}{\textbf{/pol/}}                                                                                                                                                                    & \multicolumn{3}{c|}{\textbf{\tdshort}}                                                                                                                                                     & \multicolumn{3}{c}{\textbf{Gab}}                                                                                                                                          \\ \midrule
\textbf{Entry} & {\bf Category} & \multicolumn{1}{l|}{\textbf{Clusters (\%)}} &
\textbf{Entry} & {\bf Category} & \multicolumn{1}{l|}{\textbf{Clusters (\%)}} &
\textbf{Entry} & {\bf Category} & \multicolumn{1}{l|}{\textbf{Clusters (\%)}} \\ \midrule
\href{http://knowyourmeme.com/memes/people/donald-trump}{Donald Trump}                                                                     & People                                                             & \multicolumn{1}{r|}{207 (2.2\%)}            & \href{http://knowyourmeme.com/memes/people/donald-trump}{Donald Trump}                                                      & People                                                             & \multicolumn{1}{r|}{177 (6.1\%)}            & \href{http://knowyourmeme.com/memes/people/donald-trump}{Donald Trump}                                                                 & People                                                            & 25 (5.6\%)             \\
\href{http://knowyourmeme.com/memes/happy-merchant}{Happy Merchant}                                                                   & Memes                                                              & \multicolumn{1}{r|}{124 (1.3\%)}            & \href{http://knowyourmeme.com/memes/smug-frog}{Smug Frog}                                                           & Memes                                                              & \multicolumn{1}{r|}{78 (2.7\%)}             & \href{http://knowyourmeme.com/memes/happy-merchant}{Happy Merchant}                                                                & Memes                                                             & 10 (2.2\%)             \\
\href{http://knowyourmeme.com/memes/smug-frog}{Smug Frog}                                                                        & Memes                                                              & \multicolumn{1}{r|}{114 (1.2\%)}            & \href{http://knowyourmeme.com/memes/pepe-the-frog}{Pepe the Frog}                                                      & Memes                                                              & \multicolumn{1}{r|}{63 (2.1\%)}             & \href{http://knowyourmeme.com/memes/demotivational-posters}{Demotivational Posters}                                                        & Memes                                                             & 7 (1.5\%)              \\
\href{http://knowyourmeme.com/memes/computer-reaction-faces}{Computer Reaction Faces}                                                          & Memes                                                              & \multicolumn{1}{r|}{112 (1.2\%)}            & \href{http://knowyourmeme.com/memes/feels-bad-man-sad-frog}{Feels Bad Man/ Sad Frog} & Memes                                                              & \multicolumn{1}{r|}{61 (2.1\%)}             & \href{http://knowyourmeme.com/memes/pepe-the-frog}{Pepe the Frog}                                                                  & Memes                                                             & 6 (1.3\%)              \\
\href{http://knowyourmeme.com/memes/feels-bad-man-sad-frog}{Feels Bad Man/ Sad Frog}              & Memes                                                              & \multicolumn{1}{r|}{94 (1.0\%)}             & \href{http://knowyourmeme.com/memes/make-america-great-again}{Make America Great Again}  & Memes                                                              & \multicolumn{1}{r|}{50 (1.7\%)}             & \href{http://knowyourmeme.com/memes/events/cnnblackmail}{\#Cnnblackmail}                                                               & Events                                                            & 6 (1.3\%)              \\
\href{http://knowyourmeme.com/memes/i-know-that-feel-bro}{I Know that Feel Bro}                                                             & Memes                                                              & \multicolumn{1}{r|}{90 (1.0\%)}             & \href{http://knowyourmeme.com/memes/people/bernie-sanders}{Bernie Sanders}                                                      & People                                                             & \multicolumn{1}{r|}{31 (1.0\%)}             & \href{http://knowyourmeme.com/memes/events/2016-united-states-presidential-election}{2016 US election}                                                             & Events                                                            & 6 (1.3\%)              \\
\href{http://knowyourmeme.com/memes/tony-kornheiser-s-why}{Tony Kornheiser's Why}                                                            & Memes                                                              & \multicolumn{1}{r|}{89 (1.0\%)}             & \href{http://knowyourmeme.com/memes/events/2016-united-states-presidential-election}{2016 US Election}                                                   & Events                                                             & \multicolumn{1}{r|}{27 (0.9\%)}             & \href{http://knowyourmeme.com/memes/sites/knowyourmeme}{Know Your Meme}                                                               & Sites                                                             & 6 (1.3\%)              \\
\href{http://knowyourmeme.com/memes/bait-this-is-bait}{Bait/This is Bait}                                                                & Memes                                                              & \multicolumn{1}{r|}{84 (0.9\%)}             & \href{http://knowyourmeme.com/memes/counter-signal-memes}{Counter Signal Memes}   & Memes                                                              & \multicolumn{1}{r|}{24 (0.8\%)}             & \href{http://knowyourmeme.com/memes/sites/tumblr}{Tumblr}                                                                       & Sites                                                             & 6 (1.3\%)              \\
\href{http://knowyourmeme.com/memes/events/trumpanime-rick-wilson-controversy}{\#TrumpAnime/Rick Wilson} & Events                                                             & \multicolumn{1}{r|}{76 (0.8\%)}             & \href{http://knowyourmeme.com/memes/events/cnnblackmail}{\#Cnnblackmail}                                                     & Events                                                             & \multicolumn{1}{r|}{24 (0.8\%)}             & \href{http://knowyourmeme.com/memes/cultures/feminism}{Feminism}                                                                     & Cultures                                                          & 5 (1.1\%)              \\
\href{http://knowyourmeme.com/memes/reaction-images}{Reaction Images}                                                                  & Memes                                                              & \multicolumn{1}{r|}{73 (0.8\%)}             & \href{http://knowyourmeme.com/memes/sites/knowyourmeme}{Know Your Meme}                                                     & Sites                                                              & \multicolumn{1}{r|}{20 (0.7\%)}             & \href{http://knowyourmeme.com/memes/people/barack-obama}{Barack Obama}                                                                 & People                                                            & 5 (1.1\%)              \\
\href{http://knowyourmeme.com/memes/make-america-great-again}{Make America Great Again}               & Memes                                                              & \multicolumn{1}{r|}{72 (0.8\%)}             & \href{http://knowyourmeme.com/memes/angry-pepe}{Angry Pepe}                                                         & Memes                                                              & \multicolumn{1}{r|}{18 (0.6\%)}             & \href{http://knowyourmeme.com/memes/smug-frog}{Smug Frog}                                                                     & Memes                                                             & 5 (1.1\%)              \\
\href{http://knowyourmeme.com/memes/counter-signal-memes}{Counter Signal Memes}               & Memes                                                              & \multicolumn{1}{r|}{72 (0.8\%)}             & \href{http://knowyourmeme.com/memes/demotivational-posters}{Demotivational Posters}                                             & Memes                                                              & \multicolumn{1}{r|}{18 (0.6\%)}             & \href{http://knowyourmeme.com/memes/subcultures/rwby}{rwby}                                                                         & Subcultures                                                       & 5 (1.1\%)              \\
\href{http://knowyourmeme.com/memes/pepe-the-frog}{Pepe the Frog}                                                                      & Memes                                                              & \multicolumn{1}{r|}{65 (0.7\%)}             & \href{http://knowyourmeme.com/memes/sites/4chan}{4chan}                                                              & Sites                                                              & \multicolumn{1}{r|}{16 (0.5\%)}             & \href{http://knowyourmeme.com/memes/people/kim-jong-un}{Kim Jong Un}                                                                  & People                                                            & 5 (1.1\%)              \\
\href{http://knowyourmeme.com/memes/subcultures/spongebob-squarepants}{Spongebob Squarepants}                                                            & Subcultures                                                        & \multicolumn{1}{r|}{61 (0.7\%)}             & \href{http://knowyourmeme.com/memes/sites/tumblr}{Tumblr}                                                             & Sites                                                              & \multicolumn{1}{r|}{15 (0.5\%)}             & \href{http://knowyourmeme.com/memes/murica}{Murica}                                                                       & Memes                                                             & 5 (1.1\%)              \\
\href{http://knowyourmeme.com/memes/doom-paul-its-happening}{Doom Paul its Happening}                                                          & Memes                                                              & \multicolumn{1}{r|}{57 (0.6\%)}             & \href{http://knowyourmeme.com/memes/events/gamergate}{Gamergate}                                                          & Events                                                             & \multicolumn{1}{r|}{15 (0.5\%)}             & \href{http://knowyourmeme.com/memes/events/united-airlines-passenger-removal}{UA Passenger Removal} & Events                                                            & 5 (1.1\%)              \\
\href{http://knowyourmeme.com/memes/people/adolf-hitler}{Adolf Hitler}                                                                     & People                                                             & \multicolumn{1}{r|}{56 (0.6\%)}             & \href{http://knowyourmeme.com/memes/colbertposting}{Colbertposting}                                                     & Memes                                                              & \multicolumn{1}{r|}{15 (0.5\%)}              & \href{http://knowyourmeme.com/memes/make-america-great-again}{Make America Great Again}            & Memes                                                             & 4 (0.9\%)              \\
\href{http://knowyourmeme.com/memes/sites/pol}{pol}                                                                              & Sites                                                              & \multicolumn{1}{r|}{53 (0.6\%)}             & \href{http://knowyourmeme.com/memes/donald-trumps-wall}{Donald Trump's Wall}                                                & Memes                                                              & \multicolumn{1}{r|}{15 (0.5\%)}              & \href{http://knowyourmeme.com/memes/people/bill-nye}{Bill Nye}                                                                     & People                                                            & 4 (0.9\%)              \\
\href{http://knowyourmeme.com/memes/dubs-guy-check-em}{Dubs Guy/Check'em}                     & Memes                                                              & \multicolumn{1}{r|}{53 (0.6\%)}             & \href{http://knowyourmeme.com/memes/people/vladimir-putin}{Vladimir Putin}                                                     & People                                                             & \multicolumn{1}{r|}{15 (0.5\%)}              & \href{http://knowyourmeme.com/memes/cultures/trolling}{Trolling}                                                                     & Cultures                                                          & 4 (0.9\%)              \\
\href{http://knowyourmeme.com/memes/smug-anime-face}{Smug Anime Face}                                                                  & Memes                                                              & \multicolumn{1}{r|}{51 (0.6\%)}             & \href{http://knowyourmeme.com/memes/people/barack-obama}{Barack Obama}                                                        & People                                                             & \multicolumn{1}{r|}{15 (0.5\%)}              & \href{http://knowyourmeme.com/memes/sites/4chan}{4chan}                                                                        & Sites                                                             & 4 (0.9\%)              \\
\href{http://knowyourmeme.com/memes/subcultures/warhammer-40000}{Warhammer 40000}                                                                & Subcultures                                                        & \multicolumn{1}{r|}{51 (0.6\%)}             & \href{http://knowyourmeme.com/memes/people/hillary-clinton}{Hillary Clinton}                                                    & People                                                             & \multicolumn{1}{r|}{15 (0.5\%)}              & \href{http://knowyourmeme.com/memes/cultures/furries}{Furries}                                                                      & Cultures                                                          & 3 (0.7\%)              \\
\midrule
{\bf Total} && \multicolumn{1}{r|}{\bf 1,638 (17.7\%)} &&& \multicolumn{1}{r|}{\bf 695 (23.9\%)} &&& \multicolumn{1}{r|}{\bf 121 (27.1\%)} \\
\bottomrule
\end{tabular}%
}
\caption{Top 20 KYM entries appearing in the clusters of \dspol, \td, and Gab. We report the number of clusters and their respective percentage (per community).
Each item contains a hyperlink to the corresponding entry on the KYM website.}
\label{tbl:top_entries_cluster_numbers}
\end{table*}

\descr{Top KYM entries.}
Because the majority of clusters match only one or two KYM entries (Figure~\ref{subfig:cdf_kym_entries_per_cluster}), we simplify things by giving all clusters a \emph{representative annotation} based on the most prevalent annotation given to the medoid, and, in the case of ties the average distance between all matches (see Section~\ref{subsec:pipeline}).
\emph{Thus, in the rest of this thesis, we report our findings based on the representative annotation for each cluster.}

In Table~\ref{tbl:top_entries_cluster_numbers}, we report the top 20 KYM entries with respect to the number of clusters they annotate.
These cover 17\%, 23\%, and 27\% of the clusters in \dspol, \td, and Gab, respectively, hence covering a relatively good sample of our datasets.
Donald Trump~\cite{donald_trump_meme}, Smug Frog~\cite{smug_frog_meme}, and Pepe the Frog~\cite{pepe_frog_meme} appear in the top 20 for all three communities, while the Happy Merchant~\cite{happy_merchant_meme} only in \dspol and Gab.
In particular, Donald Trump annotates the most clusters (207 in \dspol, 177 in \td, and 25 in Gab).
In fact, politics-related entries appear several times in the Table, 
e.g., Make America Great Again~\cite{maga_meme} as well as political personalities like Bernie Sanders, Barack Obama, Vladimir Putin, and Hillary Clinton.

When comparing the different communities, we observe the most prevalent categories are memes (6 to 14 entries in each community) and people (2-5).
Moreover, in \dspol, the 2nd most popular entry, related to people, is Adolf Hilter, which supports previous reports of the community's sympathetic views toward Nazi ideology~\cite{hine2016longitudinal}.
Overall, there are several memes with hateful or disturbing content (e.g., holocaust).
This happens to a lesser extent in \td and Gab: the most popular people after Donald Trump are contemporary politicians, i.e., Bernie Sanders, Vladimir Putin, Barack Obama, and Hillary Clinton.

Finally, image posting behavior in fringe Web communities is greatly influenced by real-world events.
For instance, in \dspol, we find the \#TrumpAnime controversy event~\cite{trump_anime_meme}, where a political individual (Rick Wilson) offended the alt-right community, Donald Trump supporters, and anime fans (an oddly intersecting set of interests of \dspol users).
Similarly, on \td and Gab, we find the \#Cnnblackmail~\cite{cnnblackmail_meme} event, referring to the (alleged) blackmail of the Reddit user that created the infamous video of Donald Trump wrestling the CNN.

\begin{figure*}[t]
\centering
\stackunder[1pt]{\includegraphics[height=1.5cm]{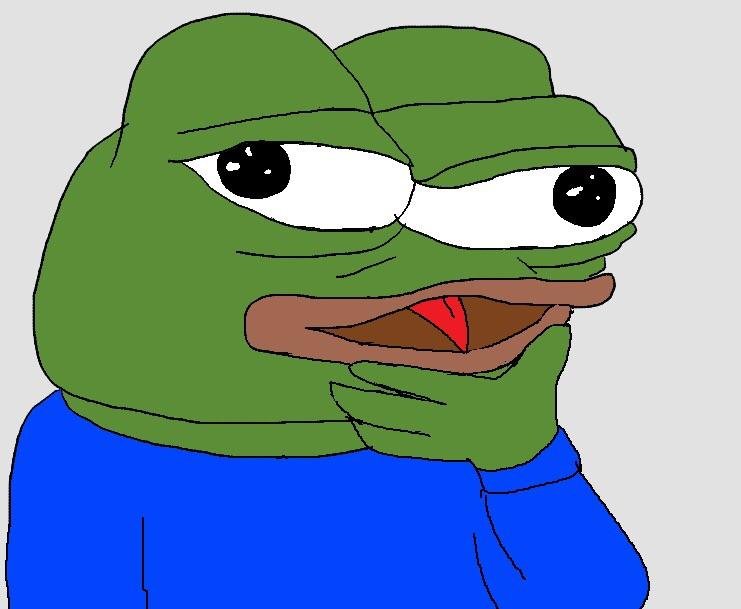}}{apustaja}~%
\stackunder[1pt]{\includegraphics[height=1.5cm]{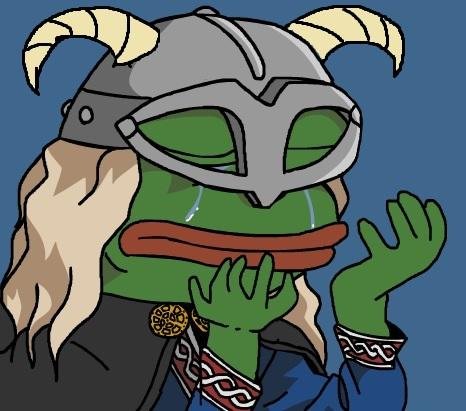}}{sad-frog}~~%
\stackunder[1pt]{\includegraphics[height=1.5cm]{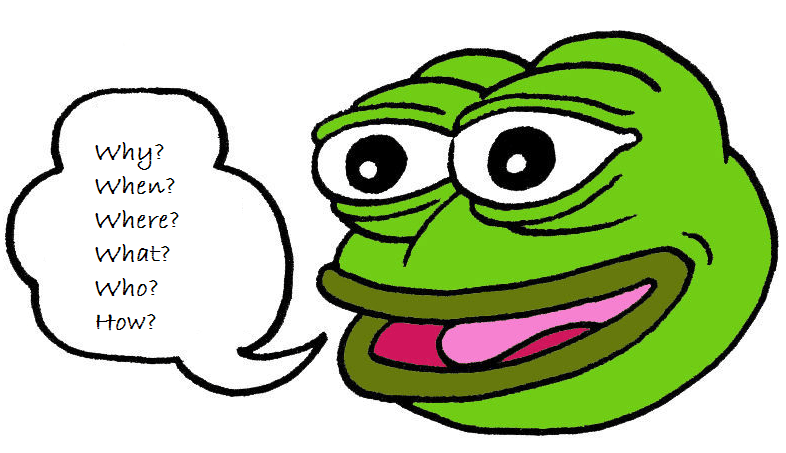}}{savepepe}~~%
\stackunder[1pt]{\includegraphics[height=1.5cm]{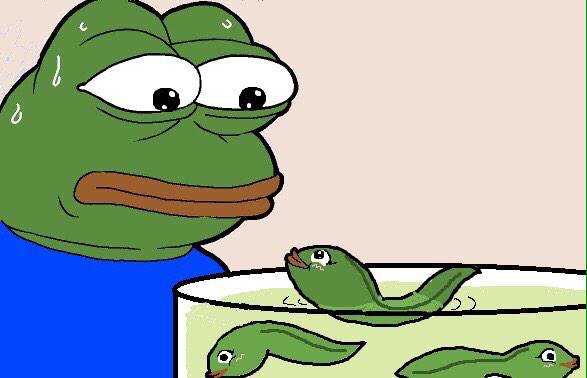}}{pepe}~~%
\stackunder[1pt]{\includegraphics[height=1.5cm]{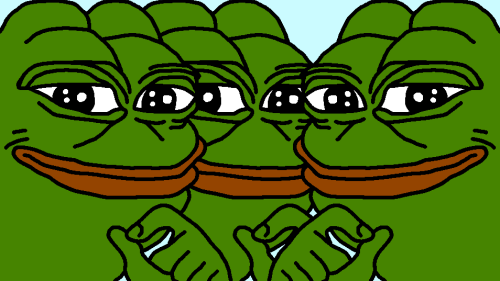}}{smug-frog-a}~~%
\stackunder[1pt]{\includegraphics[height=1.5cm]{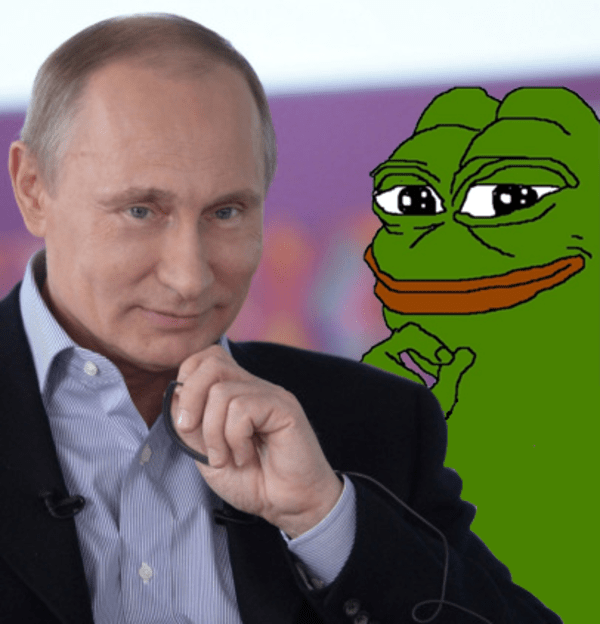}}{smug-frog-b}~~%
\stackunder[1pt]{\includegraphics[height=1.5cm]{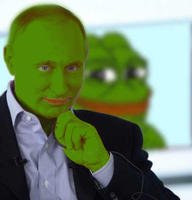}}{anti-meme}
\subfigure{\includegraphics[width=1\textwidth, trim=255 200 255 85, clip]{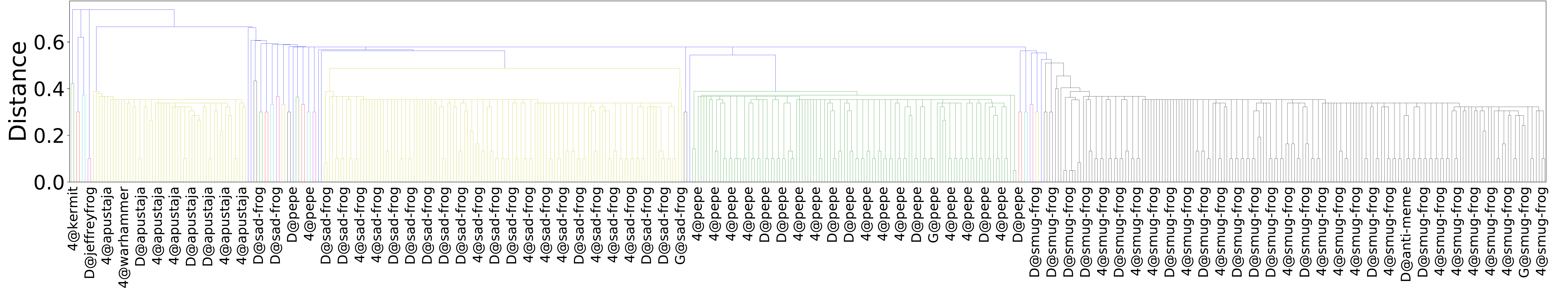}}%
\vspace*{-.5cm}
  \caption{Inter-cluster distance between all clusters with frog memes. Clusters are labeled with the origin ({\em 4} for 4chan, {\em D} for \td, and {\em G} for Gab) and the meme name. To ease  readability, we do not display all labels, abbreviate meme names, and only show an excerpt of all relationships.}
\label{fig:casestudy_frogs_dendrogram}
\end{figure*}

\subsubsubsection{Memes' Branching Nature}\label{subsection:hierarchy} 
Next, we study how memes \emph{evolve} by looking at {\em variants} across different clusters.
Intuitively, clusters that look alike and/or are part of the same meme are grouped together under the same branch of an evolutionary tree.
We use the custom distance metric introduced in Section~\ref{sec:methodology:distance}, aiming to infer the phylogenetic relationship between variants of memes.
Since there are 12.6K annotated clusters, we only report on a subset of variants. %
In particular, we focus on ``frog'' memes (e.g., Pepe the Frog~\cite{pepe_frog_meme}); as discussed later in Section~\ref{sec:analysis:annotations}, this is one of the most popular memes in our datasets.

The dendrogram in Figure~\ref{fig:casestudy_frogs_dendrogram} shows the hierarchical relationship between groups of clusters of memes related to frogs.
Overall, there are 525 clusters of frogs, belonging to 23 different memes.
These clusters can be grouped into four large categories, dominated by Apu Apustaja~\cite{apu_meme}, Feels Bad Man/Sad Frog~\cite{sad_frog_meme}, Pepe the Frog~\cite{pepe_frog_meme}, and Smug Frog~\cite{smug_frog_meme}.
The different memes express different ideas or messages: e.g., Apu Apustaja depicts a simple-minded non-native speaker using broken English, while the Feels Bad Man/Sad Frog (ironically) expresses dismay at a given situation, often accompanied with text like ``You will never do/be/have X.''
The dendrogram also shows a variant of Smug Frog ({\it smug-frog-b}) related to a variant of the Russian Anti Meme Law~\cite{russian_anti_meme} ({\it anti-meme}) as well as relationships between clusters from Pepe the Frog and Isis meme~\cite{isis_meme},
and between Smug Frog and Brexit-related clusters~\cite{brexit_meme}, as shown in Section~\ref{sec:appendix_interesting_images}.

The distance metric quantifies the similarity of any two variants of {\em different} memes; however, recall that two clusters can be close to each other even when the medoids are perceptually different (see Section~\ref{sec:methodology:distance}), as in the case of Smug Frog variants in the {\it smug-frog-a} and {\it smug-frog-b} clusters (top of Figure~\ref{fig:casestudy_frogs_dendrogram}).
Although this analysis is limited to a single ``family'' of memes, our distance metric can actually provide useful insights regarding the phylogenetic relationships of any clusters.
In fact, more extensive analysis of these relationships (through our pipeline) can facilitate the understanding of the diffusion of ideas and information across the Web, and provide a rigorous technique for large-scale analysis of Internet culture.

\begin{figure*}[t!]
\centering
\includegraphics[width=0.85\textwidth]{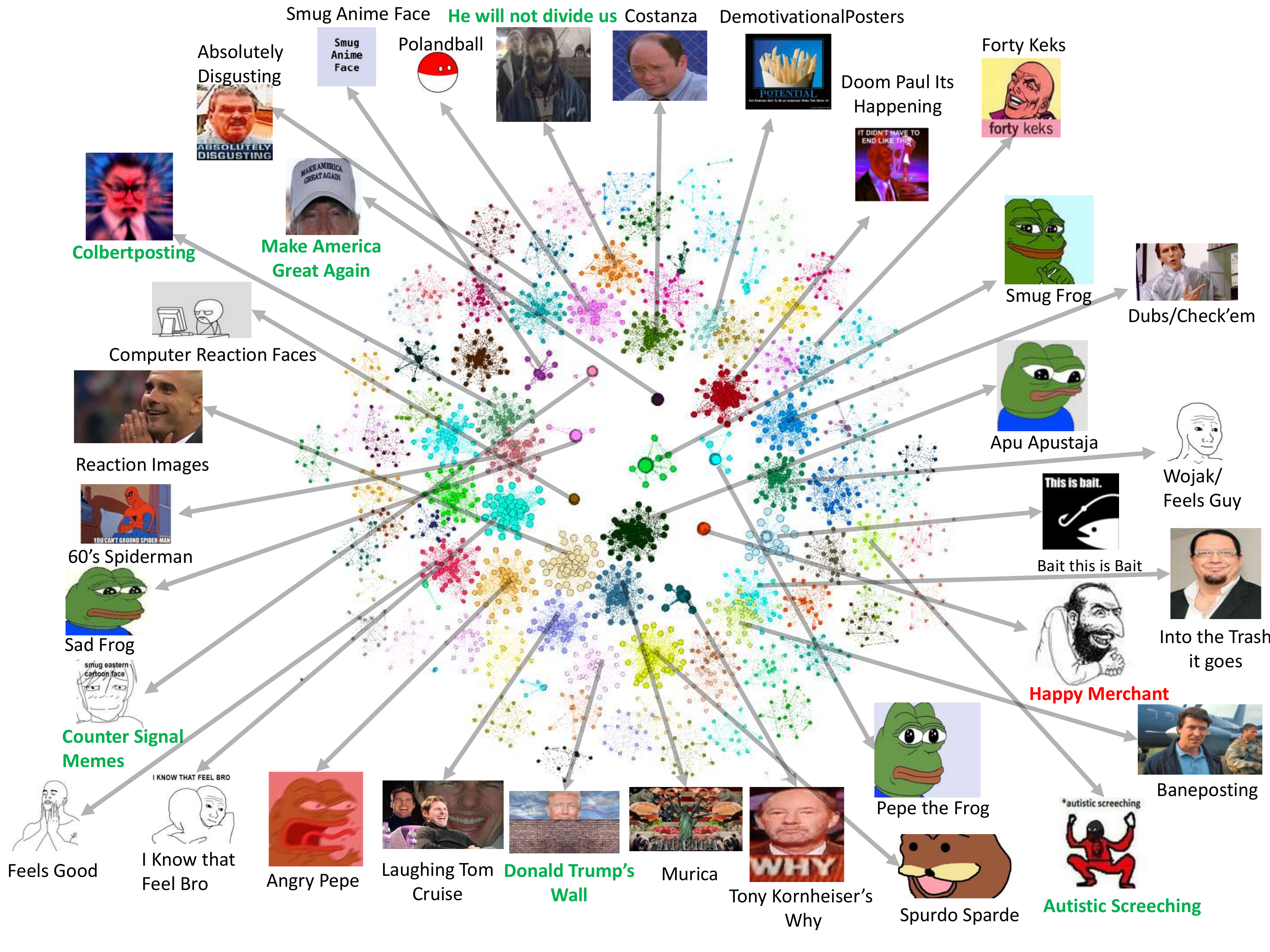}
\caption{Visualization of the obtained clusters from \dspol, \td, and Gab. Note that memes with red labels are annotated as racist, while memes with green labels are annotated as politics (see Section~\ref{sec:meme_popularity} for the selection criteria).}
\label{fig:clusters_graph}
\end{figure*}

\subsubsection{Meme Visualization}
\label{sec:analysis:clusters:visualization}
We also use the custom distance metric (see Eq.~\ref{eq:distance}) to visualize the clusters with annotations.
We build a graph $\boldsymbol{G}=(\boldsymbol{V},\boldsymbol{E})$, where $\boldsymbol{V}$ are the medoids of annotated clusters and $\boldsymbol{E}$ the connections between medoids with distance under a threshold $\kappa$.
Figure~\ref{fig:clusters_graph} shows a snapshot of the graph for $\kappa=0.45$, chosen based on the frogs analysis above.
In particular, we select this threshold as the majority of the clusters from the same meme (note coloration in Figure~\ref{fig:casestudy_frogs_dendrogram}) are hierarchically connected with a higher-level cluster at a distance close to 0.45.
To ease readability, we filter out nodes and edges that have a sum of in- and out-degree less than 10, which leaves 40\% of the nodes and 92\% of the edges.
Nodes are colored according to their KYM annotation.
NB: the graph is laid out using the OpenOrd algorithm~\cite{martin2011openord} and the distance between the components in it does not exactly match the actual distance metric.
We observe a large set of disconnected components, with each component containing nodes of primarily one color.
This indicates that our distance metric is indeed capturing the peculiarities of different memes. %
Finally, note that an interactive version of the full graph is publicly available from~\cite{memes_graph_site}. %

\subsubsection{Web Community-based Analysis}
\label{sec:analysis:annotations}
We now present a macro-perspective analysis of the Web communities through the lens of memes.
We assess the presence of different memes in each community, how popular they are, and how they evolve.
To this end, we examine the {\em posts} from all four communities (Twitter, Reddit, \dspol, and Gab) that contain {\em images} matching {\em memes} from fringe Web communities (\dspol, \td, and Gab).

\subsubsubsection{Meme Popularity}\label{sec:meme_popularity}

\descr{Memes.}
We start by analyzing clusters grouped by KYM `meme' entries, looking at the number of posts for each meme in \dspol, Reddit, Gab, and Twitter.

\begin{table*}[t]
\centering
\resizebox{\textwidth}{!}{%
\begin{tabular}{lrlrlrlr}
\toprule
\multicolumn{2}{c}{\textbf{/pol/}}                                            & \multicolumn{2}{c}{\textbf{Reddit}}                                           & \multicolumn{2}{c}{\textbf{Gab}}                                                                                           & \multicolumn{2}{c}{\textbf{Twitter}}                                        \\ \midrule
\multicolumn{1}{c}{\textbf{Entry}} & \multicolumn{1}{c|}{\textbf{Posts (\%)}} & \multicolumn{1}{c}{\textbf{Entry}} & \multicolumn{1}{c|}{\textbf{Posts (\%)}} & \multicolumn{1}{c}{\textbf{Entry}}                                              & \multicolumn{1}{c|}{\textbf{Posts (\%)}} & \multicolumn{1}{c}{\textbf{Entry}} & \multicolumn{1}{c}{\textbf{Posts(\%)}} \\ \midrule
\href{http://knowyourmeme.com/memes/feels-bad-man-sad-frog}{Feels Bad Man/Sad Frog}           & \multicolumn{1}{r|}{64,367 (4.9\%)}      & \href{http://knowyourmeme.com/memes/manningface}{Manning Face}                       & \multicolumn{1}{r|}{12,540 (2.2\%)}       & \href{http://knowyourmeme.com/memes/jesusland}{Jesusland \textbf{(P)}}                                                                       & \multicolumn{1}{r|}{454 (1.6\%)}         & \href{http://knowyourmeme.com/memes/roll-safe}{Roll Safe}                                  & 55,010 (5.9\%)                                      \\
\href{http://knowyourmeme.com/memes/smug-frog}{Smug Frog}                          & \multicolumn{1}{r|}{63,290 (4.8\%)}      & \href{http://knowyourmeme.com/memes/thats-the-joke}{That's the Joke}                    & \multicolumn{1}{r|}{7,626 (1.3\%)}       & \href{http://knowyourmeme.com/memes/demotivational-posters}{Demotivational Posters}                                                           & \multicolumn{1}{r|}{414 (1.5\%)}         & \href{http://knowyourmeme.com/memes/evil-kermit}{Evil Kermit}                                    & 50,642 (5.4\%)                                      \\
\href{http://knowyourmeme.com/memes/happy-merchant}{Happy Merchant \textbf{(R)}}                     & \multicolumn{1}{r|}{49,608 (3.8\%)}      & \href{http://knowyourmeme.com/memes/feels-bad-man-sad-frog}{Feels Bad Man/ Sad Frog}           & \multicolumn{1}{r|}{7,240 (1.3\%)}       & \href{http://knowyourmeme.com/memes/smug-frog}{Smug Frog}                                                                        & \multicolumn{1}{r|}{392 (1.4\%)}         &                                 \href{http://knowyourmeme.com/memes/arthurs-fist}{Arthur's Fist} & 37,591 (4.0\%)                                      \\
\href{http://knowyourmeme.com/memes/apu-apustaja}{Apu Apustaja}                       & \multicolumn{1}{r|}{29,756 (2.2\%)}      & \href{http://knowyourmeme.com/memes/confession-bear}{Confession Bear}                    & \multicolumn{1}{r|}{7,147 (1.3\%)}       & \href{http://knowyourmeme.com/memes/based-stickman}{Based Stickman \textbf{(P)}}                                                                  & \multicolumn{1}{r|}{391 (1.4\%)}         &   \href{http://knowyourmeme.com/memes/nut-button}{Nut Button}                               &     13,598 (1,5\%)                                 \\
\href{http://knowyourmeme.com/memes/pepe-the-frog}{Pepe the Frog}                      & \multicolumn{1}{r|}{25,197 (1.9\%)}      & \href{http://knowyourmeme.com/memes/this-is-fine}{This is Fine}                           & \multicolumn{1}{r|}{5,032 (0.9\%)}       & \href{http://knowyourmeme.com/memes/pepe-the-frog}{Pepe the Frog}                                                                   & \multicolumn{1}{r|}{378 (1.3\%)}         &  \href{http://knowyourmeme.com/memes/spongebob-mock}{Spongebob Mock}                                  &  11,136 (1,2\%)                                     \\
\href{http://knowyourmeme.com/memes/make-america-great-again}{Make America Great Again \textbf{(P)}\hspace*{-0.4cm}}            & \multicolumn{1}{r|}{21,229 (1.6\%)}      & \href{http://knowyourmeme.com/memes/smug-frog}{Smug Frog}         & \multicolumn{1}{r|}{4,642 (0.8\%)}       & \href{http://knowyourmeme.com/memes/happy-merchant}{Happy Merchant \textbf{(R)}}                                                                  & \multicolumn{1}{r|}{297 (1.1\%)}         &   \href{http://knowyourmeme.com/memes/reaction-images}{Reaction Images}                             &  9,387 (1.0\%)                                      \\
\href{http://knowyourmeme.com/memes/angry-pepe}{Angry Pepe}                         & \multicolumn{1}{r|}{20,485 (1.5\%)}      & \href{http://knowyourmeme.com/memes/roll-safe}{Roll Safe}                     & \multicolumn{1}{r|}{4,523 (0.8\%)}       & \href{http://knowyourmeme.com/memes/murica}{Murica}                                                                          & \multicolumn{1}{r|}{274 (1.0\%)}         &      \href{http://knowyourmeme.com/memes/conceited-reaction}{Conceited Reaction}  &    9,106 (1.0\%)                                  \\
\href{http://knowyourmeme.com/memes/bait-this-is-bait}{Bait this is Bait}                 & \multicolumn{1}{r|}{16,686 (1.2\%)}      & \href{http://knowyourmeme.com/memes/rage-guy-fffffuuuuuuuu}{Rage Guy}                        & \multicolumn{1}{r|}{4,491 (0.8\%)}       & \href{http://knowyourmeme.com/memes/and-its-gone}{And Its Gone}                                                                    & \multicolumn{1}{r|}{235 (0.9\%)}         &  \href{http://knowyourmeme.com/memes/expanding-brain}{Expanding Brain}                                                               & 8,701 (0.9\%)                                       \\
\href{http://knowyourmeme.com/memes/i-know-that-feel-bro}{I Know that Feel Bro}               & \multicolumn{1}{r|}{14,490 (1.1\%)}      & \href{http://knowyourmeme.com/memes/make-america-great-again}{Make America Great Again \textbf{(P)\hspace*{-0.4cm}}}             & \multicolumn{1}{r|}{4,440 (0.8\%)}       & \href{http://knowyourmeme.com/memes/make-america-great-again}{Make America Great Again \textbf{(P)}}                                                         & \multicolumn{1}{r|}{207 (0.8\%)}         & \href{http://knowyourmeme.com/memes/demotivational-posters}{Demotivational Posters}                            &                                      7,781 (0.8\%)  \\
\href{http://knowyourmeme.com/memes/cult-of-kek}{Cult of Kek}                        & \multicolumn{1}{r|}{14,428 (1.1\%)}      & \href{http://knowyourmeme.com/memes/fake-ccg-cards}{Fake CCG Cards}         & \multicolumn{1}{r|}{4,438 (0.8\%)}       & \href{http://knowyourmeme.com/memes/feels-bad-man-sad-frog}{Feels Bad Man/ Sad Frog} & \multicolumn{1}{r|}{206 (0.8\%)}         &   \href{http://knowyourmeme.com/memes/cash-me-ousside-howbow-dah}{Cash Me Ousside/Howbow Dah}                                     &      5,972 (0.6\%)                              \\
\href{http://knowyourmeme.com/memes/laughing-tom-cruise}{Laughing Tom Cruise}                & \multicolumn{1}{r|}{14,312 (1.1\%)}      &  \href{http://knowyourmeme.com/memes/confused-nick-young}{Confused Nick Young}                & \multicolumn{1}{r|}{4,024 (0.7\%)}       & \href{http://knowyourmeme.com/memes/trump-s-first-order-of-business}{Trump's First Order of Business \textbf{(P)}\hspace*{-0.4cm}}      & \multicolumn{1}{r|}{192 (0.7\%)}         &  \href{http://knowyourmeme.com/memes/salt-bae}{Salt Bae}                                                          & 5,375 (0.6\%)    \\
\href{http://knowyourmeme.com/memes/awoo}{Awoo}                               & \multicolumn{1}{r|}{13,767 (1.0\%)}     & \href{http://knowyourmeme.com/memes/daily-struggle}{Daily Struggle}                 & \multicolumn{1}{r|}{4,015 (0.7\%)}       & \href{http://knowyourmeme.com/memes/kekistan}{Kekistan}                                                                        & \multicolumn{1}{r|}{186 (0.6\%)}         &   \href{http://knowyourmeme.com/memes/feels-bad-man-sad-frog}{Feels Bad Man/ Sad Frog}                                  & 4,991 (0.5\%)                                                                           \\
\href{http://knowyourmeme.com/memes/tony-kornheiser-s-why}{Tony Kornheiser's Why}              & \multicolumn{1}{r|}{13,577 (1.0\%)}      & \href{http://knowyourmeme.com/memes/expanding-brain}{Expanding Brain}                          & \multicolumn{1}{r|}{3,757 (0.7\%)}       & \href{http://knowyourmeme.com/memes/picardia}{Picardia \textbf{(P)}}                                                                        & \multicolumn{1}{r|}{183 (0.6\%)}         &   \href{http://knowyourmeme.com/memes/math-lady-confused-lady}{Math Lady/Confused Lady}                                &  4,722 (0.5\%)                                     \\
\href{http://knowyourmeme.com/memes/picardia}{Picardia \textbf{(P)}}                           & \multicolumn{1}{r|}{13,540 (1.0\%)}      &   \href{http://knowyourmeme.com/memes/demotivational-posters}{Demotivational Posters}        & \multicolumn{1}{r|}{3,419 (0.6\%)}       & \href{http://knowyourmeme.com/memes/things-with-faces-pareidolia}{Things with Faces (Pareidolia) }                                                 & \multicolumn{1}{r|}{156 (0.5\%)}         &  \href{http://knowyourmeme.com/memes/computer-reaction-faces}{Computer Reaction Faces}
                                 & 4,720 (0.5\%)                                      \\
\href{http://knowyourmeme.com/memes/big-grin-never-ever}{Big Grin / Never Ever}              & \multicolumn{1}{r|}{12,893 (1.0\%)}      &  \href{http://knowyourmeme.com/memes/actual-advice-mallard}{Actual Advice Mallard}                  & \multicolumn{1}{r|}{3,293 (0.6\%)}       & \href{http://knowyourmeme.com/memes/serbia-strong-remove-kebab}{Serbia Strong/Remove Kebab}                                                      & \multicolumn{1}{r|}{149 (0.5\%)}         &   \href{http://knowyourmeme.com/memes/clinton-trump-duet}{Clinton Trump Duet \textbf{(P)}}
 &    3,901 (0.4\%)                                  \\
\href{http://knowyourmeme.com/memes/reaction-images}{Reaction Images}                    & \multicolumn{1}{r|}{12,608 (0.9\%)}      & \href{http://knowyourmeme.com/memes/reaction-images}{Reaction Images}                      & \multicolumn{1}{r|}{2,959 (0.5\%)}       & \href{http://knowyourmeme.com/memes/riot-hipster}{Riot Hipster}                                                                    & \multicolumn{1}{r|}{148 (0.5\%)}        &     \href{http://knowyourmeme.com/memes/kendrick-lamar-damn-album-cover}{Kendrick Lamar Damn Album Cover\hspace*{-0.4cm}}
                            &   3,656 (0.4\%)                                    \\
\href{http://knowyourmeme.com/memes/computer-reaction-faces}{Computer Reaction Faces}             & \multicolumn{1}{r|}{12,247 (0.9\%)}      &  \href{http://knowyourmeme.com/memes/handsome-face}{Handsome Face}                   & \multicolumn{1}{r|}{2,675 (0.5\%)}       & \href{http://knowyourmeme.com/memes/colorized-history}{Colorized History                                                             } & \multicolumn{1}{r|}{144 (0.5\%)}         & \href{http://knowyourmeme.com/memes/what-in-tarnation}{What in tarnation}                                    & 3,363 (0.3\%)     \\
\href{http://knowyourmeme.com/memes/wojak-feels-guy}{Wojak / Feels Guy}                  & \multicolumn{1}{r|}{11,682 (0.9\%)}      & \href{http://knowyourmeme.com/memes/absolutely-disgusting}{Absolutely Disgusting}                         & \multicolumn{1}{r|}{2,674 (0.5\%)}       &  \href{http://knowyourmeme.com/memes/the-most-interesting-man-in-the-world}{Most Interesting Man in World} & \multicolumn{1}{r|}{140 (0.5\%)}         &   \href{http://knowyourmeme.com/memes/harambe-the-gorilla}{Harambe the Gorilla}                               & 3,164 (0.3\%)  \\
\href{http://knowyourmeme.com/memes/absolutely-disgusting}{Absolutely Disgusting}              & \multicolumn{1}{r|}{11,436 (0.8\%)}      & \href{http://knowyourmeme.com/memes/pepe-the-frog}{Pepe the Frog}                    & \multicolumn{1}{r|}{2,672 (0.5\%)}       & \href{http://knowyourmeme.com/memes/chuck-norris-facts}{Chuck Norris Facts}                                                              & \multicolumn{1}{r|}{131 (0.4\%)}         &  \href{http://knowyourmeme.com/memes/i-know-that-feel-bro}{I Know that Feel Bro}                              &   3,137 (0.3\%)                                                                                                              \\
\href{http://knowyourmeme.com/memes/spurdo-sparde}{Spurdo Sparde}                      & \multicolumn{1}{r|}{9,581 (0.7\%)}       & \href{http://knowyourmeme.com/memes/i-was-only-pretending-to-be-retarded}{Pretending to be Retarded}                     & \multicolumn{1}{r|}{2,462 (0.4\%)}       & \href{http://knowyourmeme.com/memes/roll-safe}{Roll Safe}                                                                        & \multicolumn{1}{r|}{131 (0.4\%)}         & \href{http://knowyourmeme.com/memes/this-is-fine}{This is Fine}                                    & 3,094 (0.3\%)                                     \\
\midrule
{\bf Total} & \multicolumn{1}{r|}{\bf 445,179 (33.4\%)} && \multicolumn{1}{r|}{\bf 94,069 (16.7\%)} && \multicolumn{1}{r|}{\bf 4,808 (17.0\%)} && \multicolumn{1}{r|}{\bf 249,047 (26.4\%)} \\
 \bottomrule
\end{tabular}%
}
\caption{Top 20 KYM entries for memes that we find our datasets. We report the number of posts for each meme as well as the percentage over all the posts (per community) that contain images that match one of the annotated clusters. The (R) and (P) markers indicate whether a meme is annotated as racist or politics-related, respectively (see Section~\ref{sec:meme_popularity} for the selection criteria). }
\label{tbl:top_memes}
\end{table*}

In Table~\ref{tbl:top_memes}, we report the top 20 memes for each Web community sorted by the number of posts.
We observe that Pepe the Frog~\cite{pepe_frog_meme} and its variants are among the most popular memes for every platform.
While this might be an artifact of using fringe communities as a ``seed'' for the clustering, recall that the goal of this work is in fact to gain an understanding of how fringe communities disseminate memes and influence mainstream ones.
Thus, we leave to future work a broader analysis of the wider meme ecosystem.

Sad Frog~\cite{sad_frog_meme} is the most popular meme on \dspol (4.9\%), the 3rd on Reddit (1.3\%), the 10th on Gab (0.8\%), and the 12th on Twitter (0.5\%).
We also find variations like Smug Frog~\cite{smug_frog_meme}, Apu Apustaja~\cite{apu_meme}, Pepe the Frog~\cite{pepe_frog_meme}, and Angry Pepe~\cite{angry_pepe_meme}.
Considering that Pepe is treated as a hate symbol by the Anti-Defamation League~\cite{adl_pepe_frog} and that is often used in hateful or racist, this likely indicates that polarized communities like \dspol and Gab do use memes to incite hateful conversation.
This is also evident from the popularity of the anti-semitic Happy Merchant meme~\cite{happy_merchant_meme}, which depicts a ``greedy'' and ``manipulative'' stereotypical caricature of a Jew (3.8\% on \dspol and 1.1\% on Gab). 

By contrast, mainstream communities like Reddit and Twitter primarily share harmless/neutral memes, which are rarely used in hateful contexts.
Specifically, on Reddit the top memes are Manning Face~\cite{manning_face_meme} (2.2\%) and That's the Joke~\cite{thats_the_joke_meme} (1.3\%), while on Twitter the top ones are Roll Safe~\cite{roll_safe_meme} (5.9\%) and Evil Kermit~\cite{evil_kermit_meme} (5.4\%).

Once again, we find that users (in all communities) post memes to share politics-related information, possibly aiming to enhance or penalize the public image of politicians (see Section~\ref{sec:appendix_interesting_images} for an example of such memes).
For instance, we find Make America Great Again~\cite{maga_meme}, a meme dedicated to Donald Trump's US presidential campaign, among the top memes in \dspol (1.6\%), in Reddit (0.8\%), and Gab (0.8\%).
Similarly, in Twitter, we find the Clinton Trump Duet meme~\cite{clinton_trump_duet_meme} (0.4\%), a meme inspired by the 2nd US presidential debate.

\begin{table*}[t]
\centering
\resizebox{\textwidth}{!}{%
\begin{tabular}{@{}lrlrlrlr@{}}
\toprule
\multicolumn{2}{c}{\textbf{/pol/}}                                            & \multicolumn{2}{c}{\textbf{Reddit}}                                           & \multicolumn{2}{c}{\textbf{Gab}}                                              & \multicolumn{2}{c}{\textbf{Twitter}}                                        \\ \midrule
\multicolumn{1}{c}{\textbf{Entry}} & \multicolumn{1}{c|}{\textbf{Posts (\%)}} & \multicolumn{1}{c}{\textbf{Entry}} & \multicolumn{1}{c|}{\textbf{Posts (\%)}} & \multicolumn{1}{c}{\textbf{Entry}} & \multicolumn{1}{c|}{\textbf{Posts (\%)}} & \multicolumn{1}{c}{\textbf{Entry}} & \multicolumn{1}{c}{\textbf{Posts(\%)}} \\ \midrule
\href{http://knowyourmeme.com/memes/people/donald-trump}{Donald Trump}                       & \multicolumn{1}{r|}{60,611 (4.6\%)}      & \href{http://knowyourmeme.com/memes/people/donald-trump}{Donald Trump}                      & \multicolumn{1}{r|}{34,533 (6.1\%)}      & \href{http://knowyourmeme.com/memes/people/donald-trump}{Donald Trump}                       & \multicolumn{1}{r|}{1,665 (6.1\%)}       & \href{http://knowyourmeme.com/memes/people/donald-trump}{Donald Trump}                                 & 10,208 (1.3\%)                                      \\
\href{http://knowyourmeme.com/memes/people/adolf-hitler}{Adolf Hitler}                       & \multicolumn{1}{r|}{8,759  (0.6\%)}      & \href{http://knowyourmeme.com/memes/people/steve-bannon}{Steve Bannon}                       & \multicolumn{1}{r|}{3,733 (0.6\%)}       & \href{http://knowyourmeme.com/memes/people/mitt-romney}{Mitt Romney}                        & \multicolumn{1}{r|}{455 (1.7\%)}         & \href{http://knowyourmeme.com/memes/people/barack-obama}{Barack Obama}                                   & 5,187 (0.6\%)                                      \\
\href{http://knowyourmeme.com/memes/people/mike-pence}{Mike Pence}                         & \multicolumn{1}{r|}{4,738 (0.3\%)}       & \href{http://knowyourmeme.com/memes/people/stephen-colbert}{Stephen Colbert}                     & \multicolumn{1}{r|}{3,121 (0.6\%)}       & \href{http://knowyourmeme.com/memes/people/bill-nye}{Bill Nye}                           & \multicolumn{1}{r|}{370 (1.3\%)}         & \href{http://knowyourmeme.com/memes/people/chelsea-manning}{Chelsea Manning}                                   & 4,173 (0.5\%)                                      \\
\href{http://knowyourmeme.com/memes/people/jeb-bush}{Jeb Bush}                           & \multicolumn{1}{r|}{4,217 (0.3\%)}       & \href{http://knowyourmeme.com/memes/people/chelsea-manning}{Chelsea Manning}                        & \multicolumn{1}{r|}{2,261 (0.4\%)}       & \href{http://knowyourmeme.com/memes/people/adolf-hitler}{Adolf Hitler}                       & \multicolumn{1}{r|}{106 (0.4\%)}         & \href{http://knowyourmeme.com/memes/people/kim-jong-un}{Kim Jong Un}                                 & 3,271 (0.4\%)                                      \\
\href{http://knowyourmeme.com/memes/people/vladimir-putin}{Vladimir Putin}                      & \multicolumn{1}{r|}{3,218 (0.2\%)}       & \href{http://knowyourmeme.com/memes/people/ben-carson}{Ben Carson}                      & \multicolumn{1}{r|}{2,148 (0.4\%)}       & \href{http://knowyourmeme.com/memes/people/barack-obama}{Barack Obama}                       & \multicolumn{1}{r|}{104 (0.4\%)}         & \href{http://knowyourmeme.com/memes/people/anita-sarkeesian}{Anita Sarkeesian}                                  & 2,764 (0.3\%)                                     \\
\href{http://knowyourmeme.com/memes/people/alex-jones}{Alex Jones}                         & \multicolumn{1}{r|}{3,206 (0.2\%)}       & \href{http://knowyourmeme.com/memes/people/bernie-sanders}{Bernie Sanders}                     & \multicolumn{1}{r|}{1,757 (0.3\%)}       & \href{http://knowyourmeme.com/memes/people/isis-daesh}{Isis Daesh}                         & \multicolumn{1}{r|}{92 (0.3\%)}          & \href{http://knowyourmeme.com/memes/people/bernie-sanders}{Bernie Sanders}                                  & 2,277 (0.3\%)                                      \\
\href{http://knowyourmeme.com/memes/people/ron-paul}{Ron Paul}                           & \multicolumn{1}{r|}{3,116 (0.2\%)}       & \href{http://knowyourmeme.com/memes/people/ajit-pai}{Ajit Pai}                           & \multicolumn{1}{r|}{1,658 (0.3\%)}       & \href{http://knowyourmeme.com/memes/people/death-grips}{Death Grips}                        & \multicolumn{1}{r|}{91 (0.3\%)}          & \href{http://knowyourmeme.com/memes/people/vladimir-putin}{Vladimir Putin}                                   &  1,733 (0.2\%)                                      \\
\href{http://knowyourmeme.com/memes/people/bernie-sanders}{Bernie Sanders}                      & \multicolumn{1}{r|}{3,022 (0.2\%)}       & \href{http://knowyourmeme.com/memes/people/barack-obama}{Barack Obama}                          & \multicolumn{1}{r|}{1,628 (0.3\%)}       & \href{http://knowyourmeme.com/memes/people/eminem}{Eminem}                             & \multicolumn{1}{r|}{89 (0.3\%)}          & \href{http://knowyourmeme.com/memes/people/billy-mays}{Billy Mays}                                  & 1,454 (0.2\%)                                      \\
\href{http://knowyourmeme.com/memes/people/massimo-dalema}{Massimo D'alema}                     & \multicolumn{1}{r|}{2,725 (0.2\%)}       & \href{http://knowyourmeme.com/memes/people/gabe-newell}{Gabe Newell}                        & \multicolumn{1}{r|}{1,518 (0.3\%)}       & \href{http://knowyourmeme.com/memes/people/kim-jong-un}{Kim Jong Un}                       & \multicolumn{1}{r|}{87 (0.3\%)}          & \href{http://knowyourmeme.com/memes/people/adolf-hitler}{Adolf Hitler}                                  & 1,304 (0.2\%)                                      \\
\href{http://knowyourmeme.com/memes/people/mitt-romney}{Mitt Romney}                        & \multicolumn{1}{r|}{2,468 (0.2\%)}       & \href{http://knowyourmeme.com/memes/people/bill-nye}{Bill Nye}                         & \multicolumn{1}{r|}{1,478 (0.3\%)}       & \href{http://knowyourmeme.com/memes/people/ajit-pai}{Ajit Pai}                            & \multicolumn{1}{r|}{76 (0.3\%)}          & \href{http://knowyourmeme.com/memes/people/kanye-west}{Kanye West}                                  & 1,261 (0.2\%)                                      \\
\href{http://knowyourmeme.com/memes/people/chelsea-manning}{Chelsea Manning}                        & \multicolumn{1}{r|}{2,403 (0.2\%)}       & \href{http://knowyourmeme.com/memes/people/hillary-clinton}{Hillary Clinton}                        & \multicolumn{1}{r|}{1,468 (0.3\%)}       & \href{http://knowyourmeme.com/memes/people/pewdiepie}{Pewdiepie}                           & \multicolumn{1}{r|}{73 (0.3\%)}          & \href{http://knowyourmeme.com/memes/people/bill-nye}{Bill Nye}                               & 968 (0.2\%)                                      \\
\href{http://knowyourmeme.com/memes/people/hillary-clinton}{Hillary Clinton}                       & \multicolumn{1}{r|}{2,378 (0.2\%)}       & \href{http://knowyourmeme.com/memes/people/death-grips}{Death Grips}                         & \multicolumn{1}{r|}{1,463 (0.3\%)}       & \href{http://knowyourmeme.com/memes/people/bernie-sanders}{Bernie Sanders}                            & \multicolumn{1}{r|}{71 (0.3\%)}          & \href{http://knowyourmeme.com/memes/people/mitt-romney}{Mitt Romney}                                  & 923 (0.1\%)                                      \\
\href{http://knowyourmeme.com/memes/people/a-wyatt-mann}{A. Wyatt Mann}                        & \multicolumn{1}{r|}{2,110 (0.2\%)}       & \href{http://knowyourmeme.com/memes/people/adolf-hitler}{Adolf Hitler}                        & \multicolumn{1}{r|}{1,449 (0.3\%)}       & \href{http://knowyourmeme.com/memes/people/alex-jones}{Alex Jones}                           & \multicolumn{1}{r|}{70 (0.3\%)}          & \href{http://knowyourmeme.com/memes/people/filthy-frank}{Filthy Frank}                                  & 777 (0.1\%)                                      \\
\href{http://knowyourmeme.com/memes/people/ben-carson}{Ben Carson}                       & \multicolumn{1}{r|}{1,780 (0.1\%)}       & \href{http://knowyourmeme.com/memes/people/mitt-romney}{Mitt Romney}                       & \multicolumn{1}{r|}{1,294 (0.2\%)}       & \href{http://knowyourmeme.com/memes/people/hillary-clinton}{Hillary Clinton}                           & \multicolumn{1}{r|}{59 (0.2\%)}          & \href{http://knowyourmeme.com/memes/people/hillary-clinton}{Hillary Clinton}                                   & 758 (0.1\%)                                      \\
\href{http://knowyourmeme.com/memes/people/filthy-frank}{Filthy Frank}                       & \multicolumn{1}{r|}{1,598 (0.1\%)}       & \href{http://knowyourmeme.com/memes/people/eminem}{Eminem}                        & \multicolumn{1}{r|}{1,274 (0.2\%)}       & \href{http://knowyourmeme.com/memes/people/anita-sarkeesian}{Anita Sarkeesian}                           & \multicolumn{1}{r|}{54 (0.2\%)}          & \href{http://knowyourmeme.com/memes/people/ajit-pai}{Ajit Pai}                                   & 715 (0.1\%)                                      \\ \bottomrule
\end{tabular}%
}
  \caption{Top 15 KYM entries about people that we find in each of our dataset. We report the number of posts and the percentage over all the posts (per community) that match a cluster with KYM annotations.}
\label{tbl:top_people}
\end{table*}

 \begin{figure*}[t]
   \begin{minipage}[t]{1\textwidth}
 \centering
\subfigure[all memes]{\includegraphics[width=0.322\textwidth]{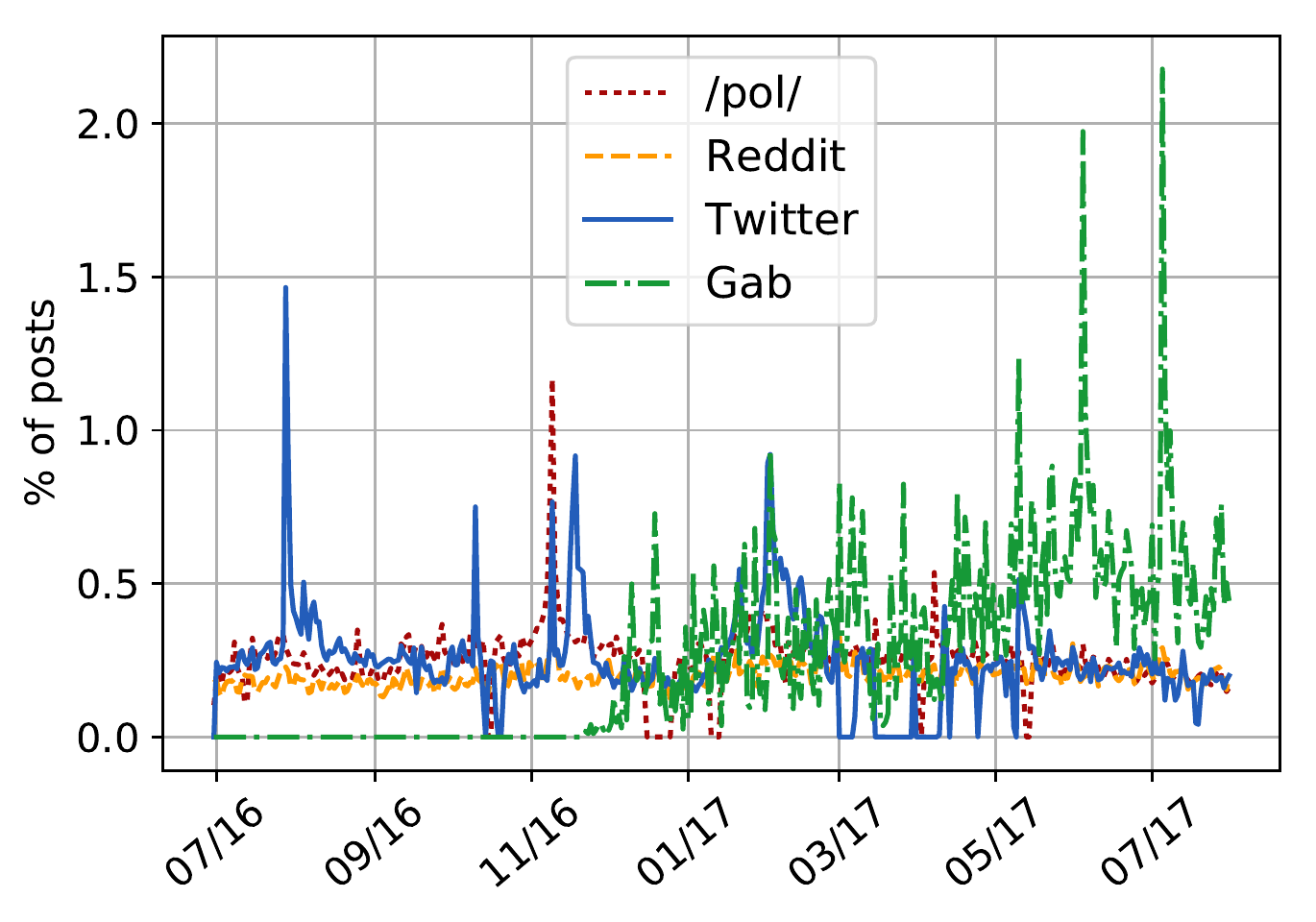}\label{temporal_all}}
\subfigure[racist]{\includegraphics[width=0.335\textwidth]{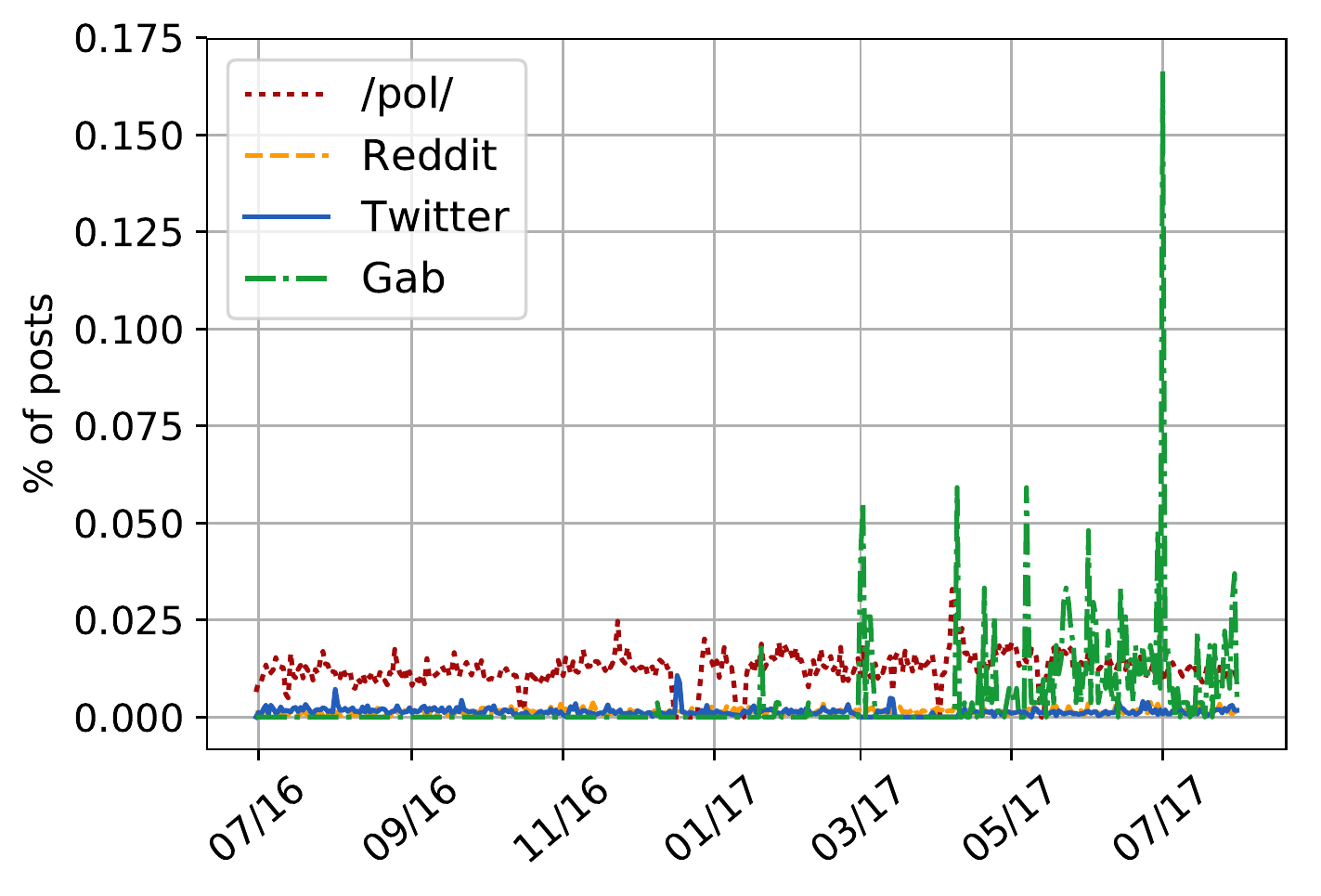}\label{temporal_racism}}
\subfigure[politics]{\includegraphics[width=0.322\textwidth]{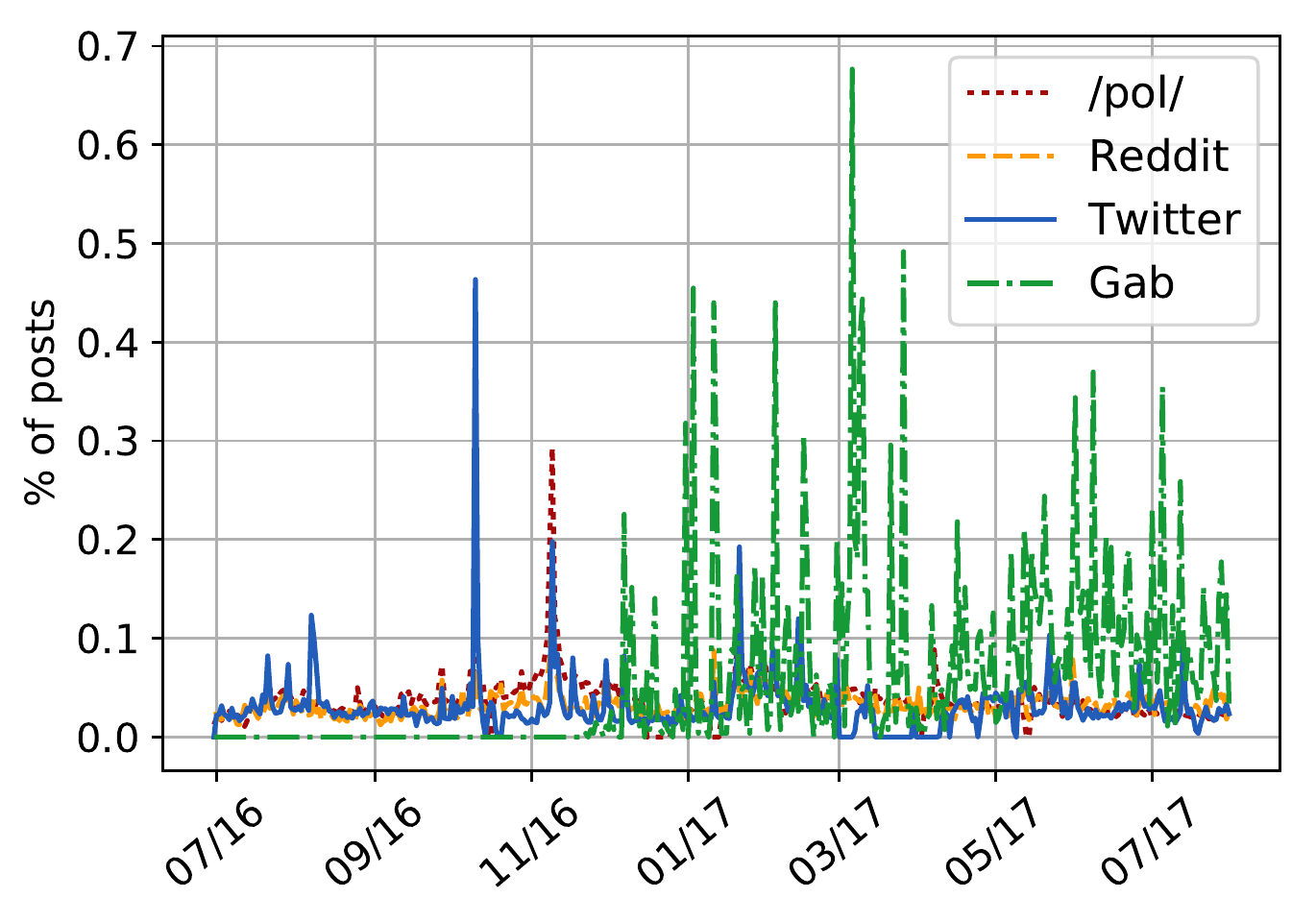}\label{temporal_politics}}
\caption{Percentage of posts per day in our dataset for all, racist, and politics-related memes.}
\label{fig:temporal_selected_memes_group}
\end{minipage}
 \centering
    \begin{minipage}[t]{0.85\textwidth}
    \centering
\subfigure[Reddit]{\includegraphics[width=0.49\textwidth]{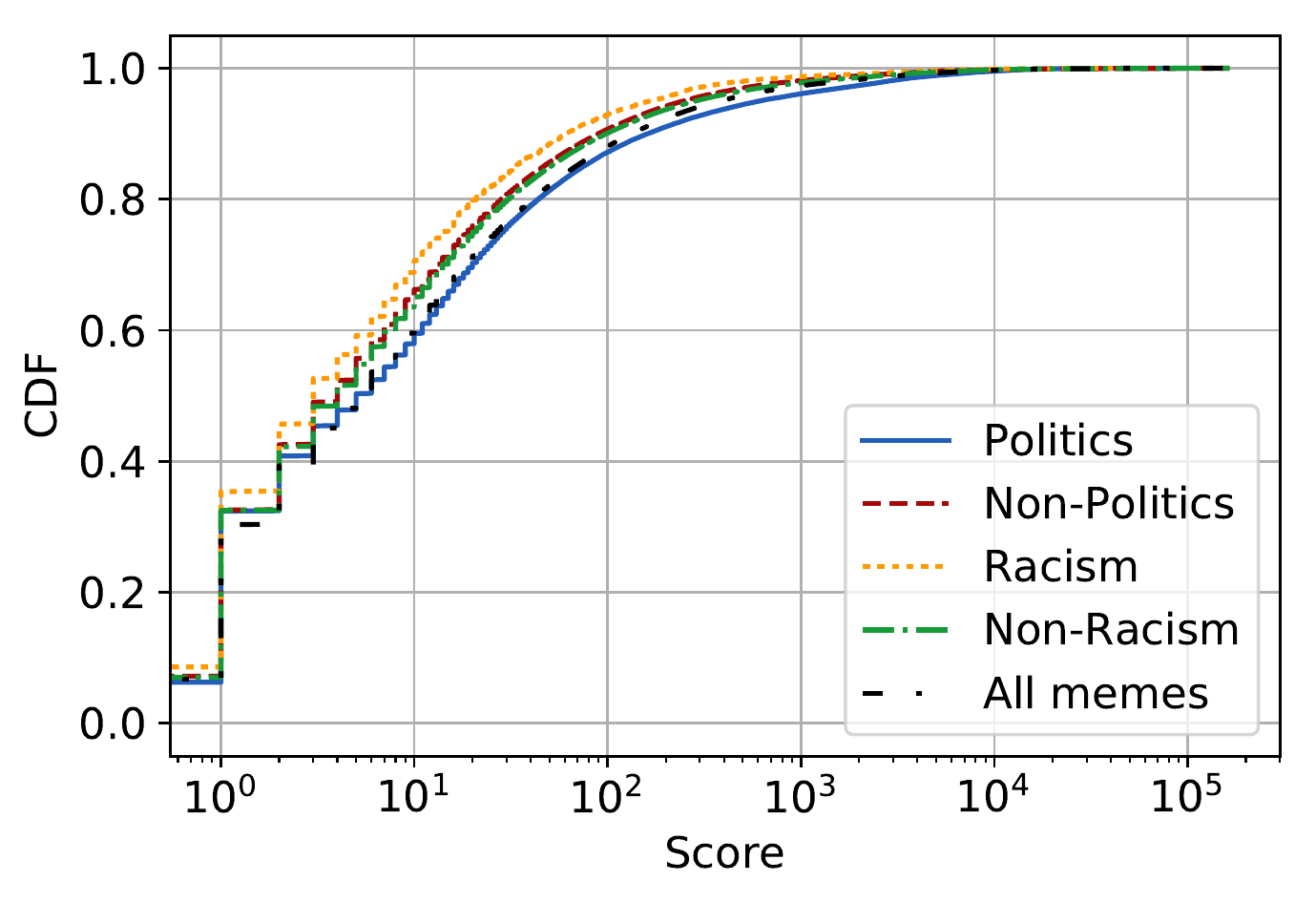}\label{scores_reddit_memes}}
\subfigure[Gab]{\includegraphics[width=0.49\textwidth]{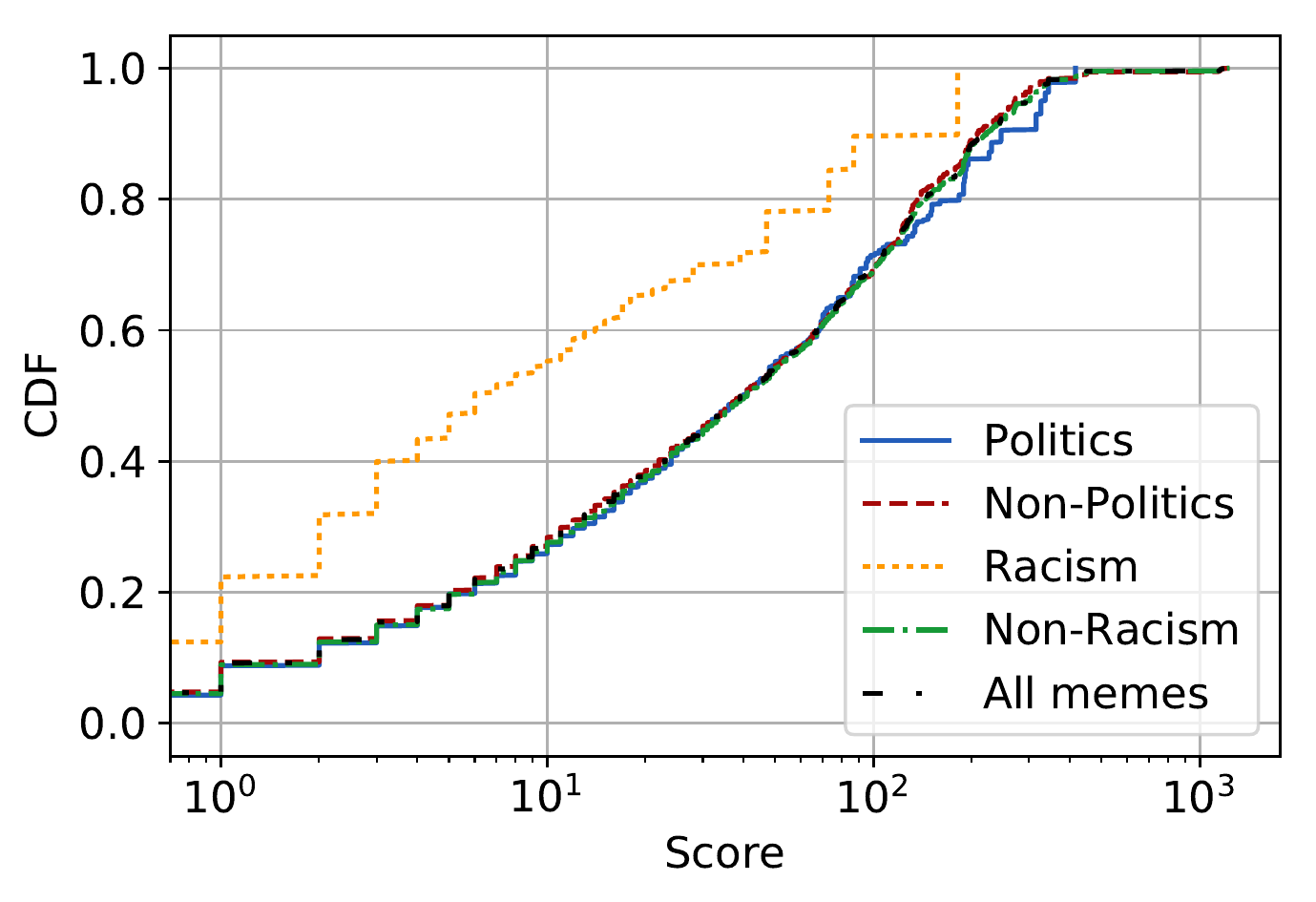}\label{scores_gab_memes}}
\caption{CDF of scores of posts that contain memes on Reddit and Gab.}
\label{fig:scores_groups}
\end{minipage}
\end{figure*}

\descr{People.}
We also analyze memes related to people (i.e., KYM entries with the people category). %
Table~\ref{tbl:top_people} reports the top 15 KYM entries in this category.
We observe that, in all Web Communities, the most popular person portrayed in memes is Donald Trump: he is depicted in 4.6\% of \dspol posts that contain annotated images, while for Reddit, Gab, and Twitter the percentages are 6.1\%, 6.1\%, and 1.3\%, respectively.
Other popular personalities, in all platforms, include several politicians.
For instance, in \dspol, we find Mike Pence (0.3\%), Jeb Bush (0.3\%), Vladimir Putin (0.2\%), while, in Reddit, we find Steve Bannon (0.6\%),  Chelsea Manning (0.6\%), and Bernie Sanders (0.3\%), in Gab, Mitt Romney (1.7\%) and Barack Obama (0.4\%), and, in Twitter, Barack Obama (0.6\%), Kim Jong Un (0.5\%), and Chelsea Manning (0.4\%).
This highlights the fact that users on these communities utilize memes to share information and opinions about politicians, and possibly try to either enhance or harm public opinion about them.
Finally, we note the presence of Adolf Hitler memes on all Web Communities, i.e., \dspol (0.6\%), Reddit (0.3\%),  Gab (0.4\%), and Twitter (0.2\%).

We further group memes into two high-level groups, racist and politics-related.
We use the {\em tags} that are available in our KYM dataset, i.e., we assign a meme to the politics-related group if it has the ``politics,'' ``2016 us presidential election,'' ``presidential election,'' ``trump,'' or ``clinton'' tags, and to the racism-related one if the tags include ``racism,'' ``racist,'' or ``antisemitism,''
obtaining 117 racist memes (4.4\% of all memes that appear in our dataset) and 556 politics-related memes (21.2\% of all memes that appear on our dataset).
In the rest of this section, we use these groups to further study the memes, and later in Section~\ref{sec:influence_estimation} to estimate influence.

\subsubsubsection{Temporal Analysis}
Next, we study the temporal aspects of posts that contain memes from \dspol, Reddit, Twitter, and Gab.
In Figure~\ref{fig:temporal_selected_memes_group}, we plot the percentage of posts per day that include memes.
For all memes (Figure~\ref{temporal_all}), we observe that \dspol and Reddit follow 
a steady posting behavior, with a peak in activity around the 2016 US elections.
We also find that memes are increasingly more used on Gab (see, e.g., 2016 vs 2017).

As shown in Figure~\ref{temporal_racism}, both \dspol and Gab include a substantially higher number of posts with racist memes, used over time with a difference in behavior: while \dspol users share them in a very steady and constant way, Gab exhibits a bursty behavior.
A possible explanation is that the former is inherently more racist, with the latter primarily reacting to particular world events.
As for political memes (Figure~\ref{temporal_politics}), we find a lot of activity overall on Twitter, Reddit, and \dspol, but with different spikes in time.
On Reddit and \dspol, the peaks coincide with the 2016 US elections.
On Twitter, we note a peak that coincides with the 2nd US Presidential Debate on October 2016.
For Gab, there is again an increase in posts with political memes after January 2017.

\begin{table*}[]
\centering
\resizebox{0.95\textwidth}{!}{%
\begin{tabular}{lrlrlr}
\hline
\multicolumn{2}{c}{\textbf{All Memes}} & \multicolumn{2}{c}{\textbf{Racism-Related Memes}} & \multicolumn{2}{c}{\textbf{Politics-Related Memes}} \\ \hline
\textbf{Subreddit} & \multicolumn{1}{r|}{\textbf{Posts (\%)}} & \textbf{Subreddit} & \multicolumn{1}{r|}{\textbf{Posts (\%)}} & \textbf{Subreddit} & \textbf{Posts (\%)} \\
\td & \multicolumn{1}{r|}{82,698 (12.5\%)} & \td & \multicolumn{1}{r|}{359 (9.3\%)} & \td & 24,343 (26.4\%) \\
AdviceAnimals & \multicolumn{1}{r|}{35,475 (5.3\%)} & AdviceAnimals & \multicolumn{1}{r|}{87 (2.2\%)} & politics & 2,751 (3.0\%) \\
me\_irl & \multicolumn{1}{r|}{15,366 (2.3\%)} & conspiracy & \multicolumn{1}{r|}{76 (2.0\%)} & EnoughTrumpSpam & 2,679 (2.9\%) \\
politics & \multicolumn{1}{r|}{8,875 (1,3\%)} & me\_irl & \multicolumn{1}{r|}{70 (1.8\%)} & TrumpsTweets & 2,363 (2.5\%) \\
funny & \multicolumn{1}{r|}{8,508 (1.3\%)} & funny & \multicolumn{1}{r|}{56 (1.4\%)} & AdviceAnimals & 1,740 (1.9\%) \\
dankmemes & \multicolumn{1}{r|}{7,744 (1,1\%)} & CringeAnarchy & \multicolumn{1}{r|}{43 (1.1\%)} & USE2016 & 1,653 (1.8\%) \\
EnoughTrumpSpam & \multicolumn{1}{r|}{6,973 (1.1\%)} & EDH & \multicolumn{1}{r|}{43 (1.1\%)} & PoliticsAll & 1,401(1.5\%) \\
pics & \multicolumn{1}{r|}{5,945 (0.9\%)} & magicTCG & \multicolumn{1}{r|}{42 (1.1\%)} & dankmemes & 881 (0.9\%) \\
AskReddit & \multicolumn{1}{r|}{5,482 (0.8\%)} & dankmemes & \multicolumn{1}{r|}{40 (1.0\%)} & pics & 877 (0.9\%) \\
HOTandTrending & \multicolumn{1}{r|}{4,674 (0.7\%)} & ImGoingToHellForThis & \multicolumn{1}{r|}{39 (1.0\%)} & me\_irl & 873 (0.9\%) \\ \hline
\end{tabular}
}
\caption{Top ten subreddits for all memes, racism-related memes, and politics-related memes.}
\label{tbl:top_subreddits_groups}
\end{table*}

\subsubsubsection{Score Analysis}
As discussed in Chapter~\ref{chapter:background}, Reddit and Gab incorporate a voting system that determines the popularity of content within the Web community and essentially captures the appreciation of other users towards the shared content.
To study how users react to racist and politics-related memes, we plot the CDF of the posts' scores that contain such memes in Figure~\ref{fig:scores_groups}.

For Reddit (Figure~\ref{scores_reddit_memes}), we find that posts that contain politics-related memes are rated highly (mean score of 224.7 and a median of 5) than posts that contain non-politics memes (mean 124.9, median 4).
On the contrary, posts that contain racist memes are rated lower (average score of 94.8 and a median of 3) than other non-racist memes (average 141.6 and median 4).
On Gab (Figure~\ref{scores_gab_memes}), posts that contain politics-related memes have a similar score as non-political memes (mean 87.3 vs 82.4).
However, this does not apply for racist and non-racist memes, as non-racist memes have over 2 times higher scores than racist memes (means 84.7 vs 35.5).

Overall, this suggests that posts that contain politics-related memes receive high scores by Reddit and Gab users, while for racist memes this applies only on Reddit.

\subsubsubsection{Sub-Communities}
Among all the Web communities that we study, only Reddit is divided into multiple sub-communities.
We now study which sub-communities share memes with a focus on racist and politics-related content.
In Table~\ref{tbl:top_subreddits_groups}, we report the top ten subreddits in terms of the percentage over all posts that contain memes in Reddit for: 1)~all memes; 2)~racist ones; and 3)~politics-related memes.

For all three groups, the most popular subreddit is \td with 12.5\%, 9.3\%, and 26.4\%, %
respectively.
Interestingly, AdviceAnimals, a general-purpose meme subreddit, is among the top-ten sub-communities also for racist and political memes, highlighting their infiltration in otherwise non-hateful communities.

Other popular subreddits for racist memes include conspiracy (2.0\%), me\_irl (1.8\%), and funny (1.4\%) subreddits.
For politics-related memes, the majority of the subreddits are related to Donald Trump, while there also are general subreddits that talk about politics, e.g., the politics (3.0\%) and the PoliticsAll subreddit (1.5\%).

\subsubsection{Take-Aways}
\noindent In summary, the main take-aways of our analysis include:\smallskip
\begin{compactenum}
\item Fringe Web communities use many variants of memes related to politics and world events, possibly aiming to share weaponized information about them (Section~\ref{sec:appendix_interesting_images} include some examples of such memes).
For instance, Donald Trump is the KYM entry with the largest number of clusters in \dspol (2.2\%), \td (6.1\%), and Gab (2.2\%).
\item \dspol and Gab share hateful and racist memes at a higher rate than mainstream communities, as we find a considerable number of anti-semitic and pro-Nazi clusters (e.g., The Happy Merchant meme~\cite{happy_merchant_meme} appears in 1.3\% of all \dspol annotated clusters and 2.2\% of Gab's, while Adolf Hitler in 0.6\% of \dspol's).
This trend is steady over time for \dspol but ramping up for Gab.
\item Seemingly ``neutral'' memes, like Pepe the Frog (or one of its variants), are used in conjunction with other memes to incite hate or influence public opinion on world events, e.g., with images related to terrorist organizations like ISIS or world events such as Brexit.
\item Our custom distance metric successfully allows us to study the interplay and the overlap of memes, as showcased by the visualizations of the clusters and the dendrogram (see Figs.~\ref{fig:casestudy_frogs_dendrogram} and \ref{fig:clusters_graph}).
\item Reddit users are more interested in politics-related memes than other type of memes.
That said, when looking at individual subreddits, we find that \td is the most active one when it comes to posting memes in general.
It is also the subreddit where most racism and politics-related memes are posted.\smallskip
\end{compactenum}

\subsection{Influence Estimation} \label{sec:influence_estimation}
So far we have studied the dissemination of memes by looking at Web communities in isolation.
However, in reality, these influence each other: e.g., memes posted on one community are often re-posted to another.
Aiming to capture the relationship between them, we use a statistical model known as Hawkes Processes~\cite{linderman2014,lindermanArxiv}, which describes how events occur over time on a collection of processes (for more details regarding Hawkes Processes see Section~\ref{sec:hawkes_background}).

\subsubsection{Influence Results}

\begin{table}[]
\centering
\small
\begin{tabular}{@{}llllll@{}}
\toprule
\dspol & Twitter & Reddit & \tdshort  & Gab   &  \\ \midrule
1,574,045               & 865,885  & 581,803 & 81,924 & 44,918 &  \\ \bottomrule
\end{tabular}
\caption{Events per community from the 12.6K clusters.} %
\label{tbl:hawkes_input}
\end{table}

We fit Hawkes models using Gibbs sampling as described in~\cite{lindermanArxiv} for the 12.6K annotated clusters; in Table~\ref{tbl:hawkes_input}, we report the total number of meme images posted to each community in these clusters.
As seen in Table~\ref{tbl:hawkes_input}, \dspol has the greatest number of memes posted, followed by Twitter and then Reddit.
In terms of total images collected (see Table~\ref{tbl:datasets_summary}), Twitter and Reddit have many more than \dspol.
However, many of the images on these communities might not be memes; additionally, because our clusters are created from the memes present on only \dspol, \td, and Gab (as these are the communities primarily of interest in this work), it is possible that there are memes on Twitter and Reddit that are not included in the clusters.
This yields an additional interesting question: how \emph{efficient} are different communities at disseminating memes?

First, we report the source of events in terms of the percent of events on the destination community.
This describes the results in terms of the data as we have collected it, e.g., it tells us the percentage of memes posted on Twitter that were caused by \dspol.
The second way we report influence is by normalizing the values by the total number of events in the source community, which lets us see how much influence each community has, relative to the number of memes they post---in other words, their efficiency.

\begin{figure}[t!]
\centering
\includegraphics[width=0.7\columnwidth]{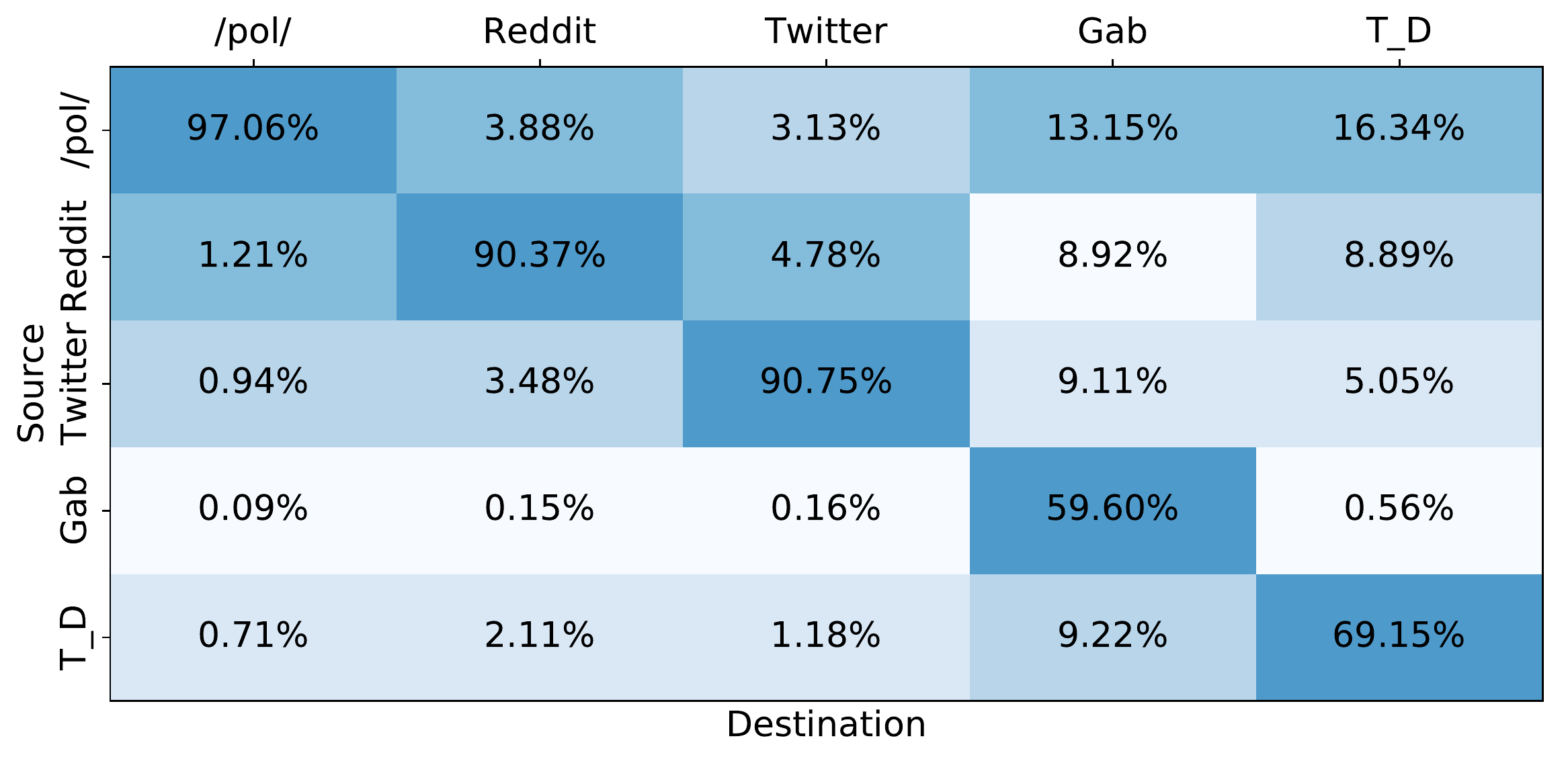}
\caption{Percent of \textit{destination} events caused by the source community on the destination community.  Colors indicate the largest-to-smallest influences per destination.}
\label{fig:hawkes_cause}
\end{figure}

We first look at the influence of all clusters together.
Figure~\ref{fig:hawkes_cause} shows the percent of events on each \emph{destination} community caused by each \emph{source} community.
The values from one community to the same community (for example, from \dspol to \dspol) include both events caused by the background rate of that community and events caused by previous events within that community; these values are the largest influence for each community.
After this, \dspol is the strongest source of influence for Reddit, \td, and Gab, but not for Twitter, which is most influenced by Reddit.
Interestingly, although Twitter has a greater number of memes posted than Reddit, it causes less influence.
Perhaps there is less original content posted directly to Twitter.

\begin{figure}[t!]
\centering
\includegraphics[width=0.7\columnwidth]{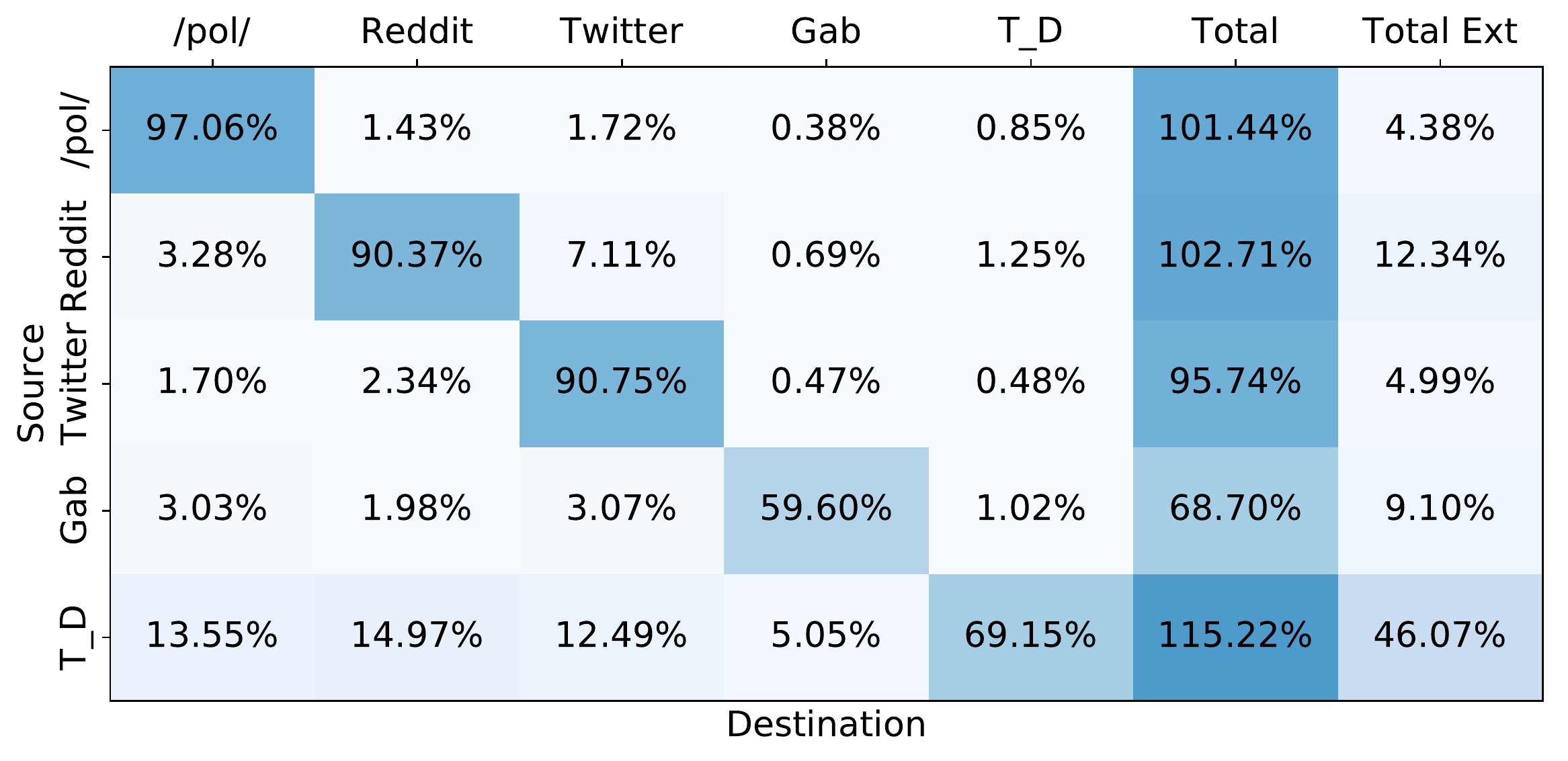}
\caption{Influence from source to destination community, normalized by the number of events in the \textit{source} community.  Columns for total influence and total external influence are shown.}
\label{fig:hawkes_cause_norm}
\end{figure}

Next, we look at the normalized influence of all clusters together.
Figure~\ref{fig:hawkes_cause_norm} shows the influence that a source community has on a destination community, normalized by the total number of memes posted on the \emph{source} community.
The values can be understood as an indication of how much influence a community has, relative to the frequency of memes posted.
For example, the influence Reddit has on Twitter is equal to 5.71\% of the total events on Reddit.
If the sum of values for a source is less than 100\%, it implies that many of the posts on the source community were caused by other communities, or that posts on the source community do not cause many posts on other communities.

There are several interesting things to note in Figure~\ref{fig:hawkes_cause_norm}.
First, \td has by far the greatest influence for the number of memes posted on it.
This is particularly apparent when looking at just external influence, where \td has more than 4 times as much influence than the rest of Reddit, the closest other community.
Memes from this community spread very well to all of the other communities.
While \dspol has a large total influence on the other communities (as seen in Figure~\ref{fig:hawkes_cause}), when normalized by its size, it has the smallest external influence: just 4.03\%.
Most of the memes posted on \dspol do not spread to other communities.
Both Gab and Twitter have a total normalized influence of less than 100\%; much less in Gab's case, although it has higher external influence.

\begin{figure}[t!]
\centering
\includegraphics[width=0.7\columnwidth]{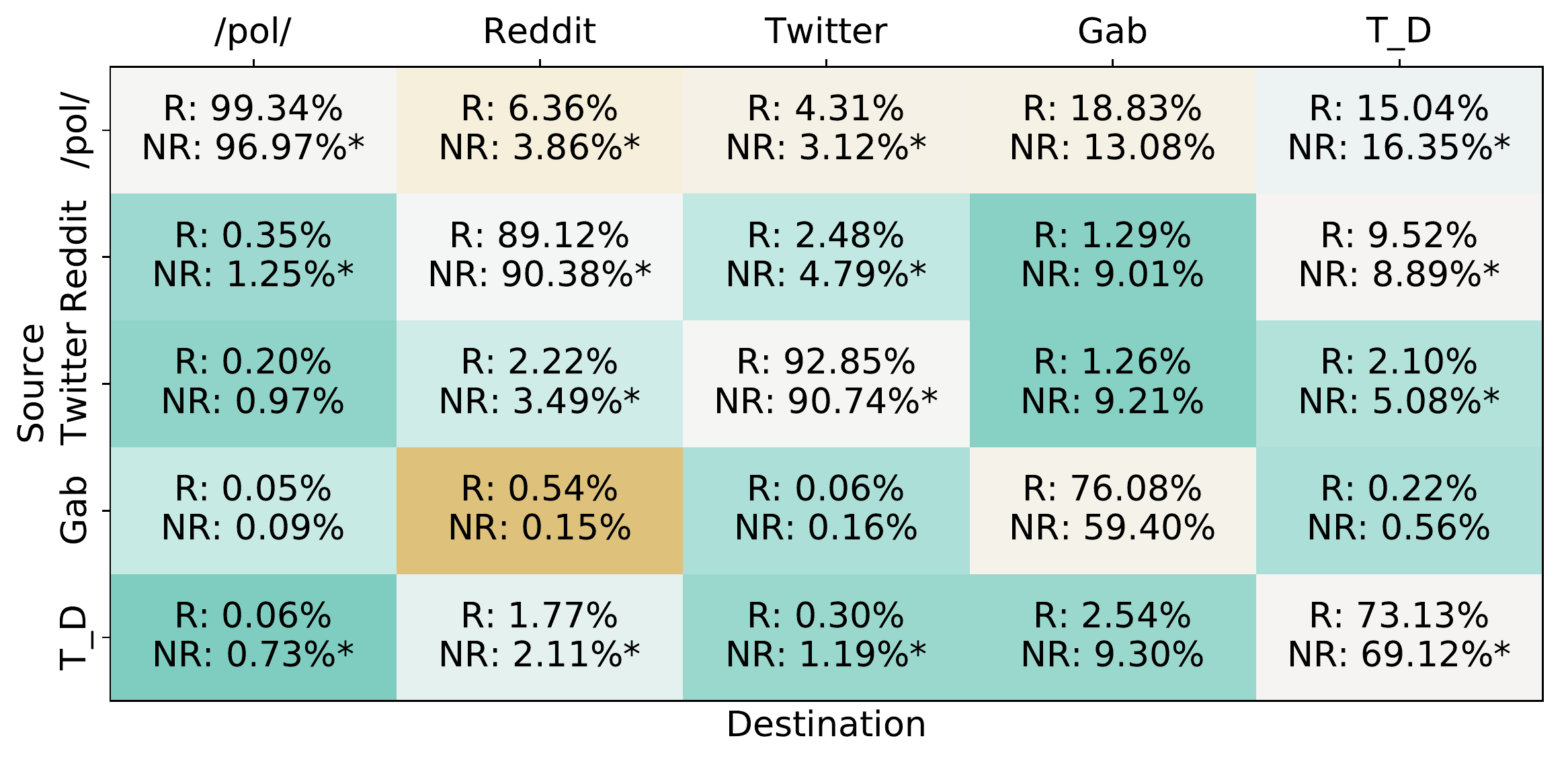}
\caption{Percent of the destination community's racist (R) and non-racist (NR) meme postings caused by the source community.  Colors indicate the percent difference between racist and non-racist.}
\label{fig:hawkes_racist_from}
\end{figure}

\begin{figure}[t!]
\centering
\includegraphics[width=0.7\columnwidth]{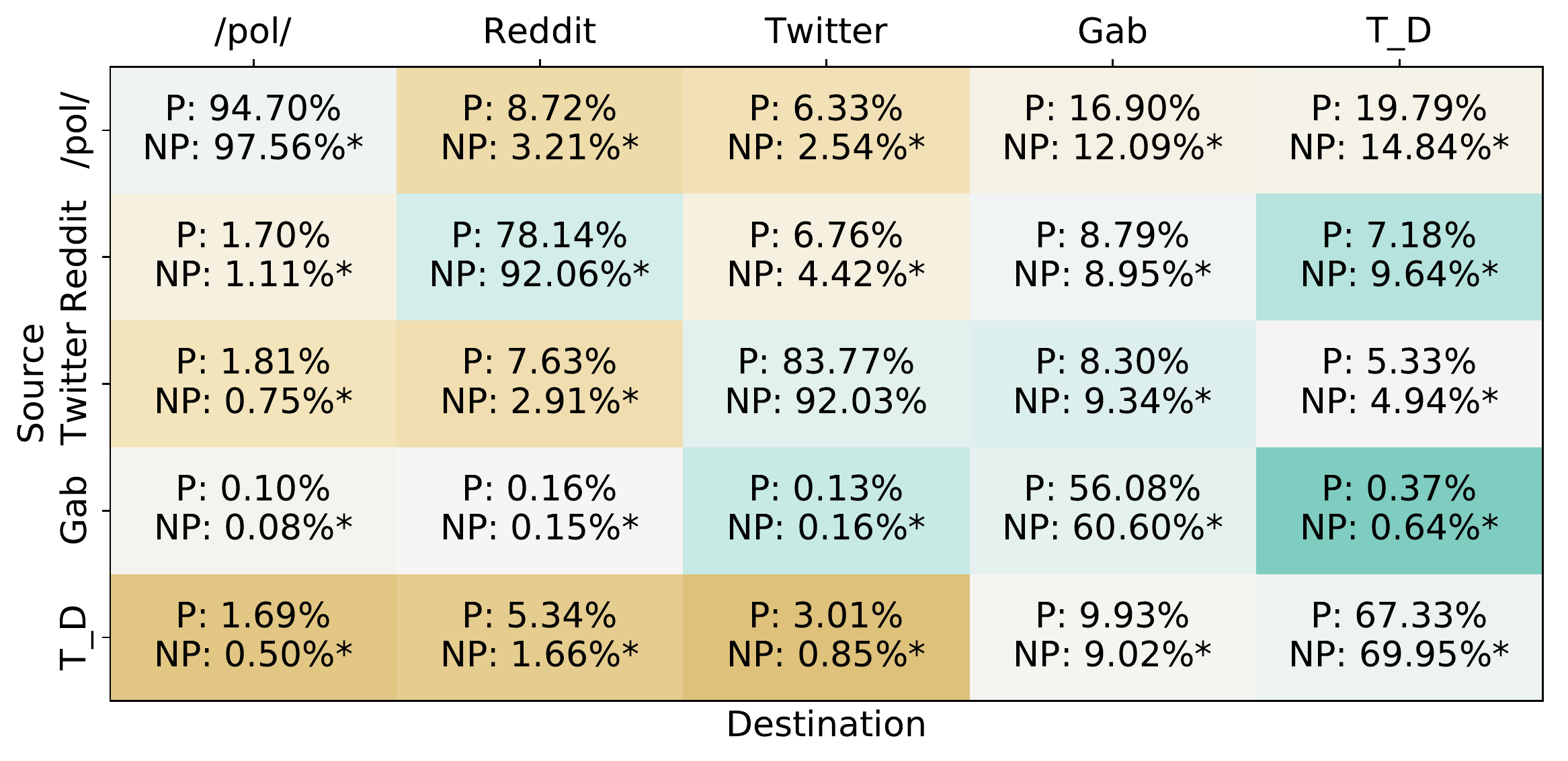}
\caption{Percent of the destination community's political (P) and non-political (NP) meme postings caused by the source community.  Colors indicate the percent difference between political and non-political.}
\label{fig:hawkes_political_from}
\end{figure}

Using the clusters identified as either racist or non-racist (see the end of Section~\ref{sec:meme_popularity}),
we compare how the communities influence the spread of these two types of content.
Figure~\ref{fig:hawkes_racist_from} shows the percentage of both the destination community's racist and non-racist meme posts caused by the source community.
We perform two-sample Kolmogorov-Smirnov tests to compare the distributions of influence from the racist and non-racist clusters; cells with statistically significant differences between influence of racist/non-racist memes (with $p {<} 0.01$) are reported with a * in the figure.
\dspol has the most \emph{total} influence for both racist and non-racist memes, with the notable exception of Twitter, where Reddit has the most the influence.
Interestingly, while the percentage of racist meme posts caused by \dspol is greater than non-racist for Reddit, Twitter, and Gab, this is \emph{not} the case for \td.
The only other cases where influence is greater for racist memes are Reddit to \td and Gab to Reddit.

When looking at political vs non political memes (Figure~\ref{fig:hawkes_political_from}), we see a somewhat different story.
Here, \dspol influences \td more in terms of political memes.
Further, we see differences in the \emph{percent} increase and decrease of influence between the two figures (as indicated by the cell colors).
For example, Twitter has a relatively larger difference in its influence on \dspol and Reddit for political and non-political memes than for racist and non-racist memes, but a smaller difference in its influence on Gab and \td.
This exposes how different communities have varying levels of influence depending on the type of memes they post.
\begin{figure}[t!]
\centering
\includegraphics[width=0.7\columnwidth]{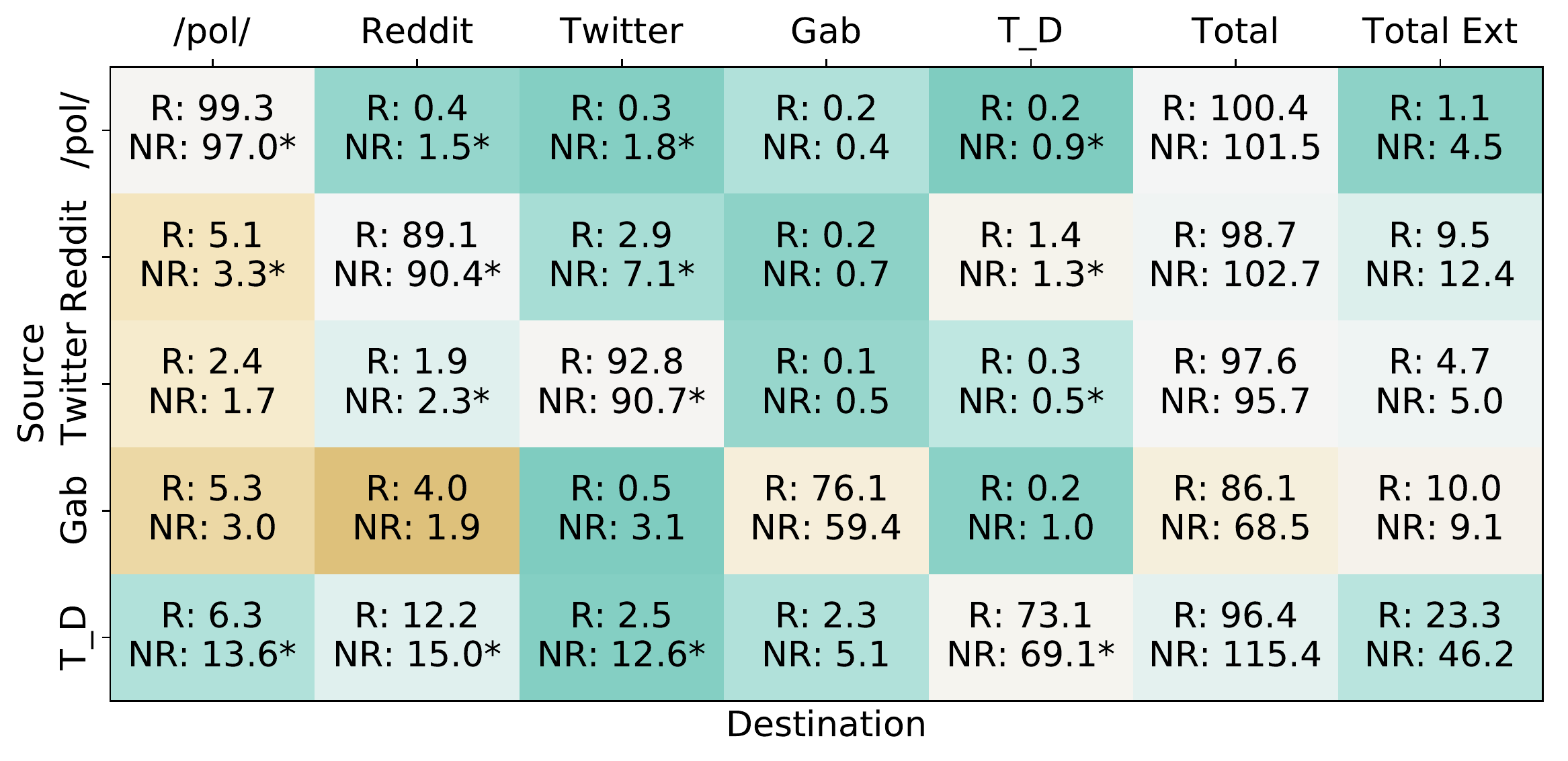}
\caption{Influence from source to destination community of racist and non-racist meme postings, normalized by the number of events in the \textit{source} community.}
\label{fig:hawkes_racist_norm}
\end{figure}

\begin{figure}[t!]
\centering
\includegraphics[width=0.7\columnwidth]{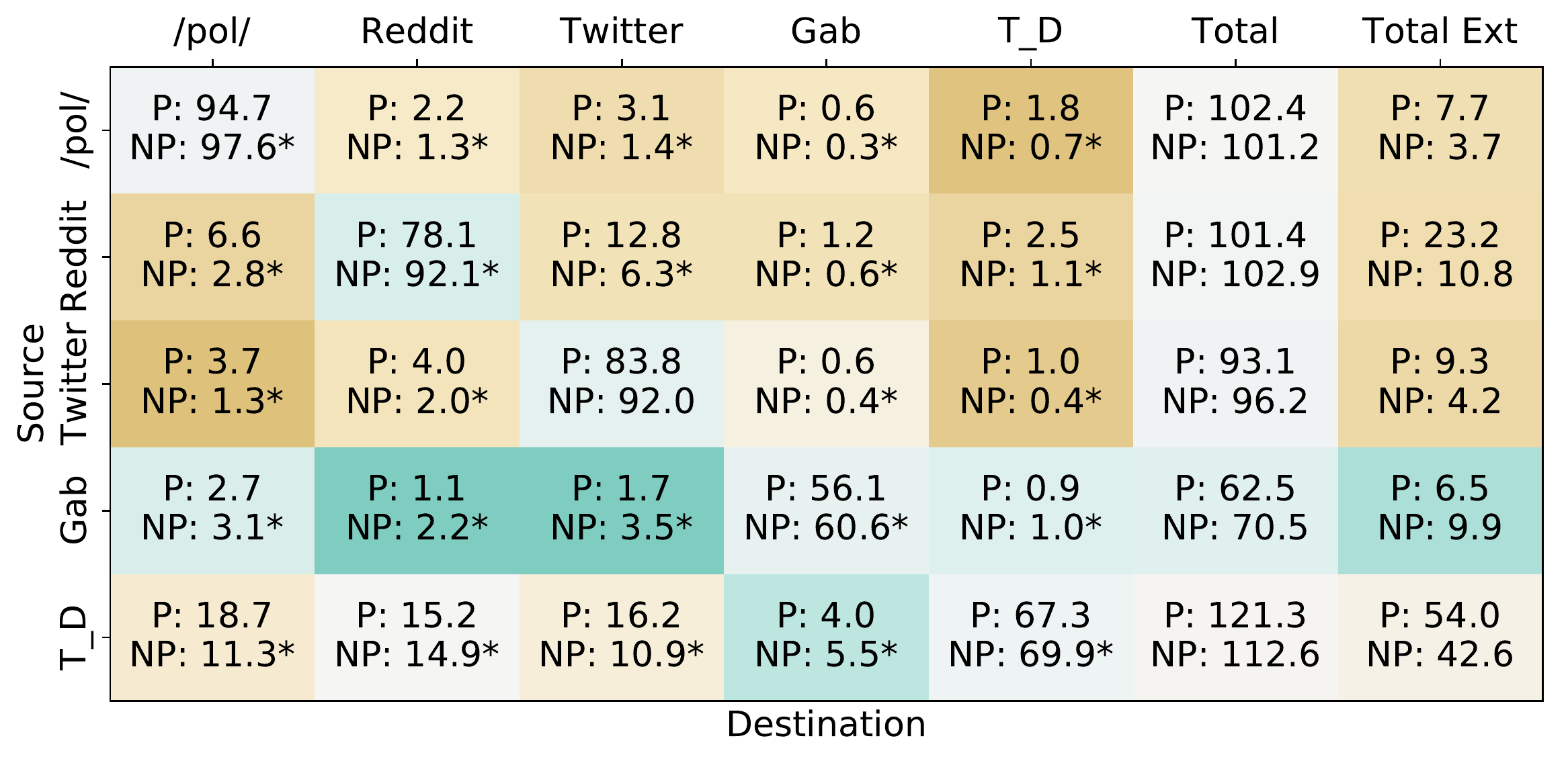}
\caption{Influence from source to destination community of political and non-political meme postings, normalized by the number of events in the \textit{source} community.}
\label{fig:hawkes_political_norm}
\end{figure}

While examining the raw influence provides insights into the meme ecosystem, it obscures notable differences in the meme posting behavior of the different communities.
To explore this, we look at the \emph{normalized} influence in Figure~\ref{fig:hawkes_racist_norm} (racist/non-racist memes) and Figure~\ref{fig:hawkes_political_norm} (political/non-political memes).
As mentioned previously, normalization reveals how \emph{efficient} the communities are in disseminating memes to other communities by revealing the \emph{per meme} influence of meme posts.
First, we note that the percent change in influence for the dissemination of racist/non-racist memes is quite a bit larger than that for political/non-political memes (again, indicated by the coloring of the cells).
More interestingly, both figures show that, contrary to the \emph{total} influence, \dspol is the \emph{least} influential when taking into account the number of memes posted.
While this might seem surprising, it actually yields a subtle, yet crucial aspect of \dspol's role in the meme ecosystem:
\dspol (and 4chan in general) acts as an evolutionary microcosm for memes.
The constant production of new content~\cite{hine2016longitudinal} results in a ``survival of the fittest''~\cite{muhinterview} scenario.
A staggering number of memes are posted on \dspol, but only the {\em best} actually make it out to other communities.
To the best of our knowledge, this is the first result quantifying this analogy to evolutionary pressure.

\descr{Take-Aways.}
There are several take-aways from our measurement of influence.
We show that \dspol is, generally speaking, the most influential disseminator of memes in terms of raw influence.
In particular, it is more influential in spreading \emph{racist} memes than non-racist one, and this difference is deeper than in any other community.
There is one notable exception: \dspol is more influential in terms of \emph{non-racist} memes on \td.
Relatedly, \dspol has generally more influence in terms of spreading political memes than other communities.
When looking at the normalized influence, however, we surface a more interesting result: \dspol is the \emph{least} efficient in terms of influence while \td is the \emph{most} efficient.
This provides new insight into the meme ecosystem: there are clearly evolutionary effects.
Many meme postings do not result in further dissemination, and one of the key components to ensuring they are disseminated is ensuring that new ``offspring'' are continuously produced.
\dspol's ``famed'' meme magic, i.e., the propensity to produce and heavily push memes, is thus the most likely explanation for \dspol's influence on the Web in general.

\subsubsection{Interesting Images} \label{sec:appendix_interesting_images}
Finally, we report some ``interesting'' examples of images from our frogs case study (see Section~\ref{subsection:hierarchy}), as well as an example of an image for enhancing/penalizing the public image of specific politicians (as discussed in Section~\ref{sec:meme_popularity}).

\begin{figure}[t]
\centering
\includegraphics[width=0.4\columnwidth]{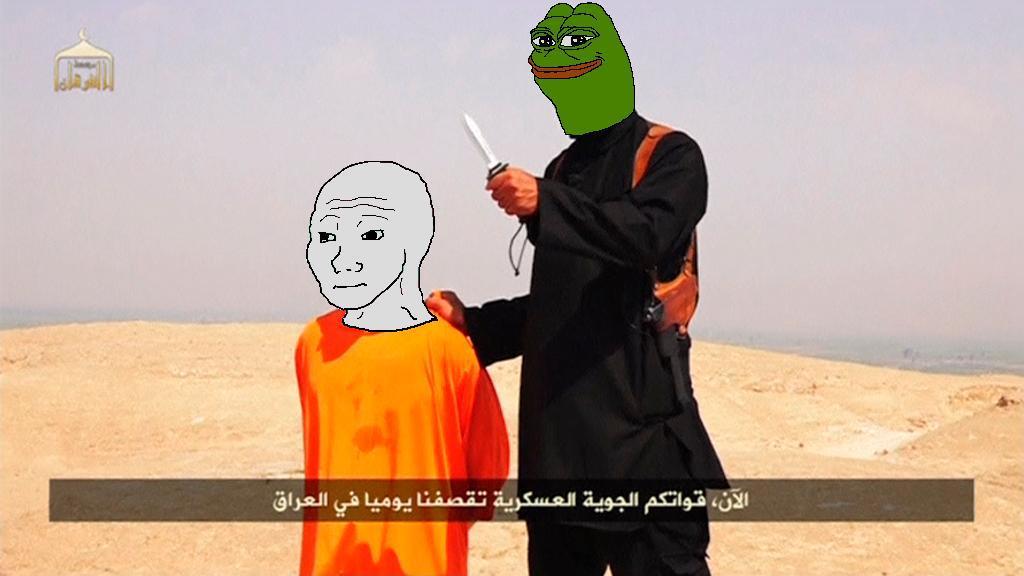}
\caption{Image that exists in the clusters that are connected with frogs and Isis Daesh.}
\label{fig:isis_pepe}
\end{figure}

\begin{figure}[t]
\centering
\includegraphics[width=0.4\columnwidth]{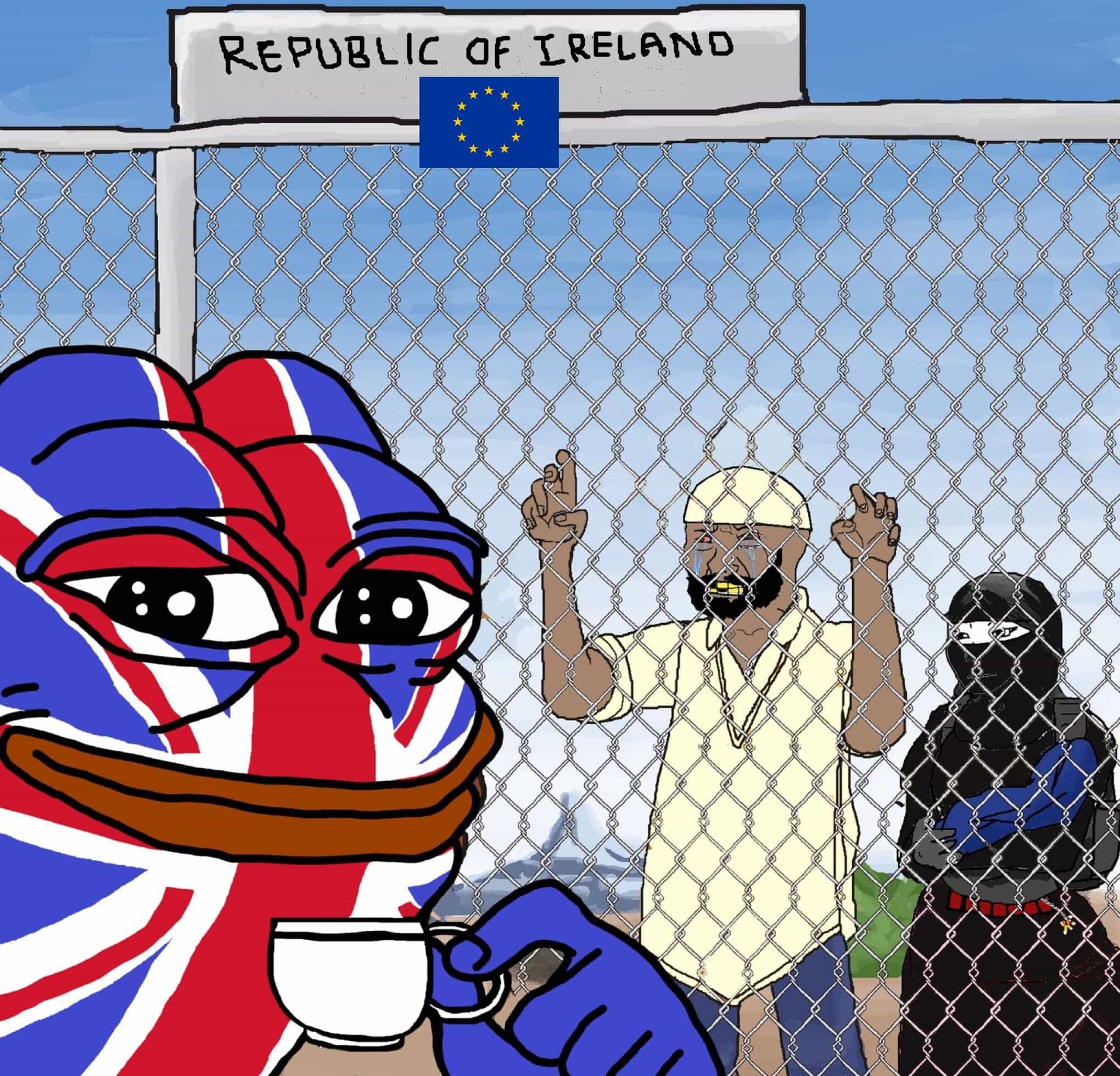}
\caption{Image that exists in the clusters that are connected with frogs and Brexit.}
\label{fig:pepe_brexit}
\end{figure}

\begin{figure}[t]
\centering
\includegraphics[width=0.4\columnwidth]{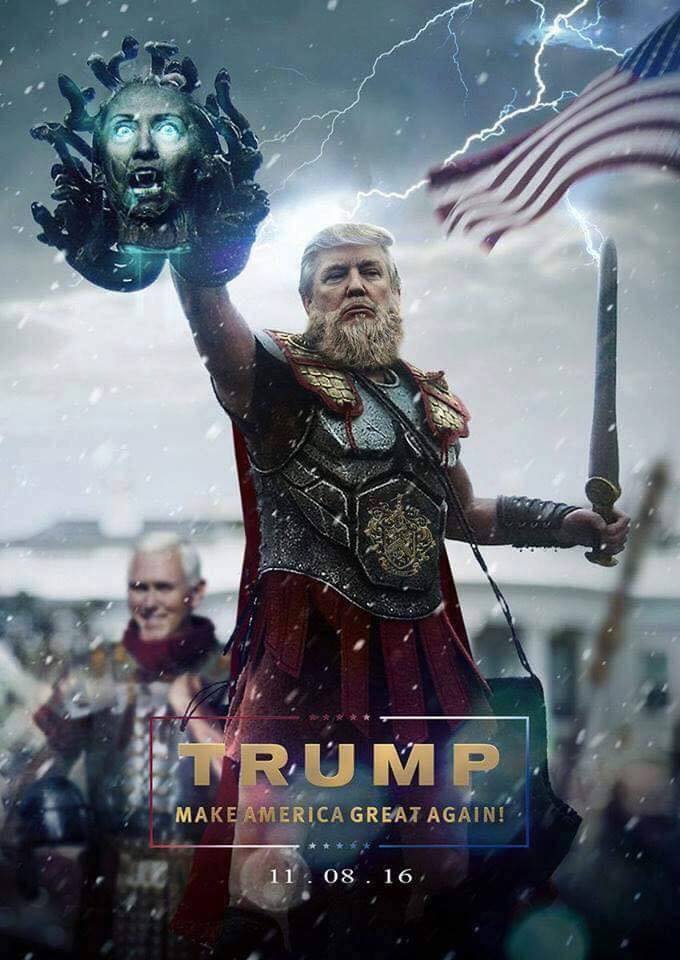}
\caption{Meme that is used for enhancing/penalizing the public image of specific politicians. Hillary Clinton is represented as Medusa, a monster, while Donald Trump is presented as Perseus (the hero who beheaded Medusa).}
\label{fig:trump_clinton_medusa}
\end{figure}

Specifically, Figure~\ref{fig:isis_pepe} shows an image connecting the Smug Frog~\cite{smug_frog_meme} and the ISIS memes~\cite{isis_meme}.
Also, Figure~\ref{fig:pepe_brexit} shows an image connecting the Smug Frog and the Brexit meme~\cite{brexit_meme}.
Finally, Figure~\ref{fig:trump_clinton_medusa} shows a graphic image found in \dspol that aims to attack the image of Hillary Clinton, while boosting that of Donald Trump.
(The image depicts Hillary Clinton as a monster, Medusa, while Donald Trump is presented as Perseus, the hero who beheaded Medusa.)

\subsection{Remarks}

In this work, we presented a large-scale measurement study of the meme ecosystem.
We introduced a novel image processing pipeline and ran it over 160M images collected from four Web communities (4chan's \dspol, Reddit, Twitter, and Gab).
We clustered images from fringe communities (\dspol, Gab, and Reddit's \td) based on perceptual hashing and a custom distance metric, annotated the clusters using data gathered from Know Your Meme, and analyzed them along a variety of axes.
We then associated images from all the communities to the clusters to characterize them through the lens of memes and the influence they have on each other.

Our analysis highlights that the meme ecosystem is quite complex, with intricate relationships between different memes and their variants.
We found important differences between the memes posted on different communities (e.g., Reddit and Twitter tend to post ``fun'' memes, while Gab and \dspol racist or political ones).
When measuring the influence of each community toward disseminating memes to other Web communities, we found that \dspol has the largest overall influence for racist and political memes, however, \dspol was the least \emph{efficient}, i.e., in terms of influence w.r.t.~the total number of memes posted,
while \td is very successful in pushing memes to both fringe and mainstream Web communities.

Our work constitutes the first attempt to provide a multi-platform measurement of the meme ecosystem, with a focus on fringe and potentially dangerous communities.
Considering the increasing relevance of digital information on world events, our study provides a building block for future cultural anthropology work, as well as for building systems to protect against the dissemination of harmful ideologies.
Moreover, our pipeline can already be used by social network providers to assist the identification of hateful content; for instance, Facebook is taking steps to ban Pepe the Frog used in the context of hate~\cite{fb-pepe}, and our methodology can help them automatically identify hateful variants.
Finally, our pipeline can be used for tracking the propagation of images from any context or other language spheres, provided an appropriate annotation dataset.

\descr{Performance.}
We also measured the time that it takes to associate images posted on Web communities to memes.
All other steps in our system are one-time batch tasks, only executed if the annotations dataset is updated.
To ease presentation, we only report the time to compare all the 74M images from Twitter (the largest dataset) against the medoids of all 12K annotated clusters:
it took about 12 days on our infrastructure, equipped with two NVIDIA Titan Xp GPUs.
This corresponds to 14ms per image, or 73 images per second.
Note that, if new GPUs are added to our infrastructure, the workload would be divided equally across all GPUs.

\chapter{Characterizing the Role of Emerging Web Communities and Services on the Information Ecosystem}\label{chapter:various_communities}
In this chapter, we study various Web communities and services, with a particular focus on understanding their role in the spread of information on the Web.
Specifically, we study Gab with the goal to understand and characterize the platform with respect to the content and users it attracts.
Also, we study Web archiving services (services that archive Web content) and how they are used by users on Twitter, Reddit, 4chan, and Gab.

\section{What is Gab?} \label{sec:gab}

\subsection{Motivation}

The Web's information ecosystem is composed of multiple communities with varying influence~\cite{zannettou2017web}.
As mainstream online social networks become less novel, users have begun to join smaller, more focused platforms.
In particular, as the former have begun to reject fringe communities identified with racist and aggressive behavior, a number of alt-right focused services have been created.
Among these emerging communities, the Gab social network has attracted the interest of a large number of users since its creation in 2016~\cite{gab_site}, a few months before the US Presidential Election.
Gab was created, ostensibly as a censorship-free platform, aiming to protect free speech above anything else.
From the very beginning, site operators have welcomed users banned or suspended from platforms like Twitter for violating terms of service, often for abusive and/or hateful behavior.
In fact, there is extensive anecdotal evidence that the platform has become the alt-right's new hub~\cite{gab_alt_right} and that it exhibits a high volume of hate speech~\cite{gab_hate_speech} and racism~\cite{gab_racism}.
As a result, in 2017, both Google and Apple rejected Gab's mobile apps from their stores because of hate speech~\cite{gab_hate_speech} and non-compliance to pornographic content guidelines~\cite{gab_apple_porn}.

In this work, we provide, to the best of our knowledge, the first characterization of the Gab social network.
We crawl the Gab platform and acquire 22M posts by 336K users over a 1.5 year period (August 2016 to January 2018).
Overall, the main findings of our analysis include:
\begin{enumerate}
\itemsep0em 
\item Gab attracts a wide variety of users, ranging from well-known alt-right personalities like Milo Yiannopoulos to conspiracy theorists like Alex Jones. We also find a number of ``troll'' accounts that have migrated over from other platforms like 4chan, or that have been heavily inspired by them.
\item Gab is predominantly used for the dissemination and discussion of world events, news, as well as conspiracy theories. Interestingly, we note that Gab reacts strongly to events related to white nationalism and Donald Trump.
\item Hate speech is extensively present on the platform, as we find that 5.4\% of the posts include hate words. This is 2.4 times higher than on Twitter, but 2.2 times lower than on 4chan's Politically Incorrect board (/pol/)~\cite{hine2016longitudinal}.
\item There are several accounts making coordinated efforts towards recruiting millennials to the alt-right.
\end{enumerate}

In summary, our analysis highlights that Gab appears to be positioned at the border of mainstream social networks like Twitter and ``fringe'' Web communities like 4chan's /pol/. 
We find that, while Gab claims to be all about free speech, this seems to be merely a shield behind which its alt-right users hide.

\begin{table*}[t]
\centering
\resizebox{0.99\textwidth}{!}{
\begin{tabular}{@{}lllllllll@{}}
\toprule
\multicolumn{3}{c}{\textbf{Followers}}                                      & \multicolumn{3}{c}{\textbf{Scores}}        & \multicolumn{3}{c}{\textbf{PageRank}}                        \\ \midrule
\textbf{Name}             & \textbf{Username} & \textbf{\#}                 & \textbf{Name}            & \textbf{Username}      & \textbf{\#}  & \textbf{Name}            & \textbf{Username}      & \textbf{PR score} \\ \midrule
Milo Yiannopoulos         & m                 & \multicolumn{1}{r|}{45,060} & Andrew Torba             & a                      &  \multicolumn{1}{r|}{819,363}    & Milo Yiannopoulos & m & 0.013655\\
PrisonPlanet              & PrisonPlanet      & \multicolumn{1}{r|}{45,059} & John Rivers              & JohnRivers             & \multicolumn{1}{r|}{606,623}     & Andrew Torba  & a & 0.012818\\
Andrew Torba              & a                 & \multicolumn{1}{r|}{38,101} & Ricky Vaughn             & Ricky\_Vaughn99        & \multicolumn{1}{r|}{496,962}   & PrisonPlanet & PrisonPlanet & 0.011762  \\
Ricky Vaughn              & Ricky\_Vaughn99   & \multicolumn{1}{r|}{30,870} & Don                      & Don                    & \multicolumn{1}{r|}{368,698 }    &Mike Cernovich & Cernovich & 0.006549\\
Mike Cernovich            & Cernovich         & \multicolumn{1}{r|}{29,081} & Jared Wyand              & JaredWyand             & \multicolumn{1}{r|}{281,798 }   & Ricky Vaughn  &Ricky\_Vaughn99 & 0.006143\\
Stefan Molyneux           & stefanmolyneux    & \multicolumn{1}{r|}{26,337} & $[omitted]$              & TukkRivers             & \multicolumn{1}{r|}{253,781}    &Sargon of Akkad & Sargonofakkad100 & 0.005823 \\
Brittany Pettibone        & BrittPettibone    & \multicolumn{1}{r|}{24,799} & Brittany Pettibone       & BrittPettibone         & \multicolumn{1}{r|}{244,025}    & $[omitted$] & d\_seaman & 0.005104 \\
Jebs                      & DeadNotSleeping   & \multicolumn{1}{r|}{22,659} & Tony Jackson             & USMC-Devildog          & \multicolumn{1}{r|}{228,370}     & Stefan Molyneux & stefanmolyneux & 0.004830\\
$[omitted]$ & TexasYankee4      & \multicolumn{1}{r|}{20,079} & [omitted]               & causticbob             & \multicolumn{1}{r|}{228,316}     & Brittany Pettibone & BrittPettibone & 0.004218\\
$[omitted]$     & RightSmarts       & \multicolumn{1}{l|}{20,042} & Constitutional Drunk     & USSANews               & \multicolumn{1}{r|}{224,261}     & Vox Day & voxday & 0.003972\\
Vox Day                   & voxday            & \multicolumn{1}{l|}{19,454} & Truth Whisper            & truthwhisper           & \multicolumn{1}{r|}{206,516}     &Alex Jones & RealAlexJones & 0.003345\\
$[omitted]$              & d\_seaman         & \multicolumn{1}{l|}{18,080} & Andrew Anglin            & AndrewAnglin           & \multicolumn{1}{r|}{203,437}  &Lauren Southern & LaurenSouthern & 0.002984  \\
Alex Jones                & RealAlexJones     & \multicolumn{1}{l|}{17,613} & Kek\_Magician            & Kek\_Magician          & \multicolumn{1}{r|}{193,819}     &Donald J Trump & realdonaldtrump & 0.002895\\
Jared Wyand               & JaredWyand        & \multicolumn{1}{l|}{16,975} & $[omitted]$           & shorty                 & \multicolumn{1}{r|}{169,167 }   &Dave Cullen &DaveCullen & 0.002824 \\
Ann Coulter               & AnnCoulter        & \multicolumn{1}{l|}{16,605} & $[omitted]$ & SergeiDimitrovicIvanov & \multicolumn{1}{r|}{169,091} & $[omitted]$ & e & 0.002648   \\
Lift                      & lift              & \multicolumn{1}{l|}{16,544} & Kolja Bonke              & KoljaBonke             & \multicolumn{1}{r|}{160,246}     & Chuck C Johnson & Chuckcjohnson & 0.002599\\
Survivor Medic            & SurvivorMed       & \multicolumn{1}{l|}{16,382} & Party On Weimerica       & CuckShamer             & \multicolumn{1}{r|}{155,021}  & Andrew Anglin & AndrewAnglin & 0.002599   \\
$[omitted]$          & SalguodNos        & \multicolumn{1}{l|}{16,124} & PrisonPlanet             & PrisonPlanet           & \multicolumn{1}{r|}{154,829}     & Jared Wyand & JaredWyand & 0.002504\\
Proud Deplorable          & luther            & \multicolumn{1}{l|}{15,036} & Vox Day                  & voxday                 & \multicolumn{1}{r|}{150,930}     & Pax Dickinson & pax & 0.002400\\
Lauren Southern           & LaurenSouthern    & \multicolumn{1}{l|}{14,827} & W.O. Cassity             & wocassity              & \multicolumn{1}{r|}{144,875}    &Baked Alaska &apple & 0.002292\\ \bottomrule
\end{tabular}
}
\caption{Top 20 popular users on Gab according to the number of followers, their score, and their ranking based on PageRank in the followers/followings network. We omit the ``screen names'' of certain accounts for ethical reasons.}
\label{tbl:top_20_users}
\end{table*}

\subsection{Dataset}

Using Gab's API, we crawl the social network using a snowball methodology. 
Specifically, we obtain data for the most popular users as returned by Gab's API and iteratively collect data from all their followers as well as their followings.
We collect three types of information: 1)~basic details about Gab accounts, including username, score, date of account creation; 2)~all the posts for each Gab user in our dataset; and 3)~all the followers and followings of each user that allow us to build the following/followers network.
Overall, we collect 22,112,812 posts from 336,752 users, between August 2016 and January 2018.

\begin{figure*}[t]
\subfigure[]{\includegraphics[width=0.32\textwidth]{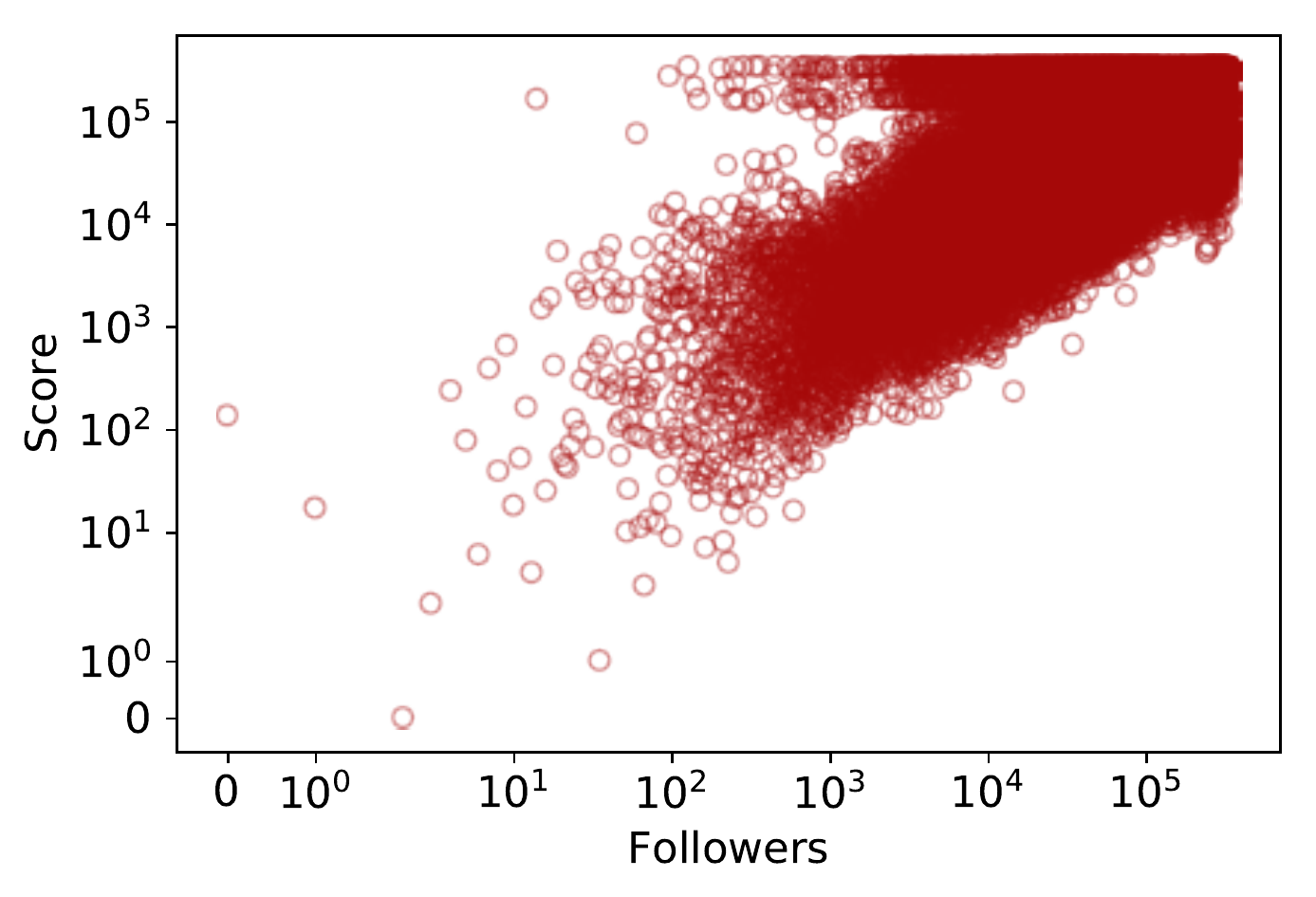}\label{subfig:followers_score}}
\subfigure[]{\includegraphics[width=0.32\textwidth]{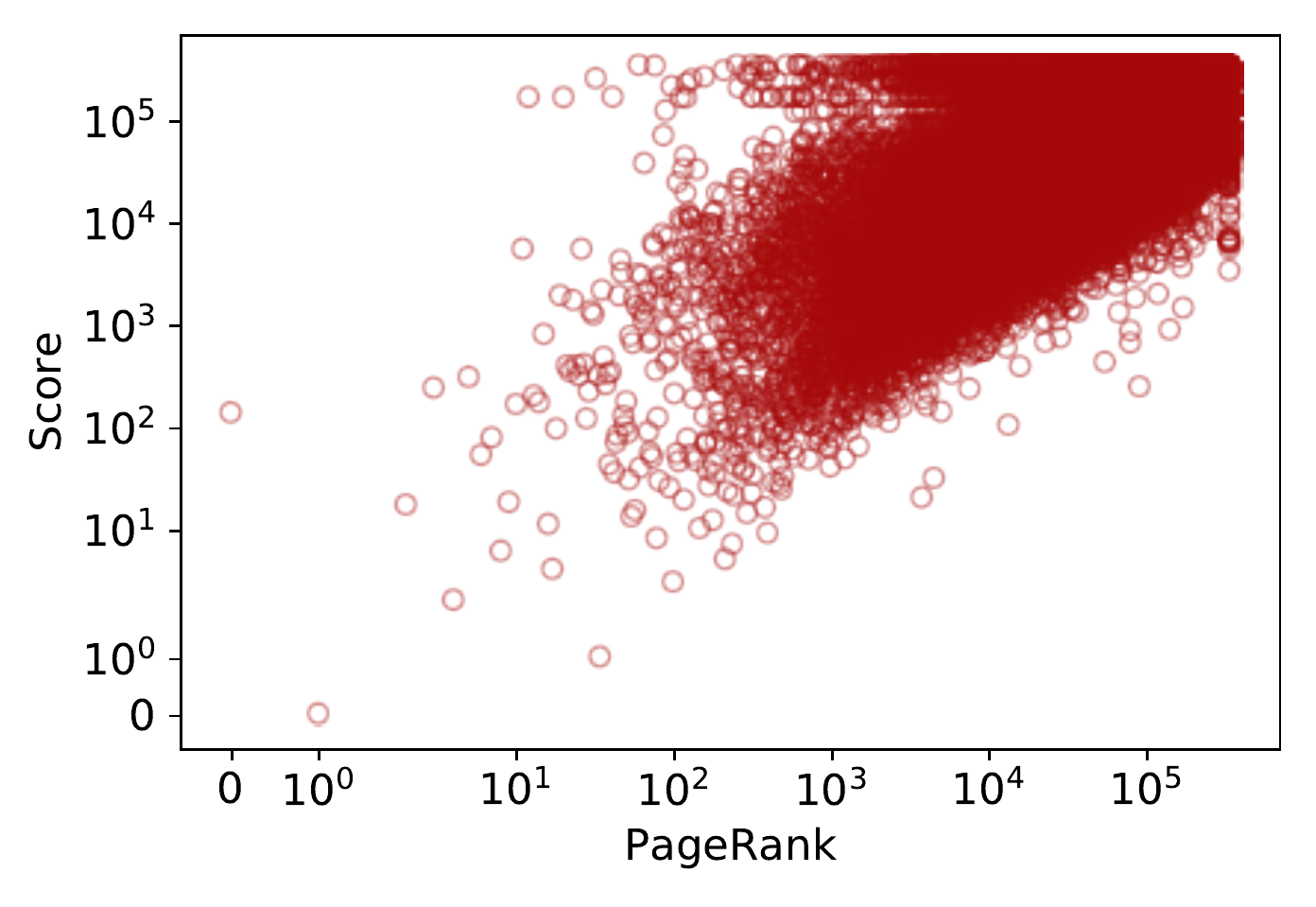}\label{subfig:pr_score}}
\subfigure[]{\includegraphics[width=0.32\textwidth]{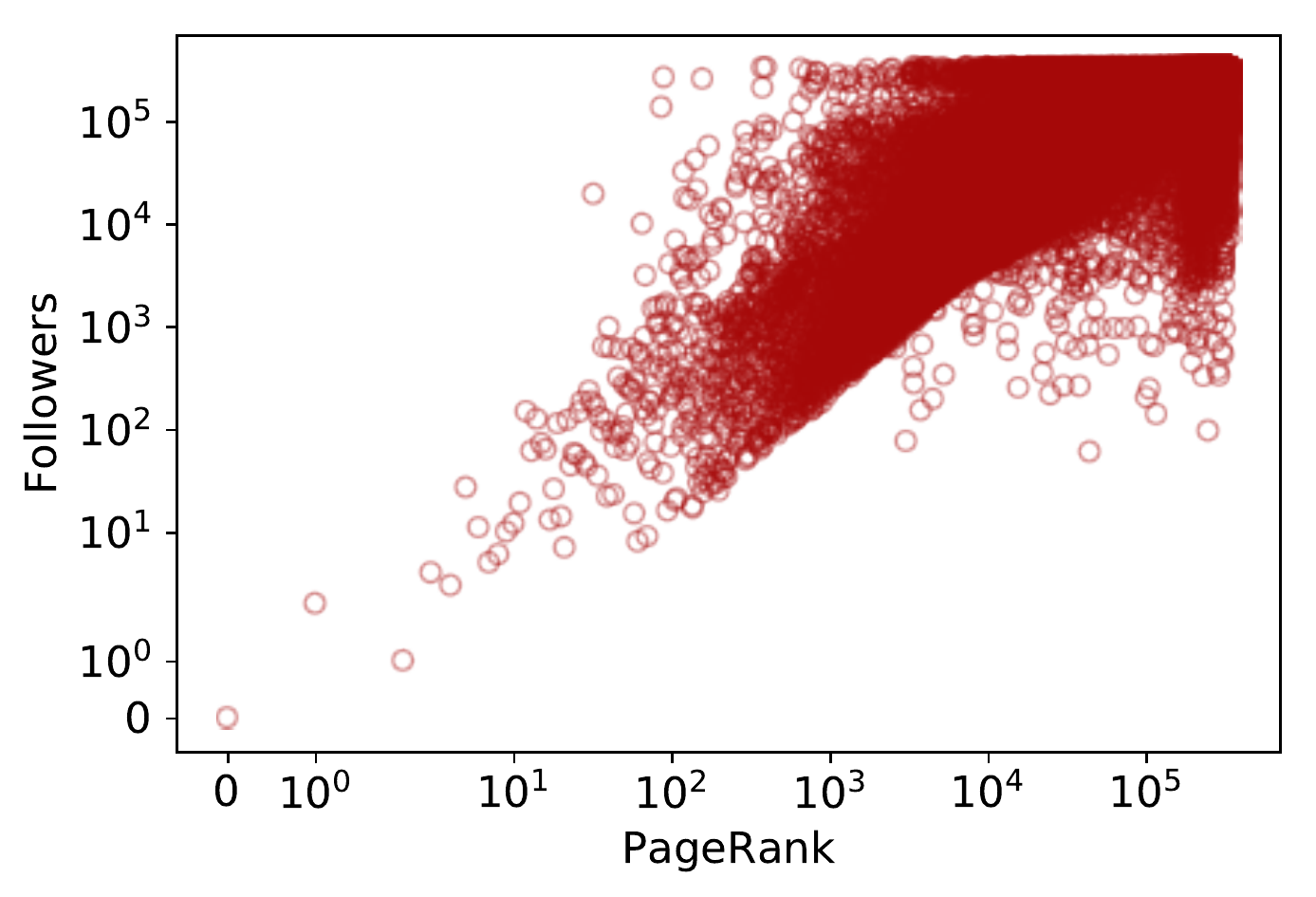}\label{subfig:pr_followers}}
\caption{Correlation of the rankings for each pair of rankings: (a) Followers - Score; (b) PageRank - Score; and (c) PageRank - Followers.}
\label{fig:correlations_scatter}
\end{figure*}

\subsection{Analysis} \label{sec:analysis}
In this section, we provide our analysis on the Gab platform. 
Specifically, we analyze Gab's user base and posts that get shared across several axes.

\subsubsection{Ranking of users}
To get a better handle on the interests of Gab users, we first examine the most popular users using three metrics: 1)~the number of followers; 2)~user account score; and 3)~user PageRank.
These three metrics provide us a good overview of things in terms of ``reach,'' appreciation of content production, and importance in terms of position within the social network.
We report the top 20 users for each metric in Table~\ref{tbl:top_20_users}.
Although we believe that their existence in Table~\ref{tbl:top_20_users} is arguably indicative of their public figure status, for ethical reasons, we omit the ``screen names'' for accounts in cases where a potential link between the screen name and the user's real life names existed \emph{and} it was unclear to us whether or not the user is a public figure.
While Twitter has many celebrities in the most popular users~\cite{kwak2010what}, Gab seems to have what can at best be described as alt-right celebrities like Milo Yiannopoulos and Mike Cernovich.

\descr{Number of followers.} The number of followers that each account has can be regarded as a metric of impact on the platform, as a user with many followers can share its posts to a large number of other users.
We observe a wide variety of different users; 1)~popular alt-right users like Milo Yiannopoulos, Mike Cernovich, Stefan Molyneux, and Brittany Pettibone; 2)~Gab's founder Andrew Torba; and 3)~popular conspiracy theorists like Alex Jones.
Notably lacking are users we might consider as counter-points to the alt-right right, an indication of Gab's heavily right-skewed user-base.

\descr{Score.} The score of each account is a metric of content popularity, as it determines the number of up-votes and down-votes that they receive from other users.
In other words, is the degree of appreciation from other users.
By looking at the ranking using the score, we observe two new additional categories of users: 1)~users purporting to be news outlets, likely pushing false or controversial information on the network like PrisonPlanet and USSANews; and 2)~troll users that seem to have migrated from or been inspired by other platforms (e.g., 4chan) like Kek\_Magician and CuckShamer.

\descr{PageRank.} We also compute PageRank on the followers/followings network and we rank the users according to the obtained score.
We use this metric as it quantifies the structural importance of nodes within a network according to its connections.
Here, we observe some interesting differences from the other two rankings.
For example, the account with username ``realdonaldtrump,'' an account reserved for Donald Trump, appears in the top users mainly because of the extremely high number of users that follow this account, despite the fact that it has no posts or score.

\descr{Comparison of rankings.} To compare the three aforementioned rankings, we plot the ranking of all the users for each pair of rankings in Fig.~\ref{fig:correlations_scatter}.
We observe that the pair with the most agreement is PageRank-Followers (Fig.~\ref{subfig:pr_followers}), followed by the pair Followers-Score (Fig,~\ref{subfig:followers_score}), while the pair with the least agreement is PageRank - Score (Fig~\ref{subfig:pr_score}.
Overall, for all pairs we find a varying degree of rank correlation.
Specifically, we calculate the Spearman's correlation coefficient for each pair of rankings; finding 0.53, 0.42, 0.26 for PageRank-Followers, Followers-Score, and PageRank-Score, respectively.
While these correlations are not terribly strong, they are significant ($p < 0.01$) for the two general classes of users: those that play an important structural role in the network, perhaps encouraging the diffusion of information, and those that produce content the community finds valuable.

\begin{table}[t]
\centering
\resizebox{0.35\columnwidth}{!}{
\begin{tabular}{@{}lrll@{}}
\textbf{Word} & \multicolumn{1}{l}{\textbf{(\%)}} & \textbf{Bigram} & \textbf{(\%)}              \\ \midrule
maga          & \multicolumn{1}{r|}{4.35\%}       & free speech     & \multicolumn{1}{r}{1.24\%} \\
twitter       & \multicolumn{1}{r|}{3.62\%}       & trump supporter & \multicolumn{1}{r}{0.74\%} \\
trump         & \multicolumn{1}{r|}{3.53\%}       & night area      & \multicolumn{1}{r}{0.49\%} \\
conservative  & \multicolumn{1}{r|}{3.47\%}       & area wanna      & \multicolumn{1}{r}{0.48\%} \\
free          & \multicolumn{1}{r|}{3.08\%}       & husband father  & \multicolumn{1}{r}{0.45\%} \\
love          & \multicolumn{1}{r|}{3.03\%}       & check link      & \multicolumn{1}{r}{0.42\%} \\
people        & \multicolumn{1}{r|}{2.76\%}       & freedom speech  & \multicolumn{1}{r}{0.41\%} \\
life          & \multicolumn{1}{r|}{2.70\%}       & hey guys        & \multicolumn{1}{r}{0.40\%} \\
like          & \multicolumn{1}{r|}{2.67\%}       & donald trump    & \multicolumn{1}{r}{0.40\%} \\
man           & \multicolumn{1}{r|}{2.49\%}       & man right       & 0.39\%                     \\
truth         & \multicolumn{1}{r|}{2.46\%}       & america great   & 0.39\%                     \\
god           & \multicolumn{1}{r|}{2.45\%}       & link contracts  & 0.35\%                     \\
world         & \multicolumn{1}{r|}{2.44\%}       & wanna check     & 0.34\%                     \\
freedom           & \multicolumn{1}{r|}{2.29\%}       & make america    & 0.34\%                     \\
right       & \multicolumn{1}{r|}{2.27\%}       & need man        & 0.34\%                     \\
american         & \multicolumn{1}{r|}{2.25\%}       & guys need       & 0.33\%                     \\
want      & \multicolumn{1}{r|}{2.23\%}       & president trump & 0.32\%                     \\
one          & \multicolumn{1}{r|}{2.20\%}       & guy sex         & 0.31\%                     \\
christian           & \multicolumn{1}{r|}{2.17\%}       & click link      & 0.30\%                     \\
time     & \multicolumn{1}{r|}{2.14\%}       & link login      & 0.30\%                     \\ \bottomrule
\end{tabular}
}
\caption{Top 20 words and bigrams found in the descriptions of Gab users.}
\label{tbl:users_description}
\end{table}

\subsubsection{User account analysis}

\descr{User descriptions.} To further assess the type of users that the platform attracts we analyze the description of each created account in our dataset.
Note that by default Gab adds a quote from a famous person as the description of each account and a user can later change it.
Although not perfect, we look for any user description enclosed in quotes with a ``--'' followed by a name, and assume it is a default quote.
Using this heuristic, we find that only 20\% of the users actively change their description from the default.
Table~\ref{tbl:users_description} reports the top words and bigrams found in customized descriptions (we remove stop words for more meaningful results).
Examining the list, it is apparent that Gab users are conservative Americans, religious, and supporters of Donald Trump and ``free speech.''
We also find some accounts that are likely bots and trying to deceive users with their descriptions; among the top bigrams there some that nudge users to click on URLs, possibly malicious, with the promise that they will get sex. 
For example, we find many descriptions similar to the following: ``\emph{Do you wanna get sex tonight? One step is left ! Click the link - $<url>$}.''
It is also worth noting that our account (created for crawling the platform) was followed by 12 suspected bot accounts between December 2017 and January 2018 without making any interactions with the platform (i.e., our account has never made a post or followed any user).

\begin{figure}[t]
\centering
\includegraphics[width=0.65\columnwidth]{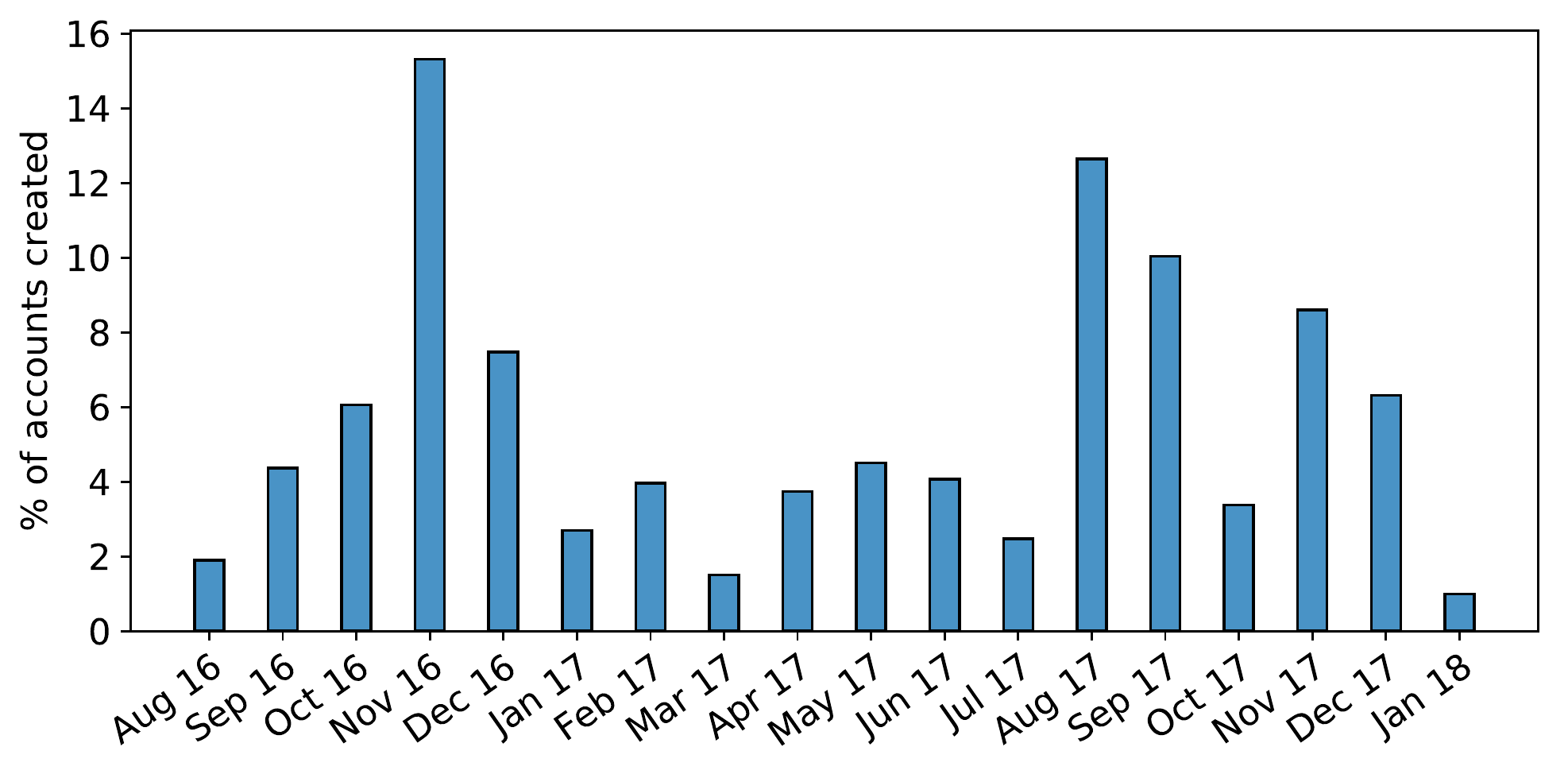}
\caption{Percentage of accounts created per month.}
\label{fig:bc_account_created}
\end{figure}

\descr{User account creation.} We also look when users joined the Gab platform. Fig.~\ref{fig:bc_account_created} reports the percentage of accounts created for each month of our dataset.
Interestingly, we observe that we have peaks for account creation on November 2016 and August 2017. 
These findings highlight the fact that Gab became popular during notable world and politics events like the 2016 US elections as well as the Charlottesville Unite the Right rally~\cite{charlotesville}.
Finally, only a small percentage of Gab's users are either pro or verified, 0.75\% and 0.5\%, respectively, while 1.7\% of the users have a private account (i.e., only their followers can see their gabs).

\begin{figure*}[t!]
\center
\subfigure[]{\includegraphics[width=0.32\textwidth]{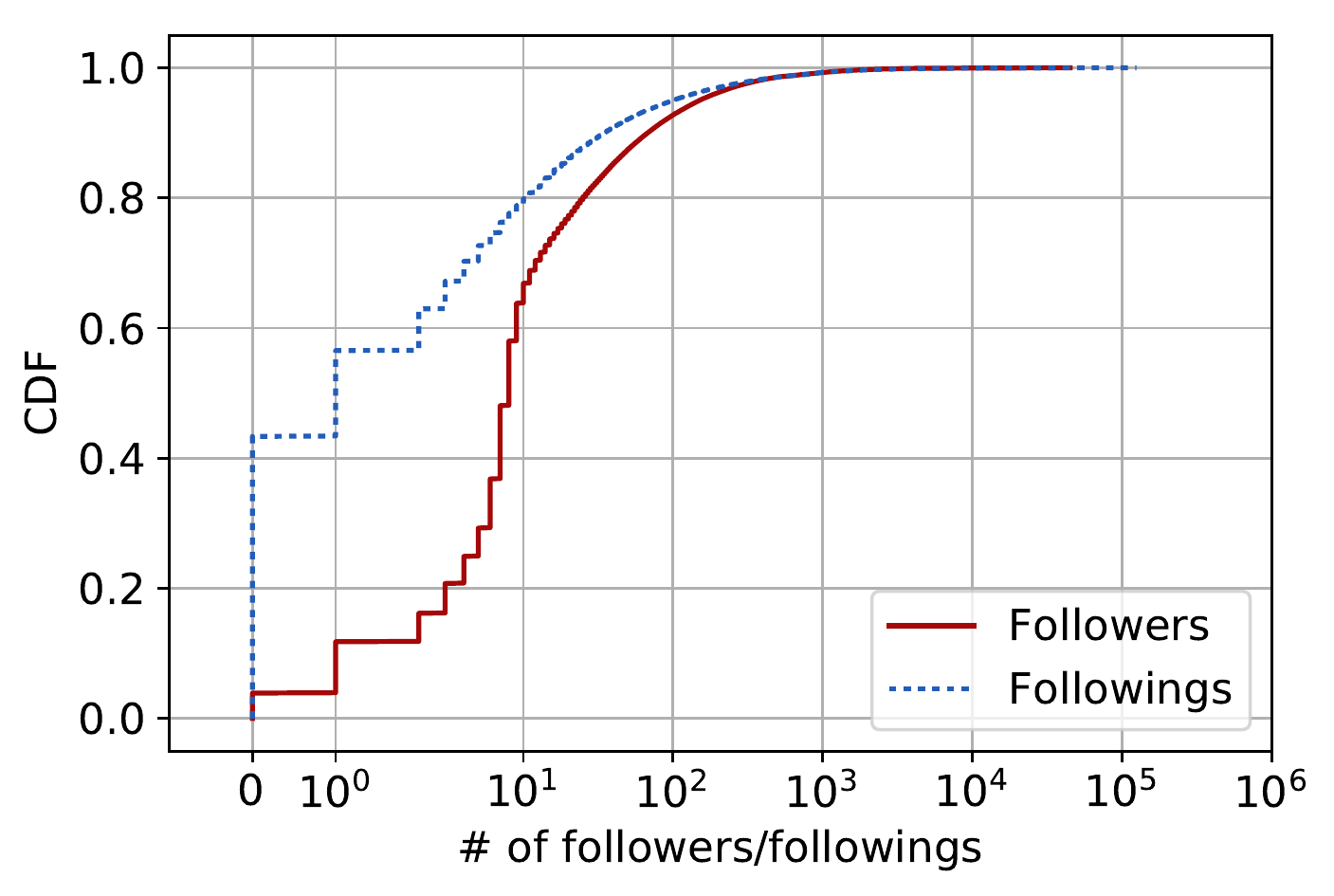}\label{fig:cdf_followers_following}}
\subfigure[]{\includegraphics[width=0.32\textwidth]{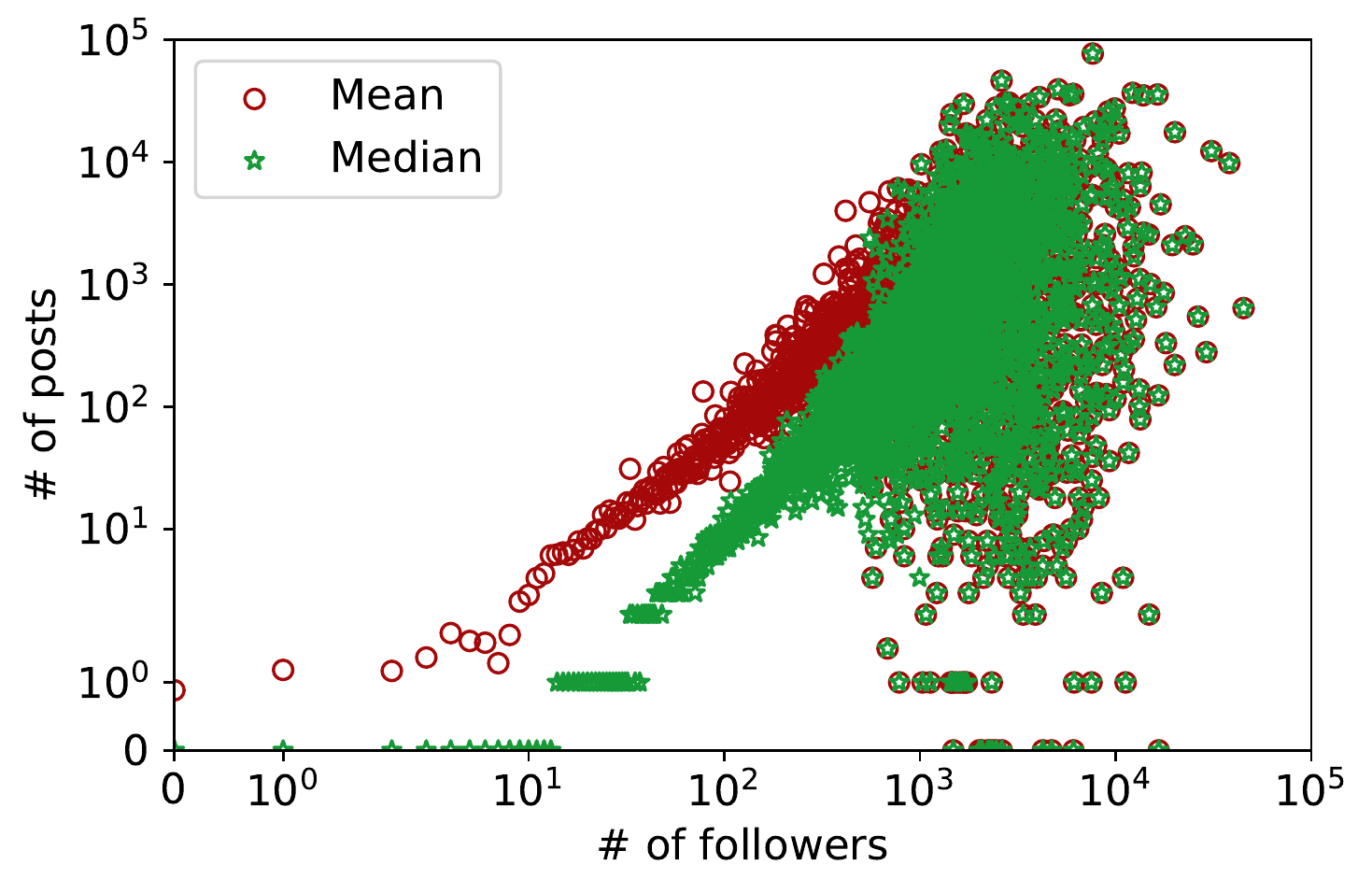}\label{subfig:scatter_followers_posts}}
\subfigure[]{\includegraphics[width=0.32\textwidth]{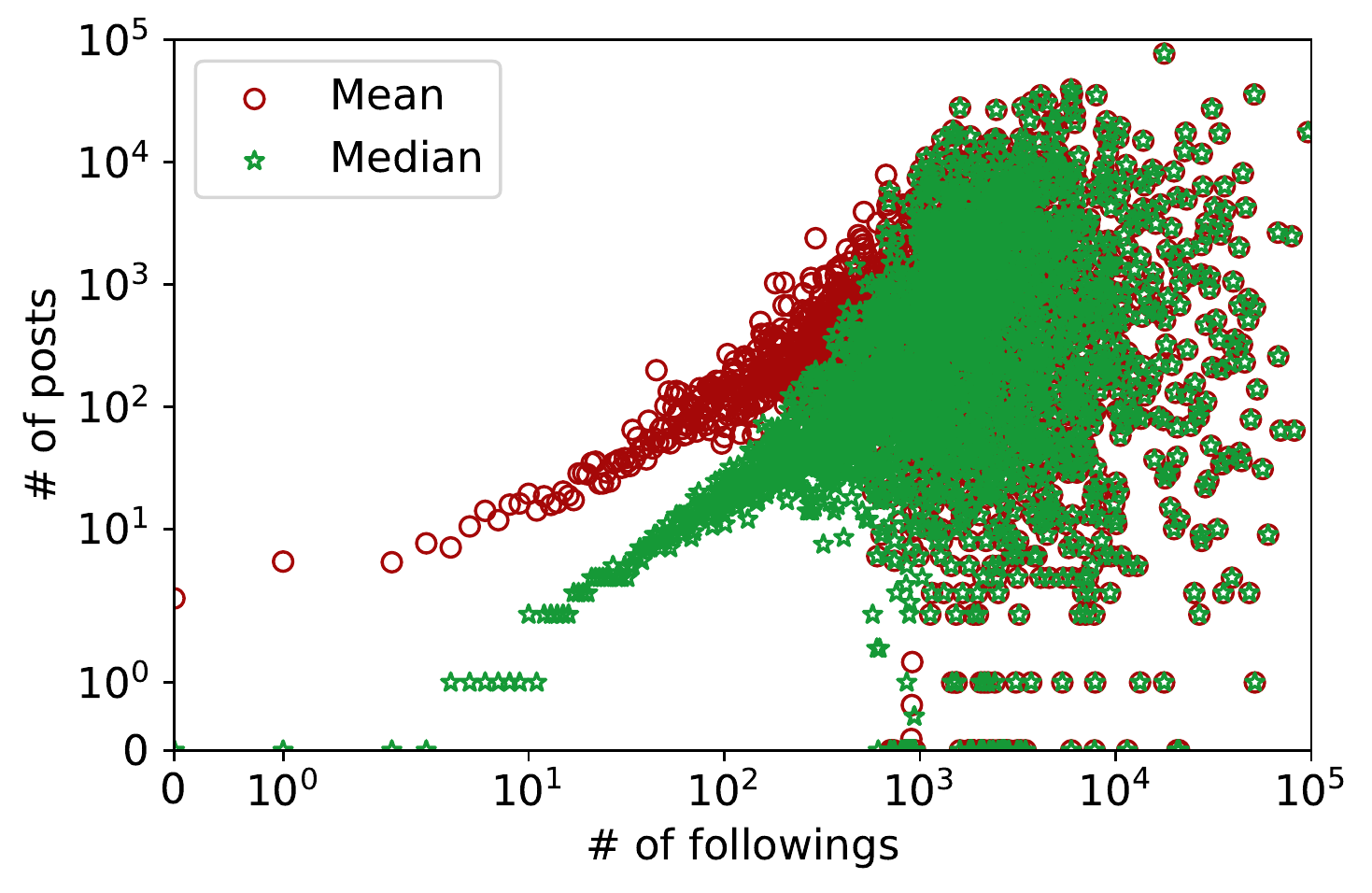}\label{subfig:scatter_following_posts}}
\caption{Followers and Following analysis (a) CDF of number of followers and following (b) number of followers and number of posts and (c) number of following and number of posts. }
\label{fig:followers_followings}
\end{figure*}

\descr{Followers/Followings.} Fig.~\ref{fig:followers_followings} reports our analysis based on the number of followers and followings for each user.
From Fig.~\ref{fig:cdf_followers_following} we observe that in general Gab users have a larger number of followers when compared with following users.
Interestingly, 43\% of users are following zero other users, while only 4\% of users have zero followers.
I.e., although counter-intuitive, most users have more followers than users they follow.
Figs.~\ref{subfig:scatter_followers_posts} and ~\ref{subfig:scatter_following_posts} show the number of followers and following in conjunction with the number of posts for each Gab user. 
We bin the data in log-scale bins and we report the mean and median value for each bin.
We observe that in both cases, that there is a near linear relationship with the number of posts and followers/followings up until around 10 followers/followings.
After this point, we see this relationship diverge, with a substantial number of users with huge numbers of posts, some over 77K.
This demonstrates the extremely heavy tail in terms of content production on Gab, as is typical of most social medial platforms.

\descr{Reciprocity.} From the followers/followings network we find a low level of reciprocity: specifically, only 29.5\% of the node pairs in the network are connected both ways, while the remaining 71.5\% are connected one way.
When compared with the corresponding metric on Twitter~\cite{kwak2010what}, these results highlight that Gab has a larger degree of network reciprocity indicating that the community is more tightly-knit, which is expected when considering that Gab mostly attracts users from the same ideology (i.e., alt-right community).

\subsubsection{Posts Analysis}
\descr{Basic Statistics.}
First, we note that 63\% of the posts in our dataset are original posts while 37\% are reposts.
Interestingly, only 0.14\%  of the posts are marked as NSFW. 
This is surprising given the fact that one of the reasons that Apple rejected Gab's mobile app is due to the share of NSFW content~\cite{gab_apple_porn}.
From browsing the Gab platform, we also can anecdotally confirm the existence of NSFW posts that are not marked as such, raising questions about how Gab moderates and enforces the use of NSFW tags by users.
When looking a bit closer at their policies, Gab notes that they use a 1964 United States Supreme Court Ruling~\cite{wikipedia_1964} on pornography that provides the famous ``I'll known it when I see it'' test.
In any case, it would seem that Gab's social norms are relatively lenient with respect to what is considered NSFW.

We also look into the languages of the posts, as returned by Gab's API. 
We find that Gab's API does not return a language code for 56\% of posts.
By looking at the dataset, we find that all posts before June 2016 do not have an associated language; possibly indicating that Gab added the language field afterwards.
Nevertheless, we find that the most popular languages are English (40\%), Deutsch (3.3\%), and French (0.14\%); possibly shedding light to Gab's users locations which are mainly the US, the UK, and Germany.

\begin{table}[t]
\centering
\small
\begin{tabular}{lrlr}
\textbf{Domain}      & \multicolumn{1}{l}{\textbf{(\%)}} & \textbf{Domain} & \multicolumn{1}{l}{\textbf{(\%)}} \\ \hline
youtube.com          & 4.22\%                            & zerohedge.com   & 0.53\%                            \\
youtu.be             & 2.67\%                            & twimg.com       & 0.53\%                            \\
twitter.com          & 1.96\%                            & dailycaller.com & 0.49\%                            \\
breitbart.com        & 1.44\%                            & t.co            & 0.47\%                            \\
bit.ly               & 0.82\%                            & ussanews.com    & 0.46\%                            \\
thegatewaypundit.com & 0.74\%                            & dailymail.co.uk & 0.46\%                            \\
kek.gg               & 0.69\%                            & tinyurl.com     & 0.44\%                            \\
imgur.com            & 0.68\%                            & wordpress.com   & 0.43\%                            \\
sli.mg               & 0.61\%                            & foxnews.com     & 0.41\%                            \\
infowars.com         & 0.56\%                            & blogspot.com    & 0.32\%                          \\ \hline 
\end{tabular}
\caption{Top 20 domains in posts and their respective percentage over all posts.}
\label{tbl:top_domains}
\end{table}

\descr{URLs.} Next, we assess the use of URLs in Gab; overall we find 3.5M unique URLs from 81K domains. 
Table~\ref{tbl:top_domains} reports the top 20 domains according to their percentage of inclusion in all posts.
We observe that the most popular domain is YouTube with almost 7\% of all posts, followed by Twitter with 2\%.
Interestingly, we note the extensive use of alternative news sources like Breitbart (1.4\%), The Gateway Pundit (0.7\%), and Infowars (0.5\%), while mainstream news outlets like Fox News (0.4\%) and Daily Mail (0.4\%) are further below.
Also, we note the use of image hosting services like Imgur (0.6\%), \url{sli.mg} (0.6\%), and kek.gg (0.7\%) and URL shorteners like \url{bit.ly} (0.8\%) and \url{tinyurl.com} (0.4\%).
Finally, it is worth mentioning that The Daily Stormer, a well known neo-Nazi web community is five ranks ahead of the most popular mainstream news source, The Hill.

\begin{table}[t]
\centering
\small
\begin{tabular}{lrlr}
\textbf{Hashtag} & \multicolumn{1}{l}{\textbf{(\%)}} & \textbf{Mention} & \multicolumn{1}{l}{\textbf{(\%)}} \\ \hline
MAGA             & \multicolumn{1}{r|}{6.06\%}       & a                & 0.69\%                            \\
GabFam           & \multicolumn{1}{r|}{4.22\%}       & TexasYankee4     & 0.31\%                            \\
Trump            & \multicolumn{1}{r|}{3.01\%}       & Stargirlx        & 0.26\%                            \\
SpeakFreely      & \multicolumn{1}{r|}{2.28\%}       & YouTube          & 0.24\%                            \\
News             & \multicolumn{1}{r|}{2.00\%}       & support          & 0.23\%                            \\
Gab              & \multicolumn{1}{r|}{0.88\%}       & Amy              & 0.22\%                            \\
DrainTheSwamp    & \multicolumn{1}{r|}{0.71\%}       & RaviCrux         & 0.20\%                            \\
AltRight         & \multicolumn{1}{r|}{0.61\%}       & u                & 0.19\%                            \\
Pizzagate        & \multicolumn{1}{r|}{0.57\%}       & BlueGood         & 0.18\%                            \\
Politics         & \multicolumn{1}{r|}{0.53\%}       & HorrorQueen      & 0.17\%                            \\
PresidentTrump   & \multicolumn{1}{r|}{0.47\%}       & Sockalexis       & 0.17\%                            \\
FakeNews         & \multicolumn{1}{r|}{0.41\%}       & Don              & 0.17\%                            \\
BritFam          & \multicolumn{1}{r|}{0.37\%}       & BrittPettibone   & 0.16\%                            \\
2A               & \multicolumn{1}{r|}{0.35\%}       & TukkRivers       & 0.15\%                            \\
maga             & \multicolumn{1}{r|}{0.32\%}       & CurryPanda       & 0.15\%                            \\
NewGabber        & \multicolumn{1}{r|}{0.28\%}       & Gee              & 0.15\%                            \\
CanFam           & \multicolumn{1}{r|}{0.27\%}       & e                & 0.14\%                            \\
BanIslam         & \multicolumn{1}{r|}{0.25\%}       & careyetta        & 0.14\%                            \\
MSM              & \multicolumn{1}{r|}{0.22\%}       & PrisonPlanet     & 0.14\%                            \\
1A               & \multicolumn{1}{r|}{0.21\%}       & JoshC            & 0.12\%                            \\ \hline
\end{tabular}
\caption{Top 20 hashtags and mentions found in Gab. We report their percentage over all posts.}
\label{tbl:top_hashtags_mentions}
\end{table}

\descr{Hashtags \& Mentions}
As discussed in Chapter~\ref{chapter:background}, Gab supports the use of hashtags and mentions similar to Twitter. 
Table~\ref{tbl:top_hashtags_mentions} reports the top 20 hashtags/mentions that we find in our dataset.
We observe that the majority of the hashtags are used in posts about Trump, news, and politics.
We note that among the top hashtags are ``AltRight'', indicating that Gab users are followers of the alt-right movement or they discuss topics related to the alt-right; ``Pizzagate'', which denotes discussions around the notorious conspiracy theory~\cite{bbc_4chan_pizzagate}; and ``BanIslam'', which indicate that Gab users are sharing their islamophobic views.
It is also worth noting the use of hashtags for the dissemination of popular memes, like the Drain the Swamp meme that is popular among Trump's supporters~\cite{vox_memes_trump}.
When looking at the most popular users that get mentioned, we find popular users related to the Gab platform like Andrew Torba (Gab's CEO with username @a).

We also note users that are popular with respect to mentions, but do \emph{not} appear in Table~\ref{tbl:top_20_users}'s lists of popular users.
For example, Amy is an account purporting to be Andrew Torba's mother. 
The user Stargirlx, who we note changed usernames three times during our collection period, appears to be an account presenting itself as a millennial ``GenZ'' young woman.
Interestingly, it seems that Amy and Stargirlx have been organizing Gab ``chats,'' which are private groups of users, for 18 to 29 year olds to discuss politics; possibly indicating efforts to recruit millennials to the alt-right community.

\begin{table}[]
\centering
\small
\begin{tabular}{lrlr}
\textbf{Topic}           & \multicolumn{1}{l}{\textbf{(\%)}} & \textbf{Category} & \multicolumn{1}{l}{\textbf{(\%)}} \\ \hline
Deutsch                  & \multicolumn{1}{r|}{2.29\%}       & News              & 15.91\%                           \\
BritFam                  & \multicolumn{1}{r|}{0.73\%}       & Politics          & 10.30\%                           \\
Introduce Yourself       & \multicolumn{1}{r|}{0.59\%}       & AMA               & 4.46\%                            \\
International News       & \multicolumn{1}{r|}{0.19\%}       & Humor             & 3.50\%                            \\
DACA                     & \multicolumn{1}{r|}{0.17\%}       & Technology        & 1.44\%                            \\
Las Vegas Terror Attack  & \multicolumn{1}{r|}{0.16\%}       & Philosophy        & 1.06\%                            \\
Hurricane Harvey         & \multicolumn{1}{r|}{0.16\%}       & Entertainment     & 1.01\%                            \\
Gab Polls                & \multicolumn{1}{r|}{0.13\%}       & Art               & 0.72\%                            \\
London                   & \multicolumn{1}{r|}{0.12\%}       & Faith             & 0.69\%                            \\
2017 Meme Year in Review & \multicolumn{1}{r|}{0.12\%}       & Science           & 0.56\%                            \\
Twitter Purge            & \multicolumn{1}{r|}{0.12\%}       & Music             & 0.52\%                            \\
Seth Rich                & \multicolumn{1}{r|}{0.11\%}       & Sports            & 0.39\%                            \\
Memes                    & \multicolumn{1}{r|}{0.11\%}       & Photography       & 0.37\%                            \\
Vegas Shooting           & \multicolumn{1}{r|}{0.11\%}       & Finance           & 0.31\%                            \\
Judge Roy Moore          & \multicolumn{1}{r|}{0.09\%}       & Cuisine           & 0.16\%                            \\ \hline
\end{tabular}
\caption{Top 15 categories and topics found in the Gab dataset}
\label{tbl:top_categories_topics}
\end{table}

\descr{Categories \& Topics.} As discussed in Chapter~\ref{chapter:background}, gabs may be part of a topic or category.
By analyzing the data, we find that this happens for 12\% and 42\% of the posts for topics and categories, respectively.
Table~\ref{tbl:top_categories_topics} reports the percentage of posts for each category as well as for the top 15 topics.
For topics, we observe that the most popular are general ``Ask Me Anything'' (AMA) topics like Deutsch (2.29\%, for German users), BritFam (0.73\%, for British users), and Introduce Yourself (0.59\%). 
Furthermore, other popular topics include world events and news like International News (0.59\%), Las Vegas shooting (0.27\%), and conspiracy theories like Seth Rich's Murder (0.11\%). 
When looking at the top categories we find that by far the most popular categories are News (15.91\%) and Politics (10.30\%).
Other popular categories include AMA 4.46\%), Humor (3.50\%), and Technology (1.44\%).

These findings highlight that Gab is heavily used for the dissemination and discussion of world events and news.
Therefore, its role and influence on the Web's information ecosystem should be assessed in the near future.
Also, this categorization of posts can be of great importance for the research community as it provides labeled ground truth about discussions around a particular topic and category.

\descr{Hate speech assessment.} As previously discussed, Gab was openly accused of allowing the dissemination of hate speech. 
In fact, Google removed Gab's mobile app from its Play Store because it violates their hate speech policy~\cite{gab_hate_speech}.
Due to this, we aim to assess the extent of hate speech in our dataset.
Using the modified Hatebase~\cite{hatebase} dictionary used by the authors of~\cite{hine2016longitudinal}, we find that 5.4\% of all Gab posts include a hate word.
In comparison, Gab has 2.4 times the rate of hate words when compared to Twitter, but less than halve the rate of hate words compared to 4chan's Politically Incorrect board (/pol/)~\cite{hine2016longitudinal}.
These findings indicate that Gab resides on the border of mainstream social networks like Twitter and fringe Web communities like 4chan's Politically Incorrect (/pol/) board. 

\begin{figure*}[t!]
\centering
\subfigure[Date]{\includegraphics[width=0.49\textwidth]{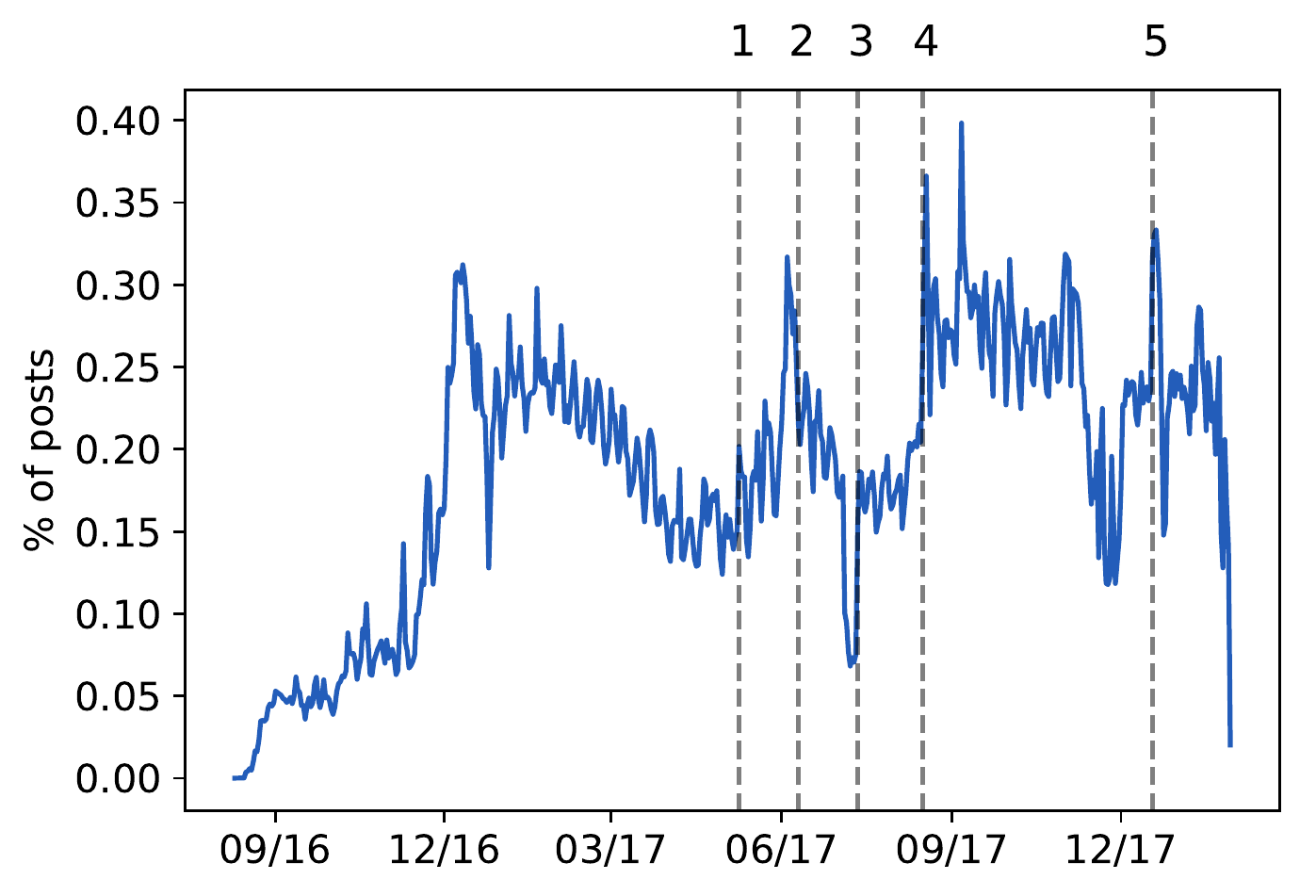}\label{subfig:counts_day}}
\subfigure[Hour of Day]{\includegraphics[width=0.49\textwidth]{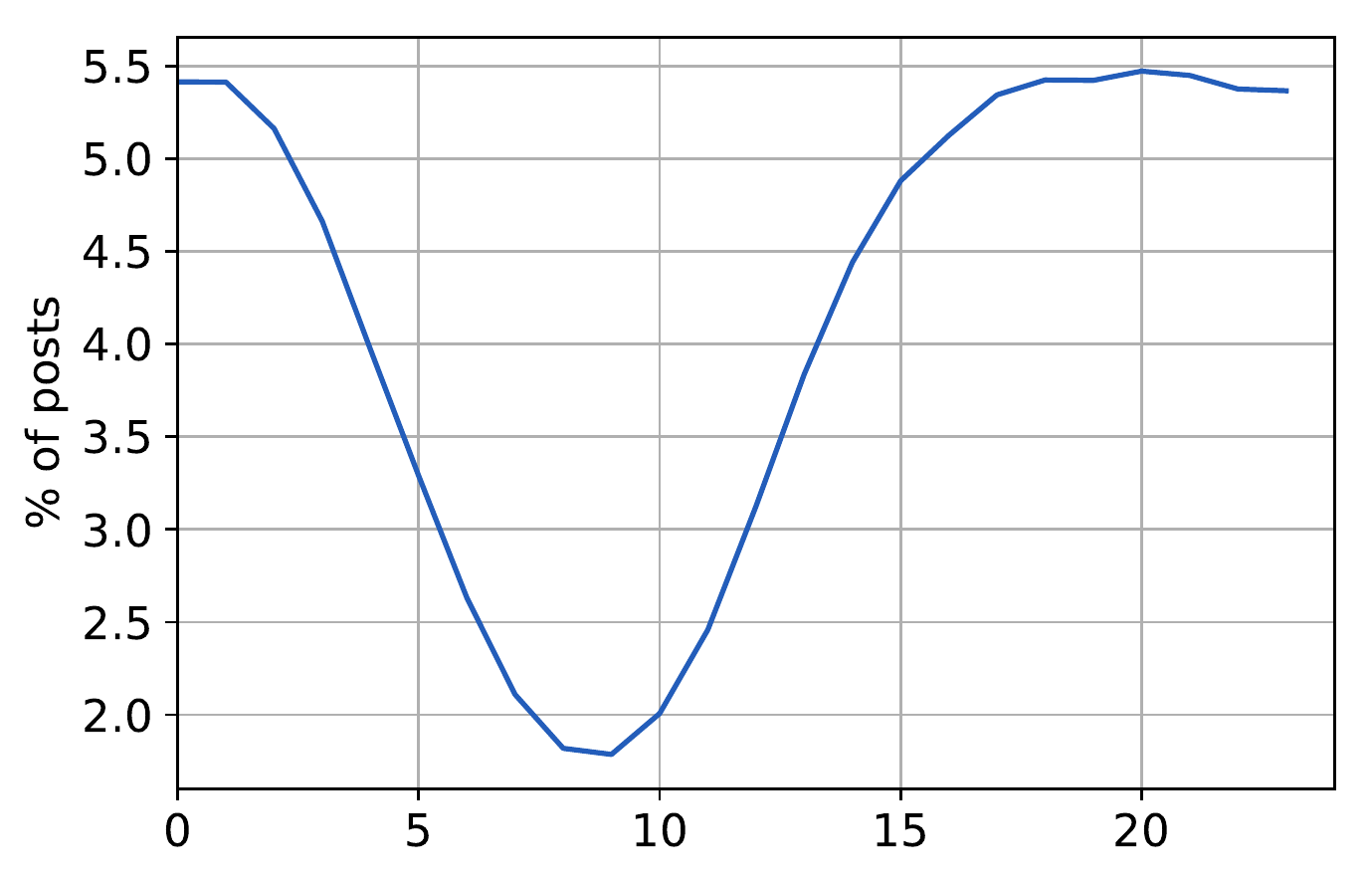}\label{subfig:counts_hour_day}}
\subfigure[Hour of Week]{\includegraphics[width=0.49\textwidth]{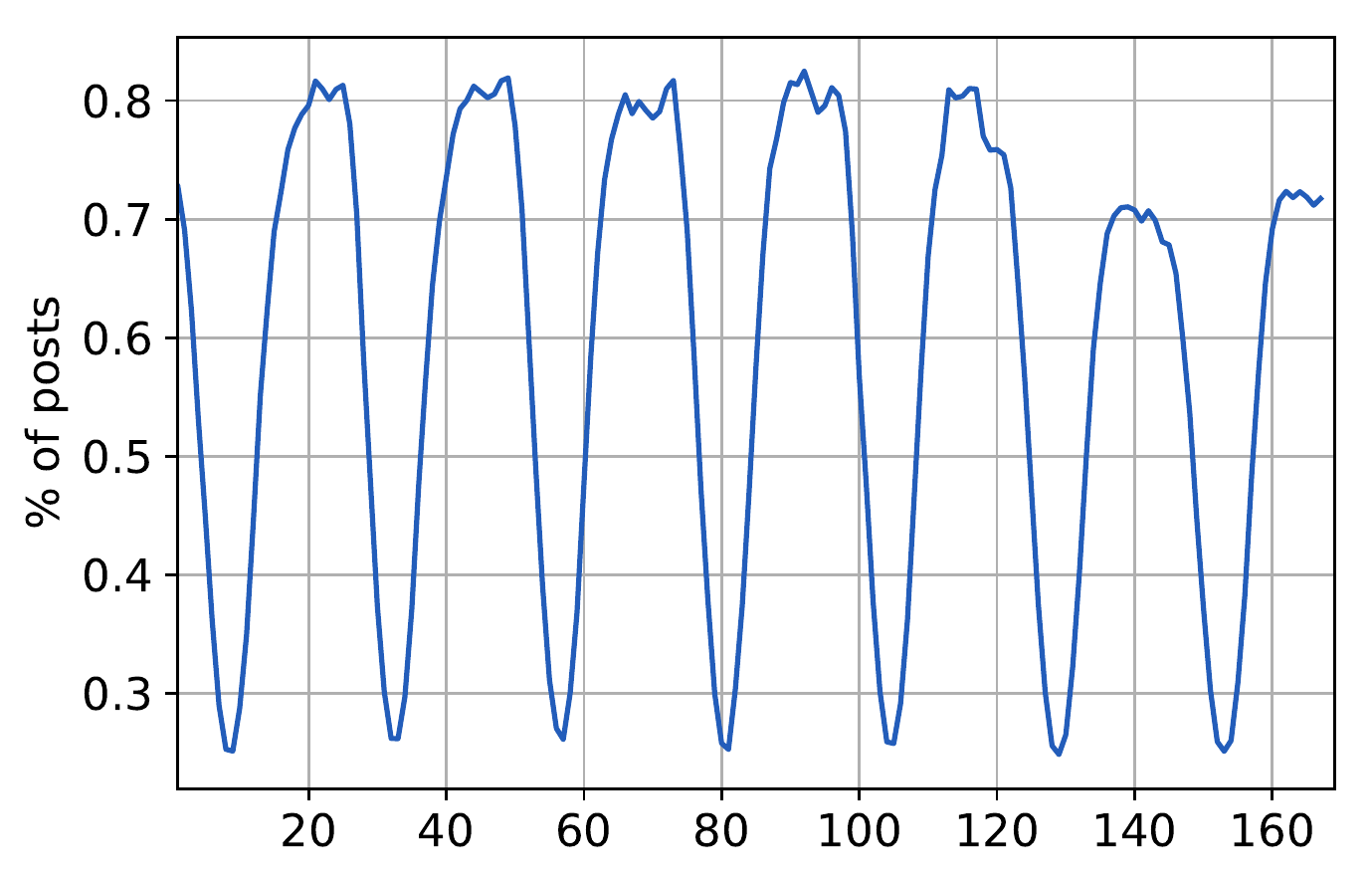}\label{subfig:counts_hour_week}}
\caption{Temporal analysis of the Gab posts (a) each day; (b) based  on hour of day and (c) based on hour of week.}
\label{fig:temporal_analysis}
\end{figure*}

\descr{Temporal Analysis.} Finally, we study the posting behavior of Gab users from a temporal point of view. Fig.~\ref{fig:temporal_analysis} shows the distribution of the Gab posts in our dataset according to each day of our dataset, as well as per hour of day and week (in UTC).
We observe that the general trend is that the number of Gab's posts increase over time (Fig.~\ref{subfig:counts_day}); this indicates an increase in Gab's popularity.
Furthermore, we note that Gab users posts most of their gabs during the afternoon and late night (after 3 PM UTC) while they rarely post during the morning hours (Fig.~\ref{subfig:counts_hour_day}).
Also, the aforementioned posting behavior follow a diurnal weekly pattern as we show in Fig.~\ref{subfig:counts_hour_week}.

To isolate significant days in the time series in Fig.~\ref{subfig:counts_day}, we perform a \emph{changepoint analysis} using the Pruned Exact Linear Time (PELT) method~\cite{changepoint}.
First, we use our knowledge of the weekly variation in average post numbers from Fig.~\ref{subfig:counts_hour_week} to subtract from our timeseries the mean number of posts for each day.
This leaves us with a mean-zero timeseries of the deviation of the number of posts per day from the daily average.
We assume that this timeseries is drawn from a normal distribution, with mean and variance that can change at a discrete number of changepoints.
We then use the PELT algorithm to maximize the log-likelihood function for the mean(s) and variance(s) of this distribution, with a penalty for the number of changepoints.
By ramping down the penalty function, we produce a ranking of the changepoints.

Examining current events around these changepoints provides insight into they dynamics that drive Gab behavior.
First, we note that there is a general increase in activity up to the Trump inauguration, at which point activity begins to decline.
When looking later down the timeline, we see an increase in activity after the changepoint marked \textbf{1} in Fig.~\ref{subfig:counts_day}.
Changepoint \textbf{1} coincides with James Comey's firing from the FBI, and the relative acceleration of the Trump-Russian collusion probe~\cite{james_comey}.

The next changepoint (\textbf{2}) coincides with the so-called ``March Against Sharia''~\cite{sharia} organized by the alt-right, with the event marked \textbf{4} corresponding to Trump's ``blame on both sides'' response to violence at the Unite the Right Rally in Charlottesville~\cite{charlotesville_blame}.
Similarly, we see a meaningful response to Twitter's banning of abusive users~\cite{twitter_purge} marked as changepoint \textbf{5}.

Changepoint \textbf{3}, occurring on July 12, 2017 is of particular interest, since it is the most extreme \emph{reduction} in activity recognized as a changepoint.
From what we can tell, this is a reaction to Donald Trump Jr. releasing emails that seemingly evidenced his meeting with a Russian lawyer to receive compromising intelligence on Hillary Clinton's campaign~\cite{donald_trump_junior}.
I.e., the disclosure of evidence of collusion with Russia corresponded to the single largest drop in posting activity on Gab.

\subsection{Remarks}

In this work, we have provided the first characterization of a new social network called Gab.
We analyzed 22M posts from 336K users, finding that Gab attracts the interest of users ranging from alt-right supporters and conspiracy theorists to trolls.
We showed that Gab is extensively used for the discussion of news, world events, and politics-related topics, further motivating the need take it into account when studying information cascades on the Web.
By looking at the posts for hate words, we also found that 5.4\% of the posts include hate words.
Finally, using changepoint analysis, we highlighted how Gab reacts very strongly to real-world events focused around white nationalism and support of Donald Trump.

\section{Understanding Web Archiving Services and their Use on Multiple Web Communities}

\subsection{Motivation}

In today's digital society, the availability and persistence of Web resources are very relevant issues.
A substantial number of URLs shared on the Web becomes unavailable after some time as websites are shutdown or redesigned in a way that does not preserve old URLs -- a phenomenon known as \emph{``link rot''}~\cite{koehler2004longitudinal}.
Moreover, content might be taken down by authorities on a legal basis, deleted by users who have shared it on social media, removed as per the ``right to be forgotten''~\cite{gdpr_right_forgotten}, etc. Overall, the ephemerality of Web content often prompts debate with respect to its impact on the availability of information, accountability, or even censorship. 

In this context, an important role is played by services like the Wayback Machine (\url{archive.org}), which \emph{proactively} archives large portions of the Web, allowing users to search and retrieve the history of more than 300 billion pages.
At the same time, {\em on-demand} archiving services like \url{archive.is} have also become popular: users can take a snapshot of a Web page by entering its URL, which the system crawls and archives, returning a permanent short URL serving as a time capsule that can be shared across the Web.

Archiving services  serve a variety of purposes beyond addressing link rot.
Platforms like \url{archive.is} are reportedly used to preserve controversial blogs and tweets that the author may later opt to delete~\cite{mondal2016forgetting}.
Moreover, %
they also reduce Web traffic toward ``source URLs'' when the original content is still accessible, thus depriving them of potential ad revenue streams (users do not visit the original site, but just the archived copy). 
In fact, anecdotal evidence has emerged that alt-right communities target outlets they disagree with by nudging their users to share 
archive URLs instead~\cite{motherboard_archive_traffic}, or discrediting them by pointing at earlier versions of articles~\cite{vice_banned_archive}.

Given the role in helping content persist, their use on social networks, as well as anecdotal evidence of their misuse in contexts where information could be weaponized~\cite{weaponized_nytimes}, archiving services are arguably impactful actors that should be thoroughly analyzed.
To this end, we aim to shed light on the Web archiving ecosystem, aiming to answer the following research questions:
How are archive URLs disseminated across popular social networks?
What kind of content gets archived, by whom and why?
Are archiving services misused in any way?

To answer these questions, we perform a large-scale quantitative analysis of Web archives, based on two data sources:
1) 21M URLs collected from the \url{archive.is} live feed, and 2) 356K \url{archive.is} plus 391K Wayback Machine URLs that were shared on four social networks: Reddit, Twitter, Gab, and 4chan's Politically Incorrect board (\dspol).

Our main findings include:
\begin{enumerate}
\item News and social media posts are the most common types of content archived, likely due to their (perceived) ephemeral and/or controversial nature. 
\item URLs of archiving services are extensively shared on ``fringe'' communities within Reddit and 4chan to preserve possibly contentious content, or to refer to it without increasing the Web traffic to the source. We also find that \dspol and Gab users favor archive.is over Wayback Machine (respectively, 15x and 16x), highlighting a particular use case in ``controversial'' online communities.
\item Web archives are exploited by users to bypass censorship policies in some communities: for instance, \dspol users post \url{archive.is} URLs to share content from 8chan and Facebook, which are banned on the platform, or to circumvent accidental censorship 
of some news sources because of substitution filters (e.g., `smh' becomes `baka', so links to \url{smh.com.au} are unusable).
\item Reddit bots are responsible for posting a very large portion of archive URLs in Reddit (respectively, 44\% and 85\% of \url{archive.is} and Wayback Machine URLs).
This is due to moderators aiming to alleviate the effects of link rot on the platform; however, this pro-active archival of content also impact traffic to archived sites originating from Reddit.

\item The\_Donald subreddit systematically targets ad revenue of news sources with conflicting ideologies:  moderation bots block URLs from 
those sites and prompt users to post archive URLs instead (some domains, e.g. \url{nydailynews.com}, have up to 46\% of their 
content censored). According to our conservative estimates, popular news sources like the Washinghton Post lose yearly approximately \textdollar 70K 
from their ad revenue because of the use of archiving services on Reddit.
\end{enumerate}

\subsection{Background}\label{sec:background}

Our analysis focuses on two popular archiving services: \url{archive.is} and the Wayback Machine (\url{archive.org}).
The former stores {\em snapshots} of Web pages upon request, while the latter is run by a 
non-profit organization (the Internet Archive) aiming to archive pages mainly through a constant crawling process.

\descr{Archive.is} offers a free, on-demand archival service of Web pages: a user 
visits the service and enters a URL to be archived. 
It also acts as a link shortener which obfuscates the source URL, by generating a 5-character URL. 
For instance, \url{http://archive.is/HVbU} shows the snapshot of Google's homepage, archived on July, 03, 2012 at 07:03:24. 

\descr{Wayback Machine.} Launched in 2001, the Wayback Machine archives a large portion of Web content, storing periodic snapshots of various pages. 
It mainly works through a proactive crawler\footnote{\url{http://crawler.archive.org/index.html}}, which visits various sites and captures 
a snapshot of the content.
However, users can also trigger information archival on demand.
When a page is archived, an archive URL is created in the following format: https://web.archive.org/web/[{\em time of archival}]/[{\em source URL}]. 
For example, the archive URL \url{https://web.archive.org/web/20100205062719/http://www.google.com/} returns the version of Google's home page on  February 5, 2010, at 06:27:19 (UTC).
In the rest of the thesis, we refer to the URLs generated by archiving services %
as {\bf\em archive URLs}, and to the archived URL as {\bf\em source URLs}. %

\smallskip
We opt to study the Wayback Machine and \url{archive.is} for a few reasons.
First of all, they are popular services: as of Jan 2018, their Alexa Global Rank is, resp., 300 and 2,920.
The Wayback Machine is actually one of the oldest initiatives, %
with about 300 billion pages archived as of 2017.
We also choose these two because of some important differences between them. The Wayback Machine is run by a  501(c)(3) non-profit organization, while  \url{archive.is} is hosted by Russian provider Hostkey (interestingly, it is only accessible via HTTP in Russia).
Moreover, the former respects robots exclusion standards (even retroactively) and generally gives website owners the right to request removal of  pages from the archive, while the latter only complies (albeit inconsistently) with DMCA take-down requests. 
Finally,  \url{archive.is} is reportedly used in ``fringe'' Web communities within 4chan and Reddit, which are known for generating~\cite{bbc_4chan_pizzagate} and incubating~\cite{guardian_reddit} fake news stories, and for their influence on the information ecosystem~\cite{zannettou2017web}.

\subsection{Datasets}\label{sec:dataset}

We now present the datasets studied in our work as well as our data collection methodology.
We perform two crawls: 1) \url{archive.is} URLs obtained from the live feed page and 
2) Wayback Machine and \url{archive.is} URLs posted on four social networks, namely, Twitter, Reddit, Gab, and 4chan's \dspol.
The resulting datasets are summarized in Table~\ref{tab:dataset}. %

\descr{\url{Archive.is} live feed.} To gather a large dataset of \url{archive.is} generated URLs, we use the live feed page 
(\url{http://archive.is/livefeed/}), which provides a view of the archive based on archival time (e.g., the first page lists URLs 
archived in the previous 10 minutes). In August 2017, we crawl the first 100K pages of the live feed,
acquiring 45.2M URLs, archived between October 7, 2015 and August 26, 2017.

Next, we visit the \url{archive.is} URLs, and scrape the content to get the archival time and the source URL.
To avoid issues for the site operators, we throttle our crawler and do not visit all 45.2M URLs. 
Instead, we randomly sample them while ensuring temporal coverage, visiting 21.5M (48\%) archive URLs, corresponding to 20.6M unique source URLs from 5.3M unique domains. %
Note that given the substantial size of our sample, which guarantees temporal coverage over almost two years, the 
resulting dataset is representative of the archive. In other words, our sampling strategy does not likely introduce substantial biases affecting our results.

\begin{table}[t]
\centering
\setlength{\tabcolsep}{0.18em} %
\resizebox{0.8\columnwidth}{!}{
\begin{tabular}{@{}llrrrrr@{}}
\toprule
\textbf{Platform}   & \textbf{Archive} & {\textbf{\#Posts with Archive}} & {\textbf{Archive}} & {\textbf{Source}} & {\textbf{Source}} & {\textbf{Filtered}} \\ 
& & {\textbf{URLs (\%all posts)}} & {\textbf{URLs}} & {\textbf{URLs}} & {\textbf{Domains}} & \\\midrule
\textbf{Live Feed} & archive.is        &                 & 21,537,554            & 20,608,834           & 5,388,112               & -                 \\ \midrule
\textbf{Reddit}     & archive.is       & 327,050 ($2.9\hspace{-0.05cm}\cdot\hspace{-0.05cm}10^{-4}\%$) & 310,392               & 291,382              & 15,994                  & 35.70\%           \\
\textbf{}           & Wayback          & 320,379 ($2.8\hspace{-0.05cm}\cdot\hspace{-0.05cm}10^{-4}\%$) & 387,081               & 343,851              & 21,124                  & 17.20\%           \\ \midrule
\textbf{/pol/}      & archive.is       & 46,912 ($1.1\hspace{-0.05cm}\cdot\hspace{-0.05cm}10^{-3}\%$)  & 36,277                & 33,824               & 3,970                   & 4.67\%            \\
\textbf{}           & Wayback          & 3,848 ($9.7\hspace{-0.05cm}\cdot\hspace{-0.05cm}10^{-5}\%$)   & 2,325                 & 2,207                & 976                     & 83.12\%           \\ \midrule
\textbf{Gab}      & archive.is       & 6,602 ($3.4\hspace{-0.05cm}\cdot\hspace{-0.05cm}10^{-4}\%$)  & 5,943              & 5,773               & 1,300                 & 5.54\%            \\
\textbf{}           & Wayback          & 478 ($5.1\hspace{-0.05cm}\cdot\hspace{-0.05cm}10^{-5}\%$)   & 361                & 349               & 240                     & 61.18\%           \\ \midrule
\textbf{Twitter}    & archive.is       & 6,750 ($3.1\hspace{-0.05cm}\cdot\hspace{-0.05cm}10^{-6}\%$)   & 3,772                 & 3,669                & 845                     & 8.23\%            \\
\textbf{}           & Wayback          & 1,905 ($9.0\hspace{-0.05cm}\cdot\hspace{-0.05cm}10^{-7}\%$)   & 1,290                 & 1,257                & 846                     & 7.49\%            \\ \bottomrule
\end{tabular}
}
\caption{Overview of our datasets: number and percentage of posts that include archive URLs, unique number of archive URLs, 
source URLs, and source domains. We also filter URLs that are malformed, unreachable, or point to resources other than Web pages.}
\label{tab:dataset}
\end{table}

\descr{Archive URLs posted on social networks.} We search for \url{archive.is} and Wayback Machine URLs on Twitter, Reddit, and \dspol, 
between Jul 1, 2016 and Aug 31, 2017, and on Gab between Aug 1, 2016--Aug 31, 2017.
We obtain the 4chan dataset from the authors of~\cite{hine2016longitudinal}, the Reddit one from \url{pushshift.io}, while, for Twitter, we rely 
on the 1\% Streaming API.\footnote{\url{https://dev.twitter.com/streaming/overview}}
For Gab, we use a snowball sampling by collecting popular users returned by Gab's API, and iteratively collecting posts for all 
their followers and users they follow.

Overall, the resulting dataset includes 50K posts from \dspol, 528K posts from Reddit, 7K posts from Gab, and about 9K tweets.
Note that we have some gaps due to failure of our data collection infrastructure, specifically, there are 70 and 13 days missing for
Twitter and \dspol, respectively.

\descr{Basic Statistics.} In Table~\ref{tab:dataset}, we report statistics from our \url{archive.is} live-feed crawl as well as the 
crawl of \url{archive.is} and Wayback Machine URLs shared on Twitter, Reddit, \dspol, and Gab. We report the number of posts with archive URLs, 
along with the percentage over the total number of posts, as well as the number of unique archive URLs, unique source URLs, unique source domains, 
and the percentage of URLs that are filtered out. %
Specifically, besides malformed URLs, we exclude, for \url{archive.is}, URLs unreachable between Aug 29 and Oct 7, 2017, while for 
Wayback Machine those pointing to types of information other than Web pages (e.g., images, videos, software, etc.).

Overall, \dspol and Gab users often share Wayback Machine URLs that point to non-Web pages: around 83\% and 61\% of the total, respectively,
suggesting that \url{archive.is} is used mostly for the dissemination of Web pages,
while Wayback Machine is preferred for other content.
Also, a high percentage of malformed \url{archive.is} URLs are shared on Reddit (35\%), due to bots trying to pro-actively archive resources but failing.
From the normalized percentages, we observe that Twitter users rarely share URLs from archiving services, while Reddit users do so from both archiving services.
On \dspol and Gab, we find 15 and 16 times, respectively, more \url{archive.is} URLs than Wayback Machine 
ones.

\subsection{Cross-Platform Analysis}
\label{sec:analysis}

In this section, we present a cross-platform analysis of archive URLs collected from the \url{archive.is} live feed, as
well as Wayback Machine and \url{archive.is} URLs shared on Twitter, Reddit, Gab, and \dspol.
We focus on understanding what kind of content gets archived, as well as the related temporal characteristics, and on 
assessing whether archived content is still available from the source.

\begin{figure*}[t]
\center
\subfigure[archive.is live feed]{\includegraphics[width=0.32\textwidth]{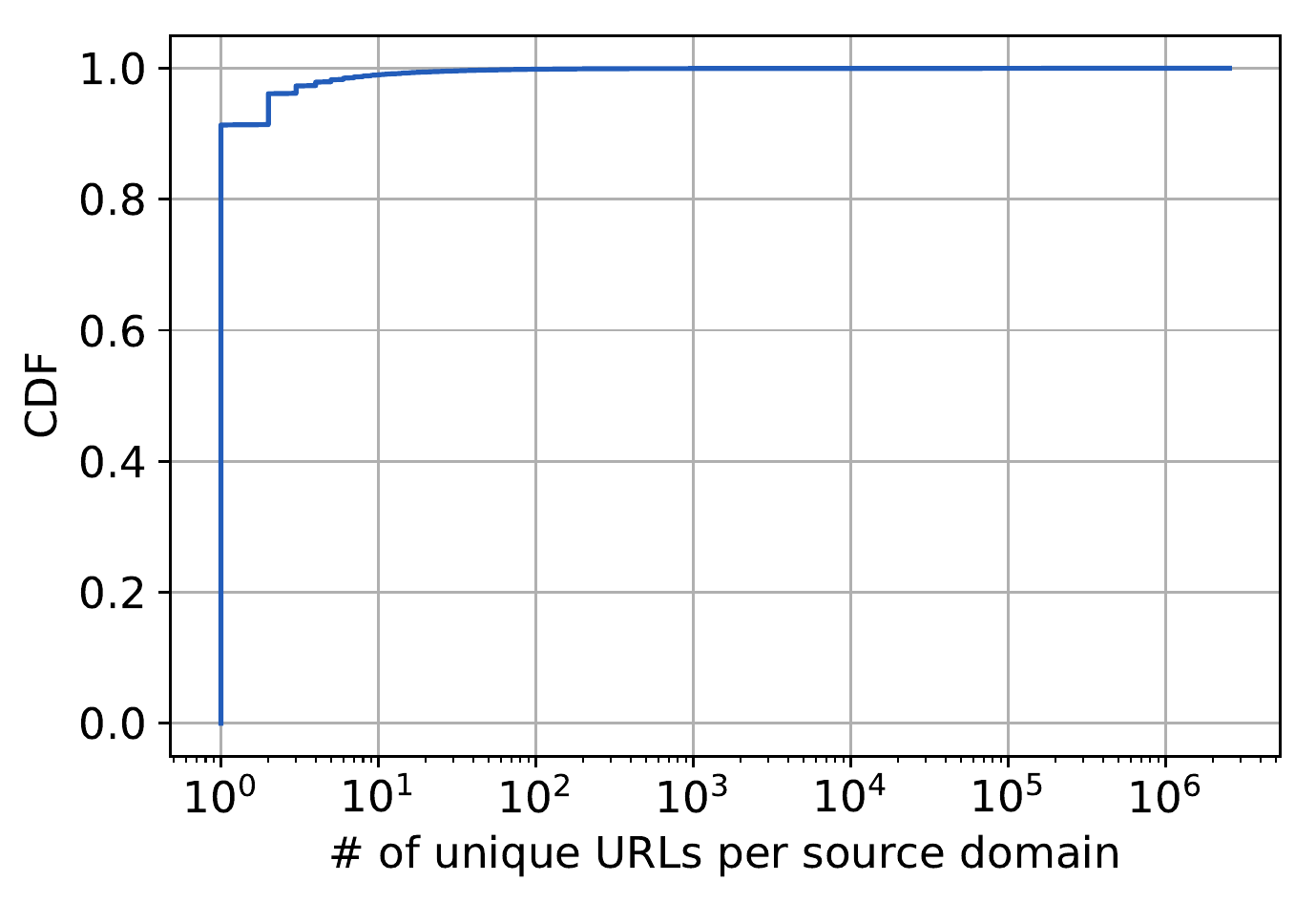}\label{fig:cdf_domain_archiveis_feed}}
\subfigure[Reddit]{\includegraphics[width=0.32\textwidth]{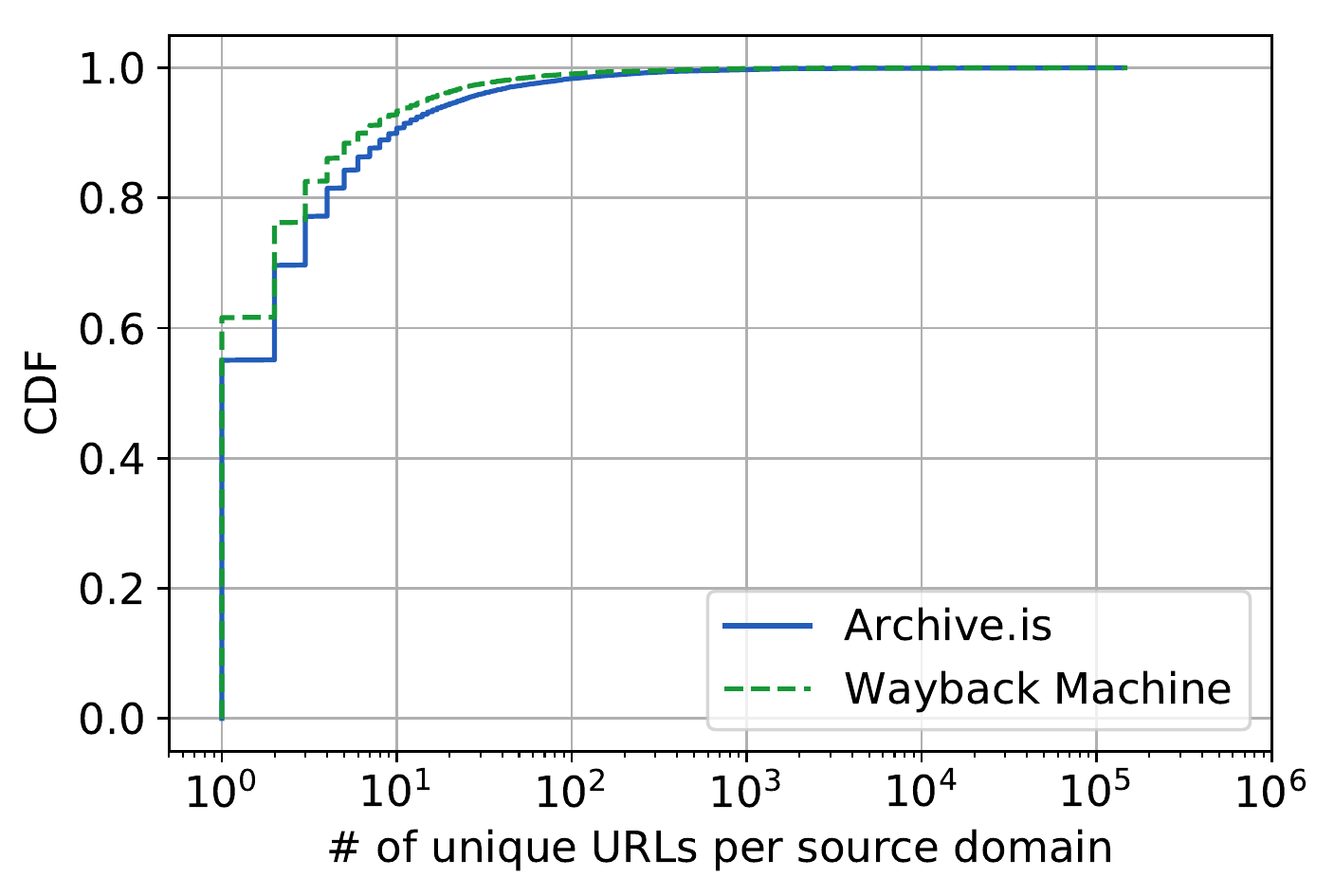}\label{subfig:cdf_domain_reddit}}
\subfigure[\dspol]{\includegraphics[width=0.32\textwidth]{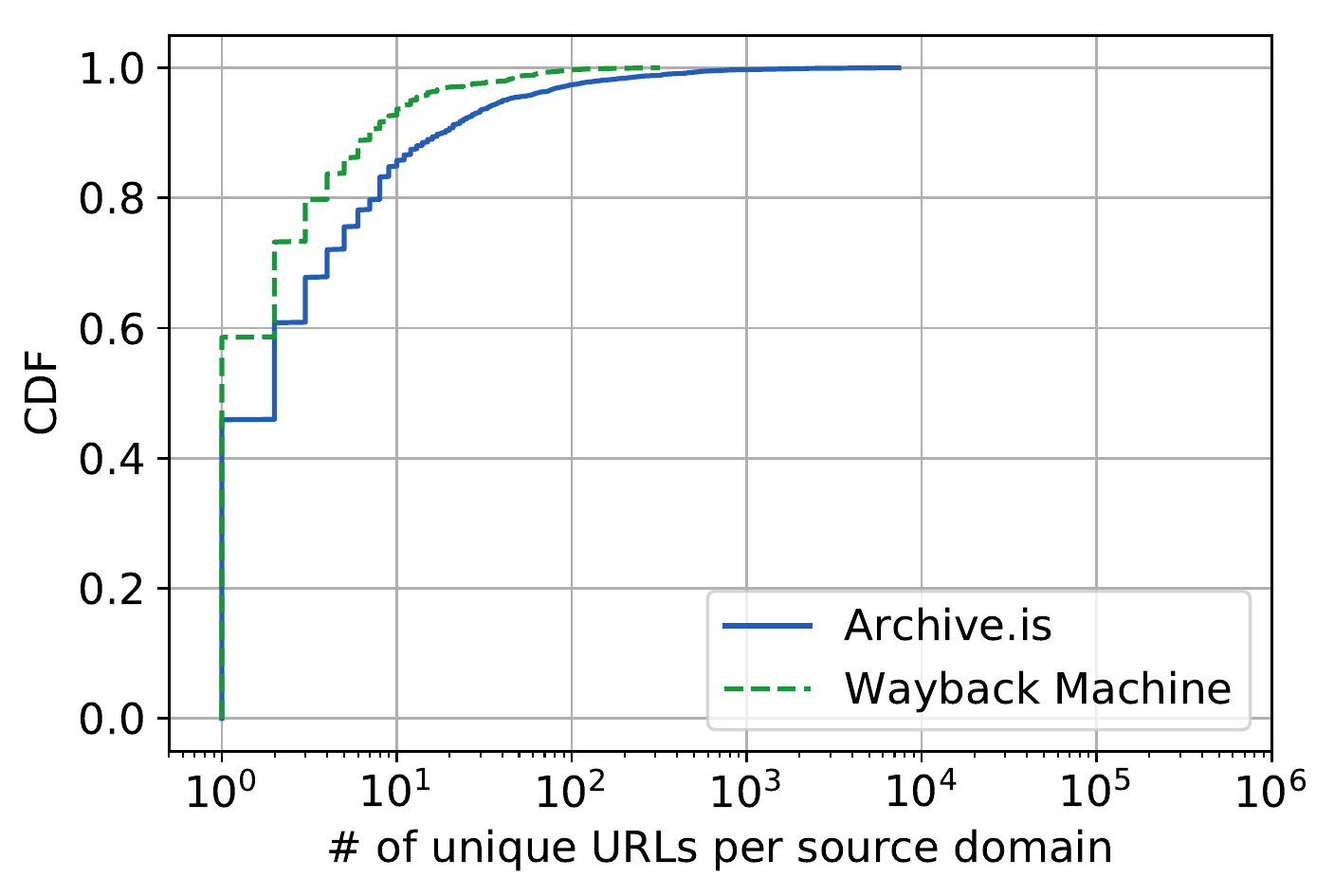}\label{subfig:cdf_domain_4chan}}
\subfigure[Twitter]{\includegraphics[width=0.32\textwidth]{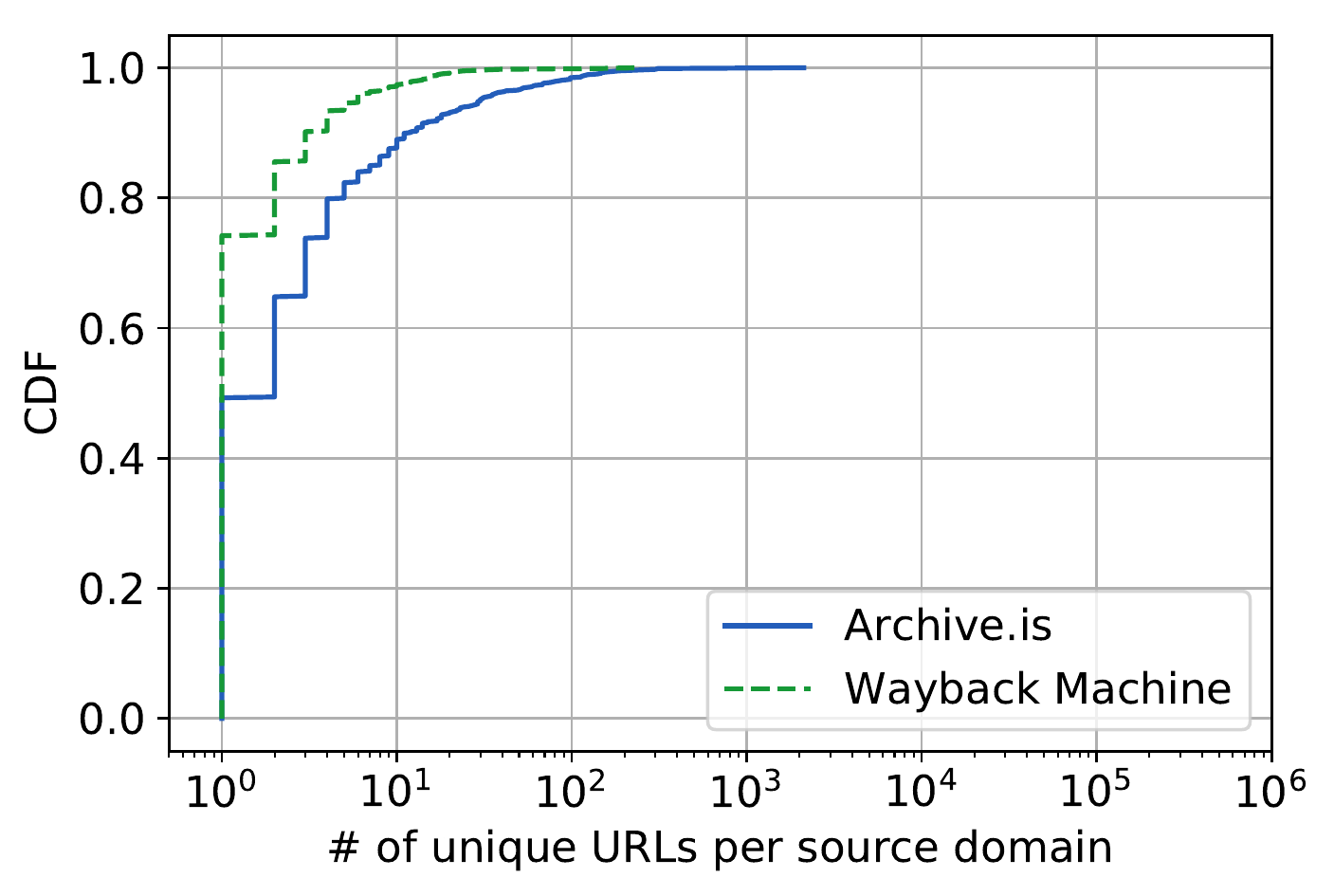}\label{subfig:cdf_domain_twitter}}
\subfigure[Gab]{\includegraphics[width=0.32\textwidth]{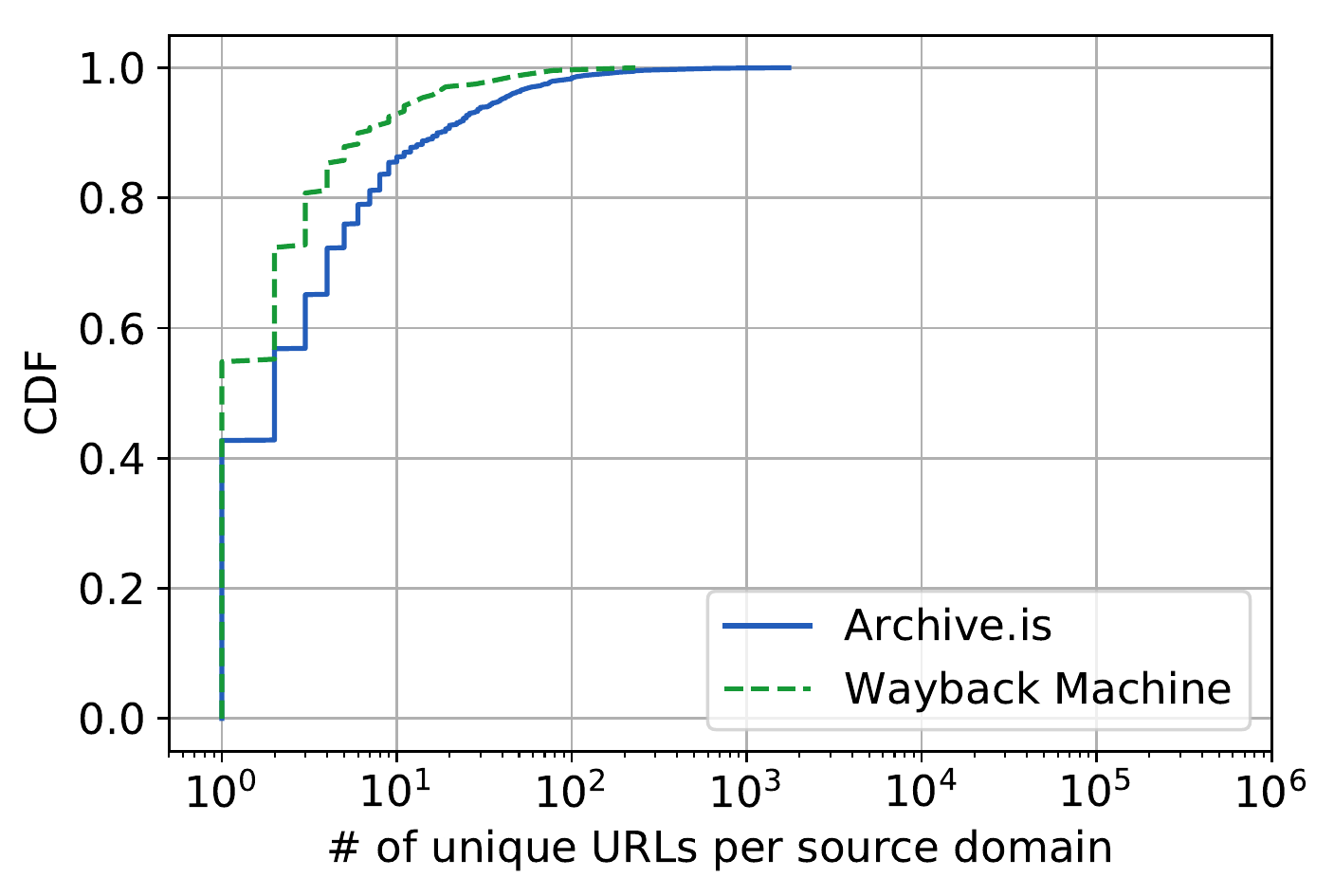}\label{subfig:cdf_domain_gab}}
\caption{CDF of the number of distinct URLs per source domain.}
\label{fig:cdf_domains}
\end{figure*}

\subsubsection{Source Domains}

\descr{Live Feed.} 
In Fig.~\ref{fig:cdf_domain_archiveis_feed}, we plot the CDF of the number of distinct URLs per domain in our \url{archive.is} live feed dataset.
The vast majority (90\%) of domains only appear once, while a few domains yield a large numbers of archive URLs -- e.g., there are 1.2M distinct 
\url{archive.is} URLs for which \url{twitter.com} is the source domain.
In Table~\ref{tbl:top_domains_archive.is}, we report the top 20 source domains as well as the top 20 domain suffixes (Sx). 
Surprisingly, the top domain (11.8\%) is actually the Wayback Machine's \url{archive.org}. %
Mainstream social networks like Twitter and Facebook are also included, likely due to their (perceived) ephemerality, i.e., users want to  preserve 
social network posts before they are removed or deleted.
As for the suffixes, we observe that common ones, such as .com and .org, are the majority, followed by domains from Germany 
(.de) and Japan (.jp) with 7\% and 5.6\% of the URLs, respectively. This suggests that a substantial portion of 
\url{archive.is}'s user base might be in Germany and Japan.

\begin{table}[t]
\centering
\footnotesize
\setlength{\tabcolsep}{0.25em} %
\resizebox{0.8\columnwidth}{!}{%
\begin{tabular}{lrrr@{\hskip 0.12cm}|@{\hskip 0.12cm}lrrr}
\toprule
\textbf{Domain} & {\bf (\%)} & \textbf{Sx} & \multicolumn{1}{r}{\bf(\%)\hspace*{0.12cm}} & \textbf{Domain} & {\bf (\%)} & \textbf{Sx} & \textbf{(\%)} \\ \midrule
\url{archive.org}     & 11.82\% & .com & 38.29\% & \url{ru-board.com}          & 0.50\% & .pl & 1.24\% \\
\url{twitter.com}     & 5.73\%  & .org & 17.64\% & \url{asstr.org}             & 0.49\% & .ch & 1.23\% \\
\url{quora.com}       & 3.18\%  & .de  & 7.02\%  & \url{ruliweb.com}           & 0.43\% & .eu & 1.01\% \\
\url{livejournal.com} & 2.17\%  & .jp  & 5.61\%  & \url{4chan.org}             & 0.40\% & .se & 0.80\% \\
\url{reddit.com}      & 1.81\%  & .net & 3.19\%  & \url{googleusercontent.com} & 0.40\% & .cz & 0.69\% \\
\url{facebook.com}    & 1.31\%  & .ru  & 3.10\%  & \url{ameblo.jp}             & 0.39\% & .br & 0.66\% \\
\url{nhk.or.jp}       & 0.78\%  & .nl  & 2.56\%  & \url{wordpress.com}         & 0.38\% & .at & 0.63\% \\
\url{youtube.com}     & 0.65\%  & .uk  & 1.51\%  & \url{yahoo.co.jp}           & 0.38\% & .es & 0.57\% \\
\url{wikipedia.org}   & 0.52\%  & .it  & 1.39\%  & \url{aaaaarg.fail}          & 0.37\% & .be & 0.55\% \\
\url{tumblr.com}      & 0.51\%  & .fr  & 1.39\%  & \url{blogspot.nl}           & 0.36\% & .ca & 0.51\% \\ \bottomrule
\end{tabular}
}
\caption{Top 20 domains and suffixes of the source URLs in the \url{archive.is} live feed dataset.}
\label{tbl:top_domains_archive.is}
\end{table}

\begin{table}[t]
\centering
\footnotesize
\resizebox{0.6\columnwidth}{!}{%
\setlength{\tabcolsep}{0.45em} %
\begin{tabular}{lrr|lrr}
\toprule
\textbf{Domain (\url{archive.is})} & \textbf{(\%)} & \multicolumn{1}{r}{\bf AF} &\textbf{Domain (Wayback)} & \textbf{(\%)} & {\bf AF} \\ \midrule
reddit.com          &      31.21\%  & {$<\hspace{-0.05cm}0.01$} & reddit.com  & 36.88\%    &  $<\hspace{-0.05cm}0.01$\\
pastebin.com       &         6.80\%  & {0.08}  & imgur.com  & 7.05\%      & $<\hspace{-0.05cm}0.01$\\
twitter.com           &     5.89\%  & {$<\hspace{-0.05cm}0.01$}  & twitter.com   &  5.19\%      & $<\hspace{-0.05cm}0.01$\\
imgur.com            &  3.02\% & {$<\hspace{-0.05cm}0.01$}  & redd.it   & 4.79\%        & $<\hspace{-0.05cm}0.01$\\
washingtonpost.com &            2.46\%    & {0.02}  & youtube.com  & 3.90\%   & $<\hspace{-0.05cm}0.01$      \\
youtube.com               & 2.33\%& {$<\hspace{-0.05cm}0.01$}  & washingtonpost.com   & 1.54\%     &   0.01  \\
redd.it                   & 2.14\%& {$<\hspace{-0.05cm}0.01$}  & youtu.be   & 1.19\%          & $<\hspace{-0.05cm}0.01$\\
nytimes.com            &     1.76\% & {0.01}  & nytimes.com   & 0.98\%   &    $<\hspace{-0.05cm}0.01$  \\
cnn.com                  & 1.64\%& {0.02}  & cnn.com   & 0.90\%         &  $<\hspace{-0.05cm}0.01$\\
wikipedia.org               & 1.37\%  & {$<\hspace{-0.05cm}0.01$}  & reddituploads.com   & 0.89\%   & 0.06       \\
huffingtonpost.com        &      0.93\%  & {0.02}  & archive.is   & 0.61\%          & $<\hspace{-0.05cm}0.01$\\
theguardian.com              &    0.78\% & {$<\hspace{-0.05cm}0.01$}  & streamable.com   & 0.61\%    &  $<\hspace{-0.05cm}0.01$      \\
googleusercontent.com      &      0.65\%      & {0.08}  & thehill.com   & 0.54\%    &   0.01     \\
politico.com                   & 0.64\%& {0.02}  & wikipedia.org   & 0.52\%           &  $<\hspace{-0.05cm}0.01$\\
wsj.com                  & 0.61\%& {0.03}  & politico.com   & 0.49\%           &  0.02\\
dailymail.co.uk       & 0.54\%& {0.01}  & theguardian.com   & 0.46\%       &    $<\hspace{-0.05cm}0.01$ \\
4chan.org                 &0.53\% & {0.16}  & rawstory.com   & 0.45\%          &  0.06\\
facebook.com            &      0.52\%  & {$<\hspace{-0.05cm}0.01$}  & huffingtonpost.com   & 0.44\%  &  $<\hspace{-0.05cm}0.01$        \\
thehill.com                &  0.43\%& {0.01}  & bbc.com   & 0.44\%       &  0.01\\
breitbart.com             & 0.40\%  & {0.01}  & kickstarter.com   & 0.37\%   &     0.02           \\ \bottomrule
\end{tabular}
}
\caption{Top 20 source domains of \url{archive.is} and Wayback Machine URLs, and archival fraction (AF), in the Reddit dataset.}
\reduce
\label{tbl:top_domains_reddit}
\end{table}
\begin{table}[t]
\centering
\footnotesize
\setlength{\tabcolsep}{0.5em} %
\resizebox{0.6\columnwidth}{!}{%
\begin{tabular}{lrr|lrr}
\toprule
\textbf{Domain (\url{archive.is})} & \textbf{(\%)} & \multicolumn{1}{r}{\bf AF} & \textbf{Domain (Wayback)} & \textbf{(\%)} & \textbf{AF} \\ \midrule
4chan.org                & 9.35\% & \multicolumn{1}{r|}{0.54} & justice4germans.com  & 7.50\%  & \textbf{0.94}    \\
theguardian.com          & 3.78\%       & \multicolumn{1}{r|}{0.13}  & chetlyzarko.com  & 3.90\% & \textbf{1.00}    \\
washingtonpost.com       & 3.70\%         & \multicolumn{1}{r|}{0.20}  & twitter.com   &  2.82\% & $<\hspace{-0.05cm}0.01$     \\
nytimes.com             & 3.46\% & \multicolumn{1}{r|}{0.16}  & dailymail.co.uk   & 2.47\%      & $<\hspace{-0.05cm}0.01$ \\
cnn.com                 & 2.78\% & \multicolumn{1}{r|}{0.14}  & revcom.us  & 2.16\%          & 0.66 \\
twitter.com             &  2.75\%   & \multicolumn{1}{r|}{0.01}  & reddit.com   & 1.98\%       & $<\hspace{-0.05cm}0.01$   \\
independent.co.uk       &  2.37\%           & \multicolumn{1}{r|}{0.13}  & tumblr.com   & 1.85\%  & 0.02        \\
breitbart.com            & 1.96\%     & \multicolumn{1}{r|}{0.08}  & thebilzerianreport.com   & 1.57\% & 0.72         \\
reddit.com               & 1.85\%  & \multicolumn{1}{r|}{0.09}  & jeffreyepsteinscience.com   & 1.55\%  & \textbf{1.00}        \\
dailymail.co.uk          & 1.72\%       & \multicolumn{1}{r|}{0.05}  & cnn.com   & 1.51\%        & $<\hspace{-0.05cm}0.01$   \\
facebook.com             & 1.69\%  & \multicolumn{1}{r|}{\textbf{0.96}}  & tdbimg.com   & 1.43\%          & \textbf{1.00} \\
huffingtonpost.com       & 1.37\%           & \multicolumn{1}{r|}{0.20}  & huffingtonpost.com   & 1.43\%  & 0.01          \\
thehill.com              & 1.21\%   & \multicolumn{1}{r|}{0.16}  & metapedia.org   & 1.22\%          & 0.04  \\
politico.com             & 1.04\%      & \multicolumn{1}{r|}{0.13}  & nytimes.com   & 1.15\%         & $<\hspace{-0.05cm}0.01$   \\
bbc.com                  & 1.01\% & \multicolumn{1}{r|}{0.08}  & washingtonpost.com   & 1.11\%        & $<\hspace{-0.05cm}0.01$    \\
8ch.net       & 0.98\% & \multicolumn{1}{r|}{\textbf{1.00}}  & theguardian.com   & 1.08\%           &  $<\hspace{-0.05cm}0.01$\\
googleusercontent.com           &  0.91\%    & \multicolumn{1}{r|}{0.59}  & independent.co.uk   & 1.08\%  & $<\hspace{-0.05cm}0.01$          \\
hypothes.is                   & 0.87\% & \multicolumn{1}{r|}{\textbf{0.98}}  & wordpress.com   & 1.06\%           & $<\hspace{-0.05cm}0.01$ \\
telegraph.co.uk               &  0.85\% & \multicolumn{1}{r|}{0.03}  & idrsolutions.com   & 1.01\%      & 0.86   \\
theatlantic.com              &  0.81\% & \multicolumn{1}{r|}{0.24}  & wikileaks.com   & 1.01\%            & $<\hspace{-0.05cm}0.01$       \\ \bottomrule
\end{tabular}
}
\caption{Top 20 source domains of \url{archive.is} and Wayback Machine URLs, and archival fraction (AF), in the /pol/ dataset.}
\label{tbl:top_domains_4chan}
\reduce
\end{table}

\begin{table}[t!]
\centering
\footnotesize
\setlength{\tabcolsep}{0.45em} %
\resizebox{0.6\columnwidth}{!}{%
\begin{tabular}{lrr|lrr}
\toprule
\textbf{Domain (\url{archive.is})} & \textbf{(\%)} & \multicolumn{1}{r}{\bf AF} & \textbf{Domain (\url{Wayback})} & \textbf{(\%)} &\textbf{\bf AF} \\ \midrule
twitter.com                 & 25.02 \% & \multicolumn{1}{r|}{$<\hspace{-0.05cm}0.01$} & justpaste.it  & 11.90 \%    & 0.02\\
facebook.com              & 3.65 \%   & \multicolumn{1}{r|}{$<\hspace{-0.05cm}0.01$}  & twitter.com  & 6.90 \%    &  0.01\\
togetter.com                & 3.58 \% & \multicolumn{1}{r|}{$<\hspace{-0.05cm}0.01$}  & dailymail.co.uk   &  1.95 \%   &  0.13  \\
seesaa.net             & 2.97 \%& \multicolumn{1}{r|}{\textbf{0.91}}  & nikkansports.com   & 1.50 \%       &  0.18\\
justpaste.it             & 2.19 \%    & \multicolumn{1}{r|}{0.01}  & mikelofgren.net  & 1.20\%       &   \textbf{1.00}\\
yahoo.co.jp              & 2.03 \%  & \multicolumn{1}{r|}{0.21}  & blogspot.com   & 1.10\%         &  0.09\\
googleusercontent.com\hspace*{-0.2cm} & 1.77 \%                 & \multicolumn{1}{r|}{\textbf{0.98}}  & whitehouse.gov   & 1.05\%     & 0.02     \\
time.com                 & 1.75 \%& \multicolumn{1}{r|}{0.01}  & journalists-in-russia.org\hspace*{-0.2cm}   & 1.00\%         &  \textbf{1.00}\\
monjiro.net               & 1.66 \%  & \multicolumn{1}{r|}{0.51}  & pcdepot.co.jp   & 0.90\%         &  \textbf{0.90}\\
pastebin.com             &   1.45 \%  & \multicolumn{1}{r|}{0.04}  & rydon.co.uk   & 0.85\%         & \textbf{1.00}\\
google.com              & 1.39 \% & \multicolumn{1}{r|}{0.01}  & yeniakit.com.tr   & 0.85\%          &  0.16\\
jimin.jp                  & 1.35 \%& \multicolumn{1}{r|}{\textbf{0.95}}  & cdse.edu   & 0.75\%            & \textbf{0.93}\\
notepad.cc              & 1.33 \%   & \multicolumn{1}{r|}{0.47}  & tetsureki.com   & 0.75\%       &    \textbf{1.00} \\
ameblo.jp                 &  1.16 \%  & \multicolumn{1}{r|}{$<\hspace{-0.05cm}0.01$}  & donaldjtrump.com   & 0.75\%      & 0.04     \\
nhk.or.jp                  & 1.16 \% & \multicolumn{1}{r|}{0.33}  & reidreport.com   & 0.75\%            & \textbf{1.00}\\
magi.md         & 1.16 \% & \multicolumn{1}{r|}{0.49}  & ameblo.cjp   & 0.70\%           &  $<\hspace{-0.05cm}0.01$\\
opensecrets.org      &  1.05 \%         & \multicolumn{1}{r|}{0.67}  & jreast.co.jp   & 0.70\%   &    \textbf{0.93}     \\
fc2.com                   & 0.99 \% & \multicolumn{1}{r|}{0.27}  & eastandard.net   & 0.65\%       &    \textbf{1.00} \\
dailyshincho.jp          &    0.93 \%    & \multicolumn{1}{r|}{\textbf{0.94}}  & yahoo.co.jp   & 0.60\%      &   0.01\\
reddit.com              &  0.89 \% & \multicolumn{1}{r|}{0.03}  & livedoor.jp   & 0.60\%                &   0.07\\ \bottomrule
\end{tabular}
}
\caption{Top 20 source domains of \url{archive.is} and Wayback Machine URLs, and archival fraction (AF), in the Twitter dataset.}
\label{tbl:top_domains_twitter}
\bigskip
\centering
\footnotesize
\setlength{\tabcolsep}{0.5em} %
\resizebox{0.6\columnwidth}{!}{%
\begin{tabular}{lrr|lrr}
\toprule
\textbf{Domain (\url{archive.is})} & \textbf{(\%)} & \multicolumn{1}{r}{\bf AF} & \textbf{Domain (Wayback)} & \textbf{(\%)} & \textbf{AF} \\ \midrule
twitter.com                & 12.28\% & \multicolumn{1}{r|}{$<\hspace{-0.05cm}0.01$} & dailymail.co.uk  & 20.98\%  & $ < 0.01 $    \\
nytimes.com          & 4.71\%       & \multicolumn{1}{r|}{0.03}  & washingtonpost.com  & 7.08\% & $ 0.01 $    \\
washingtonpost.com       & 4.17\%         & \multicolumn{1}{r|}{0.03}  & infowars.com   &  5.54\% & $<\hspace{-0.05cm}0.01$     \\
reddit.com            & 3.10\% & \multicolumn{1}{r|}{0.03}  & brandenburg.de   & 4.35\%      & 0.10 \\
googleusercontent.com                 & 2.43\% & \multicolumn{1}{r|}{0.18}  & twitter.com  & 3.63\%          & $ < 0.01$ \\
breitbart.com             &  1.82\%   & \multicolumn{1}{r|}{$ < 0.01$}  & huffingtonpost.com   & 3.08\%       & $<\hspace{-0.05cm}0.01$   \\
cnn.com       &  1.63\%           & \multicolumn{1}{r|}{0.01}  & abcnews.go.com   & 2.54\%  & $ < 0.01 $        \\
4chan.org            & 1.59\%     & \multicolumn{1}{r|}{0.07}  & salon.com   & 1.72\% & 0.01         \\
dailymail.co.uk               & 1.44\%  & \multicolumn{1}{r|}{$<\hspace{-0.05cm}0.01$}  & alexa.com   & 1.63\%  & 0.03        \\
theguardian.com          & 1.29\%       & \multicolumn{1}{r|}{$ < 0.01$}  & news.com.au   & 1.54\%        & $<\hspace{-0.05cm}0.01$   \\
wsj.com             & 1.22\%  & \multicolumn{1}{r|}{0.01}  & tu-dortmunt.de   & 1.45\%          & 0.80 \\
bbc.com       & 1.15\%           & \multicolumn{1}{r|}{0.01}  & causes.com   & 1.27\%  & 0.50          \\
huffingtonpost.com              & 1.14\%   & \multicolumn{1}{r|}{0.03}  & vigilantcitizen.com   & 1.18\%          & 0.02  \\
google.com             & 1.01\%      & \multicolumn{1}{r|}{$ < 0.01$}  & reddit.com   & 1.08\%         & $<\hspace{-0.05cm}0.01$   \\
facebook.com                  & 0.92\% & \multicolumn{1}{r|}{$ < 0.01 $}  & sahra-wagenknecht.de   & 0.99\%        & 0.78   \\
latimes.com       & 0.85\% & \multicolumn{1}{r|}{0.01}  & quillette.com   & 0.99\%           &  0.02\\
yahoo.com           &  0.81\%    & \multicolumn{1}{r|}{$ < 0.01$}  & derwesten.de   & 0.99\%  & $<\hspace{-0.05cm}0.01$          \\
dailycaller.com                   & 0.77\% & \multicolumn{1}{r|}{$ < 0.01$}  & politico.com   & 0.91\%           & $<\hspace{-0.05cm}0.01$ \\
thehill.com               &  0.74\% & \multicolumn{1}{r|}{$ < 0.01$}  & mikelofgren.net   & 0.81\%      & \textbf{0.90}   \\
wikileaks.org              &  0.73\% & \multicolumn{1}{r|}{0.01}  & alexanderhiggins.com   & 0.81\%            & 0.02       \\ \bottomrule
\end{tabular}
}
\caption{Top 20 source domains of \url{archive.is} and Wayback Machine URLs, and archival fraction (AF), in the Gab dataset.}
\label{tbl:top_domains_gab}
\reduce
\end{table}

\descr{Social Networks. } In Figs~\ref{subfig:cdf_domain_reddit}--\ref{subfig:cdf_domain_gab}, we plot the CDF of the number of URLs for each source domain in each dataset, finding that over 40\% of the source domains only appear once. %
Wayback Machine generally archives more URLs per source domain than \url{archive.is}, although for Reddit the distributions are quite similar. %
Then, in Tables~\ref{tbl:top_domains_reddit}--\ref{tbl:top_domains_gab}, we report the top 20 source domains observed on each platform, along with their {\em archival fraction} (AF), i.e., the number of times a source domain appears in an archive over the total number of times it appears in the dataset (either archived or not).

On all platforms except for Gab, the most popular domain archived through \url{archive.is} is the platform itself; e.g., archives of tweets are the most shared ones on Twitter.
This also happens for  Wayback Machine URLs, but only on Reddit.
On Reddit, this may be due to meta-subreddits focused on the preservation and discussion of dramatic happenings, e.g., flame wars and intra-Reddit conflict, that would otherwise be lost when deleted by moderators after some time.
These meta-subreddits tend to make use of bots that automatically archive drama submitted by their members.

Overall, we notice a strong presence of both mainstream (e.g., Washington Post) and alternative (e.g., Breitbart) news sources archived and shared on Reddit, \dspol, and Gab. 
Moreover, on \dspol, \url{archive.is} is often used for links to \url{hypothes.is}, a service that lets users annotate news articles,
possibly due to the fact that \dspol users often ``unravel'' conspiracy theories by researching and commenting on news articles.
On Twitter, where the footprint of archive URLs is relatively low, we find a relatively large number of Japanese domains, which might possibly indicate a stronger presence of Japanese Twitter users relying on archives.

The AFs are quite low overall, implying that archiving services disseminate a small fraction of most domains. However, on \dspol, specific domains have extremely high AFs. For instance, we find that \url{facebook.com} (AF = 0.96) and \url{8ch.net} (AF = 1.0)
are marked as spam from \dspol, and posts including links to them are rejected, a phenomenon we refer to as \textit{platform-specific censorship}.
We manually analyze other domains with high AF values, specifically, \url{hypothes.is}, \url{chetlyzarko.com}, \url{tdbming.com}, \url{justice4germans.com}, and \url{jeffreyepsteinscience.com}, without finding evidence of censorship on \dspol.
There is also ``accidental'' censorship on \dspol: for instance, the Australian newspaper \url{smh.com.au}, is affected because of a substitution filter (used for fun), which replace one word with another, as the word ``smh'' is automatically replaced on \dspol\ with ``baka.''

\subsubsection{URL Characterization}
We now proceed to characterize the type of content archived.
To this end, we extract the domain categories of source URLs using the free Virus Total API (\url{virustotal.com}), %
which we choose since it consolidates categories from multiple services including Bit Defender, TrendMicro, Alexa, etc.
Although categorization is done at domain-level, results are presented at a per-URL level (a URL is assigned the same category as its domain) in order to capture the popularity of each domain in our datasets.

\descr{Live Feed.} Due to throttling enforced by the API, we are not able to categorize all the 20.6M source URLs in our \url{archive.is} live feed dataset. %
Therefore, we first aggregate URLs into their domain, then, we follow a sampling approach using: 1) the top 100K most popular domains in our dataset, which correspond to 15M (73\%) source URLs, and 2) a sample of 121K domains drawn according to their empirical distribution in our archive datasets, resulting in 1.4M (7\%) source URLs. 

\begin{figure}[t]
\center
\hspace{-0.2cm}
\subfigure[Top 100K Domains]{\includegraphics[width=0.495\columnwidth]{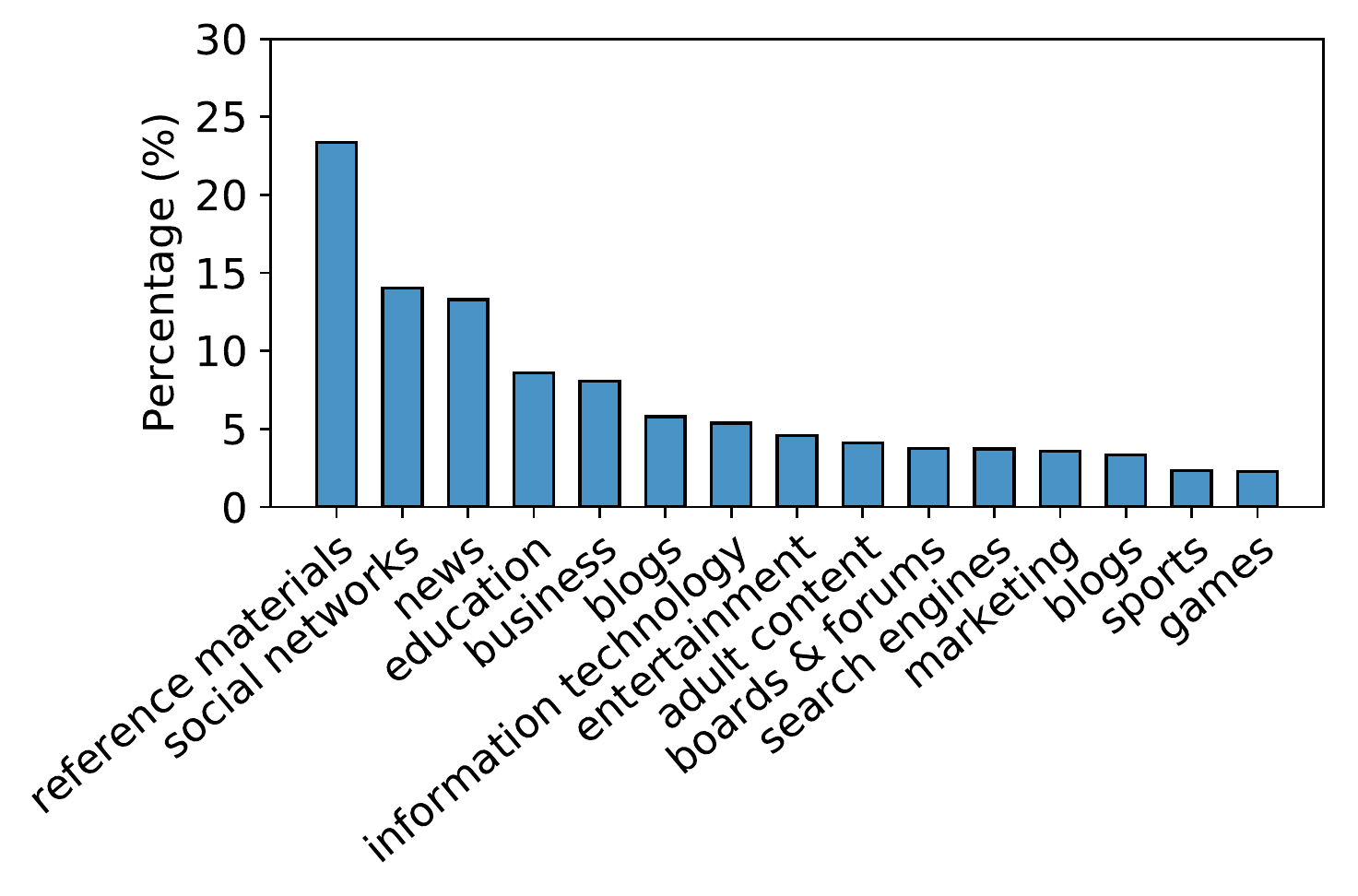}\label{subfig:bc_top_100}}
\hspace{-0.2cm}
\subfigure[Sample of 121K Domains]{\includegraphics[width=0.495\columnwidth]{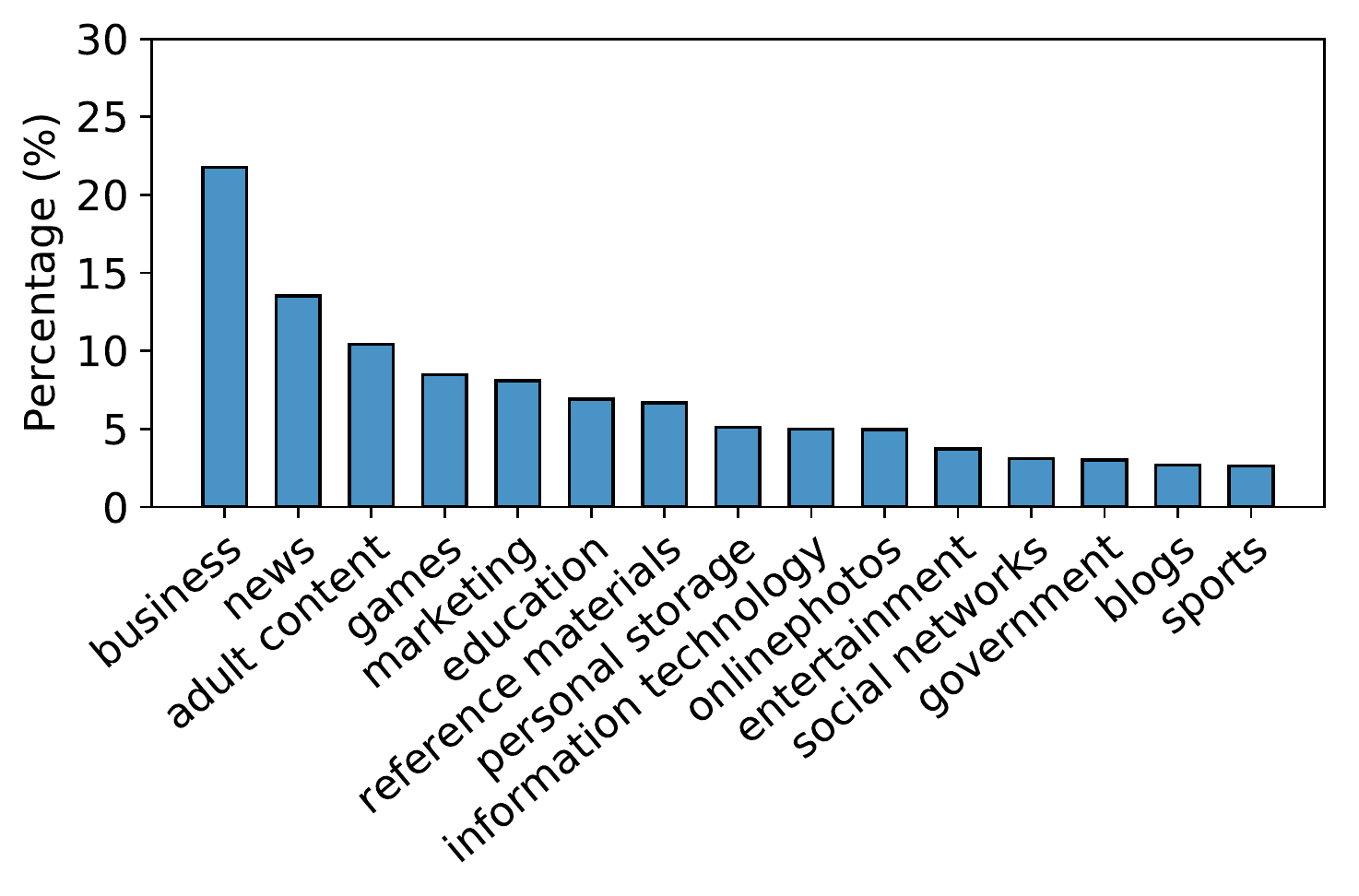}\label{subfig:bc_empirical}}
\caption{Top 15 domain categories for the \url{archive.is} live feed.}
\label{fig:bc_categorization_wild} \reduce
\end{figure}

In Fig.~\ref{fig:bc_categorization_wild}, we report the top 15 categories obtained from Virus Total for both samples.
Note that Virus Total is unable to provide a category for 1\% and 7\% of the URLs for the two sets of domains that we checked, respectively. 
From Fig.~\ref{subfig:bc_top_100}, we observe that the most popular category is Reference Materials (23\%), which is due to the fact that, as discussed earlier, many \url{archive.is} URLs archive Wayback Machine URLs.
Other popular categories include Social Networks (15\%), News Sources (14\%), Education (13\%), and Business (12\%).
Adult Content accounts for 4\% of source URLs.
Fig.~\ref{subfig:bc_empirical} shows that, for the empirically distributed sample, the top 15 categories are slightly different,
including Business (21\%), News (13\%), and Adult Content (12\%).

\begin{figure}[t]
\center
\hspace{-0.25cm}
\subfigure[Reddit]{\includegraphics[width=0.495\columnwidth]{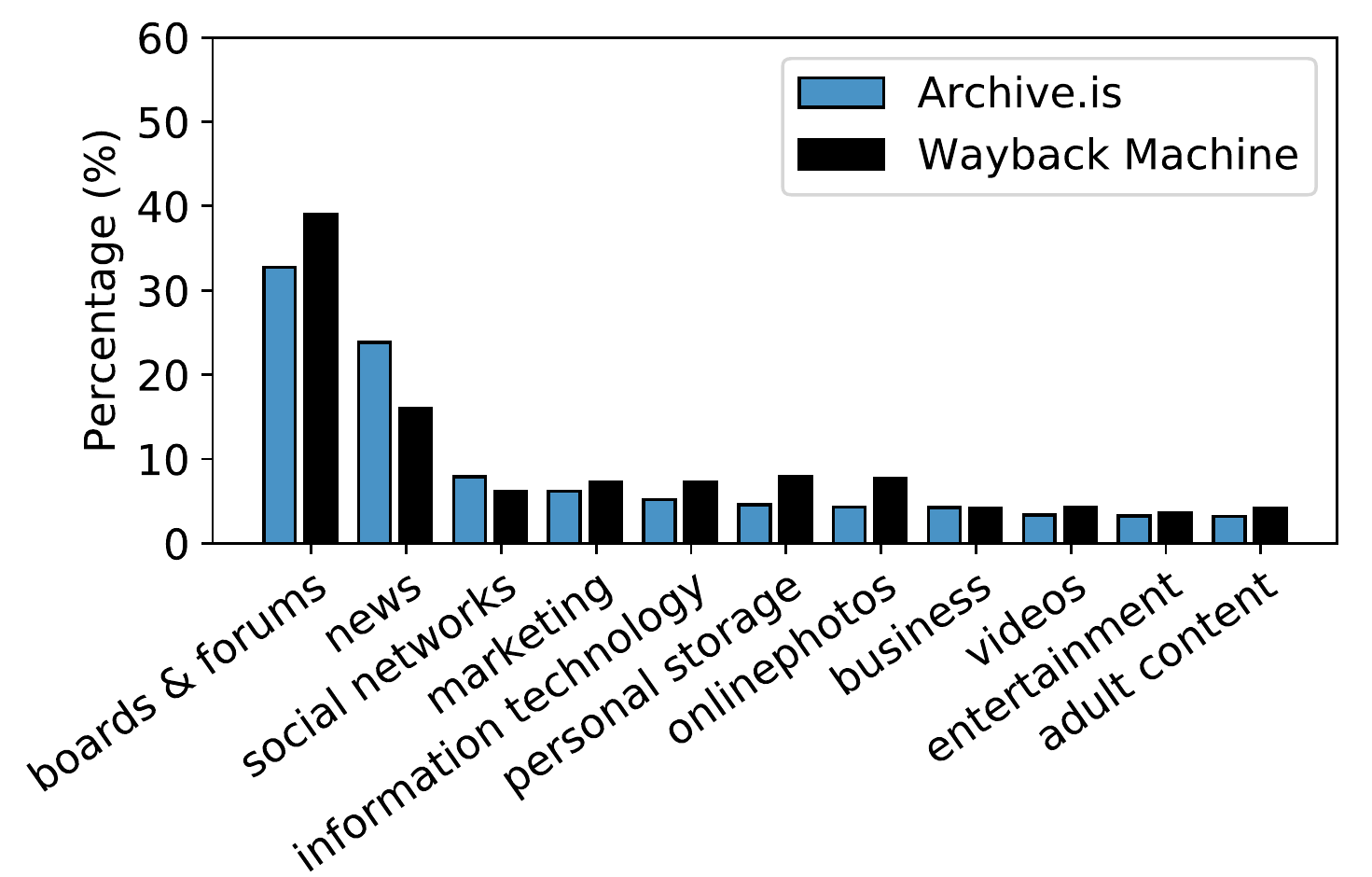}\label{subfig:bc_combined_reddit_virus}}
\hspace{-0.25cm}
\subfigure[Twitter]{\includegraphics[width=0.495\columnwidth]{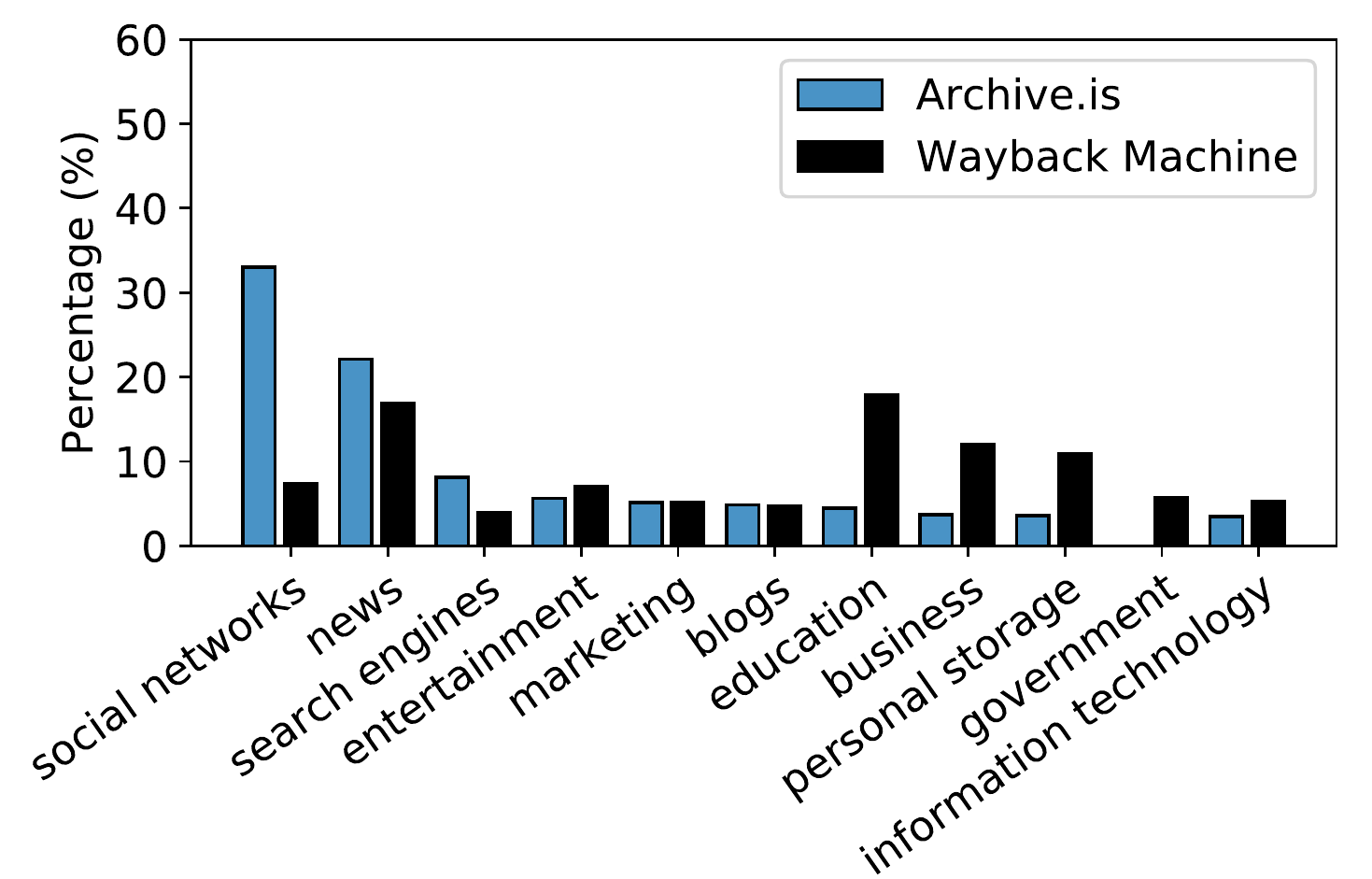}\label{subfig:bc_combined_twitter_virus}}\\
\hspace{-0.25cm}
\subfigure[\dspol]{\includegraphics[width=0.495\columnwidth]{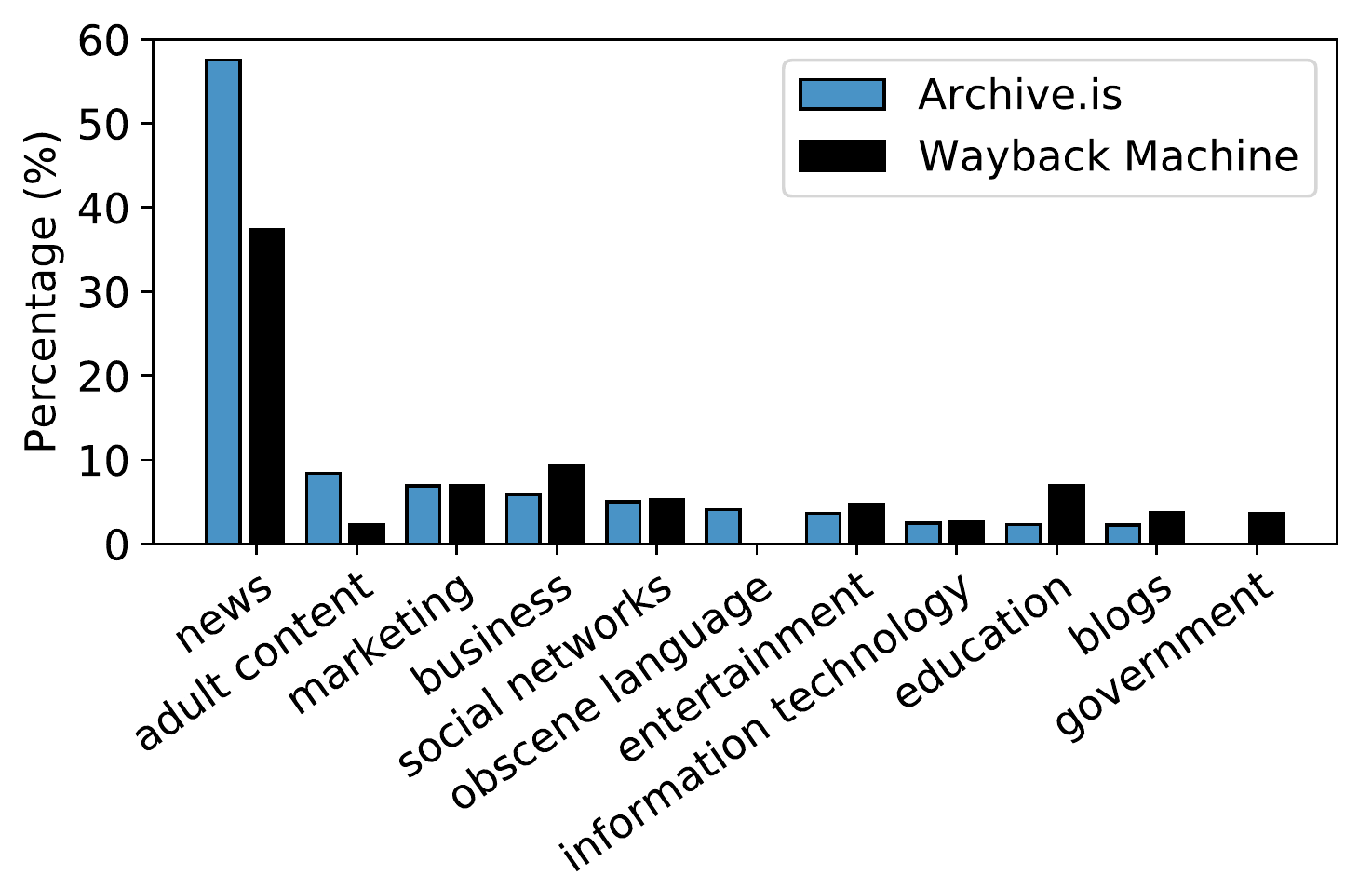}\label{subfig:bc_combined_4chan_virus}}
\hspace{-0.25cm}
\subfigure[Gab]{\includegraphics[width=0.495\columnwidth]{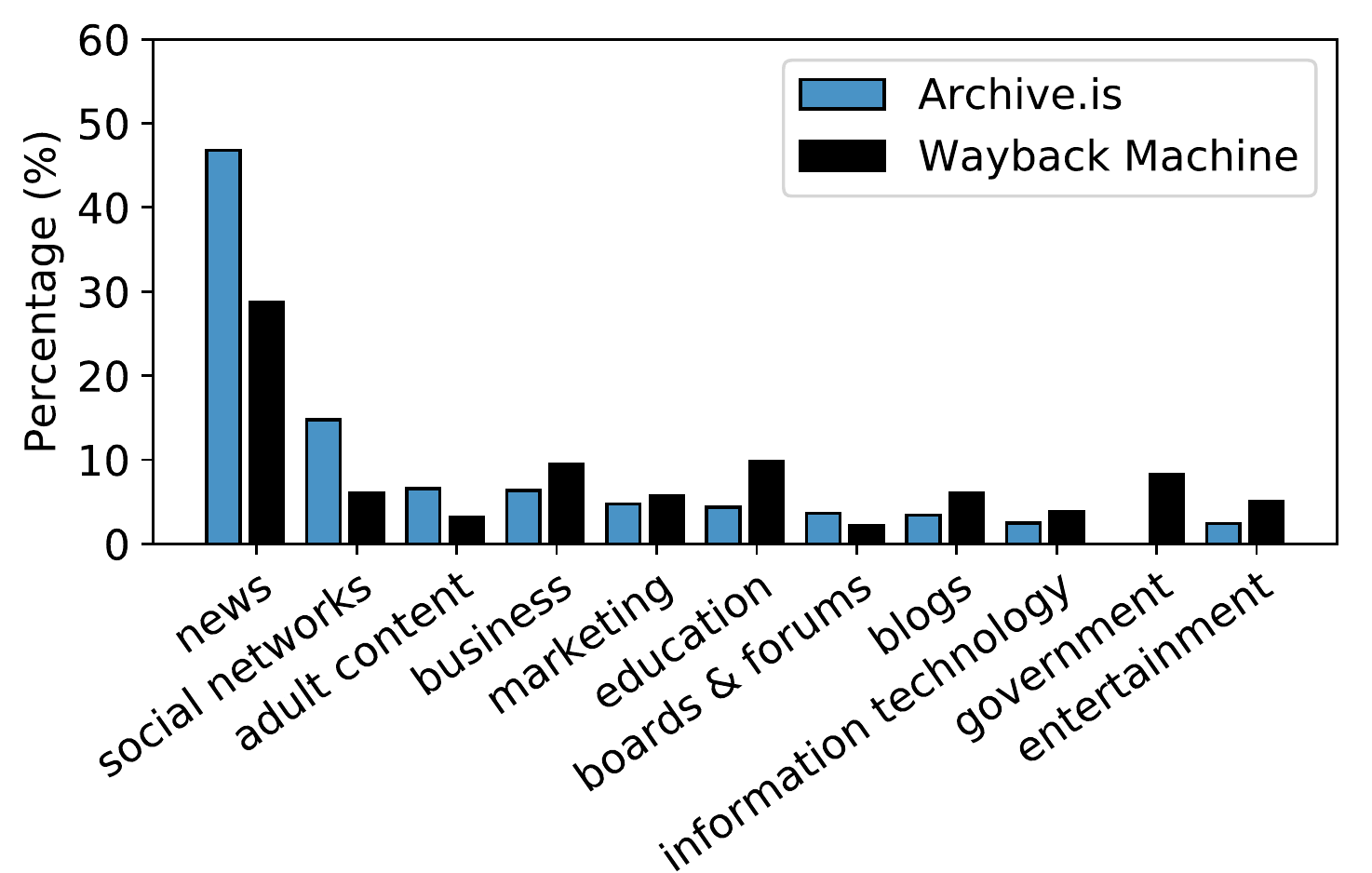}\label{subfig:bc_combined_gab_virus}}
\caption{Top domain categories for archive URLs appearing on the four social networks.}
\label{fig:top_categories}
\end{figure}

\descr{Social Networks.} Unlike the live feed dataset, we perform URL characterization for \textit{all} source URLs (aggregated by domain) found on Reddit, \dspol, Gab, and Twitter, again using the Virus Total API.
In Fig.~\ref{fig:top_categories}, we report the top categories and their corresponding percentages for both archiving services (specifically, the union of categories that appear in the top 10 categories for each service).
The Virus Total API is unable to provide a category for, on average, 1.5\% and 9\% of the \url{archive.is} and Wayback Machine 
URLs found on Reddit, \dspol, Gab, and Twitter, respectively.
Overall, both archiving services are often used to disseminate URLs from news sources, social networks, and marketing sites on all social networks.
However, there are interesting differences for the two archiving services:
Education and Government URLs appear as top categories for the Wayback Machine 
(see Fig.~\ref{subfig:bc_combined_twitter_virus},~\ref{subfig:bc_combined_4chan_virus}, and~\ref{subfig:bc_combined_gab_virus}), 
while sites that contain obscene language only for \url{archive.is} (see Fig.~\ref{subfig:bc_combined_4chan_virus}).
This suggests that the latter is used more extensively for ``questionable'' content.

Moreover, we observe that Adult Content is among the top categories for all social networks except Twitter,
while Gab and Reddit users often share archive URLs for domains related to Boards and Forums.
Also, on \dspol, \url{archive.is} is used to archive and disseminate pages with obscene language,
which is somewhat in line with previous observations~\cite{hine2016longitudinal} showing that  
\dspol conversations often include hate speech and aggressive behavior, and so \url{archive.is} URLs likely point to similar content.

\begin{figure}[t!]
\centering
\subfigure[Date]{\includegraphics[width=0.49\textwidth]{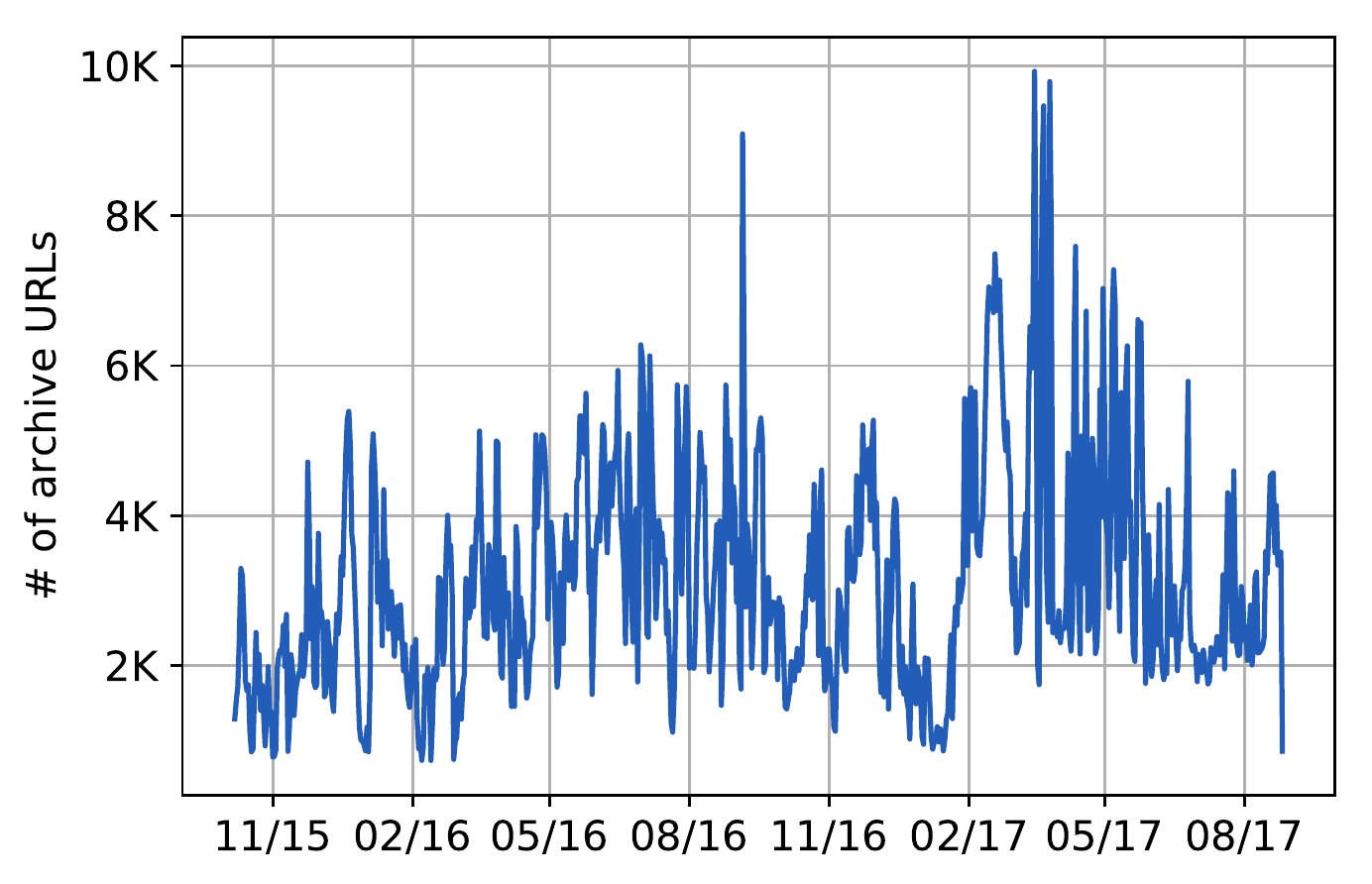}\label{subfig:counts_day}}
\subfigure[Hour of Day]{\includegraphics[width=0.49\textwidth]{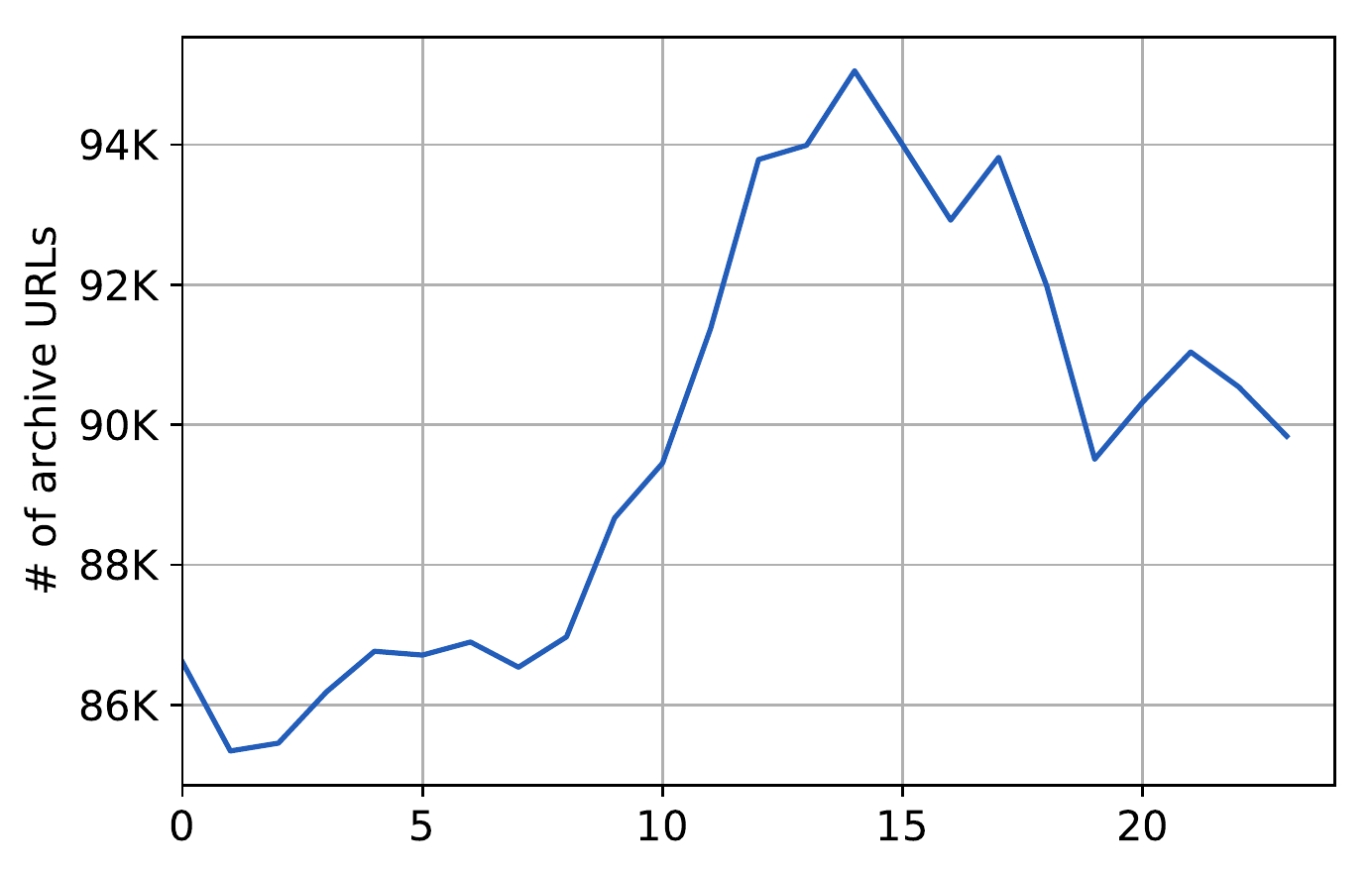}\label{subfig:counts_hour_day}}
\caption{Temporal analysis of the \url{archive.is} live feed dataset, reporting the number of URLs that are archived (a) each day and (b) based  on hour of day.}
\label{fig:temporal_analysis_capsules}
\end{figure}

\begin{figure*}[t]
\centering
\subfigure[archive.is]{\includegraphics[width=0.49\textwidth]{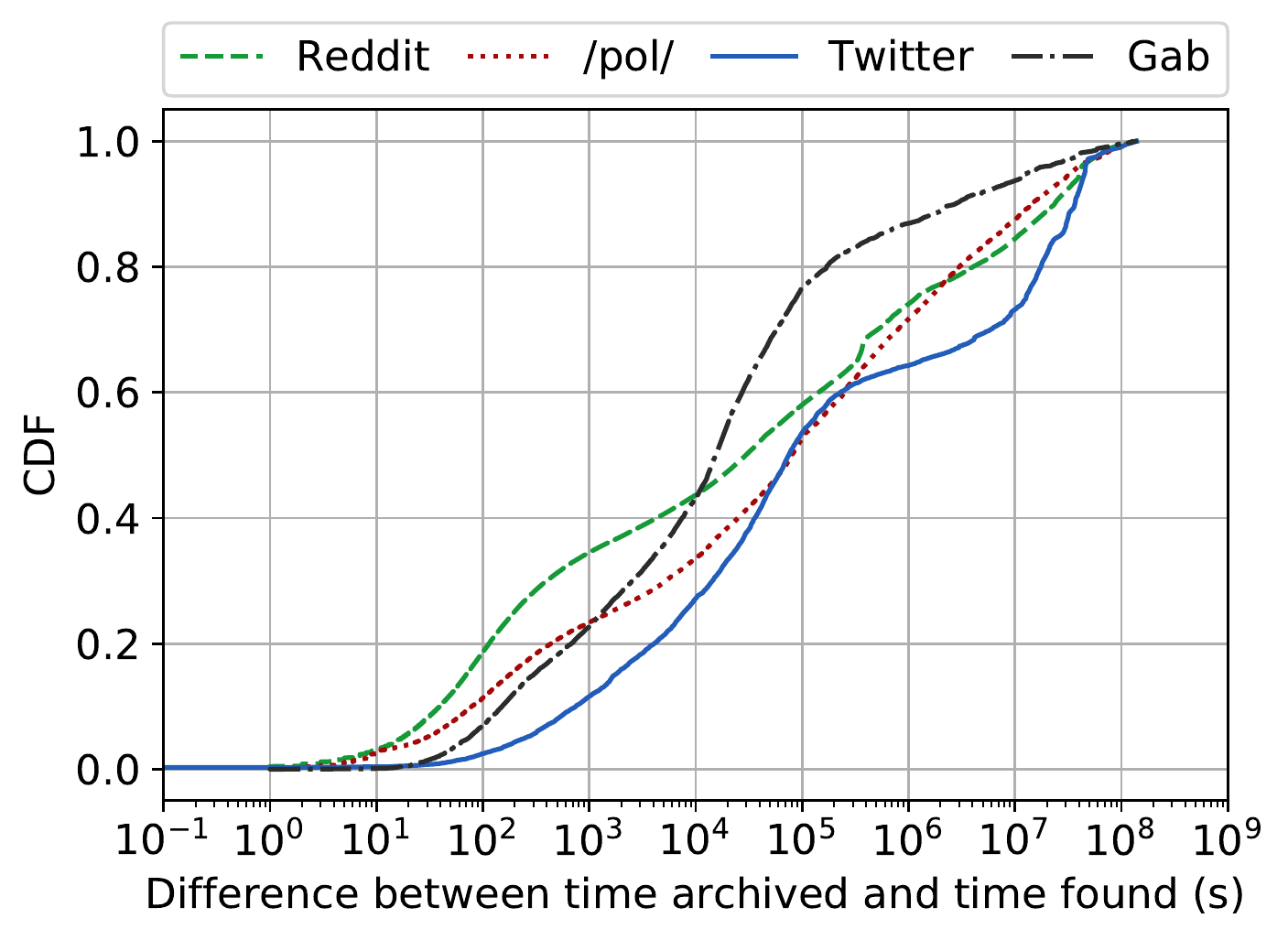}\label{subfig:cdf_time_difference_archive_is}}
\subfigure[Wayback Machine]{\includegraphics[width=0.49\textwidth]{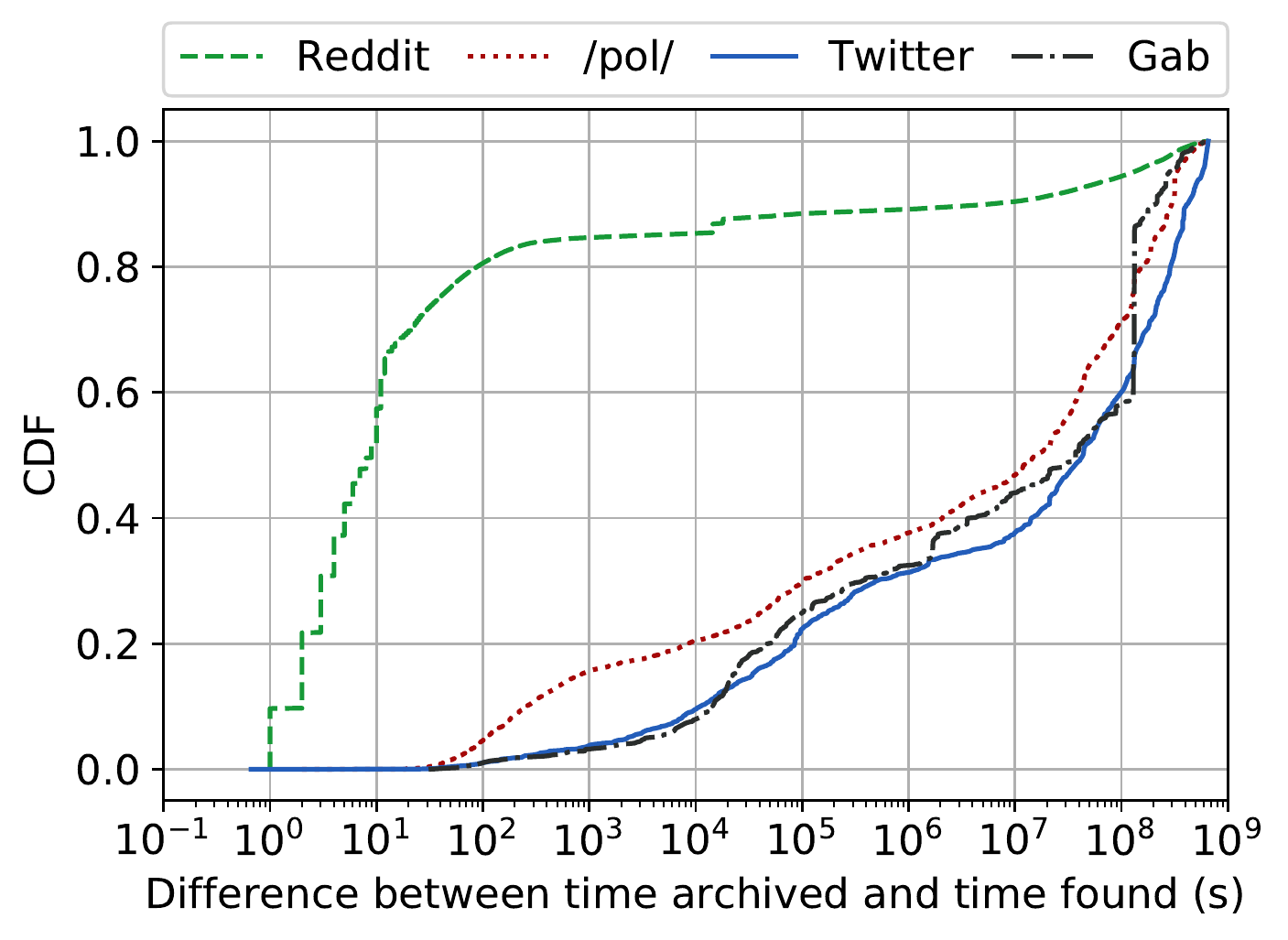}\label{subfig:cdf_time_difference_archive_org}}
\caption{CDF of the time difference between the archival time and the time appeared on each of the four platforms. (Note log scale on x-axis).}
\label{fig:cdf_time_difference}
\end{figure*}

\begin{figure*}[t]
\center
\subfigure[Reddit]{\includegraphics[width=0.24\textwidth]{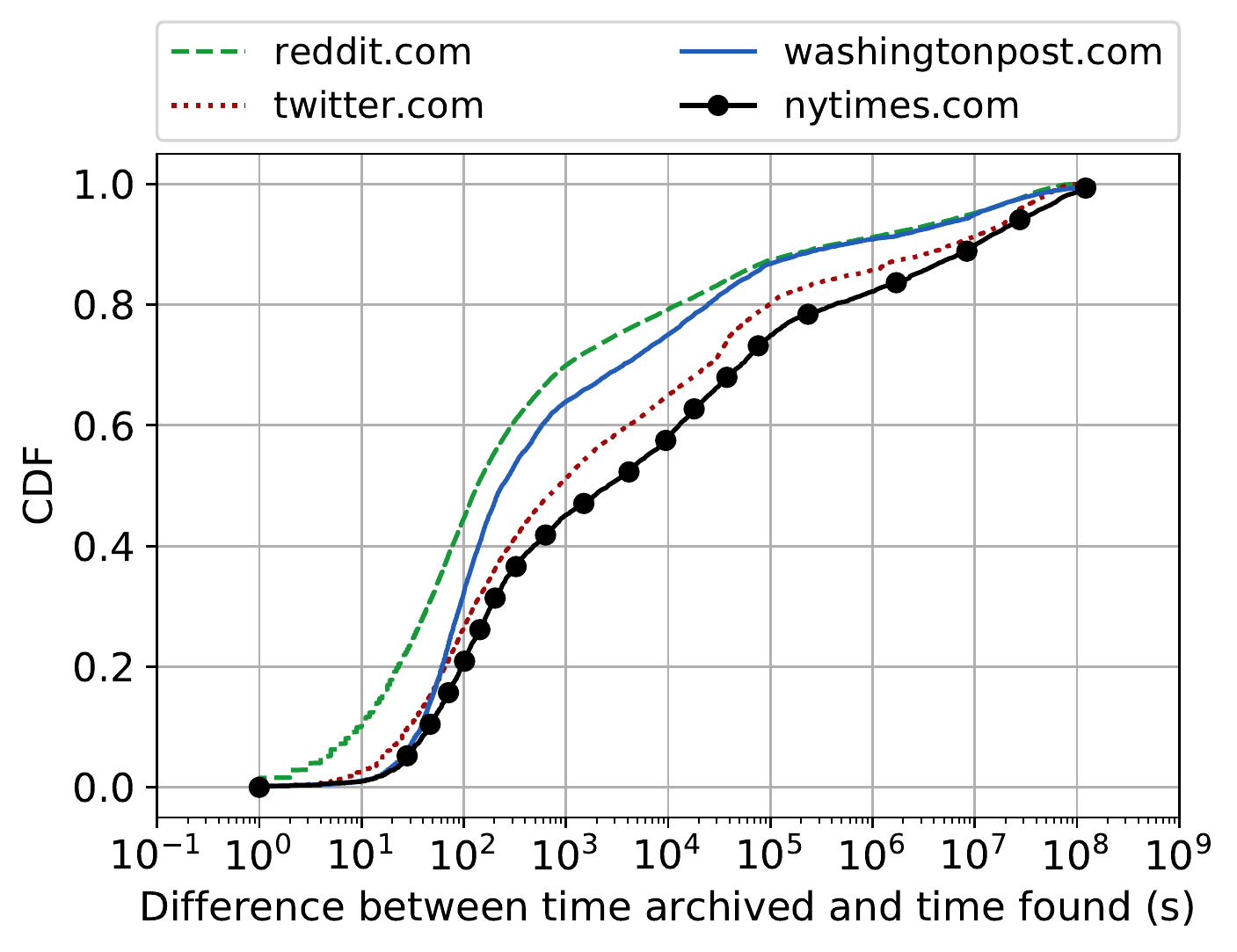}\label{subfig:cdf_time_difference_domain_archive_is_reddit}}
\subfigure[Twitter]{\includegraphics[width=0.24\textwidth]{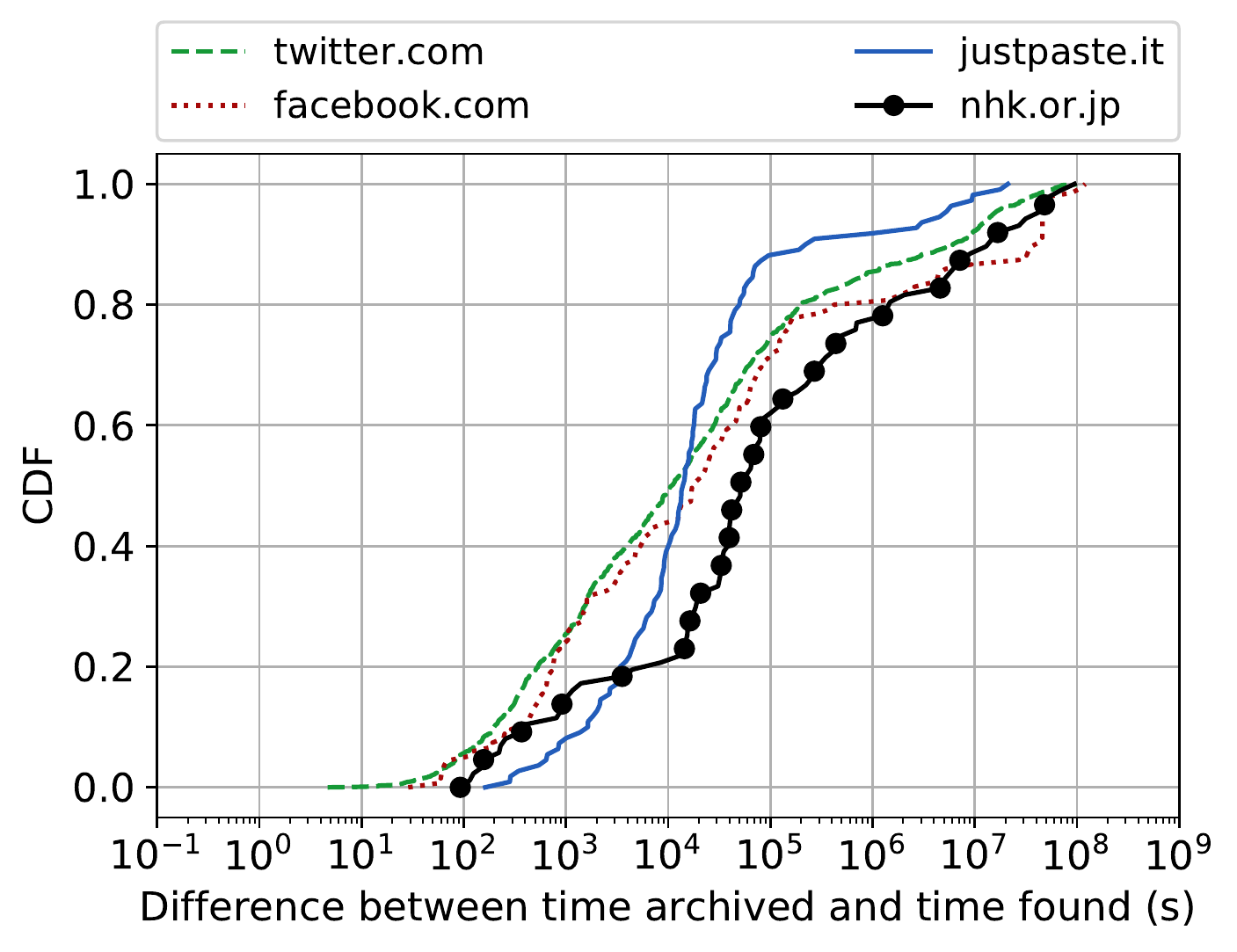}\label{subfig:cdf_time_difference_domain_archive_is_twitter}}
\subfigure[\dspol]{\includegraphics[width=0.24\textwidth]{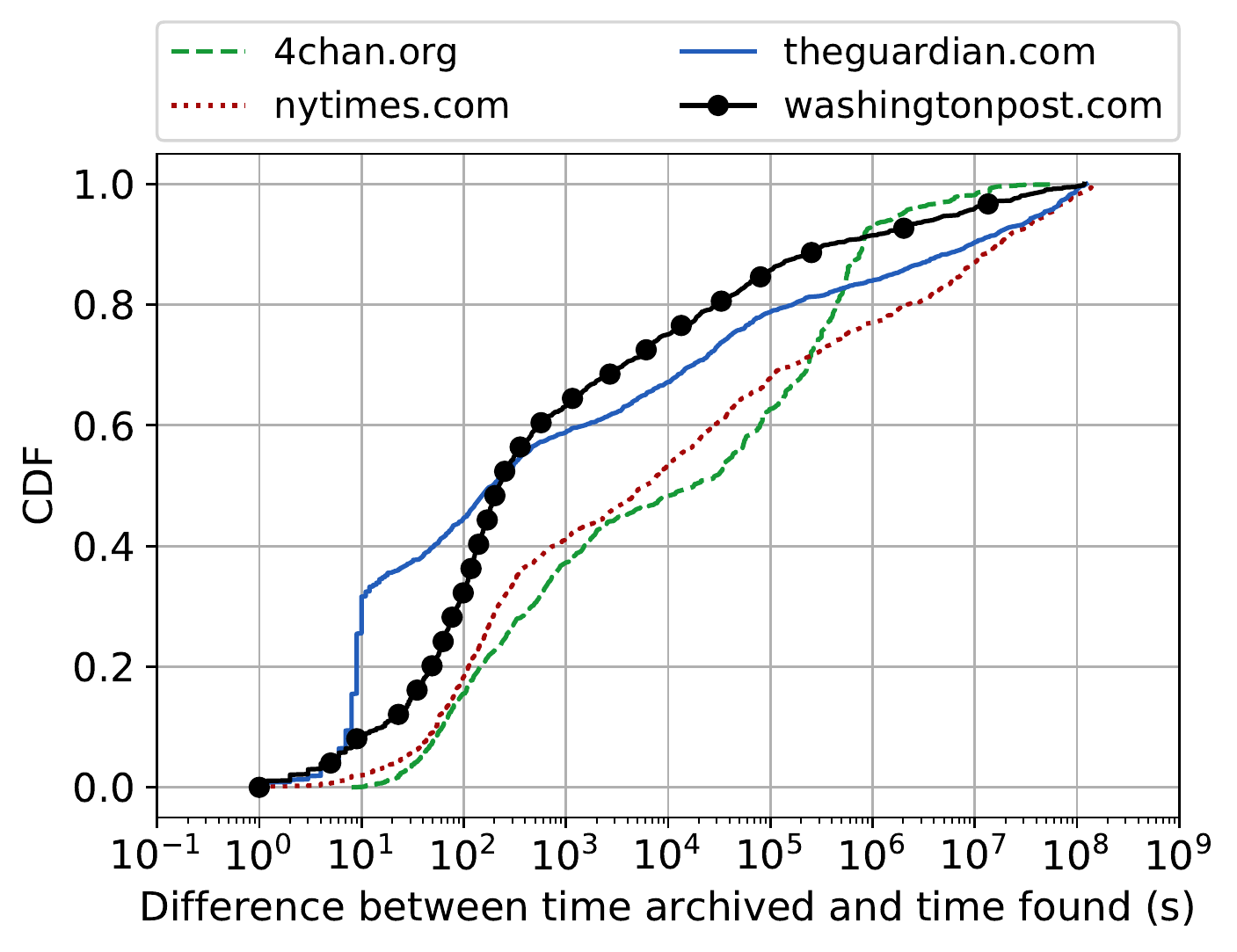}\label{subfig:cdf_time_difference_domain_archive_is_4chan}}
\subfigure[Gab]{\includegraphics[width=0.24\textwidth]{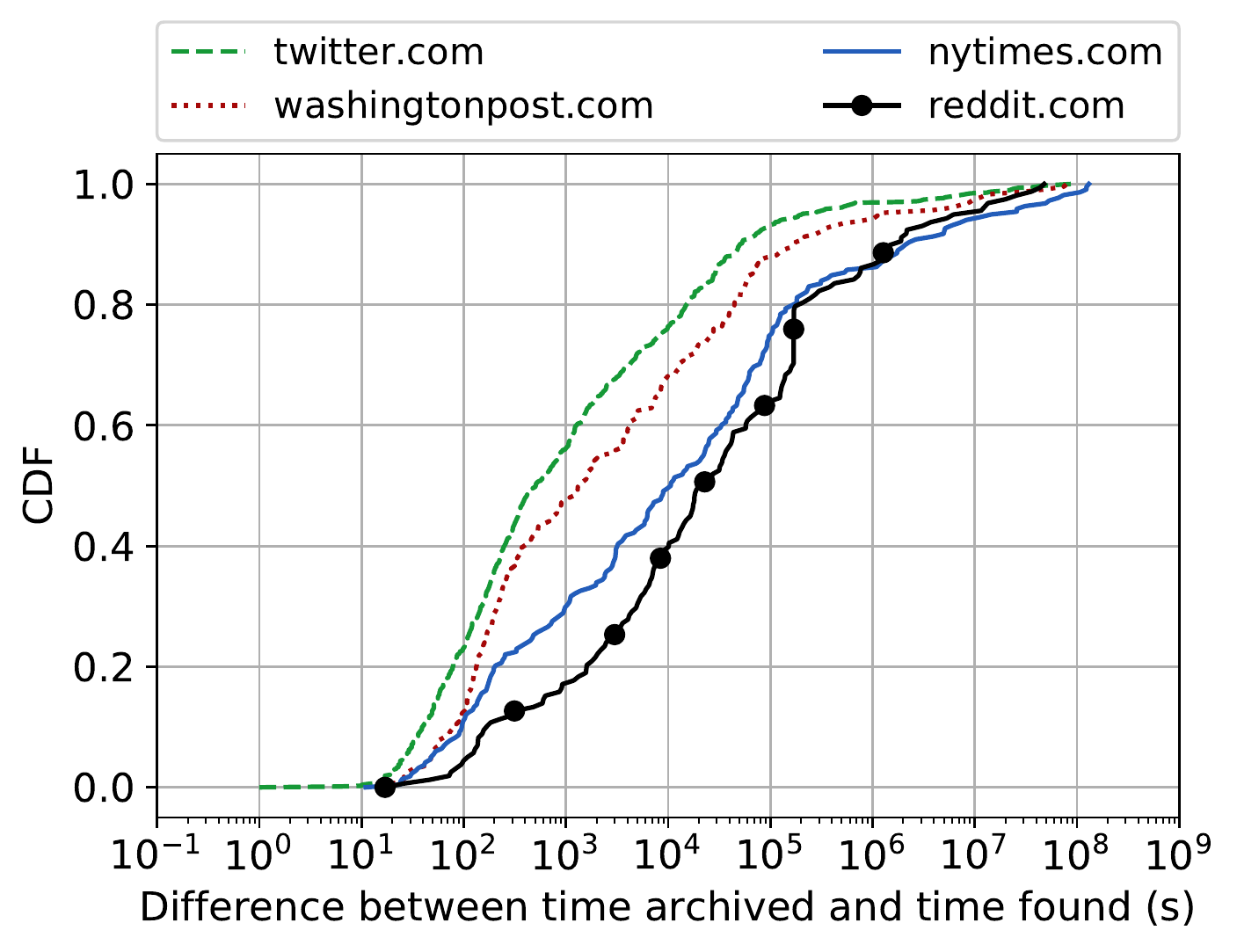}\label{subfig:cdf_time_difference_domain_archive_is_gab}}
\caption{CDF of the time difference between archival time on \url{archive.is} and appearance on social networks for the top four source domains.}
\label{fig:cdf_time_difference_domain_archive_is} 
\end{figure*}

\begin{figure*}[t]
\center
\subfigure[Reddit]{\includegraphics[width=0.24\textwidth]{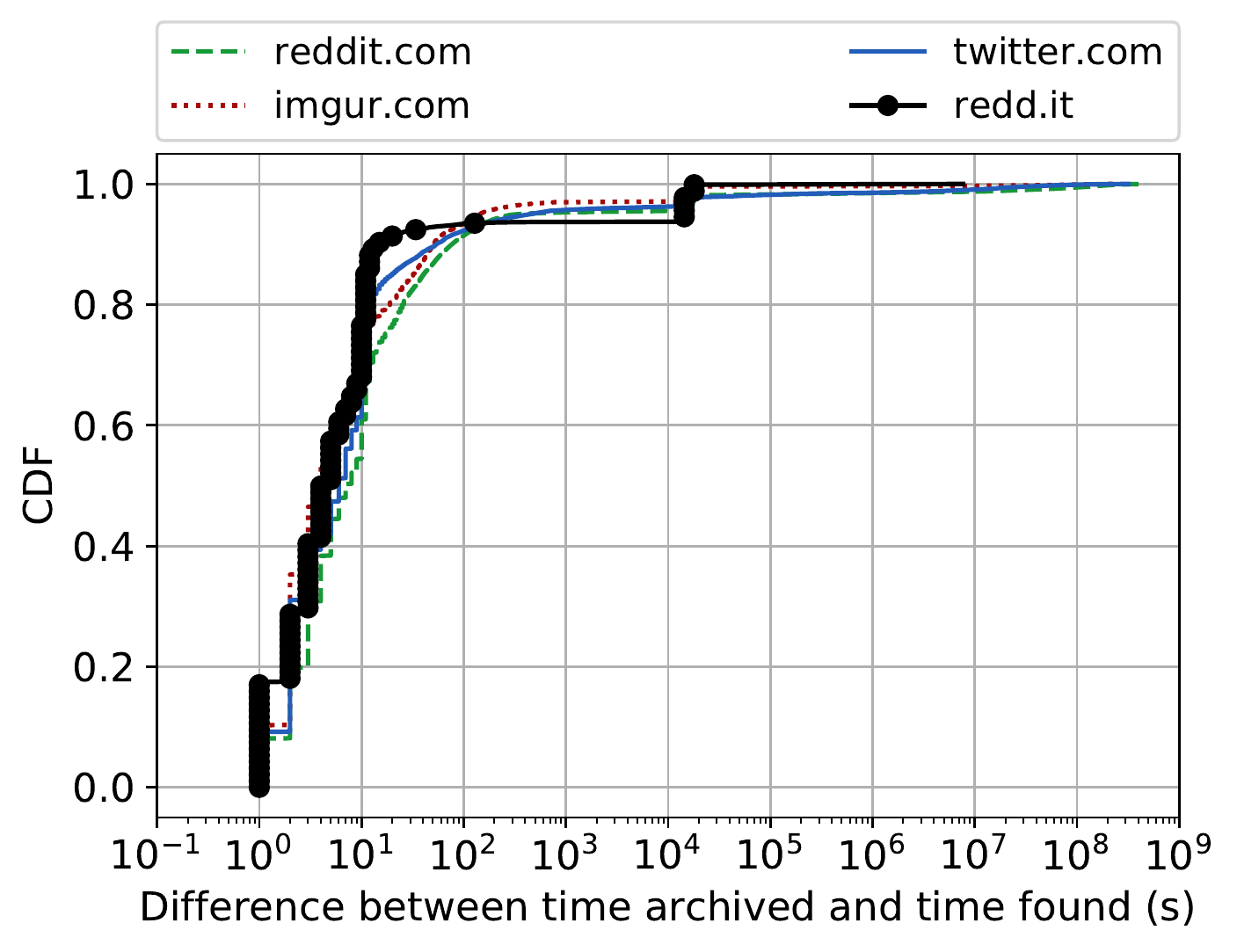}\label{subfig:cdf_time_difference_domain_archive_org_reddit}}
\subfigure[Twitter]{\includegraphics[width=0.24\textwidth]{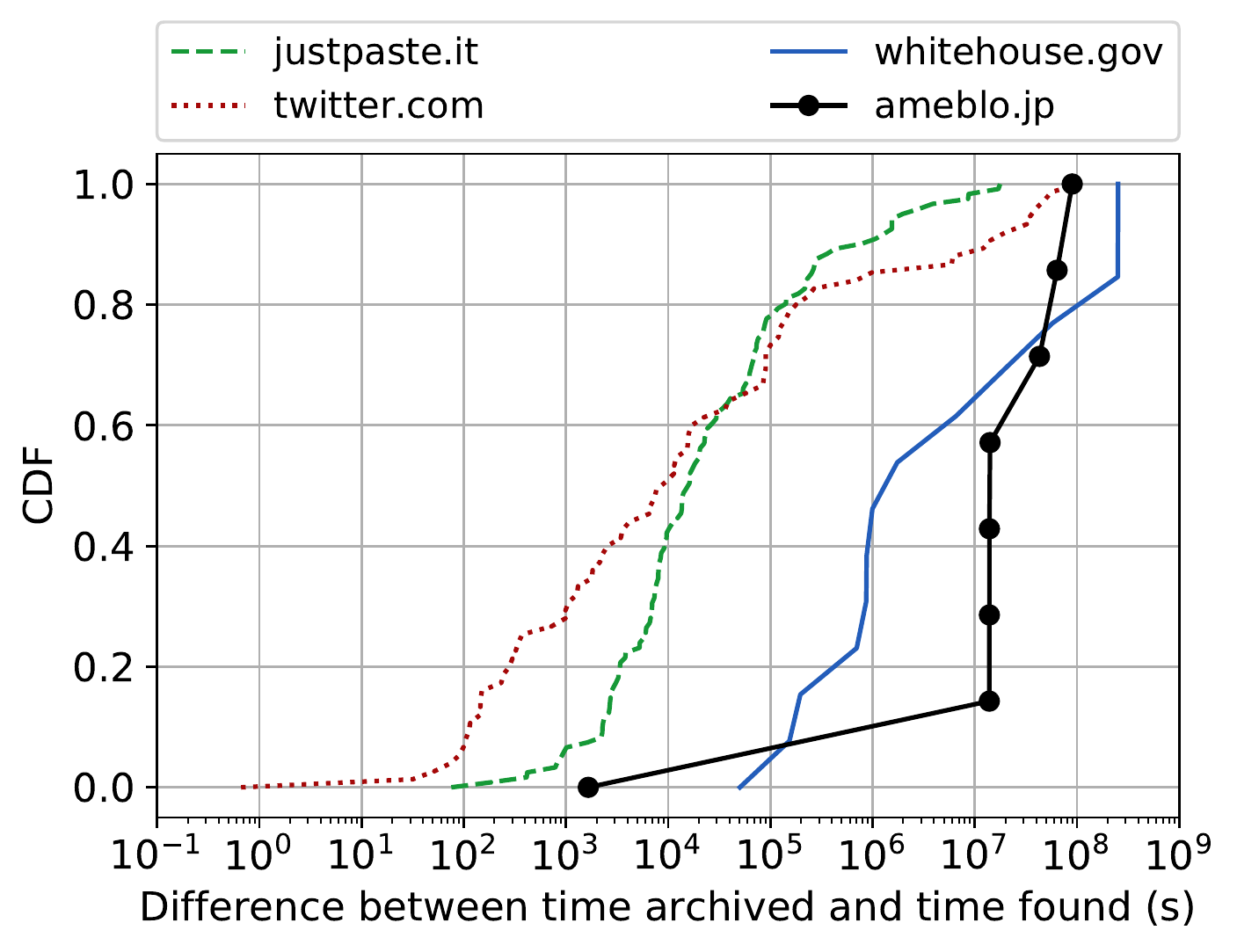}\label{subfig:cdf_time_difference_domain_archive_org_twitter}}
\subfigure[\dspol]{\includegraphics[width=0.24\textwidth]{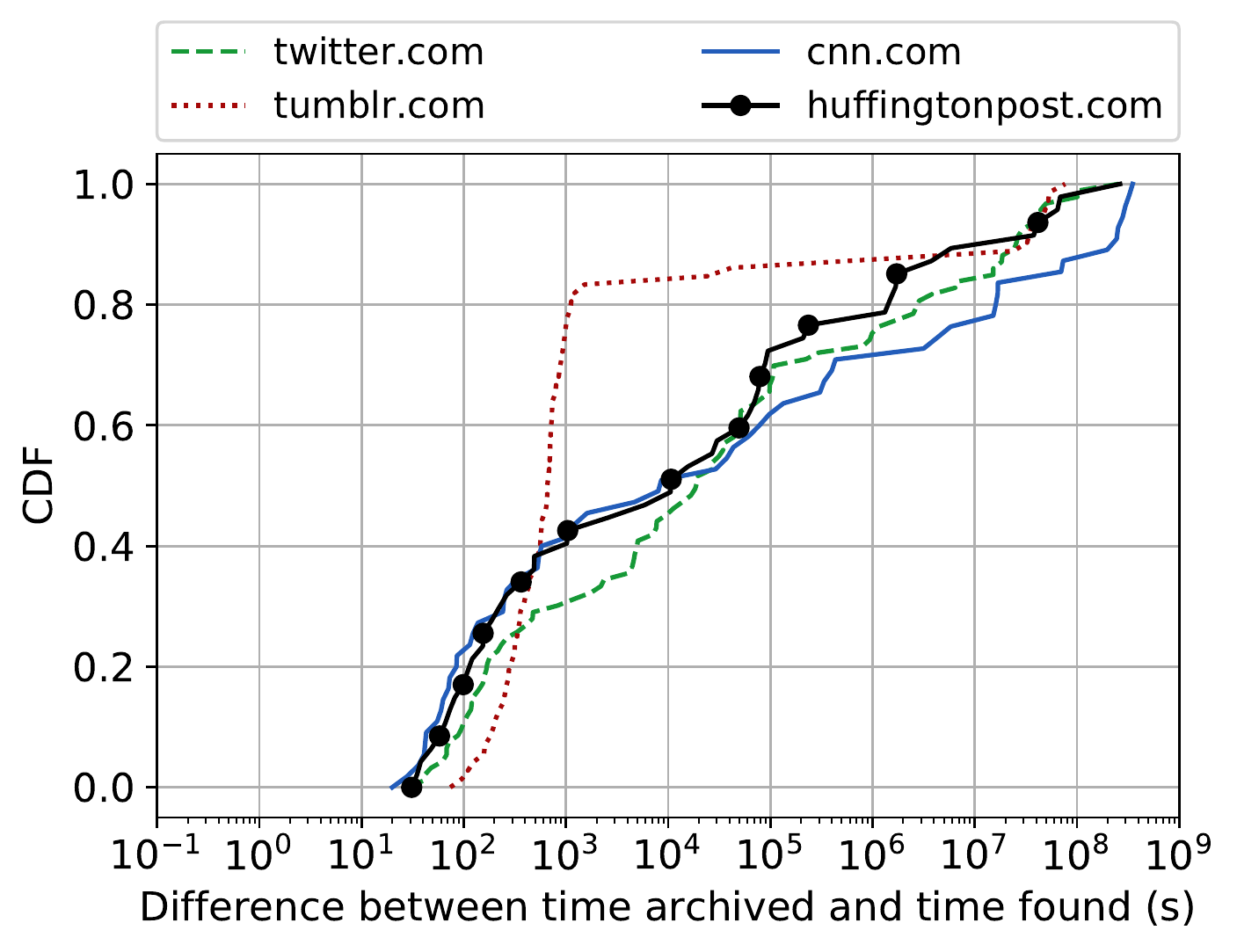}\label{subfig:cdf_time_difference_domain_archive_org_4chan}}
\subfigure[Gab]{\includegraphics[width=0.24\textwidth]{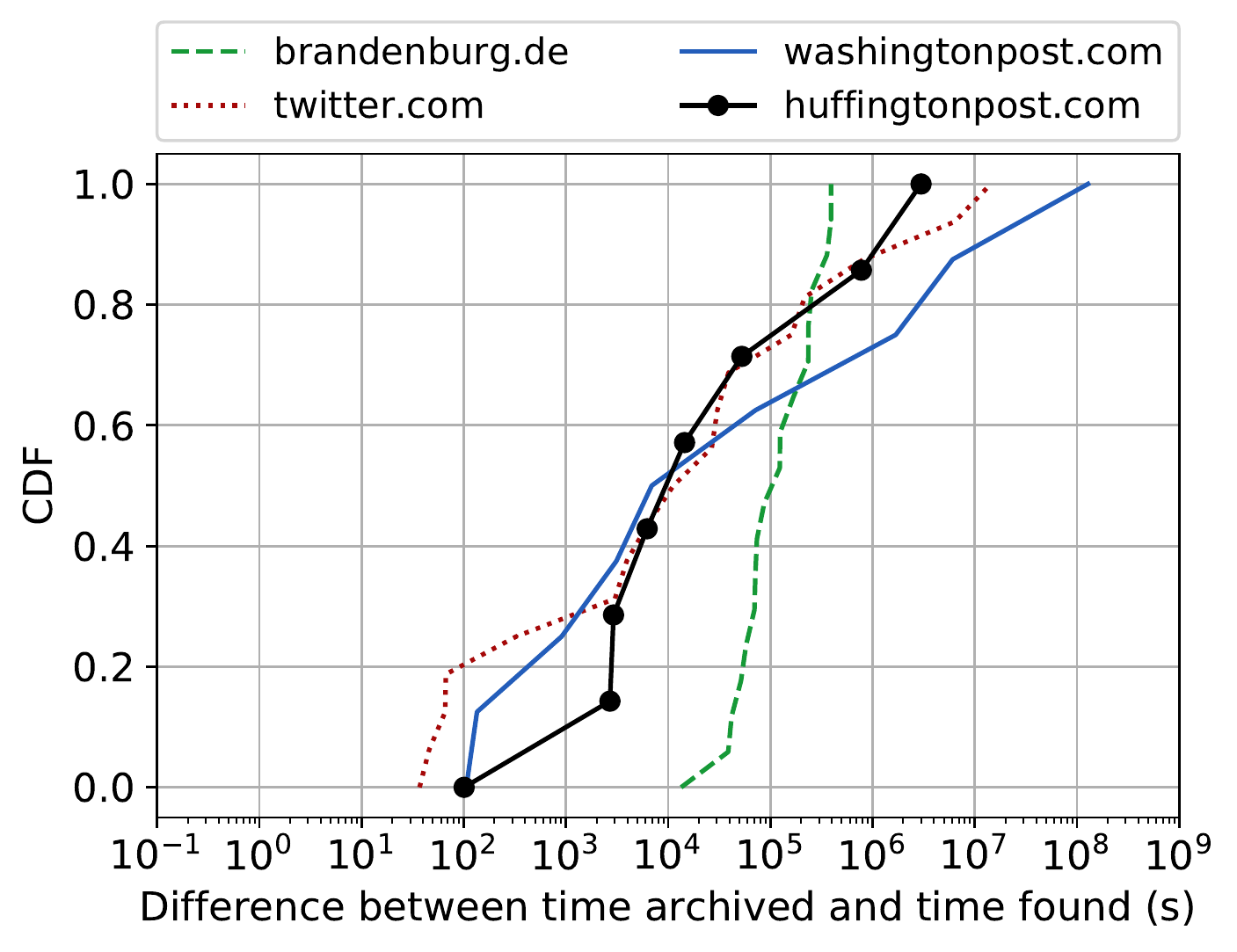}\label{subfig:cdf_time_difference_domain_archive_org_gab}}
\caption{\hspace{-0.115cm} CDF \hspace{-0.1cm} of \hspace{-0.1cm} the \hspace{-0.1cm} time \hspace{-0.1cm} difference \hspace{-0.1cm} between \hspace{-0.1cm} archival \hspace{-0.1cm} time \hspace{-0.1cm} on \hspace{-0.1cm} Wayback \hspace{-0.1cm} Machine \hspace{-0.1cm} and \hspace{-0.1cm} appearance \hspace{-0.1cm} on \hspace{-0.1cm} social \hspace{-0.1cm} networks \hspace{-0.1cm} for \hspace{-0.1cm} top \hspace{-0.1cm} four \hspace{-0.1cm} source \hspace{-0.1cm}  domains.}
\label{fig:cdf_time_difference_domain_archive_org}
\end{figure*}

\subsubsection{Temporal Dynamics}
Next, we study, from a temporal point of view, how archive URLs are created and shared on social networks. 

\descr{Live Feed.}  In Fig.~\ref{fig:temporal_analysis_capsules}, we plot the day and hour of day of the creation of the \url{archive.is} URLs.
Each day, between 1K and 10K URLs are archived (Fig.~\ref{subfig:counts_day}), mostly between 11AM and 4PM UTC time, with a peak at 2PM (Fig.~\ref{subfig:counts_hour_day}), which seems to suggest that a great number of users may be located in Europe and the US. 
According to Alexa, the top country for \url{archive.is} is the US, with 37\% of the visitors. %

\descr{Social Networks.} Next, we measure the time interval between the archiving of a URL and its appearance on one of the four social networks.
In Fig.~\ref{fig:cdf_time_difference}, we plot the CDF of these time intervals, finding that the interval between archiving and sharing times of a URL ranges from a few seconds (in which case, Reddit/4chan/Twitter/Gab users themselves might be creating the archive) to years.
Reddit is the ``fastest'' platform for Wayback Machine URLs, mainly because of bots that actively archive URLs (as we show later in this work), while for \url{archive.is} it is Gab.

We also focus on the top source domains shared via archive URLs: 
Figs~\ref{fig:cdf_time_difference_domain_archive_is}--\ref{fig:cdf_time_difference_domain_archive_org} plot the CDF of the slack time of the top four domains for \url{archive.is} and Wayback Machine URLs, respectively.
On Reddit, the top domains archived via Wayback Machine follow very similar distributions, likely due to bots, while for 
\url{archive.is} URLs, distributions vary, with the fastest domain being \url{reddit.com} itself.
On Twitter, slack times vary for URLs archived via \url{archive.is}, with the fastest domain being Twitter and the slowest \url{nhk.org.jp}. 
The same applies for the Wayback Machine, with the fastest domain being Twitter and the slowest \url{ameblo.jp}.
We also find that, on \dspol, \url{archive.is} URLs pointing to 4chan are considerably slower, suggesting that users are more interested in archiving the URL for persistence rather than sharing the content within \dspol. Based on anecdotal observations, we believe users might be archiving threads with ``evidence'' for conspiracy theories/false narratives, and using them in the future to perpetuate mis/disinformation.
This is not the case for news sources like the Washington Post or Guardian, as \dspol users might be more focused on reducing Web traffic to the source domain instead (indeed we find users explicitly mentioning this when manually examining posts).
Finally, on Gab, the faster domain is Twitter, and Reddit the slowest.

\subsubsection{Original Content Availability}
We then assess the availability of the original content that gets archived; this allow us to determine whether users are archiving URLs that are subsequently deleted. %
To this end, we make an HTTP request for each source URL in our datasets, on October 14--21, 2017 for the live feed dataset, on October 4--5, 2017 for Reddit, Twitter, \dspol datasets and on January 3, 2018 for Gab dataset.
We treat each URL as unavailable %
if we receive HTTP codes 404/410/451/5xx, or if the request times out.

\descr{Live Feed.} 
We find that 12\% of the source URLs corresponding to archive URLs on archive.is live feed are no longer available. %
Domains with most unavailable content include \url{twitter.com} (6\%), \url{nhk.or.jp} (6\%), googleusercontent.com (3\%), aaaaarg.fail (3\%), 4chan.org (3\%), and \url{8ch.net} (2\%).

\descr{Social Networks.}
In Reddit, source URLs corresponding to both \url{archive.is} and Wayback Machine are still available to a large degree (93\% and 89\% of them, respectively). 
This can be explained by the fact that Reddit bots archive URLs without considering the content.
In \dspol, the original content is available 82\% and 66\% of the times, while on Gab 87\% and 48\% for \url{archive.is} and Wayback Machine URLs, respectively.
Percentages decrease further for Twitter; 76\% and 49\% for \url{archive.is} and Wayback Machine URLs, respectively.

We also find that the top domains for which content is no longer available differ across platforms.
Except for Gab, the top unavailable domain are the social networks themselves: 10\%, 54\%, and 28\%, for Reddit, \dspol, and Twitter, respectively.
URLs from cache servers (i.e., \url{googleusercontent.com}) and Twitter are also frequently unavailable;  9\% and 10\% in Reddit,
5\% and 4\% in \dspol,  8\% and 28\% in Twitter, and 12\% and 19\% in Gab, for \url{googleusercontent.com} and Twitter, respectively.
We also note the presence of unavailable 8ch.net URLs (another ephemeral imageboard) with 5\% and 4\% on \dspol and Gab, respectively.

\subsubsection{Take-Aways}
Overall, we find that archiving services play an important role in the information ecosystem, as they are used to preserve news sources as well 
as ephemeral or controversial content. Also, users on fringe communities such as \dspol and Gab favor less popular Web archiving services 
like \url{archive.is} to archive and disseminate Web pages. %
This prompts questions as to \emph{why} less popular, and seemingly less durable, archiving services are favored by more controversial 
communities like \dspol and Gab. Although this would be out of the scope of this work, we do find one potential answer in that these 
communities also use archiving services to bypass platform-specific censorship policies. 

We also observe that temporal dynamics of how archive URLs are shared on social networks differ according to their content: for instance, on \dspol, content 
from news sources has a considerably larger time lag between first appearing on the platform and archival compared to 4chan threads.
Lastly, a non-negligible percentage of archived content is no longer available at the source; in particular, a substantial percentage of posts from social networks 
like Twitter are eventually deleted from the platform, yet remain stored in the archives.

\subsection{Social-Network-based Analysis} \label{sec:social_network_analysis}
In this section, we present a social-network-specific analysis by taking into account the fundamental differences of each platform.
We analyze the users involved in the dissemination of archive URLs as well as the content that is shared along with those URLs.
Lastly, we discuss a case study of ad revenue deprivation on Reddit.

\subsubsection{User Base}

\descr{Reddit.} Our analysis shows that archiving services are extensively used by Reddit bots.
In fact, 31\% of all \url{archive.is} URLs and 82\% of Wayback Machine URLs in our Reddit dataset are posted by a specific bot, 
namely, SnapshillBot (which is used by subreddit moderators to preserve ``drama-related'' happenings discussed earlier or just as 
a subreddit specific policy to preserve \emph{every} submission).
Other bots include AutoModerator, 2016VoteBot, yankbot, and autotldr.
We also attempt to quantify the percentage of archive URLs posted from bots, assuming that, if a username includes ``bot'' or ``auto'', it is likely a bot.
This is a reasonable strategy since Reddit bots are extensively used for moderation purposes, and do not usually try to obfuscate the fact that they are 
bots.\footnote{This is somewhat evident from the list of Reddit bots available at \url{https://www.reddit.com/r/autowikibot/wiki/redditbots}}
Using this heuristic, we find that bots are responsible for disseminating 44\% of all the \url{archive.is} and 85\% of all the Wayback Machine URLs 
that appear on Reddit between July 1, 2016 and August 31, 2017.

\begin{figure}[t]
\hspace*{-0.2cm}
\subfigure[Reddit]{\includegraphics[width=0.515\columnwidth]{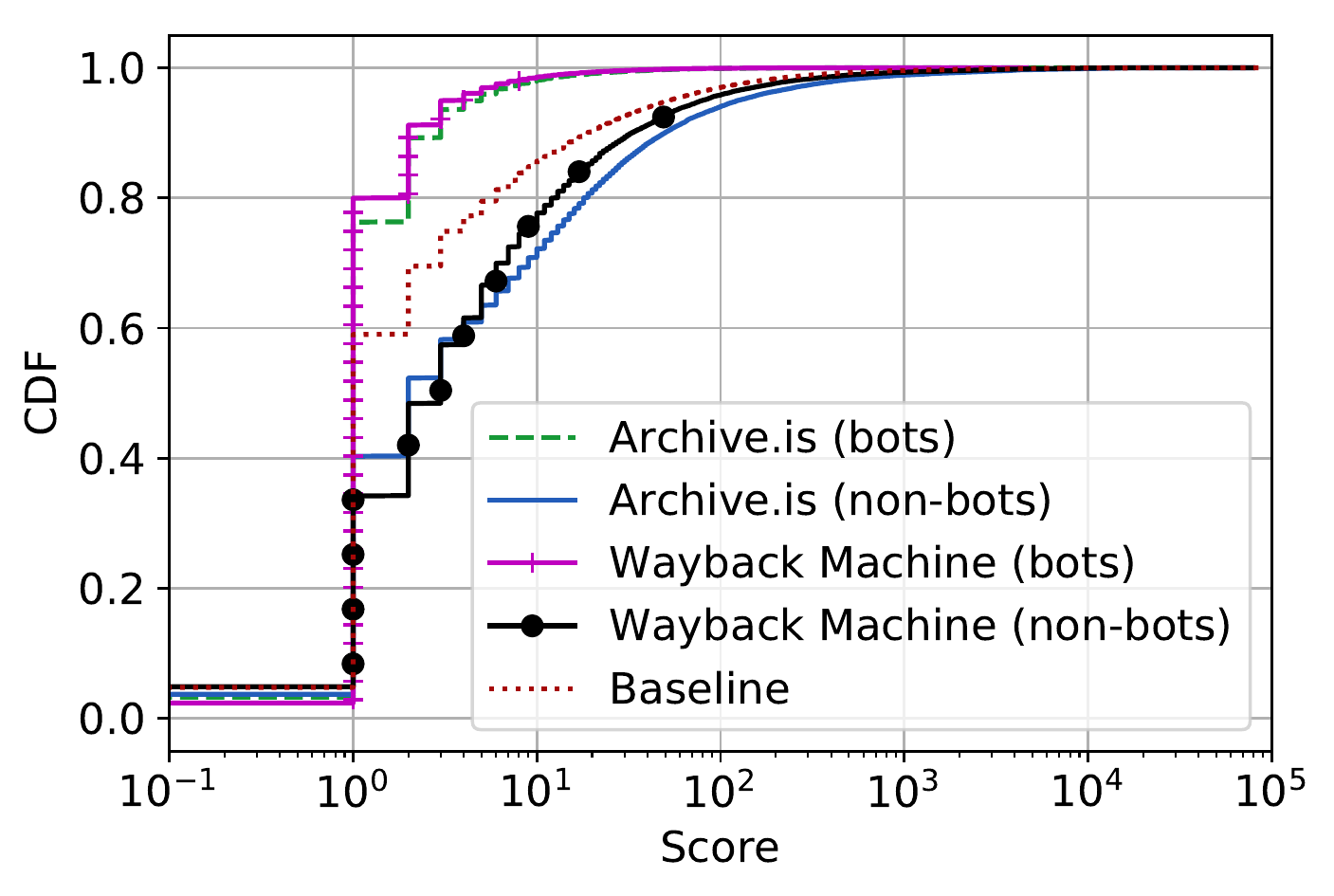}\label{subfig:cdf_scores_reddit_bots}}
\hspace*{-0.3cm}
\subfigure[Gab]{\includegraphics[width=0.515\columnwidth]{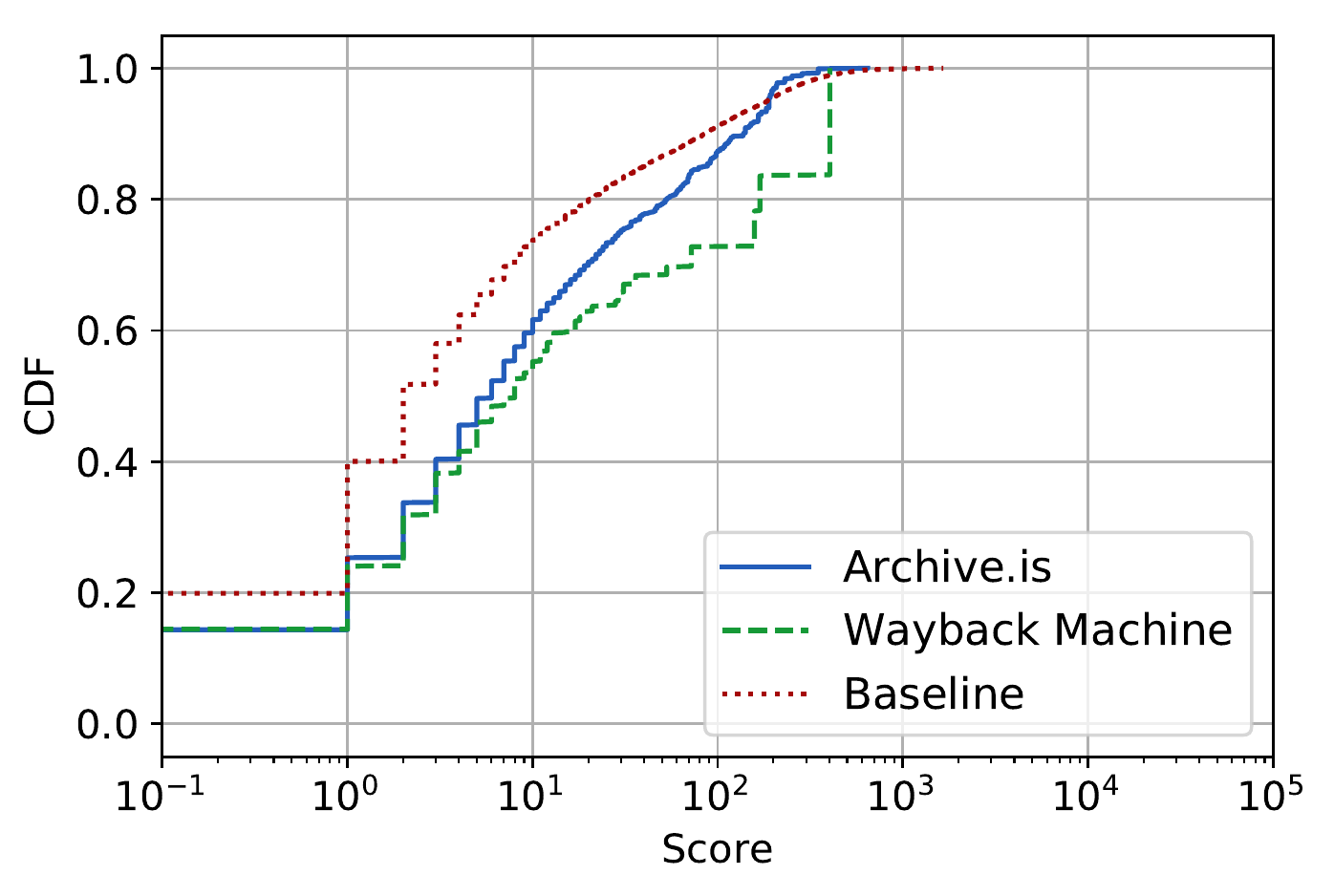}\label{subfig:cdf_scores_gab}}
\caption{CDF of the scores of posts that include \url{archive.is} and Wayback Machine URLs.}
\label{fig:score_analysis}
\end{figure}

We also use the score of each Reddit post to get an intuition of users' appreciation for posts that include archive URLs.
In Fig.~\ref{subfig:cdf_scores_reddit_bots}, we plot the CDF of the scores of posts with \url{archive.is} and Wayback Machine URLs, as 
well as all posts that contain URLs as a baseline, differentiating between bots and non-bots.
For both archiving services, posts by bots have a substantially smaller score: %
80\% of them have score of at most one, as opposed to 37\% for non-bots and 59\% of the baseline.

\begin{table}[t]
\centering
\footnotesize
\resizebox{0.6\columnwidth}{!}{%
\begin{tabular}{lr|lr}
\toprule
\textbf{Subreddit (\url{archive.is})} & \multicolumn{1}{r}{\bf (\%)} & \textbf{Subreddit (Wayback)} & \textbf{(\%)} \\ \midrule
The\_Donald                  & {24.48\%} & EnoughTrumpSpam  & 31.82\%     \\
KotakuInAction                  & {15.83\%}  & MGTOW & 7.38\%      \\
EnoughTrumpSpam                   & {12.06\%}  & SnapshillBotEx   &  7.19\%       \\
MGTOW              & {3.48\%}  & undelete   & 5.90\%        \\
undelete                & {2.74\%}  & SubredditDrama  & 5.50\%          \\
SubredditDrama                & {2.61\%}  & Drama   & 5.03\%          \\
Drama                    & {2.33\%}  & Gamingcirclejerk  & 3.47\%          \\
Gamingcirclejerk                  & {1.57\%}  & ShitAmericansSay   & 1.63\%          \\
conspiracy                  & {1.44\%}  & TopMindsOfReddit   & 1.51\%          \\
MensRights                  & {1.12\%}  & TheBluePill   & 1.25\%          \\
savedyouaclick               & {1.00\%}  & Buttcoin\_1000   & 1.15\%           \\
politics                  & {0.98\%}  &  AgainstHateSubreddits & 1.06\%            \\
DerekSmart                  & {0.76\%}  & subredditcancer   & 0.99\%            \\
ShitAmericansSay                    & {0.75\%}  & The\_Donald   & 0.95\%            \\
PoliticsAll                  & {0.72\%}  & badeconomics  & 0.75\%            \\
TopMindsOfReddit        & {0.71\%}  & ShitWehraboosSay  & 0.74\%            \\
4chan4trump                 & {0.62\%}  & shittykickstarters   & 0.71\%            \\
SnapshillBotEx                    & {0.59\%}  & jesuschristreddit   & 0.68\%            \\
Buttcoin                 & {0.56\%}  & badhistory   & 0.66\%         \\
AgainstHateSubreddits                & {0.55\%}  & politics   & 0.59\%                   \\ \bottomrule
\end{tabular}
}
\caption{Top 20 subreddits sharing \url{archive.is} and Wayback Machine URLs.}
\label{tbl:top_subreddits}
\end{table}

\descr{Reddit Sub-Communities.} We then study how specific subreddits share URLs from archiving services. 
In Table~\ref{tbl:top_subreddits}, we report the top subreddits that share the most archive URLs from \url{archive.is} and the Wayback Machine.
Among these, we find a variety of subreddits ranging from politics (e.g., EnoughTrumpSpam, The\_Donald, politics) to gaming (e.g., Gamingcirclejerk) and 
``drama-related'' communities (e.g., SubredditDrama and Drama). 
Several subreddits prefer to use \url{archive.is} rather than the Wayback Machine, e.g., KotakuInAction, which historically covers 
the GamerGate controversy~\cite{chatzakou2017hate}, The\_Donald, which discusses politics with a focus on Donald Trump, and Conspiracy, 
which focuses on various conspiracy theories. %

\descr{Gab.} On Gab, each post has a score that determines the popularity of the content. In Fig.~\ref{subfig:cdf_scores_gab}, we report the CDF of 
the scores in posts that contain \url{archive.is} and Wayback Machine URLs, between August 2016 and August 2017.
Once again, we also include a baseline, which is the scores for all the posts with URLs.
We find that posts with Wayback Machine URLs have higher scores than those with \url{archive.is} URLs, and the baseline.
Specifically, the mean score for Wayback Machine is 90, while for \url{archive.is} and the baseline the mean score is 35 and 30, respectively.
This trend mirrors the one observed on Reddit for posts not authored by bots.

\descr{/pol/.} As mentioned earlier, 4chan is an anonymous imageboard, which prevents us from performing user-level analysis.
However, we can use the flag attribute to provide a country-level estimation. %
The top country sharing archive URLs is the USA, which is in line with previous characterizations of the board~\cite{hine2016longitudinal}. 
We also find a substantial percentage of ``troll'' flags: 9\% and 5\% for \url{archive.is} and Wayback Machine, respectively.
This is somewhat surprising, since troll flags were re-introduced to \dspol on June 13, 2017, thus they were only available for about 
3 months of our 14-month dataset.

\begin{figure}[t]
\centering
\includegraphics[width=0.57\columnwidth]{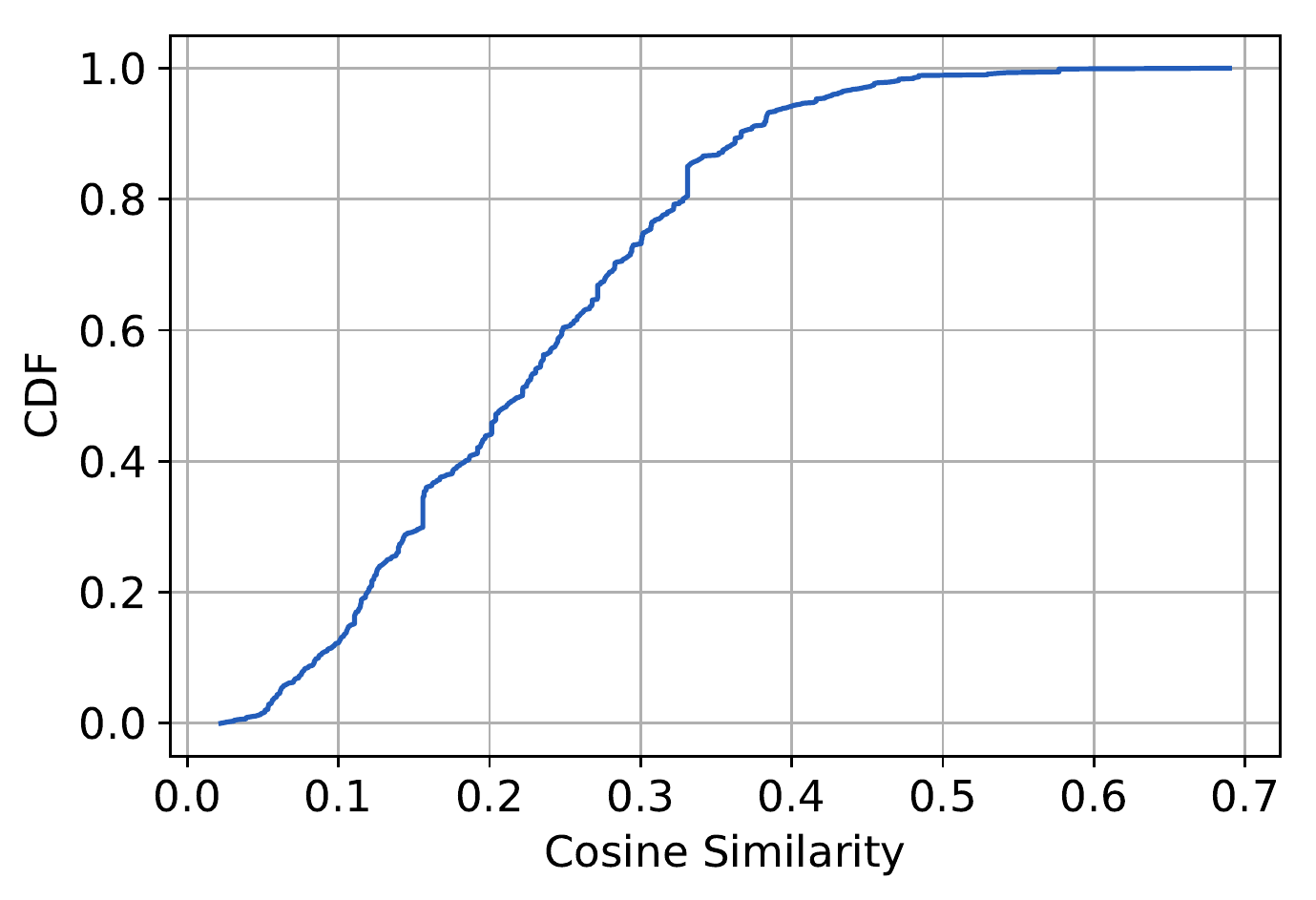}
\caption{CDF of cosine similarity of words obtained from LDA topics on Reddit and dspol threads.}
\label{fig:cdf_lda_similarity}
\end{figure}

\subsubsection{Content Analysis}
Next, we focus on the content that gets shared along with archive URLs on social platforms. We aim
to evaluate if users share the same information for a given archive URL on multiple platforms.
We do so using Latent Dirichlet Allocation (LDA)~\cite{blei2003latent}.
Before running LDA, we exclude \dspol and Reddit threads that contain less than 100 posts, 
so that the LDA can extract topics from a reasonable amount of documents. 
We then select only threads that have archive URLs appearing in {\em both} Reddit and \dspol datasets;
there are 425 such threads on \dspol and 299 on Reddit.
Next, we run LDA on all the posts within these threads and extract terms for 10 topics per thread.

In Fig.~\ref{fig:cdf_lda_similarity}, we plot the CDF of the cosine similarities on the terms extracted from LDA topics
on the two platforms when sharing the same archive URLs.
We observe that 80\% of the terms have similarity under 0.3, which is expected given the fact that the two communities discuss topics in a 
different way. By manually observing terms with high similarity scores, we find that a number of them relate 
to well-known conspiracy theories, like the Seth Rich murder~\cite{seth_rich} or Pizzagate~\cite{pizzagate}, as well as general 
discussions around politics (e.g., tensions between North Korea and the USA).
Once again, this highlights that archiving services are used to preserve content related to controversial 
stories and conspiracy theories.  

\begin{table}[t]
\centering
\footnotesize
\resizebox{0.8\columnwidth}{!}{%
\setlength{\tabcolsep}{0.55em} %
\begin{tabular}{lrr|lrr}
\toprule
\textbf{News Source} & \textbf{Count} & \multicolumn{1}{r}{(\%)} &\textbf{News Source} & \textbf{Count} & \textbf{ (\%)} \\ \midrule
washingtonpost.com          &      3,814  & {44.13\%} & change.org & 96    &  7.52\%\\
cnn.com       &         3,354  & {39.39\%}  & huffpost.com  &  62     & 13.39\%\\
nydailynews.com           &     1,070   & {46.32\%}  & fusion.net   &  60     & 44.77\%\\
huffingtonpost.com            &  978  & {43.77\%}  & cnn.it   & 58    & 44.61\% \\
nationalreview.com &            774   & {45.58\%}  & alternet.org  & 26   & 20.01\%      \\
theblaze.com               &  704  & {46.74\%}  & infostormer.com   & 16    &   27.11\%  \\
buzzfeed.com                   & 588 & {45.97\%}  & dailynewsbin.com   & 4         & 26.67\% \\
salon.com            &      373  & {44.88\%}  & todayvibes.com   & 4  &     7.27\%  \\
vice.com                  & 372 & {45.14\%}  & usanewsbets.ga   & 4       &  10.52\%\\
vox.com               & 323  & {45.23\%}  & fullycucked.com   & 1  & 1.78\%       \\
weeklystandard.com        &      253  & {46.25\%}  & northcrane.com   & 1         & 0.13\%\\
politifact.com              &    185 & {33.09\%}  &    &     &        \\ \bottomrule
\end{tabular}
}
\caption{Number and percentage of submissions deleted from The\_Donald with links to different news sources.}
\label{tbl:reddit_deleted_domains}
\end{table}

\subsubsection{Ad Revenue Deprivation}

During our experiments, we find evidence that at least one Reddit bot, AutoModerator\footnote{\url{https://www.reddit.com/r/AutoModerator/}}, is used to 
remove links to unwanted domains and nudge users to share \url{archive.is} instead. In particular, it posts:
\begin{quote}
\emph{Your submission was removed because it is from \url{cnn.com}, which has been identified as a severely anti-Trump domain. 
Please submit a cached link or screenshot when submitting content from this domain. We recommend using \url{www.archive.is} for this purpose.
}
\end{quote}

This kind of notification appears in five different subreddits that discuss mainly politics and news, specifically, 
The\_Donald, Mr\_Trump, TheNewRight, Vote\_Trump, and Republicans.
In particular, in The\_Donald, there are 13K such comments. AutoModerator blocks URLs from 23 news sources likely to be considered as 
anti-Trump by that community. In Table~\ref{tbl:reddit_deleted_domains} we report the number of submissions 
deleted for each of the sources, along with the percentage over {\em all} submissions that include that source.
Mainstream news outlets like Washington Post and CNN are the top domains that get removed from The\_Donald (3.8K and 3.3K submissions, respectively), 
and this happens slightly less than half the times (44\% and 39\% of the submissions, respectively).
Interestingly, only URLs posted via the URL submission field are censored by AutoModerator, 
but not URLs that are inserted as part of the title field.

We attempt to estimate possible ad revenue deprivation due to the practice of nudging users to share archive URLs instead of source URLs on Reddit.
We do so by providing a conservative approximation of the ad revenue loss. Since we do not have knowledge of how many times a particular URL is 
clicked, we use the up- and down-votes of a post. That is, we assume that when a user up-votes or down-votes a post, he also clicks on the URL included on the post.
This constitutes a best-effort technique as prior work shows that a substantial portion of users on Reddit do not vote~\cite{gilbert2013underprovision}, while, at 
the same time, users that do vote do not necessarily read or click on the articles~\cite{glenski2017consumers}.
That said, this approach is reasonably conservative considering the complex influence that Reddit has with respect to news 
dissemination~\cite{zannettou2017web}.

\begin{table}[]
\centering
\resizebox{0.7\columnwidth}{!}{
\setlength{\tabcolsep}{0.55em} %
\begin{tabular}{lrr|lrr}
\toprule
\textbf{ Domain} & \textbf{ Visits} & \textbf{ Loss (\$)} & \textbf{Domain} & \textbf{Visits} & \textbf{Loss (\$)} \\ \midrule
washingtonpost.com     & 79,880                  & 5,928                              & wsj.com                & 11,389                  &   845                             \\
cnn.com                & 70,483                  & 5,231                              & breitbart.com          & 11,357                  & 842                               \\
nytimes.com            & 46,442                  & 3,446                              & bbc.com                & 10,708                  & 794                               \\
huffingtonpost.com     & 27,125                  & 2,013                              & salon.com              & 10,364                  & 769                               \\
thehill.com            & 18,643                  & 1,383                               & buzzfeed.com           & 10,359                  & 768                               \\
theguardian.com        & 16,376                  & 1,215                               & foxnews.com            & 9,638                   & 715                               \\
politico.com           & 15,774                  &       1,170                         & yahoo.com              & 9,497                   & 704                              \\
dailymail.co.uk        & 14,442                  & 1,071                               & latimes.com            & 9,277                   & 688                               \\
dailycaller.com        & 12,735                  & 945                               & vox.com                & 8,976                   & 667                               \\
google.com             & 11,576                  & 859                               & washingtontimes.com    & 8,862                   & 657                               \\ \bottomrule
\end{tabular}
}
\caption{Top 20 domains with the largest ad revenue losses because of the use of archiving services on Reddit. 
We report an estimate of the average monthly visits from Reddit as well as the average monthly ad revenue loss. }
\label{tbl:revenue_losses}
\end{table}

We then calculate the potential revenue loss using only ad impressions, i.e., we conservatively estimate the revenue generated when a user 
visits the website without taking into account any potential further action (e.g., clicking on the actual ad).
To this end, we use an average Cost per 1,000 impressions (CPM) of \textdollar 24.74,
as reported by Statista\footnote{\url{https://www.statista.com/statistics/308015/online-display-cpm-usa/}}, 
while we assume an average of 3 ads per page~\cite{barford2014adscape}.
In other words, we calculate the monthly revenue loss, for each domain, based on the average CPM value as well as 
the conservative estimate of the visits using the up- and down-votes.
Overall, replacing URLs with archive URLs, as done, e.g., by the AutoModerator bot, yields an estimate of 
\textdollar 30K per month in revenue loss (for the top 20 domains in terms of views). This is detailed in Table~\ref{tbl:revenue_losses}, 
where we break down the estimate for each of the top 20 revenue-deprived domains.

On a purely pragmatic level, consider that our estimate of ad revenue deprivation is around \textdollar 70K per year for the Washington Post alone.
Although a more detailed impact analysis is out of the scope of this work, we suspect that even \textdollar 70K could have a real world effect, 
e.g., on intern budgets or even early career hires. In light of recent criticism of their credibility by President Donald 
Trump~\cite{trump-fake-news-awards}, Trump-supporting communities' deliberate use of \url{archive.is}, and the conservative 
nature of our revenue loss estimate, we believe this attack on the Fourth Estate is particularly worrying and in need of future exploration.

\subsubsection{Take-Aways}
In summary, our social-network-specific analysis shows, among other things, that moderation bots on Reddit proactively 
leverage Web archiving services to ensure that content shared on their community persists.
In particular, we find that 44\% and 85\% of \url{archive.is} and Wayback Machine URLs are shared by Reddit moderation bots.

Also, Web archiving services are extensively used for the archival and dissemination of content related to conspiracy theories 
(e.g., Pizzagate) as well as other world events related to politics (e.g., tensions between North Korea and the USA), 
thus suggesting that these services play an important role in the (false) information ecosystem and need to be taken into 
account when designing systems to detect and contain the cascade of mis/disinformation on the Web.

Finally, we find evidence that moderators from specific Reddit sub-communities force users to misuse Web archiving services 
so as to ideologically target certain news sources by depriving them of traffic and potential ad revenue. 
We also provide a best-effort conservative estimate of ad revenue loss of popular news sources showing that they can lose up to \textdollar 70K per year.

\subsection{Remarks}

This work presented a large-scale analysis of how popular Web archiving services such as \url{archive.is} and the Wayback Machine are used on social media.
Our study is based two data crawls: 1) 21M URLs, spanning almost two years, obtained from the \url{archive.is} live feed; and 
2) 356K \url{archive.is} plus 391K Wayback Machine URLs that were shared on four social networks: Reddit, Twitter, Gab, and 
4chan's Politically Incorrect board (\dspol) over 14 months. 
Among other things, we showed that these services are extensively used to archive and disseminate news, social network posts, and 
controversial content---in particular by users of fringe Web communities within Reddit and 4chan.
We also found that users not only use them to ensure persistence of Web content, but also to bypass censorship policies 
enforced on some social networks.

We uncovered evidence that certain subreddits, as well as 4chan's Politically Incorrect board (\dspol), actually nudge users to share archive URLs instead of links to news sources they perceive as having contrasting ideologies, taking away potentially hundreds of thousands of dollars in ad revenue.
Overall, our measurements illustrate the importance of archiving services in the Web's information and ad ecosystems, and the need to carefully consider 
them when studying such ecosystems.

\chapter{Towards Understanding State-Sponsored Actors}\label{chapter:trolls}
In this chapter, we study the behavior of state-sponsored actors on the Web. 
To do this, we leverage ground truth datasets released by Twitter and Reddit pertaining to Russian and Iranian trolls.
By analyzing the dataset across several axes we provide a comprehensive analysis on these actors.

Note that the methodology for detecting state-sponsored trolls employed by Twitter and Reddit is not publicly available, therefore, it is unclear on whether there are false positive or how comprehensive these datasets are (i.e., if there are still a lot of unidentified troll accounts).
Despite this fact, in this Chapter, we assume that the released datasets are high-quality ground truth datasets with negligible percentage of false positives and adequate coverage of state-sponsored trolls accounts.
Therefore, the reader should take into account that all claims and analysis made throughout this Chapter are based on these datasets and it is not clear how these claims and results will change with larger datasets or with datasets from other state-sponsored accounts (e.g., originating from other countries other than Russia and Iran).

\section[How State-Sponsored Trolls Compare to Random Users \& How Their Accounts Evolve?]{How State-Sponsored Trolls Compare to Random Users and How do Their Accounts Evolve?}

\subsection{Motivation}

Recent political events and elections have been increasingly accompanied by reports of disinformation campaigns attributed to state-sponsored actors~\cite{ferrara2017disinformation}.
In particular, ``troll farms,'' allegedly employed by Russian state agencies, have been actively commenting and posting content on social media to further the Kremlin's political agenda~\cite{independent}.
In late 2017, the US Congress started an investigation on Russian interference in the 2016 US Presidential Election, releasing the IDs of 2.7K Twitter accounts identified as Russian trolls.

Despite the growing relevance of state-sponsored disinformation, the activity of accounts linked to such efforts has not been thoroughly studied.
Previous work has mostly looked at campaigns run by bots~\cite{ferrara2017disinformation,hegelich2016are,ratkiewicz2011detecting};
however, automated content diffusion is only a part of the issue, and in fact recent research has shown that human actors are actually key in spreading false information on Twitter~\cite{starbird2017examining}.
Overall, many aspects of state-sponsored disinformation remain unclear, e.g., how do state-sponsored trolls operate? What kind of content do they disseminate? And, perhaps more importantly, how do they compare to a set of random users?

In this work, we aim to address these questions, by relying on the set of 2.7K accounts released by the US Congress as ground truth for Russian state-sponsored trolls.
From a dataset containing all tweets released by the 1\% Twitter Streaming API, we search and retrieve 27K tweets posted by 1K Russian trolls between January 2016 and September 2017.
We characterize their activity by comparing to a random sample of Twitter users.

\descr{Main findings.} Our study leads to several key observations:
\begin{enumerate} %
\item %
The main topics discussed by Russian trolls target very specific world events (e.g., Charlottesville protests) and organizations (such as ISIS), and political threads related to Donald Trump and Hillary Clinton.
\item Trolls adopt different identities over time, i.e., they ``reset'' their profile by deleting their previous tweets and changing their screen name/information.
\item Trolls exhibit significantly different behaviors compared to other (random) Twitter accounts. For instance, the locations they report concentrate in a few countries like the USA, Germany, and Russia, perhaps in an attempt to appear ``local'' and more effectively manipulate opinions of users from those countries. Also, while random Twitter users mainly tweet from mobile versions of the platform, the majority of the Russian trolls do so via the Web Client.
\end{enumerate}

\subsection{Datasets} \label{sec:datasets}

\descr{Russian trolls.} We start from the 2.7K Twitter accounts suspended by Twitter because of connections to Russia's Internet Research Agency. %
The list of these accounts was released by the US Congress as part of their investigation of the alleged Russian interference in the 2016 US presidential election, and includes both Twitter's {\em user ID} (which is a numeric unique identifier associated to the account) and the {\em screen name}.\footnote{See \url{https://democrats-intelligence.house.gov/uploadedfiles/exhibit_b.pdf}}
From a dataset storing all tweets released by the 1\% Twitter Streaming API, we search for tweets posted between January 2016 and September 2017 by the user IDs of the trolls.
Overall, we obtain 27K tweets from 1K out of the 2.7K Russian trolls. %

\descr{Baseline dataset.} We also compile a list of random Twitter users, while ensuring that the distribution of the average number of tweets per day posted by the random users is similar to the one by trolls.
To calculate the average number of tweets posted by an account, we find the first tweet posted after January 1, 2016 and retrieve the overall tweet count. %
This number is then divided by the number of days since account creation.
Having selected a set of 1K random users, we then collect all their tweets between January 2016 and September 2017, obtaining a total of 96K tweets.
We follow this approach as it gives a good approximation of posting behavior, even though
it might not be perfect, since (1) Twitter accounts can become more or less active over time, and (2) our datasets are based on the 1\% Streaming API, thus, we are unable to control the number of tweets we obtain for each account.

\begin{figure}[t]
\center
\subfigure[Hour of Day]{\includegraphics[width=0.49\columnwidth]{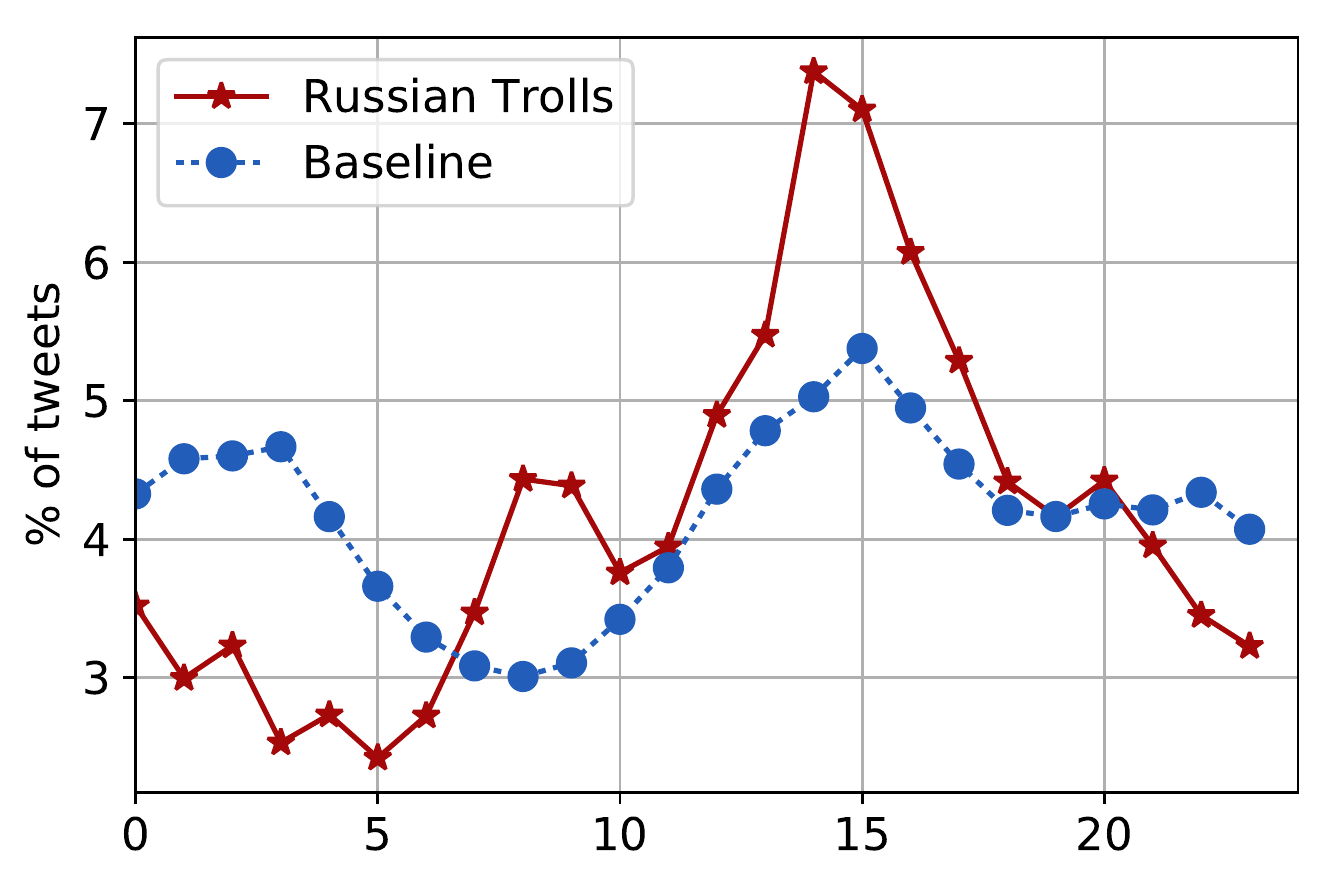}\label{subfig:cybersafety_counts_per_hour_day}}
\subfigure[Hour of Week]{\includegraphics[width=0.49\columnwidth]{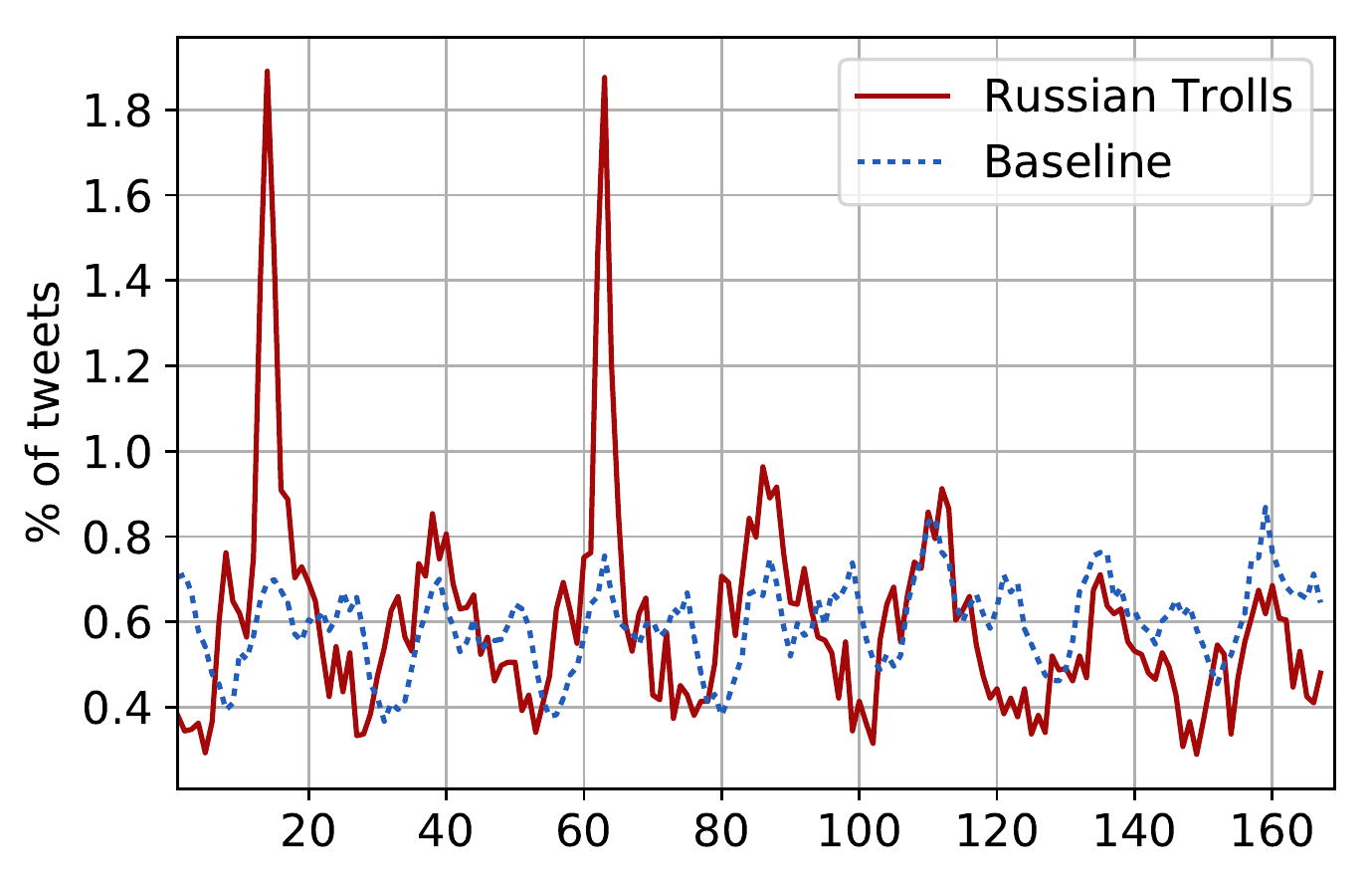}\label{subfig:cybersafety_counts_per_hour_week}}
  \caption{Temporal characteristics of tweets from Russian trolls and random Twitter users. }
\label{fig:cybersafety_temporal_analysis}
\end{figure}

\subsection{Analysis} \label{sec:analysis}
In this section, we present an in-depth analysis of the activities and the behavior of Russian trolls.
First, we provide a general characterization of the accounts and a geographical analysis of the locations they report.
Then, we analyze the content they share and how they evolved until their suspension by Twitter.
Finally, we present a case study of one specific account.

\subsubsection{General Characterization} \label{sec:general
}
\descr{Temporal analysis.}
We observe that Russian trolls are continuously active on Twitter between January, 2016 and September, 2017, with a peak of activity just before the second US presidential debate (October 9, 2016).
Fig.~\ref{subfig:cybersafety_counts_per_hour_day} shows that most tweets from the trolls are posted between 14:00 and 15:00 UTC.
In Fig.~\ref{subfig:cybersafety_counts_per_hour_week}, we also report temporal characteristics based on hour of the week,
finding that both datasets follow a diurnal pattern, while trolls' activity peaks around 14:00 and 15:00 UTC on Mondays and Wednesdays.
Considering that Moscow is three hours ahead UTC, this distribution does not rule out that tweets might actually be posted from Russia.

\begin{figure}[t!]
\centering
\includegraphics[width=0.5\columnwidth]{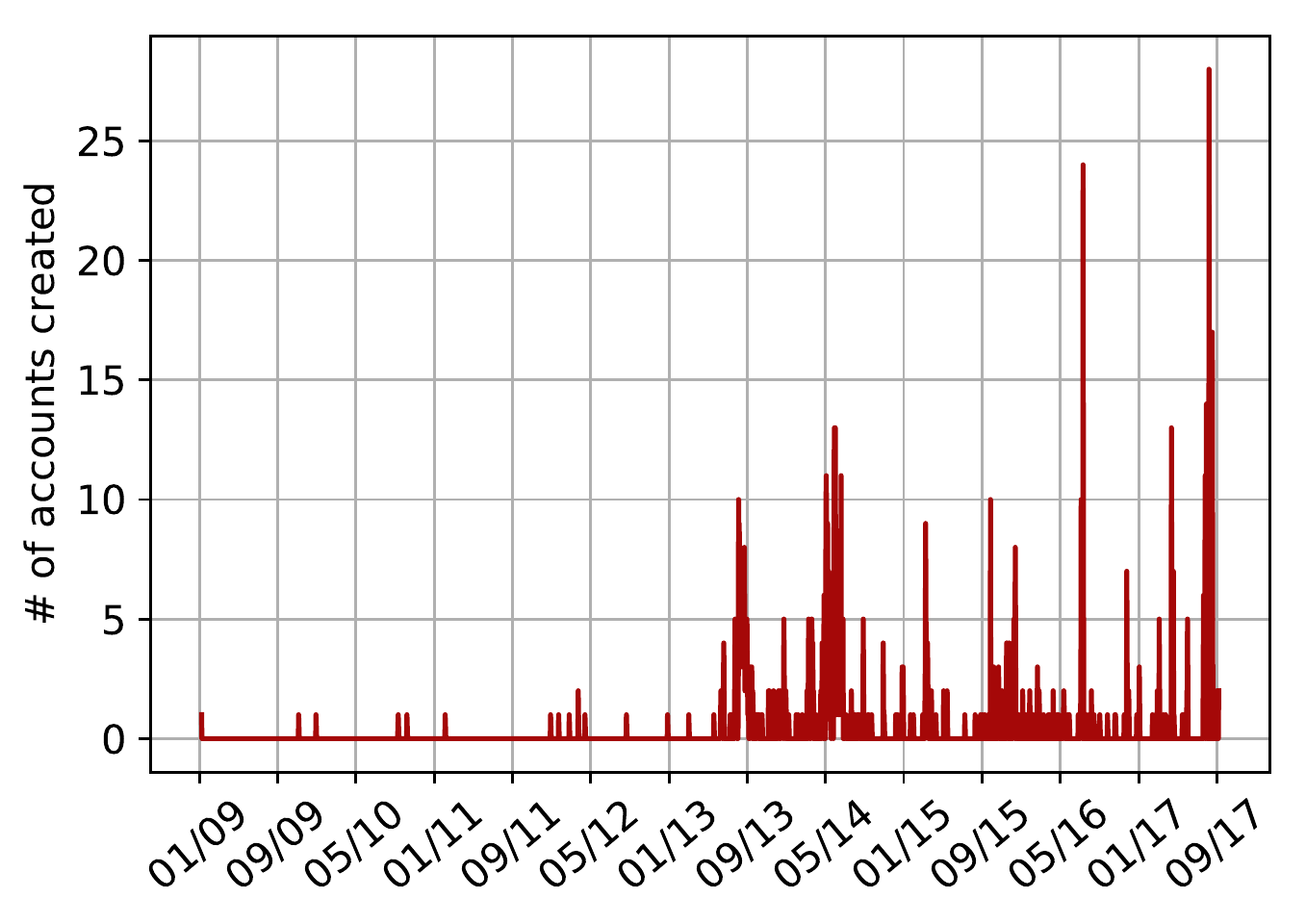}
\caption{Number of Russian troll accounts created per day.}
\label{fig:cybersafety_counts_created}
\end{figure}

\descr{Account creation.}  Next, we examine the dates when the trolls infiltrated Twitter,
by looking at the account creation dates.
From Fig.~\ref{fig:cybersafety_counts_created}, %
we observe that 71\% of them are actually created before 2016.
There are some interesting peaks, during 2016 and 2017: for instance, 24 accounts are created on July 12, 2016, approx. a week before the Republican National Convention (when Donald Trump received the nomination), %
while 28 appear on August 8, 2017, a few days before the infamous Unite the Right rally in Charlottesville. %
Taken together, this might be evidence of coordinated activities aimed at manipulating users' opinions with respect to specific events.

\descr{Account characteristics.}
We also shed light on the troll account profile information.
In Table~\ref{tbl:cybersafety_account_desc}, we report the top ten words appearing in the screen names and the descriptions of Russian trolls, as well as character 4-grams for screen names and word bigrams for profile descriptions.
Interestingly, a substantial number of Russian trolls pose as news outlets, evident from the use of the term ``news'' in both the screen name (1.3\%) and the description (10.7\%).
Also, it seems they attempt to increase the number of their followers, thus their reach of Twitter users, by nudging users to follow them (see, e.g., ``follow me'' appearing in almost 8\% of the accounts).
Finally, 10.3\% of the Russian trolls describe themselves as Trump supporters: ``trump'' and ``maga'' (Make America Great Again, one of Trump campaign's main slogans).

\begin{table}[]
\centering
\resizebox{0.8\columnwidth}{!}{
\setlength{\tabcolsep}{0.4em} %
\begin{tabular}{@{}lrlrlrlr@{}}
\toprule
\multicolumn{4}{c}{\textbf{Screen Name}}  & \multicolumn{4}{c}{\textbf{Description}} \\ %
\textbf{Word} & \textbf{(\%)} & \textbf{4-gram} & \textbf{(\%)} & \textbf{Word} & \textbf({\%)} & \textbf{Word bigram} & \textbf{(\%)} \\ \midrule
news                 & 1.3\%         & news                        & \multicolumn{1}{r|}{1.5\%}       & news                  & 10.7\%        & follow me                    & 7.8\%         \\
bote                 & 1.2\%         & line                        & \multicolumn{1}{r|}{1.5\%}       & follow                & 10.7\%        & breaking news                & 2.6\%         \\
online               & 1.1\%         & blac                        & \multicolumn{1}{r|}{1.3\%}       & conservative          & 8.1\%         & news aus                     & 2.1\%         \\
daily                & 0.8\%         & bote                        & \multicolumn{1}{r|}{1.3\%}       & trump                 & 7.8\%         & uns in                       & 2.1\%         \\
today                & 0.6\%         & rist                        & \multicolumn{1}{r|}{1.1\%}       & und                   & 6.2\%         & deiner stdt                  & 2.1\%         \\
ezekiel2517          & 0.6\%         & nlin                        & \multicolumn{1}{r|}{1.1\%}       & maga                  & 5.9\%         & die news                     & 2.1\%         \\
maria                & 0.5\%         & onli                        & \multicolumn{1}{r|}{1.0\%}       & love                  & 5.8\%         & wichtige und                 & 2.1\%         \\
black                & 0.5\%         & lack                        & \multicolumn{1}{r|}{1.0\%}       & us                    & 5.3\%         & nachrichten aus              & 2.1\%         \\
voice                & 0.4\%         & bert                        & \multicolumn{1}{r|}{1.0\%}       & die                   & 5.0\%         & aus deiner                   & 2.1\%         \\
martin               & 0.4\%         & poli                        & \multicolumn{1}{l|}{1.0\%}        & nachrichten           & 4.3\%         & die dn                       & 2.1\%         \\ \bottomrule
\end{tabular}
}
\caption{Top 10 words found in Russian troll screen names and account descriptions. We also report character 4-grams for the screen names and word bigrams for the description.}
\label{tbl:cybersafety_account_desc}
\end{table}

\begin{figure}[t!]
\center
\subfigure[]{\includegraphics[width=0.49\columnwidth]{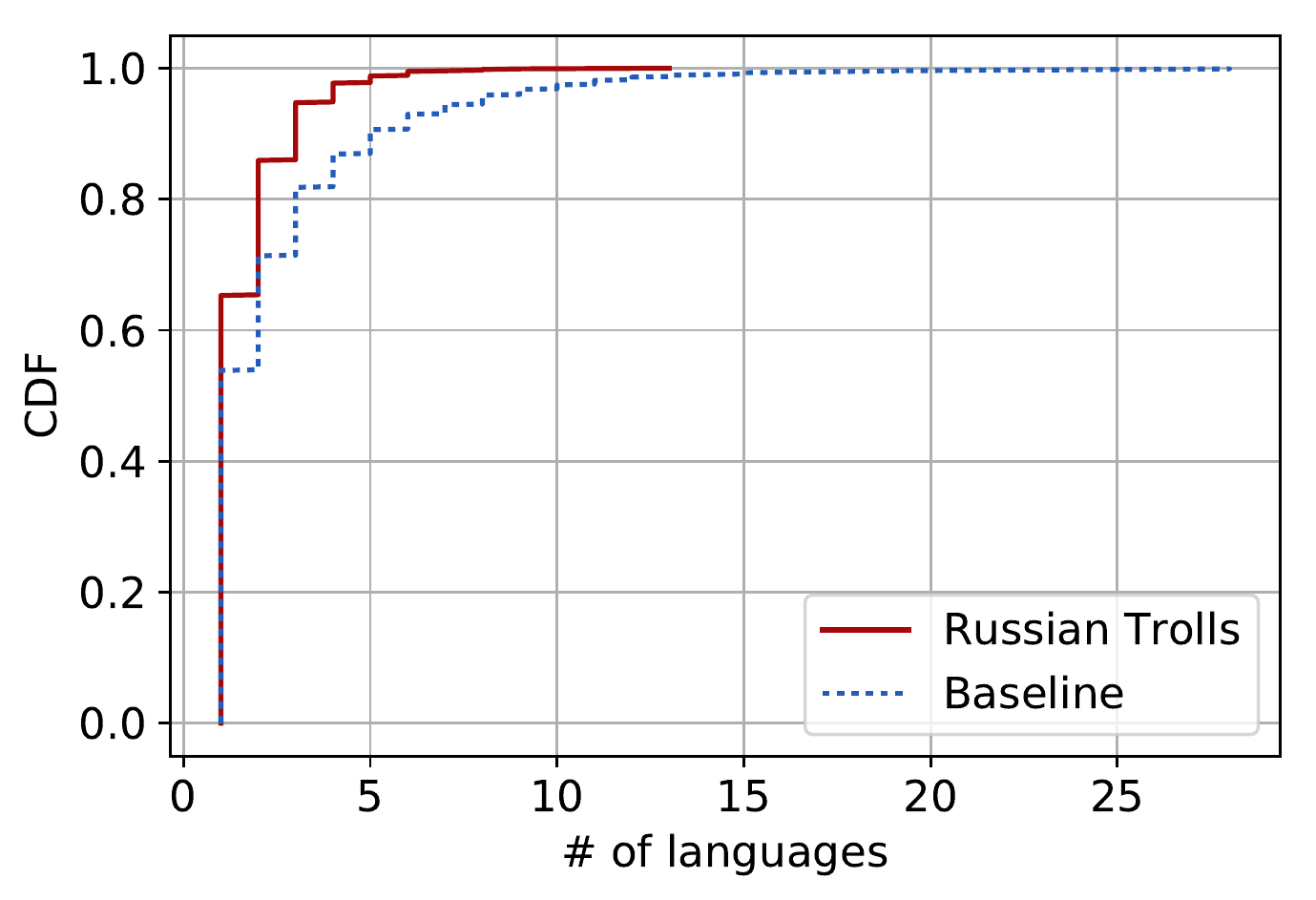}\label{subfig:cybersafety_cdf_languages_user}}
\subfigure[]{\includegraphics[width=0.49\columnwidth]{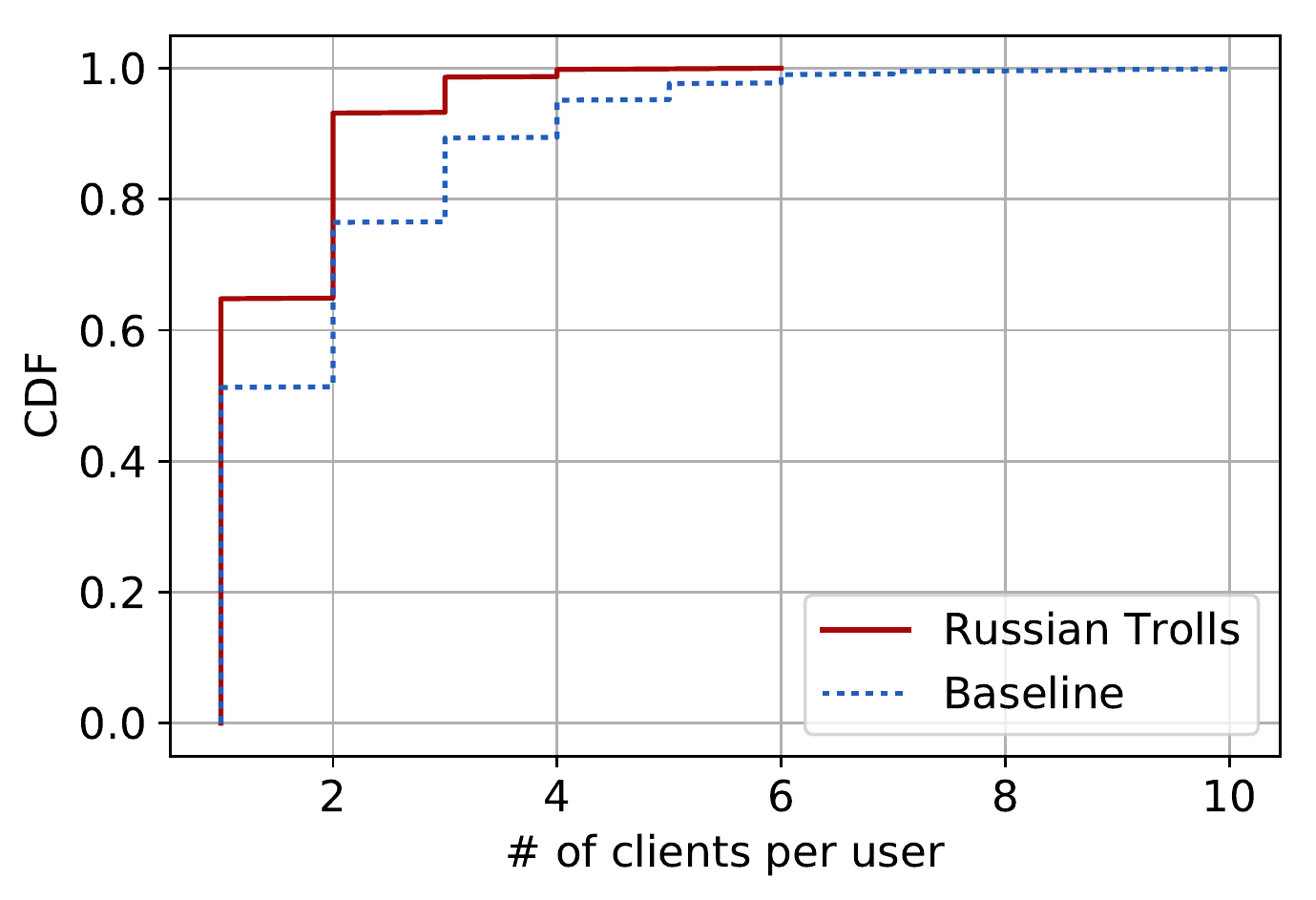}\label{subfig:cybersafety_cdf_sources_user}}
  \caption{CDF of number of (a) languages used (b) clients used per user. }
\label{fig:cybersafety_cdf_lang_sources}
\end{figure}

\descr{Language.} Looking at the language (as provided via the Twitter API) of the tweets posted by Russian trolls, we find that most of them (61\%) are in English, although a substantial portion are in Russian (27\%), and to a lesser extent in German (3.5\%).
In Fig.~\ref{subfig:cybersafety_cdf_languages_user}, we plot the cumulative distribution function (CDF) of the number of different languages for each user: 64\% of the Russian trolls post all their tweets in only one language,
compared to only 54\% for random users.
Overall, by comparing the two distributions, we observe that random Twitter users tend to use more languages in their tweets compared to the trolls.

\begin{table}[t]
\centering
\resizebox{0.7\columnwidth}{!}{
\begin{tabular}{lr|lr}
\toprule
\textbf{Client (Trolls)} & \multicolumn{1}{r}{\textbf{(\%)}} & \textbf{Client (Baseline)} & \textbf{(\%)} \\ \midrule
Twitter Web Client               & 50.1\%       & TweetDeck          & 32.6\%        \\
twitterfeed                      & 13.4\%        & Twitter for iPhone        & 26.2\%        \\
Twibble.io                            & 9.0\%         & Twitter for Android         & 22.6\%        \\
IFTTT                      & 8.6\%         &   Twitter Web Client               & 6.1\%         \\
TweetDeck                        & 8.3\%         & GrabInbox               & 2.0\%         \\
NovaPress                        & 4.6\%         & Twitter for iPad           & 1.4\%         \\
dlvr.it                          & 2.3\%         & IFTTT                      & 1.0\%         \\
Twitter for iPhone               & 0.8\%         & twittbot.net                   & 0.9\%         \\
Zapier.com                       & 0.6\%         & Twitter for BlackBerry              & 0.6\%         \\
Twitter for Android              & 0.6\%         & Mobile Web (M2)            & 0.4\%         \\ \bottomrule
\end{tabular}
}
\caption{Top 10 Twitter clients (as \% of tweets).}
\label{tbl:cybersafety_top_sources}
\end{table}

\descr{Client.} Finally, we analyze the clients used to post tweets.
We do so since previous work~\cite{egele2017towards} shows that the client used by official or professional accounts are quite different that the ones used by regular users.
Table~\ref{tbl:cybersafety_top_sources} reports the top 10 clients for both Russian trolls and baseline users.
We find the latter prefer to use Twitter clients for mobile devices (48\%) and the TweetDeck dashboard (32\%), whereas, the former mainly use the Web client (50\%).
We also assess how many different clients Russian trolls use throughout our dataset:
in Fig.~\ref{subfig:cybersafety_cdf_sources_user}, we plot the CDF of the number of clients used per user.
We find that 65\%  of the Russian trolls use only one client, 28\% of them two different clients, and the rest more than three, which is overall less than the random baseline users.

\begin{figure*}[t]
\centering
\includegraphics[width=\textwidth]{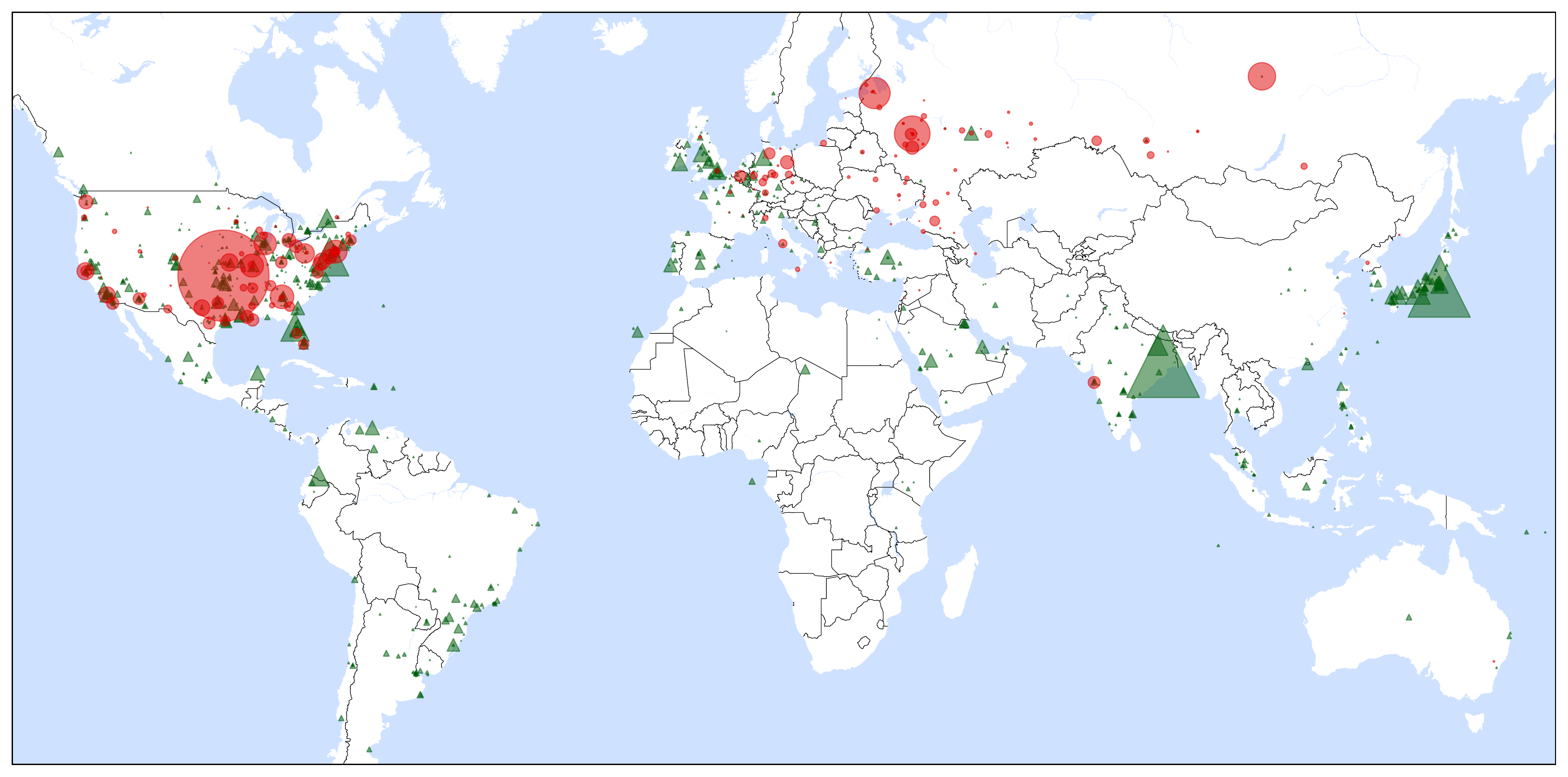}
   \caption{Distribution of reported locations for tweets by Russian trolls (red circles) and baseline (green triangles).}
\label{fig:locations_map}
\end{figure*}

\subsubsection{Geographical Analysis}

\descr{Location.} We then study users' location, relying on the self-reported location field in their profiles.
Note that users not only may leave it empty, but also change it any time they like,
so we look at locations for each tweet.
We retrieve it for 75\% of the tweets by Russian trolls, gathering 261 different entries, which we convert to a physical location using the Google Maps Geocoding API. %
In the end, we obtain 178 unique locations for the trolls,
as depicted in Fig.~\ref{fig:locations_map} (red circles).
The size of the circles on the map indicates the number of tweets that appear at each location.
We do the same for the baseline, getting 2,037 different entries, converted by the API to 894 unique locations. %
We observe that most of the tweets from Russian trolls come from locations within the USA and Russia, and some from European countries, like Germany, Belgium, and Italy.
On the other hand, tweets in our baseline are more uniformly distributed across the globe, with many tweets from North and South America, Europe, and Asia.
This suggests that Russian trolls may be pretending to be from certain countries, e.g., USA or Germany, aiming to pose as locals and better manipulate opinions.
This explanation becomes more plausible when we consider that a plurality of trolls' tweets have their location set as a generic form of ``US,'' as opposed to a specific city, state, or even region.
Interestingly, the 2nd, 3rd, and 4th most popular location for trolls to tweet from are Moscow, St. Petersburg, and a generic form of ``Russia.''
We also assess whether users change their country of origin based on the self-reported location:
only a negligible percentage (1\%) of trolls change their country, while for the baseline the percentage is 16\%.

\begin{table}[t]
\centering
\resizebox{0.7\columnwidth}{!}{
\begin{tabular}{lrlr}
\toprule
\textbf{Timezone (Trolls)}           & \multicolumn{1}{r}{\textbf{(\%)}}                    & \textbf{Timezone (Baseline)}            & \textbf{(\%)} \\ \midrule
Eastern Time  & \multicolumn{1}{r|}{38.87\%} & Athens                       & 24.41\%        \\
Pacific Time  & \multicolumn{1}{r|}{18.37\%} & Pacific Time                    & 21.41\%        \\
Volgograd                   & \multicolumn{1}{r|}{10.03\%} & London                       & 21.27\%        \\
Central Time  & \multicolumn{1}{r|}{9.43\%}  & Tokyo               & 3.83\%        \\
Moscow                      & \multicolumn{1}{r|}{8.18\%}  & Central Time                      & 3.75\%        \\
Bern                        & \multicolumn{1}{r|}{2.56\%}  & Eastern Time                      & 2.10\%        \\
Minsk                       & \multicolumn{1}{r|}{2.06\%}  & Seoul                  & 1.97\%        \\
Yerevan                     & \multicolumn{1}{r|}{1.96\%}  & Brasilia                       & 1.97\%        \\
Nairobi                     & \multicolumn{1}{r|}{1.52\%}  & Buenos Aires                         & 1.92\%        \\
Baku                        & \multicolumn{1}{r|}{1.29\%}  & Urumqi  & 1.50\%        \\ \bottomrule
\end{tabular}
}
\caption{Top 10 timezones (as \% of tweets).}
\label{tbl:top_timezones}
\end{table}

\begin{figure}[t]
\center
\subfigure[]{\includegraphics[width=0.49\columnwidth]{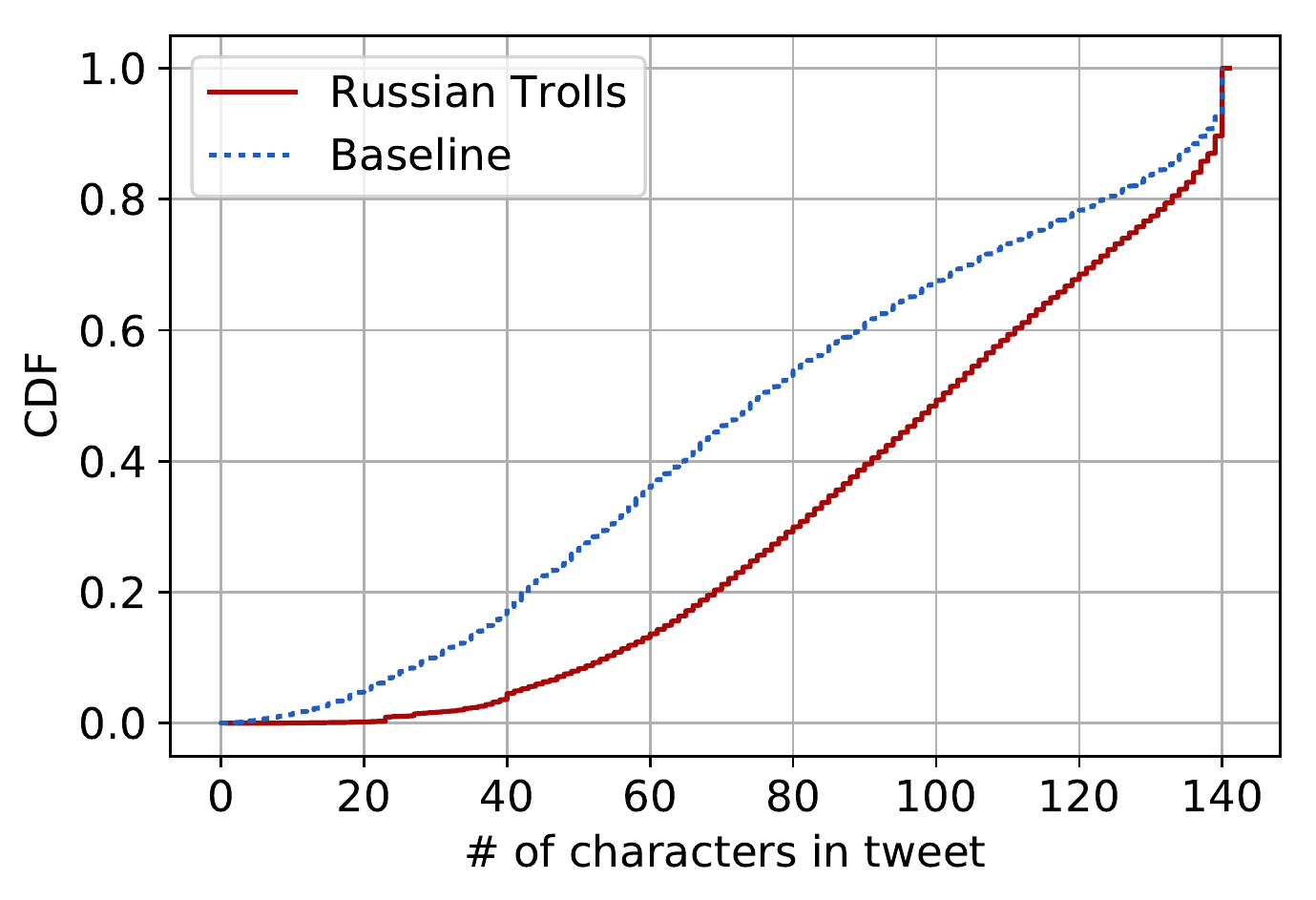}\label{subfig:cdf_characters}}
\subfigure[]{\includegraphics[width=0.49\columnwidth]{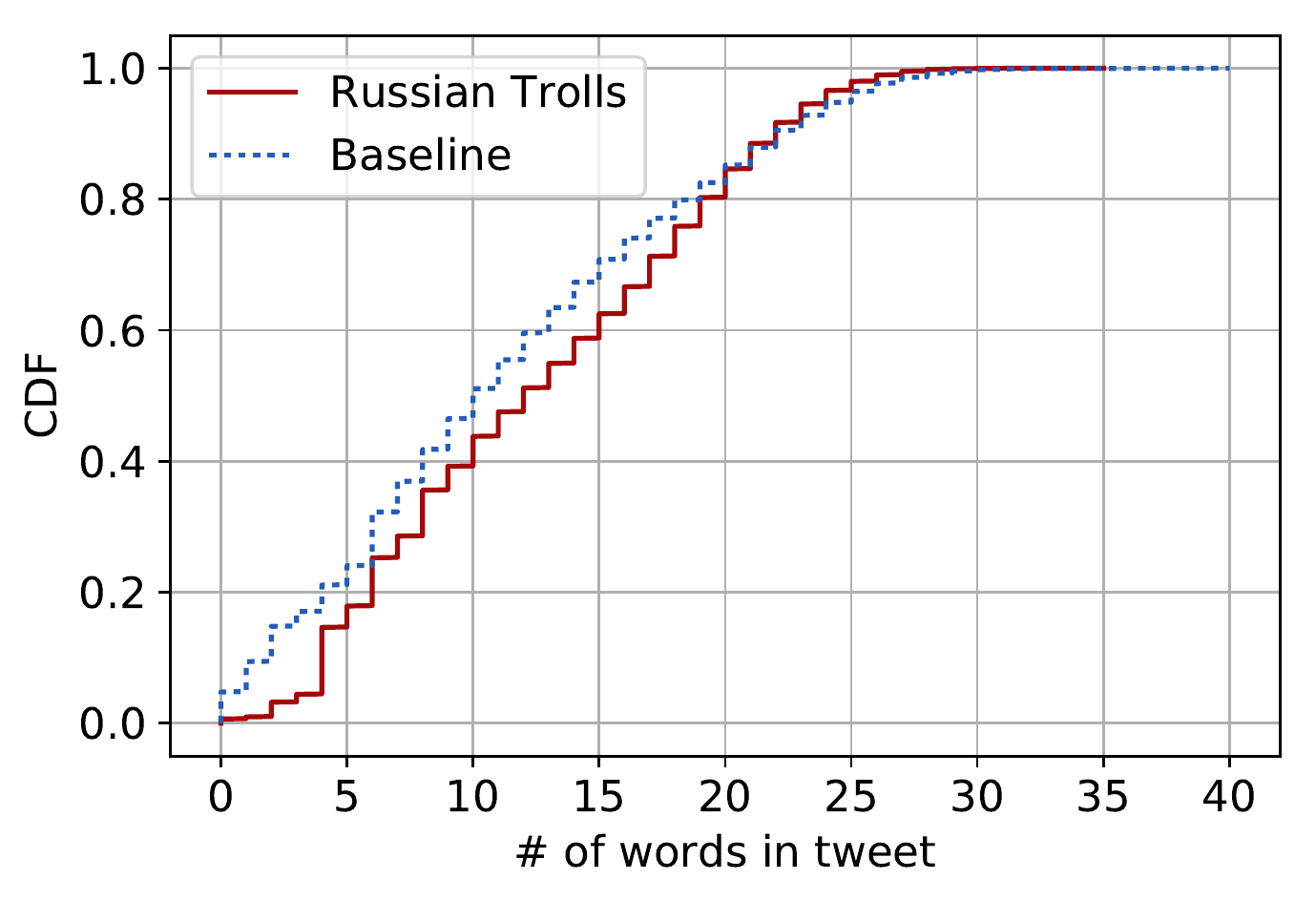}\label{subfig:cdf_words}}
  \caption{CDF of the number of (a) characters and (b) words in each tweet.}
\label{fig:cdfs_text}
\end{figure}

\descr{Timezone.} We then study the timezone chosen by the users in their account setting.
In Table~\ref{tbl:top_timezones}, we report the top 10 timezones for each dataset, in terms of the corresponding tweet volumes.
Two thirds of the tweets by trolls appear to be from US timezones, while a substantial percentage (18\%) from Russian ones. %
Whereas, the baseline has a more diverse set of timezones, which seems to mirror findings from our location analysis.

We also check whether users change their timezone settings,
finding that 7\% of the Russian trolls do so two to three times.
The most popular changes are Berlin to Bern (18 times), Nairobi to Moscow (10), and Nairobi to Volgograd (10).%
By contrast, this almost never happens for baseline accounts.

\subsubsection{Content Analysis}
\descr{Text.} Next, we quantify the number of characters and words contained in each tweet,
and plot the corresponding CDF in Fig.~\ref{fig:cdfs_text}, finding that Russian trolls tend to post longer tweets.

\descr{Media.} We then assess whether Russian trolls use images and videos in a different way than random baseline users.
For Russian trolls (resp., baseline accounts), 66\% (resp., 73\%) of the tweets include no images, 32\% (resp., 18\%) exactly one image, and 2\% (resp., 9\%) more than one.
This suggests that Russian trolls disseminate a considerable amount of information via single-image tweets.
As for videos, only 1.5\% of the tweets from Russian trolls includes a video, as opposed to 6.4\% for baseline accounts.

\begin{table}[]
\centering
\setlength{\tabcolsep}{0.2em} %
\hspace*{-0.2cm}
\resizebox{0.9\columnwidth}{!}{
\begin{tabular}{lrlrlrlr}
\toprule
\multicolumn{4}{c}{\textbf{Trolls}}                                       & \multicolumn{4}{c}{\textbf{Baseline}}       \\
\textbf{Hashtag}          & \textbf{(\%)}  & \textbf{Hashtag}           & \multicolumn{1}{l}{\textbf{(\%)}}   & \textbf{Hashtag}      & \textbf{(\%)}  & \textbf{Hashtag}      & \textbf{(\%)}  \\ \midrule
news             & 7.2\% & US                & \multicolumn{1}{r|}{0.7\%} & iHeartAwards & 1.8\% & UrbanAttires & 0.6\% \\
politics         & 2.6\% & tcot              & \multicolumn{1}{r|}{0.6\%} & BestFanArmy  & 1.6\% & Vacature     & 0.6\% \\
sports           & 2.1\% & PJNET             & \multicolumn{1}{r|}{0.6\%} & Harmonizers  & 1.0\% & mPlusPlaces  & 0.6\% \\
business         & 1.4\% & entertainment     & \multicolumn{1}{r|}{0.5\%} & iOSApp       & 0.9\% & job          & 0.5\% \\
money            & 1.3\% & top               & \multicolumn{1}{r|}{0.5\%} & JouwBaan     & 0.9\% & Directioners & 0.5\% \\
world            & 1.2\% & topNews           & \multicolumn{1}{r|}{0.5\%} & vacature     & 0.9\% & JIMIN        & 0.5\% \\
MAGA             & 0.8\% & ISIS              & \multicolumn{1}{r|}{0.4\%} & KCA          & 0.9\% & PRODUCE101   & 0.5\% \\
health           & 0.8\% & Merkelmussbleiben & \multicolumn{1}{r|}{0.4\%} & Psychic      & 0.8\% & VoteMainFPP  & 0.5\% \\
local            & 0.7\% & IslamKills        & \multicolumn{1}{r|}{0.4\%} & RT           & 0.8\% & Werk         & 0.4\% \\
BlackLivesMatter & 0.7\% & breaking          & \multicolumn{1}{l|}{0.4\%} & Libertad2016 & 0.6\% & dts          & 0.4\% \\ \bottomrule
\end{tabular}
}
\caption{Top 20 hashtags in tweets from Russian trolls and baseline users.}
\label{tbl:top_hashtags}
\end{table}

\descr{Hashtags.} Our next step is to study the use of hashtags in tweets.
Russian trolls use at least one hashtag in 32\% of their tweets, compared to 10\% for the baseline.
Overall, we find 4.3K and 7.1K unique hashtags for trolls and random users, respectively,
with 74\% and 78\% of them only appearing once.
In Table~\ref{tbl:top_hashtags}, we report the top 20 hashtags for both datasets.
Trolls appear to use hashtags to disseminate news (7.2\%) and politics (2.6\%) related content, but also use several that might be indicators of propaganda and/or controversial topics, e.g., \#ISIS, \#IslamKills, and \#BlackLivesMatter.
For instance, we find some notable examples including: 
``We just have to close the borders, `refugees' are simple terrorists \#IslamKills'' on March 22, 2016,
``\#SyrianRefugees ARE TERRORISTS from \#ISIS \#IslamKills'' on March 22, 2016, and
``WATCH: Here is a typical \#BlackLivesMatter protester:  `I hope I kill all white babes!' \#BatonRouge $<$url$>$'' on July 17, 2016. 

We also study when these hashtags are used by the trolls, finding that most of them are well distributed over time. However, there are some interesting exceptions, e.g., with \#Merkelmussbleiben (a hashtag seemingly supporting Angela Merkel) and \#IslamKills. Specifically, tweets with the former appear exclusively on July 21, 2016,
while the latter on March 22, 2016, when a terrorist attack took place at Brussels airport.
These two examples illustrate how the trolls may be coordinating to push specific narratives on Twitter.

\descr{Mentions.} We find that 46\% of trolls' tweets include {\em mentions} %
 to 8.5K unique Twitter users.
This percentage is much higher for the random baseline users (80\%, to 41K users).
Table~\ref{tbl:top_mentions} reports the 20 top mentions we find in tweets from Russian trolls and baseline users.
We find several Russian accounts, like `leprasorium' (a popular Russian account that mainly posts memes) in 2\% of the mentions, as well as popular politicians like `realDonaldTrump' (0.6\%).
The practice of mentioning politicians on Twitter may reflect an underlying strategy to mutate users' opinions regarding a particular political topic, which has been also studied in previous work~\cite{conover2011political}.

\begin{table}[]
\centering
\setlength{\tabcolsep}{0.2em} %
\hspace*{-0.2cm}
\resizebox{0.9\columnwidth}{!}{
\begin{tabular}{lrlrlrlr}
\hline
\multicolumn{4}{c}{\textbf{Trolls}}                                                     & \multicolumn{4}{c}{\textbf{Baseline}}                               \\
\textbf{Mention} & \textbf{(\%)} & \textbf{Mention} & \multicolumn{1}{l}{\textbf{(\%)}} & \textbf{Mention} & \textbf{(\%)} & \textbf{Mention} & \textbf{(\%)} \\ \hline
leprasorium      & 2.1\%         & postsovet        & \multicolumn{1}{r|}{0.4\%}        & TasbihIstighfar  & 0.3\%         & RasSpotlights    & 0.1\%         \\
zubovnik         & 0.8\%         & DLGreez          & \multicolumn{1}{r|}{0.4\%}        & raspotlights     & 0.2\%         & GenderReveals    & 0.1\%         \\
realDonaldTrump  & 0.6\%         & DanaGeezus       & \multicolumn{1}{r|}{0.4\%}        & FunnyBrawls      & 0.2\%         & TattedChanel     & 0.1\%         \\
midnight         & 0.6\%         & ruopentwit       & \multicolumn{1}{r|}{0.3\%}        & YouTube          & 0.2\%         & gemvius          & 0.1\%         \\
blicqer          & 0.6\%         & Spoontamer       & \multicolumn{1}{r|}{0.3\%}        & Harry\_Styles    & 0.2\%         & DrizzyNYC\_\_    & 0.1\%         \\
gloed\_up        & 0.6\%         & YouTube          & \multicolumn{1}{r|}{0.3\%}        & shortdancevids   & 0.2\%         & August\_Alsina\_ & 0.1\%         \\
wylsacom         & 0.5\%         & ChrixMorgan      & \multicolumn{1}{r|}{0.3\%}        & UrbanAttires     & 0.2\%         & RihannaBITCH\_   & 0.1\%         \\
TalibKweli       & 0.4\%         & sergeylazarev    & \multicolumn{1}{r|}{0.3\%}        & BTS\_twt         & 0.2\%         & sexualfeed       & 0.1\%         \\
zvezdanews       & 0.4\%         & RT\_com          & \multicolumn{1}{r|}{0.3\%}        & KylieJenner\_NYC & 0.2\%         & PetsEvery30      & 0.1\%         \\
GiselleEvns      & 0.4\%         & kozheed          & \multicolumn{1}{l|}{0.3\%}        & BaddiessNation   & 0.2\%         & IGGYAZALEAoO    & 0.1\%         \\ \hline
\end{tabular}
}
\caption{Top 20 mentions in tweets from trolls and baseline.}
\label{tbl:top_mentions}
\end{table}

\begin{figure}[t]
\centering
\includegraphics[width=0.5\columnwidth]{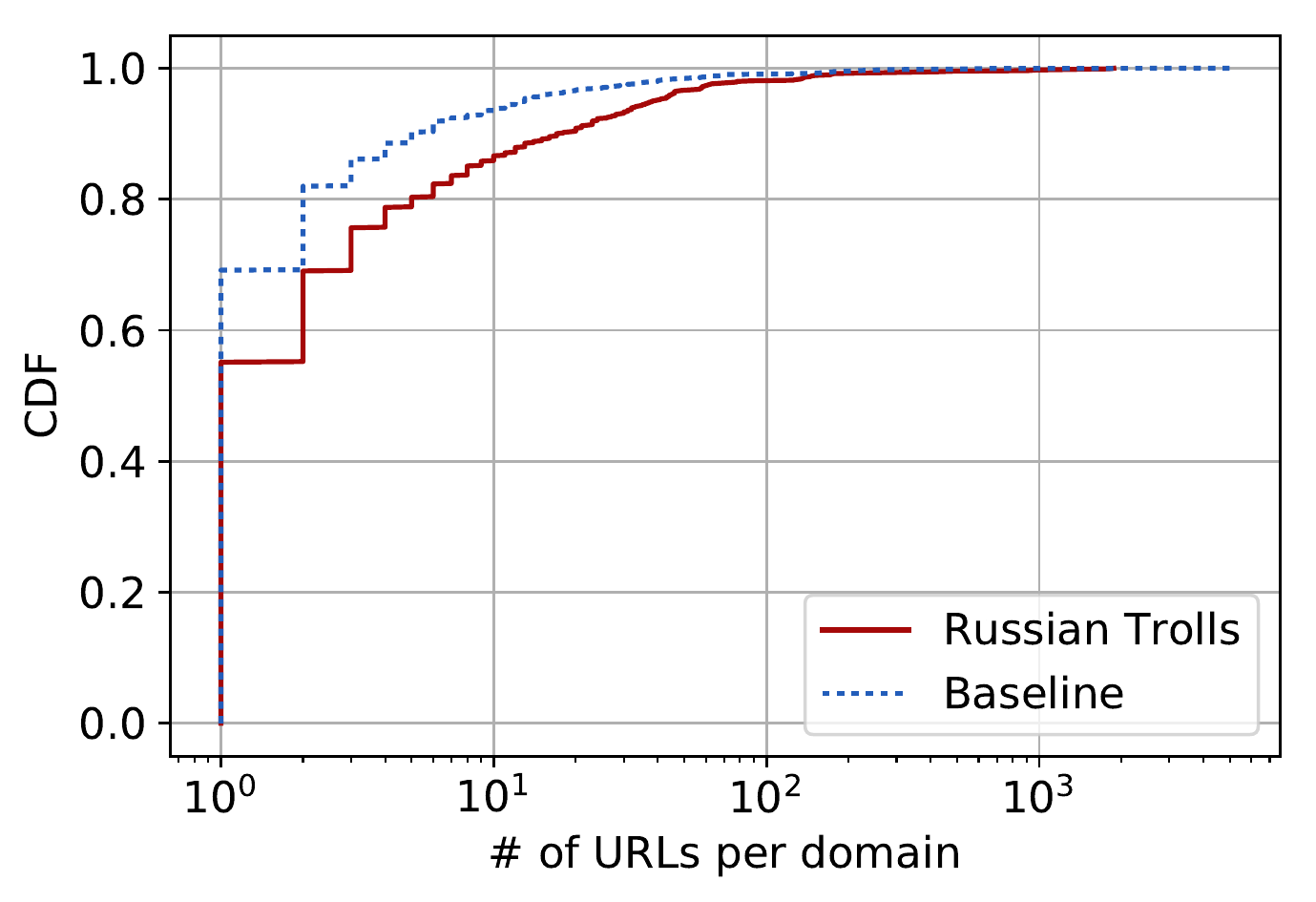}
  \caption{CDF of number of URLs per domain.}
\label{fig:cdf_domain_count}
\end{figure}

\descr{URLs.} We then analyze the URLs included in the tweets.
First of all, we note that 53\% of the trolls' tweets include at least a URL, compared to only 27\% for the random baseline.
There is an extensive presence of URL shorteners for both datasets, e.g., \url{bit.ly} (12\% for trolls and 26\% for the baseline) and \url{ift.tt} (10\% for trolls and 2\% for the baseline), therefore, in November 2017, we visit each URL to obtain the final URL after all redirections.
In Fig.~\ref{fig:cdf_domain_count}, we plot the CDF of the number of URLs per unique domain.
We observe that Russian trolls disseminate more URLs in their tweets compared to the baseline.
This might indicate that Russian trolls include URLs to increase their credibility and positive user perception; indeed, \cite{gupta2012credibility} show that adding a URL in a tweet correlates with higher credibility scores.
Also, in Table~\ref{tbl:cybersafety_top_domains}, we report the top 20 domains for both Russian trolls and the baseline.
Most URLs point to content within Twitter itself; 13\% and 35\%, respectively.
Links to a number of popular social networks like YouTube (1.8\% and 4.2\%, respectively) and Instagram (1.5\% and 1.9\%) appear in both datasets.
We also note that among the top 20 domains, there are also a number of news outlets linked from trolls' tweets, e.g., Washington Post (0.7\%), Seattle Times (0.7\%), and state-sponsored news outlets like RT (0.8\%) in trolls' tweets, but much less so from the baseline. %

\begin{table}[t]
\centering
\footnotesize
\resizebox{0.7\columnwidth}{!}{%
\begin{tabular}{lr|lr}
\toprule
\textbf{Domain (Trolls)} & \multicolumn{1}{r}{(\%)} &\textbf{Domain (Baseline)}  & \textbf{ (\%)} \\ \midrule
twitter.com     & {12.81\%} & twitter.com    &  35.51\%\\
reportsecret.com                       & {7.02\%}  & youtube.com      & 4.21\%\\
riafan.ru          & {3.42\%}  & vine.co       & 3.94\%\\
politexpert.net      & {2.10\%}  & factissues.com      & 3.24\% \\
youtube.com      & {1.88\%}  & blogspot.com.cy     & 1.92\%      \\
vk.com               & {1.58\%}  & instagram.com       &   1.90\%  \\
instagram.com             & {1.53\%}  & facebook.com           & 1.68\% \\
yandex.ru                   & {1.50\%}  & worldstarr.info     &     1.47\%  \\
infreactor.org                     & {1.36\%}  & trendytopic.info         &  1.39\%\\
cbslocal.com    & {1.35\%}  & minibird.jp         & 1.25\%\\
livejournal     & {1.35\%} & yaadlinksradio.com   &  1.24\%\\
nevnov.ru                       & {1.07\%}  & soundcloud.com      & 1.24\%\\
ksnt.com          & {1.01\%}  & linklist.me       & 1.15\%\\
kron4.com      & {0.93\%}  & twimg.com     & 1.09\% \\
viid.me      & {0.93\%}  & appparse.com     & 1.08\%      \\
newinform.com              & {0.89\%}  & cargobayy.net       &   0.88\%  \\
inforeactor.ru              & {0.84\%}  & virralclub.com          & 0.84\% \\
rt.com                & {0.81\%}  & tistory.com     &     0.50\%  \\
washigntonpost.com                      & {0.75\%}  & twitcasting.tv         &  0.49\%\\
seattletimes.com    & {0.73\%}  & nytimes.com          & 0.48\%\\ \bottomrule
\end{tabular}
}
\caption{Top 20 domains included in tweets from Russian trolls and baselines users.}
\label{tbl:cybersafety_top_domains}
\end{table}

\begin{figure}[t!]
\center
\subfigure[]{\includegraphics[width=0.49\columnwidth]{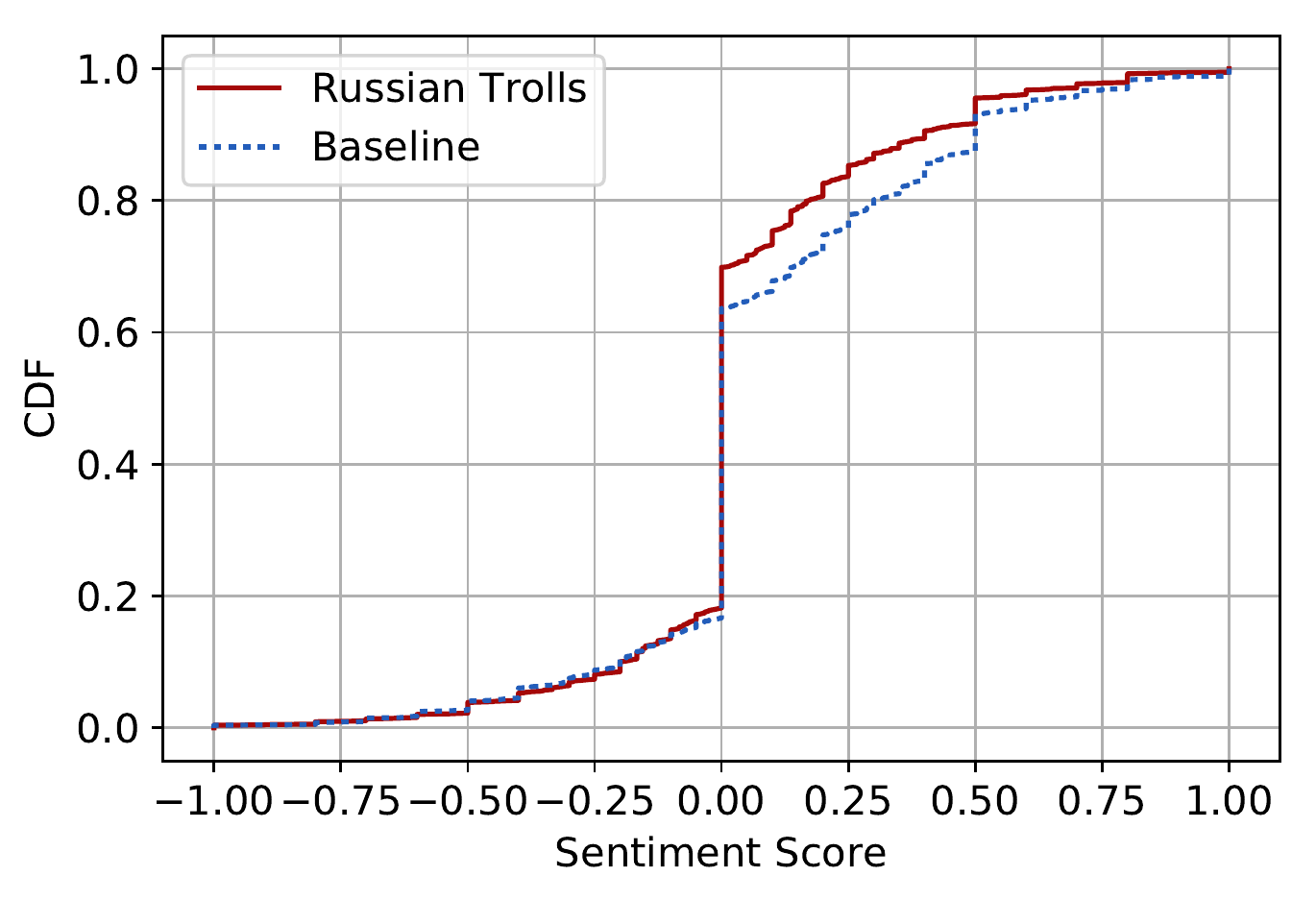}\label{subfig:cdf_sentiment}}
\subfigure[]{\includegraphics[width=0.49\columnwidth]{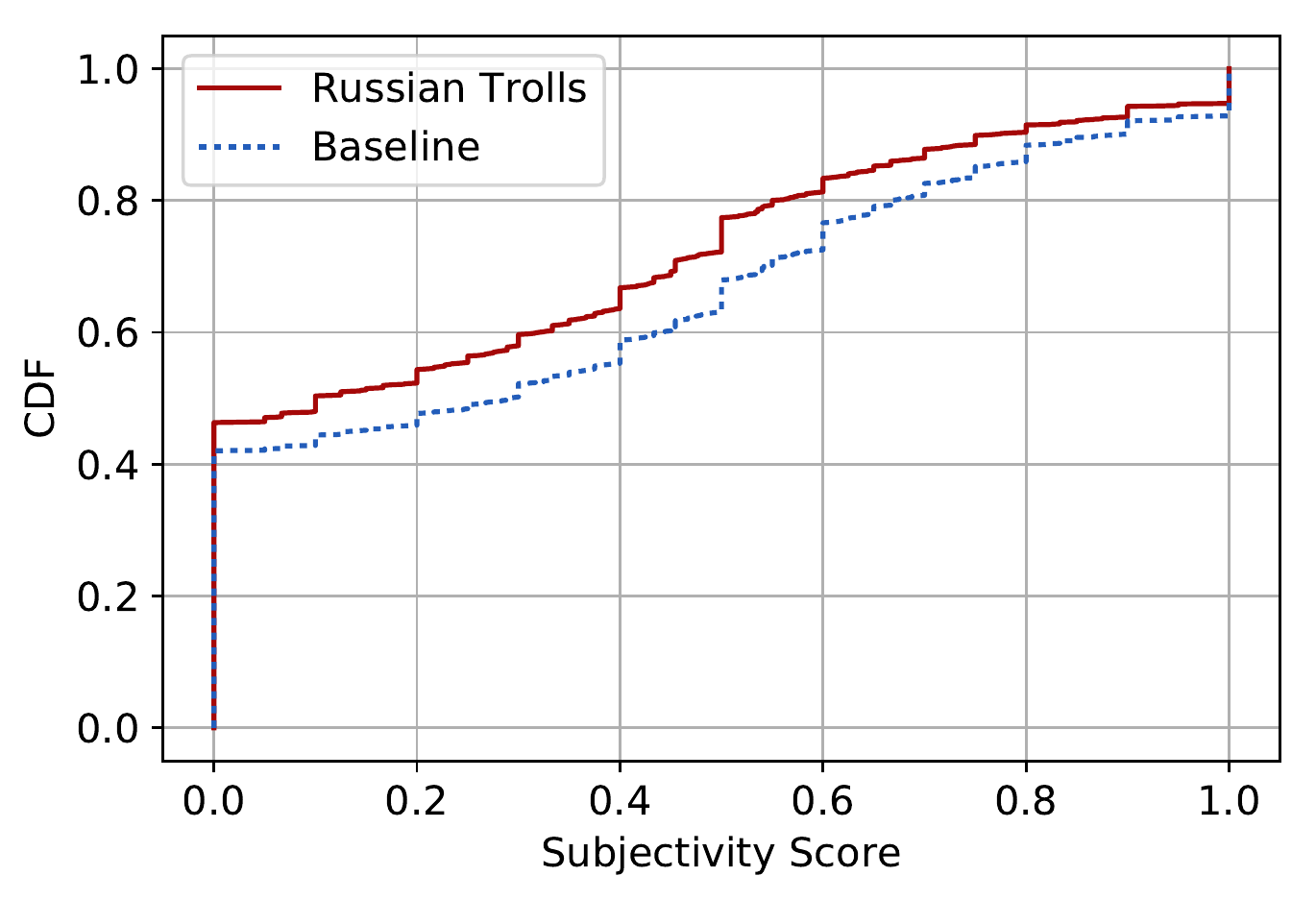}\label{subfig:cdf_subjectivity}}
\caption{CDF of sentiment and subjectivity scores for tweets of Russian trolls and random users.}
\label{fig:cdf_sentiment_subjectivity}
\end{figure}

\descr{Sentiment analysis.} Next, we assess the sentiment and subjectivity of each tweet for both datasets
using the Pattern library~\cite{smedt2012pattern}.
Fig.~\ref{subfig:cdf_sentiment} plots the CDF of the sentiment scores of tweets posted by Russian trolls and our baseline users.
We observe that 30\% of the tweets from Russian trolls have a positive sentiment, and 18\% negative.
These scores are not too distant from those of random users where 36\% are positive and 16\% negative, however, Russian trolls exhibit a unique behavior in terms of sentiment, as a two-sample Kolmogorov-Smirnov test unveils significant differences between the distributions ($p < 0.01$).
Overall, we observe that Russian trolls tend to be more negative/neutral, while our baseline is more positive.
We also compare  subjectivity scores (Fig.~\ref{subfig:cdf_subjectivity}), finding that  tweets from trolls tend to be more subjective; again, we perform significance tests revealing differences between the two distributions ($p < 0.01$).

\begin{table*}[]
\centering
\resizebox{\textwidth}{!}{
\begin{tabular}{@{}rl|rl@{}}
\toprule
{\bf Topic} & \multicolumn{1}{l}{\textbf{Terms (Trolls)}} & {\textbf{Topic}} & \textbf{Terms (Baseline)}                                                               \\ \midrule
1  & trump, black, people, really, one, enlist, truth, work, can, get                 & 1           & want, can, just, follow, now, get, see, don, love, will                                 \\
2  & trump, year, old, just, run, obama, breaking, will, news, police                 & 2           & 2016, july, come, https, trump, social, just, media, jabberduck, get                    \\
3  & new, trump, just, breaking, obamacare, one, sessions, senate, politics, york     & 3           & happy, best, make, birthday, video, days, come, back, still, little                     \\
4  & man, police, news, killed, shot, shooting, woman, dead, breaking, death          & 4           & know, never, get, love, just, night, one, give, time, can                               \\
5  & trump, media, tcot, just, pjnet, war, like, video, post, hillary                 & 5           & just, can, everyone, think, get, white, fifth, veranomtv2016, harmony, friends          \\
6  & sports, video, game, music, isis, charlottesville, will, new, health, amb        & 6           & good, like, people, lol, don, just, look, today, said, keep                             \\
7  & can, don, people, want, know, see, black, get, just, like                        & 7           & summer, seconds, team, people, miss, don, will, photo, veranomtv2016, new               \\
8  & trump, clinton, politics, hillary, video, white, donald, president, house, calls & 8           & like, twitter, https, first, can, get, music, better, wait, really                      \\
9  & news, world, money, business, new, one, says, state, 2016, peace                 & 9           & dallas, right, fuck, vote, police, via, just, killed, teenchoice, aldubmainecelebration \\
10 & now, trump, north, korea, people, right, will, check, just, playing              & 10          & day, black, love, thank, great, new, now, matter, can, much                             \\ \bottomrule
\end{tabular}
}
\caption{Terms extracted from LDA topics of tweets from Russian trolls and baseline users.}
\label{tbl:lda_topics}
\end{table*}

\descr{LDA analysis.} %
We also use the Latent Dirichlet Allocation (LDA) model to analyze
tweets' semantics.
We train an LDA model for each of the datasets and extract 10 distinct topics with 10 words,
as reported in Table~\ref{tbl:lda_topics}.
Overall, topics from Russian trolls refer to specific world events (e.g., Charlottesville) as well as specific news related to politics (e.g., North Korea and Donald Trump).
By contrast, topics extracted from the random sample are more general %
(e.g., tweets regarding birthdays).

\subsubsection{Account Evolution}

\descr{Screen name changes. }
Previous work~\cite{mariconti2017s} has shown that malicious accounts often change their screen name in order to assume different identifies.
Therefore, we investigate whether trolls show a similar behavior, as they might  change the narrative with which they are attempting to influence public opinion.
Indeed, we find that 9\% of the accounts operated by trolls change their screen name, up to 4 times during the course of our dataset.
Some examples include changing screen names from ``OnlineHouston'' to ``HoustonTopNews,'' or  ``Jesus Quintin Perez'' to ``WorldNewsPolitics,''
in a clear attempt to pose as news-related accounts.
In our baseline, we find that 19\% of the accounts changed their Twitter screen names, up to 11 times during our dataset; highlighting that changing screen names is a common behavior of Twitter users in general.

\begin{figure}[t]
\center
\subfigure[]{\includegraphics[width=0.49\columnwidth]{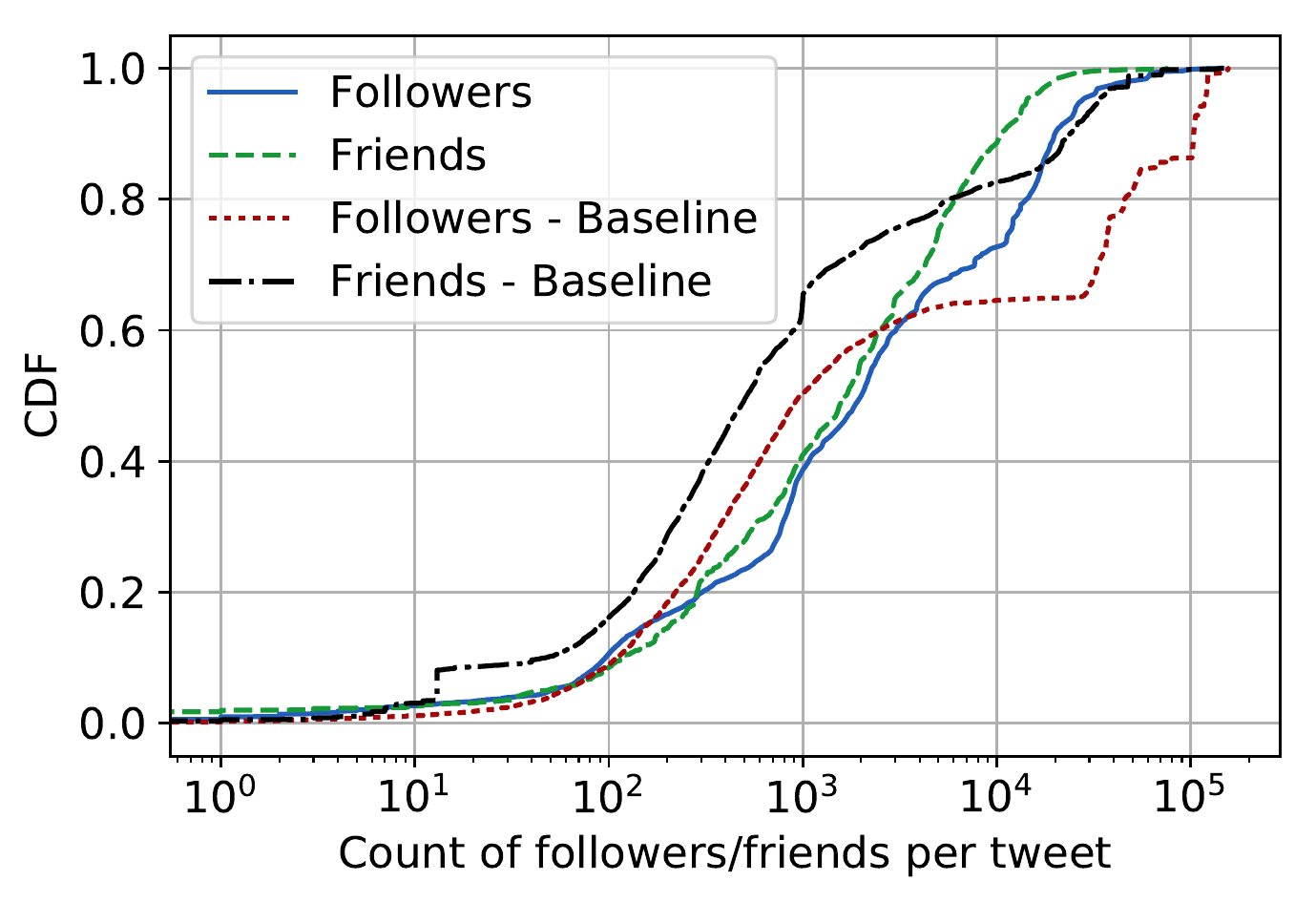}\label{subfig:cybersafety_cdf_followers}}
\subfigure[]{\includegraphics[width=0.49\columnwidth]{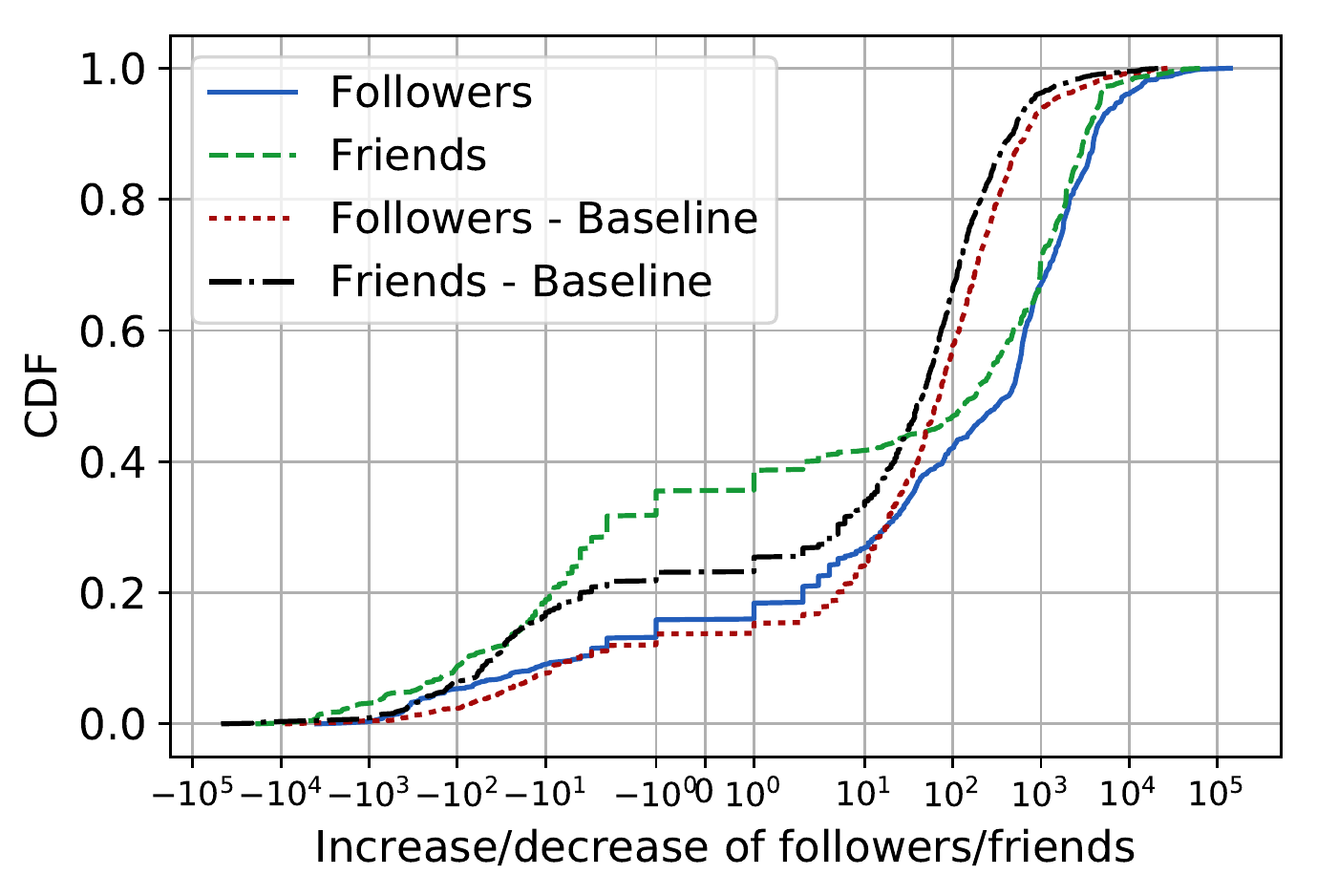}\label{subfig:cybersafety_cdf_followers_changes}}
\caption{CDF of the number of (a) followers/friends for each tweet and (b) increase in followers/friends for each user from the first to the last tweet.}
\label{fig:cdfs_domains_urls}
\end{figure}

\descr{Followers/Friends.} Next, we look at the number of followers and friends (i.e., the accounts one follows) of the Russian trolls, as this is an indication of the overall impact of a tweet.
In Fig.~\ref{subfig:cybersafety_cdf_followers}, we plot the CDF of the number of followers per tweet measured at the time of that tweet.
On average, Russian trolls have 7K followers and 3K friends, while our baseline has 25K followers and 6K friends.
We also note that in both samples, tweets reached a large number of Twitter users; at least 1K followers, with peaks up to 145K followers.
These results highlight that Russian trolls have a non-negligible number of followers, which can assist in pushing specific narratives to a much greater number of Twitter users.
We also assess the evolution of the Russian trolls in terms of the number of their followers and friends.
To this end, we get the follower and friend count for each user on their first and last tweet
and calculate the difference.
Fig.~\ref{subfig:cybersafety_cdf_followers_changes} plots the CDF of the increase/decrease of the followers and friends for each troll as well as random user in our baseline.
We observe that, on average, Russian trolls increase their number of followers and friends by 2,065 and 1,225, respectively, whereas for the baseline we observe an increase of 425 and 133 for followers and friends, respectively.
This suggests that Russian trolls work hard to increase their reachability within Twitter.

\begin{figure}[t]
\centering
\includegraphics[width=0.5\columnwidth]{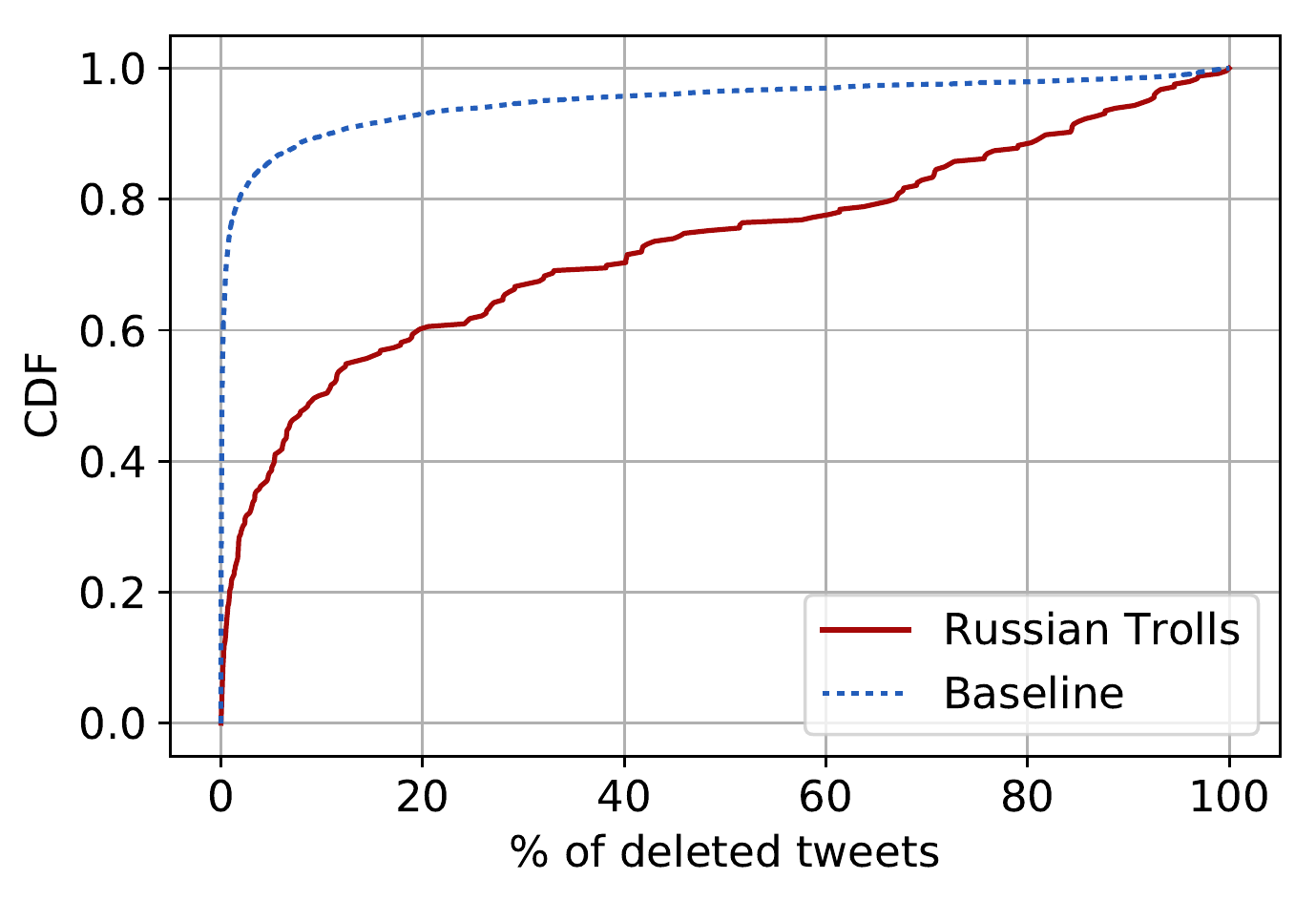}
\caption{CDF of the number of deleted tweets per observe deletion.}
\label{fig:cdf_deleted_tweets_user}
\end{figure}

\begin{figure}[t]
\centering
\includegraphics[width=0.9\columnwidth]{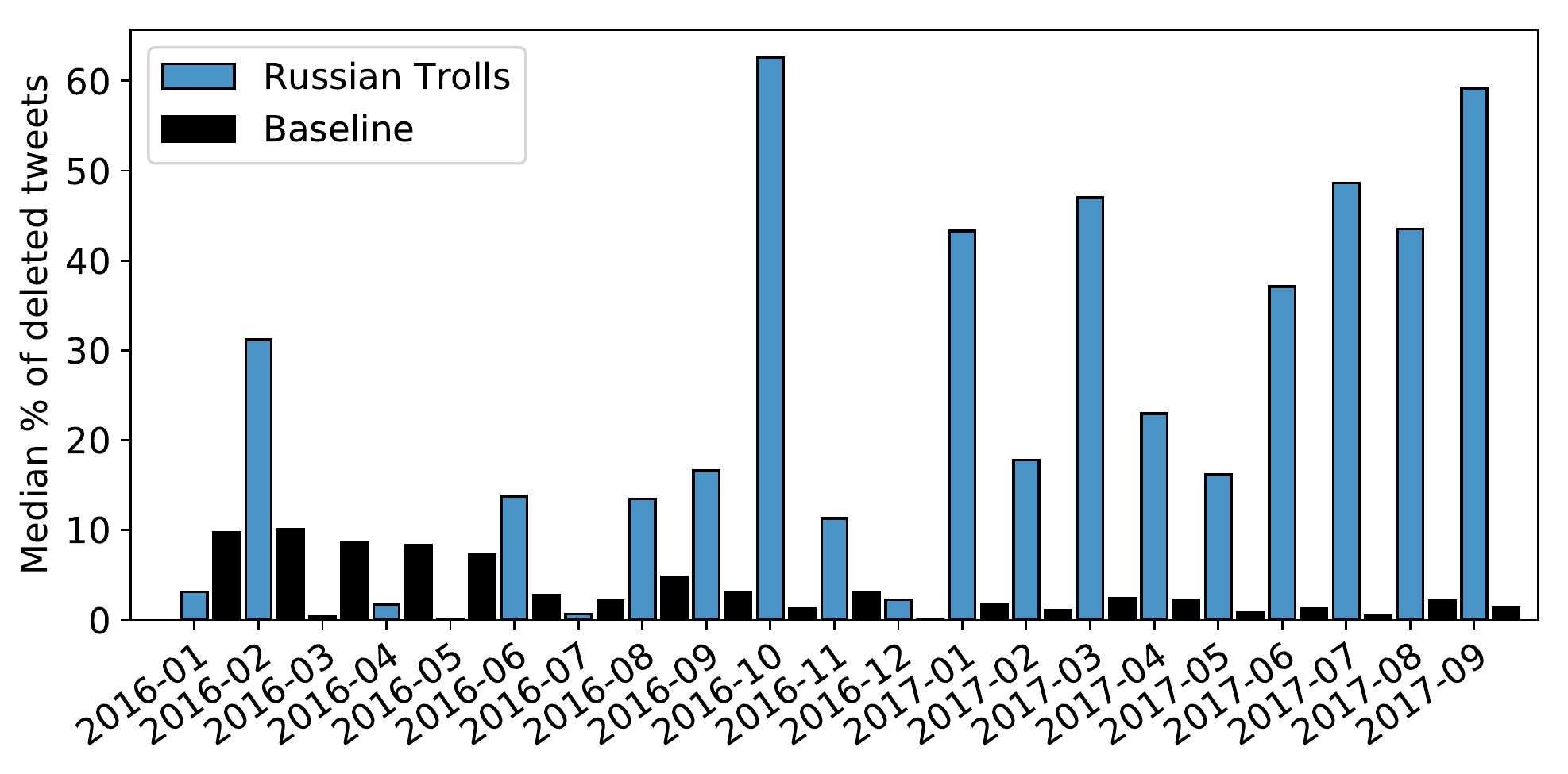}
  \caption{Average percentage of observed deletions per month. }
\label{fig:bc_deleted_tweets_month}
\end{figure}

\descr{Tweet Deletion.} Arguably, a reasonable strategy to avoid detection after posting tweets that aim to manipulate other users might be to delete them.
This is particularly useful when troll accounts change their identity and need to modify the narrative that they use to influence public opinion.
With each tweet, the Streaming API returns the total number of available tweets a user has up to that time.
Retrieving this count allows us to observe if a user has deleted a tweet, and around what period; we call this an ``observed deletion.''
Recall that our dataset is based on the 1\% sample of Twitter, thus, we can only estimate, in a conservative way, how many tweets are deleted; specifically, in between subsequent tweets, a user may have deleted and posted tweets that we do not observe. %
In Fig.~\ref{fig:cdf_deleted_tweets_user}, we plot the CDF of the number of deleted tweets per observed deletion. We observe that 13\% of the Russian trolls delete some of their tweets, with a median percentage of tweet deletion equal to 9.7\%.
Whereas, for the baseline set, 27\% of the accounts delete at least one tweet, but the median percentage is 0.1\%.
This means that the trolls delete their tweets in batches, possibly trying to cover their tracks or get a clean slate, while random users make a larger number of deletions but only a small percentage of their overall tweets, possibly because of typos.
We also report the distribution, over each month, of tweet deletions in Fig.~\ref{fig:bc_deleted_tweets_month}.
Specifically, we report the mean of the percentages for all observed deletions in our datasets.
Most of the tweets from Russian trolls are deleted in October 2016,
suggesting that these accounts attempted to get a clean slate a few months before the 2016 US elections.

\subsubsection{Case Study}

While the previous results provide a quantitative characterization of Russian trolls behavior, we believe there is value showing a concrete example of the behavior exhibited and how techniques played out.
We start on May 15, 2016, where the troll with screen name `Pen\_Air', was posing as a news account via its profile description: ``National American news.''
On September 8, 2016 as the US presidential elections approached, `Pen\_Air' became a Trump supporter, changing its screen name to `Blacks4DTrump' with a profile description of ``African-Americans stand with Trump to make America Great Again!''
Over the next 11 months, the account's tweet count grew from 49 to 642 while its follower count rose from 1.2K to \emph{nearly} 9K.
Then, around August 18, 2017, the account was seemingly repurposed.
Almost all of its previous tweets were deleted (the account's tweet count dropped to 35), it gained a new screen name (`southlonestar2'), and was now posing as a ``Proud American and TEXAN patriot! Stop ISLAM and PC. Don't mess with Texas'' according to its profile description.
When examining the accounts tweets, we see that most are clearly related to politics, featuring blunt right-wing attacks and ``talking points.''
For example, 
``Mr. Obama! Maybe you bring your girls and leave them in the bathroom with a grown man! \#bathroombill \#NObama $<$url$>$'' on May 15, 2016,
``\#HiLIARy has only two faces! And I hate both! \#NeverHillary \#Hillaryliesmatter $<$url$>$'' on May 19, 2016, and ``RT @TEN\_GOP: WikiLeaks \#DNCLeaks confirms something we all know: system is totally rigged! \#NeverHillary $<$url$>$.'' on July 22, 2016.

\subsubsection{Take-aways}

In summary, our analysis leads to the following observations.
First, we find evidence that trolls were actively involved in the dissemination of content related to world news and politics, as well as propaganda content regarding various topics such as ISIS and Islam. %
Moreover, several Russian trolls were created or repurposed a few weeks before notable world events, including the Republican National Convention meeting or the Charlottesville rally.
We also find that the trolls deleted a substantial amount of tweets in batches and overall made substantial changes to their accounts during the course of their lifespan.
Specifically, they changed their screen names aiming to pose as news outlets,
experienced significant rises in the numbers of followers and friends, etc.
Furthermore, our location analysis shows that Russian trolls might have tried to manipulate users  located in the USA, Germany, and possibly in their own country (i.e., Russia), by appearing to be located in those countries.
Finally, the fact that these accounts were active up until their recent suspension also highlights the need to develop more effective tools to detect such actors.

\subsection{Remarks}

In this work, we analyzed the behavior and use of the Twitter platform by Russian trolls during the course of 21 months.
We showed that Russian trolls exhibited interesting differences when compared with a set of random users, actively disseminated politics-related content, adopted multiple identities during their account's lifespan, and that they aimed to increase their impact on Twitter by increasing their followers.

\section[A comprehensive analysis of Russian and Iranian trolls on Twitter and Reddit]{A comprehensive analysis of Russian and Iranian trolls on Twitter and Reddit and their influence on the Web}

\subsection{Motivation}

In this work, we are motivated by the fact that many aspects of state-sponsored disinformation remain unclear, e.g., how do state-sponsored trolls operate? What kind of content do they disseminate? How does their behavior change over time? And, more importantly, is it possible to quantify the influence they have on the overall information ecosystem on the Web?

Here, we aim to address these questions, by relying on two different sources of ground truth data about state-sponsored actors. First, we use 10M tweets posted by Russian and Iranian trolls between 2012 and 2018~\cite{twitter_russian_iranians_dataset}.
Second, we use a list of 944 Russian trolls, identified by Reddit, and find all their posts between 2015 and 2018~\cite{reddit_dataset_trolls}.
We analyze the two datasets across several axes in order to understand their behavior and how it changes over time, their targets, and the content they shared.
For the latter, we leverage word embeddings to understand in what context specific words/hashtags are used and shed light to the ideology of the trolls.
Also, we use Hawkes Processes~\cite{linderman2014} to model the influence that the Russian and Iranian trolls had over multiple Web communities; namely, Twitter, Reddit, 4chan's Politically Incorrect board (\dspol)~\cite{hine2016longitudinal}, and Gab~\cite{zannettou2018gab}.

\descr{Main findings.}
Our study leads to several key observations:
\begin{enumerate}%
\item Our influence estimation results reveal that Russian trolls were extremely influential and efficient in spreading URLs on Twitter. Also, by comparing Russian to Iranian trolls, we find that Russian trolls were more efficient and influential in spreading URLs on Twitter, Reddit, Gab, but not on \dspol.
\item By leveraging word embeddings, we find ideological differences between Russian and Iranian trolls. For instance, we find that Russian trolls were pro-Trump, while Iranian trolls were anti-Trump.
\item We find evidence that the Iranian campaigns were motivated by real-world events. Specifically, campaigns against France and Saudi Arabia coincided with real-world events that affect the relations between these countries and Iran.
\item We observe that the behavior of trolls varies over time. We find substantial changes in the use of language and Twitter clients over time for both Russian and Iranian trolls. These insights allow us to understand the targets of the orchestrated campaigns for each type of trolls over time.
\item We find that the topics of discussion vary across Web communities. For example, we find that Russian trolls on Reddit were extensively discussing about cryptocurrencies, while this does not apply in great extent for the Russian trolls on Twitter.
\end{enumerate}

Finally, we make our source code publicly available~\cite{code} for reproducibility purposes and to encourage researchers to further work on understanding other types of state-sponsored trolls on Twitter (i.e., on January 31, 2019, Twitter released data related to trolls originating from Venezuela and Bangladesh~\cite{new_twitter_dataset}).

\subsection{Troll Datasets} \label{sec:datasets}

In this section, we describe our dataset of Russian and Iranian trolls on Twitter and Reddit.

\begin{table}[]
\centering
\resizebox{0.75\columnwidth}{!}{%
\begin{tabular}{llrrr}
\hline
\multicolumn{1}{c}{\textbf{Platform}} & \multicolumn{1}{c}{\textbf{Origin of trolls}} & \multicolumn{1}{c}{\textbf{\# trolls}} & \multicolumn{1}{c}{\textbf{\begin{tabular}[c]{@{}c@{}}\# trolls\\ with tweets/posts\end{tabular}}} & \multicolumn{1}{l}{\textbf{\# of tweets/posts}} \\ \hline
\textbf{Twitter} & \textbf{Russia} & 3,836 & 3,667 & 9,041,308 \\
 & \textbf{Iran} & 770 & 660 & 1,122,936 \\ \hline
\textbf{Reddit} & \textbf{Russia} & 944 & 335 & 21,321 \\ \hline
\end{tabular}
}
\caption{Overview of Russian and Iranian trolls on Twitter and Reddit. We report the overall number of identified trolls, the trolls that had at least one tweet/post, and the overall number of tweets/posts.}
\label{tbl:trolls_datasets}
\end{table}

\descr{Twitter.} On October 17, 2018, Twitter released a large dataset of Russian and Iranian troll accounts~\cite{twitter_russian_iranians_dataset}.
Although the exact methodology used to determine that these accounts were
state-sponsored trolls is unknown, based on the most recent Department of Justice indictment~\cite{trollsindictment},
the dataset appears to have been constructed in a manner that we can assume essentially no false positives, while we cannot make any postulation about false negatives.
Table~\ref{tbl:trolls_datasets} summarizes the troll dataset.

\descr{Reddit.} On April 10, 2018, Reddit released a list of 944 accounts which they determined were Russian state-sponsored trolls~\cite{reddit_dataset_trolls}.
We recover the submissions, comments, and account details for these accounts using two mechanisms:
1)~Reddit dumps provided by Pushshift~\cite{pushshift}; and 2)~crawling the user pages of those accounts. %
Although omitted for lack of space, we note that the union of these two datasets reveals some gaps in both, likely due to a combination of subreddit moderators removing posts or the troll users themselves deleting them, which would affect the two datasets in different ways.
In any case, we merge the two datasets, with Table~\ref{tbl:trolls_datasets} describing the final dataset.
Note that only about one third (335) of the accounts released by Reddit had at least one submission or comment in our dataset.
We suspect the rest were either completely missed by our data sources, or, more likely, were used as dedicated voting accounts used in an effort to push (or bury) specific content.

\subsection{Analysis} \label{sec:analysis}
In this section, we present an in-depth analysis of the activities and the behavior of Russian and Iranian trolls on Twitter and Reddit.

\subsubsection{Accounts Characteristics}

\begin{figure}[t!]
\centering
\includegraphics[width=\columnwidth]{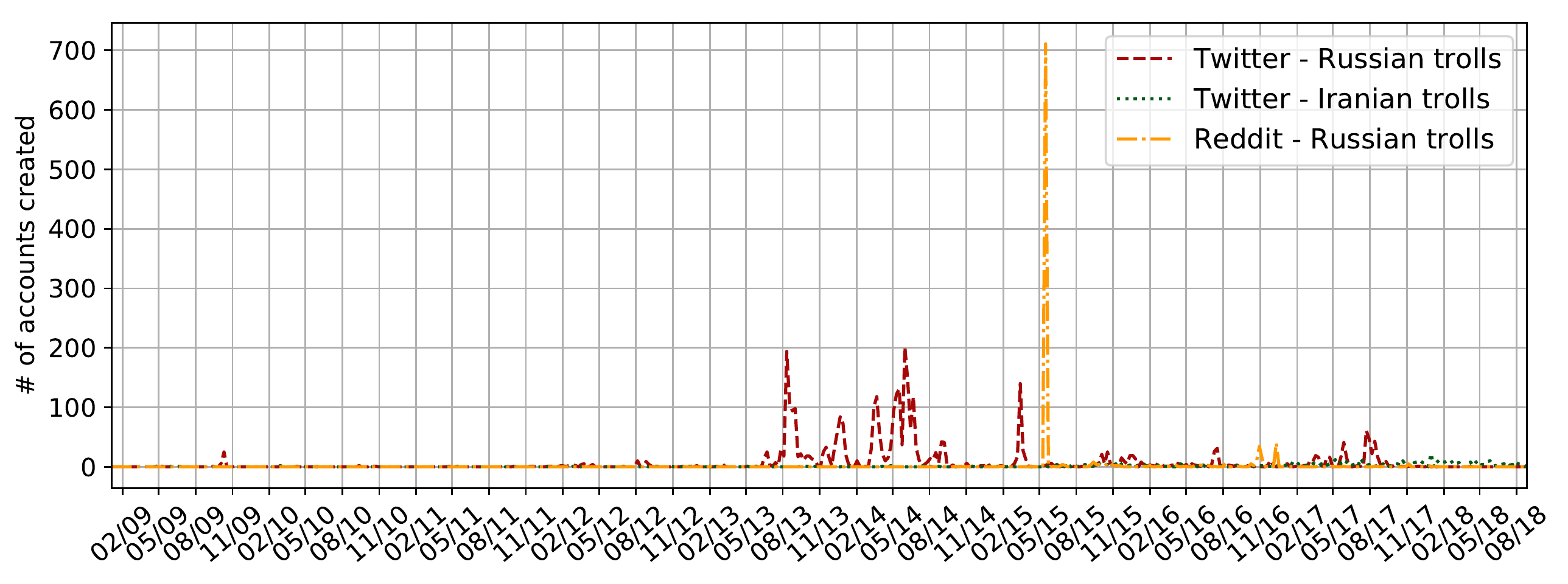}
\caption{Number of Russian and Iranian troll accounts created per week.}
\label{fig:counts_created}
\end{figure}

\begin{table}[t!]
  \centering
  \resizebox{\columnwidth}{!}{
  \setlength{\tabcolsep}{0.4em} %
\begin{tabular}{rrrrrlrl}
\hline
\multicolumn{4}{c}{\textbf{Russian troll on Twitter}} & \multicolumn{4}{c}{\textbf{Iranian trolls on Twitter}} \\ \hline
\textbf{Word} & \multicolumn{1}{l}{\textbf{(\%)}} & \textbf{Bigrams} & \multicolumn{1}{l}{\textbf{(\%)}} & \textbf{Word} & \textbf{(\%)} & \textbf{Bigrams} & \textbf{(\%)} \\ \hline
follow & 7.7\% & follow me & \multicolumn{1}{r|}{6.4\%} & journalist & 3.6\% & human rights & 1.6\% \\
love & 4.8\% & breaking news & \multicolumn{1}{r|}{0.8\%} & news & 3.2\% & independent news & 1.4\% \\
life & 4.5\% & donald trump & \multicolumn{1}{r|}{0.7\%} & independent & 2.8\% & news media & 1.4\% \\
trump & 4.4\% & lokale nachrichten & \multicolumn{1}{r|}{0.6\%} & lover (in Farsi) & 2.6\% & media organization & 1.4\% \\
conservative & 4.3\% & nachrichten aus & \multicolumn{1}{r|}{0.6\%} & social & 2.6\% & organization aim & 1.4\% \\
news & 3.4\% & hier kannst & \multicolumn{1}{r|}{0.6\%} & politics & 2.6\% & aim inspire & 1.4\% \\
maga & 3.4\% & kannst du & \multicolumn{1}{r|}{0.6\%} &  media & 2.4\% &inspire action & 1.4\% \\
{\cyr lyublyu} & 2.4\% & du wichtige & \multicolumn{1}{r|}{0.6\%} & love & 2.2\% & action likes & 1.4\% \\
will & 2.4\% & wichtige und & \multicolumn{1}{r|}{0.6\%} &  justice& 2.0\% & likes social & 1.4\% \\
proud & 2.2\% & und aktuelle & \multicolumn{1}{r|}{0.6\%} &  low (in Farsi)& 2.0\% & social justice & 1.4\% \\ \hline
\end{tabular}
  }
  \caption{Top 10 words and bigrams found in the descriptions of Russian and Iranian trolls on Twitter.}
  \label{tbl:account_desc}
\end{table}

First we explore when the accounts appeared, what they posed as, and how many followers/friends they had on Twitter.

\descr{Account Creation.} Fig.~\ref{fig:counts_created} plots the Russian and Iranian troll accounts creation dates on Twitter and Reddit.
We observe that the majority of Russian troll accounts were created around the time of the Ukrainian conflict: 80\% of have an account creation date earlier than 2016.
That said, there are some meaningful peaks in account creation during 2016 and 2017.
57 accounts were created between July 3-17, 2016, which was right before the start of the Republican National Convention (July 18-21) where Donald Trump was named the Republican nominee for President~\cite{rnc_meeting} .
Later, 190 accounts were created between July, 2017 and August, 2017, during the run up to the infamous Unite the Right rally in Charlottesville~\cite{charlotesville}.
Taken together, this might be evidence of coordinated activities aimed at manipulating users' opinions on Twitter with respect to specific events.
This is further evidenced when examining the Russian trolls on Reddit: 75\% of Russian troll accounts on Reddit were created in a single massive burst in the first half of 2015.
Also, there are a few smaller spikes occurring just prior to the 2016 US Presidential election.
For the Iranian trolls on Twitter we observe that they are much ``younger,'' with the larger bursts of account creation \emph{after} the 2016 US Presidential election.

\descr{Account Information.}
To avoid being obvious, state sponsored trolls might attempt to present a persona that masks their true nature or otherwise ingratiates themselves to their target audience.
By examining the profile description of trolls we can get a feeling for how they might have cultivated this persona.
In Table~\ref{tbl:account_desc}, we report the top ten words and bigrams that appear in profile descriptions of trolls on Twitter.
Note that we do this only for Twitter trolls as we do not have descriptions for Reddit accounts.
From the table we see that a relatively large number of Russian trolls pose as news outlets, with ``news'' (1.3\%) and ``breaking news'' (0.8\%) appearing in their description.
Further, they seem to use their profile description to more explicitly increase their reach on Twitter, by nudging users to follow them (e.g., ``follow me'' appearing in almost 6.4\% of profile descriptions).
Finally, 3.4\% of the Russian trolls describe themselves as Trump supporters: see ``trump'' (4.4\%) and ``maga'' (3.4\%) terms.
Iranian trolls are even more likely to pose as news outlets or journalists: 3.6\% have ``journalist'' and 3.2\% have ``news'' in their profile descriptions.
This highlights that accounts that pose as news outlets may in fact be accounts controlled by state-sponsored actors, hence regular users should critically think in order to assess whether the account is credible or not.

\begin{figure}[t]
\center
\subfigure[]{\includegraphics[width=0.49\columnwidth]{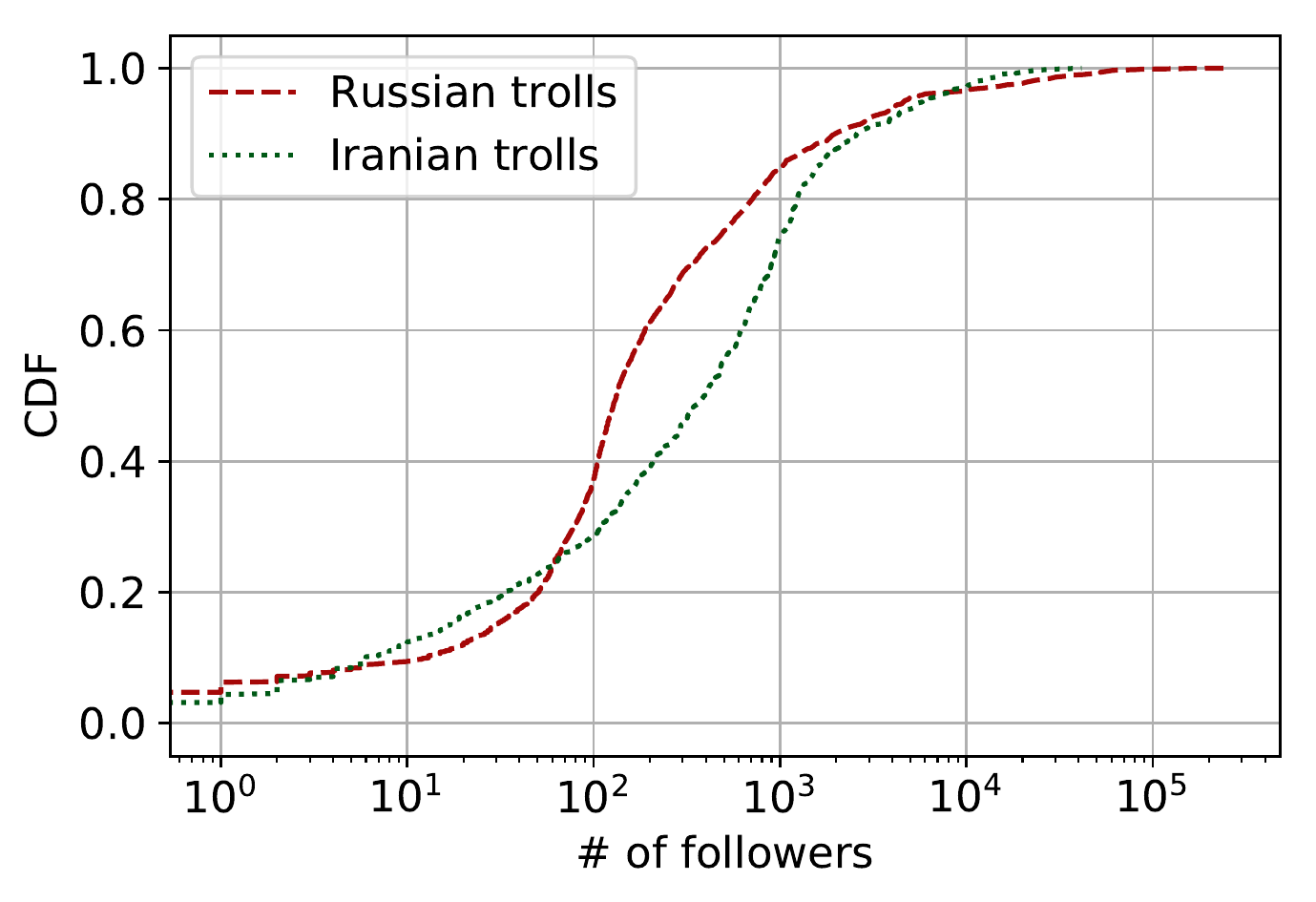}\label{subfig:cdf_followers}}
\subfigure[]{\includegraphics[width=0.49\columnwidth]{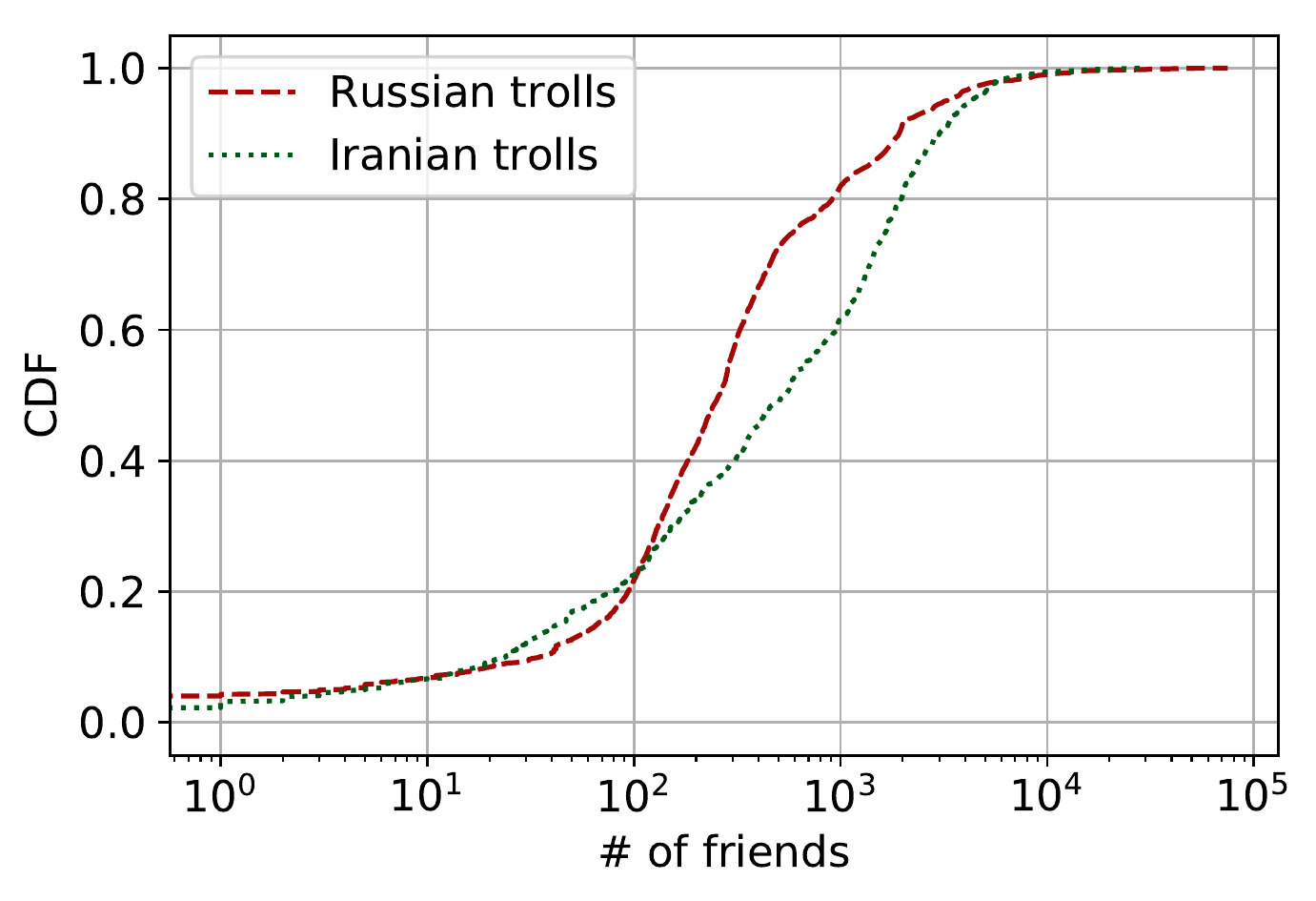}\label{subfig:cdf_followings}}
\caption{CDF of the number of a) followers and b) friends for the Russian and Iranian trolls on Twitter.}
\label{fig:cdf_followers_followings}
\end{figure}

\descr{Followers/Friends.} Fig.~\ref{fig:cdf_followers_followings} plots the CDF of the number of followers and friends for both Russian and Iranian trolls.
25\% of Iranian trolls had more than 1k followers, while the same applies for only 15\% of the Russian trolls.
In general, Iranian trolls tend to have more followers than Russian trolls (median of 392 and 132, respectively).
Both Russian and Iranian trolls tend to follow a large number of users, probably in an attempt to increase their follower count via reciprocal follows.
Iranian trolls have a median followers to friends ratio of 0.51, while Russian trolls have a ratio of 0.74.
This might indicate that Iranian trolls were more effective in acquiring followers without resorting in massive followings of other users, or perhaps that they took advantages of services that offer followers for sale~\cite{stringhini2013follow}.

\subsubsection{Temporal Analysis}

\begin{figure}[t]
\center
\subfigure[Date]{\includegraphics[width=0.97\columnwidth]{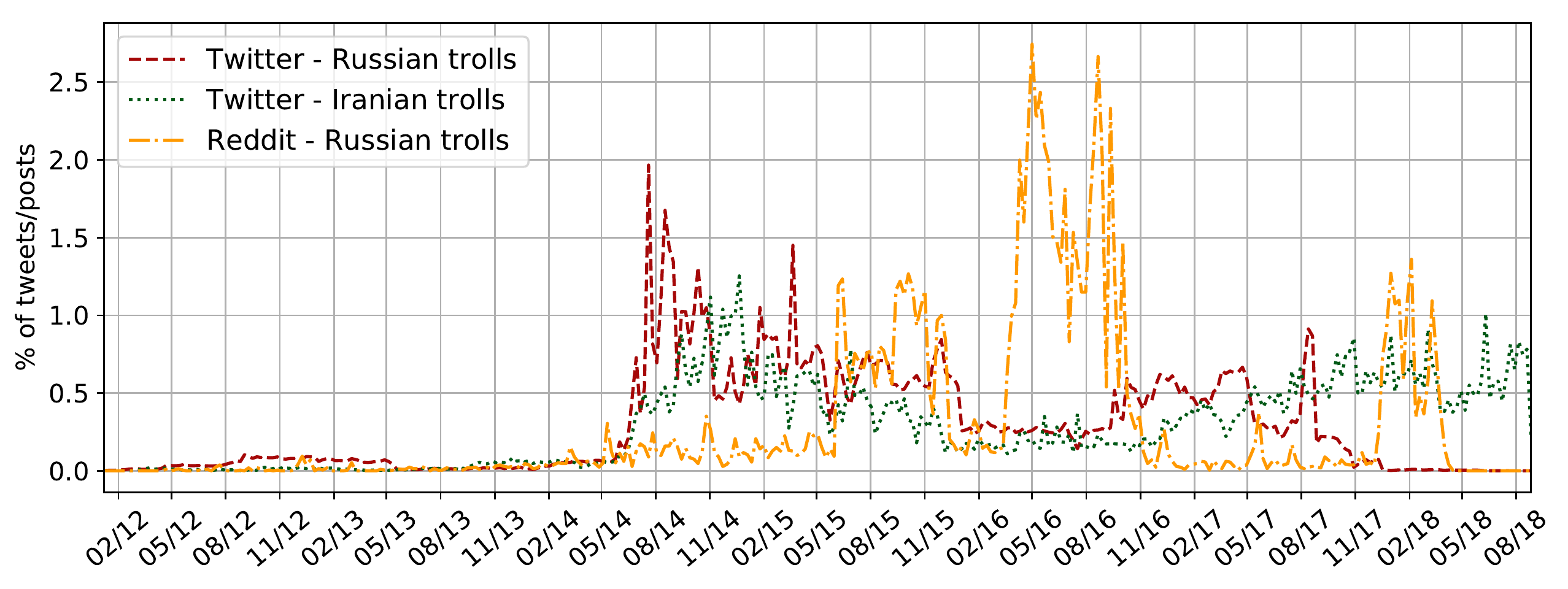}\label{subfig:counts_per_day}}
\subfigure[Hour of Day]{\includegraphics[width=0.485\columnwidth]{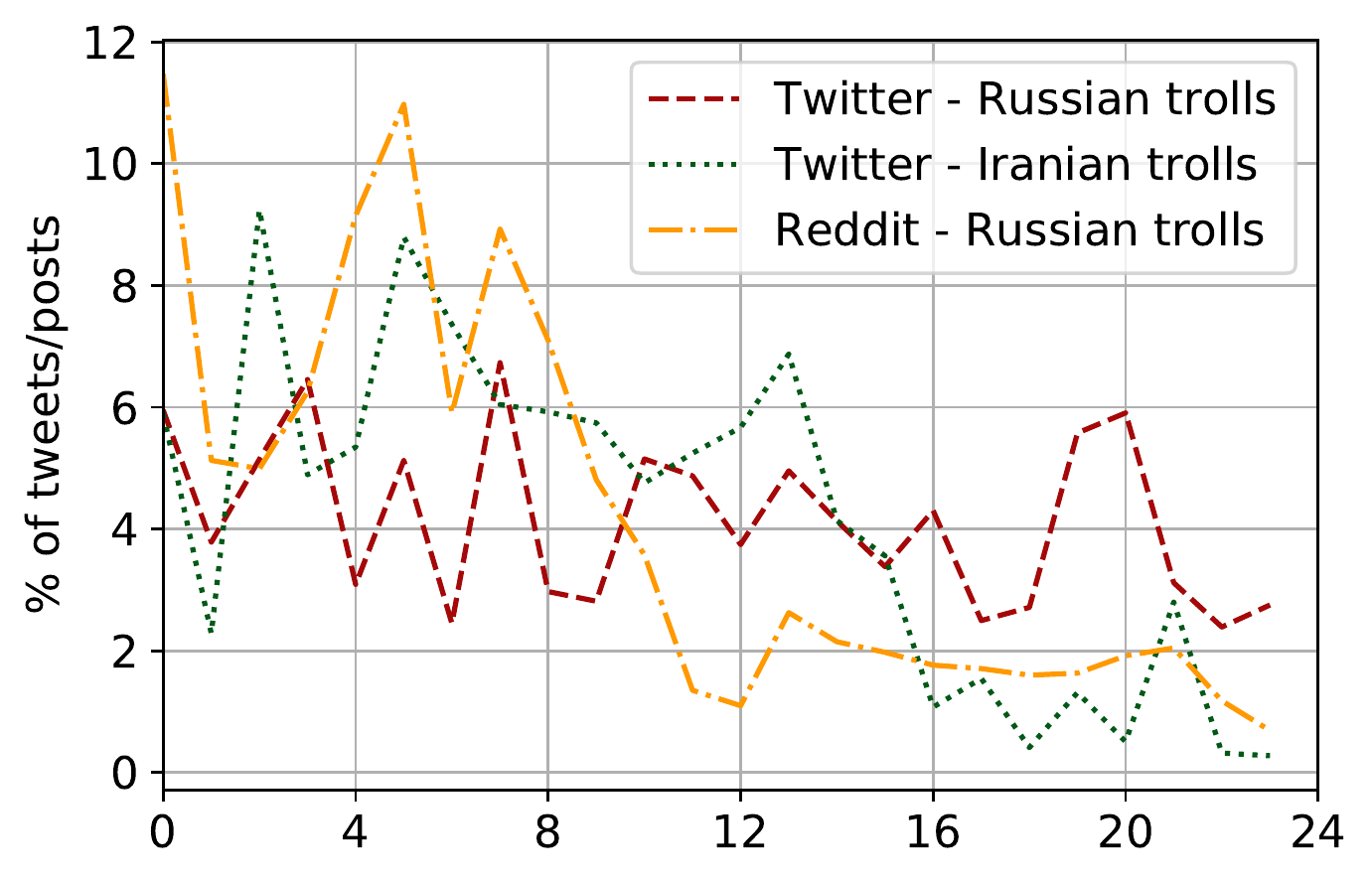}\label{subfig:counts_per_hour_day}}
\subfigure[Hour of Week]{\includegraphics[width=0.485\columnwidth]{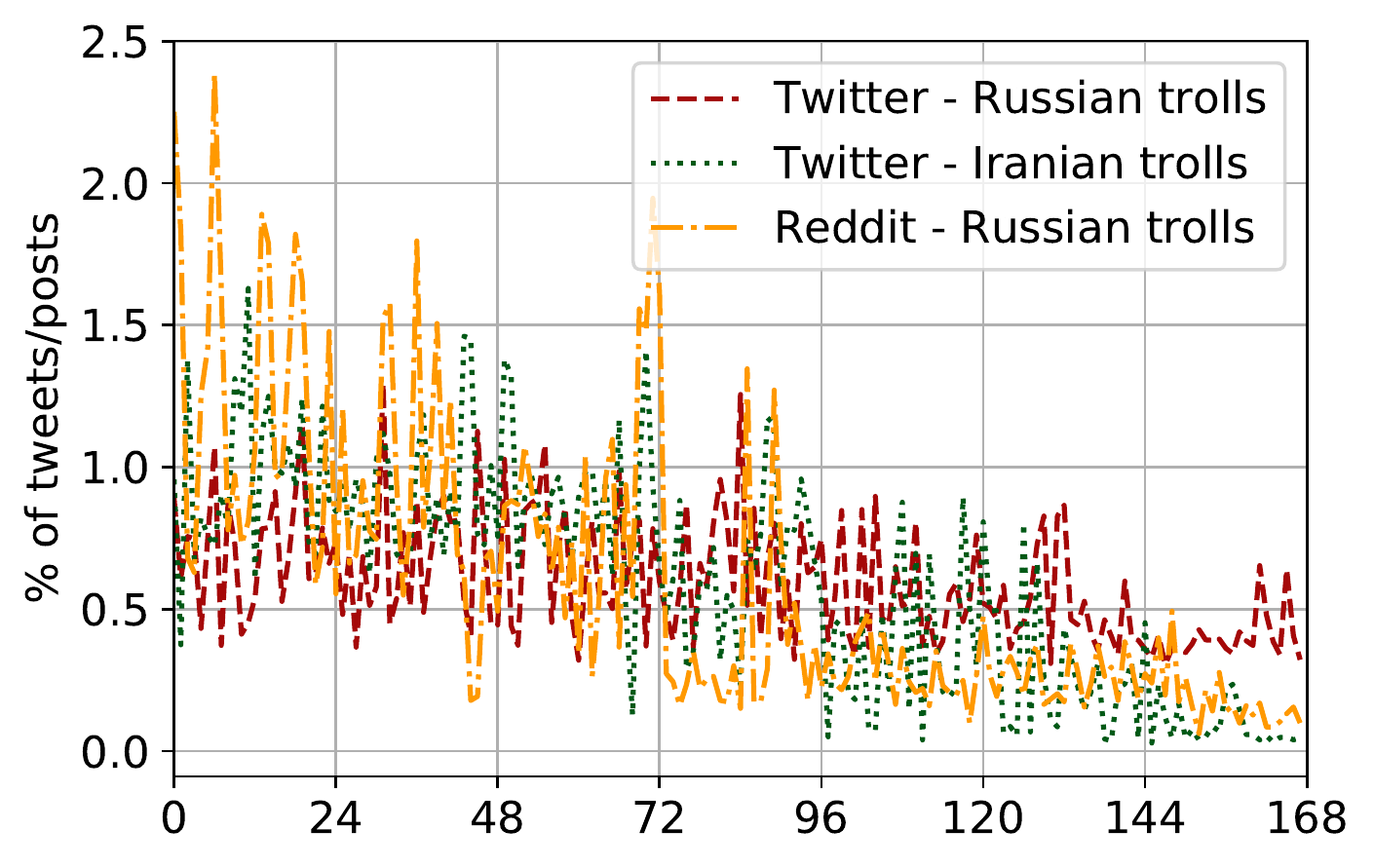}\label{subfig:counts_per_hour_week}}
  \caption{Temporal characteristics of tweets from Russian and Iranian trolls.} %
\label{fig:temporal_analysis}
\end{figure}

We next explore aggregate troll activity over time, looking for behavioral patterns.
Fig.~\ref{subfig:counts_per_day} plots the (normalized) volume of tweets/posts shared per week in our dataset.
We observe that both Russian and Iranian trolls on Twitter became active during the Ukrainian conflict.
Although lower in overall volume, there an increasing trend starts around August 2016 and continues through summer of 2017.

We also see three major spikes in activity by Russian trolls on Reddit. 
The first is during the latter half of 2015, approximately around the time that Donald Trump announced his candidacy for President.
Next, we see solid activity through the middle of 2016, trailing off shortly before the election.
Finally, we see another burst of activity in late 2017 through early 2018, at which point the trolls were detected and had their accounts locked by Reddit.

Next, we examine the hour of day and week that the trolls post.
Fig.~\ref{subfig:counts_per_hour_day} shows that Russian trolls on Twitter are active throughout the day, while on Reddit they are particularly active during the first hours of the day.
Similarly, Iranian trolls on Twitter tend to be active from early morning until 13:00 UTC.
In Fig.~\ref{subfig:counts_per_hour_week}, we report temporal characteristics based on hour of the week,
finding that Russian trolls on Twitter follow a diurnal pattern with slightly less activity during Sunday.
In contrast, Russian trolls on Reddit and Iranian trolls on Twitter are particularly active during the first days of the week, while their activity decreases during the weekend.
For Iranians this is likely due to the Iranian work week being from Sunday to Wednesday with a half day on Thursday.

\begin{figure}[t!]
\centering
\includegraphics[width=0.97\columnwidth]{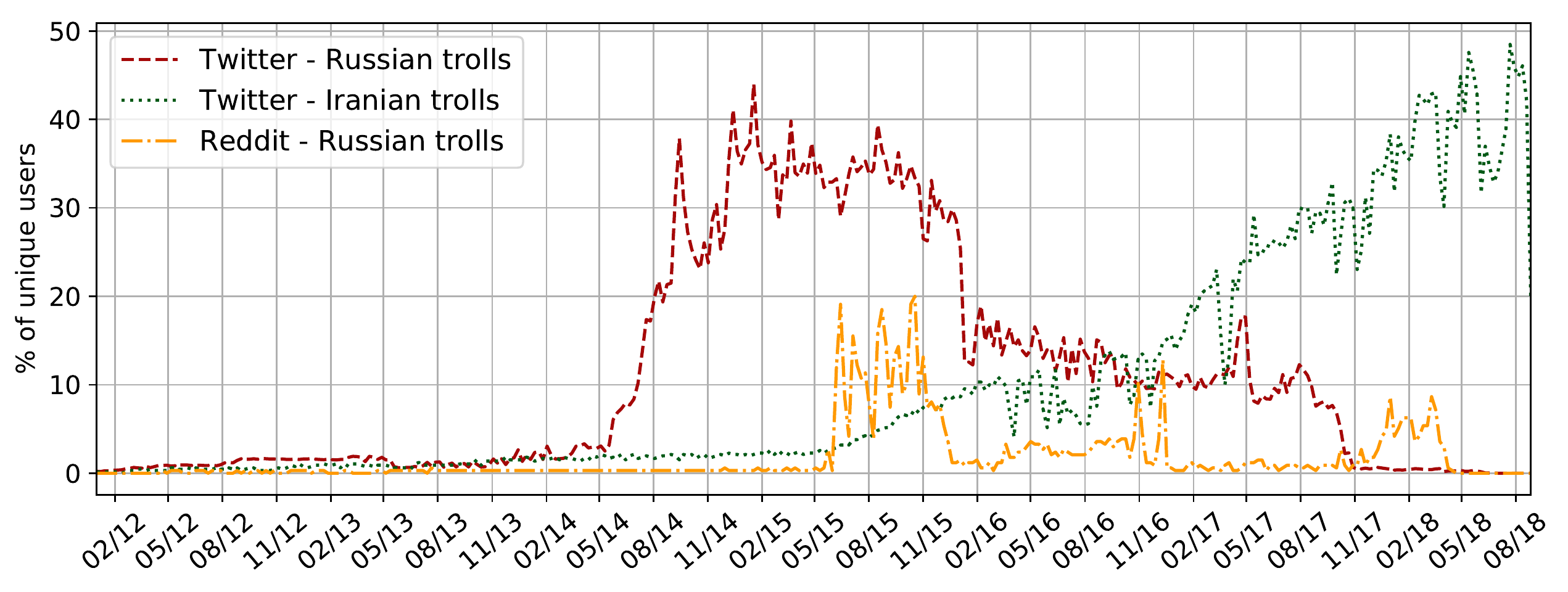}
\caption{Percentage of unique trolls that were active per week.}
\label{fig:unique_users_per_week}
\end{figure}

But are \emph{all} trolls in our dataset active throughout the span of our datasets?
To answer this question, we plot the percentage of unique troll accounts that are active per week in Fig.~\ref{fig:unique_users_per_week} from which we draw the following observations.
First, the Russian troll campaign on Twitter targeting Ukraine was much more diverse in terms of accounts when compared to later campaigns.
There are several possible explanations for this.
One explanation is that trolls learned from their Ukrainian campaign and became more efficient in later campaigns, perhaps relying on large networks of bots in their earlier campaigns which were later abandoned in favor of more focused campaigns like project Lakhta~\cite{lakhtaindictment}.
Another explanation could be that attacks on the US election might have required ``better trained'' trolls, perhaps those that could speak English more convincingly.
The Iranians, on the other hand, seem to be slowly building their troll army over time.
There is a steadily increasing number of active trolls posting per week over time.
We speculate that this is due to their troll program coming online in a slow-but-steady manner, perhaps due to more effective training.
Finally, on Reddit we see most Russian trolls posted irregularly, possibly performing other operations on the platform like manipulating votes on other posts.

\begin{figure}[t]
\centering
\includegraphics[width=0.97\columnwidth]{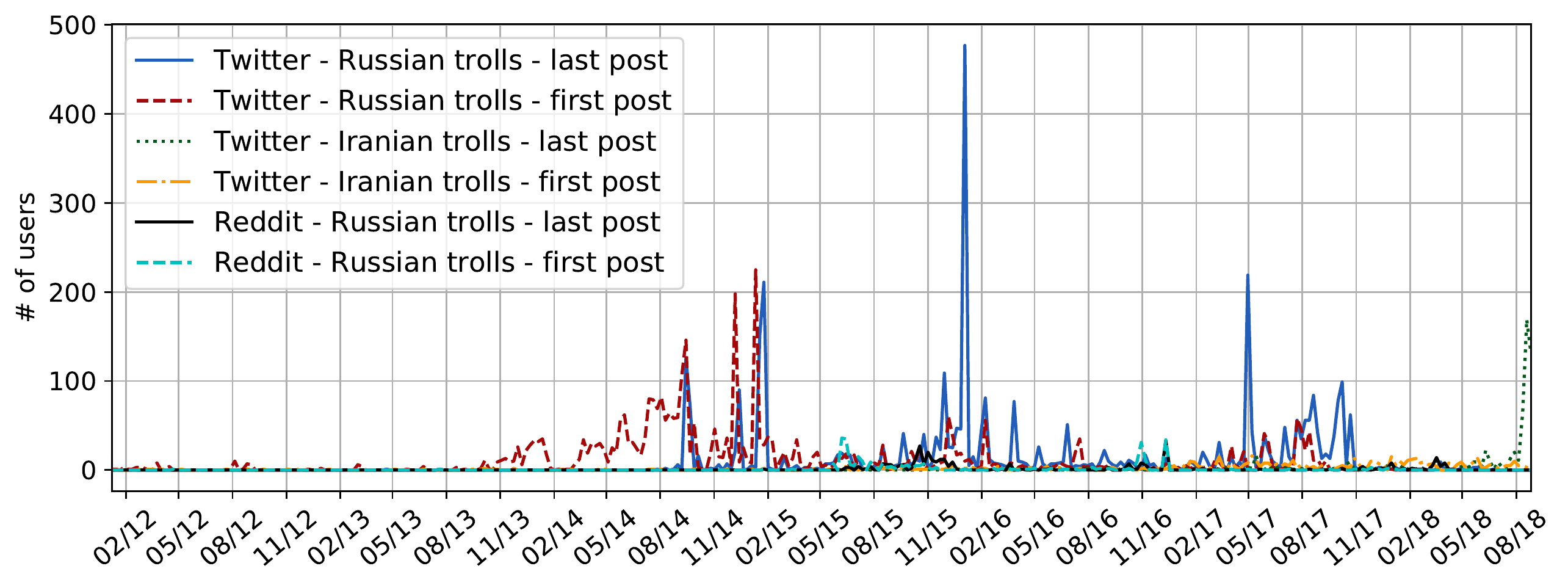}
\caption{Number of trolls that posted their first/last tweet/post for each week in our dataset.}
\label{fig:first_last_tweet}
\end{figure}

Next, we investigate the point in time when each troll in our dataset made his first and last tweet.
Fig.~\ref{fig:first_last_tweet} shows the number of users that made their first/last post for each week in our dataset, which highlights when trolls became active as well as when they ``retired.''
We see that Russian trolls on Twitter made their first posts during early 2014, almost certainly in response to the Ukrainian conflict.
When looking at the last tweets of Russian trolls on Twitter we see that a substantial portion of the trolls ``retired'' by the end of 2015.
In all likelihood this is because the Ukrainian conflict was over and Russia turned their information warfare arsenal to other targets (e.g., the USA, this is also aligned with the increase in the use of English language, see Section~\ref{sec:language}).
When looking at Russian trolls on Reddit, we do not see a substantial spike in first posts close to the time that the majority of the accounts were created (see Fig.~\ref{fig:counts_created}).
This indicates that the newly created Russian trolls on Reddit became active gradually (in terms of posting behavior).

\begin{figure}[t!]
\centering
\includegraphics[width=0.97\columnwidth]{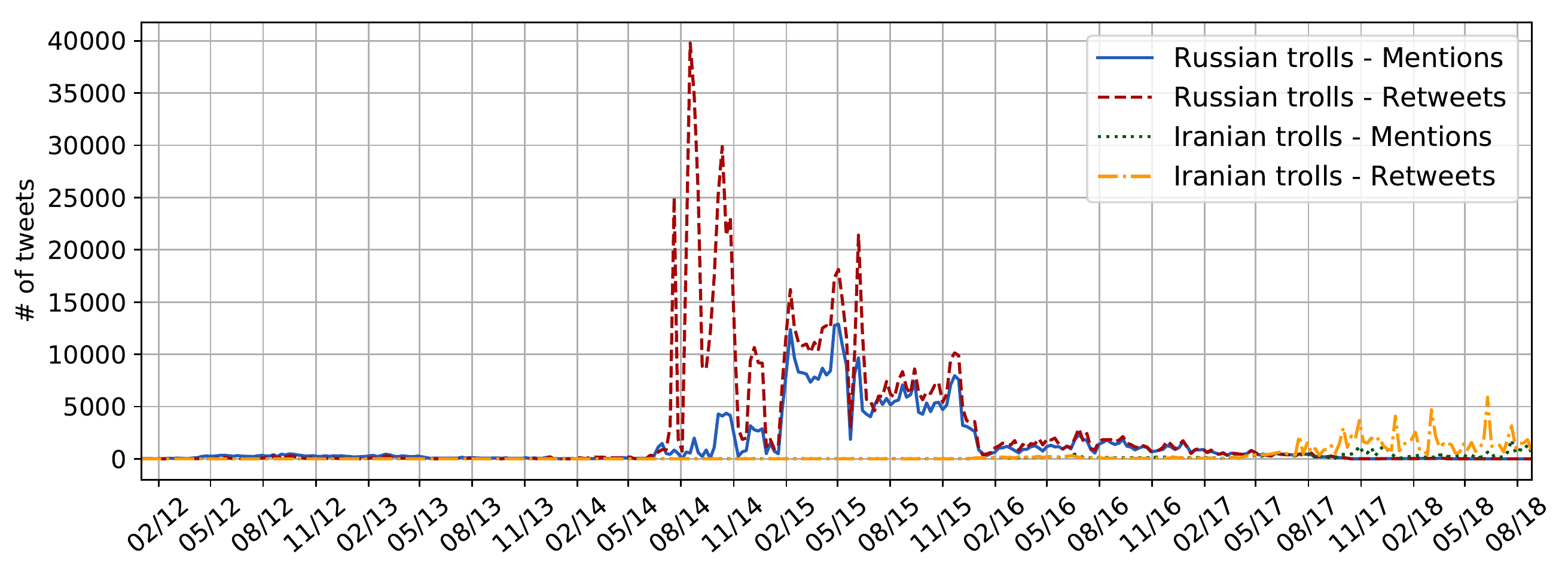}
\caption{Number of tweets that contain mentions among Russian trolls and among Iranian trolls on Twitter.}
\label{fig:mentions_among}
\end{figure}

Finally, we assess whether Russian and Iranian trolls mention or retweet each other, and how this behavior occurs over time.
Fig.~\ref{fig:mentions_among} shows the number of tweets that were mentioning/retweeting other trolls' tweets over the course of our datasets.
Russian trolls were particularly fond of this strategy during 2014 and 2015, while Iranian trolls started using this strategy after August, 2017.
This again highlights how the strategies employed by trolls adapts and evolves to new campaigns.

\subsubsection{Languages and Clients} \label{sec:language}

In this section, we study the languages that Russian and Iranian Twitter trolls posted in, as well as their Twitter clients they used to make tweets (this information is not available for Reddit).

\begin{figure}[t!]
\center
\subfigure[]{\includegraphics[width=0.485\columnwidth]{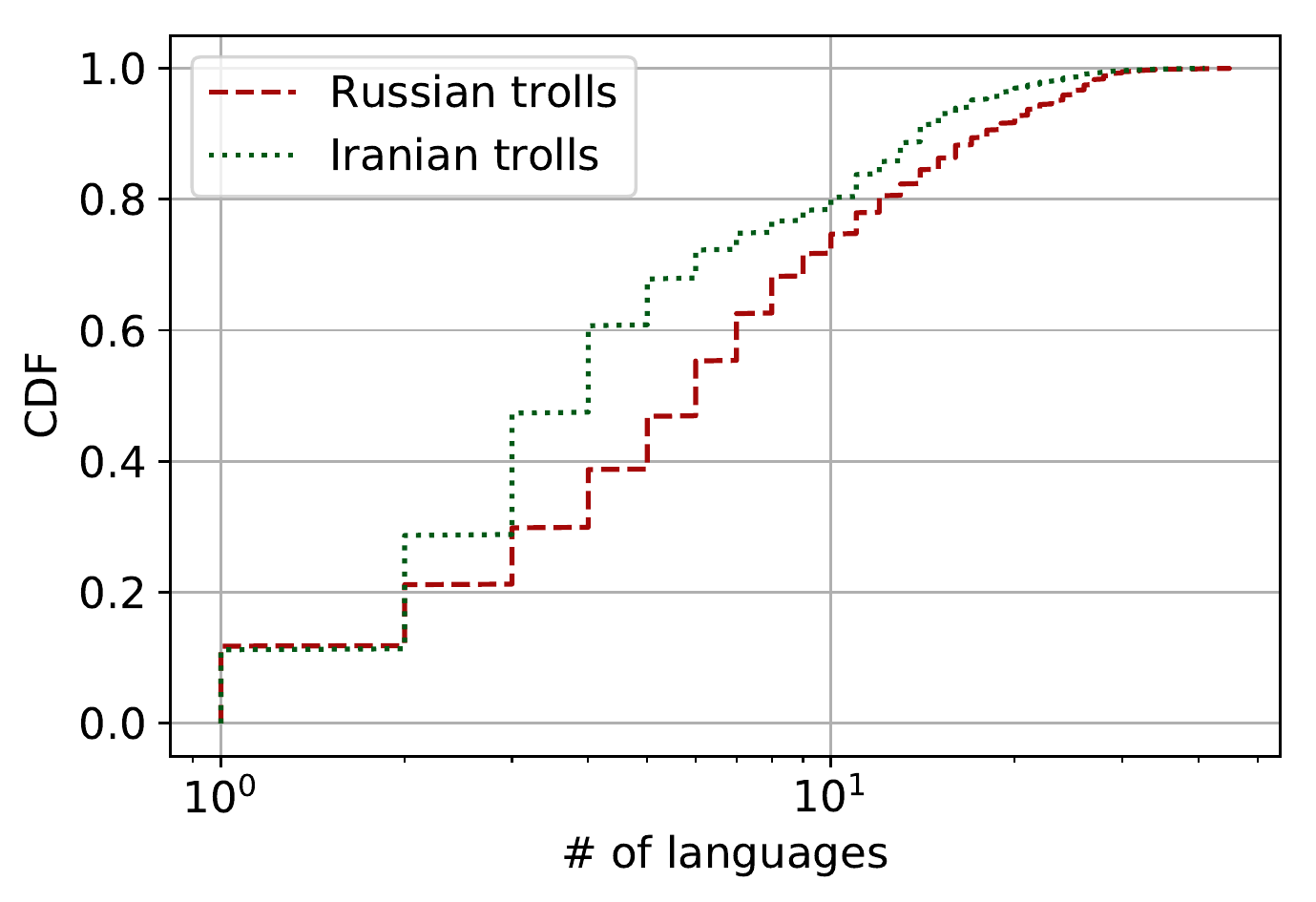}\label{subfig:cdf_languages_user}}
\subfigure[]{\includegraphics[width=0.485\columnwidth]{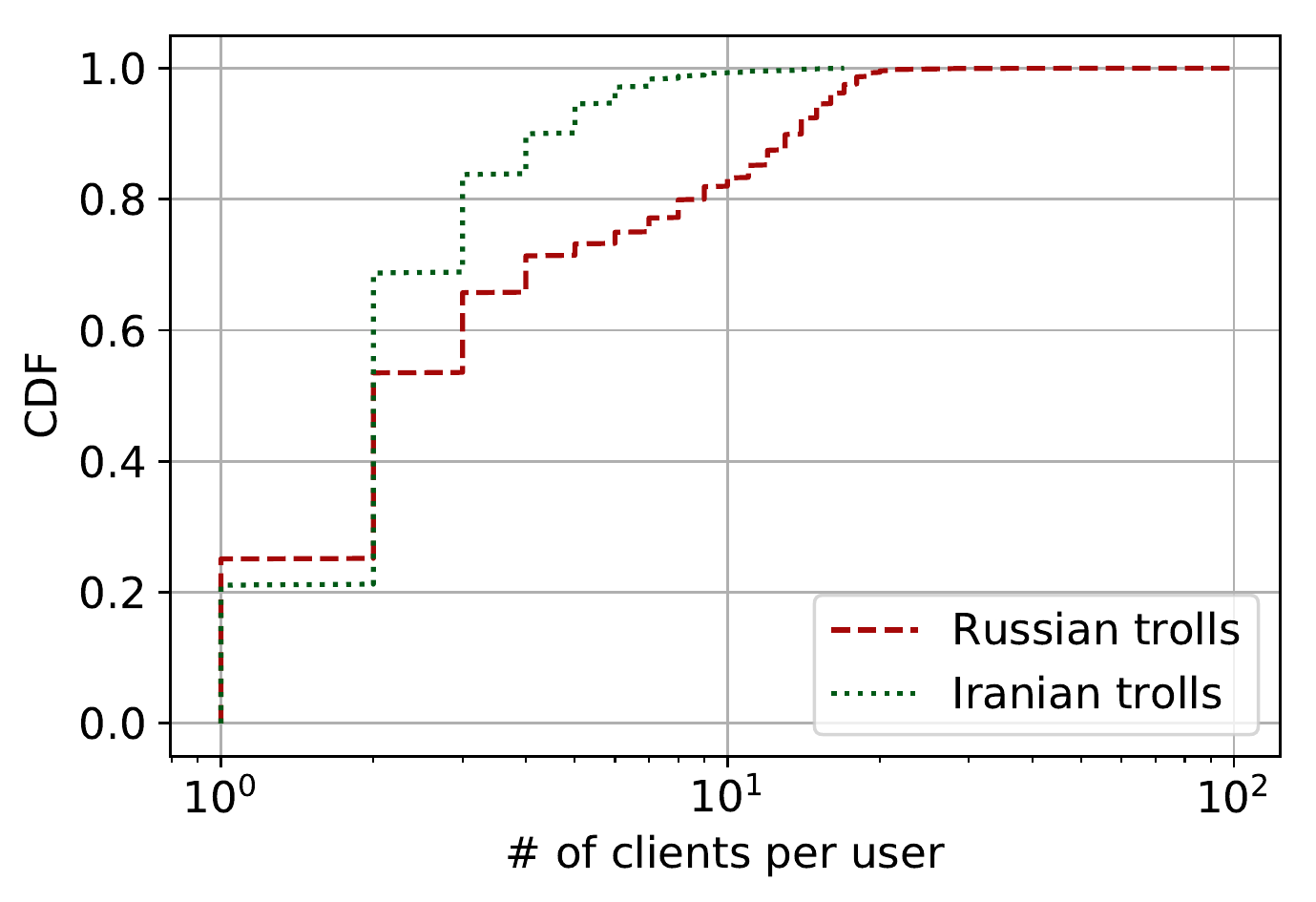}\label{subfig:cdf_sources_user}}
  \caption{CDF of number of (a) languages used (b) clients used for Russian and Iranian trolls on Twitter. }
\label{fig:cdf_lang_sources}
\end{figure}

\descr{Languages.}
First we study the languages used by trolls as it provides an indication of their targets.
The language information is included in the datasets released by Twitter.
Fig.~\ref{subfig:cdf_languages_user} plots the CDF of the number of languages used by troll accounts.
We find that 80\% and 75\% of the Russian and Iranian trolls, respectively, use more than 2 languages.
Next, we note that in general, Iranian trolls tend to use fewer languages than Russian trolls.
The most popular language for Russian trolls is Russian (53\% of all tweets), followed by English (36\%), Deutsch (1\%), and Ukrainian (0.9\%).
For Iranian trolls we find that French is the most popular language (28\% of tweets), followed by English (24\%), Arabic (13\%), and Turkish (8\%).

\begin{figure}[t!]
\center
\subfigure[Russians]{\includegraphics[width=0.49\columnwidth]{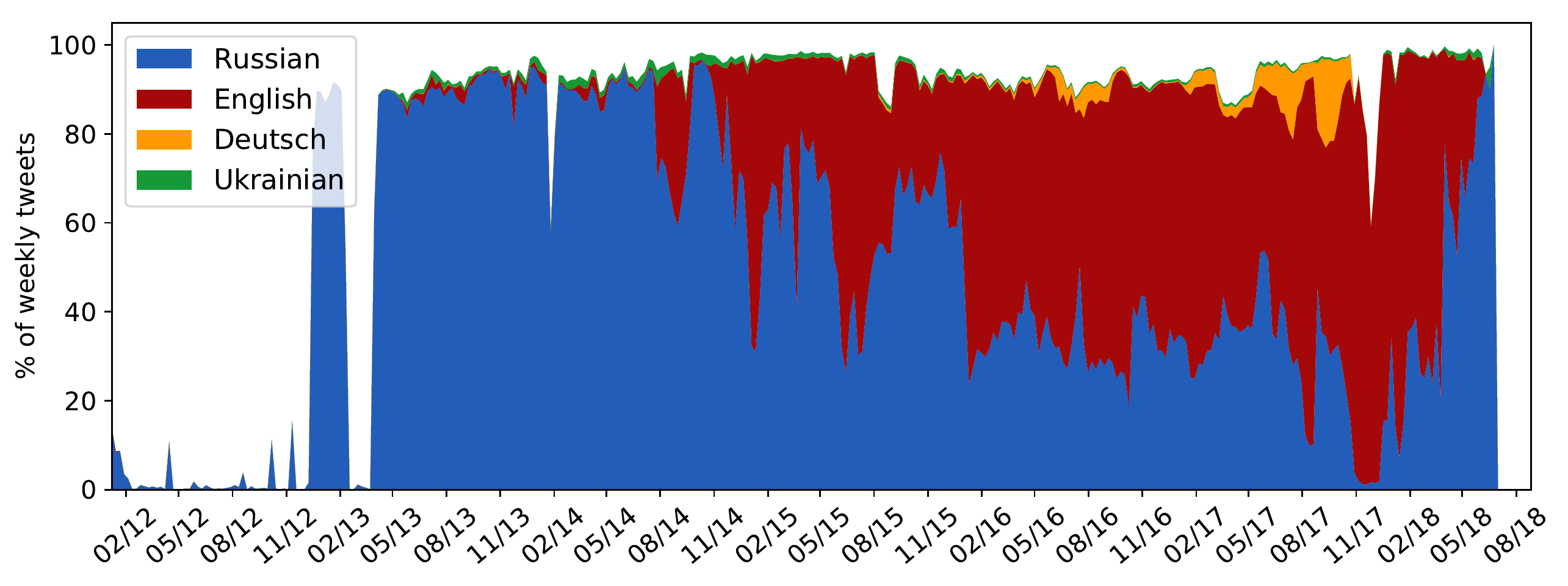}\label{subfig:russians_norm_week}}
\subfigure[Iranians]{\includegraphics[width=0.49\columnwidth]{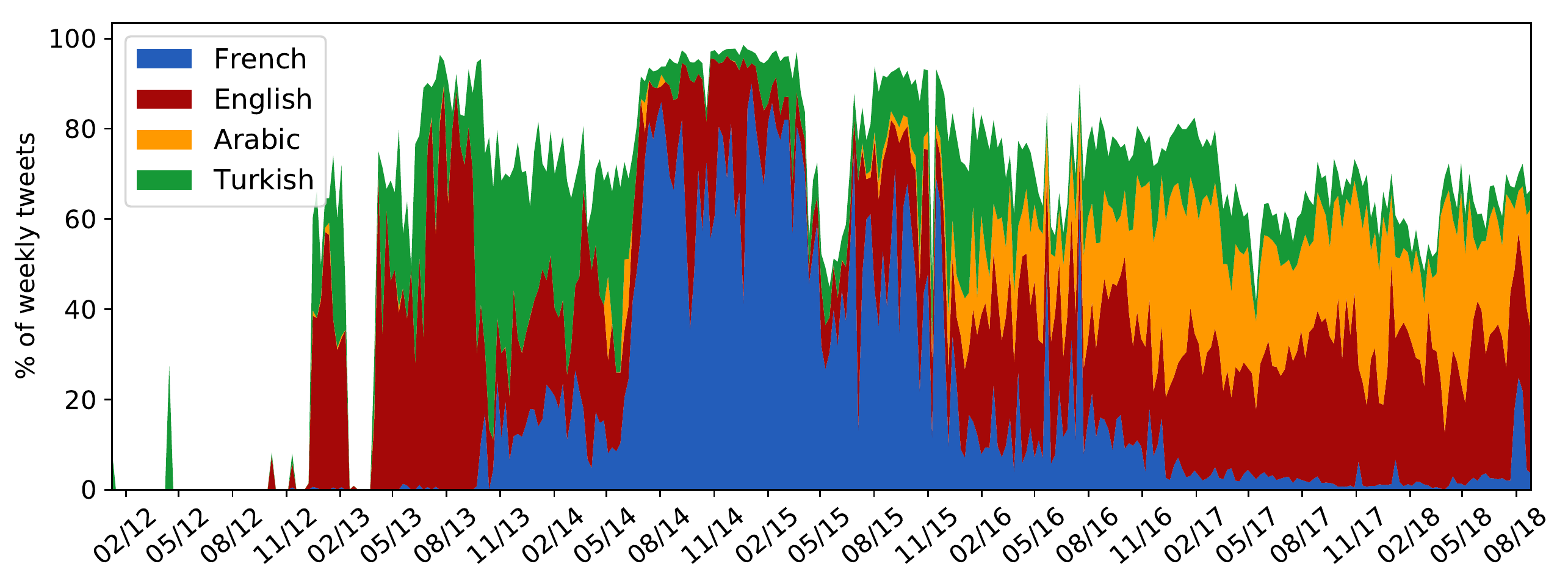}\label{subfig:iranians_norm_week}}
\subfigure[Russians]{\includegraphics[width=0.49\columnwidth]{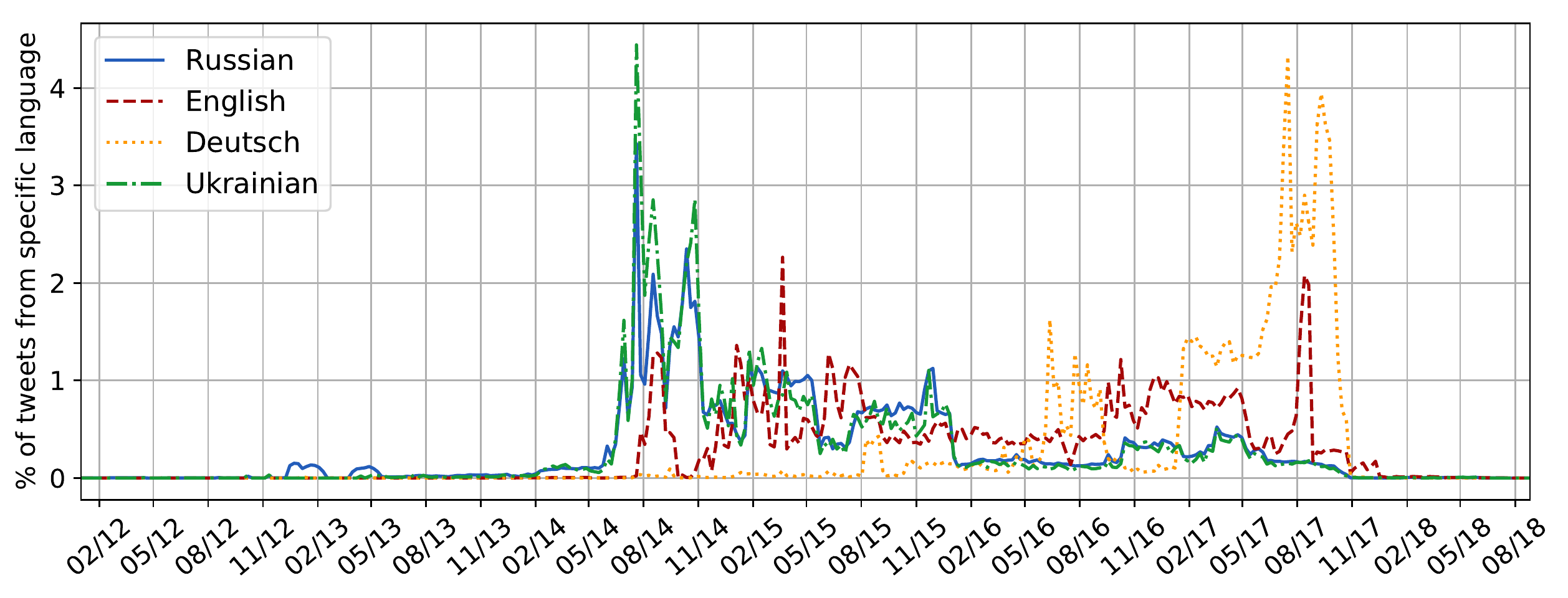}\label{subfig:russians_norm_language}}
\subfigure[Iranians]{\includegraphics[width=0.49\columnwidth]{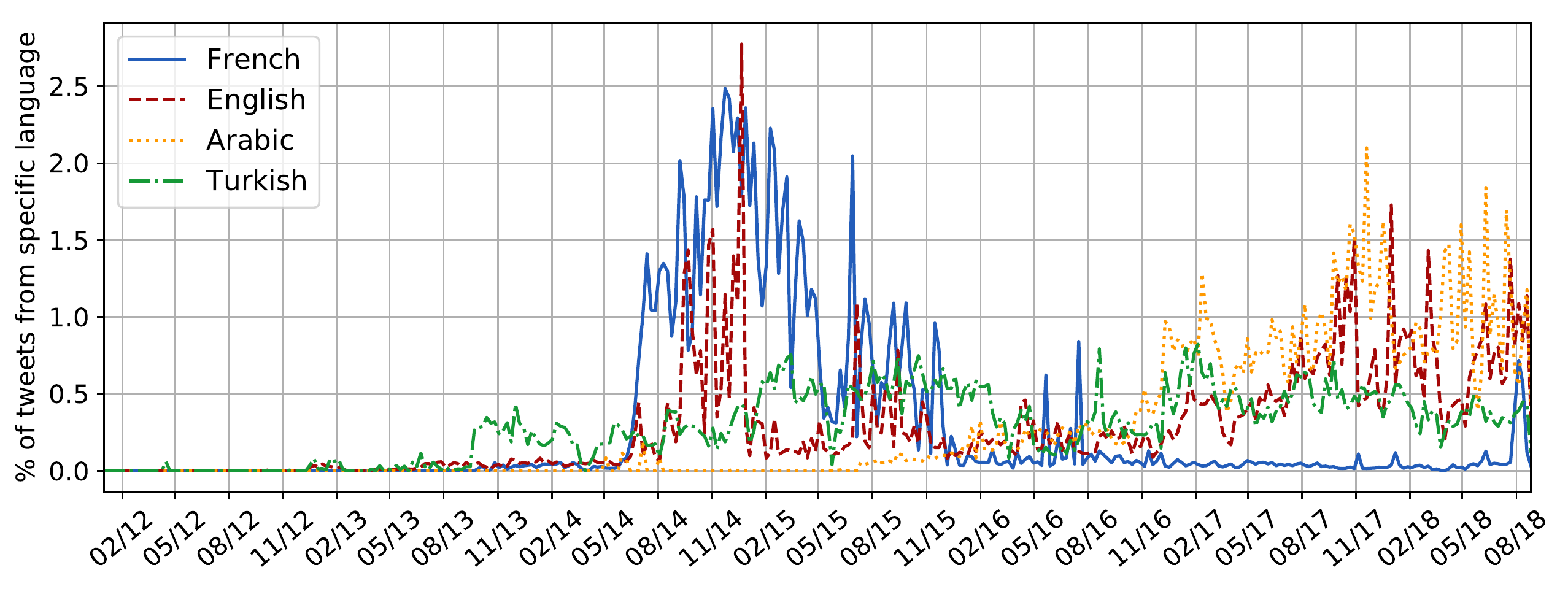}\label{subfig:iranians_norm_language}}
  \caption{Use of the four most popular languages by Russian and Iranian trolls over time on Twitter.
  (a) and (b) show the percentage of weekly tweets in each language.
  (c) and (d) show the percentage of total tweets per language that occurred in a given week.
  }
\label{fig:languages_over_time}
\end{figure}

Fig.~\ref{fig:languages_over_time} plots the use of different languages over time.
Fig.~\ref{subfig:russians_norm_week} and Fig.~\ref{subfig:iranians_norm_week} plot the percentage of tweets that were in a given language on a given week for Russian and Iranian trolls, respectively, in a stacked fashion, which lets us see how the usage of different languages changed over time relative to each other.
Fig.~\ref{subfig:russians_norm_language} and Fig.~\ref{subfig:iranians_norm_language} plot the language use from a different perspective: normalized to the overall number of tweets in a given language.
This view gives us a better idea of how the use of each particular language changed over time.
From the plots we make the following observations.
First, there is a clear shift in targets based on the campaign.
For example, Fig.~\ref{subfig:russians_norm_week} shows that the overwhelming majority of early tweets by Russian trolls were in Russian, with English only reaching the volume of Russian language tweets in 2016.
This coincides with the ``retirement'' of several Russian trolls on Twitter (see Fig~\ref{fig:first_last_tweet}).
Next, we see evidence of other campaigns, for example German language tweets begin showing up in early to mid 2016, and reach their highest volume in the latter half of 2017, in close proximity with the 2017 German Federal elections.
Additionally, we note that Russian language tweets have a huge drop off in activity the last two months of 2017.

For the Iranians, we see more obvious evidence of multiple campaigns.
For example, although Turkish and English are present for most of the timeline, French quickly becomes a commonly used language in the latter half of 2013, becoming the dominant language used from around May 2014 until the end of 2015.
This is likely due to political events that happened during this time period.
E.g., in November, 2013 France blocked a stopgap deal related to Iran's uranium enrichment program~\cite{guardian_iran}, leading to some fiery rhetoric from Iran's government (and apparently the launch of a troll campaign targeting French speakers).
As tweets in French fall off, we also observe a dramatic increase in the use of Arabic in early 2016.
This coincides with an attack on the Saudi embassy in Tehran~\cite{nytimes_iran_embassy}, the primary reason the two countries ended diplomatic relations.

When looking at the language usage normalized by the total number of tweets in that language, we can get a more focused perspective.
In particular, from Fig.~\ref{subfig:russians_norm_language} it becomes strikingly clear that the initial burst of Russian troll activity was targeted at Ukraine, with the majority of Ukrainian language tweets coinciding directly with the Crimean conflict~\cite{crimea_timeline}.
From Fig.~\ref{subfig:iranians_norm_language} we observe that English language tweets from Iranian trolls, while consistently present over time, have a relative peak corresponding with French language tweets, likely indicating an attempt to influence non-French speakers with respect to the campaign against French speakers.

\begin{figure}[t!]
\center
\subfigure[Russians]{\includegraphics[width=0.75\columnwidth]{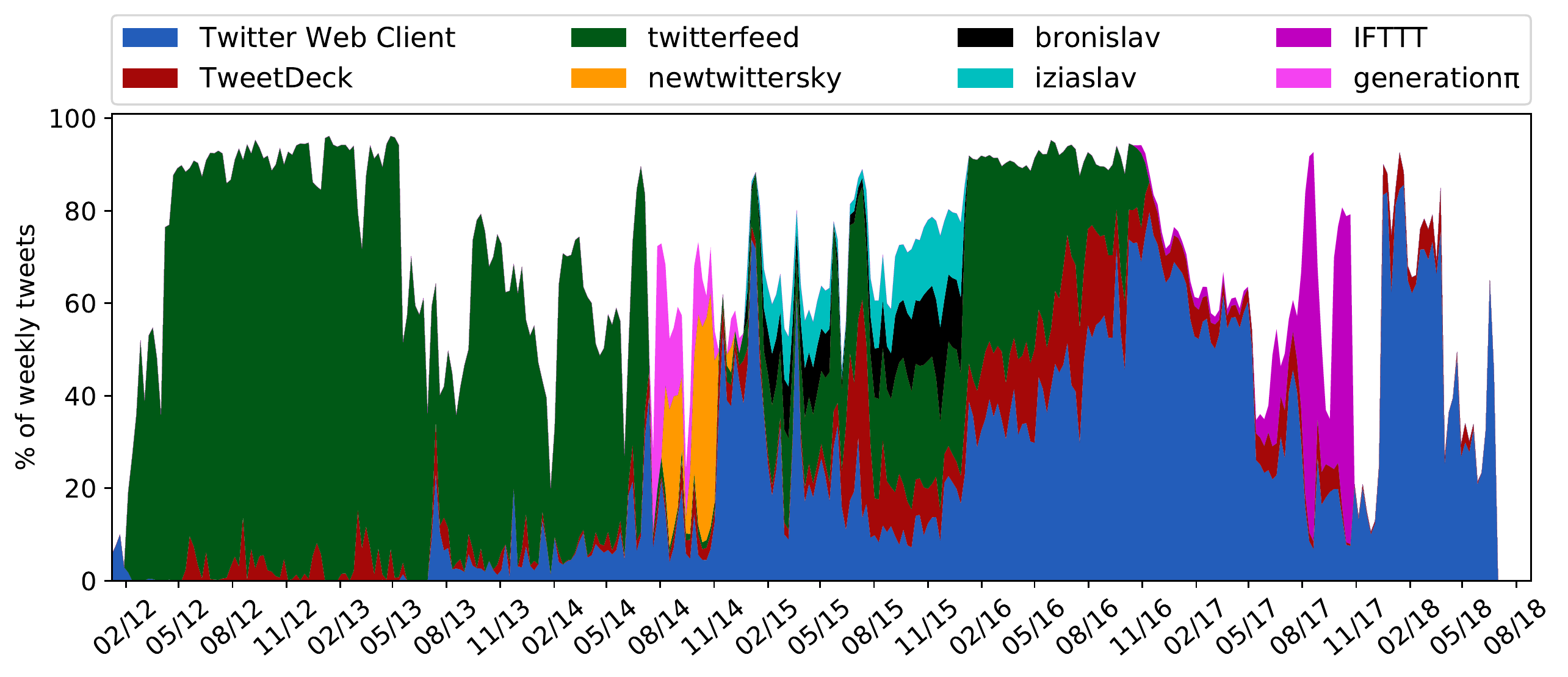}\label{subfig:russians_norm_week_clients}}
\subfigure[Iranians]{\includegraphics[width=0.75\columnwidth]{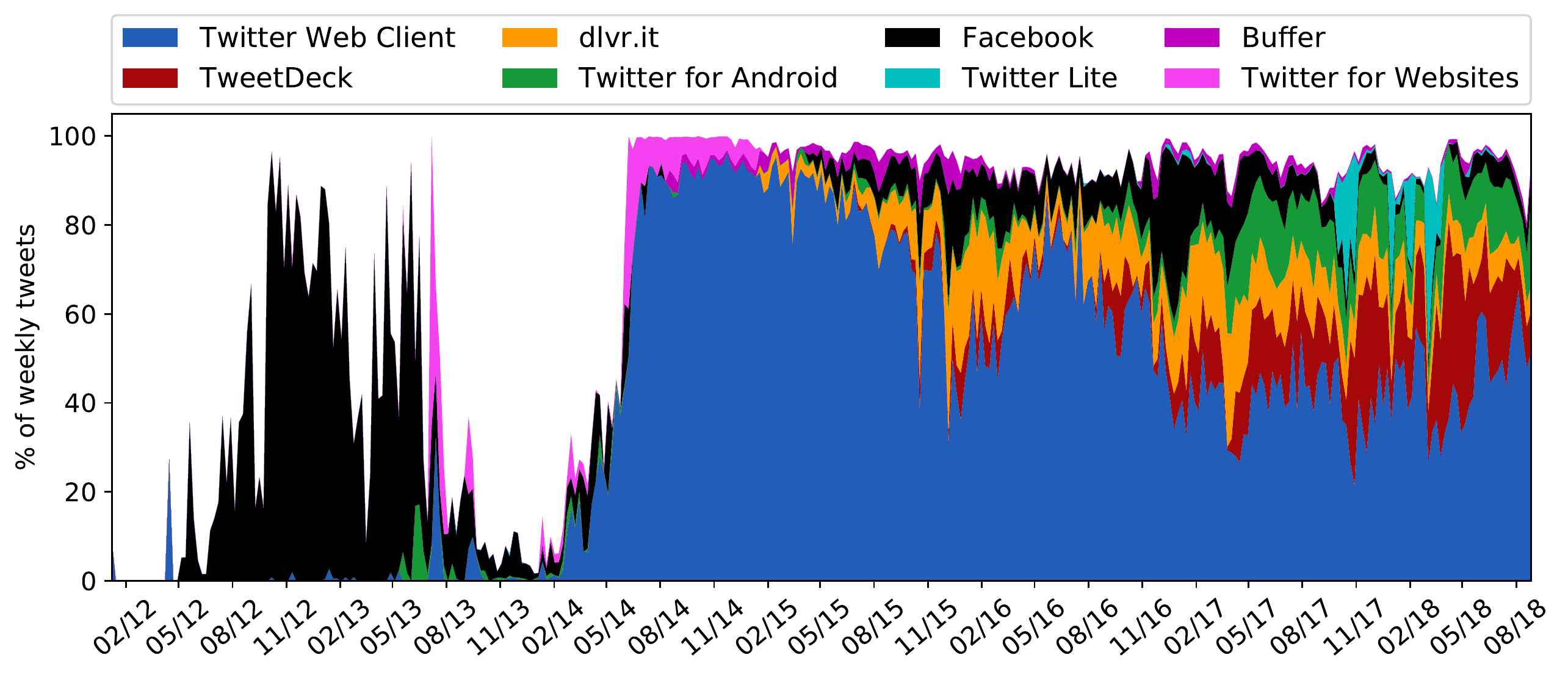}\label{subfig:iranians_norm_week_clients}}
  \caption{Use of the eight most popular clients by Russian and Iranian trolls over time on Twitter.
  }
\label{fig:clients_over_time}
\end{figure}

\descr{Client usage.}
Finally, we analyze the clients used to post tweets.
When looking at the most popular clients, we find that Russian and Iranian trolls use the main Twitter Web Client (28.5\% for Russian trolls, and 62.2\% for Iranian trolls).
This is in contrast with what normal users use: using a random set of Twitter users, we find that mobile clients make up a large chunk of tweets (48\%), followed by the TweetDeck dashboard (32\%).
We next look at how many different clients trolls use throughout our dataset:
in Fig.~\ref{subfig:cdf_sources_user}, we plot the CDF of the number of clients used per user.
25\% and 21\% of the Russian and Iranian trolls, respectively, use only one client, while in general Russian trolls tend to use more clients than Iranians.

Fig.~\ref{fig:clients_over_time} plots the usage of clients over time in terms of weekly tweets by Russian and Iranian trolls.
We observe that the Russians (Fig.~\ref{subfig:russians_norm_week_clients}) started off with almost exclusive use of the ``twitterfeed'' client.
Usage of this client drops off when it was shutdown in October, 2016.
During the Ukrainian crisis, however, we see several new clients come into the mix.
Iranians (Fig.~\ref{subfig:iranians_norm_week_clients}) started off almost exclusively using the ``facebook'' Twitter client.
To the best of our knowledge, this is a client that automatically Tweets any posts you make on Facebook, indicating that Iranians likely started with a campaign on Facebook.
At the beginning of 2014, we see a shift to using the Twitter Web Client, which only begins to decrease towards the end of 2015.
Of particular note in Fig.~\ref{subfig:iranians_norm_week_clients} is the appearance of ``dlvr.it,'' an automated social media manager, in the beginning of 2015.
This corresponds with the creation of IUVM~\cite{iuvm_about_page}, which is a fabricated ecosystem of (fake) news outlets and social media accounts created by the Iranians, and might indicate that Iranian trolls stepped up their game around that time, starting using services that allowed them for better account orchestration to run their campaigns more effectively.

\begin{figure*}[t]
\centering
\includegraphics[width=\textwidth]{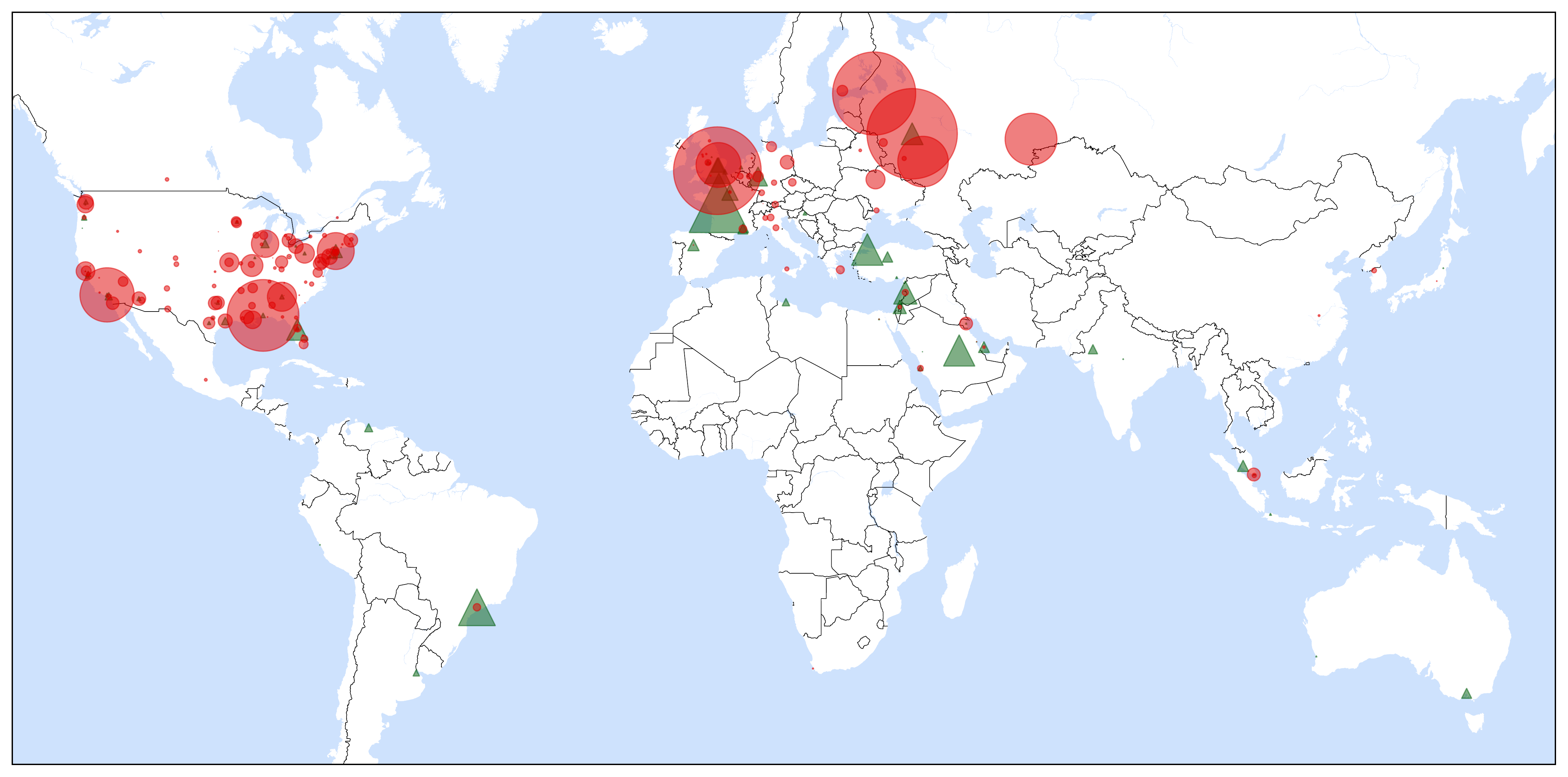}
   \caption{Distribution of reported locations for tweets by Russian trolls (100\%) (red circles) and Iranian trolls (green triangles).}
\label{fig:locations_map_agg}
\end{figure*}

\subsubsection{Geographical Analysis}

We then study users' location, relying on the self-reported location field in their profiles.
Note that this field is not required, and users are also able to change it whenever they like,
so we look at locations for each tweet.
Note that 16.8\% and 20.9\% of the tweets from Russian and Iranians trolls, respectively, do not include a self-reported location.
To infer the geographical location from the self-reported text, we use pigeo~\cite{rahimi2016pigeo}, which provides geographical information (e.g., latitude, longitude, country, etc.) given the text that corresponds to a location.
Specifically, we extract 626 self-reported locations for the Russian trolls and 201 locations for the Iranian trolls.
Then, we use pigeo to systematically obtain a geographical location (and its associated coordinates) for each text that corresponds to a location.
Fig.~\ref{fig:locations_map_agg} shows the locations inferred for Russian trolls (red circles) and Iranian trolls (green triangles).
The size of the shapes on the map indicates the number of tweets that appear on each location.
We observe that most of the tweets from Russian trolls come from locations within Russia (34\%), the USA (29\%), and some from European countries, like United Kingdom (16\%), Germany (0.8\%), and Ukraine (0.6\%).
This suggests that Russian trolls may be pretending to be from certain countries, e.g., USA or United Kingdom, aiming to pose as locals and effectively manipulate opinions.
A similar pattern exists with Iranian trolls, which were particularly active in France (26\%), Brazil (9\%), the USA (8\%), Turkey (7\%), and Saudi Arabia (7\%).
It is also worth noting that Iranians trolls, unlike Russian trolls, did not report locations from their country, indicating that these trolls were primarily used for campaigns targeting foreign countries.
Finally, we note that the location-based findings are in-line with the findings on the languages analysis (see Section~\ref{sec:language}), further evidencing that both Russian and Iranian trolls were specifically targeting different countries over time.

\subsubsection{Content Analysis}

\begin{table}[]
\centering
\resizebox{0.6\columnwidth}{!}{%
\begin{tabular}{@{}lrlr@{}}
\toprule
\multicolumn{2}{c}{\textbf{Russian trolls on Twitter}} & \multicolumn{2}{c}{\textbf{Iranian trolls on Twitter}} \\ \midrule
\textbf{Word} & \multicolumn{1}{l}{\textbf{\begin{tabular}[c]{@{}l@{}}Cosine\\ Similarity\end{tabular}}} & \textbf{Word} & \multicolumn{1}{l}{\textbf{\begin{tabular}[c]{@{}l@{}}Cosine \\ Similarity\end{tabular}}} \\ \midrule
trumparmi & \multicolumn{1}{r|}{0.68} & impeachtrump & 0.81 \\
trumptrain & \multicolumn{1}{r|}{0.67} & stoptrump & 0.80 \\
votetrump & \multicolumn{1}{r|}{0.65} & fucktrump & 0.79 \\
makeamericagreatagain & \multicolumn{1}{r|}{0.65} & trumpisamoron & 0.79 \\
draintheswamp & \multicolumn{1}{r|}{0.62} & dumptrump & 0.79 \\
trumppenc & \multicolumn{1}{r|}{0.61} & ivankatrump & 0.77 \\
@realdonaldtrump & \multicolumn{1}{r|}{0.59} & theresist & 0.76 \\
wakeupamerica & \multicolumn{1}{r|}{0.58} & trumpresign & 0.76 \\
thursdaythought & \multicolumn{1}{r|}{0.57} & notmypresid & 0.76 \\
realdonaldtrump & \multicolumn{1}{r|}{0.57} & worstpresidentev & 0.75 \\
presidenttrump & \multicolumn{1}{r|}{0.57} & antitrump & 0.74 \\ \bottomrule
\end{tabular}%
}
\caption{Top 10 similar words to ``maga'' and their respective cosine similarities (obtained from the word2vec models).}
\label{tbl:similar_to_maga}
\end{table}

\descr{Word Embeddings}
Recent indictments by the US Department of Justice have indicated that troll messaging was crafted, with certain phrases and terminology designated for use in certain contexts. 
To get a better handle on how this was expressed, we build two word2vec models on the corpus of tweets: one for the Russian trolls and one for the Iranian trolls.
To train the models, we first extract the tweets posted in English, according to the data provided by Twitter. 
Then, we remove stop words, perform stemming, tokenize the tweets, and keep only words that appear at least 500 and 100 times for the Russian and Iranian trolls, respectively.

Table~\ref{tbl:similar_to_maga} shows the top 10 most similar terms to ``maga'' for each model.
We see a marked difference between its usage by Russian and Iranian trolls.
Russian trolls are clearly pushing heavily in favor of Donald Trump, while it is the exact opposite with Iranians.

\begin{table}[]
\centering
\setlength{\tabcolsep}{0.2em} %
\hspace*{-0.2cm}
\resizebox{0.8\columnwidth}{!}{
\begin{tabular}{lrlrlrlr}
\hline
\multicolumn{4}{c}{\textbf{Russian trolls on Twitter}} & \multicolumn{4}{c}{\textbf{Iranian trolls on Twitter}} \\
\textbf{Hashtag} & \textbf{(\%)} & \textbf{Hashtag} & \multicolumn{1}{l}{\textbf{(\%)}} & \textbf{Hashtag} & \textbf{(\%)} & \textbf{Hashtag} & \textbf{(\%)} \\ \hline
news & 9.5\% & USA & \multicolumn{1}{r|}{0.7\%} & Iran & 1.8\% & Palestine & 0.6\% \\
sports & 3.8\% & breaking & \multicolumn{1}{r|}{0.7\%} & Trump & 1.4\% & Syria & 0.5\% \\
politics & 3.0\% & TopNews & \multicolumn{1}{r|}{0.6\%} & Israel & 1.1\% & Saudi & 0.5\% \\
local & 2.1\% & BlackLivesMatter & \multicolumn{1}{r|}{0.6\%} & Yemen & 0.9\% & EEUU & 0.5\% \\
world & 1.1\% & true & \multicolumn{1}{r|}{0.5\%} & FreePalestine & 0.8\% & Gaza & 0.5\% \\
MAGA & 1.1\% & Texas & \multicolumn{1}{r|}{0.5\%} & QudsDay4Return & 0.8\% & SaudiArabia & 0.4\% \\
business & 1.0\% & NewYork & \multicolumn{1}{r|}{0.4\%} & US & 0.7\% & Iuvm & 0.4\% \\
Chicago & 0.9\% & Fukushima2015 & \multicolumn{1}{r|}{0.4\%} & realiran & 0.6\% & InternationalQudsDay2018 & 0.4\% \\
health & 0.8\% & quote & \multicolumn{1}{r|}{0.4\%} & ISIS & 0.6\% & Realiran & 0.4\% \\
love & 0.7\% & Foke & \multicolumn{1}{l|}{0.4\%} & DeleteIsrael & 0.6\% & News & 0.4\% \\ \hline
\end{tabular}
}
\caption{Top 20 (English) hashtags in tweets from Russian and Iranian trolls on Twitter.}
\label{tbl:top_hashtags_russians_iranians}
\end{table}

\begin{figure*}[t!]
\centering
\subfigure[]{\includegraphics[width=0.8\textwidth]{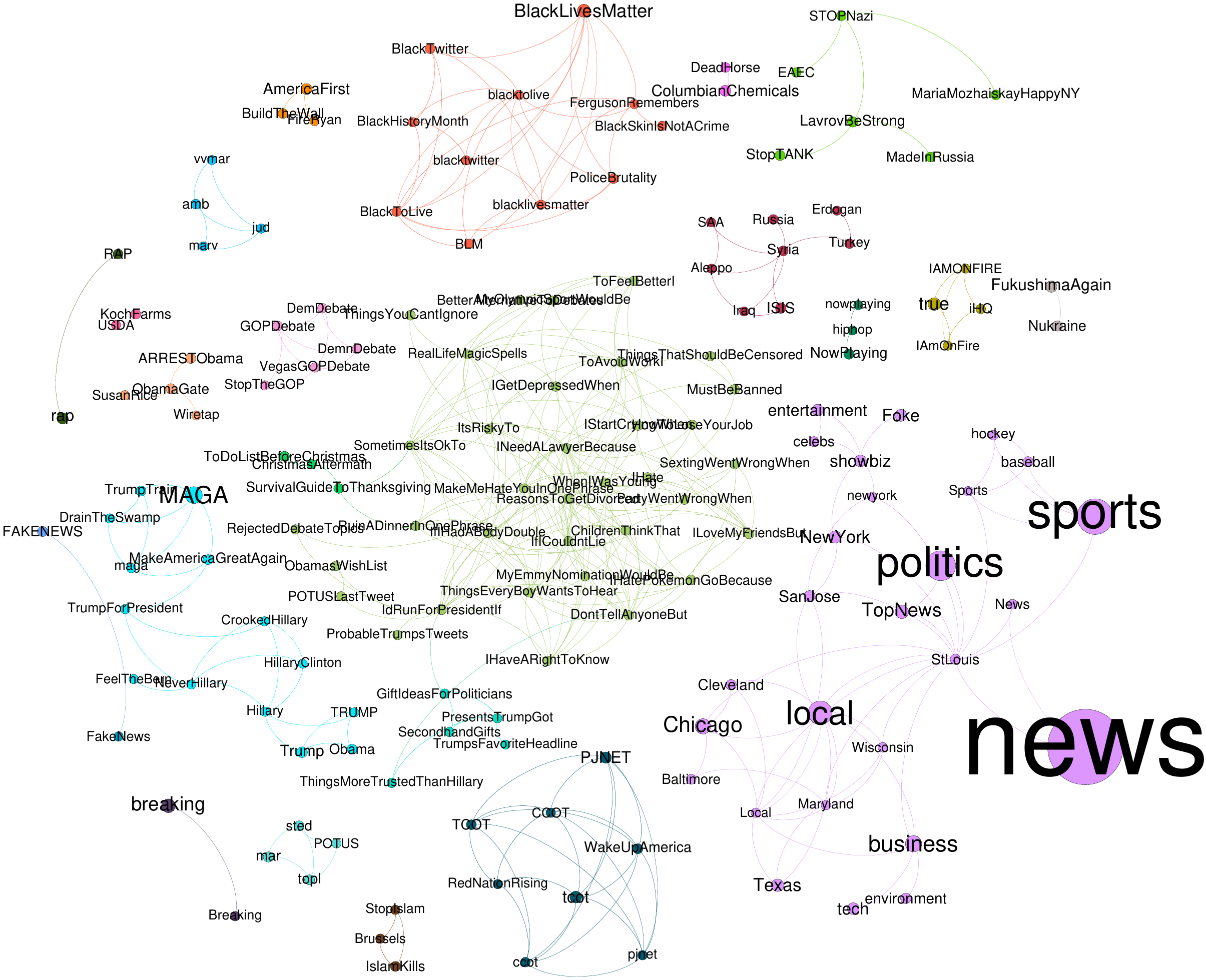}\label{fig:russians_hashtags_graph}}
\subfigure[]{\includegraphics[width=0.8\textwidth]{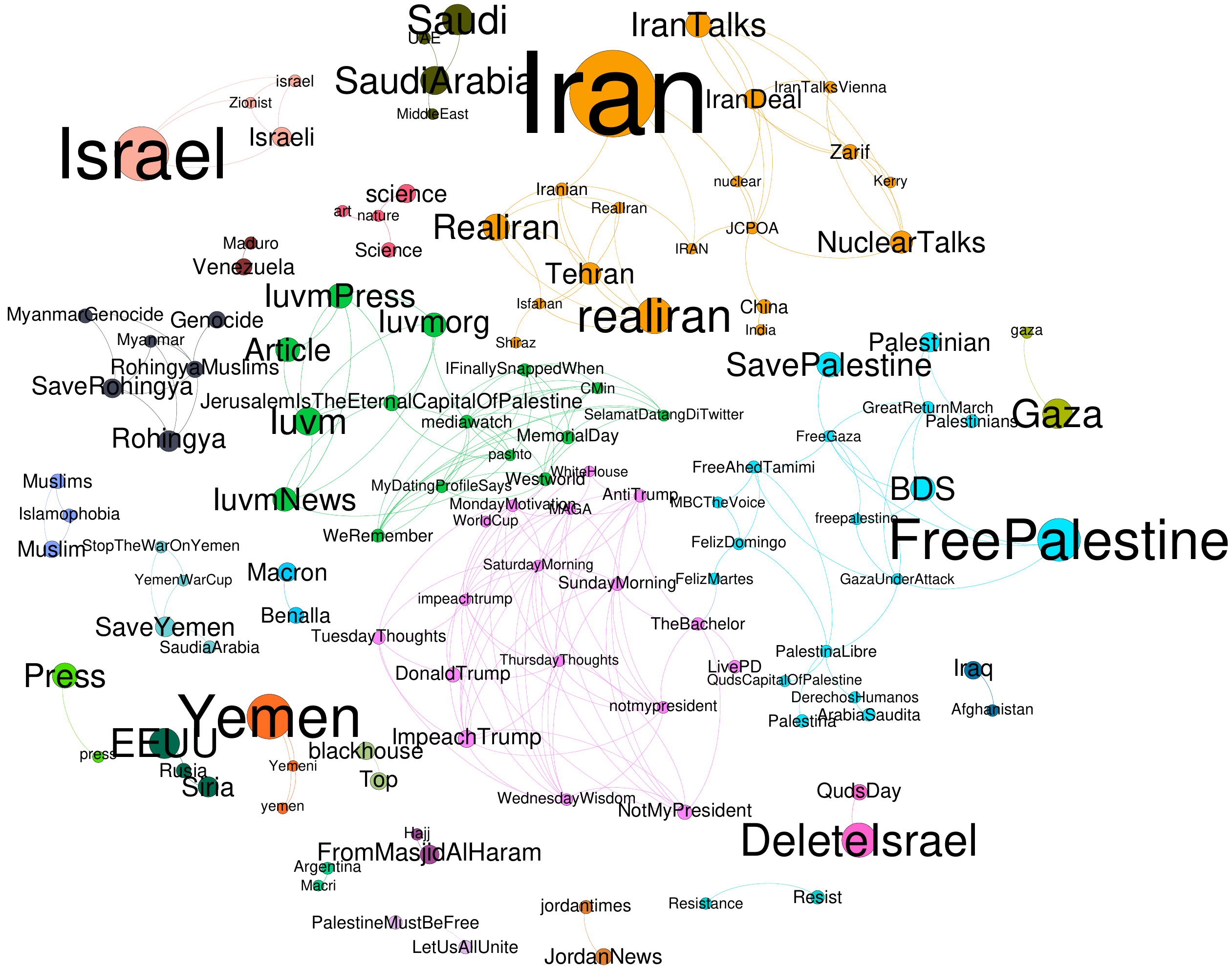} \label{fig:iranians_hashtags_graph}}
   \caption{Visualization of the top hashtags used by a) Russian trolls on Twitter (see~\cite{interactive_russians} for interactive version) and b) Iranian trolls on Twitter (see~\cite{interactive_iranians} for an interactive version).}
\label{fig:hashtags_graphs}
\end{figure*}

\descr{Hashtags.} Next, we aim to understand the use of hashtags with a focus on the ones written in English.
In Table~\ref{tbl:top_hashtags_russians_iranians}, we report the top 20 English hashtags for both Russian and Iranian trolls.
State-sponsored trolls appear to use hashtags to disseminate news (9.5\%) and politics (3.0\%) related content, but also use several that might be indicators of propaganda and/or controversial topics, e.g., \#BlackLivesMatter.
For instance, one notable example is:
``WATCH: Here is a typical \#BlackLivesMatter protester:  `I hope I kill all white babes!' \#BatonRouge $<$url$>$'' on July 17, 2016. Note that $<$url$>$ denotes a link.

Fig.~\ref{fig:hashtags_graphs} shows a visualization of hashtag usage built from the two word2vec models.
Here, we show hashgtags used in a similar context, by constructing a graph where nodes are words that correspond to hashtags from the word2vec models, and edges are weighted by the cosine distances (as produced by the word2vec models) between the hashtags.
After trimming out all edges between nodes with weight less than a threshold, based on methodology from~\cite{finkelstein2018quantitative}, we run the community detection heuristic presented in~\cite{blondel2008fast}, and mark each community with a different color.
Finally, the graph is layed out with the ForceAtlas2 algorithm~\cite{jacomy2014forceatlas2}, which takes into account the weight of the edges when laying out the nodes in 2-dimensional space.
Note that the size of the nodes is proportional to the number of times the hashtag appeared in each dataset.

We first observe that, in Fig.~\ref{fig:russians_hashtags_graph} there is a central mass of what we consider ``general audience'' hashtags (see green community on the center of the graph): hashtags related to a holiday or a specific trending topic (but non-political) hashtag.
In the bottom right portion of the plot we observe ``general news'' related categories; in particular American sports related hashtags (e.g., ``baseball'').
Next, we see a community of hashtags (light blue, towards the bottom left of the graph) clearly related to Trump's attacks on Hillary Clinton.

The Iranian trolls again show different behavior.
There is a community of hashtags related to nuclear talks (orange), a community related to Palestine (light blue), and a community that is clearly anti-Trump (pink).
The central green community exposes some of the ways they pushed the IUVM fake news network by using innocuous hashtags like ``\#MyDatingProfileSays'' as well as politically motivated ones like ``\#JerusalemIsTheEternalCapitalOfPalestine.''

\begin{figure*}[t!]
\center
\subfigure[Russian trolls - Before election]{\includegraphics[width=0.49\columnwidth]{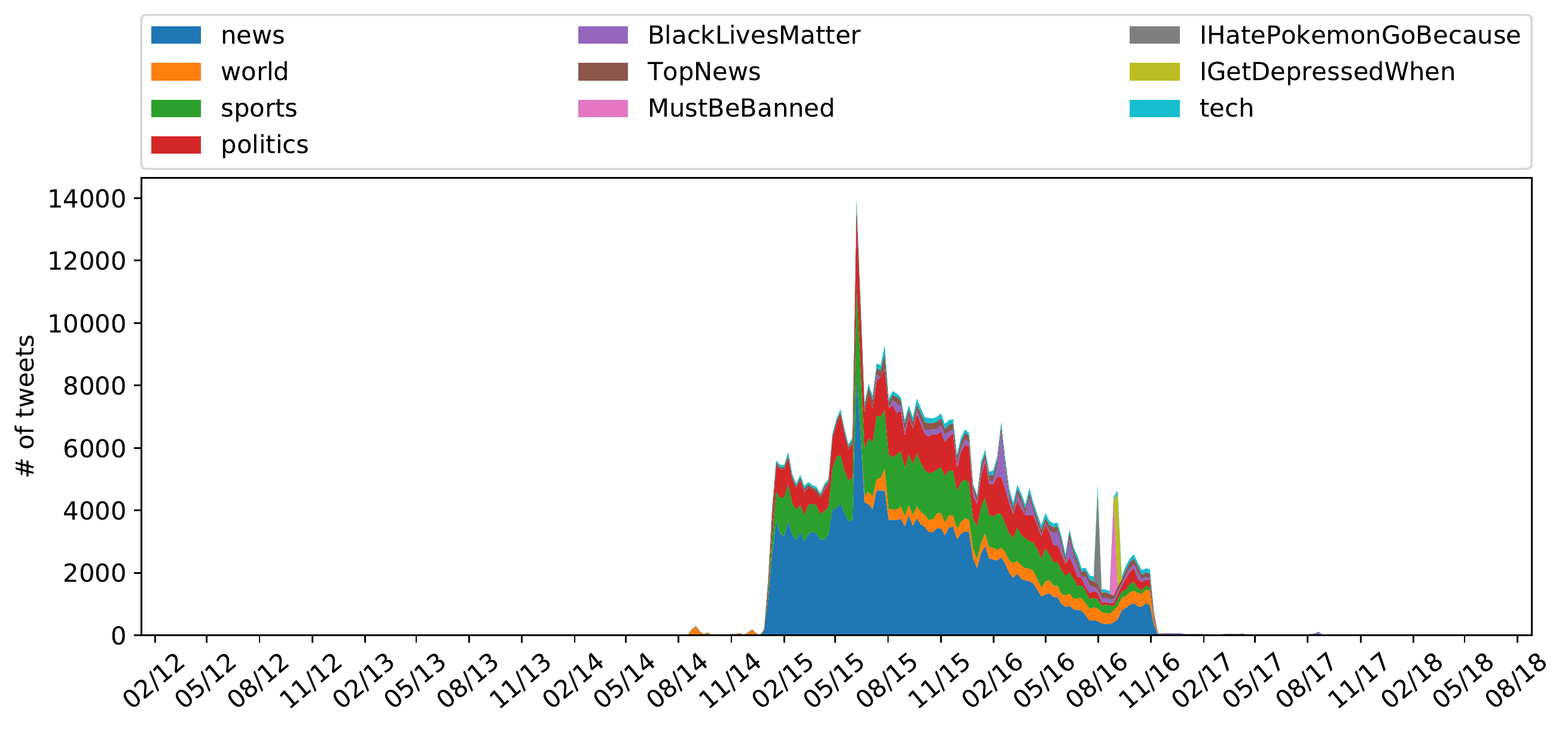}\label{subfig:russians_elections_before}}
\subfigure[Russian trolls - After election]{\includegraphics[width=0.49\columnwidth]{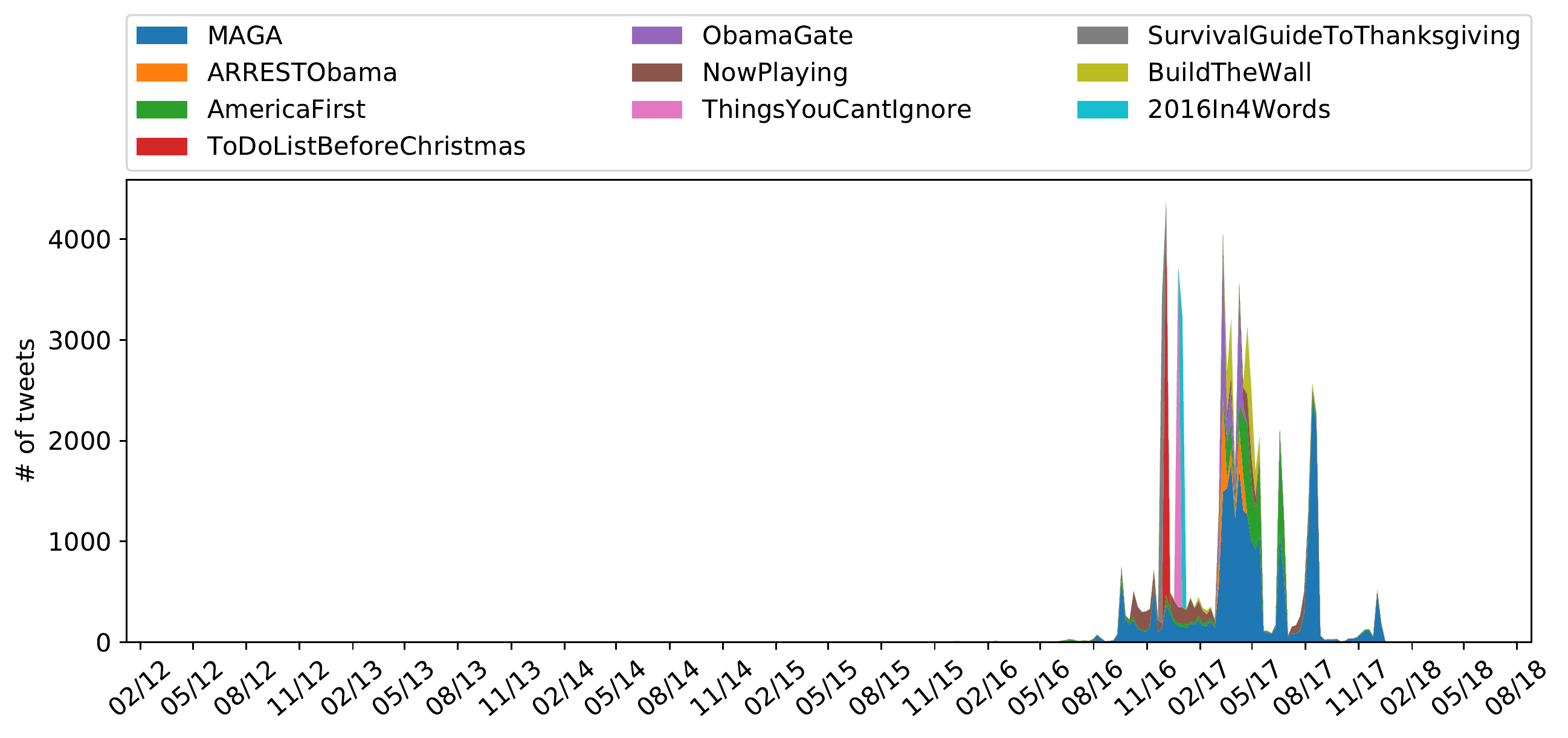}\label{subfig:russians_elections_after}}
\subfigure[Iranians trolls - Before election]{\includegraphics[width=0.49\columnwidth]{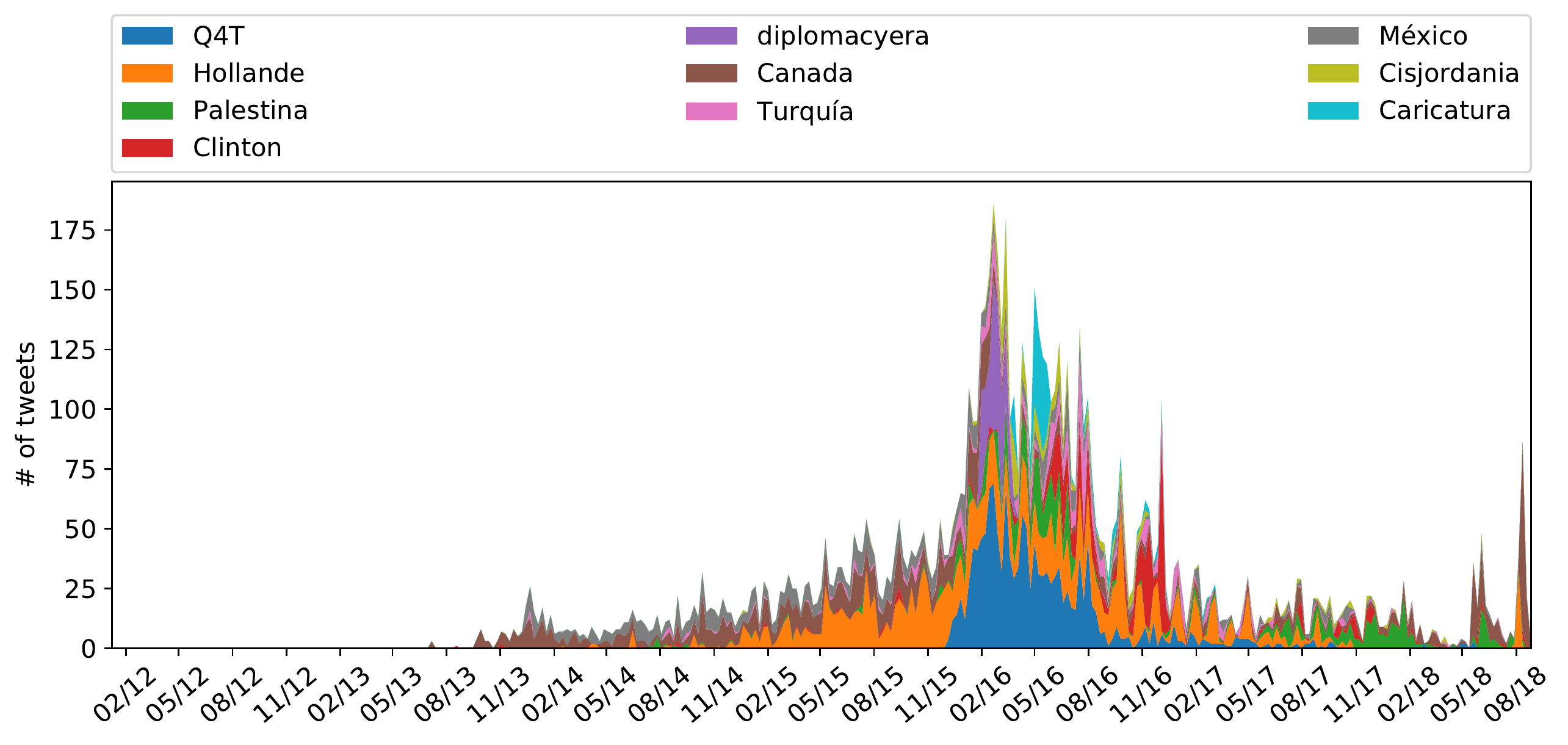}\label{subfig:iranians_elections_before}}
\subfigure[Iranians trolls - After election]{\includegraphics[width=0.49\columnwidth]{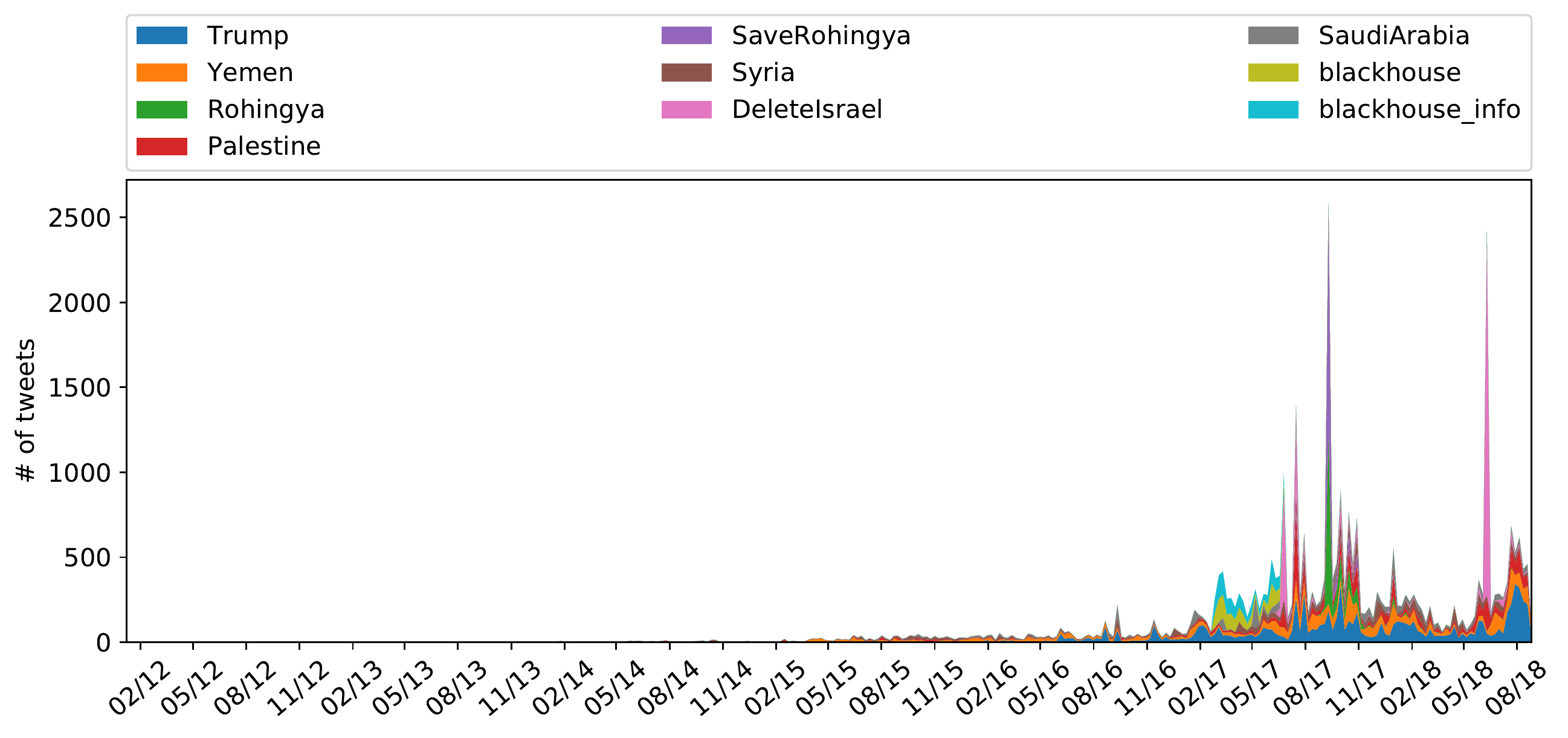}\label{subfig:iranians_elections_after}}
  \caption{ Top ten hashtags that appear a) c) substantially more times before the US elections rather than after the elections; and b) d) substantially more times after the elections rather than before. }
\label{fig:hashtags_before_after_election_russians}
\end{figure*}

We also study \emph{when} these hashtags are used by the trolls, finding that most of them are well distributed over time.
However we find some interesting exceptions.
We highlight a few of these in Fig.~\ref{fig:hashtags_before_after_election_russians}, which plots the top ten hashtags that Russian and Iranian trolls posted with substantially different rates before and after the 2016 US Presidential election.
The set of hashtags was determined by examining the relative change in posting volume before and after the election.
From the plots we make several observations.
First, we note that more general audience hashtags remain a staple of Russian trolls before the election (the relative decrease corresponds to the overall relative decrease in troll activity following the Crimea conflict).
They also use relatively innocuous/ephemeral hashtags like \#IHatePokemonGoBeacause, likely in an attempt to hide the true nature of their accounts.
That said, we also see them attaching to politically divisive hashtags like \#BlackLivesMatters around the time that Donald Trump won the Republican Presidential primaries in June 2016.
In the ramp up to the 2016 election, we see a variety of clearly political related hashtags, with \#MAGA seeing substantial peaks starting in early 2017 (higher than any peak during the 2016 Presidential campaigns).
We also see a large number of politically ephemeral hashtags attacking Obama and a campaign to push the border wall between Mexico.
In addition to these politically oriented hashtags, we again see the usage of ephemeral hashtags related to holidays.
\#SurvivalGuideToThanksgiving in late November 2016 is particularly interesting as it was heavily used for discussing how to deal with interacting with family members with wildly different view points on the recent election results.
This hashtag was exclusively used to give trolls a vector to sow discord.
When it comes to Iranian trolls, we note that, prior to the 2016 election, they share many posts with hashtags related to Hillary Clinton (see Fig.~\ref{subfig:iranians_elections_before}).
After the election they shift to posting negatively about Donald Trump (see Fig.~\ref{subfig:iranians_elections_after}).

\begin{table*}[t!]
\centering
\resizebox{\textwidth}{!}{
\begin{tabular}{rl|rl}
\hline
\textbf{Topic} & \multicolumn{1}{l|}{\textbf{Terms (Russian trolls on Twitter)}} & \textbf{Topic} & \textbf{Terms (Iranian trolls on Twitter)} \\ \hline
1 & new,  now,  music,  get,  got,  thanks,  orleans,  entertainment,  follow,  show & 1 & iran,  will,  deal,  irantalks,  iranian,  nucleartalks,  nuclear,  irandeal,  zarif,  congress \\
2 & sports,  year,  news,  old,  game,  workout,  win,  nfl,  chicago,  morning & 2 & isis,  new,  state,  fire,  blackhouse,  open,  inferno,  nation,  will,  turkish \\
3 & day,  love,  one,  foke,  today,  happy,  first,  away,  last,  time,  will,  best & 3 & yemen,  press,  front, liberty, children, saudi, isis, rohingya, school, king \\
4 & can,  don,  like,  people,  just,  know,  get,  want,  will,  never,  good,  make & 4 & isis, american, trump, sex, war, young, fbi, putin, terrorists, world \\
5 & black,  women,  great,  america,  people,  tcot,  blacklivesmatter,  read,  american,  isis & 5 & president, former, syria, obama, turkish, iraqi, russian, foreign, palestine, stop \\
6 & news,  police,  man,  local,  woman,  texas,  killed,  shooting,  chicago,  death & 6 & trump, donald, can, toonsonline, see, don, know, like, will, just \\
7 & can,  forget,  change,  wait,  book,  far,  illegal,  worst,  words,  save,  united,  done & 7 & saudi, israeli, attack, israel, days, terrorist, usa, palestinian, cia, third \\
8 & exercise,  wanna,  fight,  still,  control,  nice,  gun,  hold,  perfect,  enlist & 8 & isis, iran, first, realiran, siege, success, sydney, shame, tehran, photos \\
9 & trump,  obama,  politics,  president,  hillary,  clinton,  breaking,  just,  house,  video & 9 & saudi, united, states, isis, arabia, racist, society, structurally, oil, israel \\
10 & news,  world,  business,  health,  new,  says,  money,  tech,  water,  syria & 10 & israel, syria, police, syrian, muslim, video, people, death, trump, rights \\ \hline
\end{tabular}
}
\caption{Terms extracted from LDA topics of tweets from Russian and Iranian trolls on Twitter.}
\label{tbl:lda_topics_agg}
\end{table*}

\begin{table}[t!]
\centering
\resizebox{0.6\columnwidth}{!}{%
\begin{tabular}{rl}
\hline
\textbf{Topic} & \multicolumn{1}{l}{\textbf{Terms (Russian trolls on Reddit)}} \\ \hline
1 & police, black, man, year, cop, video, woman, shot, white, arrested \\
2 & love, one, absolutely, life, good, time, ever, wow, man, sure \\
3 & man, dog, thank, back, thing, poor, now, happy, feeling, day \\
4 & can, even, damn, cia, right, ledger, government, god, future, cap \\
5 & just, will, one, really, can, people, think, time, like, need \\
6 & like, people, just, don, looks, great, want, tie, also, tokens \\
7 & police, cop, officer, state, man, rights, obama, shooting, death, omg \\
8 & hillary, clinton, trump, new, lives, black, cute, matter, donald, recommend \\
9 & will, don, can, people, get, just, understand, buy, nothing, btc \\
10 & bitcoin, can, crypto, nice, people, try, just, tie, like, blockchain \\ \hline
\end{tabular}%
}
\caption{Terms extracted from LDA topics of posts from Russian trolls on Reddit.}
\label{tbl:lda_reddit}
\end{table}

\descr{LDA analysis.}
We also use the Latent Dirichlet Allocation (LDA) model~\cite{blei2003latent} to analyze
tweets' semantics.
We train an LDA model for each of the datasets and extract ten distinct topics with ten words,
as reported in Table~\ref{tbl:lda_topics_agg}.
While both Russian and Iranian trolls tweet about politics related topics, for Iranian trolls, this seems to be focused more on regional, and possibly even internal issues.
For example, ``iran'' itself is a common term in several of the topics, as is ``israel,'' ``saudi,'' ``yemen,'' and ``isis.''
While both sets of trolls discuss the proxy war in Syria (in which both states are involved), while the Iranian trolls have topics pertaining to Russia and Putin, the Russian trolls do not make any mention of Iran, instead focusing on more vague political topics like gun control and racism.
For Russian trolls on Reddit (see Table~\ref{tbl:lda_reddit}) we again find topics related to politics as well some topics related to discussions about cryptocurrencies (see topics 9 and 10).

\begin{figure}[t]
\centering
\includegraphics[width=0.7\columnwidth]{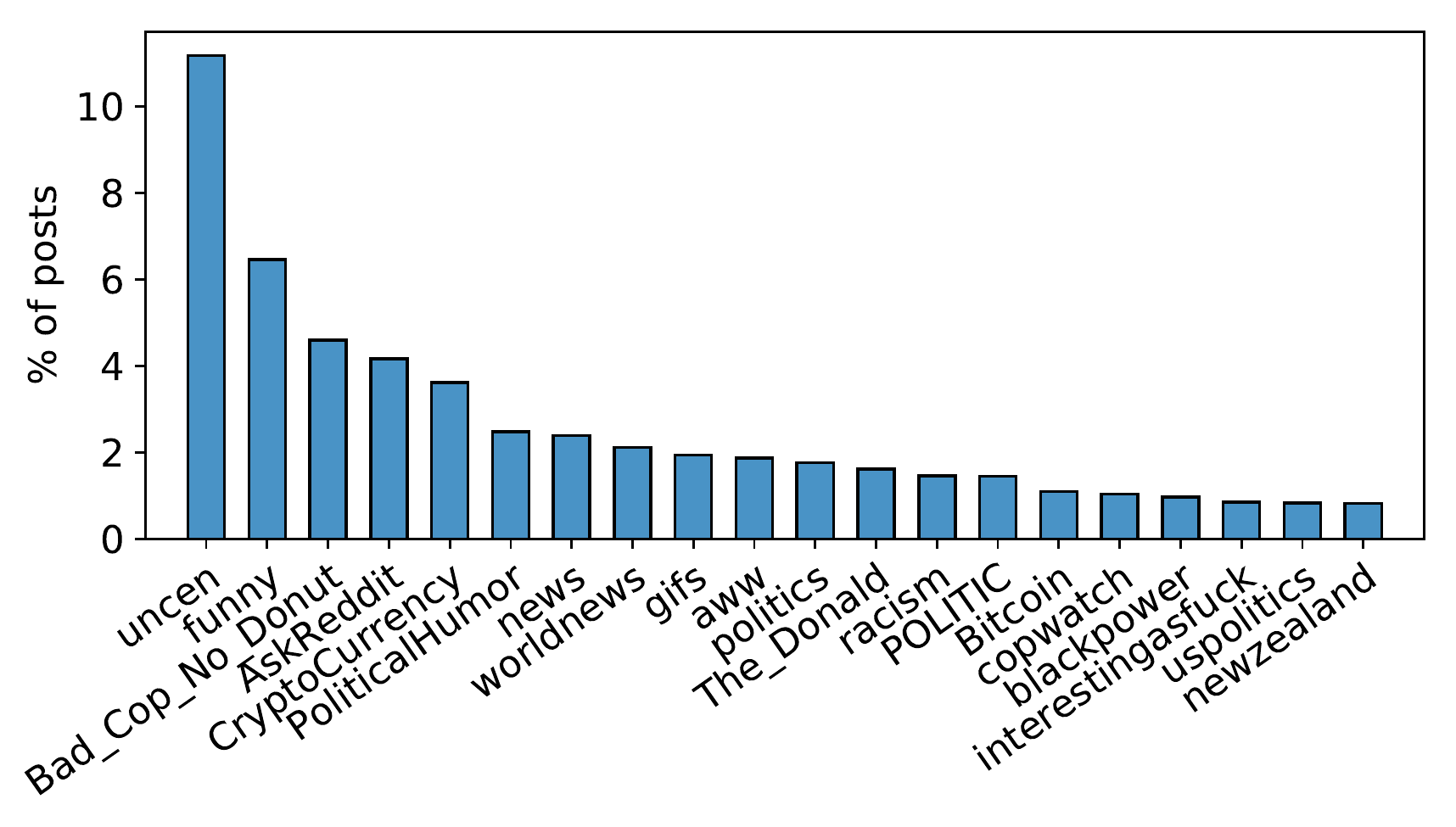}
   \caption{Top 20 subreddits that Russian trolls were active and their respective percentage of posts.}
\label{fig:top_subreddits}
\end{figure}

\descr{Subreddits.} Fig.~\ref{fig:top_subreddits} shows the top 20 subreddits that Russian trolls on Reddit exploited and their respective percentage of posts over the whole dataset.
The most popular subreddit is /r/uncen (11\% of posts), which is a subreddit created by a specific Russian troll and, via manual examination, appears to be primarily used to disseminate news articles of questionable credibility.
Other popular subreddits include general audience subreddits like /r/funny (6\%) and /r/AskReddit (4\%), likely in an attempt to obfuscate the fact that they are state-sponsored trolls in the same way that innocuous hashtags were used on Twitter.
Finally, it is worth noting that the Russian trolls were particularly active on communities related to cryptocurrencies like  /r/CryptoCurrency (3.6\%) and /r/Bitcoin (1\%) possibly attempting to influence the prices of specific cryptocurrencies.
This is particularly noteworthy considering cryptocurrencies have been reportedly used to launder money, evade capital controls, and perhaps used to evade sanctions~\cite{crypto_money_laundering,crypto_tax_evasion}.

\descr{URLs.}
We next analyze the URLs included in the tweets/posts.
In Table~\ref{tbl:top_domains}, we report the top 20 domains for both Russian and Iranian trolls.
Livejournal (5.4\%) is the most popular domain in the Russian trolls dataset on Twitter, likely due the Ukrainian campaign. 
Overall, we can observe the impact of the Crimean conflict, with essentially all domains posted by the Russian trolls being Russian language or Russian oriented.
One exception to Russian language sites is RT, the Russian-controlled propaganda outlet.
The Iranian trolls similarly post more ``localized'' domains, for example, jordan-times, but we also see them heavily pushing the IUVM fake news network.
When it comes to Russian trolls on Reddit, we find that they were mostly posting random images through Imgur (image-hosting site, 16\% of the posts), likely in an attempt to accumulate karma score.
We also note that a substantial portion of posts contained URLs to (fake) news sites linked with the Internet Research Agency like blackmattersus.com(5.7\%) and donotshootus.us (2.5\%).

\begin{table}[t]
\centering
\footnotesize
\resizebox{0.8\columnwidth}{!}{%
\begin{tabular}{rrrrrl}
\hline
\textbf{\begin{tabular}[c]{@{}r@{}}Domain (Russian \\ trolls on Twitter\end{tabular}} & \multicolumn{1}{l}{\textbf{(\%)}} & \textbf{\begin{tabular}[c]{@{}r@{}}Domain(Iranian \\ trolls on Twitter)\end{tabular}} & \multicolumn{1}{l}{\textbf{(\%)}} & \textbf{\begin{tabular}[c]{@{}r@{}}Domain (Russian \\ trolls on Reddit)\end{tabular}} & \textbf{(\%)} \\ \hline
livejournal.com & \multicolumn{1}{r|}{5.4\%} & awdnews.com & \multicolumn{1}{r|}{29.3\%} & i.imgur & 10.8\% \\
riafan.ru & \multicolumn{1}{r|}{5.0\%} & dlvr.it & \multicolumn{1}{r|}{7.1\%} & blackmattersus.com & 5.7\% \\
twitter.com & \multicolumn{1}{r|}{2.5\%} & fb.me & \multicolumn{1}{r|}{4.8\%} & imgur.com & 5.3\% \\
ift.tt & \multicolumn{1}{r|}{1.8\%} & whatsupic.com & \multicolumn{1}{r|}{4.2\%} & donotshoot.us & 2.5\% \\
ria.ru & \multicolumn{1}{r|}{1.8\%} & googl.gl & \multicolumn{1}{r|}{3.9\%} & theguardian.com & 1.0\% \\
googl.gl & \multicolumn{1}{r|}{1.7\%} & realnienovosti.com & \multicolumn{1}{r|}{2.1\%} & nytimes.com & 1.0\% \\
dlvr.it & \multicolumn{1}{r|}{1.5\%} & twitter.com & \multicolumn{1}{r|}{1.7\%} & washingtonpost.com & 0.8\% \\
gazeta.ru & \multicolumn{1}{r|}{1.4\%} & libertyfrontpress.com & \multicolumn{1}{r|}{1.6\%} & huffingtonpost.com & 0.8\% \\
yandex.ru & \multicolumn{1}{r|}{1.2\%} & iuvmpress.com & \multicolumn{1}{r|}{1.5\%} & foxnews.com & 0.8\% \\
j.mp & \multicolumn{1}{r|}{1.1\%} & buff.ly & \multicolumn{1}{r|}{1.4\%} & youtube.com & 0.8\% \\
rt.com & \multicolumn{1}{r|}{0.8\%} & 7sabah.com & \multicolumn{1}{r|}{1.3\%} & photographyisnotacrime.com & 0.7\% \\
nevnov.ru & \multicolumn{1}{r|}{0.7\%} & bit.ly & \multicolumn{1}{r|}{1.2\%} & thefreethoughtproject.com & 0.6\% \\
youtu.be & \multicolumn{1}{r|}{0.6\%} & documentinterdit.com & \multicolumn{1}{r|}{1.0\%} & butthis.com & 0.5\% \\
vesti.ru & \multicolumn{1}{r|}{0.5\%} & facebook.com & \multicolumn{1}{r|}{0.8\%} & cnn.com & 0.5\% \\
kievsmi.net & \multicolumn{1}{r|}{0.5\%} & al-hadath24.com & \multicolumn{1}{r|}{0.7\%} & dailymail.co & 0.5\% \\
youtube.com & \multicolumn{1}{r|}{0.5\%} & jordan-times.com & \multicolumn{1}{r|}{0.7\%} & rt.com & 0.5\% \\
kiev-news.com & \multicolumn{1}{r|}{0.5\%} & iuvmonline.com & \multicolumn{1}{r|}{0.6\%} & politico.com & 0.4\% \\
inforeactor.ru & \multicolumn{1}{r|}{0.4\%} & youtu.be & \multicolumn{1}{r|}{0.6\%} & truthdig.com & 0.4\% \\
lenta.ru & \multicolumn{1}{r|}{0.4\%} & alwaght.com & \multicolumn{1}{r|}{0.5\%} & nbcnews.com & 0.4\% \\
emaidan.com.ua & \multicolumn{1}{r|}{0.3\%} & ift.tt & \multicolumn{1}{r|}{0.5\%} & breitbart.com & 0.4\% \\ \hline
\end{tabular}
}
\caption{Top 20 domains included in tweets/posts from Russian and Iranian trolls on Twitter and Reddit.}
\label{tbl:top_domains}
\end{table}

\begin{table}[]
\resizebox{\columnwidth}{!}{
\begin{tabular}{@{}lrrrrrrrrr@{}}
\toprule
\multicolumn{8}{c}{\textbf{Events per community}}                                                                                                                                                                                                                                                                                                            & \multicolumn{2}{c}{\textbf{Total}}                                       \\ \midrule
\textbf{\begin{tabular}[c]{@{}c@{}}URLs\\shared by\end{tabular}} & \multicolumn{1}{c}{\textbf{\dspol}} & \multicolumn{1}{c}{\textbf{Reddit}} & \multicolumn{1}{c}{\textbf{Twitter}} & \multicolumn{1}{c}{\textbf{Gab}} & \multicolumn{1}{c}{\textbf{The\_Donald}} & \multicolumn{1}{c}{\textbf{Iran}} & \multicolumn{1}{c|}{\textbf{Russia}} & \multicolumn{1}{c|}{\textbf{Events}} & \multicolumn{1}{c}{\textbf{URLs}} \\ \hline
\textbf{Russians}                                                 & 76,155                                             & 366,319                             & 1,225,550                            & 254,016                          & 61,968                                   & 0                                 & \multicolumn{1}{r|}{151,222}         & \multicolumn{1}{r|}{2,135,230}       & 48,497                            \\
\textbf{Iranians}                                                 & 3,274                                              & 28,812                              & 232,898                              & 5,763                            & 971                                      & 19,629                            & \multicolumn{1}{r|}{0}               & \multicolumn{1}{r|}{291,347}         & 4,692                             \\
\textbf{Both}                                                     & 331                                                & 2,060                               & 85,467                               & 962                              & 283                                      & 334                               & \multicolumn{1}{r|}{565}             & \multicolumn{1}{r|}{90,002}          & 153                               \\ \bottomrule
\end{tabular}
}
\caption{Total number of events in each community for URLs shared by a) Russian trolls; b) Iranian trolls; and c) Both Russian and Iranian trolls.}
\label{tbl:hawkes}
\end{table}

\begin{figure*}[t]
\center
\subfigure[Russian trolls]{\includegraphics[width=0.49\textwidth]{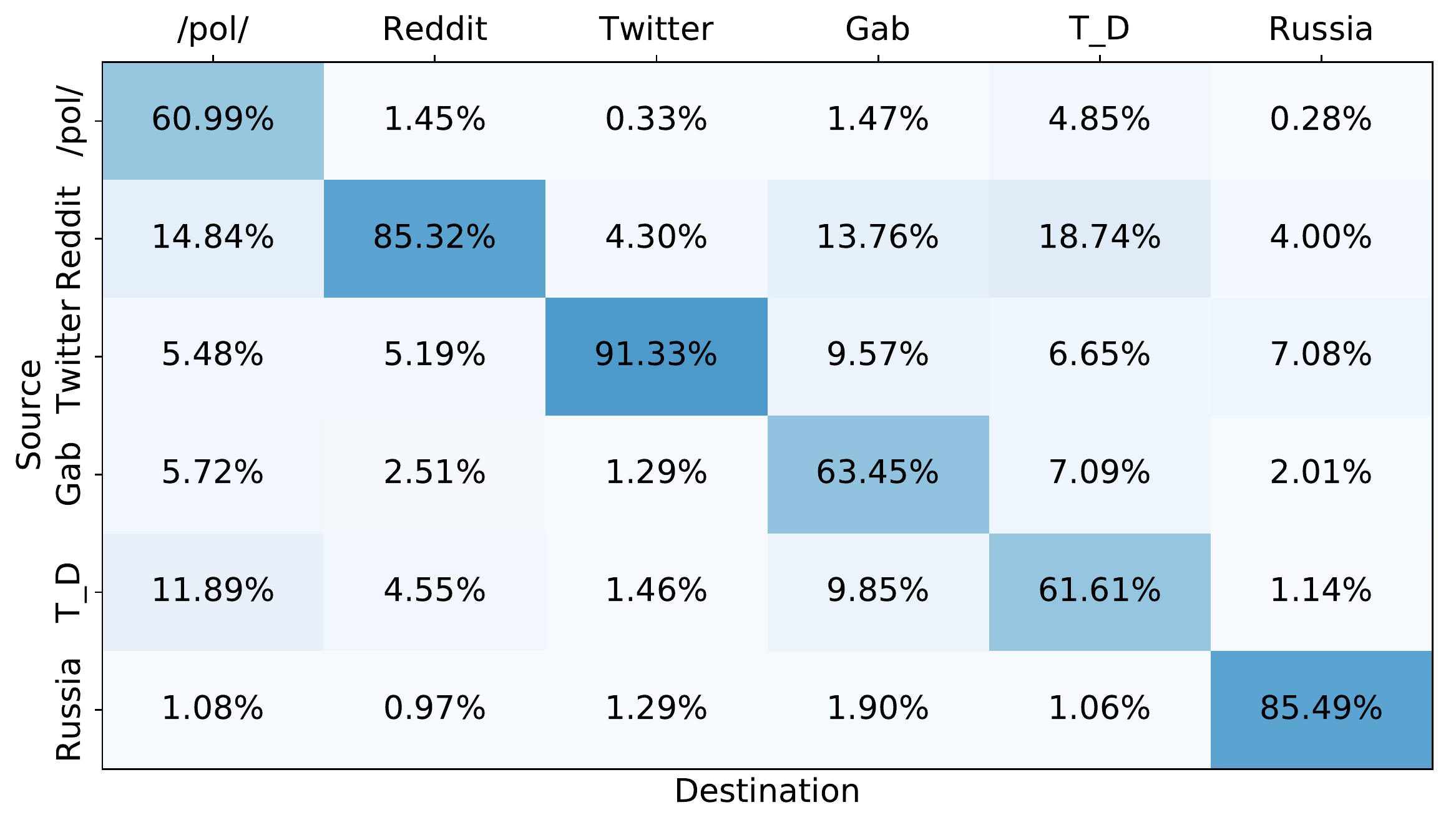}\label{subfig:raw_influence_russians}}
\subfigure[Iranian trolls]{\includegraphics[width=0.49\textwidth]{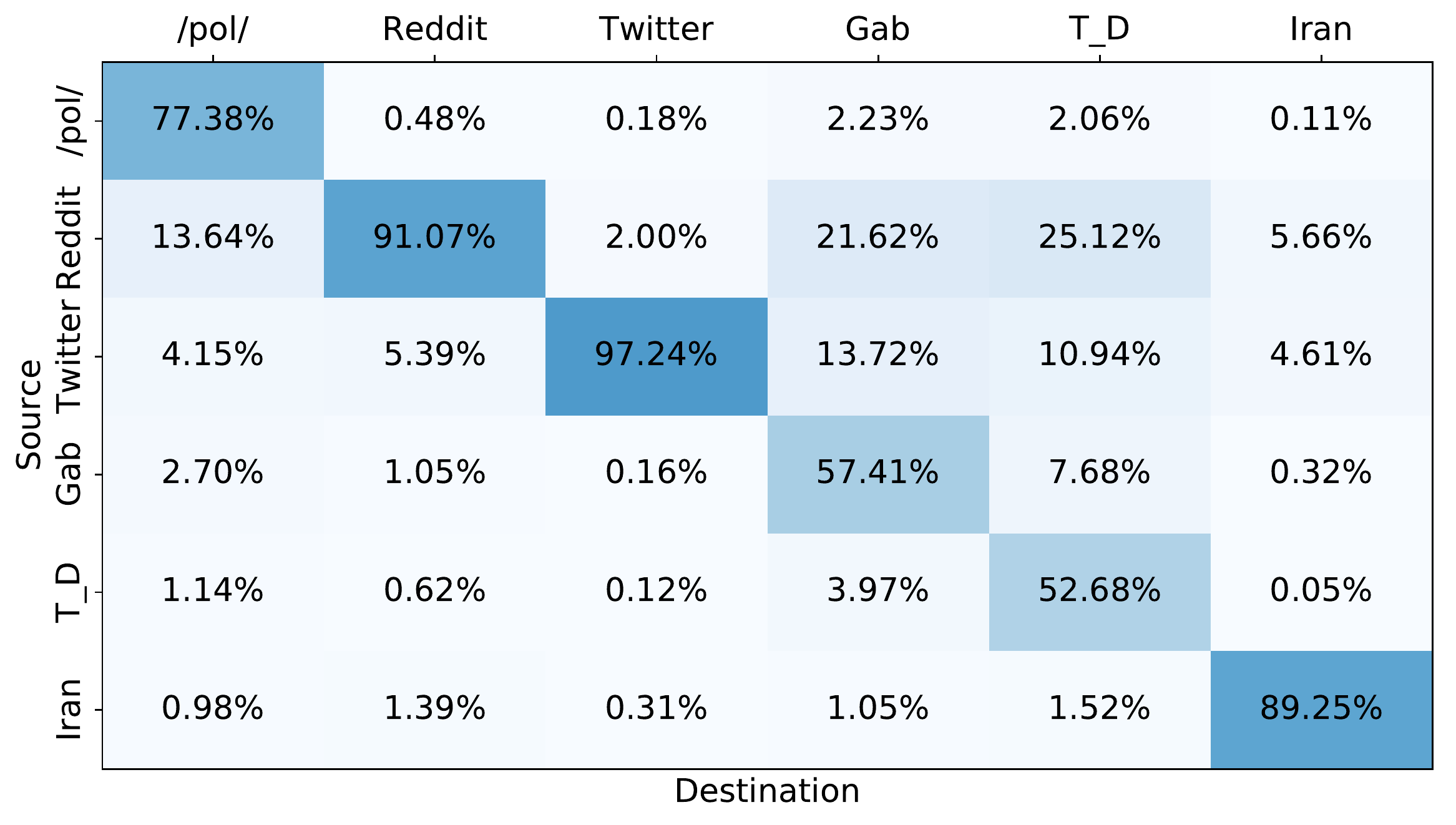}\label{subfig:raw_influence_iranians}}
\subfigure[Both]{\includegraphics[width=0.49\textwidth]{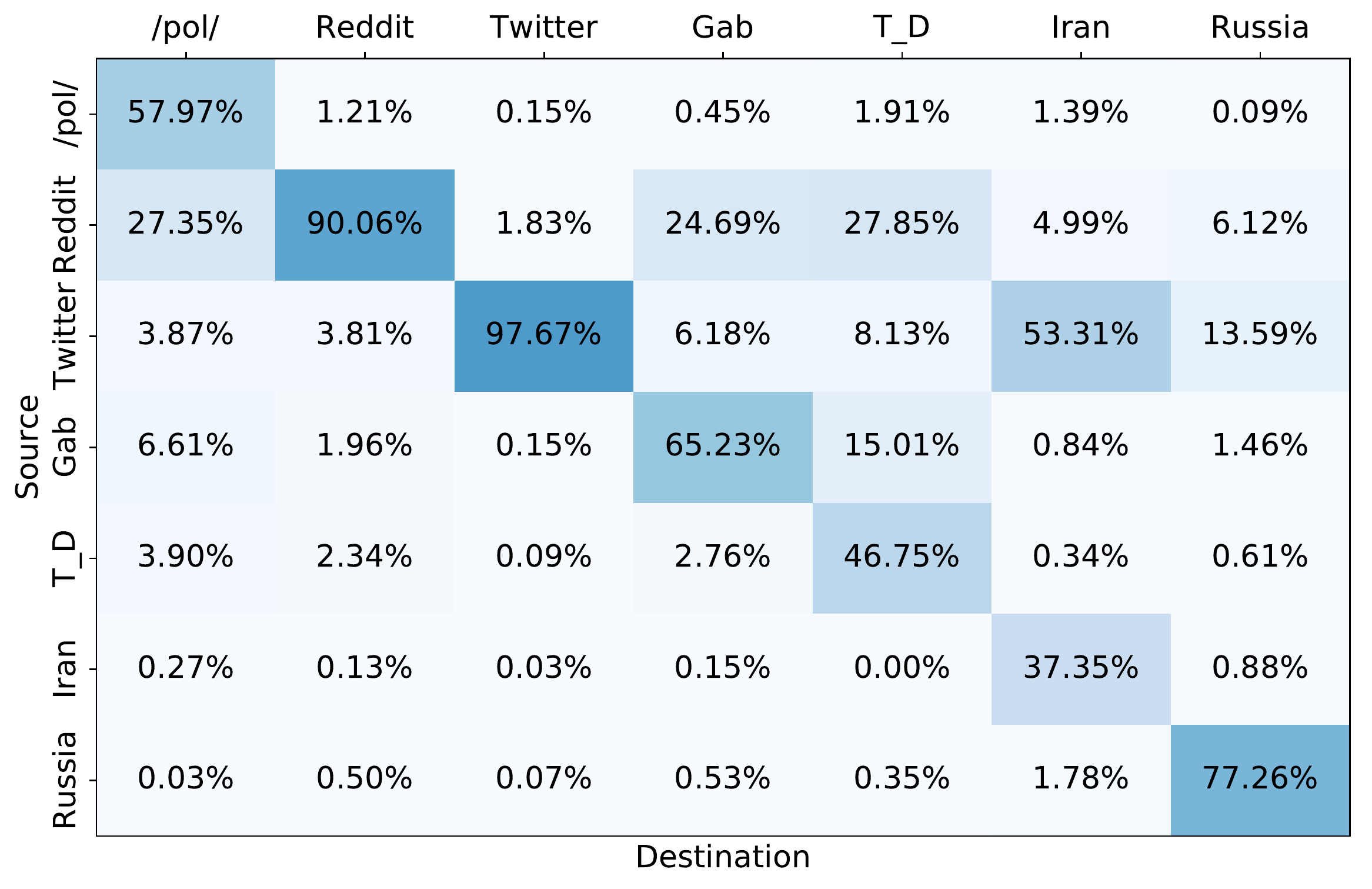}\label{subfig:raw_influence_both}}
\caption{Percent of \emph{destination} events caused by the source community to the destination community for URLs shared by a) Russian trolls; b) Iranian trolls; and c) both Russian and Iranian trolls. }
\label{fig:raw_influence}
\end{figure*}

\begin{figure*}[t]
\center
\subfigure[Russian trolls]{\includegraphics[width=0.49\textwidth]{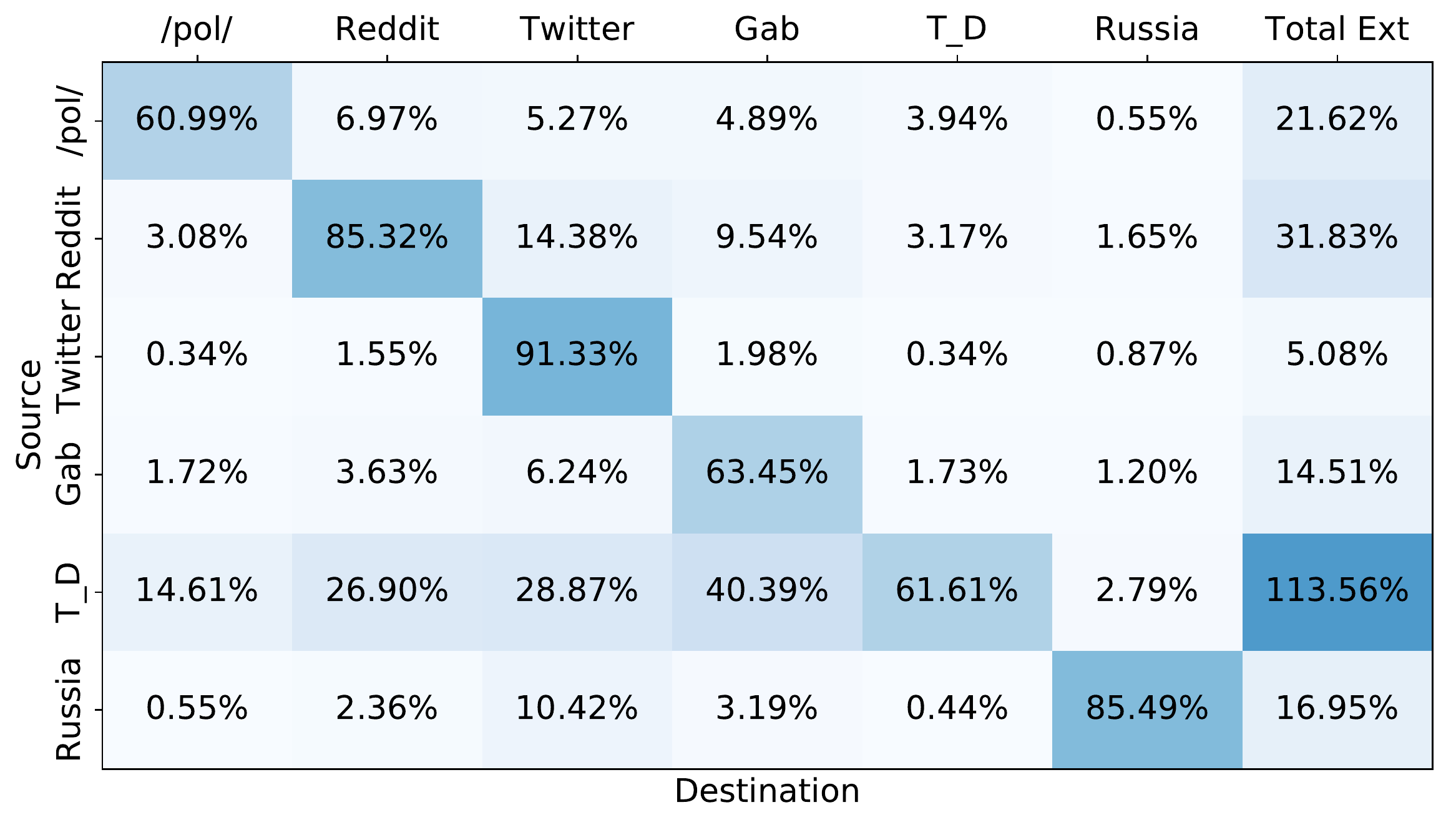}\label{subfig:norm_influence_russians}}
\subfigure[Iranian trolls]{\includegraphics[width=0.49\textwidth]{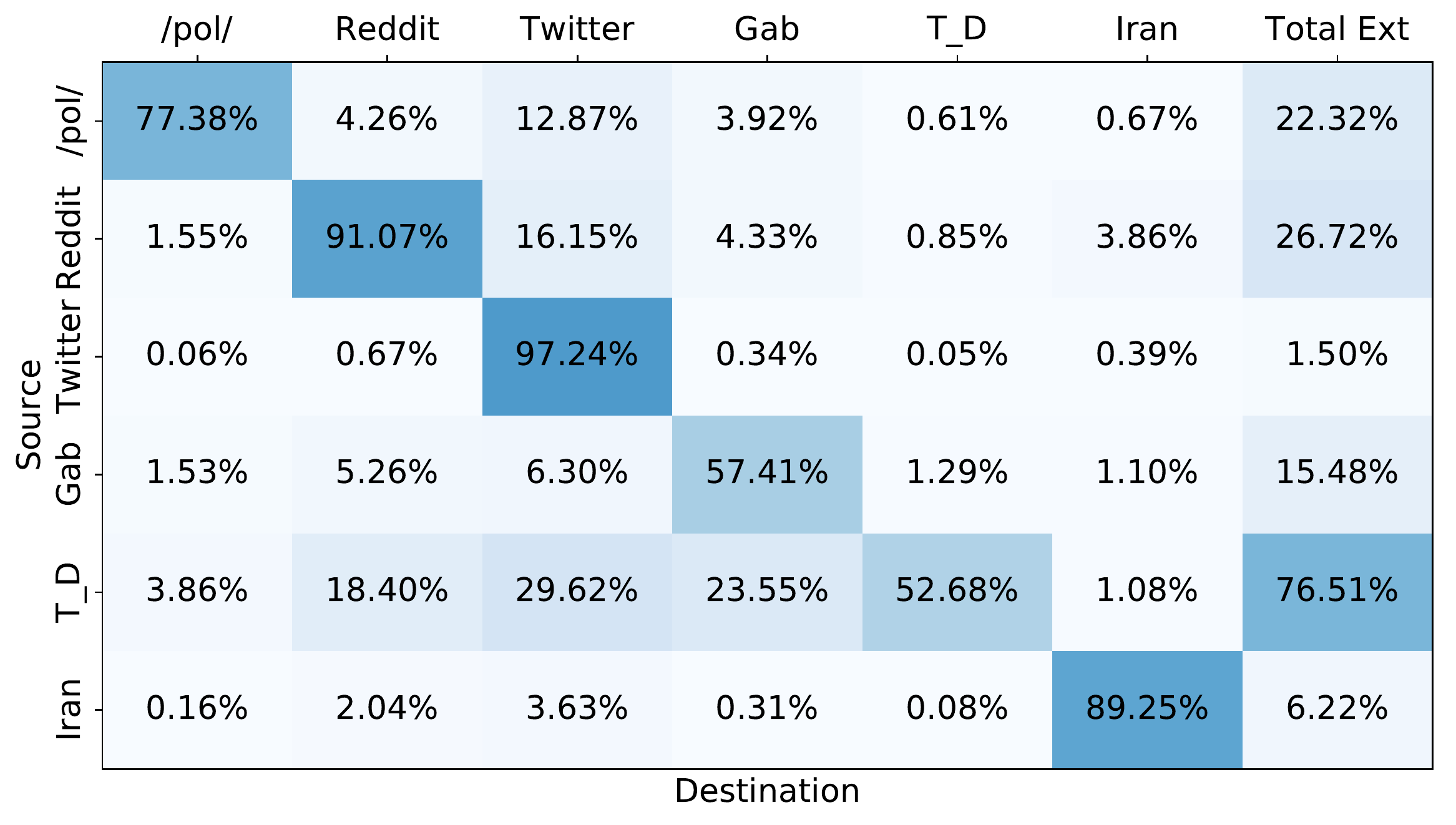}\label{subfig:norm_influence_iranians}}
\subfigure[Both]{\includegraphics[width=0.49\textwidth]{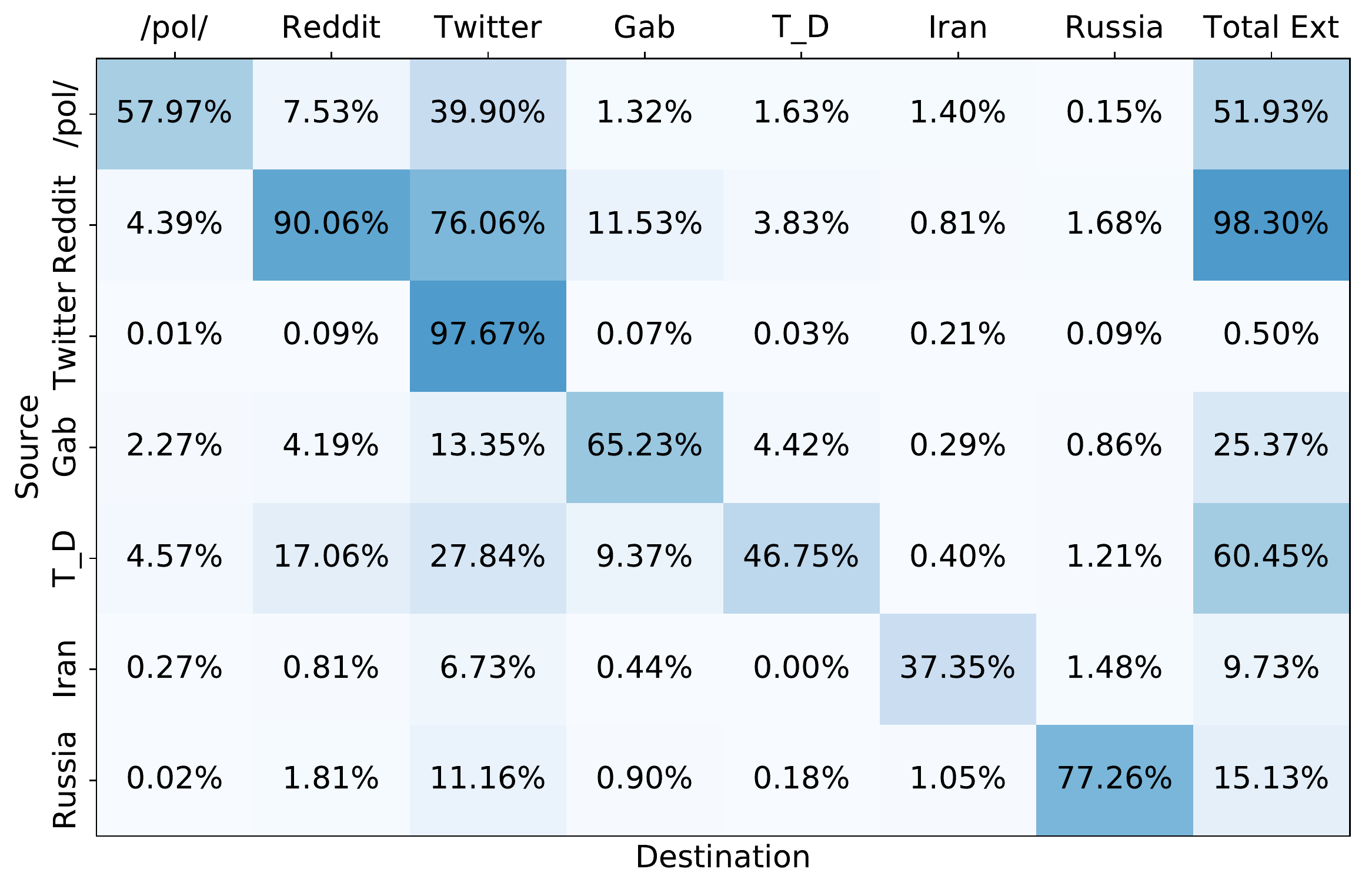}\label{subfig:norm_influence_both}}
\caption{Influence from source to destination community, normalized by the number of events in the \emph{source} community for URLs shared by a) Russian trolls; b) Iranian trolls; and c) Both Russian and Iranian trolls. We also include the total external influence of each community.}
\label{fig:norm_influence}
\end{figure*}

\subsection{Influence Estimation} \label{sec:influence}
Thus far, we have analyzed the behavior of Russian and Iranian trolls on Twitter and Reddit, with a special focus on how they evolved over time.
Allegedly, one of their main goals is to manipulate the opinion of other users and extend the cascade of information that they share (e.g., lure other users into posting similar content)~\cite{newsweek_manipulation}.
Therefore, we now set out to determine their impact in terms of the dissemination of information on Twitter, and on the greater Web.

To assess their influence, we look at three different groups of URLs: 1) URLs shared by Russian trolls on Twitter, 2) URLs shared by Iranian trolls on Twitter, and 3) URLs shared by both Russian \emph{and} Iranian trolls on Twitter.
We then find all posts that include any of these URLs in the following Web communities: Reddit, Twitter (from the 1\% Streaming API, with posts from confirmed Russian and Iranian trolls removed), Gab, and 4chan's Politically Incorrect board (\dspol).
For Reddit and Twitter our dataset spans January 2016 to October 2018, for \dspol it spans  July 2016 to October 2018, and for Gab it spans August 2016 to October 2018.\footnote{\textbf{NB:} the 4chan dataset made available by the authors of~\cite{zannettou2017web,zannettou2018origins} starts in late June 2016 and Gab was first launched in August 2016.}
We select these communities as previous work shows they play an important and influential role on the dissemination of news~\cite{zannettou2017web} and memes~\cite{zannettou2018origins}.

Table~\ref{tbl:hawkes} summarizes the number of events (i.e., occurrences of a given URL) for each community/group of users that we consider (Russia refers to Russian trolls on Twitter, while Iran refers to Iranian trolls on Twitter).
Note that we decouple The\_Donald from the rest of Reddit as previous work showed that it is quite efficient in pushing information in other communities~\cite{zannettou2018origins}.
From the table we make several observations: 1)~Twitter has the largest number of events in all groups of URLs mainly because it is the largest community and 2)~Gab has a considerably large number of events; more than \dspol and The\_Donald, which are bigger communities.

For each unique URL, we fit a statistical model known as Hawkes Processes~\cite{linderman2014,lindermanArxiv}, which allows us to estimate the strength of connections between each of these communities in terms of how likely an event -- the URL being posted by either trolls or normal users to a particular platform -- is to cause subsequent events in each of the groups.
We fit each Hawkes model using the methodology presented by~\cite{zannettou2018origins}.
In a nutshell, by fitting a Hawkes model we obtain all the necessary parameters that allow us to assess the root cause of each event (i.e., the community that is ``responsible'' for the creation of the event).
By aggregating the root causes for all events we are able to measure the influence and efficiency of each Web community we considered.

We demonstrate our results with two different metrics: 1)~the absolute influence, or percentage of events on the destination community caused by events on the source community and 2)~the influence relative to size, which shows the number of events caused on the destination platform as a percent of the number of events on the \emph{source} platform.
The latter can also be interpreted as a measure of how \emph{efficient} a community is in pushing URLs to other communities.

Fig.~\ref{fig:raw_influence} reports our results for the absolute influence for each group of URLs.
When looking at the influence for the URLs shared by Russian trolls on Twitter (Fig.~\ref{subfig:raw_influence_russians}), we find that Russian trolls were particularly influential to users from Gab (1.9\%), the rest of Twitter (1.29\%), and \dspol (1.08\%).
When looking at the communities that influenced the Russian trolls we find the rest of Twitter (7\%) followed by Reddit (4\%).
By looking at URLs shared by Iranian trolls on Twitter (Fig.~\ref{subfig:raw_influence_iranians}), we find that Iranian trolls were most successful in pushing URLs to The\_Donald (1.52\%), the rest of  Reddit (1.39\%), and Gab (1.05\%), somewhat ironic considering The\_Donald and Gab's zealous pro-Trump leanings and the Iranian trolls' clear anti-Trump leanings~\cite{flores2018mobilizing,zannettou2018gab}.
Similarly to Russian trolls, the Iranian trolls were most influenced by Reddit (5.6\%) and the rest of Twitter (4.6\%).
When looking at the URLs posted by both Russian and Iranian trolls we find that, overall, the Russian trolls were more influential in spreading URLs to the other Web communities with the exception of (again, somewhat ironically) \dspol.

But how do these results change when we normalize the influence with respect to the number of events that each community creates?
Fig.~\ref{fig:norm_influence} shows the influence relative to size for each pair of communities/groups of users.
For URLs shared by Russian trolls (Fig.~\ref{subfig:norm_influence_russians}) we find that Russian trolls were particularly efficient in spreading the URLs to Twitter (10.4\%)---which is not a surprise, given that the accounts operate directly on this platform---and Gab (3.19\%).
For the URLs shared by Iranian trolls, we again observe that were most efficient in pushing the URLs to Twitter (3.6\%), and the rest of Reddit (2.04\%).
Also, it is worth noting that in both groups of URLs The\_Donald had the highest external influence to the other platforms.
This highlights that The\_Donald is an impactful actor in the information ecosystem and is quite possibly exploited by trolls as a vector to push specific information to other communities.
Finally, when looking at the URLs shared by both Russian and Iranian trolls, we find that Russian trolls were more efficient (greater impact relative to the number of URLs posted) at spreading URLs in all the communities with the exception of \dspol, where Iranians were more efficient.

\subsection{Remarks}

In this work, we analyzed the behavior and evolution of Russian and Iranian trolls on Twitter and Reddit during the course of several years.
We shed light to the target campaigns of each group of trolls, we examined how their behavior evolved over time, and what content they disseminated.
Furthermore, we find some interesting differences between the trolls depending on their origin and the platform from which they operate.
For instance, for the latter, we find discussions related to cryptocurrencies only on Reddit by Russian trolls, while for the former we find that Russian trolls were pro-Trump and Iranian trolls anti-Trump.
Also, we quantify the influence that these state-sponsored trolls had on several mainstream and alternative Web communities (Twitter, Reddit, \dspol, and Gab), showing that Russian trolls were more efficient and influential in spreading URLs on other Web communities than Iranian trolls, with the exception of \dspol.

Our findings have serious implications for society at large.
First, our analysis shows that while troll accounts use peculiar tactics and talking points to further their agendas, these are not completely disjoint from regular users, and therefore developing automated systems to identify and block such accounts remains an open challenge.
Second, our results also indicate that automated systems to detect trolls are likely to be difficult to realize: trolls change their behavior over time, and thus even a classifier that works perfectly on one campaign might not catch future campaigns.
Third, and perhaps most worrying, we find that state-sponsored trolls have a meaningful amount of influence on fringe communities like The\_Donald, 4chan's \dspol, and Gab, and that the topics pushed by the trolls resonate strongly with these communities.
This might be due to users on these communities that sympathize with the views  the trolls aim to share (i.e., ``useful idiots'') or to unidentified state-sponsored actors on these communities.
In either case, considering recent tragic events like the Tree of Life Synagogue shootings, perpetuated by a Gab user seemingly influenced by content posted there, the potential for mass societal upheaval cannot be overstated.
Because of this, we implore the research community, as well as governments and non-government organizations to expend whatever resources are at their disposal to develop technology and policy to address this new, and effective, form of digital warfare.

\chapter{Discussion \& Conclusions}
\label{chapter:conclusions}
In this thesis, we studied several aspects of the information ecosystem on the Web.
We shed light into three main relevant lines of work:
1)~understanding the spread of information through the lens of multiple Web communities and modeling the interplay between them;
2)~characterizing emerging Web communities and services by undertaking exploratory large-scale quantitative analyses; and 
3)~understanding the behavior and impact of state-sponsored actors on the information ecosystem on the Web.
Below, for each line of work, we provide the main take-aways and possible future directions.

\section{Understanding the Spread Of Information Through The Lens Of Multiple Web Communities}

\descr{Remarks.} In this line of work, we studied the spread of news and image-based memes across four Web communities, namely, Twitter, Reddit, 4chan's \dspol and Gab.
By designing and developing a scalable processing pipeline we were able to detect and track the propagation of memes across the Web.
Then using Hawkes Processes, we modeled the interplay between the various Web communities and we quantified the influence that each community have to the other with respect to the dissemination of news and memes.
The main take-aways from this work are:  
1)~small fringe Web communities like 4chan's \dspol and The\_Donald subreddit have a surprisingly strong influence, despite their small size, to mainstream communities like Twitter; and
2)~we find important differences between the communities we study with regard to the dissemination of news and memes. For instance, for news, we find that users on different communities prefer different news sources, especially for the alternative ones, while for memes, we find that users on small fringe Web communities tend to share more memes that are likely to be used in a weaponized or hateful context.

\descr{Future Directions.}
There are several possible future directions that derive from the findings of this thesis.
First, we present a novel methodology for assessing the influence between multiple Web communities. 
This framework, based on Hawkes Processes, can be used in a lot of different domains to assess the influence between various entities.
For instance, this framework can be applied in the user-level in order to assess the influence that users of a specific community have to each other with respect to the dissemination of a specific information.
Also, by changing the notion of what a process and what an event is in the framework one can make interesting influence estimation studies.
With regard to the dissemination of news, an interesting future direction is to leverage Natural Language Processing techniques in order to understand how news articles are discussed on various Web communities and if there are important differences between the various Web communities in consideration.
With regard to the dissemination of memes there are several future directions that can be based on our developed memes processing pipeline. 
Specifically, one can leverage our pipeline to detect images pertaining to specific memes and then qualitatively analyze them in order to understand how memes are becoming weaponized and how multiple memes are combined together to deliver a specific idea. 
For instance, to study how the Pepe the Frog meme is used in conjunction with other memes with the goal to deliver a specific political message.
Another line of work, includes focusing on the detection of potential hateful and harmful memes and devising mitigation strategies that will be employed by Web communities (e.g., Twitter) in order to safeguard their users from potentially offensive content.
Finally, the developed pipeline can be used to study images that are not bounded to a specific domain (e.g., memes). 
For instance,~\cite{zannettou2019characterizing} demonstrate how our image processing pipeline can be used in conjunction with the Google Cloud Vision API to characterize the images posted by Russian trolls on Twitter.

\section{Characterizing the Role of Emerging Web Communities and Services on the Information Ecosystem}

\descr{Remarks.} In this line of work, we have explored the Gab social network as well as two Web archiving services: the Wayback Machine and \url{archive.is}, with the goal to assess their role on the information ecosystem.
We find several interesting findings when exploring these communities and services.
First, we find that Gab attracts the interest of the alt-right community as the most popular users in the platform are alt-right celebrities.
Also, we find that its users have a preference in sharing news articles from alternative news sources, while when examining the prevalence of hate speech, we find that it exhibits a high degree of hate speech.
Second, after our comprehensive analysis on the use of Web archiving services, we find that they are particularly popular in fringe Web communities for the preservation of Web content.
In addition, we find that these services are extensively used by Reddit bots to preserve content posted on specific subreddits, and that Reddit moderators ``force'' users to share archived URLs from news sources with conflicting ideology in order to penalize their ad revenue.
Overall, these findings indicate that Web archiving services are an important actor on the information ecosystem and that it should be taken into account for studies that focus on URLs.

\descr{Future Directions.} There are several future directions that can be derived from this line of work.
First, the Gab social network is still relatively unstudied when compared to other mainstream communities like Twitter, hence a lot of its aspects are unclear.
For instance, it will be interesting to study the evolution of Gab users over time and whether they are becoming more hateful/radicalized over time.
Also, it will be interesting to study whether Gab's popularity increases with purges or large bans of users from other popular Web communities like Twitter.
Furthermore, it will be interesting to study the prevalence of automated accounts within the platform and whether they are trying to promote specific talking points (in our work we find some anecdotal evidence of spam bots on Gab).
Finally, we implore the research community to qualitatively study the Gab community in order to shed light into emerging Web phenomena like hate speech, fake news, and online radicalization.

In a more broad direction, there are still a lot of Web communities for which we lack a clear understanding of what their role on the Web information ecosystem is.
For example, Web communities like Discord, WhatsApp, Mastodon, and Telegram, are relatively unstudied and it is unclear whether they contribute in the spread of false information on the Web.
A possible future direction is providing characterization of these Web communities and assessing whether campaigns are organized in such communities, especially in the ones that support private channels like Discord, WhatsApp, and Telegram.

Our work on Web archiving services points to several research avenues.
For instance, future work could better understand the role of archiving services in the dis/misinformation ecosystem, e.g., with respect to the content that gets archived and the context in which archive URLs are disseminated. 
Moreover, further work could shed light on the actors archiving specific URLs in specific contexts, as well as how much traction they get on Web communities like Twitter and Reddit. 
Finally, we believe that a deeper dive into the socio-technical and ethical implications of archiving services is warranted: they serve a crucial role in
ensuring that Web content persists, but do so without regard to (and often in spite of) the rights and consent of content producers.

\section{Towards Understanding State-Sponsored Actors}

\descr{Remarks.} In this work we provide a comprehensive exploratory analysis on the behavior of state-sponsored actors on the Web.
First, we compare the behavior of state-sponsored actors on Twitter and how they compare to a set of random users.
We find important differences between state-sponsored actors and random users ranging from the use of Twitter clients to their self-reported locations.
In addition, we provide useful insights with regard to the evolution of their accounts and how do they posed as.
Second, we provide an analysis of Russian and Iranian trolls on Twitter and Reddit.
We investigated how they evolved over time and what influence they have in other communities: namely, Reddit, Twitter, 4chan's \dspol, and Gab.
We find that the behavior and targets of these actors vary over time and that these actors were particularly influential in spreading news articles to other Web communities.
In particular, we find that the Russia state-sponsored actors were more influential in spreading news to the other communities, with the exception of \dspol where Iranian trolls were more influential.

\descr{Future Directions.} Despite providing a comprehensive overview on the behavior of state-sponsored actors on the Web, there are still several unexplored research avenues.
First, as a research community, we should develop tools to detect and mitigate campaigns organized from state-sponsored actors.
Second, there is a variety of other state-sponsored actors that are unexplored. 
For instance, Twitter detected and suspended a lot of accounts associated with Venezuela and Bangladesh governments~\cite{new_twitter_dataset}: it will be interesting to see how these actors compare with the Russian and Iranian actors presented in this thesis and whether there are meaningful differences in their behavior.
Finally, we believe that our influence estimation results show the need for more sophisticated measurements in the domains of opinion manipulation and spread of false information by state-sponsored actors.

\section{Conclusion}
In this thesis, we shed some light into the complex Web information ecosystem through the lens of multiple Web communities.
Our work reveals the need to take a cross-platform view of the information ecosystem as there are a lot of Web communities that despite their small size are particularly influential and can have real-world impact.
Also, it indicates the need to develop sophisticated tools and techniques to detect the spread of information across the Web by considering the diverse types of information (i.e., text, images, and URLs).
We argue that the aforementioned are of paramount importance for getting a more representative view of the information ecosystem, hence helping in better understanding this ecosystem as a whole.

{\footnotesize{
\bibliography{thesis}
}
\end{document}